\documentclass{book}

\usepackage{bm}
\usepackage{mathptmx}
\usepackage{helvet}
\usepackage{courier}
\usepackage{type1cm}

\usepackage{multicol}        
\usepackage[bottom]{footmisc}

\usepackage{makeidx}         
\usepackage{epsfig}
\usepackage{url}
\usepackage{graphicx}        
\usepackage[bottom]{footmisc}
\usepackage{color}

\font\aa=cmmi10 
\def\epsilon{\mbox{\aa\char'017}}

\graphicspath{{./}{./FIGURES/}{./EPS/}}
\def\av#1{\langle#1\rangle}

\def\qut#1{``#1''}
\def\CEnote#1{\bgroup \color[rgb]{0,1,1}{#1}\egroup}
\def\veci#1{\textbf{\textit{#1}}}

\def\runinhead#1{\smallskip\noindent{\bf #1 \ }}
\def\abstract#1{\noindent #1}
\def\sidecaption#1{\caption{#1}}
\def\svhline{\hline}
\def\D{\mathrm{d}}

\makeindex  

\begin{document}

\hbox{\rule{\hsize}{2pt}}
\vspace{7pt}
\begin{center}
{\huge  {\bf Complex Adaptive Dynamical}} \newline
\medskip

{\huge  {\bf Systems, a Primer\footnote{Springer 2008, second edition 2010; 
         including the solution section.}}}
\end{center}
\hbox{\rule{\hsize}{2pt}}

\vspace{5cm}
\begin{center}
{\Huge  {\bf 2008/10}}
\end{center}

\vspace{3cm}

\begin{center}
{\Large Claudius Gros \\  \ 
Institute for Theoretical Physics \\
Goethe University Frankfurt}
\end{center}
\vspace{5cm}

\frontmatter

\tableofcontents

\mainmatter
 
\setcounter{secnumdepth}{2}

\chapter{Graph Theory and Small-World Networks}
\label{chap_networks1}


\abstract{Dynamical networks constitute a very
wide class of complex and adaptive systems. Examples range from
ecological prey--predator networks to the gene expression and
protein networks constituting the basis of all living
creatures as we know it. The brain is probably the most
complex of all adaptive dynamical systems and is at the
basis of our own identity, in the form of a sophisticated neural
network. On a social level we interact through social networks, to
give a further example -- networks are ubiquitous through the domain
of all living creatures.\newline \indent A good
understanding of network theory is therefore of basic importance for
complex system theory. In this chapter we will discuss
the most important concepts of graph\footnote{Mathematicians
generally prefer the somewhat more abstract term 
\qut{graph} instead of \qut{network}.}
theory and basic realizations of possible network organizations.}

\section{Graph Theory and Real-World Networks}

\subsection{The Small-World Effect}

Six or more billion humans live on earth today and it might seem
that the world is a big place. But, as an Italian proverb says,
%
\vskip4pt
``Tutto il mondo \'e paese''\hspace{2ex} -- \hspace{2ex} 
``The world is a village''.
\vskip4pt

The network of who knows whom -- the network of acquaintances -- is
indeed quite densely webbed. Modern scientific investigations mirror
this century-old proverb.

\runinhead{Social Networks} \index{network!social}Stanley Milgram
performed a by now famous experiment in the 1960s. He distributed a
number of letters addressed to a stockbroker in Boston to a random
selection of people in Nebraska. The task was to send these letters
to the addressee (the stockbroker) via mail to an acquaintance of
the respective sender. In other words, the letters were to be sent
via a social network.

The initial recipients of the letters clearly did not know
the Boston stockbroker on a first-name basis. Their
best strategy was to send their letter to someone whom
they felt was closer to the stockbroker, socially or geographically:
perhaps someone they knew in the financial industry,
or a friend in Massachusetts.

\begin{figure}[!t]
\centering
\includegraphics{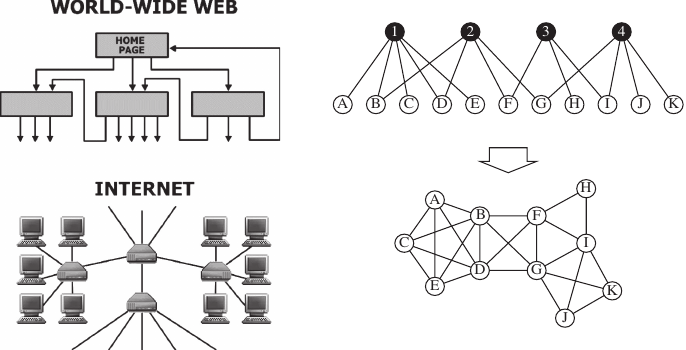}
\caption{\textit{Left}: Illustration of the network structure of the
world-wide web and of the Internet (from Albert and Barab\'asi,
2002). \textit{Right}: Construction of a graph (\textit{bottom})
from an underlying bipartite graph (\textit{top}). The {\textit{filled
circles}}{} correspond to movies and the {\textit{open circles}} to
actors {cast} in the respective movies (from
Newman, Strogatz and Watts, 2001)
         }
\label{networks1_WWW_bipartite} \index{network!internet}
\index{network!WWW} \index{network!bipartite}
\vspace*{-8pt}
\end{figure}

\runinhead{Six Degrees of Separation} About 20\% of Milgram's
letters did eventually reach their destination. Milgram found that
it had only taken an average of six steps for a letter to get from
Nebraska to Boston. This result is by now dubbed \qut{six degrees of
separation} and it is  possible to connect any two persons living on
earth via the social network in a similar number of
steps.
\begin{quotation}
{\it The Small-World Effect.\enspace}
\index{small-world!effect}The \qut{small-world effect} denotes the
result that the average distance linking two nodes belonging to the
same network can be orders of magnitude smaller than the number of
nodes {making up} the network.
\end{quotation}
\indent \index{network!actors}The small-world effect occurs in all
kinds of networks. Milgram originally examined the
networks of friends. Other examples for social nets are the network
of film actors or that of baseball players, see
Fig.~\ref{networks1_WWW_bipartite}. Two actors are linked by an edge
in this network whenever they co-starred at least once in the same
movie. In the case of baseball players the linkage
is given by the condition to have played at least once
on the same team.

\runinhead{Networks are Everywhere} Social networks are but just one
important example of a communication network. Most human
communication takes place directly among individuals. The
spreading of news, rumors, jokes and of diseases
takes place by contact between individuals. And we are 
all aware that rumors and epidemic infections can spread very
fast in densely webbed social networks.

\begin{figure}[t]
\centering
\includegraphics{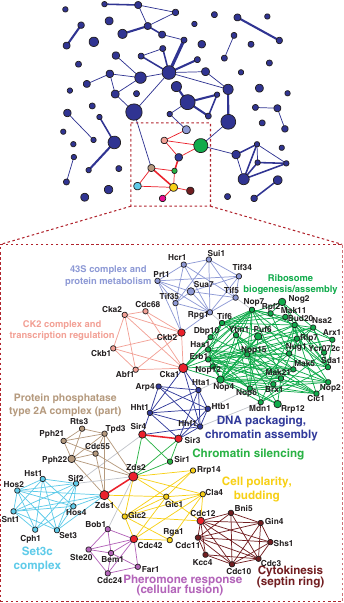}
\caption{A protein interaction network, showing a complex interplay
between highly connected hubs and communities of subgraphs with
increased densities of edges (from Palla et al., 2005)}
\label{networks1_protein} \index{network!protein interaction}
\vspace*{-8pt}
\end{figure}

\index{network!communication}Communication networks are ubiquitous.
Well known examples are the Internet and the world-wide
web, see Fig.~\ref{networks1_WWW_bipartite}. Inside
a cell the many constituent proteins form an interacting network, as
illustrated in Fig.~\ref{networks1_protein}. The same is of course
true for artificial neural networks as well as for the networks of
neurons that build up the brain. It is therefore important 
to understand the statistical properties of the
most important network classes.

\enlargethispage*{12pt} \vspace*{-12pt}
\subsection{Basic Graph-Theoretical Concepts}

We start with some basic concepts allowing
to characterize graphs and real-world networks.

\runinhead{Degree of a Vertex}
A graph is made out of vertices connected by edges.
\begin{quotation}
{\it Degree of a Vertex.\enspace}
\index{vertex!degree}The degree $k$ of the vertex 
is the number of edges linking to this node.
\end{quotation}
Nodes having a degree $k$ substantially above the
average are denoted \qut{hubs}, they are the VIPs 
of network theory.

\runinhead{Coordination Number}
\label{networks1_random_concepts}\index{random graph}The simplest
type of network is the random graph. It is characterized by only two
numbers: By the number of vertices $N$ and by the average degree
$z$, also called the coordination number.
\begin{quotation}
{\it Coordination Number.\enspace} \index{coordination number}The
coordination number $z$ is the average number of links per vertex,
viz the average degree.
\end{quotation}
A graph with an average degree $z$ has $Nz/2$ connections.
Alternatively we can define with $p$ the probability 
to find a given edge.
\begin{quotation}
{\it Connection Probability.\enspace}
\index{connection probability}
The probability that a given edge
occurs is called the connection probability~$p$. 
\end{quotation}

\begin{figure}[t]
\centering
\includegraphics{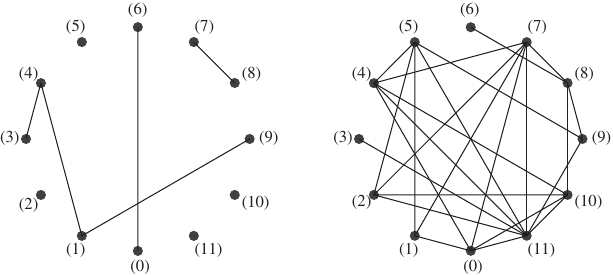}
\caption{{Random} graphs with
$N=12$ vertices and different connection probabilities $p=0.0758$
(\textit{left}) and $p=0.3788$ (\textit{right}). The three mutually
connected vertices (0,1,7) contribute to the clustering coefficient
and the fully interconnected set of sites (0,4,10,11) is a clique in
the network on the right } \label{networks1_randomGraph}
\vspace*{-10pt}
\end{figure}

\runinhead{Erd\"os--R\'enyi Random Graphs}
             \index{random graph!Erd\"os--R\'enyi}
             \index{Erd\"os--R\'enyi random graph}
We can construct a specific type of random graph simply by taking
$N$ nodes, also called vertices and by drawing $Nz/2$ lines, the
edges, between randomly chosen pairs of nodes, compare
Fig.~\ref{networks1_randomGraph}. This type of random graph is
called an \qut{Erd\"os--R\'enyi} random graph after two
mathematicians who studied this type of graph
extensively. 

Most of the following discussion will be valid for all types of
random graphs, we will explicitly state whenever we specialize to
Erd\"os--R\'enyi graphs. In Sect.~\ref{networks1_generalized_random_graphs} 
we will introduce and study other types of random graphs.

For Erd\"os--R\'enyi random graphs we have
\begin{equation}
p\ =\ {Nz\over2}{2\over N(N-1)}\ =\  {z\over N-1}
\label{networks1_p_z}
\end{equation}
for the relation between the coordination number $z$ 
and the connection probability $p$.

\runinhead{The Thermodynamic Limit} 
Mathematical graph theory is often concerned with
the thermodynamic limit.
\begin{quotation}
{\it The Thermodynamic Limit.\enspace}
\index{thermodynamic limit}The limit where the 
number of elements making up a system diverges 
to infinity is called the \qut{thermodynamic limit} in
physics.\break A \nobreak quantity is {\em extensive} if it is
proportional to the number of constituting elements, and {\em
intensive} if it scales to a constant in the thermodynamic limit.
\end{quotation}
We note that $p=p(N)\to0$ in the thermodynamic limit $N\to\infty$ for
Erd\"os--R\'enyi random graphs and
intensive $z\sim O(N^0)$, compare Eq.~(\ref{networks1_p_z}).

There are small and large real-world networks and it makes
sense only for very large networks to consider the thermodynamik
limit. An example is the network of hyperlinks.

\runinhead{The Hyperlink Network} Every web page contains links to
other web pages, thus forming a network of hyperlinks. In 1999 there
were about $N\simeq0.8\times10^9$ documents on the web, but the
average distance between documents was only about~19. The WWW
is growing rapidly; in 2007
estimates for the total number of web pages resulted in
$N\simeq(20-30)\times10^9$, with the size of the Internet backbone,
viz the number of Internet servers, being about
$\simeq0.1\times10^9$. 

\runinhead{Network Diameter and the Small-World Effect} As a first
parameter characterizing a network we discuss the diameter of a
network.
\begin{quotation}
{\it Network Diameter.\enspace}
\index{network!diameter}\index{graph!diameter}The network diameter
is the maximum degree of separation between all pairs of vertices.
\end{quotation}
\noindent For a random network with $N$ vertices and 
coordination number $z$ we have
\begin{equation}
z^D\ \approx\ N,\qquad\quad D\ \propto\ \log N/\log z~,
\label{networks1_dia_random}
\end{equation}
since any node has $z$ neighbors, $z^2$ next-nearest neighbors and
so on. The logarithmic increase in the number of degrees of
separation with the size of the network is characteristic of
small-world networks. $\log N$ increases very slowly with $N$ and
the network diameter therefore remains
small even for networks containing a large number of nodes $N$.
\begin{quotation}
{\it Average Distance.\enspace} \index{distance!average}The average
distance $\ell$ is the average of the minimal path length between
all pairs of nodes of a network. 
\end{quotation}
The average distance $\ell$ is generally 
closely related to the diameter $D$; it has the same 
scaling as the number of nodes $N$.

\begin{table}[!b]
\vspace*{12pt}
\begin{center}
\caption{The number of nodes $N$, average degree of separation
$\ell$, and
  clustering coefficient $C$, for three real-world networks.  The last
  column is the value which $C$ would take in a random graph with the same
  size and coordination number, $C_{\rm rand}=z/N$
 (from Watts and Strogatz, 1998) }
\label{networks1_wstable}\index{network!actors}
\smallskip
\begin{tabular*}{15pc}{@{}l@{\quad}l@{\quad}l@{\quad}l@{\quad}l@{\quad}@{}}
\hline\noalign{\smallskip}
Network & $N$ & $\ell$ & $C$ & $C_{\rm rand}$ \\
\noalign{\smallskip}\svhline\noalign{\smallskip}
Movie actors   & $225\,226$ & $\phantom{0}3.65$ & $0.79$ & $0.00027$ \\
Neural network & $\phantom{000}282$      & $\phantom{0}2.65$ & $0.28$ & $0.05$ \\
Power grid     & $\phantom{00}4941$     & $18.7$ & $0.08$ & $0.0005$ \\
\noalign{\smallskip}\hline\noalign{\smallskip}
\end{tabular*}
\end{center}
\vspace{-12pt}
\end{table}

\runinhead{Clustering in Networks} Real networks have strong local
recurrent connections, compare, e.g.\ the protein network
illustrated in Fig.~\ref{networks1_protein}, leading to distinct
topological elements, such as loops and clusters.
\begin{quotation}
{\it The Clustering Coefficient.\enspace}
\index{clustering!coefficient}The clustering coefficient $C$ is the
average fraction of pairs of neighbors of a node
{that} are also neighbors of each other.
\end{quotation}
 The
clustering coefficient is a normalized measure of loops of length
{3}. In a fully connected network, in which everyone
knows everyone else, $C=1$.

In a random graph a typical site has $z(z-1)/2$ pairs of neighbors.
The probability of an edge to be present between a given pair of
neighbors is $p=z/(N-1)$, see Eq.~(\ref{networks1_p_z}). The
clustering coefficient, which is just the probability of a pair of
neighbors to be interconnected {is} therefore
\begin{equation}
C_{\mathrm{rand}}\ =\ {z\over N-1}\ \approx\  {z\over N}~.
\label{network1_C_rand}
\end{equation}
It is very small for large random networks and scales to zero in the
thermodynamic limit. In Table \ref{networks1_wstable} the respective
clustering coefficients for some real-world networks and for the
corresponding random networks are listed for comparison.

\runinhead{Cliques and Communities} The clustering coefficient
measures the normalized number of triples of fully interconnected
vertices. In general, any fully connected subgraph is denoted a
clique.

\begin{figure}[t]
\centering
\includegraphics{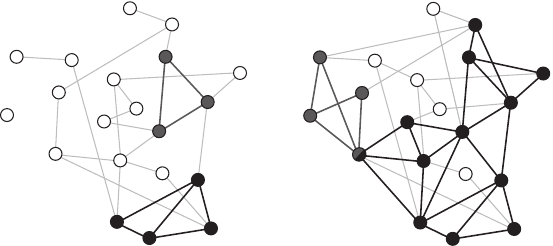}
\caption{\textit{Left}: Highlighted are three three-site cliques.
\textit{Right}: A percolating network of three-site cliques (from
Derenyi, Palla and Vicsek, 2005)} \label{networks1_cliques}
\vspace*{-6pt}
\end{figure}

\begin{quotation}
{\it Cliques.\enspace}\label{networks_def_cliques}
\index{graph!clique}\index{clique}A clique is a set of vertices for
which (a) every node is connected by an edge to every other member
of the clique and (b) no node outside the clique is connected to all
members of the clique.
\end{quotation}
 The term \qut{clique} comes from social
networks. A clique is a  group of friends where everybody
knows everybody else. The number of cliques of size $K$ in an
Erd\"os--R\'enyi graph with $N$ vertices and linking probability
$p$~is
$$
\left( \begin{array}{c} N \\ K \end{array} \right)
p^{K(K-1)/2}\left(1-p^K\right)^{N-K}~.
$$
The only cliques occurring in random graphs in the thermodynamic
limit have the size {2}, since $p=z/N$. For
{an} illustration see Fig.~\ref{networks1_cliques}.

\index{graph!community}Another term used is {\em community}. It is
mathematically not as strictly defined as \qut{clique}, it
{roughly denotes} a collection of strongly
overlapping cliques, viz of subgraphs with above-the-average
densities of edges.
%

\runinhead{Clustering for Real-World Networks}
\index{graph!clustering}Most real-world networks have a substantial
clustering coefficient, which is much greater than $O(N^{-1})$. It
is immediately evident from an inspection, for example of the
protein network presented in Fig.~\ref{networks1_protein}, that the
underlying \qut{community structure} gives rise to a high clustering
coefficient.

In Table~\ref{networks1_wstable}, we give some values of $C$,
together with the average distance $\ell$,
for three different networks:
\vskip3pt

-- the network of collaborations between movie actors

-- the neural network of the worm {\it C.~Elegans}, and

-- the Western Power Grid of the United States.

\vskip3pt

Also given in Table~\ref{networks1_wstable} are the values $C_{\rm
rand}$ {that} the clustering coefficient would have for
random graphs of the same size and coordination number. Note that
the real-world value is systematically higher than that of random
graphs. Clustering is important for real-world graphs. These are
small-world graphs, as indicated by the small values for the average
distances $\ell$ given in Table~\ref{networks1_wstable}.

Erd\"os--R\'enyi random graphs obviously do not match 
the properties of real-world networks well.
In Sect.~\ref{networks1_generalized_random_graphs}
we will discuss generalizations of random graphs
that approximate the properties of real-world
graphs much better. Before that, we will discuss some general
properties of random graphs in more detail.

\runinhead{Correlation Effects} 
\index{network!correlation effects}
The degree distribution $p_k$ captures the
statistical properties of nodes as if they where
all independent of each other. In general, the
property of a given node will however be
dependent on the properties of other nodes,
e.g.\ of its neighbors. When this happens one
speaks of \qut{correlation effects}, with the 
clustering coefficient $C$ being an example.

Another example for a correlation effect is
what one calls \qut{assortative mixing}.
\index{network!assortative mixing}
A network is assortatively correlated whenever
large-degree nodes, the hubs, tend to be mutally
interconnected and assortatively anti-correlated 
when hubs are predominantly linked to low-degree
vertices. Social networks tend to be assortatively
correlated, in agreement with the everyday 
experience that the friends of influential persons, 
the hubs of social networks, tend to be VIPs themselves.

\runinhead{Tree Graphs} \index{graph!tree}
Real-world networks typically show strong local clustering 
and loops abound. For many types of graphs commonly
considered in graph theory, like Erd\"os--R\'enyi
graphs, the clustering coefficient vanishes however 
in the thermodynamic limit, and loops become
irrelevant. One denotes a loopless graph a 
\qut{tree graph}, a concept often encountered 
in mathematical graph theory.

\runinhead{Bipartite Networks} \index{network!bipartite}Many
real-world graphs have an  underlying bipartite structure, see
Fig.~\ref{networks1_WWW_bipartite}.
\begin{quotation}
{\it Bipartite Graph.\enspace}
A bipartite graph has two kinds of vertices with
links only between vertices of unlike kinds.
\end{quotation}
Examples are networks of managers, where one kind of vertex is a
company and the other kind of {vertex} the
managers belonging to the board of directors. When eliminating one
kind of {vertex}, in this case it is customary to
eliminate the companies, one retains a social network{;}
the network of directors, as illustrated in
Fig.~\ref{networks1_WWW_bipartite}. This network has a high
clustering coefficient, as all boards of directors are mapped onto
cliques of the respective social network.

\vspace*{6pt}
\subsection{Properties of Random Graphs}
\label{networks1_properties_random_graphs}
\index{random graph!properties|textbf}

So far we have considered mostly averaged
quantities of random graphs, like the clustering coefficient
or the average coordination number $z$. We will now
develop tools allowing for a more sophisticated 
characterization of graphs.

\runinhead{Degree Distribution}
The basic description of
general random and non-random graphs is
given by the degree distribution $p_k$.
\begin{quotation}
{\it Degree Distribution.\enspace}
\index{degree distribution}If
$X_k$ is the number of vertices having the degree $k$, then
$p_k=X_k/N$ is called the degree distribution, where $N$ is the
total number of nodes.
\end{quotation}
The degree distribution is a probability distribution function
and hence normalized, $\sum_k p_k=1$.

\runinhead{Degree Distribution for Erd\"os--R\'enyi Graphs}
\index{degree distribution!Erd\"os-R\'enyi} The probability of any
node to have $k$ edges is
\begin{equation}
p_k\ =\ { N-1 \choose k}\,p^k\, (1-p)^{N-1-k}~,
\label{networks1_p_k}
\end{equation}
for an Erd\"os--R\'enyi network, where $p$ is the link connection
probability. For large $N\gg k$ we can approximate the degree
distribution $p_k$ by
\begin{equation}
p_k\ \simeq\ e^{-pN}\,{\left(pN\right)^k\over k!}
\ =\ e^{-z}\,{z^k\over k!}~,
\label{networks1_p_k_largeN}
\end{equation}
\index{Poisson distribution}where $z$ is the average coordination
number, compare Eq.~(\ref{networks1_p_z}). We have used
$$
\lim_{N\to\infty}\left(1-{x\over N}\right)^N= e^{-x},
\qquad
{ N-1 \choose k}= {(N-1)!\over k!(N-1-k)!}
\simeq {(N-1)^k\over k!}~,
$$
and $(N-1)^k p^k=z^k$, see Eq.~(\ref{networks1_p_z}). Equation
(\ref{networks1_p_k_largeN}) is a Poisson distribution with the mean
$$
\langle k\rangle \ =\ \sum_{k=0}^\infty k\, e^{-z}\,{z^k \over k!}
\ =\ z\, e^{-z} \sum_{k=1}^\infty {z^{k-1} \over (k-1)!}
\ =\ z ~,
$$
as expected.

\runinhead{Ensemble Fluctuations}
\index{ensemble!fluctuations}{In general, two specific
realizations of random graphs differ}. Their properties coincide on
the average, but not on the level of individual links. {With \qut{ensemble} one denotes} the set of
possible realizations.

In an ensemble of random graphs with fixed $p$ and $N$ the degree
distribution $X_k/N$ will be slightly different from one realization
to the next. On the average it will be given by
\begin{equation}
{1\over N} \langle X_k\rangle \ =\  p_k~.
\label{networks1_X_k}
\end{equation}

\index{ensemble!average}Here $\langle{\ldots}\rangle$
denotes the ensemble average. One can go one step further and
calculate the probability $P(X_k=R)$ that in a realization of a
random graph the number of vertices with degree $k$ equals $R$. It
is given in the large-$N$ limit by
\begin{equation}
P(X_k=R)\ =\ e^{-\lambda_k}\,{\left(\lambda_k\right)^R \over R!},
\qquad\quad \lambda_k=\langle X_k\rangle~.
\label{networks1_P_preferential_attachment}
\end{equation}
Note the similarity to Eq.~(\ref{networks1_p_k_largeN}) and
that the mean $\lambda_k=\langle X_k\rangle$ is in
general extensive while the mean $z$ of the degree distribution
(\ref{networks1_p_k_largeN}) is intensive.

\runinhead{Scale-Free Graphs} \index{graph!scale-free}Scale-free
graphs are defined by a power-law degree\break distribution
\begin{equation}
p_k\ \sim\ {1\over k^\alpha},\qquad\qquad \alpha>1~.
\label{networks1_p_k_scale_free}
\end{equation}
Typically, for real-world graphs, this scaling $\sim k^{-\alpha}$
holds only for large degrees $k$. For theoretical studies we will
mostly assume, for simplicity, that the functional dependence
{Eq.~}(\ref{networks1_p_k_scale_free}) holds for all $k$.
The power-law distribution can be normalized~if
$$
\lim_{K\to\infty}\, \sum_{k=0}^K\, p_k\ \approx\
\lim_{K\to\infty}\, \int_{k=0}^K\, p_k\
\propto\ \lim_{K\to\infty} K^{1-\alpha}\ <\ \infty~,
$$
i.e.\ when $\alpha>1$. The average degree is finite
if
$$
\lim_{K\to\infty}\, \sum_{k=0}^K k\,p_k\
\propto\ \lim_{K\to\infty} K^{-\alpha+2}\ <\ \infty~,
\qquad\quad \alpha>2~.
$$
A power-law functional relation is called scale-free, since any
rescaling $k\to a\,k$ can be reabsorbed into the normalization
constant.

Scale-free functional dependencies are also called {\em critical},
since they occur generally at the critical point of a phase
transition. We will come back to this issue recurrently in the
following chapters.

\runinhead{Graph Spectra} \index{graph!spectrum}Any graph $G$ with
$N$ nodes can be represented by a matrix encoding the topology of
the network, the adjacency matrix.
\begin{quotation}
{\it The Adjacency Matrix.\enspace}
\index{adjacency matrix}\index{matrix!adjacency}The $N\times N$ adjacency matrix
$\hat A$ has elements $A_{ij}=1$ if nodes $i$ and $j$ are connected
and $A_{ij}=0$ if they are not connected.
\end{quotation}
The adjacency matrix is symmetric and {consequently has} $N$ real eigenvalues.

\begin{quotation}
{\it The Spectrum of a Graph.\enspace}
\index{graph!spectrum}The spectrum of a graph $G$ is given by the
set of eigenvalues $\lambda_i$ of the adjacency matrix $\hat A$.
\end{quotation}
A graph with $N$ nodes has $N$ eigenvalues $\lambda_i$ and it is
useful to define the corresponding \qut{spectral density}
\begin{equation}
\rho(\lambda)\ =\ {1\over N}\sum_j \delta(\lambda-\lambda_j),
\qquad\quad \int \mathrm{d}\lambda\,\rho(\lambda)=1~,
\label{networks1_rho}
\end{equation}
where $\delta(\lambda)$ is the Dirac delta function.

\setcounter{footnote}{2}

\runinhead{Green's Function$^{\bf{2}}$}

\footnotetext{{The reader without prior experience with
Green's functions may skip the following derivation and pass
directly to the result}, namely to Eq.~(\ref{networks1_rho_self}).}
\index{Green's function} The spectral density $\rho(\lambda)$ can be
evaluated once {}the Green's function $G(\lambda)$,
\begin{equation}
G(\lambda) \ =\ {1\over N}\,
Tr\left[ {1\over \lambda-\hat A}\right]
\ =\ {1\over N} \sum_j {1\over \lambda-\lambda_j}~,
\label{networks1_G_lambda}
\end{equation}
is known. Here $Tr[{\ldots}]$ denotes the trace over
the matrix $(\lambda-\hat A)^{-1}\equiv(\lambda\,\hat{1}-\hat
A)^{-1}$, where $\hat 1$ is the identity matrix. Using the formula
$$
\lim_{\varepsilon\to0} \,{1\over \lambda-\lambda_j+i\epsilon}
\ =\ P\,{1\over \lambda-\lambda_j} \,-\, i\pi\delta(\lambda-\lambda_j)~,
$$
where $P$ denotes the principal part,\footnote{Taking the principal
part signifies that one has to consider {the positive and the negative contributions to the
$1/\lambda$ divergences carefully}.} we find the relation
\begin{equation}
\rho(\lambda) \ =\ -{1\over \pi}
\lim_{\varepsilon\to0} Im G(\lambda+i\varepsilon)~.
\label{networks1_rel_G_rho}
\end{equation}

\runinhead{The Semi-Circle Law} \index{law!semi-circle}The graph
spectra can be evaluated for random matrices for the case of small
link densities $p=z/N$, where $z$ is the average connectivity.
Starting from a random site we can connect on the average to $z$
neighboring sites and from there on to $z-1$ next-nearest
neighboring sites, and so on:
\begin{equation}
G(\lambda) \ =\ {1\over \lambda -
{z\over \lambda - {z-1\over \lambda -{z-1\over \lambda - \dots}}}}
\ \approx\  {1\over \lambda - z\,G(\lambda)}~,
\label{networks1_G_self_retracing}
\end{equation}
where we have approximated $z-1\approx z$ in the last step. Equation
(\ref{networks1_G_self_retracing}) is also called the
\qut{self-retracting path approximation} and can be derived by
evoking a mapping to {} Green's function of a particle
moving along the vertices of the graph. \index{self-retracting!path
approximation} It constitutes a self-consistency equation for
$G=G(\lambda)$, with the solution \index{self-consistency
condition!spectral density}
$$
G^2-{\lambda\over z}G+{1\over z}=0,
\qquad\quad
G = {\lambda\over 2z} -
\sqrt{{\lambda^2\over 4z^2} -{1\over z} }~,
$$
since $\lim_{\lambda\to\infty}G(\lambda)=0$. The spectral density
{Eq.~}(\ref{networks1_rel_G_rho}) then takes the form
\begin{equation}
\rho(\lambda) \ =\ \left\{
\begin{array}{cl}
\sqrt{4z-\lambda^2}/(2\pi z) & \mbox{if}\ \lambda^2<4z \\
0 & \mbox{if}\ \lambda^2>4z
\end{array}
\right.
\label{networks1_rho_self}
\end{equation}
of a half-ellipse also known as \qut{Wigner's law}, or the
\qut{semi-circle law}. \index{Wigner's law}\index{law!Wigner's}
\index{semi-circle law}\index{law!semi-circle}

\runinhead{Loops and the Clustering Coefficient}
\index{clustering!coefficient!loops}\index{loop!network} The total
number of triangles, viz the overall number of loops of length
3 in a network is $C(N/3)(z-1)z/2$, where $C$ is
the clustering coefficient. This number is related to the adjacency
matrix via
\begin{eqnarray*}
C {N\over 3} {z(z-1)\over 2} & =&
\mbox{number\ of\ triangles} \\
& =& {1\over 6}\sum_{i_1,i_2,i_3} A_{i_1i_2} A_{i_2i_3} A_{i_3 i_1}
\ =\ {1\over 6} \mathrm{Tr}\left[A^3\right]~,
\end{eqnarray*}
since three sites $i_1$, $i_2$ and $i_3$ are interconnected only
when the respective entries of the adjacency matrix are unity. The
sum of the right-hand side of above relation is also denoted a
\qut{moment} of the graph spectrum. The factors $1/3$ and $1/6$ on
the left-hand {side} and on the right-hand side account
for overcountings.

\runinhead{Moments of the Spectral Density}
\index{graph!spectrum!moments} The graph spectrum is directly
related to certain topological features of a graph via its moments.
The $l$th moment of $\rho(\lambda)$ is given~by
\begin{eqnarray}
\label{networks1_rho_moment} \int \mathrm{d}\lambda\,
\lambda^l\rho(\lambda) & = & {1\over N}\sum_{j=1}^N
\left(\lambda_j\right)^l \nonumber\\ & =& {1\over N}{\rm Tr}\left[A^l\right]
\ =\ {1\over N}\sum_{i_1,i_2,\dots,i_l} A_{i_1i_2}
A_{i_2i_3}\cdot\cdot\cdot A_{i_l i_1}~,
\end{eqnarray}
as one can see from Eq.~(\ref{networks1_rho}). The $l$th moment of
$\rho(\lambda)$ is therefore equivalent to the number of closed
paths of length $l$, the number of all paths of length $l$ returning
to the starting point.

\runinhead{Graph Laplacian}
\index{graph!Laplacian}
Consider a function $f(x)$. The first and
second derivatives are given by
$$
{d\over dx}f(x) \ =\ {f(x+\Delta x)-f(x)\over \Delta x},
\qquad
{d^2\over dx^2}f(x) \ =\ {f(x+\Delta x)+f(x-\Delta x)-2f(x)\over \Delta x^2}~,
$$
in the limit $\Delta x\to 0$. Consider now a function
$f_i$, $i=1,...,N$ on a graph with $N$ sites. One defines
the graph Laplacian $\hat \Lambda$ via
\begin{equation}
\Lambda_{ij}\ = \ \left(\sum_j A_{ij}\right)\delta_{ij}-A_{ij}
\ =\ \left\{
\begin{array}{ccl}
k_i &\quad & i=j \\
-1  &\quad & i\ \mathrm{and}\ j\ \mathrm{connected} \\
0   &\quad & \mathrm{otherwise}
\end{array}
            \right.~,
\label{network1_graph_lapacian}
\end{equation}
where the $\Lambda_{ij}=(\hat \Lambda)_{ij}$ are the elements of the
Laplacian matrix, $A_{ij}$ the adjacency matrix,
 and where $k_i$ is the degree of vertex $i$.
$\hat \Lambda$ corresponds, apart from a sign convention, to
a straightforward generalization of the usual
Laplace operator. To see this, just apply the Laplacian
matrix $\Lambda_{ij}$ to a graph-function $\mathbf{f}=(f_1,...,f_N)$.

Alternatively one defines by
\begin{equation}
L_{ij}\ = \ \left\{
\begin{array}{ccl}
1 &\quad & i=j \\
-1/\sqrt{k_i\,k_j} &\quad & i\ \mathrm{and}\ j\ \mathrm{connected} \\
0 &\quad & \mathrm{otherwise}
\end{array}
            \right.~,
\label{network1_normalized_graph_lapacian}
\end{equation}
the \qut{normalized graph Laplacian}, where
$k_i=\sum_j A_{ij}$ is the degree
of vertex $i$. The
eigenvalues of the normalized graph
Laplacian have a straightforward interpretation
in terms of the underlying graph topology.
\index{graph!Laplacian!normalized}

\runinhead{Eigenvalues of the Normalized Graph Laplacian}
Of interest are the eigenvalues $\lambda_l$, $l=0,..,(N-1)$
of the normalized graph Laplacian.
\begin{itemize}
\item[--] The normalized graph Laplacian is positive semidefinite,
$$
0\ =\ \lambda_0\ \le\ \lambda_1 \ \le\ \dots \le \lambda_{N-1}\ \le\ 2~.
$$
\item[--] The lowest eigenvalue $\lambda_0$ is always zero,
          corresponding to the eigenfunction
\begin{equation}
\mathbf{e}(\lambda_0) \ =\ {1\over\sqrt{C}}\left(
\sqrt{k_1},\sqrt{k_2},\dots,\sqrt{k_N}\right)~,
\label{network1_normalized_laplacian_e_0}
\end{equation}
where $C$ is a normalization constant and where the $k_i$
are the respective vertex-degrees.

\item[--] The degeneracy of $\lambda_0$ is given by the number
          of disconnected subgraphs contained in the network.
          The eigenfunctions of $\lambda_0$ then vanish
          on all subclusters beside one, where it has the
          functional form (\ref{network1_normalized_laplacian_e_0}).

\item[--] The largest eigenvalue $\lambda_{N-1}$ is $\lambda_{N-1}=2$,
          if and only if the network is bipartite. Generally, a
          small value of $2-\lambda_{N-1}$ indicates that the graph is
          nearly bipartite.

\item[--] The inequality
$$
\sum_l \lambda_l \ \le\ N
$$
         holds generally. The equality holds for connected
         graphs, viz when $\lambda_0$ has degeneracy one.
\end{itemize}

\runinhead{Examples of Graph Laplacians}
The eigenvalues of the normalized graph Laplacian can be given
analytically for some simple graphs.

\begin{itemize}
\item For a complete graph (all sites are mutually
          interconnected), containing $N$ sites,
          the eigenvalues are
$$
\lambda_0=0,\qquad \lambda_l= N/(N-1),\qquad (l=1,...,N-1)~.
$$
\item For a complete bipartite graph (all sites of
      one subgraph are connected to all other sites
      of the other subgraph) the eigenvalues are
$$
\lambda_0=0,\qquad \lambda_{N-1}=2,\qquad
\lambda_l= 1,\qquad (l=1,...,N-2)~.
$$
The eigenfunction for $\lambda_{N-1}=2$ has the form
\begin{equation}
\mathbf{e}(\lambda_{N-1}) \ =\ {1\over\sqrt{C}}\big(
\underbrace{\sqrt{k_A},\dots,\sqrt{k_A}}_{\mathrm{A\ sublattice}},
\underbrace{-\sqrt{k_B},\dots,-\sqrt{k_B}}_{\mathrm{B\ sublattice}} \big)~.
\label{network1_normalized_laplacian_e_2}
\end{equation}
Denoting with $N_A$ and $N_B$ the number of sites in
two sublattices $A$ and $B$, with $N_A+N_B=N$, the
degrees $k_A$ and $k_B$ of vertices belonging to
sublattice A and B respectively are $k_A=N_B$ and
$k_B=N_A$ for a complete bipartite lattice.
\end{itemize}

A densely connected graph will therefore have many eigenvalues
close to unity. For real-world graphs one may therefore plot
the spectral density of the normalized graph Laplacian in order
to gain an insight into its overall topological properties.
The information obtained from the spectral density of the
adjacency matrix and from the normalized graph Laplacian
are distinct.

\enlargethispage*{12pt}
\section{Generalized Random Graphs}
\label{networks1_generalized_random_graphs}
\index{random graph!generalized|textbf}
\index{model!configurational}
\index{configurational model}

The most random of all graphs are
Erd\"os--R\'enyi graphs. One can relax the
degree of randomness somewhat and construct
random networks having an arbitrarily
given degree distribution. This procedure
is also denoted \qut{configurational model}.

\subsection{Graphs with Arbitrary Degree Distributions}
\label{networks1_arbitrary_degree_distributions}
\index{degree distribution!arbitrary}
In order to generate random graphs that have
non-Poisson degree distributions we may choose a
specific set of degrees.
\begin{quotation}
{\it The Degree Sequence.\enspace}
\index{degree!sequence}A degree sequence is a specified set
$\{k_i\}$ of the degrees for the vertices $i=1\ldots N$.
\end{quotation}

\begin{figure}[t]
\centering
\includegraphics{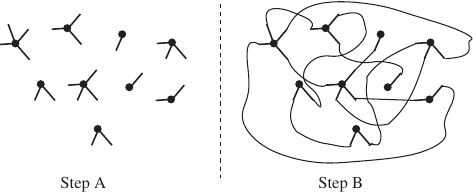}
\caption{Construction procedure of a random network with
{nine} vertices and degrees $X_1=2$, $X_2=3$, $X_3=2$,
$X_4=2$. In step A the vertices with the desired number of stubs
(degrees) are constructed. In step B the stubs are connected
randomly} \label{construct_random_net}
\end{figure}

\runinhead{Construction of Networks with Arbitrary Degree
Distribution} \index{random graph!generalized!construction} The
degree sequence can be chosen in such a way that the fraction of
vertices having degree $k$ will tend to the desired degree
distribution
$$
p_k,\qquad\quad N\to\infty
$$
in the thermodynamic limit. The network can then be constructed
in the following way:
\begin{enumerate}
\item Assign $k_i$  \qut{stubs} (ends of edges emerging from a vertex)
      to every vertex $i=1,\dots,N$.
\item {Iteratively choose} pairs of stubs at random and join them
      together to make complete edges.
\end{enumerate}
When all stubs have been used up, the resulting graph is a random
member of the ensemble of graphs with the desired degree sequence.
Figure~\ref{construct_random_net} illustrates the construction
procedure.

\runinhead{{The} Average Degree and Clustering}
\index{degree!average}
The mean number of neighbors is the coordination number
$$
z\ =\ \langle k\rangle\ =\ \sum_k k\,p_k~.
$$
The probability that one of the second neighbors of a
given vertex is also a first neighbor, scales as $N^{-1}$
for random graphs, regardless of the degree distribution,
and hence can be ignored in the limit $N\to\infty$.

\runinhead{Degree Distribution of Neighbors} \index{degree
distribution!of neighbors}Consider a given vertex $A$ and a vertex
$B$ {that} is a neighbor of $A$, i.e.\ $A$ and $B$
are linked by an edge.

We are now interested in the degree distribution for vertex $B$, viz
in the degree distribution of a neighbor vertex of $A$, where $A$ is
an arbitrary vertex of the random network with degree distribution
$p_k$. As a first step we consider the average degree of a neighbor
node.

A high-degree vertex has more edges connected to it.
There is then a higher chance that any given edge on the graph
will be connected to it, with this chance being directly
proportional to the degree of the vertex. Thus the probability
distribution of the degree of the vertex to which an edge
leads is proportional to $kp_k$ and not just to~$p_k$.

\runinhead{Excess Degree Distribution}
\index{excess degree distribution}
\index{degree distribution!excess}
When we are interested in determining the size of loops or the size
of connected components in a random graph, we are normally
interested not in the complete degree of the vertex reached by
following an edge from $A$, but in the number of edges emerging from
such a vertex {that} do not lead back to $A$,
because the latter contains all information about the number of
second neighbors of $A$.

The number of new edges emerging from $B$ is just the
degree of $B$ minus one and its correctly normalized
distribution is therefore
\begin{equation}
q_{k-1}\ =\ { k\,p_k\over \sum_j jp_j},
\qquad\quad
q_k\ =\ {(k+1) p_{k+1}\over\sum_j j p_j}~,
\label{networks1_def_q_k}
\end{equation}
since $kp_k$ is the degree distribution of a neighbor. The
distribution $q_k$ of the outgoing edges of a neighbor vertex
is also denoted \qut{excess degree distribution}.
The average number of outgoing edges of a neighbor vertex is then
\begin{eqnarray} \nonumber
\sum_{k=0}^\infty k q_k
&=& {\sum_{k=0}^\infty k(k+1)p_{k+1}\over\sum_j j p_j}
\ =\ {\sum_{k=1}^\infty (k-1)kp_k\over\sum_j j p_j} \\
&=& {\langle k^2 \rangle-\langle k \rangle\over\langle k \rangle}~.
\label{networks1_avqk}
\end{eqnarray}

\runinhead{Number of Next-Nearest Neighbors} 
We denote with
$$
z_m,\qquad\quad z_1=\langle  k\rangle\equiv z
$$
the average number of $m$-nearest neighbors. Equation
(\ref{networks1_avqk}) gives the average number of vertices two
steps away from the starting vertex $A$ via a particular neighbor
vertex. Multiplying this by the mean degree of $A$, namely
$z_1\equiv z$, we find that the mean number of second neighbors
$z_2$ of a vertex is\enlargethispage*{12pt}
\begin{equation}
z_2\ =\ \langle k^2\rangle - \langle k\rangle~.
\label{networks1_z2}
\end{equation}

\runinhead{${\bf{z_2}}$ for the Erd\"os--R\'enyi graph}
The degree distribution of an Erd\"os--R\'enyi graph is the Poisson
distribution, $p_k= e^{-z}\,{z^k/ k!}$, see
Eq.~(\ref{networks1_p_k_largeN}). We obtain for the average
number of second neighbors, Eq.~(\ref{networks1_z2}),
\begin{eqnarray*}
z_2& =& \sum_{k=0}^\infty\, k^2e^{-z}{z^k\over k!}\, -\, z
\ =\ z e^{-z}\sum_{k=1}^\infty\, (k-1+1){z^{k-1}\over (k-1)!}\, -\, z
\\ & =& z^2 \  =\ \langle k\rangle^2~.
\end{eqnarray*}
The mean number of second neighbors of a vertex in an
Erd\"os--R\'enyi random graph is just the square of the mean number
of first neighbors. This is a special case however. For most degree
distributions, Eq.~(\ref{networks1_z2}) will be dominated by the
term $\langle k^2\rangle$, so the number of second neighbors is
roughly the mean square degree, rather than the square of the mean.
For broad distributions these two quantities can be very different.

\runinhead{Number of Far Away Neighbors} The average number of edges
emerging from a second neighbor, and not leading back to where we
{came} from, is also given by
{Eq.~}(\ref{networks1_avqk}), and indeed this is true at
any distance $m$ away from vertex $A$. The average number of
neighbors at a distance $m$ is then
\begin{equation}
z_m\ =\ {\langle k^2\rangle-\langle k\rangle\over\langle k\rangle} \,
z_{m-1}\ =\ {z_2\over z_1}\, z_{m-1}~,
\end{equation}
where $z_1\equiv z=\langle k\rangle$ and $z_2$ are
given by Eq.~(\ref{networks1_z2}).
Iterating this relation we find
\begin{equation}
z_m\ =\ \biggl[{z_2\over z_1}\biggr]^{m-1} z_1~.
\label{networks1_z_m}
\end{equation}

\runinhead{{The Giant} Connected Cluster}
\index{giant connected!cluster}Depending on whether $z_2$ is greater
than $z_1$ or not, {Eq.~} (\ref{networks1_z_m})
will either diverge or converge exponentially as $m$ becomes large:
\begin{equation}
\lim_{m\to\infty} z_m \ =\ \left\{
\begin{array}{cl}
\infty & \mbox{\ if\ }\ z_2>z_1 \\
0 & \mbox{\ if\ }\  z_2<z_1
\end{array}
\right. ~,
\label{networks1_lim_z_m}
\end{equation}
$z_1=z_2$ is the percolation point.
In the second case the total number of neighbors
$$
\sum_m z_m\  =\ z_1\sum_{m=1}^\infty
\biggl[{z_2\over z_1}\biggr]^{m-1}
\ =\ {z_1\over 1-z_2/z_1} \ =\ {z_1^2\over z_1-z_2}
$$
is finite even in the thermodynamic limit,
in the first case it is infinite. The network
decays, for $N\to\infty$,
into non-connected components when
the total number of neighbors is finite.

\vspace*{6pt}
\begin{quotation}
{\it The Giant Connected Component.\enspace}
\index{giant connected!component}When the largest cluster of a
graph encompasses a finite fraction of all vertices, in the
thermodynamic limit, it is said to form a giant connected component
(GCC).
\end{quotation}
If the total number of neighbors is infinite,
then there must be a giant connected component. When the
total number of neighbors is finite, there can
be no\break GCC.

\runinhead{{The} Percolation Threshold}
\index{percolation!threshold}When a system has two or more
{possibly} macroscopically different states, one
speaks of a phase transition.
\begin{quotation}
{\it Percolation Transition.\enspace}
\index{percolation!transition}When the structure of an evolving
graph goes from a state in which two (far away) sites are on the
average connected/not connected one speaks of a percolation
transition.
\end{quotation}
This phase transition occurs precisely at the point where $z_2=z_1$. Making
use of Eq.~(\ref{networks1_z2}),
$z_2 = \langle k^2\rangle - \langle k\rangle$,
we find that this condition is equivalent to
\begin{equation}
\langle k^2\rangle-2\langle k\rangle \ =\ 0,
\qquad\quad
\sum_{k=0}^\infty k(k-2) p_k\  =\  0~.
\label{networks1_mrcondition}
\end{equation}

\looseness1 We note that, because of the factor $k(k-2)$, vertices of degree
zero and degree two do not contribute to the sum. The number of
vertices with degree zero or two {therefore affects} neither the phase transition nor the
existence of the giant \nobreak component.
\begin{itemize}
\item[--] Vertices of degree zero are not connected to any other node, they do not contribute to the network topology.

\item[--] Vertices of degree two act as intermediators between two other nodes.
    Removing vertices of degree two does not change the topological structure of a graph.
\end{itemize}
One can therefore remove (or add) vertices of degree two or zero
without affecting the existence of the giant component.

\runinhead{Clique Percolation} \index{percolation!of cliques}Edges
correspond to cliques with $Z=2$ sites (see page
\pageref{networks_def_cliques}). The percolation transition can then
also be interpreted as a percolation of cliques having size two
and larger. It is then
clear that the concept of percolation can be generalized to that of
percolation of cliques with $Z$ sites, see
Fig.~\ref{networks1_cliques} for an illustration.

\runinhead{{The Average} Vertex--Vertex Distance}
\index{distance!average!below percolation}Below the percolation
threshold the average vertex--vertex distance $\ell$ is finite and
the graph decomposes into an infinite number of disconnected
subclusters.
\begin{quotation}
{\it Disconnected {\it Subclusters}.\enspace}
A
disconnected subcluster or subgraph constitutes a subset of vertices
for which (a) there is at least one path in between all pairs of
nodes making up the subcluster and (b) there is no path between a
member of the subcluster and any out-of-subcluster vertex.
\end{quotation}

Well above the percolation transition, $\ell$ is given
approximately by the condition $z_\ell\simeq N$:
\begin{equation}
\log(N/z_1)\ =\ (\ell-1)\log(z_2/z_1),
\qquad\quad
\ell\ =\ {\log (N/z_1)\over\log (z_2/z_1)} + 1~,
\label{networks1_ell}
\end{equation}
using Eq.~(\ref{networks1_z_m}). For the special case of the
Erd\"os--R\'enyi random graph, for which $z_1=z$ and
$z_2=z^2${,} this expression reduces to the standard
formula (\ref{networks1_dia_random}),
$$
\ell\ =\ { \log N - \log z\over \log z} \,+\,1
    \ =\ { \log N\over \log z}~.
$$

\runinhead{{The Clustering} Coefficient of
Generalized Random Graphs} \index{random
graph!generalized!clustering coefficient}
\index{clustering!coefficient!random graph}The clustering
coefficient $C$ denotes the probability that two neighbors $i$ and
$j$ of a particular vertex $A$ have stubs that do
interconnect. The probability that two given stubs are connected is
$1/(zN-1)\approx 1/zN$, since $zN$ is the total number of stubs. We
then have, compare Eq.~(\ref{networks1_avqk}),
\begin{eqnarray}\nonumber
C& =& {\av{k_i k_j}_q\over Nz}
 \ =\ {\av{k_i}_q\av{k_j}_q\over Nz}
 \ =\ {1\over Nz}\,\left[\sum_k k q_k\right]^2 \\[6pt]
 & =& {1\over Nz}\,\left[{\av{k^2}-\av{k}\over\av{k}}\right]^2
 \ =\ {z\over N}\,\left[{\av{k^2}-\av{k}\over\av{k}^2}\right]^2~,
\label{networks1_crg}
\end{eqnarray}
since the distributions of two neighbors $i$ and $j$
are statistically independent. The notation $\av{...}_q$
indicates that the average is to be take with respect to
the excess degree distribution $q_k$,
as given by Eq.\ (\ref{networks1_def_q_k}).

The clustering coefficient vanishes in the thermodynamic limit
$N\to\infty$, as expected. {However,} it may have a
very big leading coefficient, especially for degree distributions
with fat tails. The differences listed in
Table~\ref{networks1_wstable}, between the measured clustering
coefficient $C$ and the value $C_{\mathrm{rand}}=z/N$ for
Erd\"os--R\'enyi graphs, are partly due to the fat tails in the
degree distributions $p_k$ of the corresponding networks.

\vspace*{-12pt}
\subsection{Probability Generating Function Formalism}
\label{networks1_generating function}
\index{probability generating function|textbf}

Network theory is about the statistical properties of graphs.
A very powerful method from probability theory is the
generating function formalism, which we will discuss now
and apply later on.

\runinhead{Probability Generating Functions} \index{probability
generating function!degree distribution} We define by
\begin{equation}
G_0(x)\ =\ \sum_{k=0}^\infty\, p_k x^k
\label{networks1_def_G_0}
\end{equation}
the {\sl generating function} $G_0(x)$
for the probability distribution $p_k$. The generating
function $G_0(x)$ contains all information present in
$p_k$. We can recover $p_k$ from $G_0(x)$ simply
by differentiation:
\begin{equation}
p_k\ =\ {1\over k!} {\mathrm{d}^k G_0\over \mathrm{d}
x^k}\bigg|_{x=0}~. \label{networks1_derivatives}
\end{equation}
One says that the function $G_0$ \qut{generates} the probability
distribution~$p_k$.

\runinhead{{The Generating} Function for
Degree Distribution of Neighbors} \index{probability generating
function!degree distribution!of neighbors}We can also define a
generating function for the distribution~$q_k$,
Eq.~(\ref{networks1_def_q_k}), of the other edges leaving a vertex
{that we} reach by following an edge in the graph:
\begin{eqnarray}\nonumber
G_1(x)& =& \sum_{k=0}^\infty q_k x^k
      \ =\ {\sum_{k=0}^\infty (k+1) p_{k+1}x^k\over\sum_j j p_j}
      \ =\ {\sum_{k=0}^\infty k p_k x^{k-1}\over\sum_j j p_j}
  \\  & =& {G_0'(x)\over z}~,
\label{networks1_def_G_1}
\end{eqnarray}
where $G_0'(x)$ denotes the first derivative of $G_0(x)$ with
respect to its argument.

\runinhead{Properties of Generating Functions} \index{probability
generating function!properties} Probability generating functions
have a couple of important properties:
\begin{enumerate}
\item{Normalization:} The distribution $p_k$ is normalized and hence
\begin{equation}
G_0(1)\ =\ \sum_k p_k\ =\ 1~.
\label{networks1_normalization}
\end{equation}
\item{Mean:} A simple differentiation
\begin{equation}
G'_0(1)\ =\ \sum_k k\, p_k\ =\ \av{k} {}
\label{networks1_mean}
\end{equation}
 yields the average degree $\av{k}$.
\item{Moments:}  The $n$th moment $\av{k^n}$ of the distribution $p_k$ is given
by\vspace*{1.5pt}
\begin{equation}
\av{k^n}\ =\ \sum_k k^n p_k\ = \
  \biggl[ \biggl(x {\mathrm{d}\over \mathrm{d} x}\biggr)^{\!n} G_0(x) \biggr]_{x=1}~.
\label{networks1_moments}\vspace*{1.5pt}
\end{equation}
\end{enumerate}

\runinhead{The Generating Function for Independent Random Variables}
Let us assume that we have two random variables. As an example we
consider two dice. Throwing the two dice are two independent random
events. The joint probability to obtain $k=1,\dots,6$ with the first
die and $l=1,\dots,6$ with the second dice is $p_k\,p_l$. This
probability function is generated by\vspace*{1.5pt}
$$
\sum_{k,l}\, p_k p_l\, x^{k+l} \ =\ \left(\sum_{k}\, p_k\,  x^{k}
\right) \left(\sum_{l}\, p_l\,  x^{l} \right)~,\vspace*{1.5pt}
$$
i.e.\ by the product of the individual generating functions.
This is the reason why generating functions are so useful in
describing combinations of independent random events.

As an application consider $n$ randomly chosen vertices. The sum
$\sum_i k_i$ of the respective degrees has a cumulative degree
distribution, which is generated by
$$ \Big[\,G_0(x)\,\Big]^n~.
$$
\runinhead{The Generating Function of the Poisson Distribution}
\index{probability generating function!Poisson distribution}As an
example we \hbox{consider} the Poisson distribution
$p_k=e^{-z}\,z^k/k!$, see
Eq.~(\ref{networks1_p_k_largeN}), with $z$ being the
average degree. Using
Eq.~(\ref{networks1_def_G_0}) we obtain
\begin{equation}
G_0(x)\ =\ e^{-z} \sum_{k=0}^\infty\, {z^k\over k!}\, x^k
      \ =\ e^{z(x-1)}~.
\end{equation}

This is the generating function for the Poisson distribution.  The
generating function $G_1(x)$ for the excess degree distribution
$q_k$ is, see Eq.~(\ref{networks1_def_G_1}),
\begin{equation}
G_1(x)\ =\ {G_0'(x)\over z}\ =\ e^{z(x-1)}~.
\end{equation}
Thus, for the case of the Poisson distribution we have, as
expected, $G_1(x)=G_0(x)$.

\runinhead{Further Examples of Generating Functions}
\index{probability generating function!examples} As a second
example, consider a graph with an exponential degree \nobreak
distribution:
\begin{equation}
p_k\ =\ (1 - e^{-1/\kappa})\, e^{-k/\kappa},
\qquad \quad
\sum_{k=0}^\infty p_k\ =\ {1 - e^{-1/\kappa}\over 1 - e^{-1/\kappa}}
\  =\ 1~,
\label{expdist}
\end{equation}
where $\kappa$ is\enlargethispage{-12pt} a constant. The generating function for this
distribution is
\begin{equation}
G_0(x)\ =\ (1 - e^{-1/\kappa}) \sum_{k=0}^\infty e^{-k/\kappa} x^k
      \ =\ {1 - e^{-1/\kappa}\over 1 - xe^{-1/\kappa}}~,
\end{equation}
and
\begin{equation}
z\ =\ G_0'(1) \ =\ { e^{-1/\kappa} \over 1-e^{-1/\kappa}},
\qquad\quad
G_1(x)\ =\ {G_0'(x)\over z} \ =\
\biggl[ {1 - e^{-1/\kappa}\over 1 - xe^{-1/\kappa}} \biggr]^2~.
\end{equation}

As a third example, consider a graph in which all vertices have degree~0,
1, 2, or~3 with probabilities $p_0\ldots p_3$.  Then the generating
functions take the form of simple polynomials
\begin{eqnarray}
G_0(x) &=& p_3 x^3 + p_2 x^2 + p_1 x + p_0,\\
G_1(x) &=& q_2 x^2 + q_1 x + q_0
        =  {3 p_3 x^2 + 2 p_2 x + p_1\over 3 p_3 + 2 p_2 + p_1}~.
\end{eqnarray}
\runinhead{Stochastic Sum of Independent Variables}
Let's assume we have random variables $k_1,\ k_2, \dots$,
each having the same generating functional $G_0(x)$.
Then 
$$
G_0^{\,2}(x),\qquad
G_0^{\,3}(x),\qquad
G_0^{\,4}(x),\qquad \dots
$$
are the generating functionals for 
$$
k_1+k_2,\qquad
k_1+k_2+k_3,\qquad
k_1+k_2+k_3+k_4, \qquad \dots~.
$$
Now consider that the number of times $n$ this 
stochastic prozess is executed is distributed 
as $p_n$. As an example consider throwing a
dice several times, with a probablity $p_n$ of
throwing exacly $n$ times. The
distribution of the results obtained is then
generated by
\begin{equation}
\sum_n p_n G_0^{\,n}(x)\ =\ G_N\left(G_0(x)\right),
\qquad\quad
G_N(z) = \sum_n p_n z^n~.
\label{networks1_sum_random_variables}
\end{equation}
We will make use of this relation further on.

\vspace*{3pt}
\subsection{Distribution of Component Sizes}
\label{networks1_distribution_sizes}
\index{distribution!component sizes|textbf}

\runinhead {The Absence of Closed Loops}
\index{loop!absence} We consider here a network below the
percolation transition and are interested in the distribution of the
sizes of the individual subclusters. The calculations will
crucially depend on the fact that the
generalized random graphs considered here do not have
{any} significant clustering nor any closed\break loops.
\begin{quotation}
{\it Closed Loops.\enspace}
\index{loop!closed}\index{closed loops}A
set of edges linking vertices
$$
i_1 \to i_2 \ \dots\ i_n \to i_1
$$ is called a closed loop of length $n$.
\end{quotation}
In physics jargon, all finite components are {\sl tree-like}. The
number of closed loops of length 3 corresponds to the clustering
coefficient $C$, viz to the probability that two of your friends are
also friends of each other. For random networks
$C=[\av{k^2}-\av{k}]^2/(z^3N)$, see Eq.~(\ref{networks1_crg}), tends
to zero as $N\to\infty$.

\begin{figure}[t]
\begin{center}
\includegraphics{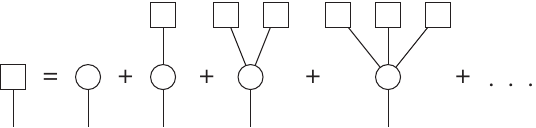}
\end{center}
\caption{Graphical representation of the self-consistency
         Eq.~(\ref{networks1_self_con_H_1})
         for the generating function
         $H_1(x)$, represented by the \textit{box}. A single vertex is
         represented by a \textit{circle}. The subcluster connected to
         an incoming vertex can be either a single vertex or
         an arbitrary number of subclusters of the same type
         connected to the first vertex
         (from Newman et al., 2001)
        }
\label{network_sum_components}
\vspace*{-12pt}
\end{figure}

\runinhead{Generating Function for the Size Distribution of
Components} \index{probability generating function!graph
components}We define by
$$
H_1(x)\ =\ \sum_m h_m^{(1)}x^m
$$
the generating function that generates the distribution of cluster
sizes {containing} a given vertex $j$, which
is linked to a specific incoming edge, see
Fig.~\ref{network_sum_components}. That is, $h_m^{(1)}$ is the
probability that the such-defined cluster contains $m$ nodes.

\runinhead{Self-Consistency Condition for
$\textit{\textbf{H}}_1\textit{\textbf{(x)}}$}
\index{self-consistency condition!graph component sizes}We note
{the following:}
\begin{enumerate}
\item The first\enlargethispage{6pt} vertex $j$ belongs to the subcluster with probability 1,
      its generating function is $x$.
\item The probability that the vertex $j$ has $k$ outgoing stubs
      is $q_k$.
\item At every stub outgoing from vertex $j$ there is a subcluster.
\item The total number of vertices consists
of those generated by $H_1(x)$ plus the starting vertex.
\end{enumerate}
The number of outgoing edges $k$ from vertex $j$ is described by the
distribution function $q_k$, see Eq.~(\ref{networks1_def_q_k}). The
total size of the $k$ clusters is generated by $[H_1(x)]^k$, as a
consequence of the multiplication property of generating functions
discussed in Sect.~\ref{networks1_generating function}. The
self-consistency equation for the total number of vertices reachable
is then
\begin{equation}
H_1(x)\ =\ x\, \sum_{k=0}^\infty\, q_k\, [H_1(x)]^k\ =\ x\, G_1(H_1(x))~,
\label{networks1_self_con_H_1}
\end{equation}
where we have made use of Eqs.~(\ref{networks1_def_G_1}) and
(\ref{networks1_sum_random_variables}).


\runinhead{The Embedding Cluster Distribution Function} 
\index{probability generating
function!embedding clusters}The quantity {that} we
actually want to know is the distribution of the sizes of the
clusters to which the entry vertex belongs. We note
{that}
\begin{enumerate}
\item The number of edges emanating from a randomly chosen vertex
      is distributed according to the degree distribution $p_k$.
\item Every edge leads to a cluster whose size is generated by $H_1(x)$.
\end{enumerate}

\noindent The size of a complete component is thus generated by
\begin{equation}
H_0(x)\ =\ x\, \sum_{k=0}^\infty p_k\, [H_1(x)]^k\ =\ x\, G_0(H_1(x))~,
\label{networks1_def_H_0}
\end{equation}
where the prefactor $x$ corresponds to the generating function of
the starting vertex. The complete distribution of component sizes is
given by solving Eq.~(\ref{networks1_self_con_H_1})
self-consistently for $H_1(x)$ and then substituting the result into
Eq.~(\ref{networks1_def_H_0}).

\runinhead{{The} Mean Component Size}
\index{mean!component size}The calculation of $H_1(x)$ and $H_0(x)$
in closed form is not possible. We are, however, interested only in
the first moment, viz the mean component size, see
Eq.~(\ref{networks1_mean}).

\enlargethispage*{12pt}

The component size distribution is generated by $H_0(x)$,
Eq.~(\ref{networks1_def_H_0}), and hence the mean component size
below the percolation transition is
\begin{eqnarray} \nonumber
\av{s}& =& H_0'(1) =
\Big[\, G_0(H_1(x)) + x\, G_0'(H_1(x))\, H_1'(x)
\,\Big]_{x=1} \\[3pt]
 & =& 1 + G_0'(1) H_1'(1)~,
\label{networks1_avs1}
\end{eqnarray}
where we have made use of the normalization
$$G_0(1)\ =\ H_1(1)\ =\ H_0(1) \ =\ 1~.$$

\noindent of generating functions, see
Eq.~(\ref{networks1_normalization}). The value of $H_1'(1)$ can be
calculated from Eq.~(\ref{networks1_self_con_H_1}) by
differentiating:
\begin{eqnarray} \label{networks1_H_1prime}
H_1'(x)& =& G_1(H_1(x))\, +\, x\,G_1'(H_1(x))\,H_1'(x), \\[3pt]
H_1'(1)& =& {1\over1-G_1'(1)}~. \nonumber
\end{eqnarray}
Substituting this into (\ref{networks1_avs1}) we find
\begin{equation}
\av{s}\ =\ 1 + {G_0'(1)\over1-G_1'(1)}~.
\label{networks1_avs2}
\end{equation}
We note that
\begin{eqnarray} \label{networks1_G_0_1prime}
G_0'(1) &=& \sum_k k\, p_k\  =\  \av{k}\  =\  z_1,\\[3pt]
G_1'(1) &=& {\sum_k k(k-1) p_k\over\sum_k k p_k}
        \  =\ {\av{k^2}-\av{k}\over\av{k}}\ =\ {z_2\over z_1}~,
\nonumber
\end{eqnarray}
where we have made use of Eq.~(\ref{networks1_z2}).
Substitution into (\ref{networks1_avs2}) then gives
the average component size below the transition as
\begin{equation}
\av{s}\ =\ 1 + {z_1^2\over z_1-z_2}~. \label{networks1_s_scaling}
\end{equation}
This expression has a divergence at $z_1=z_2$. The 
mean component size diverges at the percolation 
threshold, compare
Sect.~\ref{networks1_generalized_random_graphs}, 
and the giant connected component forms.

\section{Robustness of Random Networks}
\label{networks1_robustness}
\index{random graph!robustness|textbf}
\index{robustness!random networks|textbf}
Fat tails in the degree distributions $p_k$ of real-world
networks (only slowly decaying with large $k$) increase
the robustness of the network. That is, the network
retains functionality even when a certain number of vertices
or edges is removed. The Internet remains functional, to give
an example, even when a substantial number of
Internet routers have failed.

\runinhead{Removal of Vertices}
\index{vertex!removal}We consider a graph model in which each
vertex is either \qut{active} or \qut{inactive}. Inactive vertices
are nodes that have either been removed, or are present but
non-functional. We denote {by}
$$
b(k)\ =\ b_k
$$
the probability that a vertex is active. The probability can be,
in general, a function of the degree $k$.
The generating function
\begin{equation}
F_0(x)\ =\ \sum_{k=0}^{\infty}\, p_k\, b_k\, x^k,
\qquad\quad
F_0(1)\ =\ \sum_k\, p_k\,b_k\ \le\ 1~,
\label{networks1_def_F_0}
\end{equation}
generates the probabilities that a vertex has degree $k$ and
is present. The normalization $F_0(1)$ is
equal to the fraction of all vertices that are present.
By analogy with Eq.~(\ref{networks1_def_G_1}) we define by
\begin{equation}
F_1(x)\ =\ \frac{\sum_k\, k\,p_k\, b_k\, x^{k-1}}{\sum_k\, k\,p_k}
\  =\  \frac{F_0'(x)}{z}
\label{networks_F_1}
\end{equation}
the (non-normalized) generating function for
the degree distribution of neighbor sites.

\runinhead{Distribution of Connected Clusters}
The distribution of the sizes of connected clusters
reachable from a given vertex, $H_0(x)$,
or from a given edge, $H_1(x)$, is generated
respectively by the normalized functions
\begin{eqnarray} \nonumber
H_0(x)& =& 1\, -\, F_0(1)\, +\, x F_0(H_1(x)),\qquad H_0(1) = 1, \\
H_1(x)& =& 1\, -\, F_1(1)\, +\, x F_1(H_1(x)),\qquad H_1(1) = 1~,
\label{networks1_H_0_H_1}
\end{eqnarray}
which are logical equivalents of Eqs.~(\ref{networks1_self_con_H_1}) and
(\ref{networks1_def_H_0}).

\runinhead{Random Failure of Vertices}
{First} we consider the case of random failure of
vertices. In this case, the probability

$$
b_k\ \equiv\ b\ \leq 1, \qquad F_0(x)=b\,G_0(x), \qquad
F_1(x)=b\,G_1(x)
$$
of a vertex being present is independent of the
degree~$k$ and just equal to a constant~$b$,
which means that

\begin{equation}
H_0(x) = 1 - b + b x G_0(H_1(x)),\qquad H_1(x) = 1 - b + b x
G_1(H_1(x)), \label{networks1_siteperc}
\end{equation}

\noindent where $G_0(x)$ and $G_1(x)$ are the standard generating
functions for the degree of a vertex and of a
neighboring vertex, Eqs.~(\ref{networks1_def_G_0})
and (\ref{networks1_def_G_1}). This implies that the
mean size of a cluster of connected and present vertices is

$$
\av{s}\ =\ H_0'(1) = b\, +\, b G_0'(1)\, H_1'(1)
      \ =\ b\, +\, {b^2 G_0'(1) \over 1-bG_1'(1)}
      \ =\ b\left[1+\frac{bG_0'(1)}{1-bG_1'(1)}\right]~,
$$

\noindent where we have followed the derivation presented in
Eq.~(\ref{networks1_H_1prime}) in order to obtain
$H_1'(1)=b/(1-bG_1'(1))$. With
{Eq.~}(\ref{networks1_G_0_1prime}) for
$G_0^\prime(1)=z_1=z$ and $G_1^\prime(1)=z_2/z_1$ we obtain the
generalization

\begin{equation}
\av{s}\ =\ b + {b^2z_1^2\over z_1-bz_2}
\end{equation}

\noindent of {Eq.~}(\ref{networks1_s_scaling}). The model has a
phase transition at the critical value of~$b$
\begin{equation}
b_c\ =\ \frac{z_1}{z_2}\ =\ \frac{1}{G_1'(1)}~.
\end{equation}
If the fraction $b$ of the vertices present in the network is
smaller than the critical fraction $b_c$, then
there will be no giant component. This is the point at which the network
ceases to be functional in terms of connectivity. When there is no giant
component, connecting paths exist only within small isolated groups of
vertices, but no long-range connectivity exists. For a communication
network such as the Internet, this would be fatal.

For networks with fat {tails,
however, we expect} that the number of next-nearest neighbors $z_2$
is large compared to the number of nearest neighbors $z_1$ and that
$b_c$ is consequently small{.} The network is robust as
one would need to take out a substantial fraction of the nodes
before it would fail.

\runinhead{Random Failure of Vertices in Scale-Free Graphs}
\index{robustness!scale-free graphs}
\index{graph!scale-free!robustness}We consider a pure power-law
degree distribution
$$ p_k\ \sim\  {1\over k^{\alpha}},\qquad\quad
\int {\mathrm{d}k\over k^\alpha}\ < \ \infty, \qquad\quad \alpha>1~,
$$
see Eq.~(\ref{networks1_p_k_scale_free}) and also
Sect.~\ref{networks1_scale_free}. The first two moments are
$$
z_1\ = \langle k\rangle \ \sim\ \int \mathrm{d}k\, (k/
k^\alpha),\qquad\quad \langle k^2\rangle\ \sim\ \int \mathrm{d}k\,
(k^2/ k^\alpha)~.
$$
Noting that the number of next-nearest neighbors $z_2=\langle
k^2\rangle-\langle k\rangle$, Eq.~(\ref{networks1_z2}), we can
identify three regimes:
\begin{itemize}
\item[--]\underline{$1<\alpha\le2$}: $z_1\to\infty$, $z_2\to\infty$\\
     $b_c=z_1/z_2$ is arbitrary in the thermodynamic limit $N\to\infty$.
\item[--]\underline{$2<\alpha\le3$}: $z_1<\infty$, $z_2\to\infty$\\
     $b_c=z_1/z_2\to0$  in the thermodynamic limit. Any number of vertices
     can be randomly removed with the network remaining above the
     percolation limit. The network is extremely robust.
\item[--]\underline{$3<\alpha$}: $z_1<\infty$, $z_2<\infty$\\
     $b_c=z_1/z_2$ can acquire any value and the network has
     normal robustness.
\end{itemize}

\runinhead{Biased Failure of Vertices}
What happens when one sabotages the most important sites of a
network? This is equivalent {to} removing vertices in
decreasing order of their degrees, starting with the highest degree
vertices. The probability that a given node is active
{then takes} the form
\begin{equation}
b_k\ =\ \theta(k_c-k)~,
\end{equation}
where $\theta(x)$ is the Heaviside step function
\begin{equation}
\theta(x)\ =\ \biggl\lbrace \begin{array}{ll}
            0             & \mbox{for $x<0$}\\
            1 \qquad\null & \mbox{for $x\ge0$}
            \end{array}~.
\end{equation}
This corresponds to setting the upper limit of the sum in
Eq.~(\ref{networks1_def_F_0}) to $k_c$.

Differentiating Eq.~(\ref{networks1_H_0_H_1}) with respect to $x$
yields
$$
H_1'(1)\ =\ F_1(H_1(1)) \,+\, F_1'(H_1(1))\, H_1'(1),
\qquad\quad
H_1'(1)\ =\ {F_1(1)\over 1-F_1'(1)}~,
$$
as $H_1(1)=1$. The phase transition occurs when $F_1'(1)=1$,
\begin{equation}
{\sum_{k=1}^\infty k(k-1) p_k b_k\over\sum_{k=1}^\infty k p_k} \ =\
{\sum_{k=1}^{k_c} k(k-1) p_k \over\sum_{k=1}^\infty k p_k} \ =\
1~, \label{networks1_biase_transition}
\end{equation}
where we used the definition Eq.\ (\ref{networks_F_1}) for
$F_1(x)$.

\begin{figure}[t]
 \centering
\includegraphics{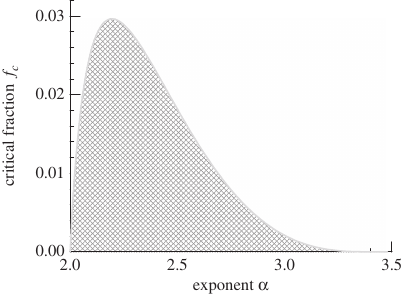}
\caption{The critical fraction $f_c$ of
vertices,
  Eq.~(\ref{networks1_def_f_c}). Removing
  a fraction greater than $f_c$ of highest degree
  vertices from a scale-free network,
  with a power-law degree distribution
  $p_k\sim k^{-\alpha}$ drives the network
  below the percolation limit. For a
  smaller loss of highest degree vertices (\textit{shaded area})
  the giant connected component remains intact
  (from Newman, 2002)}
\label{networks1_criticalFraction}
\vspace*{12pt}
\end{figure}

\runinhead{Biased Failure of Vertices for Scale-Free Networks}
Scale-free networks have a power-law degree distribution,
$p_k\propto k^{-\alpha}$. We can then rewrite
Eq.~(\ref{networks1_biase_transition}) as
\begin{equation}
H_{k_c}^{(\alpha-2)}\, -\, H_{k_c}^{(\alpha-1)}\ =\ H_{\infty}^{(\alpha-1)}~,
\label{networks1_def_H_k_c}
\end{equation}
where $H_n^{(r)}$ is the $n$th harmonic number of order~$r$:
\begin{equation}
H_n^{(r)}\ =\ \sum_{k=1}^n {1\over k^r}~.
\end{equation}
The number of vertices present is $F_0(1)$, see
Eq.~(\ref{networks1_def_F_0}), or $F_0(1)/\sum_k p_k$, since the
degree distribution $p_k$ is normalized. If we remove a certain
fraction $f_c$ of the vertices we reach the transition determined by
Eq.~(\ref{networks1_def_H_k_c}):
\begin{equation}
f_c\ =\ 1\, -\, {F_0(1) \over \sum_k p_k} \ =\
1\, -\, {H_{k_c}^{(\alpha)}\over H_\infty^{(\alpha)}}~.
\label{networks1_def_f_c}
\end{equation}
It is impossible to determine $k_c$ from (\ref{networks1_def_H_k_c})
and (\ref{networks1_def_f_c}) to get $f_c$ in closed form. One can,
however, solve Eq.~(\ref{networks1_def_H_k_c}) numerically for $k_c$
and substitute it into {Eq.~}(\ref{networks1_def_f_c}).
The results are shown in Fig.~\ref{networks1_criticalFraction}, as a
function of {the} exponent $\alpha$. The network is very
susceptible with respect to a biased removal of highest-degree
vertices.
\begin{itemize}
\item[--] A removal of more than about 3\% of the highest
      degree vertices {always leads} to a destruction of the giant
      connected component. Maximal robustness is achieved for
       $\alpha\approx 2.2$, which is actually close to
       the exponents measured in some real-world networks.
\item[--]Networks with $\alpha<2$ have
      no finite mean, $\sum_k k/k^2\to\infty$,
      and therefore make little sense physically.
\item[--]Networks with $\alpha>\alpha_c=3.4788{\ldots}$ have no
      giant connected component. The critical exponent
      $\alpha_c$ is given by the percolation condition
      $H_\infty^{(\alpha-2)}= 2 H_\infty^{(\alpha-1)}$,
      see Eq.~(\ref{networks1_mrcondition}).
\end{itemize}

\section{Small-World Models}
\label{networks1_small-world_models}\index{model!small-world
network|textbf}\index{small-world!graph|textbf}Random graphs and
random graphs with arbitrary degree distribution show no clustering
in the thermodynamic limit, in contrast to real-world networks. It
is therefore important to find methods to generate graphs
that have a finite clustering coefficient and, at
the same time, the small-world property.

\begin{figure}[t!]
\centering
\includegraphics{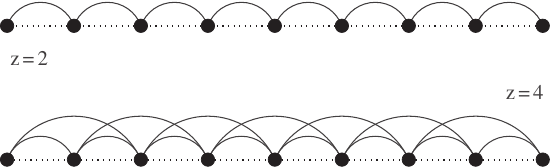}
\caption{Regular linear graphs with connectivities $z=2$
(\textit{top}) and $z=4$ (\textit{bottom})} \label{networks1_1D}
\vspace*{14pt}
\end{figure}

\runinhead{Clustering in Lattice Models}
\index{clustering!lattice models}Lattice models and random graphs
are two extreme cases of network models. In Fig.~\ref{networks1_1D}
we illustrate a simple one-dimensional lattice with
connectivity $z=2,4$. We consider periodic boundary conditions, viz
the chain wraps around itself in a ring. We then can
calculate the clustering coefficient $C$ exactly.

\begin{itemize}
\item[--]{{The One-Dimensional} Lattice}:
The number of clusters can be easily counted. One finds
\begin{equation}
C \ =\ {3(z-2)\over4(z-1)}~,
\label{networks1_C_1d}
\end{equation}
which tends to $3/4$ in the limit of large $z$.

\item[--]{Lattices with Dimension $d$}:
Square or cubic lattices have dimension $d=2,3$, respectively. The
clustering coefficient for general dimension $d$ is
\begin{equation}
C\ =\ {3(z-2d)\over4(z-d)}~,
\label{networks1_gend}
\end{equation}
which generalizes Eq.~(\ref{networks1_C_1d}). We note that the
clustering coefficient tends to $3/4$ for $z\gg2d$ for regular
hypercubic lattices in all dimensions.
\end{itemize}

\begin{figure}[t]
\centering
\includegraphics{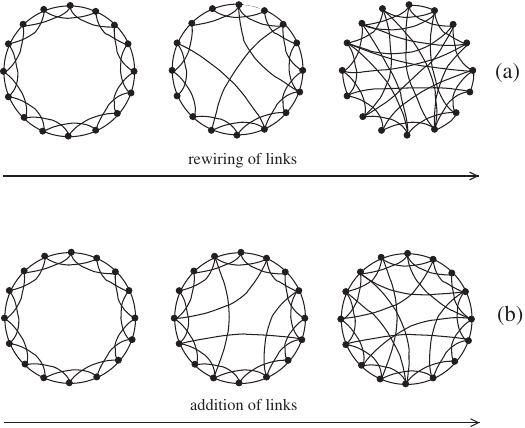}
\caption{Small-world networks in which the crossover from a
  regular lattice to a random network is realized.
  (\textbf{a}) The original Watts--Strogatz model with the rewiring of links.
  (\textbf{b}) The network with the addition of shortcuts
  (from Dorogovtsev and Mendes, 2002)
  }
\label{networks1_model}
\vspace*{-18pt}
\end{figure}

\runinhead{Distances in Lattice Models}\index{distance!lattice
model}Regular lattices do not show the small-world effect. A
regular hypercubic lattice in $d$ dimensions with linear size $L$
has $N=L^d$ vertices. The average vertex--vertex distance increases
as $L$, or equivalently as
$$
\ell\ \approx\ N^{1/d}~.
$$

\runinhead{The Watts and Strogatz Model}
\index{Watts--Strogatz model}\index{model!Watts--Strogatz}Watts
and Strogatz have proposed a small-world model
that interpolates smoothly between a regular
lattice and an Erd\"os--R\'enyi random graph. The 
construction starts with a one-dimensional lattice, see
Fig.~\ref{networks1_model}(a). One goes through all the
links of the lattice and rewires the link with some \hbox{probability
$p$.}
\begin{quotation}
{\it Rewiring Probability.\enspace}
\index{probability!rewiring}We
move one end of every link with the probability $p$ to a new
position chosen at random from the rest of the lattice.
\end{quotation}
For small $p$ this process produces a graph {that}
is still mostly regular but has a few connections
that stretch long distances across the lattice as
illustrated in Fig.~\ref{networks1_model}(a). The average \nobreak
coordination number of the lattice is by construction still the
initial degree $z$. The number of neighbors of any particular vertex
can, however, be greater or smaller than
$z$.

\runinhead{The Newman and Watts Model} \index{Newman--Watts
model}\index{model!Newman--Watts}  A variation of the
Watts--Strogatz model has been suggested by Newman and Watts.
Instead of rewiring links between sites as in
Fig.~\ref{networks1_model}(a), extra links, also called
\qut{shortcuts}, are added between pairs of sites chosen at random,
but no links are removed from the underlying lattice, see
Fig.~\ref{networks1_model}(b). This model is somewhat easier to
analyze than the original Watts and Strogatz model, because it is
not possible for any region of the graph to become disconnected from
the rest, whereas this can happen in the original model.

The small-world models illustrated in Fig.~\ref{networks1_model},
have an intuitive justification for social networks. Most people are
friends with their immediate neighbors. Neighbors on the same
street, people that they work with or their relatives.
{However,} some people are also friends with a few far
away persons. Far away in a social sense, like people in other
countries, people from other walks of life, acquaintances from
previous eras of their lives, and so forth.  These long-distance
acquaintances are represented by the long-range links in the
small-world models illustrated in Fig.~\ref{networks1_model}.

\begin{figure}[t]
\centering
\sidecaption[t]
\includegraphics{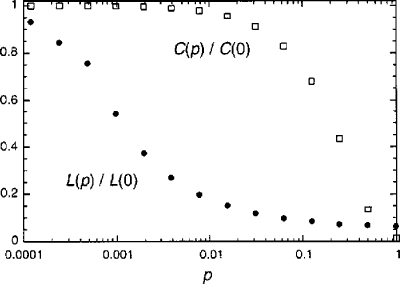}
\caption{The clustering coefficient $C(p)$ and the average
         path length $L(p)$, as a function of the rewiring
         probability for the Watts and Strogatz model,
         compare Fig.~\ref{networks1_model}
         (from Watts and Strogatz, 1998)
} \label{networks1_Watts_Strogatz_results}
\end{figure}

\runinhead{Properties of the Watts and Strogatz Model}
In Fig.~\ref{networks1_Watts_Strogatz_results} the clustering
coefficient and the average path length are shown as a function of
the rewiring probability $p$. The key result is that
there is a parameter range, say $p\approx 0.01-0.1$, where the
network still has a very high clustering
coefficient and already a small average path length, as observed in
real-world networks. Similar results hold for the Newman--Watts
model.

\section{Scale-Free Graphs}
\label{networks1_scale_free}\index{scale-free!graph|textbf}
\index{graph!scale-free|textbf}

\runinhead{Evolving Networks} \index{network!evolving}Most
real-world networks are {\sl open}, i.e.\ they are formed by the
continuous addition of new vertices to the system. The number of
vertices, $N$, increases throughout the lifetime of the network, as
{}it is the case for the WWW, which grows exponentially
by the continuous addition of new web pages. The small world
networks discussed in Sect.~1.4 are, however, constructed for a
fixed number of nodes $N$, growth is not considered.

\runinhead{Preferential Connectivity}
\index{connectivity!preferential}\index{preferential!connectivity}
Random network models assume that the probability that two vertices
are connected is random and uniform. In contrast, most real networks
exhibit the \qut{rich-get-richer} phenomenon.
\begin{quotation}
{\it Preferential Connectivity.\enspace}
When the probability for a
new vertex to connect to any of the existing nodes is not uniform
for an open network we speak of preferential connectivity.
\end{quotation}

\enlargethispage*{15pt}

A newly created web page, to give an example, will include links to
well-known sites with a quite high probability. Popular web pages
will therefore have both a high number of incoming links and a high
growth rate for incoming links. {The growth} of
vertices in terms of edges is therefore in general not uniform.

\begin{figure}[t]
\centering
\includegraphics{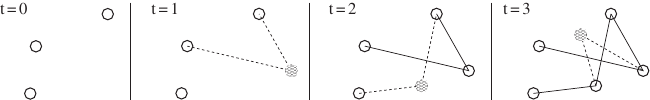}
\caption{Illustration of the preferential attachment model
         for an evolving network. At $t=0$ the system
         consists of $m_0=3$ isolated vertices. At every
         time step a new vertex ({\it shaded circle}) is added, which is connected
         to $m=2$ vertices, preferentially to the vertices
         with high connectivity, determined by the rule
         Eq.~(\ref{networks1_preferential_attachment})}
\label{networks1_scaleFree_models}
\vspace*{-12pt}
\end{figure}

\runinhead{Barab\'asi--Albert Model}
\index{graph!scale-free!construction}We start with $m_0$ unconnected
vertices. The preferential attachment growth process can then be
carried out in two steps:
\begin{itemize}
\item[--]{Growth}:
   At every time step we add a new vertex and $m\leq m_0$ stubs.
\item[--]{Preferential Attachment}:
\index{preferential!attachment}
    We connect the $m$ stubs to vertices already present
    with the probability
\begin{equation}
\Pi(k_i)\ =\ k_i/\sum_j k_j~,
\label{networks1_preferential_attachment}
\end{equation}
viz we have chosen the attachment probability $\Pi(k_i)$ to be
linearly proportional to the number of links already present. Other
functional dependencies for $\Pi(k_i)$ are of course possible, but
not considered here.
\end{itemize}
After $t$ time steps this model leads to a network with $N=t+m_0$
vertices and $mt$ edges, see Fig.~\ref{networks1_scaleFree_models}.
We will now show that the preferential rule leads to a scale-free
degree distribution
\begin{equation}
p_k\ \sim\ k^{-\gamma}
\qquad\quad \gamma > 1 ~,
\label{networks1_P_scale_free}
\end{equation}
with $\gamma=3$. The relation
Eq.~(\ref{networks1_preferential_attachment}) is
valid for the case we consider here, large degrees $k_i$.
For numerical simulations one should use
$\Pi(k_i)\propto (k_i+1)$.

\runinhead{Time-Dependent Connectivities}
\index{connectivity!time-dependent}\index{mean-field
theory!scale-free evolving nets}The time dependence of the degree
of a given vertex can be calculated analytically using a mean-field
approach. We are interested in vertices with large degrees
$k${;} the scaling relation
Eq.~(\ref{networks1_P_scale_free}) is\enlargethispage{6pt} defined asymptotically for the
limit $k\to\infty$. We may therefore assume $k$ to be continuous:
\begin{eqnarray} \nonumber
\Delta k_i(t)& \equiv & k_i(t+1)-k_i(t) \ \approx\
\frac{\partial k_i}{\partial t}\\
& =& A\, \Pi(k_i)
\ =\ A\,\frac{k_i}{\sum_{j=1}^{m_0+t-1}k_j}~,
\label{networks1_scale_free_eq}
\end{eqnarray}
where $\Pi(k_i)=k_i/\sum_j k_j$ is the attachment probability.
The overall number of new links is proportional to
a normalization constant $A$, which is hence
determined by the sum rule
$$
\sum_i\Delta k_i(t)\ \equiv\ m \ =\ A\, {\sum_i k_i\over\sum_j k_j}\ =\ A~,
$$
where the sum runs over the already existing nodes.
At every time step $m$ new edges are attached to
the existing links. The total number of connectivities
is then $\sum_j k_j=2m(t-1)$.
We thus obtain
\begin{equation}
\frac{\partial k_i}{\partial t}\ =\
{m k_i\over 2m (t-1)} \ =\ \frac{k_i}{2(t-1)}
\ \approx\ \frac{k_i}{2t}~.
\label{networks1_dot_k}
\end{equation}
Note that Eq.~(\ref{networks1_scale_free_eq}) is not well defined
for $t=1$, since there are no existing edges present in the system.
{In principle
preferential attachment needs} some starting connectivities to work.
We have therefore set $t-1\approx t$ in Eq.~(\ref{networks1_dot_k}),
since we are only interested in the long-time behaviour.

\begin{figure}[t]
\centering
\includegraphics{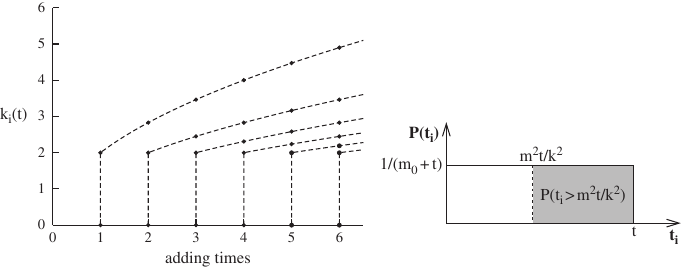}
\caption{\textit{Left}: Time evolution of the connectivities for
vertices with adding times $t=1,2,3,\dots$ and $m=2$, following
Eq.~(\ref{networks1_sqrt}).  \textit{Right}: {The} integrated probability, $P(k_i(t)<k)\ =\ P(t_i>{t
m^2/k^2})$, see Eq.~(\ref{networks1_prob_integrate})
\label{networks1_prop_distribution}}
\end{figure}

\runinhead{Adding Times} {Equation} (\ref{networks1_dot_k}) can be
easily solved taking into account that every vertex $i$ is
characterized by the time $t_i=N_i-m_0$ {that} it was added to the
system with $m=k_i(t_i)$ initial links:\vspace*{3pt}
\begin{equation}
k_i(t)\ =\ m\left(\frac {t}{t_i}\right)^{0.5}, \quad\qquad t_i \
=t\, m^2/k_i^2~. \label{networks1_sqrt}\vspace*{3pt}
\end{equation}
Older nodes, i.e.\ those with smaller $t_i$, increase their
connectivity faster than the younger vertices, viz those with bigger
$t_i$, see Fig.~\ref{networks1_prop_distribution}. For social
networks this mechanism is dubbed the rich-gets-richer phenomenon.

The number of nodes $N(t)=m_0+t$ is identical to the number of
adding times,\vspace*{3pt}
$$
t_1,\dots,t_{m_0}\ =\ 0,\qquad\quad t_{m_0+j}\ = j,\quad\quad
j=1,2,\dots~,\vspace*{3pt}
$$
where we have defined the initial $m_0$ nodes to have adding times
zero.

\runinhead{Integrated Probabilities} Using (\ref{networks1_sqrt}),
the probability that a vertex has a connectivity $k_i(t)$ smaller
than a certain $k$, $P(k_i(t)<k)$ can be written as\vspace*{3pt}
\begin{equation}
P(k_i(t)<k)\ =\ P\left(t_i>{{m^2t}\over{k^2}}\right)~. \label{networks1_prob_integrate}\vspace*{3pt}
\end{equation}
The adding times are uniformly distributed, compare
Fig.~\ref{networks1_prop_distribution}, and the probability $P(t_i)$
to find an adding time $t_i$ is then 
\begin{equation}
P(t_i) \ =\  \frac{1}{m_0+t}, \label{networks1_prob_adding_times}
\end{equation}
just the inverse of the total number of adding times, which
coincides with the total number of nodes. $P(t_i> {{m^2t}/{k^2}})$
is therefore the cumulative number of adding times $t_i$ larger than
$m^2 t/k^2$, multiplied with the probability $P(t_i)$
{(Eq.~(\ref{networks1_prob_adding_times}))}
to add a new node:
\begin{equation}
P\left(t_i> {{m^2t}\over{k^2}}\right)
\ =\left(t- {{m^2t}\over{k^2}}\right){1\over m_0+t}~.
\label{networks1_cumm_P}
\end{equation}
%


\runinhead{Scale-Free Degree Distribution}

\index{degree distribution!scale-free}\index{scale-free!degree
distribution}The degree distribution $p_k$ then follows from
Eq.~(\ref{networks1_cumm_P}) via a simple differentiation,
\begin{equation}
p_k\ =\ \frac{\partial P(k_i(t)<k)}{\partial k}
\ =\ \frac{\partial P(t_i > {{m^2t}/{k^2}})}{\partial k}
\ =\ \frac{2m^2t}{m_0+t}\, \frac{1}{k^3}
\label{networks1_cube}~,
\end{equation}
in accordance with Eq.~(\ref{networks1_P_scale_free}). The degree
distribution Eq.~(\ref{networks1_cube}) has a well defined limit
$t\to\infty$, approaching a stationary distribution.
\index{stationary distribution!scale-free evolving graph}We note
that $\gamma=3$, {which is} independent of the number $m$
of added links per new site. This result indicates that growth and
preferential attachment play an important role for the occurrence of
a power-law scaling in the degree distribution. To verify that both
ingredients are really necessary, we {now
investigate} a variant of above model.

\runinhead{Growth with Random Attachment}
\index{random!attachment} We examine then whether growth alone can
result in a scale-free degree distribution. We assume random
instead of preferential attachment. The growth equation for the
connectivity $k_i$ of a given node $i$, compare
Eqs.~(\ref{networks1_scale_free_eq}) and
(\ref{networks1_prob_adding_times}), then takes the form
\begin{equation}
\frac{\partial k_i}{\partial t}\ =\ {m\over m_0+(t-1)}~.
\label{networks1_scale_free_random}
\end{equation}
The $m$ new edges are linked randomly at time $t$ to the $(m_0+t-1)$
nodes present at the previous time step. Solving {Eq.~}
(\ref{networks1_scale_free_random}) for $k_i$, with the initial
condition $k_i(t_i)=m$, we obtain
\begin{equation}
\label{networks1_evol_a}
k_i\ =\ m\Big[\, \ln(m_0+t-1)\, -\, \ln(m_0+t_i-1)+1\, \Big]~,
\end{equation}
{which is} a logarithmic increase with time. The
probability that vertex $i$ has connectivity $k_i(t)$ smaller than
$k$ is then
\begin{eqnarray}
P(k_i(t)<k) & = & P\left(t_i>(m_0+t-1)\exp\left(1-\frac k m\right)-m_0+1\right)
\nonumber \\
& =& \left[t  - {(m_0+t-1)\exp\left(1-\frac km\right)-m_0+1}\right]\frac 1 {m_0+t}~,
\label{networks1_diff}
\end{eqnarray}
where we assumed that we add the vertices uniformly in time to the
system. Using
$$
p_k\ =\ \frac{\partial P(k_i(t)<k)}{\partial k}
$$
and assuming long times, we find
\begin{equation}
p_k\ =\ {1\over m} e^{1-k/m}
\ =\ \frac{e}{m}\,\exp\left(-\frac{k}{m}\right)~.
\label{networks1_p_k_exp_no_grows}
\end{equation}
{Thus for a growing network with random
attachment we find a characteristic degree}
\begin{equation}
k^* \ =\ m~,
\end{equation}
{which is} identical to half of the average connectivities
of the vertices in the system, since $\left<k\right>=2m$. Random
attachment does not lead to a scale-free degree distribution. Note
that $p_k$ in Eq.~(\ref{networks1_p_k_exp_no_grows}) is not properly
normalized, {nor} in Eq.\
(\ref{networks1_cube}), since we used a large-$k$ approximation
during the respective derivations.

\runinhead{Internal Growth with Preferential Attachment}
The original preferential attachment model yields a degree
distribution $p_k\sim k^{-\gamma}$ with $\gamma=3$. Most social
networks such as the WWW and the Wikipedia network, {however,} have exponents $2<\gamma<3$, with the exponent
$\gamma$ being relatively close to 2. It is also observed that new
edges are mostly added in between existing nodes, albeit with
(internal) preferential attachment.

We can then generalize the preferential attachment
model discussed above in the following way:
\begin{itemize}
\item[--]{Vertex Growth:}
    At every time step a new vertex is added.
\item[--]{Link Growth:}
    At every time step $m$ new edges are added.
\item[--]{External Preferential Attachment:}
    With probability $r\in[0,1]$ any one of the $m$ new edges
    is added between the new vertex and an existing
    vertex $i$, which is selected with a probability
    $\propto \Pi(k_i)$, see Eq.~(\ref{networks1_preferential_attachment}).
\item[--]{Internal Preferential Attachment:}
    With probability $1-r$ any one of the $m$ new edges
    is added in between two existing
    vertices $i$ and $j$, which are selected with a
    probability $\propto \Pi(k_i)\, \Pi(k_j)$.
\end{itemize}
The model reduces to the original preferential attachment
model in the limit $r\to1$. The scaling exponent $\gamma$
can be evaluated along the lines used above for the
case $r=1$. One finds
\begin{equation}
p_k\ \sim {1\over k^\gamma}, \qquad\quad
\gamma \ =\ 1\,+\, {1\over 1-r/2}~.
\label{network1_scaleFree_r}
\end{equation}
The exponent $\gamma=\gamma(r)$ interpolates smoothly between
{2 and 3}, with $\gamma(1)=3$ and
$\gamma(0)=2$. For most real-world graphs $r$ is quite
small{;} most links are added internally. Note,
{however, that the average connectivity
$\langle k\rangle = 2m$ remains constant}, since one new vertex is
added for $2m$ new stubs.

\vspace*{-6pt}

\enlargethispage{12pt}

\addcontentsline{toc}{section}{Exercises} 
\section*{Exercises}

{\sc Bipartite Networks}
\begin{list}{}
\item Consider $i=1,\dots,9$ managers sitting
      on the boards of six companies with
      (1,9), (1,2,3), (4,5,9), (2,4,6,7), (2,3,6)
      and (4,5,6,8) being the respective board
      compositions. Draw the graphs for the managers
      and companies, by eliminating from the bipartite 
      manager/companies graph one type of nodes.
      Evaluate for both networks the average degree $z$,
      the clustering coefficient $C$ and the 
      graph diameter $D$.
\end{list}

\hspace*{-12pt}{\sc Degree Distribution}
\begin{list}{}
\item Online network databases can be found on the Internet.
  Write a program and evaluate for a network of your
  choice the degree distribution $p_k$, the
  clustering coefficient $C$ and compare it with
  the expression (\ref{networks1_crg}) for
  a generalized random net with the same $p_k$.
\end{list}

\hspace*{-12pt}{\sc Ensemble Fluctuations}
\begin{list}{}
\item Derive Eq.~(\ref{networks1_P_preferential_attachment}) 
for the distribution of ensemble
fluctuations. In {the case of difficulties Albert and
Barab\'asi (2002) can be consulted}. Alternatively, check
Eq.~(\ref{networks1_P_preferential_attachment}) numerically.
\end{list}

\hspace*{-12pt}{\sc Self-Retracing Path Approximation}
\begin{list}{}
\item Look at {Brinkman and Rice (1970)} and prove
Eq.~(\ref{networks1_G_self_retracing}). This derivation is only
suitable for {readers with a solid training in physics}.
\end{list}

\hspace*{-12pt}{\sc Probability Generating Functions}
\begin{list}{}
\item Prove that the variance $\sigma^2$ of a probability
distribution $p_k$ with a generating functional
$G_0(x)=\sum_k p_k\,x^k$ and average $\langle k\rangle$
is given
by $\sigma^2=G_0^{\prime\prime}(1)+\langle k\rangle -\langle k\rangle^2$.

Consider now a cummulative process, compare
Eq.~(\ref{networks1_sum_random_variables}), generated
by $G_C(x)=G_N(G_0(x))$. Calculate the mean and the
variance of the cummulative process and discuss the
result.
\end{list}

\hspace*{-12pt}{\sc Clustering Coefficient}
\begin{list}{}
\item Prove Eq.~(\ref{networks1_C_1d}) for the
clustering coefficient of one-dimensional lattice graphs.
Facultatively, generalize this formula to a
$d$-dimensional lattice with links along the main
axis.
\end{list}

\hspace*{-12pt}{\sc Scale-Free Graphs}
\begin{list}{}
\item Write a program {that} implements preferential
attachments and calculate the resulting degree distribution $p_k$.
If you are adventurous, try alternative functional dependencies for
the attachment probability $\Pi(k_i)$ instead of the linear
assumption (\ref{networks1_preferential_attachment}).
\end{list}

\index{SIRS model!on a network}
\hspace*{-12pt}{\sc Epidemic Spreading in Scale-Free Networks}
\begin{list}{}
\item Consult \qut{R.~Pastor-Satorras and A.~Vespigiani,
{\sl Epidemic spreading in scale-free networks},
Physical Review Letters, Vol. 86, 3200 (2001)}, and solve a
simple molecular-field approach to the SIS model for the
spreading of diseases in scale-free networks by using the
excess degree distribution discussed in
Sect.~\ref{networks1_arbitrary_degree_distributions}, where
S and I stand for susceptible and infective
individuals respectively.
\end{list}

\index{SIRS model!on a network}
\hspace*{-12pt}{\sc Epidemic Outbreak in the Configurational Model}
\begin{list}{}
\item Consult \qut{M.E.J.~Newman,
{\sl Spread of epidemic disease on networks},
Physical Review E, Vol. 66, 16128 (2002)}, and solve the
SIR model for the spreading of diseases in social networks
by a generalization of the techniques discussed in
Sect.~\ref{networks1_robustness}, where S, I and R
stand for susceptible, infective and removed
individuals respectively.
\end{list}

\enlargethispage{12pt}

\vspace*{4pt}


\def\refer#1#2#3#4#5#6{\item{\frenchspacing\sc#1}\hspace{4pt}
                       #2\hspace{8pt}#3 {\it\frenchspacing#4} {\bf#5}, #6.}
\def\bookref#1#2#3#4{\item{\frenchspacing\sc#1}\hspace{4pt}
                     #2\hspace{8pt}{\it#3}  #4.}

\addcontentsline{toc}{section}{Further Reading} 
\section*{Further Reading}
\markboth{\thechapter\enspace Graph Theory and Small-World Networks}{Further Reading}

For further studies several books (Watts, 1999; Dorogovtsev and
Mendes, 2003; Caldarelli, 2007) and review articles
(Albert and Barab\'asi, 2002; Dorogovtsev and Mendes, 2002)
on general network theory are recommended.

The interested reader might {delve into} some of
the original literature on, e.g. the original Watts and Strogatz
(1998) small-world model, the Newman and Watts (1999) model, the
mean-field solution of the preferential attachment model\break
(Barab\'asi et al., 1999), the formulation of the concept
of clique percolation (Derenyi et al., 2005), an early
study of the WWW (Albert et al., 1999), a recent
study of the time evolution of the Wikipedia network (Capocci et
al., 2006), a study regarding the community structure of real-world
networks (Palla et al., 2005), the notion of assortative mixing 
in networks (Newman, 2002) or the mathematical basis of graph
theory (Erd\"os and R\'enyi, 1959). A good starting point is
Milgram's (1967) account of his by now famous experiment, which led
to the law of ``six degrees of separation'' (Guare, 1990).

{\baselineskip=15pt
\begin{list}{}{\leftmargin=2em \itemindent=-\leftmargin%
\itemsep=3pt \parsep=0pt \small}

\refer{Albert, R., Barab\'asi, A.-L.}{2002}{Statistical mechanics of
complex networks.}{Review of Modern Physics} {74}{47--97}

\refer{Albert, R., Jeong, H., Barab\'asi, A.-L.}{1999}{Diameter of
the
  world-wide web.}{Nature}{401}{130--131}

\refer{Barabasi, A.L., Albert, R., Jeong, H.}{1999}{Mean-field
theory for scale-free random networks.}{Physica
A}{272}{173--187}
\refer{Brinkman, W.F., Rice, T.M.}{1970}
  {Single-particle excitations in magnetic insulators.}
  {\textit{Physical Review B}} {\bf 2}{1324--1338}
\bookref{Caldarelli, G.}{2007}{Scale-Free Networks:
Complex Webs in Nature and Technology.}
{Oxford University Press Oxford}
\refer{Capocci, A. { et al.}}{2006} {Preferential attachment in the
growth of social networks:
 The internet encyclopedia Wikipedia.}
{Physical Review E}{74}{036116}
\refer{Derenyi, I., Palla, G., Vicsek, T.}{2005} {Clique percolation
in random networks.}{Physical Review Letters} {94}{160202}
\refer{Dorogovtsev, S.N., Mendes, J.F.F.}{2002} {Evolution of
  networks.}{Advances in Physics}{51}{1079--1187}
\bookref{Dorogovtsev, S.N., Mendes, J.F.F.}{2003}{Evolution of
  Networks. From Biological Nets to the Internet and WWW.}
  {Oxford University Press Oxford}
\refer{Erd\"os, P., R\'enyi, A.}{1959}{On random graphs.}
      {Publications Mathematicae\/}{6}{290--297}
\bookref{Guare, J.}{1990}{Six Degrees of Separation: A play.}
{Vintage New York}
\refer{Milgram, S.}{1967}{The small world problem.}{Psychology
  Today\/}{2}{60--67}
\refer{Moukarzel, C.F.}{1999}{Spreading and shortest paths in
systems
  with sparse long-range connections.}{Physics Review
  E\/}{60}{6263--6266}
\bookref{Newman, M.E.J.}{2002}{Random Graphs as Models of Networks.}
{\url{http://arxiv.org/abs/cond-mat/0202208}}

\refer{Newman, M.E.J.}{2002}{Assortative mixing in networks.}
{Physical Review Letters}{89}{208701}

\refer{Newman, M.E.J., Strogatz, S.H., Watts, D.J.}{2001}
 {Random graphs with arbitrary degree distributions and their applications.}
 {Physical Review E}{64}{026118}

\refer{Newman, M.E.J., Watts, D.J.}{1999}{Renormalization group
  analysis of the small world network model.}{Physics Letters
  A}{263}{341--346}

\refer{Palla, G., Derenyi, I., Farkas, I., Vicsek, T.}{2005}
{Uncovering the overlapping community structure of complex
 networks in nature and society.}{Nature}{435}{814--818}
\bookref{Watts, D.J.}{1999}{Small Worlds: The Dynamics of Networks
Between Order and Randomness.} {Princeton University Press, Princeton}
\refer{Watts, D.J., Strogatz, S.H.}{1998}{Collective dynamics of
  small world networks.}{Nature\/}{393}{440--442}
\end{list}
\par}

 

\chapter{Chaos, Bifurcations and Diffusion}
\label{chap_chaos1}


\abstract{Complex system theory deals with dynamical systems
containing very large numbers of variables. It extends dynamical
system theory, which deals with dynamical systems containing a few
variables. A good understanding of dynamical
systems theory is therefore a prerequisite when
studying complex systems.
\newline
\indent In this chapter we introduce important concepts, like
regular and irregular behavior, attractors and Lyapunov exponents,
bifurcation, and deterministic chaos from the realm of dynamical
system theory. A short introduction to dissipative and
stochastic, viz noisy systems is given further on,
together with two important examples out of noise-controlled
dynamics, namely stochastic escape and stochastic\break resonance.
\newline
\indent Most of the chapter will be devoted to ordinary differential
equations, the traditional focus of dynamical system theory, venturing
however towards the end into the intricacies of time-delayed dynamical
systems.
}


\section{Basic Concepts of Dynamical Systems Theory}
\label{chaos_introduction}
\index{dynamical system!basic concepts|textbf}

Dynamical systems theory deals with the
properties of coupled differential equations, determining the time
evolution of a few, typically a handful of variables. Many
interesting concepts have been developed and we will present a short
overview covering the most important phenomena.

\begin{figure}[t]
\centering
\includegraphics{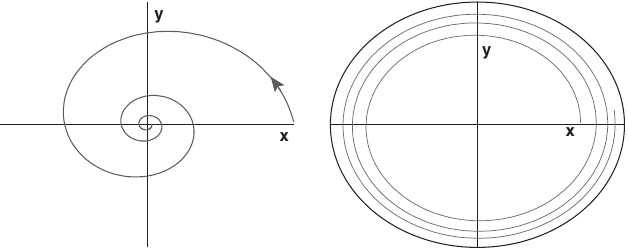}
\vspace*{0.0cm} \caption{The solution of the non-linear rotator
{equations } (\ref{chaos_x_y_r}) and (\ref{chaos_dot_r})
for $\Gamma<0$ (\textit{left}) and $\Gamma>0$ (\textit{right})}
\label{chaos_fig_sprials}\vspace*{-8pt}
\end{figure}

\runinhead{Fixpoints and Limiting Cycles}
\index{fixpoint}\index{limiting cycle}\index{cycle!limiting} We
start by discussing an elementary non-linear rotator, just to
illustrate some procedures {that are} typical for
dynamical systems theory. We consider a two-dimensional system ${\bf
x}=(x,y)$. Using {the} polar coordinates
\index{coordinates!polar}
\begin{equation}
x(t)\ =\ r(t)\cos(\varphi(t)), \qquad\quad y(t)\ =\
r(t)\sin(\varphi(t))~, \label{chaos_x_y_r}
\end{equation}
\noindent we assume that the following non-linear differential
equations:
\begin{equation}
\dot r\ =\ (\Gamma-r^2)\,r,
\qquad\quad \dot\varphi\ =\ \omega
\label{chaos_dot_r}
\end{equation}
govern the dynamical behavior. {The typical}
orbits $(x(t),y(t))$ are illustrated in
Fig.~\ref{chaos_fig_sprials}. The limiting behavior of
Eq.~(\ref{chaos_dot_r}) is
\begin{equation}
\lim_{t\to\infty}
\left[
\begin{array}{c} x(t) \\ y(t) \end{array}
\right]
\ =\
\left\{
\begin{array}{ccr}
\left[\begin{array}{c} 0 \\ 0 \end{array}\right] &\quad& \Gamma<0 \\
\left[\begin{array}{c} r_c\cos(\omega t) \\ r_c\sin(\omega t)
       \end{array}\right] &\quad& r_c^2=\Gamma>0
\end{array}
\right. ~.
\label{chaos_r_t_infty}
\end{equation}
In the first case, $\Gamma<0$, we have a stable
fixpoint; in the second case, $\Gamma>0$, the dynamics
{approaches} a limiting cycle.

\begin{quotation}
{\it Bifurcation. \enspace} \index{bifurcation} When a dynamical
system, described by a set of parameterized differential equations,
changes qualitatively, as a function of an external parameter, the
nature of its long-time limiting behavior in terms of fixpoints or
limiting cycles{,} one speaks of a bifurcation.
\end{quotation}
The dynamical system (\ref{chaos_x_y_r}) and (\ref{chaos_dot_r})
shows a bifurcation at $\Gamma=0$. A fixpoint turns into a limiting
cycle at $\Gamma=0$, and one denotes this specific type
of bifurcation as a
\qut{Hopf bifurcation}\index{Hopf bifurcation}.

\runinhead{Stability of Fixpoints}
\index{fixpoint!stability}
The dynamics of orbits close to a fixpoint or a 
limiting orbit determines its stability. 
\begin{quotation}
{\it Stability Condition. \enspace} 
A fixpoint is stable (unstable) if nearby
orbits are attracted (repelled) by the fixpoint, 
and metastable if the distance does not change. 
\end{quotation}
The stability of fixpoints is
closely related to their Lyapunov exponents, see 
Sect.~\ref{sect_chaos_logistic_map}.

One can examine the stability of a fixpoint ${\bf x}^*$ by linearizing 
the equation of motions for ${\bf x}\approx{\bf x}^*$. For the fixpoint
$r^*=0$ of Eq.~(\ref{chaos_dot_r}) we find
$$
\dot r\ =\ \left(\Gamma-r^2\right)r \ \approx\ \Gamma r
\qquad\quad r\ll1~,
$$
and $r(t)$ decreases (increases) for $\Gamma<0$ ($\Gamma>0$).
For a $d$-dimensional system ${\bf x}=(x_1,\ \dots,\ x_d)$ 
the stability of a fixpoint ${\bf x}^*$ is determined by
calculating the $d$ eigenvalues of the linearized equations
of motion. The system is stable if all eigenvalues are
negative and unstable if at least one eigenvalue is positive.

\runinhead{First-Order Differential Equations} \index{differential
equation!first-order} Let us consider the third-order differential
equation
\begin{equation}
{{d}^3\over \mathrm{d}t^3} x(t) \ =\ f(x,\dot x, \ddot x)~.
\label{chaos_x3dot_f}
\end{equation}
Using
\begin{equation}
x_1(t) = x(t), \qquad
x_2(t) = \dot x(t), \qquad
x_3(t) = \ddot x(t)~,
\label{chaos_x123}
\end{equation}
we can rewrite (\ref{chaos_x3dot_f}) as
a first-order differential equation:
$$
\frac{\D}{{\D}t} \left[\begin{array}{c} x_1 \\ x_2 \\
x_3
\end{array}\right] \ =\ \left[\begin{array}{c} x_2 \\ x_3 \\
f(x_1,x_2,x_3) \end{array}\right]~.
$$
\runinhead{Autonomous Systems} It is then generally true that one
can reduce any set of coupled differential equations to a set of
first-order differential equations {by} introducing an
appropriate number of additional variables. We {}
therefore consider in the following only first-order, ordinary
differential equations such as
\begin{equation}
{\mathrm{d} {\bf x}(t) \over \mathrm{d}t}\ =\ {\bf f}({\bf x}(t)),
\qquad\quad {\bf x},{\bf f} \in {\rm I\!R}^d, \qquad\quad
t\in[-\infty,+\infty]~, \label{chaos_ode}
\end{equation}
\index{dynamics!continuous time}when time is continuous, or,
equivalently, maps such as
\begin{equation}
{\bf x}(t+1)\ =\ {\bf g}({\bf x}(t)), \qquad\quad {\bf x},{\bf g}
\in {\rm I\!R}^d, \qquad\quad t=0,1,2,\ldots \label{chaos_maps}
\end{equation}
\index{dynamics!discrete time}when time is discrete. An evolution
equation of type Eq.~(\ref{chaos_ode}) is denoted
\qut{autonomous},\index{dynamical
system!autonomous}\index{autonomous dynamical system} since it does
not contain an explicit time dependence. A system of type $\dot{\bf
x}={\bf f}(t,{\bf x})$ is dubbed \qut{non-autonomous}.

\begin{quotation}
{\it The Phase Space.\enspace}
\index{phase space} One denotes by \qut{phase space} the space
spanned by all allowed values of the variables entering the set of
first-order differential equations defining the dynamical system.
\end{quotation}
The phase space depends on the representation. For a two-dimensional
system $(x,y)$ the phase space is just ${\rm I\!R}^2$, but in
{the} polar coordinates Eq.~(\ref{chaos_x_y_r}) it is
$$
\Big\{\,
(r,\varphi)\,\Big|\, r\in[0,\infty],\, \varphi\in[0,2\pi[
\,\Big\}~.
$$

\runinhead{Orbits and Trajectories} \index{orbit} A particular
solution ${\bf x}(t)$ of the dynamical system\break
Eq.~(\ref{chaos_ode}) can be visualized as a \qut{trajectory}, also
denoted  \qut{orbit}, in phase space. Any orbit is uniquely
determined by the set of \qut{initial conditions}, ${\bf x}(0)\equiv
{\bf x}_0$, since we are dealing with first-order differential
equations.

\begin{figure}[t]
\centering \sidecaption[t]
\includegraphics{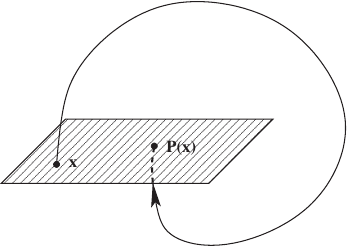}
\caption{{The} Poincar\'e map $\ {\bf
x}\to {\bf P}({\bf x})$} \label{chaos_map_Poincare}
\end{figure}\vspace*{12pt}

\runinhead{The Poincar\'e Map} \index{Poincar\'e
map}\index{map!Poincar\'e} It is difficult to illustrate graphically
the motion of ${\bf x}(t)$ in $d$ dimensions. Our retina as well as
our print media {are} two-dimensional and it is
therefore convenient to consider a plane $\Sigma$ in $ {\rm I\!R}^d$
and the points ${\bf x}^{(i)}$ of {the} intersection of
an orbit $\gamma$ with $\Sigma$, see Fig.~\ref{chaos_map_Poincare}.

For the purpose of illustration let us consider the plane
$$
\Sigma\ =\ \{\,(x_1,x_2,0,\dots,0)\,|\, x_1,x_2\in{\rm I\!R}\, \}
$$
and the sequence of intersections (see
Fig.~\ref{chaos_map_Poincare})
$$
{\bf x}^{(i)}\ =\ (x_1^{(i)},x_2^{(i)},0,\dots,0),\qquad\quad
(i=1,2,\ldots)
$$
which define the {\em Poincar\'e map}
$$
{\bf P}:\;\; {\bf x}^{(i)}\ \mapsto\ {\bf x}^{(i+1)}~.
$$
The Poincar\'e map is therefore a discrete map of
{the type of} Eq.~(\ref{chaos_maps}), which can be
constructed for continuous-time dynamical systems like
Eq.~(\ref{chaos_ode}). The Poincar\'e map is very useful, since we
can print and analyze it directly. A periodic orbit, to give an
example, would show up in the Poincar\'e map as the identity
mapping.

\runinhead{Constants of Motion and Ergodicity} We mention
{here} a few general concepts from the theory of
dynamical systems.

\begin{itemize}
\item[--]{{The Constant} of Motion}: \index{constant of motion}A function $F({\bf x})$ on phase space ${\bf x}=(x_1,\dots,x_d)$ is
called a  \qut{constant of motion} or a \qut{conserved quantity} if
it is conserved under the time evolution of the dynamical system,
i.e.\ when
$$
{\mathrm{d}\over \mathrm{d}t}F({\bf x}(t)) \ =\ \sum_{i=1}^d
\left({\partial\over\partial x_i}F({\bf x})\right) \dot x_i(t) \
\equiv\ 0
$$
holds for all times $t$. In many mechanical systems the
energy is a conserved quantity.
\item[--]{Ergodicity}:
\index{dynamical system!ergodic}A dynamical system in which orbits
come {arbitrarily} close to any allowed point in
{the} phase space, irrespective of the initial condition,
is called ergodic.

\enlargethispage*{12pt}

\hspace*{12pt}All conserving systems of classical mechanics, obeying
Hamiltonian dynamics, are ergodic. The ergodicity of a mechanical
system is closely related to \nobreak \qut{Liouville's theorem},
which will be discussed in Sect.~\ref{chaos_strange_attractors}.

\hspace*{12pt}Ergodicity holds only modulo conserved quantities, as
{}is the case for the energy in many mechanical
systems. Then, only points in {the} phase space having
the same energy as the trajectory considered are approached
arbitrarily close.
\item[--]{{Attractors}}:
\index{attractor}A bounded region in phase space to which orbits
with certain initial conditions come
{arbitrarily} close is called an attractor.

\hspace*{10pt} Attractors can be isolated points (fixpoints),
limiting cycles or more complex objects.
\item[--]{{The Basin} of Attraction}:
\index{attractor!basin}\index{basin of attraction}The set of initial
conditions {that} leads to orbits
{approaching a certain
attractor arbitrarily closely} is called the basin of attraction.
\end{itemize}
It is clear that ergodicity and attractors
are mutually exclusive: An ergodic system cannot
have attractors and a dynamical system with one
or more attractors cannot be ergodic.

\begin{figure}[t]
\centering
\includegraphics{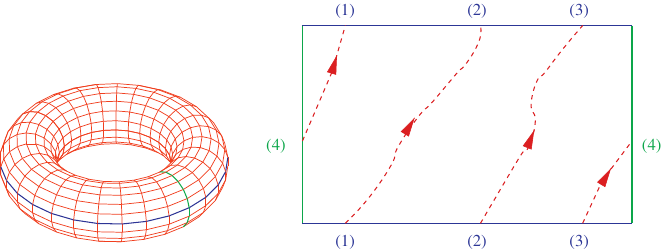}
\caption{\label{KAM_tori} {A} KAM-torus.
\textit{Left}: The torus can be cut along two lines
(\textit{vertical/horizontal}) and unfolded.  \textit{Right}: A
closed orbit on the unfolded torus with $\omega_1/\omega_2=3/1$. The
numbers indicate points {that} coincide after
refolding (periodic boundary conditions)
        }
\index{KAM!torus}\index{orbit!closed}
\end{figure}

\runinhead{Mechanical Systems and Integrability} \index{dynamical
system!mechanical} A dynamical system of type
$$
\ddot x_i=f_i({{\veci x}}, \dot{\veci x}), \qquad\quad i=1,\ldots ,f
$$
is denoted {a} \qut{mechanical system} since all
equations of motion in classical mechanics are of this form, e.g.\
Newton's law. $f$ is called the degree of freedom and a mechanical
system can be written as a set of coupled first-order differential
equations with $2f$ variables
$$
(x_1\dots x_f,v_1\dots v_f),
\qquad\quad  v_i=\dot x_i,
\qquad\quad  i=1,\dots,N
$$
constituting the phase space, with ${\bf v}=(v_1,\ldots ,v_f)$ being
denoted the generalized velocity.\index{dynamical system!integrable}
A mechanical system is {\em integrable} if there are
$\alpha=1,\ldots,f$ independent constants of motion $F_\alpha(\veci
x, \dot{\veci x})$ with
$$
{\textit{d}\over \mathrm{d}t}F_{\alpha}(\veci x, \dot{\veci x})\ =\
0,\qquad\quad \alpha=1,\dots,f~.
$$
The motion in the $2f$-dimensional phase space $\ (x_1\dots
x_f,v_1\dots v_f)\ $ is then restricted to an $f$-dimensional
subspace, {which is} an $f$-dimensional torus, see
Fig.~\ref{KAM_tori}.

An example {of} an integrable mechanical system is the
Kepler problem, viz the motion of the earth around the sun.
Integrable systems, {however, are} very rare,
but they constitute important reference points for the understanding
of more general dynamical systems. A classical example of a
non-integrable mechanical system is the three-body problem, viz the
combined motion of earth, moon and sun around each other.


\runinhead{The KAM Theorem} \index{KAM!theorem}\index{theorem!KAM}
Kolmogorov, Arnold and Moser (KAM) have examined the question of
what happens to an integrable system when it is perturbed. Let us
consider a {two}-dimensional torus, as illustrated in
Fig.~\ref{KAM_tori}. The orbit wraps around the torus with
frequencies $\omega_1$ and $\omega_2$, respectively. A key quantity
is the ratio of revolution frequencies $\omega_1/\omega_2$; it might
be rational or irrational.

We remember that any irrational number $\ r\ $ may be approximated
with arbitrary accuracy by a sequence of quotients\vspace*{3pt}
$$
{m_1\over s_1},\ {m_2\over s_2},\ {m_3\over s_3},\ \ldots
\qquad\qquad s_1<s_2<s_3<\ldots\vspace*{3pt}
$$
with ever larger denominators $s_i$. A number $r$ is \qut{very
irrational} when it is difficult to approximate $r$ by such a series
of rational numbers, viz when very large denominators $s_i$ are
needed to achieve a certain given accuracy $\ |r-m/s|$.

The KAM theorem states that orbits with
rational ratios of revolution frequencies
$\omega_1/\omega_2$ are the most unstable
under a perturbation of an integrable system
and that tori are most stable when this
ratio is very irrational.

\runinhead{Gaps in {the Saturn Rings}} A
spectacular example {of} the instability of rational
KAM-tori are the gaps in the rings of the planet Saturn.

The time a particle orbiting in Cassini's gap (between the
{A-ring} and the B-ring, $r= 118\ 000$~km) would need
around {} Saturn is exactly half the time the
\qut{shepherd-moon} Mimas needs to orbit Saturn. The quotient of the
revolving frequencies is\break $\ 2:1$.  Any particle orbiting in
Cassini's gap is therefore unstable against the perturbation caused
by Mimas and it is consequently thrown out of its orbit.

\vspace*{7pt}
\section{The Logistic Map and Deterministic Chaos}
\label{sect_chaos_logistic_map}
\index{map!logistic|textbf}\index{logistic map|textbf}
\index{chaos!deterministic|textbf}\index{deterministic chaos|textbf}

\runinhead{Chaos} \looseness1The notion of \qut{chaos} plays an
important role in dynamical systems theory. A chaotic system is
defined as a system {that} cannot be predicted
within a given numerical accuracy. At first sight this seems to be a
surprising concept, since differential equations of type
Eq.~(\ref{chaos_ode}), which do not contain any noise or randomness,
are perfectly deterministic. Once the starting point is known, the
resulting trajectory can be calculated for all times. Chaotic
behavior can arise nevertheless, due to an exponential sensitivity
to the initial conditions.

\begin{quotation}
{\it Deterministic Chaos.\enspace}
A deterministic dynamical system {that} shows
exponential sensibility of the time development on the initial
conditions is called chaotic.
\end{quotation}
This means that a very small change in the initial condition can
blow up even after a short time. When considering real-world
applications, when models need to be determined from measurements
containing inherent errors and limited accuracies, an exponential
sensitivity can result in unpredictability. A well known example is
the problem of long-term weather prediction.

\runinhead{The Logistic Map}

\enlargethispage{12pt}

One of the most cherished models in the field of deterministic chaos
is the logistic map of the interval $\ [0,1]\ $ onto
itself:\vspace*{3pt}
\begin{equation}
x_{n+1}\ =\ f(x_n)\ \equiv\ r\,x_n\,(1-x_n),
\qquad\quad x_n\in[0,1],
\qquad\quad r\in[0,4]~,
\label{eq_chaos_logistic_map}
\end{equation}
%

\begin{figure}[t]
\centering
\includegraphics{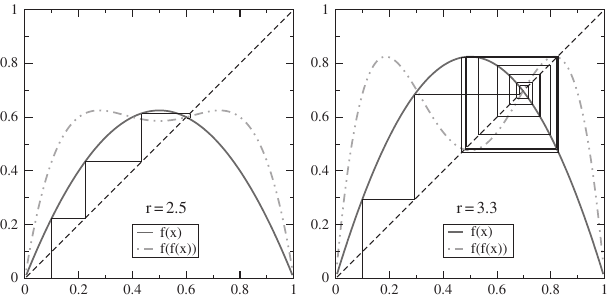}
\caption{Illustration of the logistic map $f(x)$ (\textit{thick
solid line}) and of the iterated logistic map $f(f(x))$
(\textit{thick dot-dashed line}) for $r=2.5$ (\textit{left}) and
$r=3.3$ (\textit{right}). Also shown is an iteration of $f(x)$,
starting from $x=0.1$ (\textit{thin solid line}) Note, that the
fixpoint $f(x)=x$ is stable/unstable for $r=2.5$ and $r=3.3$,
respectively. The orbit is attracted to a fixpoint of $f(f(x))$ for
$r=3.3$, corresponding to a cycle of period {2} for
$f(x)$ } \label{fig_chaos_logistic}
\end{figure}
\noindent where we have used the notation $x(t+n)=x_n$. The logistic map is
illustrated in Fig.~\ref{fig_chaos_logistic}. The logistic map shows,
despite its apparent simplicity, an infinite series of bifurcations
and a transition to chaos.

\runinhead{Biological Interpretation} We may consider
{}$\ x_n\in[0,1]\ $ {as standing}
for the population density of a reproducing species in the year $n$.
In this case the factor $\ r(1-x_n)\in[0,4]\ $ is the number of
{offspring} per year, which is limited in the
case of high population densities $x\to1$, when resources become
scarce. The classical example is that of a herd of reindeer on an
island.

Knowing the population density $\ x_n\ $ in a given year $\ n\ $ we
may predict via Eq.~(\ref{eq_chaos_logistic_map}) the population density
for all subsequent years exactly{;} the system is
deterministic. {Nevertheless the population shows} irregular behavior
for certain values of $r$, which one calls ``chaotic''.

\enlargethispage*{12pt}

\runinhead{Fixpoints of the Logistic Map} \index{fixpoint!logistic
map} We start considering the fixpoints of $f(x)$:
$$
x\ =\ rx(1-x)\qquad\Longleftrightarrow
\qquad x=0\quad {\rm or}\quad  1=r(1-x)~.
$$
The non-trivial fixpoint is then
\begin{equation}
1/r\ =\ 1-x,\quad\qquad x^{(1)}\ =\ 1-1/r,
\quad\qquad r_1<r,
\quad\qquad r_1=1~.
\label{chaos_x(1)}
\end{equation}
It occurs only for $\ r_1<r $, with $\ r_1=1$,
due to the restriction
$\ x^{(1)}\in [0,1]$.

\runinhead{Stability of the Fixpoint} \index{fixpoint!logistic
map!stability} We examine the stability of $x^{(1)}$ against
perturbations by linearization of Eq.~(\ref{eq_chaos_logistic_map}),
using
$$
y_n\ =\ x_n-x^{(1)},\quad\qquad
x_n\ =\ x^{(1)}+y_n,\quad\qquad
|y_n|\ \ll\ 1~.
$$
We obtain
\begin{eqnarray*}
x^{(1)}+y_{n+1}& =& r(x^{(1)}+y_{n})
(1-x^{(1)}-y_{n})  \\
& =& rx^{(1)} (1-x^{(1)}-y_{n}) +
ry_n(1-x^{(1)}-y_{n})~.
\end{eqnarray*}
Using the fixpoint condition $x^{(1)}=f(x^{(1)})$ and neglecting
terms $\sim y_n^2$, we obtain
$$
y_{n+1}\ =\ -rx^{(1)}y_n +r y_n(1-x^{(1)}) \ = \ r(1-2x^{(1)})\,y_n,
$$
and, using Eq.~(\ref{chaos_x(1)}), we find
\begin{equation}
y_{n+1}\ =\ r(1-2(1-1/r))\, y_n
       \ =\ (2-r)\,y_n \ =\ (2-r)^{n+1}\,y_0~.
\label{stabil_1}
\end{equation}
The perturbation $y_n$ increases/decreases in magnitude for
$|2-r|>1$ and $|2-r|<1$, respectively. Noting that $r\in[1,4]$, we
find
\begin{equation}
|2-r|<1\quad\quad \Longleftrightarrow \quad\quad
\fbox{\parbox{3cm}{$$
r_1<r<r_2
                   $$}}
\qquad\quad
\begin{array}{c}
r_1=1 \\ r_2=3
\end{array}
\label{chaos_range_1}
\end{equation}
for the region of stability of $x^{(1)}$.

\runinhead{Fixpoints of Period 2} \index{fixpoint!logistic
map!period {2}} For $\ r>3\ $ a fixpoint of period
{2} appears, {which} is a fixpoint of
the iterated function
$$
f(f(x))\ =\ rf(x)(1-f(x))\ =\ r^2x(1-x)(1-rx(1-x)).
$$
The fixpoint equation $\ x=f(f(x))\ $ leads to
the cubic equation
\begin{eqnarray}\nonumber
1& =& r^2(1-rx+rx^2) -r^2x(1-rx+rx^2), \\
0& =&
r^3x^3 - 2r^3x^2 + (r^3+r^2) x + 1-r^2~.
\label{kubisch}
\end{eqnarray}
In order to find the roots of Eq.~(\ref{kubisch}) we use the fact
that $\ x=x^{(1)}=1-1/r\ $ is a stationary point of both $f(x)$
and $f(f(x))$, see Fig.~\ref{fig_chaos_logistic}. We
divide (\ref{kubisch}) by the root $\ (x-x^{(1)})=(x-1+1/r) $:
$$
(r^3x^3 - 2r^3x^2 + (r^3+r^2) x + 1-r^2):(x-1+1/r)\ = \qquad
$$
$$
\qquad\qquad\qquad r^3x^2 - (r^3+r^2)x + (r^2+r)~.
$$
The two new fixpoints of $\ f(f(x))\ $
are therefore the roots of
$$
x^2 - \left(1+{1\over r}\right)x + \left({1\over r}+{1\over r^2}\right)
\ =\ 0~.
$$
We obtain
\begin{equation}
x_\pm^{(2)}\ =\
{1\over2}\left(1+{1\over r}\right) \pm
\sqrt{
{1\over4}\left(1+{1\over r}\right)^2
-\left({1\over r}+{1\over r^2}\right)
     }~.
\label{chaos_x(2)}
\end{equation}

\begin{figure}[t]
\centering
\includegraphics{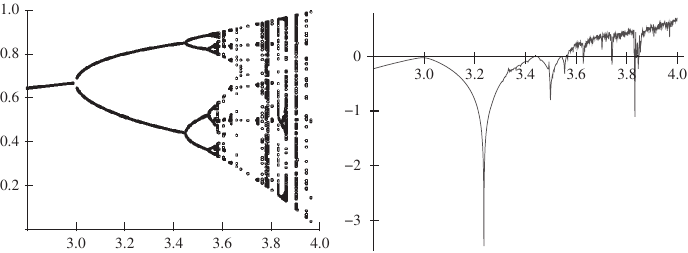}
\vspace*{0.0cm} \caption{The fixpoints of the (iterated) logistic
map (\textit{left}) and the corresponding maximal Lyapunov exponents
(\textit{right}), see Eq.~(\ref{chaos1_exponent_Lyapunov_maximal}),
both as a function of the parameter $r$. Positive Lyapunov exponents
$\lambda$ indicate chaotic behavior} \label{chaos_bifurcation}
\end{figure}

\runinhead{Bifurcation} \index{logistic
map!bifurcation}\index{bifurcation!logistic map} {We have two fixpoints for
$r>3$ and only one fixpoint for $r<3$.} What happens for
$r=3$?
%
$$
x_\pm^{(2)}(r=3)\ =\ {1\over2}{3+1\over 3}
\pm\sqrt{{1\over4}\left({3+1\over3}\right)^2
-\left({3+1\over9}\right)}
$$
$$
\quad=\ {2\over3}\ =\ 1-{1\over 3}\ =\ x^{(1)}(r=3)~.
$$
At $r=3$ the fixpoint splits into two, see
Fig.~\ref{chaos_bifurcation},  a typical {\em
bifurcation}.

\runinhead{More Bifurcations} We may now carry out a stability
analysis for $\ x_\pm^{(2)}$, just as we did for $x^{(1)}$. We find
a critical value $r_3>r_2$ such that
\begin{equation}
x_\pm^{(2)}(r)\ \ {\rm stable}\quad\quad \Longleftrightarrow
\quad\quad \fbox{\parbox{3cm}{$$ r_2<r<r_3.
                   $$}}
\label{chaos_range_2}
\end{equation}
Going further on one finds an $r_4$ such that there are four
fixpoints of period {4}, that is of $\ f(f(f(f(x))))
$, for $\ r_3<r<r_4$. In general there are critical values $\ r_n\ $
and $\ r_{n+1}\ $ such that there are
$$
2^{n-1}\ \ {\rm fixpoints}\ x^{(n)}\ {\rm of\ period }\ 2^{n-1}
\qquad \Longleftrightarrow \qquad \fbox{\parbox{4cm}{$$
r_{n}<r<r_{n+1}.
                   $$}}
$$
The logistic map therefore shows iterated bifurcations.
{This, however, is}  not yet chaotic
behavior.

\runinhead{Chaos in the Logistic Map} \index{logistic map!chaos}
\index{chaos!logistic map} The critical $r_n$ for doubling of the
period converge:
$$\lim_{n\to\infty} r_{n}\ \rightarrow\  r_\infty,
\qquad\quad r_\infty=3.5699456\ldots\
$$
There are consequently no stable fixpoints of $f(x)$
or of the iterated logistic map in the region
$$
r_\infty\ <\ r\ <\ 4~.\vspace*{3pt}
$$
In order to characterize the {sensitivity} of
Eq.~(\ref{eq_chaos_logistic_map}) with respect to the initial condition,
we consider two slightly different starting populations $\ x_1\ $
and $\ x_1^\prime$:
$$
x_1-\ x_1^\prime \ =\ y_1,
\qquad\quad |y_1|\ \ll\ 1~.
$$
The key question is then whether the difference in
populations
$$ y_m\ =\ x_m-\ x_m^\prime\
$$
is still small after $m$ iterations. Using $\ x_1^\prime=x_1-y_1$ we
find for $\ m=2\ $\vspace*{3pt}
\begin{eqnarray*}
y_2& =& x_2-x_2^\prime \ =\
rx_1(1-x_1) - rx_1^\prime(1-x_1^\prime) \\
& =& rx_1(1-x_1) - r(x_1-y_1)(1-(x_1-y_1))  \\
& =& rx_1(1-x_1) - rx_1(1-x_1+y_1) +ry_1(1-x_1+y_1) \\
& =& -rx_1y_1 +ry_1(1-x_1+y_1)~.\vspace*{3pt}
\end{eqnarray*}
Neglecting the term $\ \sim y_1^2\ $ we obtain
$$
y_2\ =\ -rx_1y_1 +ry_1(1-x_1)
   \ =\ r(1-2x_1)\,y_1 \
\equiv\ {d f(x)\over dx}\Big|_{x=x_1} y_1
\equiv\
\epsilon\, y_1~.
$$
For $\ |\epsilon|<1$ the map is stable, as two initially different
populations close in with time passing. For $\ |\epsilon|>1\ $ they
diverge; the map is ``chaotic''.

\runinhead{Lyapunov Exponents} \index{Lyapunov exponent}
\index{exponent!Lyapunov} We define via
\begin{equation}
|\epsilon|\ =\ {\rm e}^{\lambda},
\qquad\quad
\lambda \ =\ \log\left|d f(x)\over dx\right|
\label{chaos1_exponent_Lyapunov}
\end{equation}
the {\em Lyapunov exponent} $\ \lambda=\lambda(r)\ $:
$$
\lambda<0\ \Leftrightarrow\ \hbox{\rm stability},\qquad\qquad
\lambda>0\ \Leftrightarrow\ \hbox{\rm instability}~.
$$
For positive  Lyapunov exponents the time development is
exponentially sensitive to the initial conditions and shows chaotic
features. This is indeed observed in nature, e.g.\ for populations
of reindeer on isolated islands, as well as for the logistic map for
$r_\infty<r<4$, compare Fig.~\ref{chaos_bifurcation}.

\vspace*{3pt} \runinhead{Maximal Lyapunov Exponent}
\index{exponent!Lyapunov!maximal} 
\index{exponent!Lyapunov!global} 
The Lyapunov exponent, as defined
by Eq.~(\ref{chaos1_exponent_Lyapunov}) provides a description of
the short time behavior. For a corresponding characterization of the
long time dynamics one defines the\vspace*{3pt} \qut{maximal
Lyapunov exponent}
\begin{equation}
\lambda^{(max)}\ =\
\lim_{n\gg 1}\,{1\over n}\log\left|d f^{(n)}(x)\over dx\right|,
\qquad\quad
f^{(n)}(x)\ =\ f(f^{(n-1)}(x))~.
\label{chaos1_exponent_Lyapunov_maximal}
\end{equation}
Using Eq.~(\ref{chaos1_exponent_Lyapunov}) for the
short time evolution we can decompose $\lambda^{(max)}$
into an averaged sum of short time Lyapunov exponents. We
leave this as an exercise to the reader, $\lambda^{(max)}$
is also denoted the \qut{global Lyapunov exponent}.

One needs to select advisedly the number of iterations $n$ in 
Eq.~(\ref{chaos1_exponent_Lyapunov_maximal}). On one side
$n$ should be large enough such that short-term fluctuations
of the Lyapunov exponent are averaged out. The available 
phase space is however generically finite, for the logistic map
$y\in[0,1]$, and two initially close orbits cannot diverge
ad infinitum. One needs hence to avoid phase-space restrictions,
evaluating $\lambda^{(max)}$ for large but finite numbers
of iterations $n$.

\runinhead{Routes to Chaos} \index{chaos!routes to chaos} The
chaotic regime $\ r_\infty<r<4\ $ of the logistic map connects to
the regular regime $\ 0<r<r_\infty\ $ with increasing period
doubling. One speaks of a \qut{route to chaos via period-doubling}.
The study of chaotic systems is a wide field of research and a
series of routes leading from regular to chaotic behavior have been
found. Two important alternative routes to chaos are:
\begin{itemize}
\item[--] {{The Intermittency} route to chaos.}\\
      The trajectories are almost periodic{; they are} interdispersed with regimes
      of irregular behaviour. The occurrence of these irregular bursts
      increases until the system becomes irregular.
\item[--] {{Ruelle--Takens--Newhouse} route to chaos.}\\
      A strange attractor appears in a dissipative system after
      two (Hopf) bifurcations. As a function of an external parameter
      a fixpoint evolves into a limiting cycle (Hopf bifurcation), which
      then turns into a limiting torus{, which subsequently} turns into a strange
      attractor.
\end{itemize}

\section{Dissipation and Adaption}
\label{chaos_dissipation_adaption}

{In the preceding sections, we discussed
deterministic dynamical systems},  viz systems for which the time
evolution can be computed exactly, at least in principle, once the
initial conditions are known. We now turn to \qut{stochastic
systems}, i.e.\ dynamical systems {that} are
influenced by noise and fluctuations.

\subsection{Dissipative Systems and Strange Attractors}
\label{chaos_strange_attractors}
\index{dissipative system|textbf}
\index{dynamical system!dissipative|textbf}

\runinhead{Friction and Dissipation}
\index{friction}\index{dissipation} Friction plays an important role
in real-world systems. One speaks also of \qut{dissipation} since
energy is dissipated away by friction in physical systems.

The total energy{, however, is} conserved in
nature and friction{ then} just stands for a
transfer process of energy{;} when energy is
transferred from a system we observe, like a car on a motorway
with the engine {turned off}, to a system not
under observation, such as the surrounding air. In this case the
combined kinetic energy of the car and the thermal energy of the
air body is constant{;} the air heats up a little bit
while the car slows down.

\runinhead{The Mathematical Pendulum} \index{mathematical
pendulum}\index{oscillator!mathematical} As an example we consider
the damped \qut{mathematical pendulum}
\begin{equation}
\ddot{\phi}\,+\,\gamma\, \dot \phi\, +\,\omega_0^2\,\sin\phi\ =\ 0~,
\label{chaos_oscill_damped}
\end{equation}
which describes a pendulum with a rigid bar, capable of turning over
completely, with $\phi$ corresponding to the angle between the bar
and the vertical. The mathematical pendulum reduces to the damped
harmonic oscillator for small $\phi\approx\sin\phi$, which is
damped/critical/overdamped for $\gamma<2\omega_0$,
$\gamma=2\omega_0$ and $\gamma>2\omega_0$.

\enlargethispage*{12pt}

\runinhead{Normal Coordinates} \index{coordinates!normal}
Transforming the damped mathematical pendulum
Eq.~(\ref{chaos_oscill_damped}) to a set of coupled first-order
differential equations via $x=\phi$ and $\dot\phi=y$ one
gets\enlargethispage{-24pt}

\begin{equation}
\begin{array}{rcl}
\dot x &=& y \\
\dot y &=& -\gamma y-\omega_0^2\sin x~.
\end{array}
\label{chaos_oscill_damped_normal}
\end{equation}

The phase space is $ \veci x \in {\rm I\!R}^2,$ with $\veci
x=(x,y)$. For all $\gamma>0$ the motion approaches one of the
equivalent global fixpoints $(2\pi n,0)$ for $t\to\infty$ and
$n\in{\rm Z}$.

\begin{figure}[t]
\centering
\includegraphics{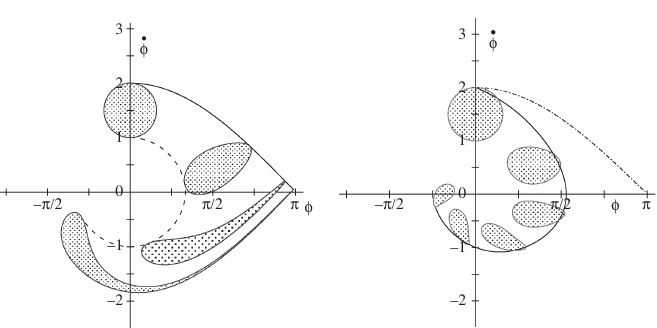}
\vspace*{0.0cm} \caption{Simulation of the mathematical pendulum
$\ddot\phi = -\sin(\phi)-\gamma\dot\phi$. The \textit{shaded
regions} illustrate the evolution of the phase space volume for
consecutive times, starting with $t=0$ ({\it top}). \textit{Left}:
Dissipationless case $\gamma=0$. The energy
$E=\dot\phi^2/2-\cos(\phi)$ is conserved as well as the phase space
volume (Liouville's theorem). The \textit{solid/dashed lines} are
the trajectories for $E=1$ and $E=-0.5$, respectively.
\textit{Right}: Case $\gamma=0.4$. Note the contraction of the phase
space volume} \label{chaos_fig_Liouville_Pendel}
\end{figure}

\runinhead{Phase Space Contraction} \index{phase space!contraction}
\index{dissipative system!phase space contraction} Near an attractor
the phase space contracts. We con-\break sider a three-dimensional
phase space $(x,y,z)$ for illustrational purposes. The
\hbox{quantity}
$$
\Delta V(t) \ =\
\Delta x(t) \Delta y(t) \Delta z(t) \ =\
(x(t)-x'(t))\, (y(t)-y'(t))\, (z(t)-z'(t))
$$
corresponds to a small volume of phase space. Its time evolution is
given by\vspace*{3pt}
$$
{\mathrm{d}\over \mathrm{d}t} \Delta V \ =\ \Delta \dot x \Delta y
\Delta z + \Delta x \Delta \dot y \Delta z + \Delta x \Delta y
\Delta \dot z~,\vspace*{3pt}
$$
or\vspace*{3pt}
\begin{equation}
{\Delta \dot V \over \Delta x\Delta y\Delta z} \ =\
{\Delta \dot x\over \Delta x} +
{\Delta \dot y\over \Delta y} +
{\Delta \dot z\over \Delta z}
\ =\ \vec\nabla\cdot\dot{\veci x}~.
\label{chaos_measure_phase_space}
\end{equation}

The time evolution of {the} phase space is illustrated in
Fig.~\ref{chaos_fig_Liouville_Pendel} for the case of the
mathematical pendulum. An initially simply connected volume of
{the} phase space {thus remains}
under the effect of time evolution, but it might undergo substantial
deformations.

\begin{quotation}
{\it Dissipative and Conserving Systems.\enspace}
\index{dissipative system!vs.\ conserving}\index{conserving system}
\index{dynamical system!dissipative}\index{dynamical
system!conserving} A dynamical system is dissipative, if its phase
space volume contracts continuously, $\vec\nabla\cdot\dot{\veci
x}<0$, for all $\veci x(t)$. The system is said to be conserving if
{the} phase space volume is a constant of motion, viz if
$\vec\nabla\cdot\dot{\veci x}\equiv 0$.
\end{quotation}
Mechanical systems, i.e.\ systems described by Hamiltonian
mechanics, are all conserving in {the} above sense. One
denotes this result from classical mechanics {as}
\qut{Liouville's theorem}.\index{theorem!Liouville}
\index{Liouville's theorem}

Mechanical systems in general have bounded and non-bounded orbits,
depending on the energy. The planets run through bounded orbits
around the sun, to give an example, but some comets leave the solar
system for {ever} on unbounded trajectories. One
can easily deduce from Liouville's theorem, i.e.\ from phase space
conservation, that bounded orbits are ergodic. {This comes arbitrarily close} to every
point in phase space having the identical conserved energy.

\vspace*{6pt}

\runinhead{Examples} Dissipative systems are a special class of
dynamical systems. Let us consider a few examples:
\begin{itemize}
\item[--]
For the damped mathematical pendulum
Eq.~(\ref{chaos_oscill_damped_normal}) we find
$$
{\partial \dot x\over\partial x} \ =\ 0, \qquad {\partial\dot
y\over\partial y} \ =\ {\partial[-\gamma y-\omega_0^2\sin
x]\over\partial y} \ =\ -\gamma \qquad \vec\nabla\cdot\dot{\veci x}
\ =\ -\gamma\ <\ 0~.
$$
The damped harmonic oscillator is consequently dissipative. It has a
single fixpoint $(0,0)$ and the basis of attraction is the full
phase space (modulo $2\pi$). Some examples of trajectories and phase
space evolution are illustrated in
Fig.~\ref{chaos_fig_Liouville_Pendel}.

\item[--]
For the non-linear rotator defined by Eq.~(\ref{chaos_dot_r}) we
have
\begin{equation}
{\partial \dot r\over\partial r} +
{\partial \dot \varphi\over\partial \varphi}  \ =\ \Gamma - 3r^2
\ =\ \left\{
\begin{array}{rcl}
<0 &\mbox{for}& \Gamma<0 \\
<0 &\mbox{for}& \Gamma>0\ \ \mbox{and} \ \ r>r_c/\sqrt3 \\
>0 &\mbox{for}& \Gamma>0\ \ \mbox{and} \ \ 0<r<r_c/\sqrt3
\end{array}
  \right.~ ,
\label{chaos1_grad_dot_x_NLR}
\end{equation}
\end{itemize}
where $r_c=\sqrt\Gamma$ is the radius of the limiting cycle when
$\Gamma>0$. The system might either dissipate or {take up energy, which is typical} behavior of
\qut{adaptive systems} as we will discuss further in
Sect.~\ref{chaos_adaptive_systems}. Note that the phase space
contracts both close to the fixpoint, for $\Gamma<0$, and close to
the limiting cycle, for $\Gamma>0$.

\vspace*{6pt}

\runinhead{Phase Space Contraction and Coordinate Systems}
\index{phase space!contraction} The time development of a small
phase space volume, Eq.~(\ref{chaos_measure_phase_space}), depends on
the coordinate system chosen to represent the variables. As an
example we reconsider the non-linear rotator defined by
Eq.~(\ref{chaos_dot_r}) in terms of the Cartesian coordinates
$x=r\cos\varphi$ and $y=r\sin\varphi$.

The respective infinitesimal phase space volumes are
related via the Jacobian,
$$
\mathrm{d}x\,\mathrm{d}y \ =\ r\,\mathrm{d}r\,\mathrm{d}\varphi~,
$$
and we find
$$
{\dot{\Delta V} \over \Delta V} \ =\
{\dot r\Delta r\Delta\varphi +
 r\dot\Delta r\Delta\varphi +
 r\Delta r\dot\Delta\varphi
\over r\Delta r\Delta\varphi} \ =\
{\dot r\over r} +
{\partial \dot r\over\partial r} +
{\partial \dot \varphi\over\partial \varphi}
\ =\ 2\Gamma - 4r^2 ~,
$$
compare Eqs.~(\ref{chaos_dot_r}) and (\ref{chaos1_grad_dot_x_NLR}).
The amount and even the sign of {the phase
space} contraction can depend on the choice of the coordinate
system.

\runinhead{The Lorenz Model}
\index{model!{Lorenz}}
\index{Lorenzmodel@Lorenz model} A
rather natural question is the possible existence of attractors with
less regular behaviors, i.e. {which are} different from
stable fixpoints, periodic or quasi-periodic motion. For this
question we examine the Lorenz model
\begin{eqnarray}
{\mathrm{d}x \over \mathrm{d}t} & = & -\sigma (x - y), \nonumber\\
{\mathrm{d}y \over \mathrm{d}t} & = & -xz+rx-y, \label{chaos_lorenz}\\
{\mathrm{d}z \over \mathrm{d}t} & = & xy-bz~. \nonumber
\end{eqnarray}
The classical values are
$\sigma=10$ and $b=8/3$, with $r$ being the
control variable.

\runinhead{Fixpoints of the Lorenz Model} \index{fixpoint!Lorenz
model} A trivial fixpoint is $(0,0,0)$. The non-trivial fixpoints
are
$$
\begin{array}{rclrcl}
0 & = & -\sigma (x - y),        &             x& = & y , \\
0 & = & -xz+rx-y,  &\quad \qquad z& =& r-1, \\
0 & = & xy-bz ,    &        x^2=y^2&=&b\,(r-1)~.
\end{array}
$$
It is easy to see by linear analysis that the fixpoint $(0,0,0)$ is
stable for $r < 1$. For $r > 1$ it becomes unstable and two new
fixpoints appear:
\begin{equation}
C_{+,-}\ =\ \left(\pm \sqrt{b(r-1)}, \pm \sqrt{b(r-1)}, r-1\right)~.
\label{chaos_fixedpts}
\end{equation}
These are stable for $r < r_c = 24.74$ ($\sigma=10$ and $b=8/3$).
For $r>r_c$ the behavior becomes more complicated and generally
non-periodic.

\runinhead{Strange Attractors} \index{strange attractor}
\index{attractor!strange} One can show, that the Lorenz model has
positive Lyapunov exponents for $r>r_c$. It is chaotic with
sensitive dependence on the initial conditions. The Lorenz model is
at the same time dissipative, since
\begin{equation}
{\partial \dot x \over \partial x} +
{\partial \dot y \over \partial y} +
{\partial \dot z \over \partial z} \ =\ -(\sigma+1+b) \ < \ 0,
\qquad\quad \sigma>0,\ b>0 ~.
\label{chaos_divlorenz}
\end{equation}
The attractor of the Lorenz system therefore cannot be a smooth
surface. Close to the attractor the phase space contracts. At the
same time two nearby orbits are repelled due to the positive
Lyapunov exponents. One finds a self-similar structure for the
Lorenz attractor with a fractal dimension $2.06\;\pm\;0.01$. Such a
structure is called a {\em strange attractor}.

The Lorenz model has an important historical
relevance in the development of chaos theory and is now considered a
paradigmatic example of a chaotic system.

\begin{figure}[!t]
\centering
\includegraphics{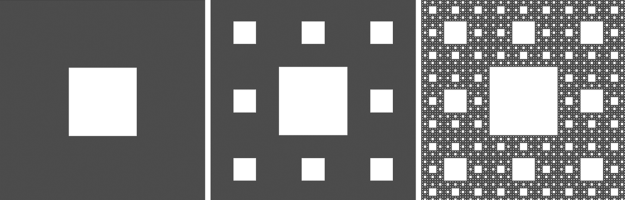}
\caption{The Sierpinski carpet and its iterative construction}
\label{chaos_sierpinski}
\end{figure}

\runinhead{Fractals}\index{fractal} Self-similar structures are
called fractals. Fractals can be defined by recurrent geometric
rules{;} examples are the Sierpinski triangle and carpet
(see Fig.~\ref{chaos_sierpinski}) and the Cantor set. Strange
attractors are normally {\em multifractals}, i.e.\ fractals with
non-uniform self-similarity.

\vspace*{6pt}

\runinhead{{The Hausdorff} Dimension}
\index{Hausdorff dimension}\index{dimension!Hausdorff} An important
notion in the theory of fractals is the \qut{Hausdorff dimension}.
We consider a geometric structure defined by a set of points in $d$
dimensions and the number $N(l)$ of $d$-dimensional spheres of
diameter $l$ needed to cover this set. If $N(l)$ scales like
\begin{equation}
N(l)\ \propto\ l^{-D_H},
\qquad\quad\mbox{for}\qquad
l\ \to\ 0~,
\label{chaos_Hausdorff}
\end{equation}
then $D_H$ is called the Hausdorff dimension of the set.
Alternatively we can rewrite
{Eq.~}(\ref{chaos_Hausdorff}) as
\begin{equation}
{N(l)\over N(l')} \ =\  \left({l\over l'}\right)^{-D_H},
\qquad\quad
D_H\ =\ -{ \log[N(l)/N(l')] \over \log[l/l'] }~,
\label{chaos_Hausdorff_fractal}
\end{equation}
which is useful for self-similar structures (fractals).

The {$d$-dimensional} spheres necessary
to cover a given geometrical structure will generally overlap. The
overlap does not affect the value of the fractal dimension as long
as the degree of overlap does not change qualitatively with
decreasing\break diameter~$l$.

\enlargethispage*{12pt}
 \runinhead{{The Hausdorff} Dimension of
the Sierpinski Carpet} \index{Hausdorff dimension!of Sierpinski
carpet}\index{Sierpinski carpet} For the Sierpinski carpet we
increase the number of points $N(l)$ by a factor of 8, compare
Fig.~\ref{chaos_fig_seesawWater}, when we decrease the length scale
$l$ by a factor of 3 (see Fig.~\ref{chaos_sierpinski}):
$$
D_H\ \to\ -{ \log[8/1] \over \log[1/3] }
\ =\ {\log 8\over \log 3} \ \approx\ 1.8928.
$$

\enlargethispage*{18pt}
\subsection{Adaptive Systems}
\label{chaos_adaptive_systems}
\index{dynamical system!adaptive}
\index{adaptive system}

\runinhead{Adaptive Systems} \index{adaptive
{systems}} A general complex system is neither
fully conserving nor fully dissipative. Adaptive systems will have
periods where they take up energy and \hbox{periods} where they give
energy back to the environment. An example is the non-linear rotator
of Eq.~(\ref{chaos_dot_r}), see also
Eq.~(\ref{chaos1_grad_dot_x_NLR}).

In general one affiliates with the term \qut{adaptive system} the
notion of complexity and adaption. Strictly speaking any dynamical
system is adaptive if $\ \nabla\cdot\dot \veci x\ $ may take both
positive and negative values. {In practice, however, it is usual to} reserve the term
adaptive system to dynamical systems showing a certain complexity,
such as emerging behavior.

\begin{figure}[t]
\centering
\includegraphics{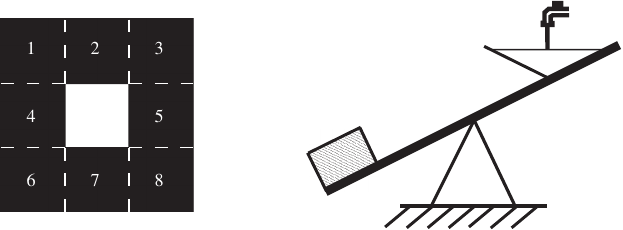}
\caption{\textit{Left}: The fundamental unit of the Sierpinski
carpet, compare Fig.~\ref{chaos_sierpinski}, contains eight squares
{that can be covered by discs of an appropriate diameter}.
 \textit{Right}: The seesaw with a water container at one end{;} an
example of an oscillator {that
takes up/disperses} takes up/disperses energy periodically}
\label{chaos_fig_seesawWater}\vspace{12pt}
\end{figure}

\runinhead{The Van der Pol Oscillator} \index{van der Pol
oscillator} \index{oscillator!van der Pol} Circuits or mechanisms
built for the purpose of controlling an engine or machine are
intrinsically adaptive. An example is the van der Pol oscillator,
\begin{equation}
\ddot x\,-\,\epsilon(1-x^2)\dot x\,+\,x \ =\ 0,
\qquad\qquad
\begin{array}{rcl}
\dot x &=& y\\
\dot y &=& \epsilon(1-x^2)y -x
\end{array}
\label{chaos_van_der_pol}
\end{equation}
where $\epsilon>0$ and where we have used {the} phase
space variables $\veci x=(x,y)$. We evaluate the time evolution
$\vec\nabla\cdot\dot{\veci x}$ of the phasespace volume,\vspace{3pt}
$$
\vec\nabla\cdot\dot{\veci x}\ =\ +\epsilon\,(1-x^2)~.\vspace{3pt}
$$
The oscillator takes up/dissipates energy for $x^2<1$ and $x^2>1$,
respectively. A simple mechanical example for a system with similar
properties is illustrated in\break Fig.~\ref{chaos_fig_seesawWater}

\runinhead{Secular Perturbation Theory} \index{van der Pol
oscillator!secular perturbation theory} \index{perturbation
theory!secular} We consider a perturbation expansion in $\epsilon$.
The solution of Eq.~(\ref{chaos_van_der_pol}) is\vspace{-3pt}
\begin{equation}
x_0(t)\ =\ a\,e^{i(\omega_0t+\phi)}\,+\, c.c.,
\qquad\quad \omega_0\ =\ 1~,
\label{chaos_harmonic_sol}\vspace{3pt}
\end{equation}
for $\epsilon=0$. We note that the   amplitude $a$ and phase $\phi$
are arbitrary in Eq.~(\ref{chaos_harmonic_sol}). The perturbation
$\epsilon(1-x^2)\dot x$ might change, in principle, also the given
frequency $\omega_0=1$ by an amount $\propto\epsilon$. In order to
account for this \qut{secular perturbation} we make the ansatz\vspace{3pt}
\begin{equation}
x(t)\ =\ \left[ A(T)e^{it}+A^*(T)e^{-it}\right] \,+\epsilon x_1\,
+\, \cdots, \qquad\quad A(T)\ =\ A(\epsilon t)~,
\label{chaos_van_der_pol_ansatz}\vspace{3pt}
\end{equation}
which differs from the usual expansion $x(t)\to x_0(t)+\epsilon
x'(t)+\cdots$ of the full solution $x(t)$ of a dynamical system with
respect to a small parameter $\epsilon$.

\runinhead{Expansion} From Eq.~(\ref{chaos_van_der_pol_ansatz}) we
find to the order $O(\epsilon^1)$\vspace{3pt}
\begin{eqnarray*}
x^2& \approx& A^2e^{2it} \,+\, 2|A|^2 \,+\,(A^*)^2e^{-2it}
\,+\,2\epsilon x_1\left[ Ae^{it}+Ae^{-it}\right] \\
\epsilon(1-x^2) & \approx &
\epsilon(1-2|A|^2)-\epsilon\left[A^2e^{2it}+(A^*)^2e^{-2it}\right]~,
\end{eqnarray*}
\begin{eqnarray*}
\dot x& \approx& \left[\left(\epsilon A_T+iA\right)e^{it}
+c.c.\right]\,+\,\epsilon\,\dot x_1,
\qquad\quad A_T\ = {\partial A(T)\over\partial T} \\
\epsilon(1-x^2)\dot x & = &
\epsilon(1-2|A|^2)\left[iAe^{it}-iA^*e^{-it}\right] \\
&-&\epsilon\left[A^2e^{2it}+(A^*)^2e^{-2it}\right]
 \left[iAe^{it}-iA^*e^{-it}\right]
\end{eqnarray*}
and
\begin{eqnarray*}
\ddot x& =& \left[\left(\epsilon^2 A_{TT}+2i\epsilon A_T
-A\right)e^{it}+c.c.\right] \,+\,\epsilon\,\ddot x_1 \\ &\approx&
\left[ \left(2i\epsilon A_T -A\right)e^{it}+c.c.\right]
\,+\,\epsilon\ddot x_1 ~.
\end{eqnarray*}
Substituting these expressions into Eq.~(\ref{chaos_van_der_pol}) we
obtain in {the} order $O(\epsilon^1)$
\begin{equation}
\ddot x_1\,+\, x_1 \ =\
\left(-2iA_T+iA-i|A|^2A\right)e^{it}\, -\, iA^3e^{3it}\,+\,c.c.~.
\label{chaos_perturbation_epsilon}
\end{equation}
\runinhead{The Solvability Condition}
Equation (\ref{chaos_perturbation_epsilon}) is identical
to a driven harmonic oscillator, which will be discussed
in Chap.~\ref{chap_synchro1} in more detail. The time\break
dependencies
$$
\sim\,e^{it}\qquad\quad\mbox{and}
\qquad\quad \sim\,e^{3it}
$$
of the two terms on the right-hand side of
Eq.~(\ref{chaos_perturbation_epsilon}) are proportional to the
unperturbed frequency $\omega_0=1$ and to $3\omega_0$, respectively.

The term $\sim e^{it}$ is therefore exactly at resonance and would
induce a diverging response $x_1\to\infty$, in contradiction to the
perturbative assumption made by ansatz
(\ref{chaos_van_der_pol_ansatz}). Its prefactor must therefore
vanish:\vspace*{3pt}
\begin{equation}
A_T\ =\ {\partial A\over \partial T}\ =\ {1\over
2}\left(1-|A|^2\right)A, \qquad\quad {\partial A\over \partial t}\
=\ {\epsilon\over 2}\left(1-|A|^2\right)A~,
\label{chaos_solvability_condition}
\end{equation}

\noindent where we have used $T=\epsilon t$. The solubility
condition Eq.~(\ref{chaos_solvability_condition}) can be written as
$$
\dot a\, e^{i\phi}\,+\,i\dot\phi\, a\,e^{i\phi} \ =\
{\epsilon\over 2}\left(1-a^2\right) a\,e^{i\phi}
$$
in phase-magnitude representation $A(t)=a(t)e^{i\phi(t)}$,
or\vspace*{3pt}
\begin{equation}
\begin{array}{rcl}
\dot a &=& \epsilon\left(1-a^2\right) a/2, \\
\dot\phi &\sim & O(\epsilon^2)~.
\end{array}
\label{chaos_solvability_condition_phase_magnitude}\vspace*{3pt}
\end{equation}
The system takes up energy for $a<1$ and the amplitude $a$ increases
until the saturation limit $a\to1$, the conserving point. For $a>1$
the system dissipates energy to the environment and the amplitude
$a$ decreases, approaching unity for $t\to\infty$, just as we
discussed in connection {with} Eq.~(\ref{chaos_dot_r}).

\begin{figure}[t]
\centering
\includegraphics{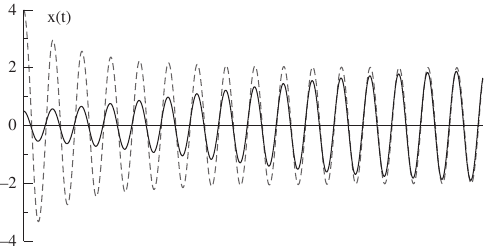}
\caption{The solution of the van der Pol oscillator,
         Eq.~(\ref{chaos_van_der_pol}),
         for small $\epsilon$ and two different initial conditions.
         Note the self-generated amplitude stabilization
        }
\label{chaos1_fig_vanDerPol_x_t}\vspace*{9pt}
\end{figure}

The solution $x(t)\approx 2\,a\cos(t)$, compare\enlargethispage*{-1pt}
Eqs.~(\ref{chaos_van_der_pol_ansatz}) and
(\ref{chaos_solvability_condition_phase_magnitude}), of the van der
Pol equations {therefore
constitutes} an amplitude-regulated oscillation, as illustrated in
Fig.~\ref{chaos1_fig_vanDerPol_x_t}. This behavior was the technical
reason for {historical development
of the} control systems {that} are described by the
van der Pol equation (\ref{chaos_van_der_pol}).

\runinhead{Li\'enard Variables}
\index{variable!Li\'enard}\index{Li\'enard variables}\index{van der
Pol oscillator!Li\'enard variables} For large $\epsilon$ it is
convenient to define, compare Eq.~(\ref{chaos_van_der_pol}),
with\vspace*{3pt}
\begin{equation}
\epsilon\,{\mathrm{d}\over \mathrm{d}t} Y(t) \ =\ \ddot
x(t)\,-\,\epsilon\left(1-x^2(t)\right)\dot x(t) \ =\ -x(t)
\label{chaos_Lienard_var}\vspace*{3pt}
\end{equation}
or\vspace*{3pt}
$$
\epsilon \dot Y\ =\ \ddot X\,-\,\epsilon\left(1-X^2\right)\dot X,
\qquad\quad X(t)\ =\ x(t),\vspace*{3pt}
$$
the Li\'enard variables $X(t)$ and $Y(t)$. Integration of $\dot Y$
with respect to $t$ yields\vspace*{3pt}
$$
\epsilon Y \ =\  \dot X\,-\,\epsilon\left(X-{X^3\over
3}\right)~,\vspace*{3pt}
$$
where we have set the integration constant to zero. We obtain,
together with Eq.~(\ref{chaos_Lienard_var}),
\begin{equation}
\begin{array}{rcl}
\dot X &=& c\,\Big(Y-f(X)\Big)\\
\dot Y &=& -X/c
\end{array}
\qquad\quad
f(X)\ =\ X^3/3-X~,
\label{chaos_Lineard_van_der_Pol}
\end{equation}
where we have set $c\equiv\epsilon$, as we are now interested in the
case $c\gg1$.

\enlargethispage{-6pt}

\begin{figure}[t]
\centering
\includegraphics{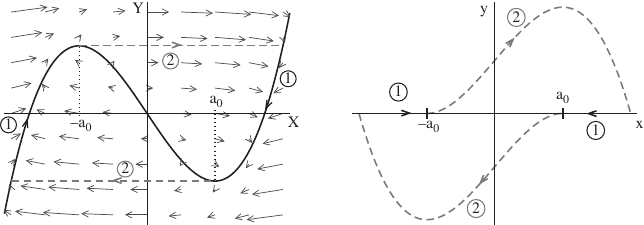}
\caption{Van der Pol oscillator for a large driving
         $c\equiv\epsilon$.
\textit{Left}: The relaxation oscillations
      with respect to the Li\'enard variables
      Eq.~(\ref{chaos_Lineard_van_der_Pol}).
      The \textit{arrows} indicate the flow $(\dot X,\dot Y)$, for
      $c=3$, see Eq.\ (\ref{chaos_Lineard_van_der_Pol}).
      Also shown is the $\dot X=0$
      isocline $Y= -X+X^3/3$ (\textit{solid line}) and the limiting cycle,
      which includes the \textit{dashed line with an arrow} and part
      of the isocline.
\textit{Right}: The limiting cycle in terms of the original
variables
       $(x,y)=(x,\dot x)=(x,v)$. Note that $X(t)=x(t)$
} \label{chaos_fig_relax_oscil}
\end{figure}

\runinhead{Relaxation Oscillations} \index{relaxation oscillator!van
der Pol} We discuss the solution of the van der Pol oscillator
Eq.~(\ref{chaos_Lineard_van_der_Pol}) for a large driving $c$
graphically, compare Fig.~\ref{chaos_fig_relax_oscil}, by
considering the flow $(\dot X,\dot Y)$ in phase space $(X,Y)$.  For
$c\gg1$ there is a separation of time\break scales,\index{time scale
separation!van der Pol oscillator}
$$
(\dot X,\dot Y) \ \sim\ (c,1/c),
\qquad\quad
\dot X \gg \dot Y~,
$$

\noindent which leads to the following dynamical behavior:
\begin{itemize}
\item[--] Starting at a general $(X(t_0),Y(t_0))$ the orbit
      develops very fast $\sim c$ and nearly {horizontally}
      until it hits the \qut{isocline}\footnote{The term isocline stands for \qut{equal slope}
                 in ancient Greek.}\index{isocline!van der Pol}
\begin{equation}
      \dot X\ =\ 0, \qquad\quad Y\ =\ f(X)\ =\ -X+X^3/3~.
\label{chaos_X_isocline}
\end{equation}
\item[--] Once the orbit is close to the $\dot X=0$
  isocline $Y= -X+X^3/3$ the motion slows down
  and it develops slowly, with a velocity $\sim 1/c$
  close-to (but not exactly on) the isocline
  {(Eq.~(\ref{chaos_X_isocline}))}.
\item[--] Once the slow motion reaches one of the two local
          extrema $X=\pm a_0=\pm 1$ of the isocline,
          it cannot follow the isocline any more and
          makes a rapid transition towards the other branch of
          the $\dot X=0$ isocline, with $Y\approx$ const.
          Note, that \hbox{trajectories} may cross the isocline vertically,
          e.g.\ right at the extrema $\dot Y|_{X=\pm1}=\mp1/c$
          is small but finite.
\end{itemize}

\looseness2 The orbit therefore relaxes rapidly
towards a limiting oscillatory trajectory, illustrated in
Fig.~\ref{chaos_fig_relax_oscil}, with the time needed to perform a
whole oscillation depending on the relaxation constant
$c$; therefore the term \qut{relaxation oscillation}.
Relaxation oscillators represent an important class of
cyclic attractors, allowing to model systems going through
several distinct and well characterized phases during
the course of one cycle. We will discuss relaxation oscillators
further in Chap.~\ref{chap_synchro1}.

\vspace*{6pt}
\section{Diffusion and Transport}
\label{chaos_diffusion}
\index{diffusion|textbf}
\index{transport|textbf}

\runinhead{Deterministic vs.\ Stochastic Time Evolution}
\index{dynamical system!deterministic}\index{dynamical
system!stochastic} So far we have discussed some concepts and
examples of deterministic dynamical systems, governed by sets of
coupled differential equations without noise or randomness.
{At the other extreme} are diffusion
processes for which the random process dominates the dynamics.

Dissemination of information through social networks is one of many
examples where diffusion processes plays a paramount role. The
simplest model of diffusion is the Brownian motion,
{which is} the erratic movement of grains suspended in
liquid observed by the botanist Robert Brown as early as
{}1827. Brownian motion became the prototypical example
of a stochastic process after the seminal works of Einstein and
Langevin at the beginning of the 20th century.

\enlargethispage{-12pt}


\vspace*{6pt}
\subsection{Random Walks, Diffusion and L\'evy Flights}
\label{chaos_diffusion_random_walk}
\index{random!walk|textbf}
\index{walk!random|textbf}

\runinhead{One-Dimensional Diffusion} \index{diffusion!one
dimensional} We consider the random walk of a particle along a line,
with {the} equal probability $1/2$ to move left/right at every time
step. The \nobreak probability\vspace*{3pt}
$$
p_t(x), \qquad\quad x=0,\pm 1,\pm 2, \ldots, \qquad\quad
t=0,1,2,\ldots\vspace*{3pt}
$$
to find the particle at time $t$ at position $x$ obeys the master
equation\vspace*{3pt}
\begin{equation}
p_{t+1}(x)\ =\ {1 \over 2}\, p_{t}(x-1)\, +\, {1 \over 2}\,
p_{t}(x+1)~. \label{chaos_master_walk}\vspace*{3pt}
\end{equation}
In order to obtain the limit of continuous time and space, we
introduce explicitly the steps $\Delta x$ and $\Delta t$ in space
and time, and write
\begin{equation}
{p_{t+\Delta t}(x)-p_{t}(x) \over \Delta t}\ = \
{(\Delta x)^2 \over 2 \Delta t}\,
{p_{t}(x+\Delta x)+p_{t}(x-\Delta x)-2 p_{t}(x) \over
(\Delta x)^2} \, .
\label{chaos_master_new}
\end{equation}
Now, taking the limit $\Delta x, \Delta t \to 0$ in such a way
that $(\Delta x)^2/(2\Delta t)$ remains finite,
we obtain the diffusion equation
\index{diffusion!equation}
\index{equation!diffusion}
\begin{equation}
{\partial p(x,t) \over \partial t}\ =\
D\, {\partial^2 p(x,t) \over \partial x^2} \quad\qquad
D= {(\Delta x)^2\over 2\Delta t}~.
\label{chaos_diffusion_eq}
\end{equation}
\runinhead{Solution of the Diffusion Equation} The solution to
Eq.~(\ref{chaos_diffusion_eq}) is readily obtained~as\footnote{Note:
$\int e^{-x^2/a}\D x=\sqrt{a\pi}$ and
$\lim_{a\to0} \exp(-x^2/a)/\sqrt{a\pi}=\delta(x)$.}
\begin{equation}
p(x,t)\ =\ {1 \over \sqrt{4 \pi D t}}
\exp \left(-{x^2 \over 4 D t}\right),
\qquad\quad \int_{-\infty}^\infty \mathrm{d}x\, \rho(x,t)
\ =\ 1~, \label{chaos_diffusion_sol}
\end{equation}
for the initial condition $\rho(x,t=0)=\delta(x)$.
From Eq.~(\ref{chaos_diffusion_sol}) one concludes that the variance
of the displacement follows diffusive behavior, i.e.
\begin{equation}
\langle x^2(t) \rangle\ =\ 2D\,t~,
\qquad\quad
\bar x \ =\ \sqrt{\langle x^2(t) \rangle}\ =\ \sqrt{2D\,t}~.
\label{chaos_diffusion_time}
\end{equation}
\vskip2pt

\begin{figure}[t]
\centerline{
\includegraphics{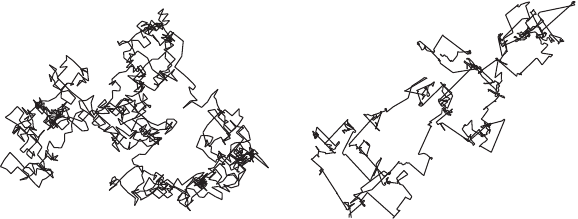}}
\caption{Examples of random walkers with scale-free distributions
         $\sim |\Delta x|^{1+\beta}$ for the real-space jumps, see
         Eq.~(\ref{chaos1_Levy_distributions}).
\textit{Left}: $\beta=3$, which falls into the universality class
      of standard Brownian motion.
\textit{Right}: $\beta = 0.5$, a typical L\'evy flight. Note the
       occurrence of longer-ranged jumps in conjunction
       with local walking
        }\vspace*{-8pt}
\label{chaos_random_walk_fig}
\end{figure}

\index{transport!diffusive} Diffusive transport is characterized by
transport sublinear in time in contrast to ballistic transport with
\index{transport!ballistic} $x=vt$, as illustrated in
Fig.~\ref{chaos_random_walk_fig}.

\enlargethispage*{-18pt}

\begin{figure}[t]
\centering
\includegraphics{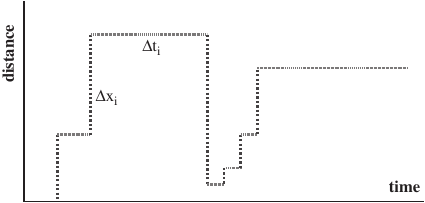}
\caption{A random walker with distributed waiting times $\Delta t_i$
and jumps $\Delta x_i$ may become a generalized L\'evy flight }
\label{chaos_fig_Levey_flights}\vspace*{-6pt}
\end{figure}

\runinhead{L\'evy Flights} \index{L\'evy flight} We can generalize
the concept of a random walker, which is at the basis of ordinary
diffusion, and consider a random walk with distributions $p(\Delta
t)$ and $p(\Delta x)$ for waiting times $\Delta t_i$ and jumps
$\Delta x_i$, at every step $i=1,\ 2,\ldots $ of the walk, as
illustrated in
Fig.~\ref{chaos_fig_Levey_flights}. One may assume scale-free
distributions
\begin{equation}
p(\Delta t) \ \sim\ {1\over (\Delta t)^{1+\alpha}},
\qquad\quad
p(\Delta x) \ \sim\ {1\over (\Delta x)^{1+\beta}},
\qquad\quad \alpha,\beta > 0~.
\label{chaos1_Levy_distributions}
\end{equation}
%

\setcounter{table}{0}
\begin{table}[!b]
\vspace*{-8pt}\centering
 \caption{The four regimes of a generalized
walker with distribution functions,
Eq.~(\ref{chaos1_Levy_distributions}), characterized by scalings
$\propto (\Delta t)^{-1-\alpha}$ and $\propto (\Delta x)^{-1-\beta}$
for the waiting times $\Delta t$ and jumps $\Delta x$, as depicted
in Fig.~\ref{chaos_fig_Levey_flights}
         \label{chaos1_table_Levy}}
\begin{tabular*}{19pc}{@{}l@{\quad}l@{\quad}l@{\quad}l@{\quad}@{}}
\hline\noalign{\smallskip}
\phantom{\large |$^|$}\hspace*{-3.5ex} $\alpha>1\ $ &
\  $\beta>2$\  &\  $\bar x\sim \sqrt{t}$\  &\  Ordinary diffusion \\
\phantom{\large |$^|$}\hspace*{-3.5ex} $\alpha>1\ $ &
\  $0<\beta<2$\  &\  $\bar x\sim t^{1/\beta}$\  &\ L\'evy flights \\
\phantom{\large |$^|$}\hspace*{-3.5ex} $0<\alpha<1\ $ &
\  $\beta>2$\  &\  $\bar x\sim t^{\alpha/2}$\  &\ Subdiffusion \\
\phantom{\large |$^|$}\hspace*{-3.5ex} $0<\alpha<1\ $ &
\  $0<\beta<2$\  &\  $\bar x\sim t^{\alpha/\beta}$\  &\ Ambivalent processes \\
\noalign{\smallskip}\hline\noalign{\smallskip}
\end{tabular*}
\index{diffusion!ordinary}\index{diffusion!subdiffusion}
{\vspace*{-1pc}}
\end{table}

If $\alpha>1$ (finite mean waiting time) and $\beta>2$ (finite
variance), nothing special happens. In this case the central
limiting theorem for well behaved distribution \nobreak functions is
valid for the {spatial} component and one obtains standard Brownian
diffusion. Relaxing {the} above conditions one finds four regimes:
normal Brownian diffusion, \qut{L\'evy flights}, fractional Brownian
motion, also denoted \qut{subdiffusion} and generalized L\'evy
flights termed \qut{ambivalent processes}. Their respective scaling
laws are listed in Table \ref{chaos1_table_Levy} and two examples
are shown in Fig.~\ref{chaos_random_walk_fig}.

L\'evy flights occur in a wide range of processes,
such as in the flight patterns of wandering albatrosses or in human
travel habits, which seem to be characterized by a generalized
L\'evy flight with $\alpha,\beta\approx 0.6$.

\enlargethispage{18pt}

\runinhead{Diffusion Within Networks}
\index{diffusion!within networks} \index{network!diffusion}
Diffusion occurs in many circumstances. We consider here the
case of diffusion within a network, such as the diffusion
of information within social networks. This is an interesting
issue as the control of information is important for
achieving social influence and prestige. We will however
neglect in the following the creation of new information,
which is clearly relevant for real-life applications.

Consider a network of $i=1,\ldots,N$ vertices connected by edges
with weight $W_{ij}$, corresponding to the elements of the weighted
adjacency matrix. We denote by
$$
\rho_i(t), \qquad\quad \sum_{i=1}^N\,\rho_i(t) \ =\ 1
$$
the density of information present at time $t$ and vertex $i$.

\runinhead{Flow of Information} \index{flow!of information} The
information flow can then be described by the master equation
\begin{equation}
\rho_i(t+\Delta t)\ =\ \rho_i(t)\, +\, J_i^{(+)}(t)\Delta t
                         \, -\, J_i^{(-)}(t)\Delta t~,
\label{chaos_info_flow}
\end{equation}
where $J_i^{(\pm)}(t)$ denotes the density of information
entering~($+$) and leaving~($-$) vertex $i$ per time interval
$\Delta t$, given by\vspace*{4pt}
$$
J_i^{(+)}(t)\ =\ \sum_j {W_{ij}\over \sum_k W_{kj}}\,\rho_j(t),
\qquad\quad
J_i^{(-)}(t)\ =\ \sum_j {W_{ji}\over \sum_k W_{ki}}\,\rho_i(t)
            \ =\ \rho_i(t)~.\vspace*{4pt}
$$

\noindent Introducing the time step $\Delta t=1$ and the expressions
for $J_i^{(\pm)}(t)$ into Eq.~(\ref{chaos_info_flow}) we find\vspace*{4pt}
\begin{equation}
{\rho_i(t+\Delta t)-\rho_i(t)\over \Delta t} \ =\
{\partial\over \partial t} \rho_i(t) \ =\
\sum_j T_{ij}\, \rho_j(t)\, -\, \rho_i(t)~,
\label{chaos_master-eq-info}\vspace*{4pt}
\end{equation}
where we have performed the limit $\Delta t\to 0$ and
defined
$$
T_{ij} \ =\  \frac{W_{ij}}{\sum_k W_{kj}}~.
$$
This equation can easily be cast into the following matrix form:
\begin{equation}
{\partial\over \partial t}\vec{\rho}(t) \ =\
{\bf D}\,\vec{\rho}(t),
\qquad\quad
D_{ij}\ =\  T_{ij} - \delta_{ij}~,
\label{chaos_eq_ME5}
\end{equation}
where $\vec\rho=(\rho_1,\ldots ,\rho_N)$. It resembles the diffusion
equation (\ref{chaos_master_new}), so we may denote ${\bf
D}=\left(D_{ij}\right)$ {as} the diffusion matrix (or
operator). Physically, Eq.~(\ref{chaos_master-eq-info}) means that
${\bf T}=\left(T_{ij}\right)$ transfers (propagates) the energy
density $\vec{\rho}(t)$ one step forward in time. Due to this
property, ${\bf T}$ has been termed the \qut{transfer matrix}.
\index{matrix!transfer}\index{transfer matrix!diffusion of information}

\enlargethispage{18pt}

\vspace*{6pt}
\runinhead{The Stationary State} \index{stationary
solution!information flow} When no new information is created we may
expect the distribution of information to settle into a stationary
state
$$
{\partial \rho_i(t)\over \partial t}\ \to\ 0,
\qquad\quad
\rho_i(t)\ \to\ \rho_i(\infty)~.
$$
Formally, the stationary state corresponds to the unitary eigenvalue
of $\bf T$, see Eq.~(\ref{chaos_master-eq-info}). Here we assume
\begin{equation}
\rho_i(\infty)\ \propto\ \sum_j\,W_{ji}~,
\label{synchro_ansatz_rho}
\end{equation}
in Eq.~(\ref{chaos_master-eq-info}):
\begin{equation}
\sum_j{W_{ij}\over\sum_k W_{kj}} \sum_k W_{kj} \ =\
\sum_l\, W_{li},
\qquad\quad
\sum_j\,W_{ij}\ =\
\sum_l\, W_{li}~.
\label{chaos_network_info_stationary_state}
\end{equation}
Consequently, a global steady state has the form of the
ansatz (\ref{synchro_ansatz_rho}) when the weight of incoming links
$\sum_j\,W_{ij}$ equals the weight of outgoing links $\sum_l\,
W_{li}$ for every vertex $i$. That is if there are no sinks or
sources for information. The condition
Eq.~(\ref{chaos_network_info_stationary_state}) is fulfilled for
symmetric weight matrices with $W_{ij}=W_{ji}$.
The information density is proportional to the
vertex degree, $\rho_i(\infty)\propto k_i$, when
the $W_{ij}$ reduces to the adjacency matrix.

\subsection{The Langevin Equation and Diffusion}
\label{chaos_langevin_diffusion}

\runinhead{Diffusion as a Stochastic Process}
\index{diffusion!stochastic process} Langevin proposed to describe
the diffusion of a particle by the stochastic differential equation
\index{Langevin equation} \index{equation!Langevin}
\begin{equation}
m\,\dot v\ =\ -m\,\gamma\,v\,+\, \xi(t),
\quad\quad <\xi(t)>=0,
\quad\quad <\xi(t)\xi(t')>=Q\delta(t-t'),
\label{chaos1_eq_langevin}
\end{equation}
where $v(t)$ is the velocity of the particle and
$m>0$ its mass.
\begin{description}
\item[(i)] The term $-m \gamma v$ on the
     right-hand-side of Eq.~(\ref{chaos1_eq_langevin})
     corresponds to a damping term, the friction being
     proportional to $\gamma>0$.\index{friction!damping term}

\item[(ii)] $\xi(t)$ is a stochastic variable, viz noise.
            The brackets $<\ldots >$ denote ensemble averages, i.e.\
            averages over different noise realizations.\index{stochastic!variable}
\item[(iii)] As {\em white noise} (in contrast to
             {\em colored noise}) one denotes noise
             with {a}
             flat power spectrum (as white light), viz
             $\ <\xi(t)\xi(t')>\propto \delta(t-t')$.\index{noise!white}\index{noise!colored}
\item[(iv)] The constant $Q$ is a measure for the strength of
            the noise.
\end{description}
\runinhead{Solution of the Langevin Equation} \index{Langevin
equation!solution} Considering a specific noise realization
$\xi(t)$, one finds\vspace*{4pt}
\begin{equation}
v(t)\ =\ v_0\,e^{-\gamma t} \,+\, {e^{-\gamma t}\over m}\,\int_0^t
\mathrm{d}t'\, e^{\gamma t'}\, \xi(t') \label{chaos1_sol_Langevin}\vspace*{4pt}
\end{equation}
for the solution of the Langevin Eq.~(\ref{chaos1_eq_langevin}),
where $v_0\equiv v(0)$.

\runinhead{Mean Velocity} \index{mean!velocity} For the ensemble
average $<v(t)>$ of the velocity one finds\index{ensemble!average}
\begin{equation}
<v(t)> \ =\ v_0\,e^{-\gamma t}\,+\, {e^{-\gamma t}\over m}\,\int_0^t
\mathrm{d}t'\, e^{\gamma t'}\, \underbrace{<\xi(t')>}_{0}
 =\ v_0\,e^{-\gamma t}~.
\label{chaos1_av_v_langevin}
\end{equation}
The average velocity decays exponentially to
zero.

\enlargethispage{-24pt}

\runinhead{Mean Square Velocity} For the ensemble average $<v^2(t)>$
of the velocity squared one finds
\begin{eqnarray*}
<v^2(t)> & =& v_0^2\,e^{-2\gamma t}\,+\,
{2\,v_0\,e^{-2\gamma t}\over m}\,\int_0^t
\mathrm{d}t'\, e^{\gamma t'}\, \underbrace{<\xi(t')>}_{0} \\
&\quad +& {e^{-2\gamma t}\over m^2}\, \int_0^t \mathrm{d}t'\, \int_0^t
\mathrm{d}t''\, e^{\gamma t'}\, e^{\gamma t''}\,
\underbrace{<\xi(t')\xi(t'')>}_{Q\,\delta(t'-t'')} \\
&=& v_0^2\,e^{-2\gamma t}\,+\, {Q\,e^{-2\gamma t}\over m^2}\,
\underbrace{\int_0^t \mathrm{d}t'\,e^{2\gamma t'}}_{\left(e^{2\gamma
t}-1\right)/(2\gamma)}
\end{eqnarray*}


\noindent and finally
\begin{equation}
<v^2(t)> \ =\ v_0^2\,e^{-2\gamma t}\,+\,
{Q\over 2\,\gamma\,m^2}\,
\left(1-e^{-2\gamma t}\right)~.
\label{chaos1_av_vv_langevin}
\end{equation}
For long times the average squared velocity
\begin{equation}
\lim_{t\to\infty}<v^2(t)> \ =\
{Q\over 2\,\gamma\,m^2}
\label{chaos1_av_vv_langevin_t_infty}
\end{equation}
becomes, as expected, independent of the initial velocity $v_0$.
Equation (\ref{chaos1_av_vv_langevin_t_infty}) shows explicitly that
the dynamics is driven exclusively by the stochastic process
$\propto Q$ for long time scales.


 \runinhead{{The Langevin}
Equation and Diffusion} \index{Langevin equation!diffusion} The
Langevin equation is formulated in terms of the particle velocity.
In order to make connection with the time evolution of a real-space
random walker, Eq.~(\ref{chaos_diffusion_time}), we multiply the
Langevin equation (\ref{chaos1_eq_langevin}) by $x$ and take the
ensemble average:
\begin{equation}
\vspace{2pt} < x\,\dot v>\ =\ -\gamma <x\,v>\,+\, {1\over
m}<x\,\xi>~. \label{chaos1_eq_langevin_xv}
\end{equation}
We note that
$$
x\,v = x\,\dot x = {\mathrm{d}\over \mathrm{d}t}{x^2\over 2},
\qquad\quad x\,\dot v = x\,\ddot x = {\mathrm{d}^2\over
\mathrm{d}t^2}{x^2\over 2}-\dot x^2, \qquad\quad <x\xi>=x<\xi>=0~.
$$

\noindent We then find for Eq.~(\ref{chaos1_eq_langevin_xv})
$$
{\mathrm{d}^2\over \mathrm{d}t^2}{<x^2>\over 2}-<v^2> \ =\
-\gamma\,{\mathrm{d}\over \mathrm{d}t}{<x^2>\over 2}
$$
or
\begin{equation}
\vspace{2pt} {\mathrm{d}^2\over
\mathrm{d}t^2}<x^2>\,+\,\gamma\,{\mathrm{d}\over \mathrm{d}t}<x^2> \
=\ 2<v^2> \ =\ {Q\over \gamma m^2},
\label{chaos1_eq_langevin_diffusion} \vspace{2pt}
\end{equation}
where we have used the long-time result
Eq.~(\ref{chaos1_av_vv_langevin_t_infty}) for $<v^2>$. The solution
of Eq.~(\ref{chaos1_eq_langevin_diffusion}) is
\begin{equation}
\vspace{2pt} <x^2> \ =\ \left[\gamma t-1+e^{-\gamma
t}\right]\,{Q\over\gamma^3m^2}~. \label{chaos1_langevin_diffusion}
\vspace{2pt}
\end{equation}
For long times we find
\begin{equation}
\lim_{t\to\infty}<x^2>\ =\ {Q\over\gamma^2m^2}\,t \ \equiv\ 2D\,t,
\qquad\quad
D= {Q\over 2\gamma^2m^2}
\label{chaos1_rel_noise_diffusion}
\end{equation}
diffusive behavior, compare Eq.~(\ref{chaos_diffusion_time}). This
shows that diffusion is microscopically due to a stochastic process,
since $D\propto Q$.

\vspace*{-1.3pc}
\section{Noise-Controlled Dynamics}
\label{chaos_noise_dynamics}
\index{dynamical system!noise-controlled}

\runinhead{Stochastic Systems}\index{stochastic!system}
\index{dynamical system!stochastic}\index{noise!stochastic system} A
set of first-order differential equations with a stochastic term is
generally denoted {a} \qut{stochastic system}. The
Langevin equation (\ref{chaos1_eq_langevin}) discussed in
Sect.~\ref{chaos_langevin_diffusion} is a prominent example. The
stochastic term corresponds quite generally to noise. Depending on
the {circumstances}, noise might be very
important for the long-term dynamical behavior{. Some examples of this are as follows}:
\begin{itemize}
\item[--]{Neural Networks}:
  Networks of interacting neurons are responsible for
  the cognitive information processing in the brain.
  They must remain functional also in the presence of noise and
  need to be stable as stochastic systems.
  In this case the introduction of a noise term to the
  evolution equation should not change
  the dynamics qualitatively. This postulate should be valid
  for the vast majorities of biological networks.
\item[--]{Diffusion}:
  The Langevin equation reduces, in the absence of noise,
  to a damped motion without an external driving force, with
  $v=0$ acting as a global attractor. The stochastic term
  is therefore essential in the long-time limit, leading
  to diffusive behavior.
\item[--]{Stochastic Escape and Stochastic Resonance}:
\index{stochastic escape}\index{stochastic resonance}A particle
trapped in a local\break minimum may escape this
  minimum by a noise-induced diffusion process{;} a phenomenon
  called \qut{stochastic escape}. Stochastic escape in a driven
  bistable system leads to an even more subtle consequence
  of noise-induced dynamics, the \qut{stochastic resonance}.
\end{itemize}

\vspace*{-8pt}
\subsection{Stochastic Escape}
\label{chaos_stochastic_escape}

\runinhead{Drift Velocity} \index{drift velocity} We generalize the
Langevin equation
 (\ref{chaos1_eq_langevin}) and consider an external
potential $V(x)$,
\begin{equation}
m\,\dot v\ =\ -m\,\gamma\,v\,+\,F(x) \,+\, \xi(t), \qquad\quad
F(x)=-V'(x)=-{\mathrm{d}\over \mathrm{d}x}V(x)~,
\label{chaos1_eq_langevin_F}
\end{equation}
where $v,m$ are the velocity and the mass of the particle,
$<\xi(t)>=0$ and $<\xi(t)\xi(t')>=Q\delta(t-t')$. In the absence of
damping ($\gamma=0$) and noise ($Q=0$),
Eq.~(\ref{chaos1_eq_langevin_F}) reduces to Newton's law.
\index{Newton's law}\index{equation!Newton}

We consider for a moment a constant force $F(x)=F$ and the absence
of noise, $\xi(t)\equiv0$. The system {then
reaches} an equilibrium for $t\to\infty$ when relaxation and
force cancel each other:
\begin{equation}
m\,\dot v_D\ =\ -m\,\gamma\,v_D\,+\,F\ \equiv\ 0,
\quad\qquad
v_D\ =\ {F\over \gamma m}~.
\label{chaos1_eq_drift_v}
\end{equation}
$v_D$ is called the \qut{drift velocity}. A typical example is the
motion of electrons in a metallic wire. An applied voltage, which
leads an electric field along the wire, induces an electrical
current (Ohm's law).\index{Ohm's law}\index{law!Ohm}
{This} results in the drifting electrons being
continuously accelerated by the electrical field, while bumping into
lattice imperfections or colliding with the lattice vibrations,
{i.e.} the phonons.

\runinhead{{The Fokker--Planck} Equation}
\index{Fokker--Planck equation}\index{equation!Fokker--Planck} We
consider now an ensemble of particles diffusing in an external
potential, and denote with $P(x,t)$ the density of particles at
location $x$ and time $t$. Particle number conservation defines the
particle current density $J(x,t)$ via the continuity equation
\index{equation!continuity}\index{continuity equation}
\index{Fokker--Planck equation!particle current}
\begin{equation}
{\partial P(x,t)\over\partial t}\,+\,
{\partial J(x,t)\over\partial x}\ =\ 0.
\label{chaos1_continuity_equation}
\end{equation}
There are two contributions, $J_{v_D}$ and $J_\xi$,
to the total particle current density, $J=J_{v_D}+J_\xi$,
induced by the diffusion and by the stochastic motion
respectively. We derive these two contributions in two steps.

In a first step we consider with $Q=0$ the absence of noise
in Eq.~(\ref{chaos1_eq_langevin_F}).
The particles then move uniformly with the drift velocity
$v_D$ in the stationary limit, and the current density is
$$
J_{v_D} = v_D\,P(x,t)~.
$$
In a second step we set the force to zero, $F=0$, and
derive the contribution $J_\xi$ of the noise term $\sim\xi(t)$
to the particle current density. For this purpose
we rewrite the diffusion equation (\ref{chaos_diffusion_eq})
$$
{\partial P(x,t) \over \partial t}\ =\
D\, {\partial^2 P(x,t) \over \partial x^2} \ \equiv\
-{\partial J_\xi(x,t) \over \partial x}
\quad\qquad
{\partial P(x,t) \over \partial t}
+{\partial J_\xi(x,t) \over \partial x}
\ =\ 0
$$
as a continuity equation, which allows us
to determine the functional form of $J_\xi$,
\begin{equation}
J_\xi\ =\  -D{\partial P(x,t)\over \partial x}~.
\label{chaos_diffusion_eq_current}
\end{equation}
Using the relation $D=Q/(2\gamma^2m^2)$,
see Eq.~(\ref{chaos1_rel_noise_diffusion}),
and including the drift term we find
\begin{equation}
J(x,t) \ =\ v_D\,P(x,t)\,-\,D\,{\partial P(x,t)\over \partial x}
\ =\ {F\over\gamma m}\,P(x,t)\,-\,
{Q\over 2\gamma^2m^2}\,{\partial P(x,t)\over \partial x}
\label{chaos1_J_total}
\end{equation}
for the total current density $J=J_{v_D}+J_\xi$. 
Using expression (\ref{chaos1_J_total}) 
for the total particle current density in
(\ref{chaos1_continuity_equation}) one obtains
the \qut{Fokker--Planck} or \qut{Smoluchowski} equation
\begin{equation}
{\partial P(x,t)\over\partial t} \ =\
-{\partial v_D\,P(x,t)\over\partial x}\,+\,
{\partial^2 D\, P(x,t)\over \partial x^2}
\label{chaos1_Fokker_Planck}
\end{equation}
for the density distribution $P(x,t)$.

\runinhead{The Harmonic Potential}

\begin{figure}[t]
\centering
\includegraphics{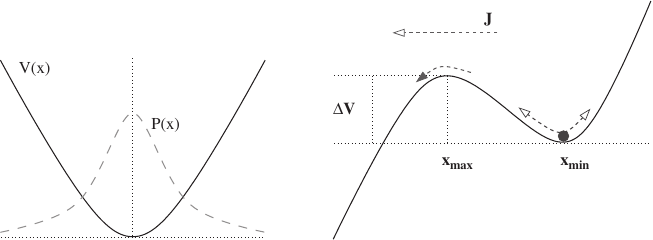}
\caption{\textit{Left}: Stationary distribution $P(x)$ of diffusing
particles in a harmonic potential $V(x)$.  \textit{Right}:
Stochastic escape from a local minimum, with $\Delta
V=V(x_{\max})-V(x_{\min})$ being the potential barrier height and
$J$ the escape current} \label{chaos_stochastic_escape_fig}\vspace*{6pt}
\end{figure}

\index{potential!harmonic}\index{Fokker--Planck equation!harmonic
potential} We consider the harmonic confining potential
$$
V(x)\ =\ {f\over 2}\,x^2,
\qquad\quad
F(x)\ =\ -f\,x~,
$$
and a stationary density distribution,
\index{stationary solution!density distribution}
$$
{\mathrm{d} P(x,t) \over \mathrm{d}t}\ =\ 0
\qquad\Longrightarrow\qquad {\mathrm{d} J(x,t) \over \mathrm{d}x}\
=\ 0~.
$$
Expression (\ref{chaos1_J_total}) yields then the differential
equation
$$
{\mathrm{d}\over \mathrm{d}x}\left[ {f\,x\over \gamma\,m} + {Q\over
2\gamma^2m^2}{\mathrm{d}\over {\mathrm{d}}x} \right]\,P(x) \ =\ 0 \
=\ {\mathrm{d}\over \mathrm{d}x}\left[ \beta f x + {\mathrm{d}\over
{\mathrm{d}}x} \right]\,P(x),
$$
%
${\textrm{with}}\ \beta = {2\gamma m/Q}$ and where for the
stationary distribution function
$P(x)=$\break$\lim_{t\to\infty}P(x,t)$. The
system is confined and the steady-state current
vanishes consequently. We find
\begin{equation}
\framebox{$\displaystyle
P(x)\ =\ A\,e^{-\beta {f\over 2} x^2}
\ =\ A\,e^{-\beta V(x)}
         $}
\qquad\quad
A=\sqrt{f\gamma m\over \pi Q},
\label{networks2_sol_P_x}
\end{equation}
where the prefactor is determined by the normalization condition
$\int \mathrm{d}x P(x)=1$. The density of diffusing particles in a
harmonic trap is Gaussian-distributed,\index{Gaussian distribution}
\index{distribution!Gaussian} see
Fig.~\ref{chaos_stochastic_escape_fig}.

\enlargethispage{6pt}

\runinhead{The Escape Current} \index{Fokker--Planck equation!escape
current} \index{current!escape}\index{escape!current} We now
consider particles in a local minimum, as depicted in
Fig.~\ref{chaos_stochastic_escape_fig}, with a typical
potential having a functional form like
\begin{equation}
V(x)\ \sim \ -x\,+\,x^3~.
\label{chaos1_potential_stoch_escape}
\end{equation}
Without noise, the particle will oscillate around
the local minimum eventually coming
to a standstill $x\to x_{\min}$ under
the influence of friction.

With noise, the particle will have a small but
finite probability
$$
\propto e^{-\beta \Delta V}, \qquad\quad \Delta V =
V(x_{\max})-V(x_{\min})
$$
to reach the next saddlepoint, where $\Delta V$ is the potential
difference between the saddlepoint and the local minimum, see
Fig.~\ref{chaos_stochastic_escape_fig}. The solution
Eq.~(\ref{networks2_sol_P_x}) for the stationary particle
distribution in a confining potential $V(x)$ has a vanishing
total current $J$. For non-confining potentials, like
Eq.\ (\ref{chaos1_potential_stoch_escape}), the particle
current $J(x,t)$ never vanishes. Stochastic escape occurs
when starting with a density of diffusing particles
close the local minimum, as illustrated in
Fig.~\ref{chaos_stochastic_escape_fig}. The escape current
will be nearly constant whenever the escape probability is
small. In this case the escape current will be proportional
to the probability a particle has to reach the saddlepoint,
$$
J(x,t)\Big|_{x=x_{\max}} \ \propto\
e^{-\beta\,[V(x_{\max})-V(x_{\min})]}~,
$$
when approximating the functional dependence of $P(x)$ with
that valid for the harmonic potential,
Eq.~(\ref{networks2_sol_P_x}).

\runinhead{Kramer's Escape} \index{escape!Kramer's}\index{Kramer's
escape} \index{probability!stochastic escape} When the escape
current is finite, there is a finite probability per unit of time
for the particle to escape the local minima, the {\em Kramer's
escape rate}~$r_K$,
\begin{equation}
r_K \ =\ {\omega_{\max}\omega_{\min}\over 2\pi\,\gamma} \,
\mbox{exp}\left[-\beta\, (V(x_{\max})-V(x_{\min}))\right]~,
\label{networks2_escape_rate}
\end{equation}
where the prefactors $\omega_{\min}=\sqrt{|V''(x_{\min})|/m}$ and
$\omega_{\max}=\sqrt{|V''(x_{\max})|/m}$ can be derived from a more
detailed calculation, and where $\beta=2\gamma m/Q$.

\runinhead{Stochastic Escape in Evolution} Stochastic escape occurs
in many real-world systems. Noise allows the system to escape from a
local minimum {where it would otherwise remain stuck for eternity}.

As an example, we mention stochastic escape from a local fitness
maximum (in evolution fitness is to be maximized) by random
mutations {that} play the role of noise. These
issues will be discussed in more detail in
Chap.~\ref{chap_evolution1}.

\enlargethispage{6pt}

\vspace*{-6pt}

\subsection{Stochastic Resonance}
\label{chaos_stochastic_resonance}
\index{stochastic resonance|textbf}

\begin{figure}[t]
\centering
\includegraphics{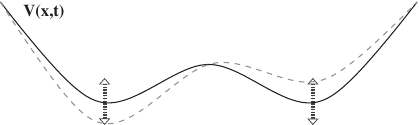}
\caption{The driven double-well potential, $V(x)-A_0\cos(\Omega
t)x$,
         compare Eq.~(\ref{chaos1_eq_driven_double_well}).
         The driving force is small enough to retain the two local
         minima}
\label{chaos_fig_driven_double_well}
\end{figure}

\runinhead{The Driven Double-Well Potential} We consider
diffusive dynamics in a driven double-well potential, see
Fig.~\ref{chaos_fig_driven_double_well},
\index{potential!double-well}
\begin{equation}
\dot x \ =\ -V'(x)+A_0\cos(\Omega t) + \xi(t),
\qquad\quad
V(x)= -{1\over 2}x^2+{1\over 4}x^4~.
\label{chaos1_eq_driven_double_well}
\end{equation}
{The following is to be remarked}:
\begin{itemize}
\item[--]
  Equation (\ref{chaos1_eq_driven_double_well}) corresponds
  to the Langevin equation  (\ref{chaos1_eq_langevin_F})
  in the limit of very large damping, $\gamma\gg m$,
  keeping $\gamma m\equiv 1$ constant (in dimensionless
  units).\index{friction!large damping}
\item[--] The potential in Eq.~(\ref{chaos1_eq_driven_double_well})
      is in normal form, which one can always achieve
      by rescaling the variables appropriately.
\item[--] The potential $V(x)$ has two minima $x_0$ at
$$
-V'(x)\ =\ 0 \ =\ x-x^3 \ =\ x(1-x^2),
\qquad\quad x_0=\pm1~.
$$
The local maximum $x_0=0$ is unstable.
\index{periodic driving}
\item[--] We assume that the periodic driving $\propto A_0$
      is small enough, such that the effective
      potential $V(x)-A_0\cos(\Omega t)x$ retains
      two minima at all times, compare
      Fig.~\ref{chaos_fig_driven_double_well}.
\end{itemize}

\begin{figure}[t]
\centering
\includegraphics{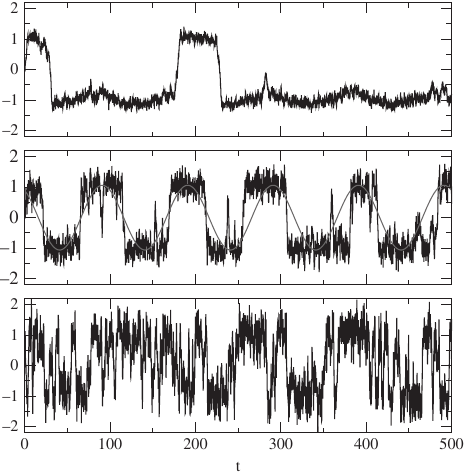}
\caption{Example trajectories $x(t)$ for the
         driven double-well potential.
         The strength and the period of the driving potential
         are $A_0=0.3$ and $2\pi/\Omega=100$, respectively. The
         noise level $Q$ is 0.05, 0.3 and 0.8 (\textit{top/middle/bottom}),
         see Eq.~(\ref{chaos1_eq_driven_double_well})
        }
\label{chaos_fig_stochastic_resonance_transient_states}\vspace{-7pt}
\end{figure}

\runinhead{Transient State Dynamics} \index{transient state
dynamics!stochastic resonance} The system will stay close to one of
the two minima, $x\approx \pm1$, for most of the time when both
$A_0$ and the noise strength are weak, see
Fig.~\ref{chaos_fig_stochastic_resonance_transient_states}.
This is an instance of \qut{transient state dynamics},
which will be discussed in more detail in Chap.~\ref{chap_cogSys1}.
The system switches between a set of preferred states.

\begin{figure}[t]
\centering
\includegraphics{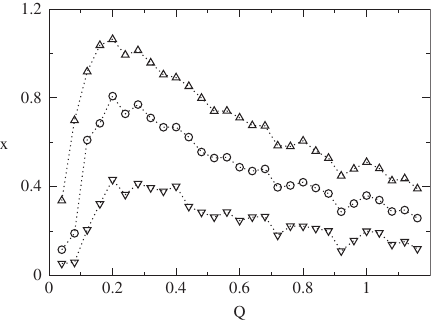}
\caption{The gain $\bar x$, see Eq.~(\ref{chaos2_x_bar}),
 as a function of noise level $Q$. The strength of
 the driving amplitude $A_0$ is 0.1, 0.2 and 0.3
 (\textit{bottom/middle/top}
 {curves}), see Eq.~(\ref{chaos1_eq_driven_double_well}) and
 {the period} $2\pi/\Omega=100$.
 The response $\bar x$ is very small for vanishing noise
 $Q=0$, when the system performs only small-amplitude
 oscillations in one of the local minima
} \label{chaos_fig_stochResonanz_response}
\end{figure}

\runinhead{Switching Times} \index{stochastic resonance!switching
times} An important question is then: How often does the system
switch between the two preferred states $x\approx 1$ and $x\approx
-1$? There are two time scales present:
\begin{itemize}
\item[--] In the absence of external driving, $A_0\equiv0$,
      the transitions are noise driven and irregular,
      with the average switching time given by
      Kramer's lifetime $T_K=1/r_K$, see
      Fig.~\ref{chaos_fig_stochastic_resonance_transient_states}.
      The system is translational invariant with respect
      to time and the ensemble averaged expectation value
$$
< x(t) > \ =\ 0
$$
      {therefore vanishes} in the absence of
      an external {force}.
\item[--] When $A_0\ne0$ the external {force}
      induces a reference time and a non-zero
      response $\bar x$,
\begin{equation}
< x(t) > \ =\ \bar x\,\cos(\Omega t-\bar\phi)~,
\label{chaos2_x_bar}
\end{equation}
      which follows the time evolution of the
      driving potential with a certain phase shift $\bar\phi$,
      see Fig.~\ref{chaos_fig_stochResonanz_response}.
\end{itemize}

\enlargethispage{12pt}

\runinhead{The Resonance Condition} \index{stochastic
resonance!resonance condition} When the time scale $2T_K=2/r_K$ to
switch {back and forth} due to the
stochastic process equals the period $2\pi/\Omega$, we expect a
large response $\bar x$, see
Fig.~\ref{chaos_fig_stochResonanz_response}. The time-scale matching
condition\vspace{-2pt}
$$
{2\pi\over\Omega}\ \approx\ {2\over r_K}\vspace{-2pt}
$$
depends on the noise-level $Q$, via
Eq.~(\ref{networks2_escape_rate}), for the Kramer's escape rate
$r_K$. The response $\bar x$ first increases with rising $Q$ and
then becomes smaller again, for otherwise constant parameters, see
Fig.~\ref{chaos_fig_stochResonanz_response}. Therefore the name
\qut{stochastic resonance}.

\runinhead{Stochastic Resonance and the Ice Ages} \index{stochastic
resonance!ice ages} The average temperature $T_e$ of the earth
differs by about $\Delta T_e\approx 10^\circ\mbox{C}$ in between a
typical ice age and the interglacial periods. Both states of the
climate are locally stable.
\begin{itemize}
\item[--]{{The Ice} Age}:
    The large ice covering increases the albedo of the earth
    and a larger part of sunlight is reflected back to space.
    The earth remains cool.
\item[--]{{The Interglacial Period}}:
    The ice covering is small and a larger portion of the
    sunlight is absorbed by the oceans and land{. The} earth
    remains warm.
\end{itemize}
A parameter of the orbit of {the planet} earth, the
eccentricity, varies slightly with a period $T=2\pi/\Omega\approx
10^5\,\mbox{years}$. The intensity of the incoming radiation from
the sun therefore varies with the same period. {Long-term} climate changes can therefore be modeled by a
driven two-state system, i.e.\ by
Eq.~(\ref{chaos1_eq_driven_double_well}). The driving force, viz the
variation of the energy flux the earth receives from the sun, is
however very small. The increase in the amount of incident sunlight
is too weak to pull the earth out of an ice age into an interglacial
period or vice versa. Random climatic fluctuation, like variations
in the strength of the gulf stream, are needed to finish the job.
The alternation of ice ages with interglacial periods may therefore
be  modeled as a stochastic resonance phenomenon.

\enlargethispage{12pt}

\runinhead{Neural Networks and Stochastic Resonance}
\index{stochastic resonance!neural network}\index{neural
network!stochastic resonance} Neurons are driven bistable devices
operating in a noisy environment. It is therefore not surprising
that stochastic resonance may play a role for certain neural network
setups with undercritical driving.

\runinhead{Beyond Stochastic Resonance}
Resonance phenomena generally occur when two frequencies,
or two time scales, match as a function of some control
parameter. For the case of stochastic resonance
these two time scales correspond to the period of the
external driving and to the average waiting time for
the Kramer's escape respectively, with the later depending
directly on the level of the noise. The phenomenon is
denoted as \qut{stochastic resonance} since one of the
time scales involved is controlled by the noise.

\index{coherence resonance}
One generalization of this concept is the one of
\qut{coherence resonance}. In this case one has a
dynamical system with two internal time scales $t_1$
and $t_2$. These two time scales will generally be
affected to a different degree by an additional
source of noise. The stochastic term may therefore
change the ratio $t_1/t_2$, leading to internal
resonance phenomena.

\section{Dynamical Systems with Time Delays}
\label{sect_chaos_time_delays}
\index{dynamical system!time delays|textbf}
\index{time delays!dynamical system|textbf}

The dynamical systems we have considered so far
all had instantaneous dynamics, being of the
type
\begin{eqnarray}
\label{chaos1_instantaneous_dynamics}
{d\over dt} y(t) &=& f(y(t)), \qquad  t>0 \\
\qquad\quad y(t=0) &=& y_0~,
\nonumber
\end{eqnarray}
when denoting with $y_0$ the initial condition.
This is the simplest case:
one dimensional (a single dynamical variable only),
autonomous ($f(y)$ is not an explicit function
of time) and deterministic (no noise).

\runinhead{Time Delays}
In many real-world applications the couplings between
different sub-systems and dynamical variables is
not instantaneous. Signals and physical interactions
need a certain time to travel from one subsystem to
the next. Time delays are therefore encountered commonly
and become important when the delay time $T$ becomes
comparable with the intrinsic time scales of the
dynamical system. We consider here the simplest
case, a noise-free one-dimensional dynamical system
with a single delay time,
\begin{eqnarray}
\label{chaos1_dynamics_time_delay}
{d\over dt} y(t) &=& f(y(t),y(t-T)), \qquad  t>0 \\[6pt]
\qquad\quad y(t) &=& \phi(t), \qquad\qquad\qquad t\in[-T,0] ~.
\nonumber
\end{eqnarray}
Due to the delayed coupling we need now to specify
an entire initial function $\phi(t)$. Differential
equations containing one or more time delays need
to be considered very carefully, with the
time delay introducing an additional dimension to
the problem. We will discuss here a few illustrative
examples.

\runinhead{Linear Couplings}
We start with the linear differential equation
\begin{equation}
{d\over dt}y(t) \ =\ -a\,y(t)\,-\,b\,y(t-T),
\qquad\quad a,b>0~.
\label{chaos1_time_delay_linear}
\end{equation}
The only constant solution for $a+b\ne0$ is
the trivial state $y(t)\equiv 0$.
The trivial solution is stable in the
absence of time delays, $T=0$, whenever
$a+b>0$. The question is now, whether a
finite $T$ may change this.

We may expect the existence of a certain critical
$T_c$, such that $y(t)\equiv0$ remains stable for
small time delays $0\le T<T_c$. In this case
the initial function $\phi(t)$ will affect the
orbit only transiently, in the long run the
motion would be damped out, approaching
the trivial state asymptotically for $t\to\infty$.

\begin{figure}[t]
\centering
\includegraphics{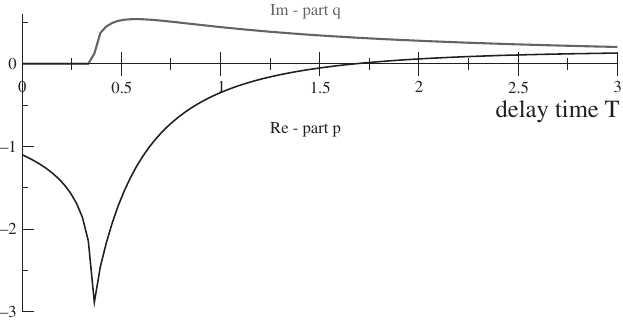}
\caption{The solution $\hbox{e}^{(p+iq)t}$ of the time-delayed
system, Eq.\ (\ref{chaos1_time_delay_linear}), for $a=0.1$ and
$b=1$. The state $y(t)\equiv0$ become unstable whenever $p>0$. $q$
is given in units of $\pi$
        }
\label{chaos_fig_delayTime_sol}
\end{figure}

\runinhead{Hopf Bifurcation}
Trying our luck with the usual exponential
ansatz, we find\vspace*{2pt}
$$
\lambda \ =\ -a-b\hbox{e}^{-\lambda T},
\qquad\quad
y(t)=y_0\hbox{e}^{\lambda t},
\qquad\quad
\lambda=p+iq~.\vspace*{2pt}
$$
Separating into a real and imaginary part
we obtain\vspace*{2pt}
\begin{equation}
\begin{array}{rcl}
p+a &=& -b \hbox{e}^{-pT}\cos(qT), \\[6pt]
q &=& \phantom{-}b \hbox{e}^{-pT}\sin(qT).
\end{array}~
\label{chaos1_eq_time_eigen_square}\vspace*{2pt}
\end{equation}
For $T=0$ the solution is $p=-(a+b)$, $q=0$,
as expected, and the trivial solution $y(t)\equiv0$
is stable. A numerical solution is shown
in Fig.~\ref{chaos_fig_delayTime_sol} for
$a=0.1$ and $b=1$. The crossing point
$p=0$ is determined by\vspace*{2pt}
\begin{equation}
a \ = \ -b\cos(qT),
\qquad\quad
q \ =\  b \sin(qT)~.
\label{chaos1_p=0_solution}\vspace*{2pt}
\end{equation}
The first condition in
Eq.\ (\ref{chaos1_p=0_solution})
can be satisfied only for $a<b$. Taking the
squares in Eq.\ (\ref{chaos1_p=0_solution})
and eliminating $qT$ one has
$$
q \ =\ \sqrt{b^2-a^2},
\qquad\quad
T\ \equiv\ T_c \ =\  \arccos(-a/b)/q~.
$$
\index{Hopf bifurcation}

One therefore has a Hopf bifurcation at $T \ =\ T_c$ and the trivial
solution becomes unstable for $T>T_c$. For the case $a=0$ one has
simply $q=b$, $T_c=\pi/(2b)$. Note, that there is a Hopf bifurcation
only for $a<b$, viz whenever the time-delay dominates, and that $q$
becomes non-zero well before the bifurcation point, compare
Fig.~\ref{chaos_fig_delayTime_sol}. One has therefore a region of
damped oscillatory behavior with $q\ne0$ and
$p<0$.\enlargethispage*{12pt}

\runinhead{Discontinuities} For time-delayed differential equations
one may specify an arbitrary initial function $\phi(t)$ and the
solutions may in general show discontinuities in their derivatives,
as a consequence. As an example we consider the case $a=0$, $b=1$ of
Eq.\ (\ref{chaos1_time_delay_linear}), with a non-zero constant
initial function,\vspace*{-6pt}
\begin{equation}
\vspace*{-6pt} {d\over dt}y(t) \ =\ -y(t-T), \qquad\quad
\phi(t)\equiv 1~. \label{chaos1_time_delay_discontinuity}
\end{equation}
The solution can be evaluated simply by stepwise integration,
$$
y(t)-y(0)\ =\ \int_0^t dt'\dot y(t')
         \ =\ -\int_0^t dt' y(t'-T)
         \ =\ -\int_0^t dt'
         \ =\ -t,
\qquad 0<t<T~.
$$
The first derivative in consequently discontinuous at
$t=0$,
$$
\lim_{t\to 0-} {d\over dt} y(t) \ =\ 0,
\qquad\quad
\lim_{t\to 0+} {d\over dt} y(t) \ =\ -1~.
$$
For larger times, $T<t<2T$, one finds
$$
y(t)-y(T) = -\int_T^t dt' y(t'-T)
          = \int_T^t dt' \big[t'-1\big]
          = {t^2-T^2\over 2} - (t-T)~,
$$
and the second derivative has a discontinuity
at $t=T$.

\runinhead{Dependence on Initial Function}
The solution of ordinary differential equations
is determined by their initial condition and
different initial conditions lead to distinct
trajectories (injectivity). This is not necessarily
the case anymore in the presence of time delays.
We consider
\begin{equation}
{d\over dt}y(t) \ =\ y(t-T)\big(y(t)-1\big),
\qquad\quad \phi(t=0) \ =\ 1~.
\label{chaos1_time_delay_injectivity}
\end{equation}
For any $\phi(t)$ with $\phi(0)=1$ the solution
is $y(t)\equiv 1$ for all $t\in[0,\infty]$.

\runinhead{Non-Constant Time Delays}
Thinks may become rather weird when the
time delays are not constant anymore.
Consider
\begin{equation}
\begin{array}{rcl}
{d\over dt}y(t) & =& y\big(t-|y(t)|-1\big) \,+\, {1\over 2},
\qquad\quad t >0, \\[8pt]
\phi(t) & = & \left\{
\begin{array}{rcl}
0 & \ & -1<t<0\\
1 & \ & \phantom{-1<}t<-1
\end{array}
              \right.
~.
\end{array}
\label{chaos1_time_delay_non_constant}
\end{equation}
It is easy to see, that both functions
$$
y(t) \ =\ {t\over 2}, \qquad\quad
y(t) \ =\ {3t\over 2}, \qquad\quad
t\in[0,2]~,
$$
are solutions of Eq.\ (\ref{chaos1_time_delay_non_constant}),
with appropriate continuations for $t>2$. Two different
solutions of the same differential equation and
identical initial conditions, that cannot happen for
ordinary differential equations. It is evident, that
especial care must be taken when examining dynamical
systems with time delays numerically.


\section*{Exercises}
\addcontentsline{toc}{section}{Exercises} \markright{Exercises}
\begin{list}{}
\item \hspace*{-15pt}{\sc {The Lorenz Model}} \\
Perform the stability analysis of the fixpoint $(0,0,0)$ and of
$C_{+,-}\ =\break \ (\pm \sqrt{b(r-1)}, \pm \sqrt{b(r-1)}, r-1)$ for the
Lorenz model
{Eq.~(\ref{chaos_lorenz})} with
$r,b>0$. Discuss the difference between the dissipative case and the
ergodic case $\sigma=-1-b$, see Eq.~(\ref{chaos_divlorenz}).
\item \hspace*{-15pt}{\sc {The Poincar\'e Map}} \\
For the Lorenz model
{Eq.~(\ref{chaos_lorenz})} with
$\sigma=10$ and $\beta=8/3$, evaluate numerically the Poincar\'e map
for (a) $r=22$ (regular regime) and the plane $z=21$ and (b) $r=28$
(chaotic regime) and the plane $z=27$.
\item \hspace*{-15pt}{\sc {The Hausdorff Dimension}} \\
Calculate the Hausdorff dimension of a straight line and
of the Cantor set,
\index{Cantor set}
which is generated by removing consecutively
the middle-1/3 segment of a line having a given initial length.
\item \hspace*{-15pt}{\sc The Driven Harmonic Oscillator} \\
Solve the driven, damped harmonic oscillator
$$
\ddot x\,+\,\gamma\, \dot x\, +\,\omega_0^2\,x\ =\ \epsilon\cos(\omega t)
$$
in the long-time limit. Discuss the behavior close to the
resonance $\omega\to\omega_0$.
\item \hspace*{-15pt}{\sc Continuous-Time Logistic Equation}\\
Consider the continuous-time logistic equation
$$
\dot y(t) \ =\  \alpha y(t)\Big[1-y(t)\Big]~.
$$
(A) Find the general solution and (B) compare to the
logistic map Eq.~(\ref{eq_chaos_logistic_map})
for discrete times $t=0,\ \Delta t,\ 2\Delta t,\ ..$.
\item \hspace*{-15pt}{\sc Information Flow in Networks} \\
{Choose a not-too-big social network} and examine numerically
the flow of information, Eq.~(\ref{chaos_info_flow}), through the
network. Set the weight matrix $W_{ij}$ identical to the adjacency
matrix $A_{ij}$, with entries being either unity or zero. Evaluate
the steady-state distribution of information and plot the result as
a function of vertex degrees.
\item \hspace*{-15pt}{\sc Stochastic Resonance} \\
Solve the driven double-well problem
Eq.~(\ref{chaos1_eq_driven_double_well}) numerically and try to
reproduce
Figs.~\ref{chaos_fig_stochastic_resonance_transient_states} and
\ref{chaos_fig_stochResonanz_response}.
\item \hspace*{-15pt}{\sc Delayed Differential Equations} \\
The delayed Eq.~(\ref{chaos1_time_delay_linear}) allows for
harmonically oscillating solutions for certain sets of
parameters $a$ and $b$. Which are the conditions? 
Speciallize then for the case $a=0$.
\item \hspace*{-15pt}{\sc Car-Following Model} \\
A car moving with velocity $\dot x(t)$ follows another
car driving with velocity $v(t)$ via
\begin{equation}
\ddot x(t+T) \ =\ \alpha (v(t)-\dot x(t)),
\qquad\quad \alpha > 0~,
\label{chaos1_car_following_model}
\end{equation}
with $T>0$ being the reaction time of the driver. Prove
the stability of the steady-state solution for a constant
velocity $v(t)\equiv v_0$ of the preceding car.
\end{list}


\def\refer#1#2#3#4#5#6{\item{\frenchspacing\sc#1}\hspace{4pt}
                       #2\hspace{8pt}#3 {\it \frenchspacing#4} {\bf#5}, #6.}
\def\bookref#1#2#3#4{\item{\frenchspacing\sc#1}\hspace{4pt}
                     #2\hspace{8pt}{\it#3}  #4.}

\addcontentsline{toc}{section}{Further Reading} 
\section*{Further Reading}
\markboth{\thechapter\enspace Chaos, Bifurcations and Diffusion}{Further Reading}

For further studies we refer to introductory texts for dynamical
system theory \hbox{(Katok} and Hasselblatt, 1995), classical
dynamical systems (Goldstein, 2002), chaos (Schuster and Just, 2005;
Devaney, 1989; Gutzwiller, 1990, Strogatz, 1994), stochastic
systems (Ross, 1982; Lasota and Mackey, 1994) and
differential equations with time delays (Erneux, 2009).
Other textbooks on complex and/or adaptive systems are those by Schuster
(2001) and Boccara (2003). For an alternative approach to complex
system theory via Brownian agents consult Schweitzer (2003).

The interested reader may want to study some selected 
subjects in more depth, such as the KAM theorem 
(Ott, 2002), relaxation oscillators (Wang, 1999),
stochastic resonance (Benzit et al., 1981; Gammaitoni et al., 1998), 
coherence resonance (Pikovsky and Kurths, 1997),
L\'evy flights (Metzler and Klafter,
2000), the connection of L\'evy flights to the patterns of wandering
albatrosses (Viswanathan et al., 1996), human traveling (Brockmann,
Hufnagel and Geisel, 2006) and diffusion of information in networks
(Eriksen et al., 2003).

The original literature provides more insight, such as the
seminal works of Einstein (1905) and Langevin (1908) on Brownian
motion or the first formulation and study of the Lorenz (1963)
model.

{\baselineskip=15pt
\begin{list}{}{\leftmargin=2em \itemindent=-\leftmargin%
\itemsep=3pt \parsep=0pt \small}
\refer{Benzit, R., Sutera, A.,  Vulpiani, A.}{1981} {The mechanism
of stochastic resonance.}{Journal of Physics A} {14}{L453--L457}
\refer{Brockmann, D., Hufnagel, L.,  Geisel, T.}{2006}
 {The scaling laws of human travel.}{Nature}{439}{462}
\bookref{Boccara, N.}{2003}{Modeling Complex Systems.}{Springer,
Berlin}
\bookref{Devaney, R.L.}{1989} {An Introduction to Chaotic Dynamical
Systems.}{Addison-Wesley{, Reading, MA}}
\refer{Einstein, A.}{1905}{\"Uber die von der molekularkinetischen
Theorie der W\"arme geforderte\break Bewegung von in ruhenden
Fl\"ussigkeiten suspendierten Teilchen.} {Annalen der
Physik}{17}{549}
\refer{Eriksen, K.A., Simonsen, I., Maslov, S.,  Sneppen, K.}{2003}
   {Modularity and extreme edges of the internet.}
   {Physical Review Letters}{90}{148701}
\bookref{Erneux, T.}{2009}{Applied Delay Differential Equations.}
{Springer, New York}  
\refer{Gammaitoni, L., H\"anggi, P., Jung, P.,  Marchesoni,
F.}{1998} {Stochastic resonance.}{Review of Modern
Physics}{70}{223--287}
\bookref{Goldstein, H.}{2002}{Classical Mechanics.} {3rd Edition,
Addison-Wesley, Reading, MA}
\bookref{Gutzwiller, M.C.}{1990} {Chaos in Classical and Quantum
Mechanics.}{Springer, New York}
\bookref{Katok, A., Hasselblatt, B.}{1995} {Introduction to the
Modern Theory of Dynamical Systems.} {Cambridge University Press,
Cambridge}

\refer{Langevin, P.}{1908}{Sur la th\'eorie du mouvement brownien.}
{Comptes Rendus}{146}{530--532}

\bookref{Lasota, A.,  Mackey, M.C.}{1994} {Chaos, Fractals, and
Noise -- Stochastic Aspects of\break Dynamics.} {Springer, New
York}

\refer{Lorenz, E.N.}{1963}{Deterministic nonperiodic flow.} {Journal
of the Atmospheric Sciences}{20}{130--141}
\refer{Metzler, R., Klafter J.}{2000}
       {The random walk's guide to anomalous diffusion:
       a fractional dynamics approach}{Physics Reports}{339}{1}
\bookref{Ott, E.}{2002}{Chaos in Dynamical Systems.}
 {Cambridge University Press, Cambridge}

\refer{Pikovsky, A.S., Kurths, J.}{1997}
{Coherence resonance in a noise-driven excitable system}
{Physical Review Letters}{78}{775}
\bookref{Ross, S.M.}{1982}{Stochastic Processes.} {Wiley, New
York}
\bookref{Schuster, H.G.}{2001}{Complex Adaptive Systems.}{Scator,
Saarbr\"ucken}

\bookref{Schuster, H.G.,  Just, W.}{2005}{Deterministic Chaos.} {4th.
Edition, Wiley-VCH{, New York}}

\bookref{Schweitzer, F.}{2003}{Brownian Agents and Active Particles:
Collective Dynamics in the Natural and Social Sciences.}
{Springer, New York}

\def\CEnote#1{\bgroup \color[rgb]{0,1,1}{#1}\egroup}
\bookref{Strogatz, S.H}{1994}{Nonlinear Systems and Chaos.} {Perseus
Publishing, Cambridge, MA}

\refer{Viswanathan, G.M., Afanasyev, V., Buldyrev, S.V.,
       Murphy, E.J., Prince, P.A.,  Stanley, H.E.}{1996}
      {L\'evy flight search patterns of wandering albatrosses.}
      {\it Nature}{\bf 381}{413}
\bookref{Wang, D.L.}{1999}{\em{Relaxation oscillators and
networks}.} {In J.G. Webster (ed.), {\em Encyclopedia of Electrical
and Electronic Engineers}, pp.\ 396--405, Wiley, New York}
\end{list}
\par}

 

\chapter{Complexity and Information Theory}
\label{chap_complex1}


\abstract{What do we mean when by saying that
a given system shows \qut{complex behavior}, 
can we provide precise measures for the degree 
of complexity?  This chapter offers an account 
of several common measures of complexity and the 
relation of complexity to predictability and emergence.
\newline
\indent 
The chapter starts with a self-contained
introduction to information theory and statistics.
We will learn about probability distribution 
functions, the law of large numbers and the central
limiting theorem. We will then discuss
the Shannon entropy and the mutual information, 
which play central roles both in the context 
of time series analysis and as starting points
for the formulation of quantitative measures of
complexity. This chapter then concludes with
a short overview over generative approaches
to complexity.
}


\section{Probability Distribution Functions}
\label{complex_PDFs}
\index{information theory!basic concepts|textbf}
\index{probability distribution|textbf}

Statistics is ubiquitous in everyday life and we are
used to chat, e.g., about the probability that our child
will have blue or brown eyes, the chances to win a lottery
or those of a candidate to win the presidential 
elections. Statistics is also ubiquitous in all realms
of the sciences and basic statistical concepts are
used throughout these lecture notes\footnote{In
\index{probability distribution!Bayesian} some areas,
like the neurosciences or artificial intelligence, the term
\qut{Bayesian} is used for approaches using statistical methods,
in particular in the context of hypothesis building,  
when estimates of probability distribution functions 
are derived from observations.}.

\runinhead{Variables and Symbols}
\index{stochastic variable}
Probability distribution functions may
be defined for continuous or discrete variables
as well as for sets of symbols,
$$
x\,\in\,[0,\infty],
\qquad \quad
x_i\,\in\,\{1,2,3,4,5,6\},
\qquad \quad
\alpha\,\in\,\{\mbox{blue},\mbox{brown},\mbox{green}\}~.
$$
E.g.\ we may define with $p(x)$ the probability distribution
of human life expectancy $x$, with $p(x_i)$ the chances to
obtain $x_i$ when throwing a dice or with $p(\alpha)$
the probability to meet somebody having eyes of color $\alpha$.
Probabilities are in any case positive definite and the
respective PDF\footnote{PDF is a commonly used abbreviation 
for \qut{probability distribution function}.} normalized,
\index{probability distribution!PDF}
$$
p(x),\ p(x_i),\ p(\alpha)\ \ge\ 0,
\qquad\quad
\int_0^\infty p(x)\,dx \ =\ 1 \ = \ \sum_{\alpha}\, p(\alpha),
\qquad \dots~.
$$
The notation used for a given variable will indicate
in the following its nature, i.e.\ whether it is
a continuous or discrete variable, or denoting a symbol.

\runinhead{Continuous vs.\ Discrete Stochastic Variables}
\index{stochastic variable!continuous}
\index{stochastic variable!discrete}
When discretizing a stochastic variable, e.g.\ when
approximating an integral by a Riemann sum,
\begin{equation}
\int_0^\infty p(x)\,dx \ \approx\ 
\sum_{i=0}^\infty\, p(x_i)\,\Delta x,
\qquad\quad
x_i\,=\,\Delta x\,(0.5+i)~,
\label{complex1_PDF_continuous_discrete}
\end{equation}
the resulting discrete PDF $p(x_i)$ is not
any more normalized; the properly normalized
discrete PDF is $p(x_i)\Delta x$. Note, that
both notations $p_i$ and $p(x_i)$ are used 
for discrete distribution 
functions\footnote{The expression $p(x_i)$ is therefore
context specific and can denote both a properly
normalized discrete PDF as well as the value of
a continuous probability distribution function.}.

\runinhead{Mean, Median and Standard Deviation}
\index{probability distribution!mean}
\index{probability distribution!average}
\index{probability distribution!expectation value}
\index{probability distribution!standard deviation}
The average $\langle x\rangle$, denoted also by $\bar x$,
and the standard deviation $\sigma$ are
given by
\begin{equation}
\langle x\rangle \ =\ \int x\, p(x)\, dx,
\qquad\quad
\sigma^2\ =\ \int \left(x-\bar x\right)^2p(x)\, dx~.
\label{complex1_def_mean_sigma}
\end{equation}
One also calls $\bar x$ the expectation value
or just the mean, and $\sigma^2$ the 
variance\footnote{In formal texts on statistics 
and information theory the notation $\mu=E(X)$ is
often used for the mean $\mu$, the expectation
value $E(X)$ and a random variable $X$, where
$X$ represents the abstract random variable,
whereas $x$ denotes its particular value and
$p_X(x)$ the probability distribution.}.
\index{probability distribution!variance}
\index{probability distribution!median}
For everyday life situations the median $\tilde x$,
\begin{equation}
\int_{x<\tilde x} p(x)\,dx\ =\ \frac{1}{2}\ =\
\int_{x>\tilde x} p(x)\,dx~,
\label{complex1_def_median}
\end{equation}
is somewhat more intuitive than the mean. We have
a 50\% chance to meet somebody being smaller/taller 
than the median height.

\begin{figure}[t]
\centering
\includegraphics[width=0.48\textwidth]{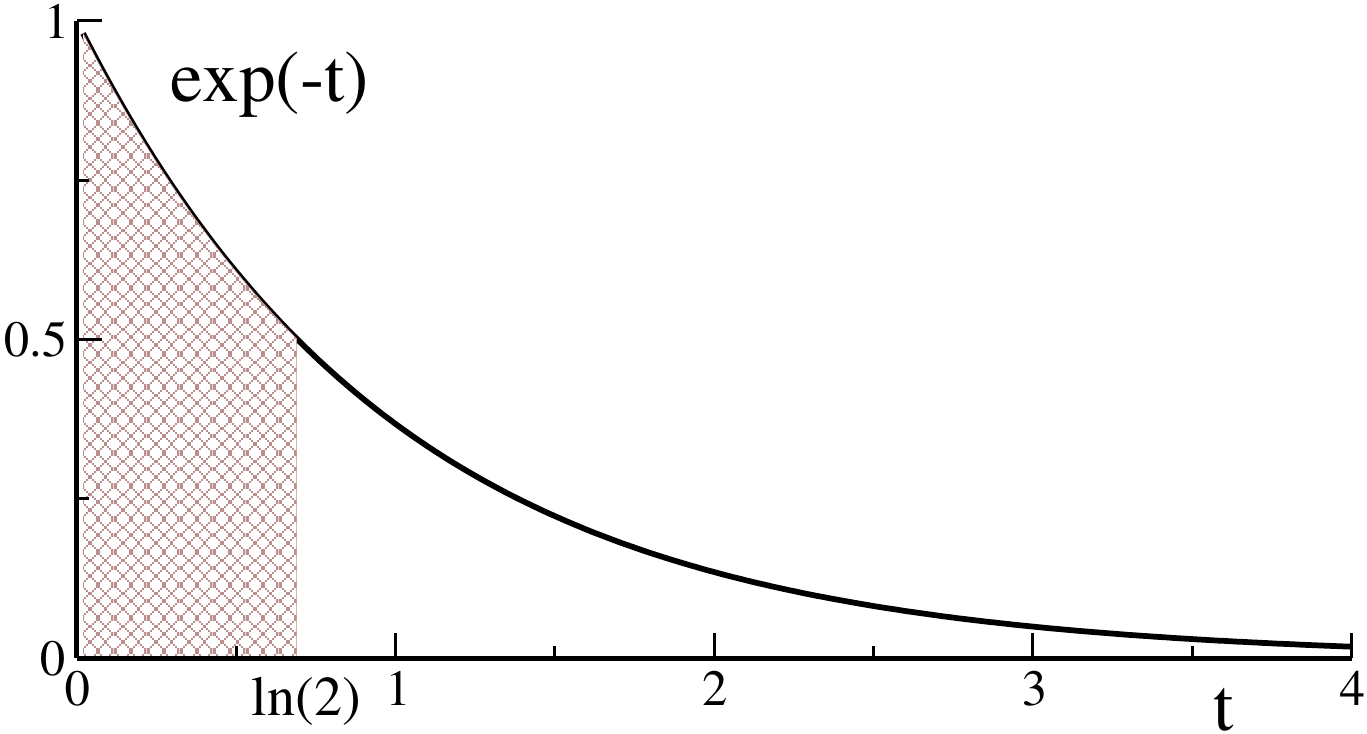}\hspace{1ex}
\includegraphics[width=0.48\textwidth]{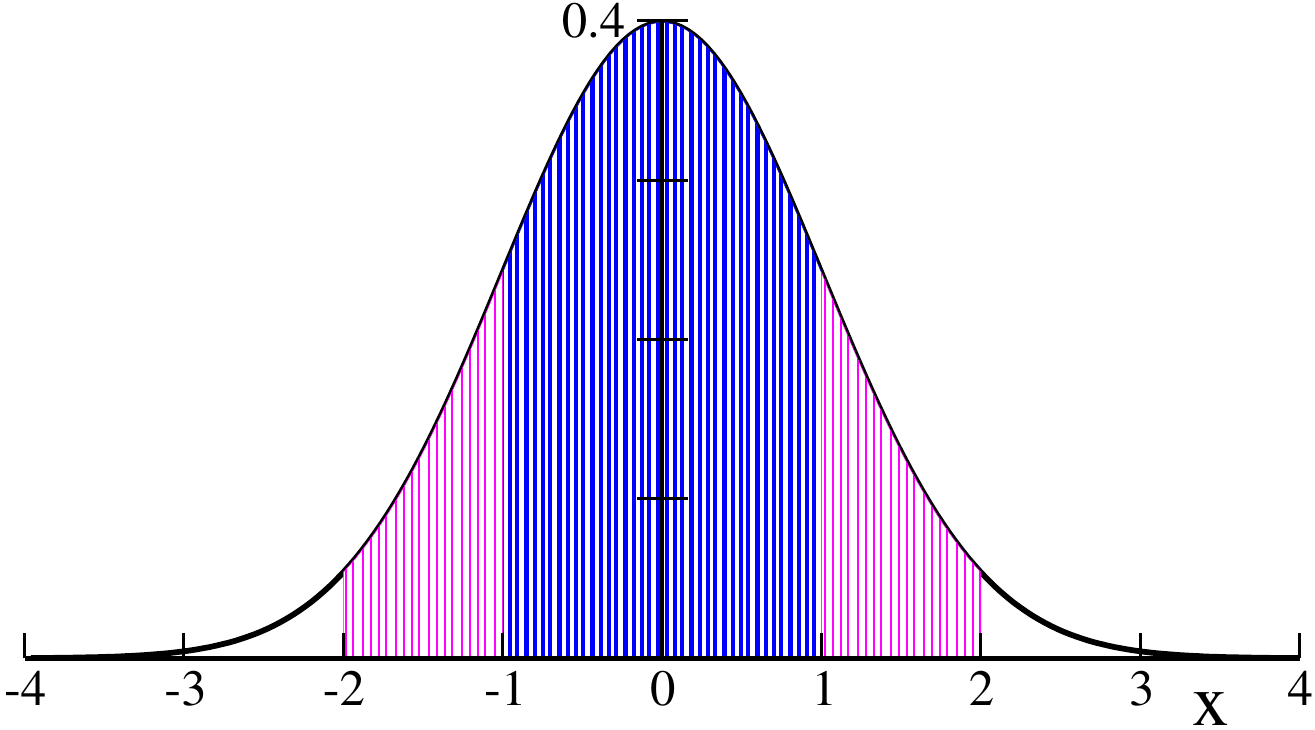}
\vspace*{0.0cm} \caption{\textit{Left}: The exponential distribution
$\,\exp(-t/T)/T$, for an average waiting time $T=1$.
The shaded area, $t\in[0,\ln(2)]$, is 1/2, where $\ln(2)$
is the median. \textit{Right}: The normal distribution
$\exp(-x^2/2)/\sqrt{2\pi}$ having a standard deviation $\sigma=1$.
The probability to draw a result within one/two standard deviations
of the mean ($x\in[-1,1]$ and $x\in[-2,2]$ respectively, 
shaded regions), is $68\%$ and $95\%$.
}
\label{complex_fig_PDF}
\end{figure}

\runinhead{Exponential Distribution}
\index{probability distribution!exponential}
Let us consider, as an illustration, the exponential distribution, 
which describes, e.g.\ the distribution of waiting times
for radioactive decay,
\begin{equation}
p(t) = {1\over T}\,\mathrm{e}^{-t/T},
\quad\qquad
\int_0^\infty p(t)\,dt\ =\ 1~,
\label{complex1_exponential}
\end{equation}
with the mean waiting time
$$
\langle t\rangle \ =\ 
{1\over T}\int_0^\infty t\,\mathrm{e}^{-t/T}\,dt\ =\ 
{t\over T}\,\mathrm{e}^{-t/T}\big|_{0}^{\infty}\, +\,
\int_0^\infty \mathrm{e}^{-t/T}\,dt\ =\ T~.
$$
The median $\tilde t$ and the standard deviation $\sigma$
are evaluated readily as
$$
\tilde t \ =\ T\,\mathrm{ln}(2),
\qquad\quad
\sigma\ =\ T~.
$$
In 50\% of times we have to wait less
than $\tilde t\approx0.69\,T$, which is 
smaller than our average waiting time $T$,
compare Fig.~\ref{complex_fig_PDF}.

\runinhead{Standard Deviation and Bell Curve}
\index{probability distribution!standard deviation}
The standard deviation $\sigma$ measures the 
size of the fluctuations around the mean.
The standard deviation is especially 
important for the \qut{Gaussian distribution}
\index{probability distribution!Gaussian}
\begin{equation}
p(x)\ =\ {1\over \sigma\sqrt{2\pi}}\,
\mathrm{e}^{-{(x-\mu)^2\over 2\sigma^2}},
\qquad\quad
\langle x\rangle \ =\ \mu,
\qquad\quad
\langle (x-\bar x)^2\rangle \ =\ \sigma^2~,
\label{complex1_PDF_gaussian}
\end{equation}
\index{probability distribution!Bell curve}
\index{probability distribution!normal}
also denoted \qut{Bell curve}, or \qut{normal
distribution}. Bell curves are ubiquitous
in daily life, characterizing cumulative processes
(see Sect.~\ref{complex1_subsec_law_large_numbers}).

The Gaussian falls off rapidly with distance from
the mean $\mu$, compare Fig.~\ref{complex_fig_PDF}.
The probability to draw a value within $n$
standard deviation of the mean, viz the
probability that $x\in[\mu-n\sigma,\mu+n\sigma]$,
is $68\%,\,95\%,\,99.7\%$ for $n=1,\,2,\,3$.
Note, that these numbers are valid
only for the Gaussian, not for a general PDF.

\runinhead{Probability Generating Functions}
\index{probability distribution!generating function}
We recall the basic properties of the
generating function
\begin{equation}
G_0(x)\ =\ \sum_{k}\, p_k\,x^k~,
\label{complex1_def_G_0}
\end{equation}
introduced in Sect.~\ref{networks1_generating function},
for the probability distribution $p_k$ of a discrete
variable $k=0,1,2,..$, namely
\begin{equation}
G_0(1)\ =\ \sum_{k}\, p_k \ =\ 1,
\qquad\quad
G_0'(1)\ =\ \sum_{k}\,k\,p_k \ =\ 
\langle k\rangle\ \equiv\ \bar k
\label{complex1_G_0_norm_mean}
\end{equation}
for the normalization and the mean
$\langle k\rangle$ respectively.
The second moment
$\langle k^2\rangle$
\begin{equation}
\langle k^2\rangle \ =\ 
\sum_k\, k^2\, p_k\, x^k\Big|_{x=1}\ =\ 
\frac{d}{dx} \left(x\,G_0'(x)\right)\Big|_{x=1} 
\label{complex1_G_0_second_moment}
\end{equation}
allows to express the standard deviation $\sigma$ as
\begin{eqnarray}
\nonumber
\sigma^2\ =\ 
\langle (k-\bar k)^2\rangle\ =\ 
\langle k^2\rangle\,-\,\bar k^2& =& 
\frac{d}{dx} \left(x\,G_0'(x)\right)\Big|_{x=1} 
\,-\,\left(G_0'(1)\right)^2 \\
&=&  G_0''(1)\,+\,G_0'(1)-\left(G_0'(1)\right)^2~.
\label{complex1_G_0_standard_deviation}
\end{eqnarray}
The importance of probability generating functions
lies in the fact that the distribution for
the sum $k=\sum_i k_i$ of independent stochastic 
variables $k_i$ is generated by the product of 
the generating functions $G_0^{(i)}(x)$ of the 
respective individual processes $p_{k_i}^{(i)}$, viz
$$
G_0(x)\ =\ \sum_k p_k x^k\ =\
\prod_i G_0^{(i)}(x),
\qquad\quad
G_0^{(i)}(x)\ =\ \sum_{k_i} p_{k_i}^{(i)} x^{k_i}~,
$$
see Sect.~\ref{networks1_generating function}
for further details and examples.

\runinhead{Bayesian Theorem}
\index{theorem!Bayes}
\index{Bayes!theorem}
Events and processes may have dependencies upon each other.
A physician will typically have to know, to give an example,
the probability that a patient has a certain illness, 
given that the patient shows a specific symptom.
\begin{quotation}
{\it Conditional Probability.\enspace}
\index{probability distribution!conditional}
The probability that an event $x$ occurs, given that
an event $y$ has happened, is denoted 
\qut{conditional probability} $p(x|y)$.
\end{quotation}
Throwing a dice twice, the probability that the
first throw resulted in a 1, given that the total
result was $4=1+3=2+2=3+1$, is 1/3. Obviously,
\begin{equation}
p(x) \ =\ \int p(x|y)\,p(y)\,dy
\label{complex1_PDF}
\end{equation}
holds. The probability distribution of throwing 
$x$ in the first throw and $y$ in the second throw 
is determined, on the other hand, by the joint distribution 
$p(x,y)$.
\begin{quotation}
{\it Joint Probability Distribution.\enspace}
\index{probability distribution!joint}
The probability of events $x$ and $y$ occurring is
given by the \qut{joint probability} $p(x,y)$.
\end{quotation}
Note, that $\int p(x,y)dx dy=1$. The self-evident relation
\begin{equation}
p(x,y) \ =\ p(x|y)\, p(y)
\label{complex1_theorem_Bayes}
\end{equation}
is denoted \qut{Bayes' theorem}. As a corollary of
Eq.\ (\ref{complex1_theorem_Bayes}),
$p(y|x)p(x)=p(x|y)p(y)$ holds.

\subsection{The Law of Large Numbers}
\label{complex1_subsec_law_large_numbers}
\index{law!large numbers|textbf}

Throwing a dice many times and adding up the
results obtained, the resulting average
will be close to $3.5\,N$, where $N$
is the number of throws. This is the typical
outcome for cumulative stochastic processes\footnote{Please 
take note of the difference between a cumulative 
stochastic process, when adding the results of 
individual trials, and the \qut{cumulative PDF}
\index{probability distribution!cumulative}
$F(x)$ defined by $F(x)=\int_{-\infty}^x p(x')dx'$.}.
\begin{quotation}
{\it Law of Large Numbers.\enspace}
Repeating $\,N\,$ times a stochastic process with mean
$\,\bar x\,$ and standard deviation $\,\sigma\,$,
the mean and the standard
deviation of the cumulative result will 
approach $\,\bar x\, N\,$ and $\,\sigma\sqrt{N}\,$
respectively in the thermodynamic limit $\,N\to\infty$.
\end{quotation}
The law of large numbers implies, that one 
obtains $\bar x$ as an averaged result, with
a standard deviation $\sigma/\sqrt{N}$ for the
averaged process. One needs to square the 
number of trials in order to improve
accuracy by a factor of two.

\runinhead{Proof}
For a proof of the law of large numbers
we consider a discrete process $p_k$ described by the
generating functional $G_0(x)$. This is not really
a restriction, since PDFs of continuous variables
can be discretized with arbitrary accuracy.
The cumulative stochastic process is then 
characterized by a generating functional
$$
G_0^N(x), \qquad\quad
\bar k^{(N)}\ =\ {d\over dx} G_0^N(x)\Big|_{x=1} \ =\
N\, G_0^{N-1}(x)\,G_0'(x)\Big|_{x=1} \ =\ N\,\bar k
$$
and the mean $\bar k^{(N)}=N\bar k$ respectively. For the
standard deviation $\sigma^{(N)}$ of the cumulative
process we use Eq.\ (\ref{complex1_G_0_standard_deviation}),
\begin{eqnarray} \nonumber
\left(\sigma^{(N)}\right)^2 & =& 
\frac{d}{dx} \left(x\,{d\over dx}G_0^N(x)\right)\Big|_{x=1}  
\,-\, \left(N\bar k\right)^2\\ \nonumber
& =& {d\over dx}\big(
x\,N\,G_0^{N-1}(x)\,G_0'(x)\big)\Big|_{x=1}  
\,-\, N^2\left(G_0'(1)\right)^2 \\ \nonumber
&=&
NG_0'(1)\,+\, N(N-1)\left(G_0'(1)\right)^2\,+\,NG_0''(1)
\,-\,N^2\,\left(G_0'(1)\right)^2 \\
&=& N\left(
G_0''(1)+G_0'(1)-\left(G_0'(1)\right)^2\right)
\ \equiv\ N\,\sigma^2~,
\label{eq_complex_law_large_numbers}
\end{eqnarray}
and obtain the law of large numbers.

\runinhead{Central Limiting Theorem}
\index{central limiting theorem}
\index{theorem!central limiting}
The law of large numbers tells us, that the
variance $\sigma^2$ is additive for
cumulative processes, not the 
standard deviation $\sigma$. The 
\qut{central limiting theorem} then
tells us, that the limiting distribution
function is a Gaussian.
\begin{quotation}
{\it Central Limiting Theorem.\enspace}
Given $i=1,\dots,N$ independent random variables $x_i$,
distributed with mean $\mu_i$ and standard deviations
$\sigma_i$. The cumulative distribution $x=\sum_i x_i$
is then described, for $N\to\infty$, by a Gaussian
with mean $\mu=\sum_i \mu_i$ and variance
$\sigma^2=\sum_i\sigma_i^2$.
\end{quotation}
In most cases one is not interested in the cumulative 
result, but in the averaged one, which is obtained
by rescaling of variables
$$
y\ =\ x/N,\quad\qquad 
\bar\mu\ =\ \mu/N,\qquad\quad
\bar\sigma\ =\ \sigma/N,
\qquad\quad
p(y)\ =\ {1\over \bar\sigma\sqrt{2\pi}}\,
\mathrm{e}^{-{(y-\bar\mu)^2\over 2\bar\sigma^2}}~.
$$
The rescaled standard deviation scales with
$1/\sqrt{N}$. To see this, just consider
identical processes with $\sigma_i\equiv\sigma_0$,
$$
\bar\sigma\ =\ {1\over N}\sqrt{\sum_i\sigma_i^2}\ =\
{\sigma_0\over\sqrt{N}}~,
$$
in accordance with the law of large numbers.

\runinhead{Is Everything Boring Then?}
One might be tempted to draw the conclusion
that systems containing a large number of
variables are boring, since everything seems 
to average out. This is actually not the case,
the law of large numbers holds only for
statistically independent processes. 
Subsystems of distributed complex systems 
are however dynamically dependent and 
these dynamical correlations may lead to
highly non-trivial properties in the
thermodynamic limit.

\begin{figure}[t]
\centering
\includegraphics[width=0.6\textwidth]{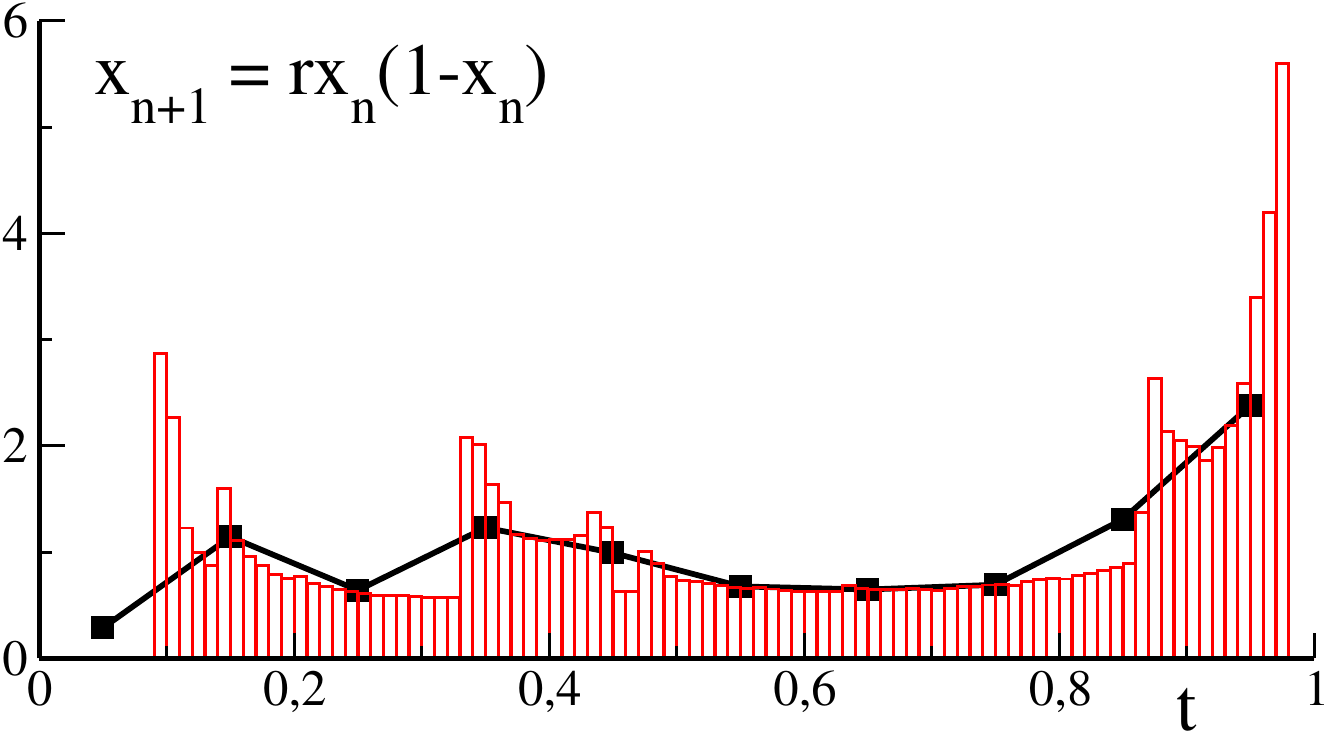}\hspace{2ex}
\includegraphics[width=0.35\textwidth]{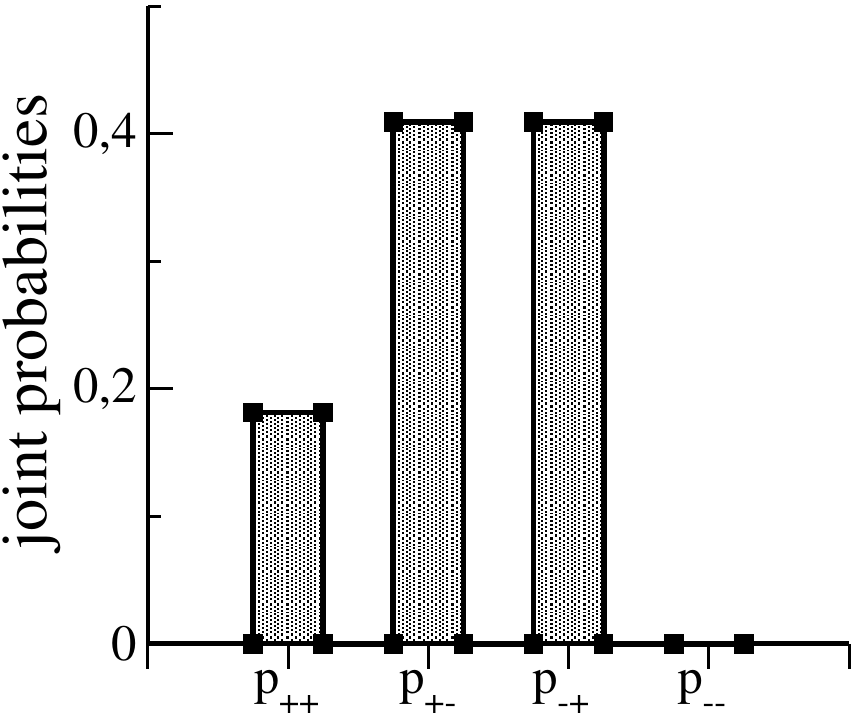}
\vspace*{0.0cm} \caption{
For the logistic map with $r=3.9$ and $x_0=0.6$, two statistical
analyses of the time series $x_n$, $n=0,\dots,N$, with $N=10^6$.
\textit{Left}: The distribution $p(x)$ of the $x_n$.
Plotted is $N_{bin}p(x)/N$, for $N_{bin}=10/100$ bins 
(curve with square symbols and open vertical bars respectively).
The data is plotted at the midpoints of the respective bins.
\textit{Right}: The joint probabilities $p_{\pm\pm}$, as defined by 
Eq.\ (\ref{complex1_joint_PDF_symbol}), of consecutive increases/decreases
of the $x_n$. The probability $p_{--}$ that the data decreases 
consecutively twice vanishes.
}
\label{complex_fig_PDF_logistic}
\end{figure}

\subsection{Time Series Characterization}
\label{complex1_subsec_time_series_characterization}
\index{time series|textbf}

In many cases one is interested in
estimating the probability distribution 
functions for data generated by some
known or unknown process, like the
temperature measurements of a weather
station. It is important, when doing so,
to keep a few caveats in mind.

\runinhead{Binning of Variables}
Here we will be dealing mainly with
the time series of data generated by
dynamical systems. As an example we
consider the logistic map, compare
Sect.\ \ref{sect_chaos_logistic_map},
\begin{equation}
x_{n+1}\ =\ f(x_n)\ \equiv\ r\,x_n\,(1-x_n),
\qquad\quad x_n\in[0,1],
\qquad\quad r\in[0,4]~.
\label{eq_complex_logistic_map}
\end{equation}
The dynamical variable is continuous
and in order to estimate the probability
distribution of the $x_n$ we need to
bin the data. In Fig.~\ref{complex_fig_PDF_logistic}
the statistics of a time series in the
chaotic regime, for $r=3.9$, is given.

One needs to select the number of bins $N_{bin}$ 
and, in general, also the positions and the widths 
of the bins. When the data is not uniformly distributed
one may place more bins in the region of interest,
generalizing the relation 
(\ref{complex1_PDF_continuous_discrete})
through $\Delta x\to\Delta x_i$, with the
$\Delta x_i$ being the width of the individual
bins.

For our illustrative example see 
Fig.~\ref{complex_fig_PDF_logistic}, we have selected
$N_{bin}=10/100$ equidistant bins. The data is
distributed over more bins, when $N_{bin}$ 
increases. In order to make the PDFs for
different number of bins comparable one needs 
to rescale them with $N_{bin}$, as it has
been done for the data shown in 
Fig.~\ref{complex_fig_PDF_logistic}.

The selection of the binning procedure is in
general a difficult choice. Fine structure
will be lost when $N_{bin}$ is too low, but
statistical noise will dominate for a too
large number of bins.

\runinhead{Symbolization}
\index{time series!symbolization}
One denotes by \qut{symbolization} the
construction of a finite number of symbols
suitable for the statistical characterization
of a given time series\footnote{For continuous-time 
data, as for an electrocardiogram, an additional 
symbolization step is necessary, the discretization of 
time. Here we consider however only discrete-time series.}.
The binning procedure discussed above is a commonly 
used symbolization procedure.

For a further example of a symbolization 
procedure we denote with $\delta_t=\pm1$,
\begin{equation}
\delta_t\ =\ \mathrm{sign}(x_t-x_{t-1}) \ =\
\left\{
\begin{array}{rcl}
1 &\quad& x_t>x_{t-1} \\
-1 &\quad& x_t<x_{t-1} 
\end{array}
\right.
\label{complex1_symbol_delta}
\end{equation}
the direction of the time development. 
The consecutive development of the
$\delta_t$ may then be encoded in higher-level
symbolic stochastic variables. E.g.\ one might
be interested in  the joint probabilities
\begin{equation}
\begin{array}{rclcrcl}
p_{++} &=& \langle p(\delta_t=1,\delta_{t-1}=1) \rangle_t&\qquad&
p_{+-} &=& \langle p(\delta_t=1,\delta_{t-1}=-1) \rangle_t\\
p_{-+} &=& \langle p(\delta_t=-1,\delta_{t-1}=1) \rangle_t&\qquad&
p_{--} &=& \langle p(\delta_t=-1,\delta_{t-1}=-1) \rangle_t 
\end{array}~,
\label{complex1_joint_PDF_symbol}
\end{equation}
where $p_{++}$ gives the probability that the
data increases at least twice consecutively, etc., and
where $\langle\dots\rangle_t$ denotes the time average.
In Fig.~\ref{complex_fig_PDF_logistic} the values for the
joint probabilities $p_{\pm\pm}$ are given for a selected
time series of the logistic map in the chaotic regime. 
The data never decreases twice consecutively, $p_{--}=0$,
a somewhat unexpected result. 

There are many possible symbolization procedures and the
procedure used to analyze a given time series determines 
the kind of information one may hope to extract, as evident 
from the results illustrated in Fig.~\ref{complex_fig_PDF_logistic}.
The selection of the symbolization procedures needs to be 
given attention, and will be discussed further in
Sect.~\ref{complex1_subsec_information_content_of_real_world_time_series}.

\runinhead{Self Averaging}
\index{time series!self averaging}
A time series produced by a dynamical system depends on the
initial condition and so will generally also the statistical
properties of the time series. As an example we consider the
XOR series\footnote{Remember, that 
$\mathrm{XOR}(0,0)=0=\mathrm{XOR}(1,1)$ and
$\mathrm{XOR}(0,1)=1=\mathrm{XOR}(1,0)$.} 
\index{time series!XOR}
\begin{equation}
\sigma_{t+1} \ =\ \mathrm{XOR}(\sigma_t,\sigma_{t-1}),
\qquad\quad \sigma_t=0,\,1~.
\label{complex1_time_series_XOR}
\end{equation}
The four initial conditions $00$, $01$, $10$ and $11$ give
rise to the respective time series
\begin{equation}
\begin{array}{c}
\dots0000000\underline{00} \qquad\quad
\dots1011011\underline{01} \\
\dots1101101\underline{10} \qquad\quad
\dots0110110\underline{11}
\end{array}
\label{complex1_XOR_series}
\end{equation}
where time runs from right to left and where we have underlined
the initial condition $\sigma_1$ and $\sigma_0$. The typical
time series, occurring for 75\% of the initial conditions, 
is $..011011011011..$, with $p(0)=1/3$ and $p(1)=2/3$ for
the probability to find a $0/1$. When averaging over all
four initial conditions, we have on the other hand
$(2/3)(3/4)=1/2$ for the probability to find a $1$. Then
$$
p(1) \ =\ \left\{
\begin{array}{rcl}
2/3 &\quad&\mathrm{typical}\\
1/2 &\quad&\mathrm{average}
\end{array}~.
\right.
$$
When observing a single time series we are likely 
to obtain the typical probability, analyzing many 
time series will result on the other hand
in the average probability.
\begin{quotation}
{\it Self Averaging.\enspace}
\index{time series!self averaging}
When the statistical properties of a time series generated by
a dynamical process are independent of the respective
initial conditions, one says the time series is
\qut{self averaging}.
\end{quotation}
The XOR series is not self averaging and one can generally not
assume self averaging to occur. An inconvenient situation whenever
only a single time series is available, as it is the case for 
most historical data, e.g.\ of past climatic conditions.

\runinhead{XOR Series with Noise}
Most real-world processes involve a certain degree 
of noise and one may be tempted to assume, that 
noise could effectively restart the dynamics, 
leading to an implicitly averaging over initial 
conditions. This assumption is not generally 
valid but works out for XOR process with noise,
\begin{equation}
\sigma_{t+1} \ =\ \left\{ 
\begin{array}{rcl}
\mathrm{XOR}(\sigma_t,\sigma_{t-1}) &\quad&
\mathrm{probability}\ 1-\xi \\
\neg\,\mathrm{XOR}(\sigma_t,\sigma_{t-1}) &\quad&
\mathrm{probability}\ \xi 
\end{array}
\right.
\qquad\quad 0\le\xi\ll1~.
\label{complex1_XOR_noise}
\end{equation}
For low level of noise, $\xi\to0$, the time series 
$$
\dots000000\underline{0}011011011\underline{0}10110110110
\underline{1}1101101101\underline{1}00000000\dots
$$
has stretches of regular behavior interseeded by four types 
of noise induced dynamics (underlined, time running from 
right to left). Denoting with $p_{000}$ and $p_{011}$ 
the probability of finding regular dynamics of type
$..000000000..$ and $..011011011..$ respectively,
we find the master equation
\begin{equation}
\dot p_{011} \ = \ \xi p_{000}\,-\,\xi p_{011}/3 \ =\ 
-\dot p_{000} 
\label{complex1_XOR_noise_master}
\end{equation}
for the noise-induced transition probabilities. 
In the stationary case $p_{000}=p_{011}/3$ for 
the XOR process with noise, the same ratio one 
would obtain for the deterministic XOR series averaged
over the initial conditions. 

The introduction of noise generally introduces a complex dynamics
akin to the master Eq.\ (\ref{complex1_XOR_noise_master}) and it
is generally not to be expected that the time series becomes such
self-averaging. A simple counter example is the OR time series;
we leave its analysis to the reader.

\runinhead{Time Series Analysis and Cognition}
\index{cognitive system!time series analysis}
Time series analysis is a tricky business whenever the fundamentals
of the generative process are unknown, e.g.\ whether noise is
important or not. This is however the setting in which cognitive
systems, see Chap.~\ref{chap_cogSys1}, are operative.
Our sensory organs, eyes and ears, provide us with a continuous 
time series encoding environmental information. Performing an
informative time series analysis is paramount for surviving.

\section{Entropy and Information}
\label{complex_entropy_information}
\index{entropy|textbf}

\index{law!second, of thermodynamics}
Entropy is a venerable concept from physics encoding 
the amount of disorder present in a thermodynamic system 
at a given temperature. The \qut{Second Law of Thermodynamics} 
states, that entropy can only increase in an isolated 
(closed) system. The second law has far reaching 
consequences, e.g.\ determining the maximal efficiency
of engines and power plants, and philosophical implications for
our understanding of the fundamentals underpinning the nature of
life as such.

\runinhead{Entropy and Life}
\index{entropy!life}
Living organisms have a body and such create ordered structures
from basic chemical constituents. Living beings therefore decrease
entropy locally, in their bodies, seemingly in violation of the
second law. In reality, the local entropy depressions are created
on the expense of corresponding entropy increases in the environment,
in agreement with the second law of thermodynamics. All living
beings need to be capable of manipulating entropy.

\runinhead{Information Entropy and Predictability}
\index{Shannon entropy!predictability} 
\index{Shannon entropy} 
\index{entropy!Shannon} 
\index{entropy!information} 
Entropy is also a central concept in information theory,
where it is commonly denoted \qut{Shannon entropy} or
\qut{information entropy}. In this context one is interested
in the amount of information encoded by a sequence of symbols
$$
\dots \sigma_{t+2},\ \sigma_{t+1},\
\sigma_t,\ \sigma_{t-1},\ \sigma_{t-2},\ \dots~,
$$
e.g.\ when transmitting a message. Typically, in everyday
computers, the $\sigma_t$ are words of bits. Let us
consider two time series of bits, e.g.\
\begin{equation}
\dots 101010101010\dots, \quad\qquad
\dots 1100010101100\dots~.
\label{complex1_time_series_entropy_examples}
\end{equation}
The first example is predictable, from the perspective of
a time-series, and ordered, from the perspective of an
one-dimensional alignment of bits. The second example is
unpredictable and disordered respectively.

Information can be transmitted through a time series of symbols 
only when this time series is not predictable. Talking to a friend,
to illustrate this statement, we will not learn anything new when
capable of predicting his next joke. We have therefore
the following two perspectives,
$$
\mathrm{high\ entropy} \ \hat{=}\
\left\{
\begin{array}{rcl}
\mathrm{large\ disorder} &\quad& \mathrm{physics}\\
\mathrm{high\ information\ content} &\quad& \mathrm{information\ theory}\\
\end{array}
\right. ~,
$$
and vice versa. Only seemingly disordered sequences of 
symbols are unpredictable and thus potential carriers 
of information. Note, that the predictability of
a given time series, or its degree of disorder, may 
not necessarily be as self evident as in above 
example, Eq.\ (\ref{complex1_time_series_entropy_examples}),
depending generally on the analysis procedure used, see
Sect.~\ref{complex1_subsec_information_content_of_real_world_time_series}.

\runinhead{Extensive Information}
In complex system theory, as well as in physics,
we are often interested in properties of systems 
composed of many subsystems.
\begin{quotation}
{\it Extensive and Intensive Properties.\enspace}
For systems composed of $N$ subsystems a property
is denoted \qut{extensive} if it scales as $O(N^1)$
and \qut{intensive} when it scales with $O(N^0)$.
\end{quotation}
A typical extensive property is the mass, a typical
intensive property the density. When lumping together 
two chunks of clay, their mass adds,
but the density does not change. 

One demands, both in physics and in information theory,
that the entropy should be an extensive quantity. The
information content of two independent transmission channels
should be just the sum of the information carried by
the two individual channels.

\begin{figure}[t]
\centering
\includegraphics[width=0.44\textwidth]{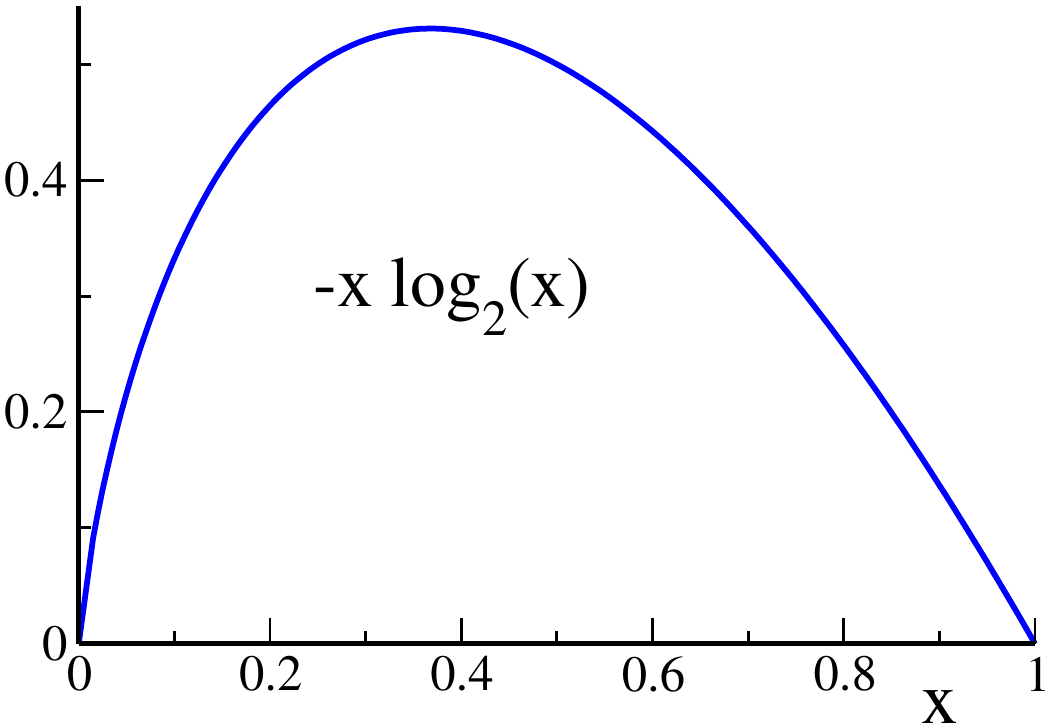}\hspace{1ex}
\includegraphics[width=0.44\textwidth]{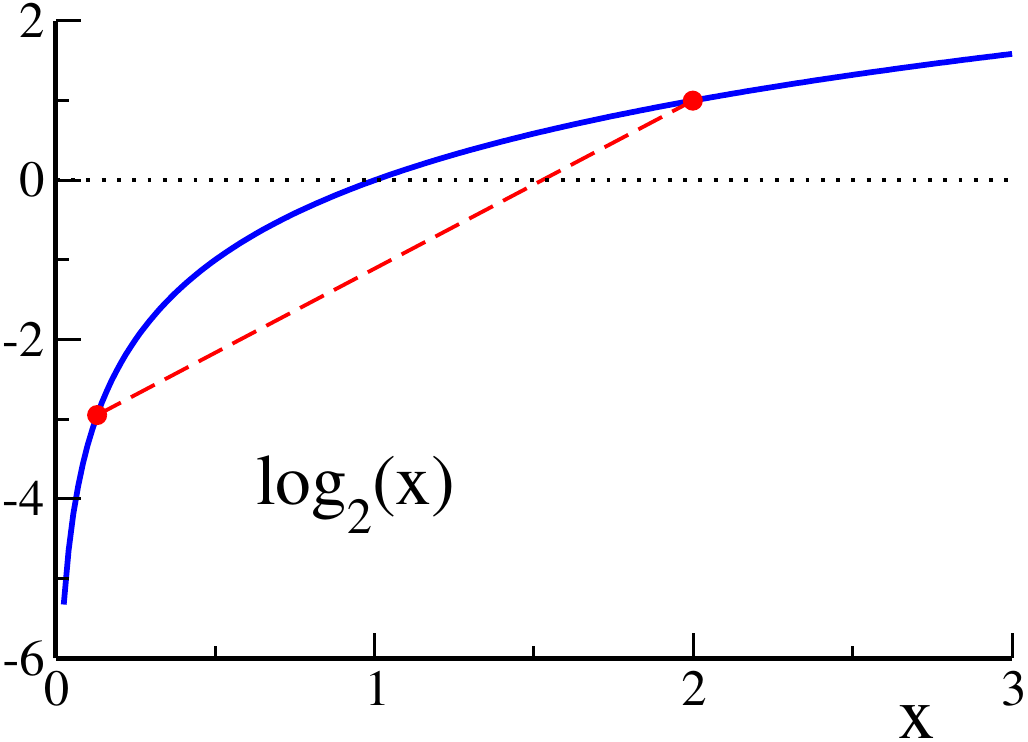}
\vspace*{0.0cm} \caption{\textit{Left}: Plot of
$-x\,\log_2(x)$. \textit{Right}: The logarithm $\log_2(x)$
(full line) is concave, every cord (dashed line) lies
below the graph.
}
\label{complex_entropy_plots_log}
\end{figure}

\runinhead{Shannon Entropy}
\index{Shannon entropy|textbf}
The Shannon entropy $H[p]$ is defined by
\begin{equation}
H[p] \ =\ -\sum_{x_i} p(x_i)\,\log_b(p(x_i))
 \ =\ -\langle\,\log_b(p)\,\rangle,
\qquad\quad
H[p] \ge 0 ~,
\label{complex1_Shannon_entropy}
\end{equation}
where $p(x_i)$ is a normalized discrete probability 
distribution function and where the brackets 
in $H[p]$ denote the functional
dependence\footnote{A function $f(x)$ is a 
function of a variable $x$; a functional $F[f]$
is, on the other hand, functionally dependent on a
function $f(x)$. In formal texts on information 
theory the notation $H(X)$ is often used for the 
Shannon entropy and a random variable $X$ with
probability distribution $p_X(x)$.}.
Note, that $-p\log(p)\ge0$ for $0\le p\le 1$, 
see Fig.~\ref{complex_entropy_plots_log},
the entropy is therefore strictly positive.

$b$ is the base of the logarithm used in
Eq.\ (\ref{complex1_Shannon_entropy}). Common values of $b$ are 2, 
Euler's number $e$ and $10$. The corresponding units of entropy
are then termed \qut{bit} for $b = 2$, 
\qut{nat} for $b = e$ and \qut{digit} for $b = 10$.
In physics the natural logarithm is always used and
there is an additional constant (the Boltzmann constant
$k_B$) in front of the definition of the entropy. Here we
will use $b=2$ and drop in the following the index $b$.

\runinhead{Extensiveness of the Shannon Entropy}
The $\log$-dependence in the definition of the information
entropy in Eq.\ (\ref{complex1_Shannon_entropy}) is necessary
for obtaining an extensive quantity. To see this,
let us consider a system composed of two independent
subsystems. The joint probability distribution is 
multiplicative, 
$$
p(x_i,y_j)\ =\ p_X(x_i)p_Y(y_j), \qquad\quad
\log(p(x_i,y_j)) \ =\ \log(p_X(x_i))\,+\, \log(p_Y(y_j))~.
$$
The logarithm is the only function which maps a
multiplicative input onto an additive output.
Consequently, 
\begin{eqnarray*}
H[p] & =& -\sum_{x_i,y_j}p(x_i,y_j)\log(p(x_i,y_j)) \\
& =& -\sum_{x_i,y_j} p_X(x_i)p_Y(y_j)\,\Big[\,\log(p_X(x_i))+\log(p_Y(y_j))\,\Big] \\
& =& -\sum_{x_i} p_X(x_i)\,\sum_{y_j} p_Y(y_j)\log(p_Y(y_j))
\, -\, \sum_{y_j} p_Y(y_j)\,\sum_{x_i} p_X(x_i)\log(p_X(x_i)) \\
&=& H[p_Y]\,+\, H[p_X]~,
\end{eqnarray*}
as necessary for the extensiveness of $H[p]$.
Hence the $\log$-dependence in 
Eq.\ (\ref{complex1_Shannon_entropy}).

\runinhead{Degrees of Freedom}
\index{entropy!vs.\ degrees of freedom}
We consider a discrete system
with $x_i\in[1,\dots,n]$, having
$n$ \qut{degrees of freedom} in
physics' slang. If the probability 
of finding any value is equally likely,
as it is the case for a thermodynamic system
at infinite temperatures, the entropy is
\begin{equation}
H\ =\ -\sum_{x_i} p(x_i)\log(p(x_i)) \ =\ 
-n{1\over n}\log(1/n) \ =\ \log(n)~,
\label{complex1_entropy_degrees_freedom}
\end{equation}
a celebrated result. The entropy grows
logarithmically with the number of 
degrees of freedom.

\runinhead{Shannon's Source Coding Theorem}
So far we have shown, that 
Eq.\ (\ref{complex1_Shannon_entropy}) 
is the only possible definition, modulo 
renormalizing factors, for an extensive
quantity depending exclusively on the probability
distribution. The operative significance of the
entropy $H[p]$ in terms of informational content 
is given by Shannon's theorem.
\begin{quotation}
{\it Source Coding Theorem.\enspace}
\index{theorem!source coding}
Given a random variable $x$ with a PDF $p(x)$ 
and entropy $H[p]$. The cumulative entropy $NH[p]$ 
is then, for $N\to\infty$, a lower bound for the 
number of bits necessary when trying to compress $N$ 
independent processes drawn from $p(x)$.
\end{quotation}
If we compress more, we will lose information, the
entropy $H[p]$ is therefore a measure of information content.

\runinhead{Entropy and Compression}
Let's make an example. Consider we have words made out of
the four letter alphabet $A$, $B$, $C$ and $D$. Suppose, that
these four letters would not occur with the same probability, 
the relative frequencies being
$$
p(A)={1\over2},\qquad
p(B)={1\over4},\qquad
p(C)={1\over8}=p(D)~.
$$
When transmitting a long series of words using this alphabet
we will have the entropy
\begin{eqnarray}
\nonumber
H[p] & =& {-1\over2}\log(1/2)
\,-\,{1\over4}\log(1/4)
\,-\,{1\over8}\log(1/8)
\,-\,{1\over8}\log(1/8) \\
&=& {1\over2}\,+\,{2\over4}\,+\,
    {3\over8}\,+\,{3\over8}
\ =\ 1.75~,
\label{complex1_example_encoding_entropy}
\end{eqnarray}
since we are using the logarithm with base $b=2$.
The most naive bit encoding,
$$
A\to 00,\qquad
B\to 01,\qquad
C\to 10,\qquad
D\to 11~,
$$
would use exactly two bit, which is larger than the
Shannon entropy. An optimal encoding would be,
on the other hand,
\begin{equation}
A\to 1,\qquad
B\to 01,\qquad
C\to 001,\qquad
D\to 000~,
\label{complex1_example_encoding_optimal}
\end{equation}
leading to an average length of words transmitted
of
\begin{equation}
p(A)\,+\,2p(B)\,+\,3p(C)\,+\,3p(D) \ =\ 
{1\over2}\,+\,{2\over4}\,+\,{3\over 8}\,+\,{3\over8} \ =\ 1.75~,
\label{complex1_example_encoding_word_length}
\end{equation}
which is the same as the information entropy $H[p]$.
The encoding given in Eq.\ (\ref{complex1_example_encoding_optimal})
is actually \qut{prefix-free}. When we read the words
from left to right, we know where a new word starts and
stops,
$$
110000010101\qquad\longleftrightarrow\qquad
AADCBB~,
$$ 
without ambiguity. Fast algorithms for optimal,
or close to optimal encoding are clearly of importance
in the computer sciences and for the compression of 
audio and video data.

\runinhead{Discrete vs.\ Continuous Variables}
When defining the entropy we have considered
hitherto discrete variables. The information entropy
can also be defined for continous variables.
We should be careful though, being aware
that the transition from continuous to discrete
stochastic variables, and vice versa, is slightly 
non-trivial, compare 
Eq.\ (\ref{complex1_PDF_continuous_discrete}):
\begin{eqnarray}
\nonumber
H[p]\Big|_{\mathrm{con}} & =&
-\int p(x)\log(p(x))\,dx\ \approx\
\sum_i p(x_i)\log(p(x_i)))\,\Delta x \\
&=& -\sum_i p_i\log(p_i/\Delta x) \ =\ -\sum_i p_i\log(p_i)
    \,+\,\sum_i p_i\log(\Delta x) 
\nonumber \\
&=& H[p]\Big|_{\mathrm{dis}} \,+\, \log(\Delta x)~,
\label{complex1_entropy_continuous_discrete}
\end{eqnarray}
where $p_i=p(x_i)\Delta x$ is here the properly normalized
discretized PDF, compare Eq.\ (\ref{complex1_PDF_continuous_discrete}). 
The difference $\log(\Delta x)$ between the continuous-variable entropy 
$H[p]\big|_{\mathrm{con}}$ and the discretized version
$H[p]\big|_{\mathrm{dis}}$ diverges for $\Delta x\to 0$,
the transition is discontinuous. 

\runinhead{Entropy of a Continuous PDF}
From Eq.\ (\ref{complex1_entropy_continuous_discrete})
it follows, that the Shannon entropy 
$H[p]\big|_{\mathrm{con}}$ can be negative
for a continous probability distribution function. As 
an example consider the flat distribution
$$
p(x) \ =\ \left\{
\begin{array}{rcl}
1/\epsilon &\quad& \mathrm{for}\ x\in[0,\epsilon]\\
0&\quad& \mathrm{otherwise}
\end{array}
\right.,
\qquad\quad \int_0^{\epsilon} p(x)\,dx\,=\,1
$$
in the small interval $[0,\epsilon]$,
with the entropy
$$
H[p]\Big|_{\mathrm{con}} \ =\ -\int_0^{\epsilon}
{1\over\epsilon}\log(1/\epsilon)\,dx \ =\ \log(\epsilon)\ <\ 0,
\qquad\quad \mbox{for}\quad \epsilon\,<\,1~.
$$
The absolute value of the entropy is hence not meaningful 
for continous PDFs, only entropy differences. 
$H[p]\big|_{\mathrm{con}}$ is therefore also 
referred-to as \qut{differential entropy}.
\index{entropy!differential}

\runinhead{Maximal Entropy Distributions}
Which kind of distributions maximize entropy, 
viz information content? Remembering that
$$
\lim_{p\to0,1} p\log(p)\ =\ 0,
\qquad\quad
\log(1)\ =\ 0~,
$$
see Fig.~\ref{complex_entropy_plots_log},
it is intuitive that a flat distribution might
be optimal. This is indeed correct in the absence
of any further constraints. We consider three cases.
\begin{itemize}
\item[--] No constraint: we need to maximize
\begin{equation}
H[p] \ =\ \int f(p(x))\,dx, \qquad\quad
f(p) \ =\ -p\log(p)~,
\label{complex1_H_f}
\end{equation}
where the notation used will turn out useful
later on. Maximizing a functional like $H[p]$ is
a typical task of variational calculus. One
considers with
$$
p(x) \ =\ p_{opt}(x)\,+\,\delta p(x),
\qquad\quad
\delta p(x)\ \mbox{arbitrary}
$$
a general variation of $\delta p(x)$ around the
optimal function $p_{opt}(x)$. At optimality, the
dependence of $H[p]$ on the variation $\delta p$
should be stationary,
\begin{equation}
0\ \equiv\ \delta H[p] \ =\ 
\int f'(p)\,\delta p\,dx, \qquad\quad
0\ =\ f'(p)~,
\label{complex1_variational_calculus}
\end{equation}
where $f'(p)=0$ follows from the fact that
$\delta p$ is an arbitrary function. For
$f(p)=-p\log(p)$ we find then with
\begin{equation}
f'(p)\ =\ -\log(p) -1\ =\ 0,
\quad\quad
p(x)\ =\ \mathrm{const.}
\label{complex1_maximal_entropy_free}
\end{equation}
the expected flat distribution.

\item[--] Fixed mean: next we consider the
entropy maximization under the constraint of
fixed average $\mu$,
\begin{equation}
\mu\ =\ \int x\,p(x)\,dx~.
\label{complex1_PDF_maximal_entropy_fixed_mu_condition}
\end{equation}
This condition can be enforced by a Lagrange
parameter $\lambda$ via
$$
f(p)\ =\ -p\log(p)\,-\,\lambda xp~.
$$
The stationary condition $f'(p)=0$ then leads to
\begin{equation}
f'(p) \ =\ -\log(p)-1-\lambda x\ =\ 0,
\qquad\quad
p(x)\ \propto\ 2^{-\lambda x}\ \sim\ \mathrm{e}^{-x/\mu}
\label{complex1_maximal_entropy_mean}
\end{equation}
the exponential distribution, see Eq.\ (\ref{complex1_exponential}),
with mean $\mu$. The Lagrange parameter $\lambda$ needs to
be determined such that the condition of fixed mean,
Eq.\ (\ref{complex1_PDF_maximal_entropy_fixed_mu_condition}),
is satisfied. For a support $x\in[0,\infty]$, as assumed
above, we have $\lambda\log_e(2)=1/\mu$.

\item[--] Fixed mean and variance: Lastly we consider the
entropy maximization under the constraint of
fixed average $\mu$ and variance $\sigma^2$,
\begin{equation}
\mu\ =\ \int x\,p(x)\,dx,
\qquad\quad
\sigma^2\ =\ \int (x-\mu)^2\,p(x)\,dx~.
\label{complex1_PDF_maximal_entropy_fixed_mu_sigma_condition}
\end{equation}
We leave it to the reader to show that the entropy is
the maximal for a Gaussian.
\end{itemize}
%

\subsection{Information Content of a Real-World Time Series}
\label{complex1_subsec_information_content_of_real_world_time_series}

The Shannon entropy is a very powerful concept in
information theory. The encoding rules are typically 
known in information theory, there is no ambiguity 
regarding the symbolization procedure
(see Sect.~\ref{complex1_subsec_time_series_characterization})
to employ when receiving a message via some technical
communication channel. This is however not any more 
the case, when we are interested in determining the 
information content of real-world processes, 
e.g.\ the time series of certain financial data or 
the data stream produced by our sensory organs.

\runinhead{Symbolization and Information Content}
The result obtained for the information content of a 
real-world time series $\{\sigma_t\}$ depends in 
general on the symbolization procedure used. Let us 
consider, as an example, the first time series of 
Eq.\ (\ref{complex1_time_series_entropy_examples}),
\begin{equation}
\dots 101010101010\dots~.
\label{complex1_time_series_01}
\end{equation}
When using a one-bit symbolization procedure, we have
$$
p(0) = {1\over2} = p(1), \qquad\quad
H[p]\ =\ -2\,{1\over2}\log(1/2) \ =\ 1~,
$$
as expected. If, on the other hand, we use a two-bit
symbolization, we find
$$
p(00) = p(11) = p(01) = 0,\qquad
p(10) = 1,\qquad\quad
H[p] \ =\ -\log(1) \ =\ 0~.
$$
When two-bit encoding is presumed, the time series is
predictable and carries no information. This seems
intuitively the correct result and the question is:
Can we formulate a general guiding principle which
tells us which symbolization procedure would yield
the more accurate result for the information content
of a given time series?

\runinhead{The Minimal Entropy Principle}
\index{entropy!principle of minimal}
The Shannon entropy constitutes a lower bound for
the number of bits, per symbol, necessary when
compressing the data without loss of information.
Trying various symbolization procedures, the symbolization
procedure yielding the lowest information entropy then
allows us to represent, without loss of information, 
a given time series with the least number of bits.

\begin{quotation}
{\it Minimal Entropy Principle.\enspace}
The information content of a time series with unknown
encoding is given by the minimum (actually the 
infimum) of the Shannon entropy over all 
possible symbolization procedures.
\end{quotation}
The minimal entropy principle then gives us a
definite answer with respect to the information
content of the time series given in
Eq.\ (\ref{complex1_time_series_01}).
We have seen that at least one symbolization procedure 
yields a vanishing entropy and one cannot get
a lower value, since $H[p]\ge0$. This is the expected
result, since $..01010101..$ is predictable.

\runinhead{Information Content of a Predictable Time Series}
Note, that a vanishing information content $H[p]=0$
only implies that the time series is strictly 
predictable, not that it is constant. One therefore
needs only a finite amount of information to encode
the full time series, viz for arbitrary lengths $N\to\infty$.
When the time series is predictable, the information
necessary to encode the series is intensive and not 
extensive.

\runinhead{Symbolization and Time Horizons}
The minimal entropy principle is rather
abstract. In practice one may not be able than to
try out more than a handful of different symbolization
procedures. It is therefore important to gain an 
understanding of the time series at hand.

An important aspect of many time series is the intrinsic
time horizon $\tau$. Most dynamical processes have certain
characteristic time scales and memories of past states
are effectively lost for times exceeding these intrinsic
time scales. The symbolization procedure used should
therefore match the time horizon $\tau$

This is what happened when analyzing the time series 
given in Eq.\ (\ref{complex1_time_series_01}), for 
which $\tau=2$. A one-bit symbolization procedure
implicitly presumes that $\sigma_t$ and
$\sigma_{t+1}$ are statistically independent and
such missed the intrinsic time scale $\tau=2$,
in contrast to the two-bit symbolization procedure.

\subsection{Mutual Information}
\label{complex_mutual_information}
\index{mutual information|textbf}
\index{information!mutual|textbf}

We have been considering so far the statistical
properties of individual stochastic processes as well
as the properties of cumulative processes generated by 
the sum of stochastically independent random variables. 
In order to understand complex systems we need to develop 
tools for the description of a large number of interdependent
processes. As a first step towards this direction we consider
in the following the case of two stochastic processes, which 
may now be statistically correlated.

\runinhead{Two Channels - Markov Process}
We start by considering an illustrative example of
two correlated channels $\sigma_t$ and $\tau_t$, with
\begin{equation}
\sigma_{t+1}\ =\ XOR(\sigma_t,\tau_t),
\qquad
\tau_{t+1}\ =\ \left\{
\begin{array}{rcl}
XOR(\sigma_t,\tau_t) &\quad& \mathrm{probability}\ 1-\xi \\
\neg XOR(\sigma_t,\tau_t) &\quad& \mathrm{probability}\ \xi 
\end{array}
\right.~.
\label{complex1_two_channels}
\end{equation}
\index{Markov!property} 
This dynamics has the \qut{Markov property}, the
value for the state $\{\sigma_{t+1},\tau_{t+1}\}$ 
depends only on the state at the previous time step,
viz on $\{\sigma_{t},\tau_{t}\}$.
\begin{quotation}
{\it Markov Process.\enspace}
\index{Markov!process} A discrete-time memory-less dynamical 
process is denoted a \qut{Markov process}. The likelihood of 
future states depends only on the present state, and not on 
any past states.
\end{quotation}
\index{Markov!chain}
When the state space is finite, as in our example, the
term \qut{Markov chain} is also used. 
We will not adhere here to the distinction which is
sometimes made between discrete and 
continuous time, with Markov processes 
being formulated for discrete time and 
\qut{master equations} describing stochastic 
processes for continuous time. 
\index{master equation} \index{equation!master}

\runinhead{Joint Probabilities}
A typical time series of the Markov chain
specified in Eq.\ (\ref{complex1_two_channels})
looks like
$$
\begin{array}{rcl}
\dots\sigma_{t+1}\sigma_t\dots &\quad :\quad &     
0\,0\,0\,1\,0\,0\,0\,0\,0\,0\,1\,0\,1\,0\dots \\
\dots\tau_{t+1}\tau_t\dots &\quad :\quad &         
0\,0\,0\,1\,\underline{1}\,0\,0\,0\,0\,0\,1\,\underline{1}\,1\,1\dots
\end{array}
~,
$$
where we have underlined instances of noise-induced
transitions. For $\xi=0$ the stationary state is
$\{\sigma_t,\tau_t\}=\{0,0\}$ and therefore fully correlated.
We now calculate the joint probabilities $p(\sigma,\tau)$ for 
general values of noise $\xi$, using the transition
probabilities
$$
\begin{array}{rcl}
 p_{t+1}(0,0) &= & (1-\xi) \left[ p_t(1,1)+p_t(0,0)\right] \\
 p_{t+1}(1,1) &= & (1-\xi) \left[ p_t(1,0)+p_t(0,1)\right] 
\end{array},
\qquad
\begin{array}{rcl}
 p_{t+1}(1,0) &= & \xi \left[ p_t(0,1)+p_t(1,0)\right] \\
 p_{t+1}(0,1) &= & \xi \left[ p_t(0,0)+p_t(1,1)\right] 
\end{array}~,
$$
for the ensemble averaged joint probability distributions
$p_t(\sigma,\tau)=\langle p(\sigma_t,\tau_t)\rangle_{ens}$,
where the average $\langle ..\rangle_{ens}$ denotes
the average over an ensemble of time series. For
the solution in the stationary case 
$p_{t+1}(\sigma,\tau)=p_t(\sigma,\tau)\equiv p(\sigma,\tau)$ 
we use the normalization
$$
p(1,1)\,+\,p(0,0)\,+\,p(1,0)\,+\,p(0,1) \ =\ 1~.
$$
We find 
$$
p(1,1)\,+\,p(0,0)\ = 1-\xi,
\qquad\quad
p(1,0)\,+\,p(0,1)\ = \xi~,
$$
by adding the terms $\propto(1-\xi)$ and $\propto\xi$
respectively. It then follows immediately
\begin{equation}
\begin{array}{rcl}
 p(0,0) &= & (1-\xi)^2 \\
 p(1,1) &= & (1-\xi)\xi
\end{array},
\quad\quad
\begin{array}{rcl}
 p(1,0) &= & \xi^2 \\
 p(0,1) &= & \xi(1-\xi)
\end{array}~.
\label{complex1_two_channels_master_joint_PDF}
\end{equation}
For $\xi=1/2$ the two channels become 100\% uncorrelated,
as the $\tau$-channel is then fully random.
The dynamics of the Markov process given in
Eq.\ (\ref{complex1_two_channels}) is self averaging
and it is illustrative to verify the result for the
joint PDF, 
Eq.\ (\ref{complex1_two_channels_master_joint_PDF}),
by a straightforward numerical simulation.

\begin{figure}[t]
\centering
\includegraphics[width=0.70\textwidth]{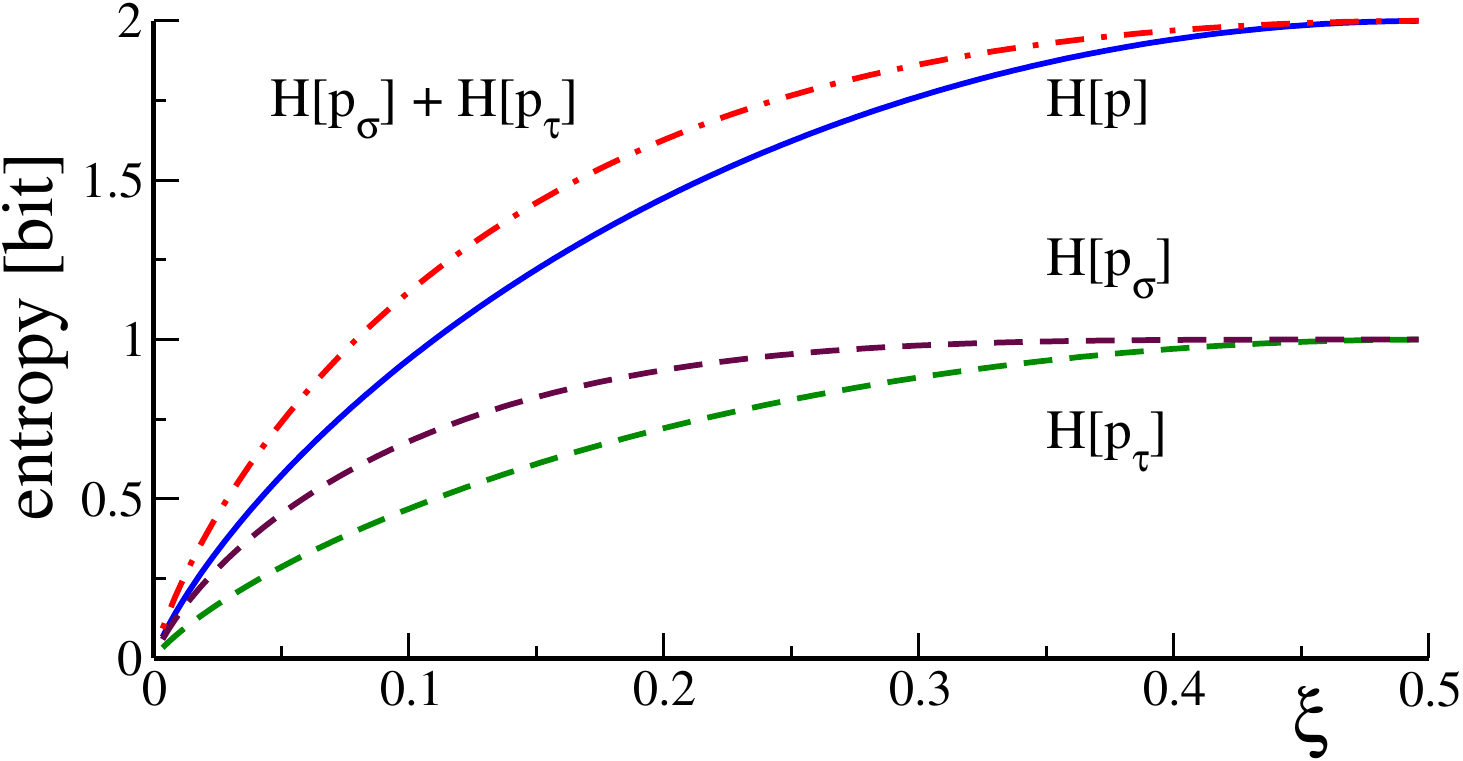}
\vspace*{0.0cm} \caption{For the two-channel XOR-Markov chain
$\{\sigma_t,\tau_t\}$ with noise $\xi$, 
see Eq.\ (\ref{complex1_two_channels}),
the entropy $H[p]$ of the combined process 
(full line, Eq.\ (\ref{complex1_two_channels_combined_entropy})),
of the individual channels (dashed lines,
Eq.\ (\ref{complex1_two_channels_individual_entropies})),
$H[p_\sigma]$ and $H[p_\tau]$, and of the sum of the
joint entropies (dot-dashed line). Note the positiveness of
the mutual information, $I(\sigma,\tau)=H[p_\sigma]+H[p_\tau]-H[p]>0$. 
}
\label{complex_fig_entropy}
\end{figure}

\runinhead{Entropies}
Using the notation
$$
p_\sigma(\sigma') \ =\ \sum_{\tau'}p(\sigma',\tau'),
\qquad\quad
p_\tau(\tau') \ =\ \sum_{\sigma'}p(\sigma',\tau')
$$
for the \qut{marginal PDFs} 
\index{probability distribution!marginal}
$p_\sigma$ and $p_\tau$, we find from 
Eq.\ (\ref{complex1_two_channels_master_joint_PDF})
\begin{equation}
\begin{array}{rcl}
p_\sigma(0)&=&1-\xi \\
p_\sigma(1)&=&\xi
\end{array},\qquad\quad
\begin{array}{rcl}
p_\tau(0)&=&1-2\xi(1-\xi) \\
p_\tau(1)&=&2\xi(1-\xi)
\end{array}
\label{complex1_two_channels_master_mariginal_PDF}
\end{equation}
for the PDFs of the two individual channels. We may
now evaluate both the entropies of the individual
channels, $H[p_\sigma]$ and $H[p_\tau]$, the
\qut{marginal entropies}, viz
\index{entropy!marginal}
\begin{equation}
H[p_\sigma] \ =\ -\langle\log(p_\sigma)\rangle,\quad\qquad
H[p_\tau] \ =\ -\langle\log(p_\tau)\rangle~,
\label{complex1_two_channels_individual_entropies}
\end{equation}
as well as the entropy of the combined process,
termed \qut{joint entropy},
\index{entropy!joint}
\begin{equation}
H[p]\ =\  -\sum_{\sigma',\tau'} p(\sigma',\tau')
\log(p(\sigma',\tau'))~.
\label{complex1_two_channels_combined_entropy}
\end{equation}
In Fig.~\ref{complex_fig_entropy} the respective entropies are
plotted as a function of noise strength $\xi$. Some observations:
\begin{itemize}
\item In the absence of noise, $\xi=0$, both the 
individual channels as well as the combined process are
predictable and all three entropies, $H[p]$, $H[p_\sigma]$
and $H[p_\tau]$, vanish consequently. 
\item For maximal noise $\xi=0.5$, the information content
of both individual chains is one bit and of the combined 
process two bits, implying statistical independence.
\item For general noise strengths $0<\xi<0.5$, the two channels are
statistically correlated. The information content of the
combined process $H[p]$ is consequently smaller than the
sum of the information contents of the individual channels,
$H[p_\sigma]+H[p_\tau]$.
\end{itemize}

\runinhead{Mutual Information}
The degree of statistical dependency of two channels
can be measured by comparing the joint entropy with
the respective marginal entropies.
\begin{quotation}
{\it Mutual Information.\enspace}
\index{mutual information} \index{information!mutual}
For two stochastic processes $\sigma_t$ and $\tau_t$ the difference
\begin{equation}
I(\sigma,\tau)\ =\ H[p_\sigma]+H[p_\tau] - H[p]
\label{complex1_def_mutual_information}
\end{equation}
between the sum of the marginal entropies $H[p_\sigma]+H[p_\tau]$
and the joint entropy $H[p]$ is denoted 
\qut{mutual information} $I(\sigma,\tau)$.
\end{quotation}
When two dynamical processes become correlated,
information is lost and this information loss 
is given by the mutual information. Note, that
$I(\sigma,\tau)=I[p]$ is a functional of
the joint probability distribution $p$ only,
the marginal PDFs $p_\sigma$ and $p_\tau$ being
themselves functionals of $p$.

\runinhead{Positiveness}
We will now discuss some properties of the mutual information,
considering the general case of two stochastic processes
described by the joint PDF $p(x,y)$ and the respective
marginal PDFs $p_X(x)=\int p(x,y) dy$, $p_Y(y)=\int p(x,y) dx$.

The mutual information
\begin{equation}
I(X,Y) \ =\ \langle\log(p)\rangle\,-\,
            \langle\log(p_X)\rangle\,-\,
            \langle\log(p_Y)\rangle\qquad\quad
I(X,Y) \ \ge 0~,
\label{complex1_mutual_information_positive}
\end{equation}
is strictly positive. Rewriting
the mutual information as
\begin{eqnarray}
\label{complex1_mutal_info_one_log}
I(X,Y)
 &= & \int p(x,y)\Big[\log(p(x,y)) - \log(p_X(x))
     -\log(p_Y(y))\Big]  dx\,dy\\
 &= & \int p(x,y)\log\left({p(x,y)\over p_X(x)p_Y(y)}\right) dx\,dy 
\ = \ -\int p\log\left({p_Xp_Y\over p}\right) dx\,dy~,
\nonumber
\end{eqnarray}
we can easily show that $I(X,Y)\ \ge\ 0
$ 
follows from the concaveness of the logarithm,
see Fig.~\ref{complex_entropy_plots_log},
\begin{equation}
\log(p_1 x_1+p_2 x_2) \ \ge\ 
p_1\log(x_1) + p_2\log(x_2),
\qquad\quad
\forall x_1,x_2\in[0,\infty]~,
\label{complex1_log_concave}
\end{equation}
and $p_1,p_2\in[0,1]$, with $p_1+p_2=1$; any cord
of a concave function lies below the graph.
We can regard $p_1$ and $p_2$ as the coefficients 
of a distribution function and generalize, 
$$
p_1\delta(x-x_1)+p_2\delta(x-x_2)
\quad\longrightarrow\quad p(x)~,
$$
where $p(x)$ is now a generic, properly normalized PDF. The
concaveness condition, Eq.\ (\ref{complex1_log_concave}),
then reads 
\begin{equation}
\log\left(\int p(x)\,x\,dx)\right) \ \ge\ 
\int p(x)\log(x)\,dx~,
\qquad\quad
\varphi\left(\langle x\rangle\right) \ \ge\ 
\langle\, \varphi(x)\,\rangle~,
\label{complex1_jensens_inequality}
\end{equation}
\index{Jensen inequality}
the \qut{Jensen inequality}, which holds for 
any concave function $\varphi(x)$. This inequality 
remains valid when substituting $x\to p_X p_Y/p$ 
for the argument of the logarithm\footnote{For a 
proof consider the generic substitution $x\to q(x)$
and a transformation of variables $x\to q$ 
via $dx=dq/q'$, with $q'=dq(x)/dx$, for the integration 
in Eq.\ (\ref{complex1_jensens_inequality}).}. 
We then obtain for the mutual 
information, Eq.\ (\ref{complex1_mutal_info_one_log}),
\begin{eqnarray*}
I(X,Y) &= & -\int p\log\left({p_Xp_Y\over p}\right) dx\,dy 
 \ \ge \ -\log\left(\int pp_Xp_Y/p\, dx\,dy\right) \\
 & = & -\log\left(\int p_X(x)\,dx \int p_Y(y)\,dy \right) 
 \ =\ -\log(1)\ =\ 0~,
\end{eqnarray*}
viz $I(X,Y)$ is non-negative. Information can only be loost
when correlating two previously independent processes.

\runinhead{Conditional Entropy}
There are various ways to rewrite the mutual
information, using Bayes theorem
$p(x,y)=p_|(x|y)p_Y(y)$ between the
joint PDF $p(x,y)$, the conditional
PDF $p_|(x|y)$ and the marginal PDF $p_Y(y)$, e.g.\
\begin{eqnarray*}
I(X,Y) & =& \left\langle\log\left({p\over p_X p_Y}\right)\right\rangle
       \ =\ \int p(x,y)\,\log\left({p(x|y)\over p_X(x)}\right)\,dx\,dy \\
       &\equiv& H[p_X] \,-\, H[p_|]~,
\end{eqnarray*}
\index{entropy!conditional}
where we have defined the \qut{conditional entropy} 
\begin{equation}
H[p_|]\ =\ -\langle\,\log(p_|)\,\rangle \ =\
-\int p(x,y)\log(p_|(x|y))\,dx\,dy~.
\label{complex1_def_conditional_entropy}
\end{equation}
The conditional entropy is positive for discrete
processes, since
$$
-p(x_i,y_j)\log(p_|(x_i|y_j))\ =\ -p_|(x_i|y_j)p_Y(y_j)\log(p_|(x_i|y_j)) 
$$
is positive, as $-p_|\log(p_|)\ge0$ in 
the interval $p_|\in[0,1]$, compare Fig.~\ref{complex_entropy_plots_log}
and Eq.\ (\ref{complex1_entropy_continuous_discrete})
for the change-over from continous to discrete variables.
Several variants of the conditional entropy may
be used to extend the statistical complexity
measures discussed in
Sect.~\ref{complex1_subsec_Complexity_Predictability}.

\runinhead{Kullback-Leibler Divergence}
\index{Kullback-Leibler divergence}
The mutual information, Eq.\ (\ref{complex1_mutal_info_one_log}),
is a special case of the \qut{Kullback-Leibler Divergence}
\begin{quotation}
{\it Kullback-Leibler Divergence.\enspace}
Given two probability distribution functions $p(x)$ and
$q(x)$ the functional
\begin{equation}
K[p;q] \ =\ \int p(x)\log\left({p(x)\over q(x)}\right)dx
\ \ge \ 0
\label{complex1_def_Kullback_Leibler_divergence}
\end{equation}
is a non-symmetric measure of the difference between
$p(x)$ and $q(x)$.
\end{quotation}
The Kullback-Leibler divergence $K[p;q]$ is also denoted
\qut{relative entropy}\index{entropy!relative}
and the proof for $K[p;q]\ge0$ is analogous to
the one for the mutual information given above. 
The Kullback-Leibler divergence vanishes for
$p(x)\equiv q(x)$.

\begin{figure}[t]
\centering
\includegraphics[width=0.44\textwidth]{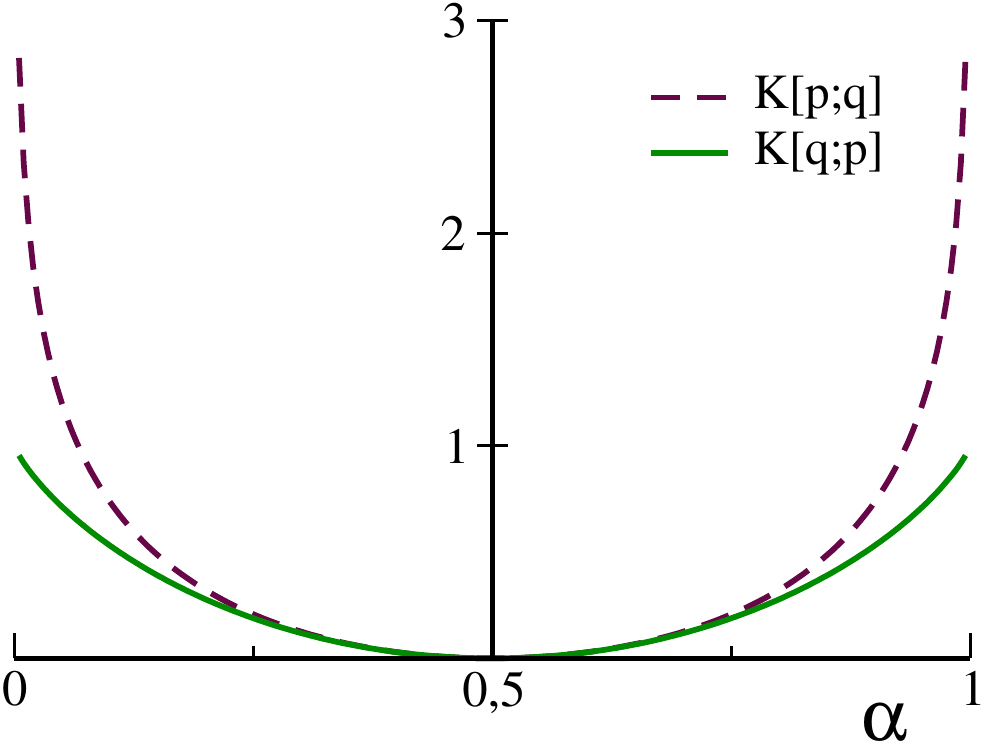}
\hspace{1ex}
\includegraphics[width=0.44\textwidth]{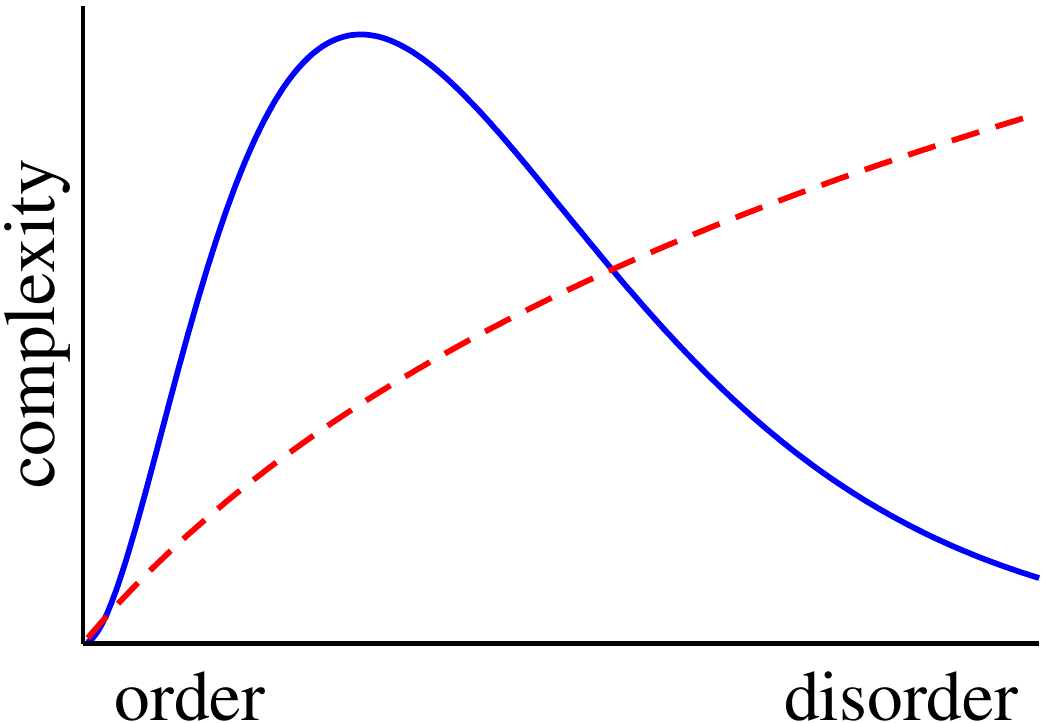}
\vspace*{0.0cm} \caption{\textit{Left}: For the two
PDFs $p$ and $q$ parametrized by $\alpha$, 
see Eq.\ (\ref{complex1_p_q_example_KL}),
the respective Kullback-Leibler divergences $K[p;q]$ (dashed line) and
$K[q;p]$ (full line). Note the maximal asymmetry for $\alpha\to0,1$,
where $\lim_{\alpha\to0,1}K[p;q]=\infty$.
\textit{Right}: The degree of complexity (full line)
should be minimal both in the fully ordered and the fully
disordered regime. For some applications it may however 
be meaningful to consider complexity measures maximal 
for random states (dashed line).
}
\label{complex_complex_plot}
\end{figure}

\runinhead{Example}
As a simple example we consider two distributions,
$p(\sigma)$ and $q(\sigma)$,
for a binary variable $\sigma=0,1$, 
\begin{equation}
p(0)\,=\,1/2\,=\,p(1),
\qquad\quad
q(0)\,=\,\alpha,\quad\qquad
q(1)\,=\,1-\alpha ~,
\label{complex1_p_q_example_KL}
\end{equation}
with $p(\sigma)$ being flat and $\alpha\in[0,1]$.
The Kullback-Leibler divergence,
\begin{eqnarray*}
K[p;q] & =& \sum_{\sigma=0,1} p(\sigma)\log\left({p(\sigma\over q(\sigma)}\right)
       \ =\ {-1\over2}\log(2\alpha)\,-\, {1\over2}\log(2(1-\alpha)) \\
       & =& -\log(4(1-\alpha)\alpha)\,/\,2\ \ge\ 0~,
\end{eqnarray*}
is unbounded, since $\lim_{\alpha\to0,1} K[p;q]\to\infty$.
Interchanging $p\leftrightarrow q$ we find
\begin{eqnarray*}
K[q;p] & =& \alpha\log(2\alpha) \,+\, (1-\alpha)\log(2(1-\alpha)) \\
& =& \log(2)\,+\,\alpha\log(\alpha) \,+\, (1-\alpha)\log(1-\alpha)\ \ge\ 0~,
\end{eqnarray*}
which is now finite in the limit $\lim_{\alpha\to0,1}$.
The Kullback-Leibler divergence is highly asymmetric,
compare Fig.~\ref{complex_complex_plot}.

\section{Complexity Measures}
\label{complex_measures}
\index{complexity!measure|textbf}

Can we provide a single measure, or a
small number of measures, suitable
for characterizing the \qut{degree of
complexity} of any dynamical system at
hand? This rather philosophical question
has fascinated researchers for decades and
no definitive answer is known. 

The quest of complexity measures touches many
interesting topics in dynamical system theory and
has led to a number of powerful tools suitable for
studying dynamical systems, the original goal
of developing a one-size-fit-all measure for
complexity seems however not anymore a scientifically
valid target. Complex dynamical systems can 
show a huge variety of qualitatively different 
behaviors, one of the reasons why complex system 
theory is so fascinating, and it is not appropriate
to shove all complex systems into a single basket
for the purpose of measuring their degree of complexity
with a single yardstick.

\runinhead{Intuitive Complexity}
The task of developing a mathematically well defined
measure for complexity is handicapped by the lack of 
a precisely defined goal. In the following we will 
discuss some selected prerequisites and constraints 
one may postulate for a valid complexity measure. In the 
end it is, however, up to our intuition for deciding
whether these requirements are appropriate or not.

An example of a process one may intuitively 
attribute a high degree of complexity are the 
intricate spatio-temporal patterns generated 
by the forest fire model discussed 
in Sect.~\ref{automata1_sec_cellular_automata},
and illustrated in Fig.~\ref{automata_fig_fires},
with perpetually changing fronts of fires 
burning through a continuously regrowing forest.

\runinhead{Complexity vs.\ Randomness}
\index{complexity!vs.\ randomness}
A popular proposal for a complexity measure
is the information entropy $H[p]$, see
Eq.\ (\ref{complex1_Shannon_entropy}).
It vanishes when the system is regular, which
agrees with our intuitive presumption that
complexity is low when nothing happens. The 
entropy is however maximal for random dynamics,
as shown in Fig.~\ref{complex_fig_entropy}.

It is a question of viewpoints to which extend
one should consider random systems as complex,
compare Fig.~\ref{complex_complex_plot}.
For some considerations, e.g.\ when dealing with
\qut{algorithmic complexity} (see 
Sect.~\ref{complex1_subsec_algorithmic_generative_complexity})
it makes sense to attribute maximal complexity
degrees to completely random sets of objects.
In general, however, complexity measures should
be concave and minimal for regular behavior
as well as for purely random sequences.

\runinhead{Complexity of Multi-Component Systems}
\index{complexity!intensive vs.\ extensive}
Complexity should be a positive quantity, like
entropy. Should it be, however, extensive or intensive?
This is a difficult and highly non-trivial question
to ponder.

Intuitively one may demand complexity to be intensive,
as one would not expect to gain complexity when considering 
the behavior of a set of $N$ independent and identical 
dynamical systems. On the other side we cannot rule out
that $N$ strongly interacting dynamical systems could
show more and more complex behavior with an increasing 
number of subsystems, e.g.\ we consider intuitively the
global brain dynamics to be orders of magnitude more complex 
than the firing patterns of the individual neurons.

There is no simple way out of this quandary when 
searching for a single one-size-fits-all complexity measure.
Both intensive and extensive complexity measures have their
areas of validity.

\runinhead{Complexity and Behavior}
\index{complexity!behavior}
The search for complexity measures is not just an 
abstract academic quest. As an example consider
how bored we are when our environment is repetitive, 
having low complexity, and how stressed when the complexity 
of our sensory inputs is too large. There are indeed 
indications that a valid behavioral strategy for highly 
developed cognitive systems may consist in optimizing 
the degree of complexity. Well defined complexity measures 
are necessary in order to quantify this intuitive 
statement mathematically. 

\subsection{Complexity and Predictability}
\label{complex1_subsec_Complexity_Predictability}

Interesting complexity measures can be constructed
using statistical tools, generalizing concepts like
information entropy and mutual information. We will
consider here time series generated from a finite
set of symbols. One may, however, interchange the time 
label with a space label in the following, whenever 
one is concerned with studying the complexity of 
spatial structures.

\runinhead{Stationary Dynamical Processes}
As a prerequisite we need stationary dynamical
processes, viz dynamical processes which do not
change their behavior and their statistical
properties qualitatively over time. In practice this
implies that the time series considered, as generated
by some dynamical system, has a finite time horizon $\tau$.
The system might have several time scales $\tau_i\le\tau$,
but for large times $t\gg\tau$ all correlation functions
need to fall off exponentially, like the autocorrelation
function defined in Sect.~\ref{section_criticality_dynamical_systems}.
Note, that this assumption may break down for
critical dynamical systems, which are characterized, as discussed
in Chap.~\ref{chap_automata1}, by dynamical and 
statistical correlations decaying only slowly, with 
an inverse power of time.

\runinhead{Measuring Joint Probabilities}
For times $t_0,\, t_1,\, ..$, a set of symbols $X$,
and a time series containing $n$ elements,
\begin{equation}
x_{n},\,x_{n-1},\,\dots,\,x_2,\,x_1,
\qquad\quad
x_i \,=\,x(t_i),
\qquad\quad
x_i \,\in\,X
\label{complex1_time_series_discrete}
\end{equation}
we may define the joint probability distribution
\begin{equation}
p_n\,:\qquad\quad
p(x_n,\dots,x_1)~.
\label{complex1_joint_PDF_n}
\end{equation}
The joint probability $p(x_n,\dots,x_1)$ is not
given a priori. It needs to be measured from an
ensemble of time series. This is a very demanding
task as $p(x_n,\dots,x_1)$ has $\left(N_s\right)^n$ 
components, with $N_s$ being the number of
symbols in $X$.

It clearly makes no sense to consider joint
probabilities $p_n$ for time differences
$t_n\gg\tau$, the evaluation of joint probabilities
exceeding the intrinsic time horizon $\tau$ is a
waste of effort. In practice finite values of $n$
are considered, taking subsets of length $n$ of a 
complete time series containing normally a vastly 
larger number of elements. This is an admissible 
procedure for stationary dynamical processes.

\runinhead{Entropy Density}
\index{entropy!density}
We recall the definition of the Shannon entropy
\begin{equation}
H[p_n] \ =\ -\sum_{x_{n},..,x_1\in X}
p(x_n,\dots,x_1)\, \log(p(x_n,\dots,x_1)) 
\ \equiv\ -\langle\, \log({p_n})\,\rangle_{p_n}~,
\label{complex1_def_entropy_n}
\end{equation}
which needs to be measured for an ensemble of
time series of length $n$ or greater. Of interest is the
entropy density in the limit of large times,
\begin{equation}
h_\infty \ =\ \lim_{n\to\infty} {1\over n}\,H[p_n]~, 
\label{complex1_def_entropy_density}
\end{equation}
which exists for stationary dynamical processes with
finite time horizons. The entropy density is the mean
number of bits per time step needed for encoding 
the time series statistically.

\begin{figure}[t]
\centering
\includegraphics[width=0.60\textwidth]{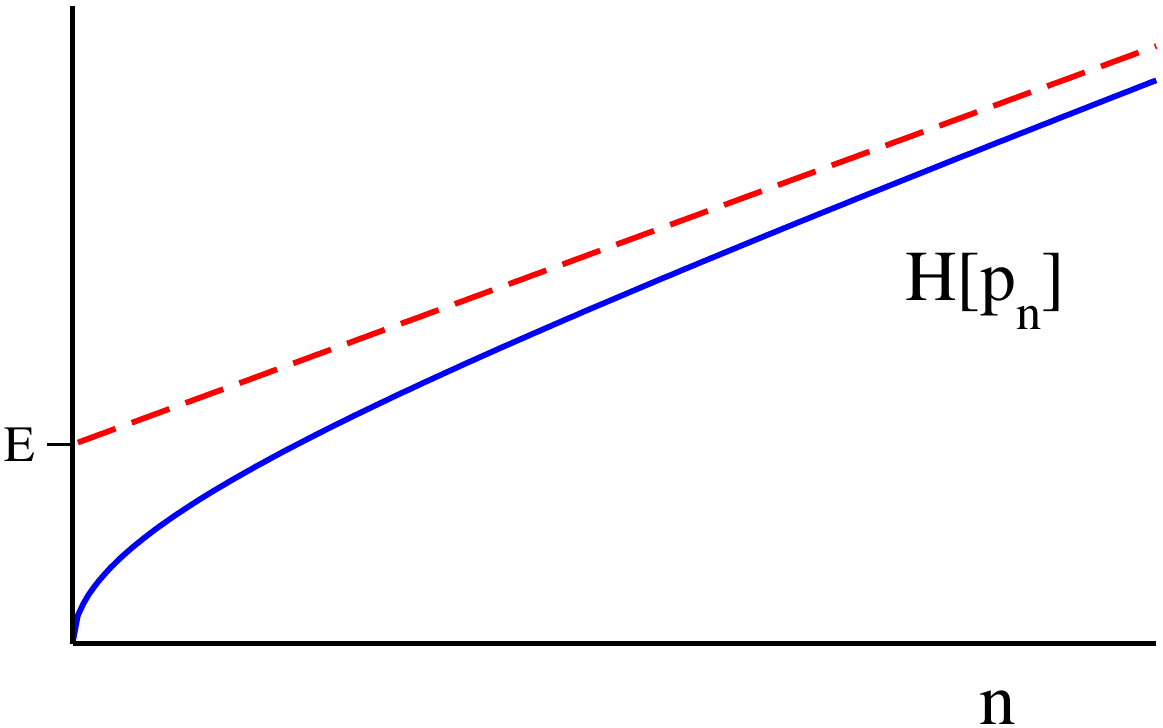}
\vspace*{0.0cm} \caption{The entropy (full line) 
$H[p_n]$ of a time series of length $n$ increases monotonically, 
with the limiting slope (dashed line) $h_\infty$. For large
$n\to\infty$ the entropy $H[p_n]\approx E+h_\infty n$, with
the excess entropy $E$ given by the intercept of asymptote
with the $y$-axis.
}
\label{complex_graphs_excess_entropy}
\end{figure}

\runinhead{Excess Entropy}
\index{entropy!excess}
\index{excess entropy}
\index{complexity!excess entropy}
We define the \qut{excess entropy} $E$ as
\begin{equation}
E \ =\ \lim_{n\to\infty} \big( H[p_n]\,-\,n\,h_\infty\big) \ \ge \ 0~.
\label{complex1_def_excess_entropy}
\end{equation}
The excess entropy is just the
non-extensive part of the entropy, it is the
coefficient of the term $\propto n^0$ when expanding 
the entropy in powers of $1/n$,
\begin{equation}
H[p_n] \ =\ n\,h_\infty\, +\, E\, +\, O(1/n),
\qquad\quad
n\,\to\,\infty~,
\label{complex1_excess_entropy_intensive}
\end{equation}
compare Fig.~\ref{complex_graphs_excess_entropy}.
The excess entropy $E$ is positive as long as
$H[p_n]$ is concave as a function of $n$ (we leave
the proof of this statement as an exercise to the reader), 
which is the case for stationary dynamical processes.
For practical purposes one may approximate
the excess entropy via
\begin{equation}
h_\infty \ =\ \lim_{n\to\infty} h_n,\qquad\quad
h_n\ =\ H[p_{n+1}]\,-\, H[p_{n}]~,
\label{complex1_h_infty_h_n}
\end{equation}
since $h_\infty$ corresponds to the asymptotic slope
of $H[p_n]$, compare Fig.~\ref{complex_graphs_excess_entropy}.
\begin{itemize}
\item One may also use Eqs.\ (\ref{complex1_h_infty_h_n}) 
and (\ref{complex1_def_conditional_entropy}) for
rewriting the entropy density $h_n$ in terms of an 
appropriately generalized conditional entropy.
\item Using Eq.~(\ref{complex1_excess_entropy_intensive})
we may rewrite the excess entropy as
$$
\sum_n\left[ {H[p_n]\over n}-h_\infty\right]~.
$$
In this form the excess entropy is known as
the \qut{effective measure complexity} (EMC)
or \qut{Grassberger entropy}.
\index{complexity!Grassberger}
\index{complexity!EMC}
\end{itemize}

\runinhead{Excess Entropy and Predictability}
The excess entropy vanishes both for a random and
for an ordered system. For a random system
$$
H[p_n] \ =\ n\, H[p_X] \ \equiv \ n\,h_\infty~,
$$
where $p_X$ is the marginal probability. The excess
entropy, Eq.\ (\ref{complex1_def_excess_entropy})
vanishes consequently. For an example of 
a system with ordered states we consider 
the dynamics
$$
\dots 000000000000000 \dots,
\qquad\quad
\dots 111111111111111 \dots ~,
$$
for a binary variable, occurring with
probabilities $\alpha$ and $1-\alpha$ respectively.
This kind of dynamics is the natural output of
logical AND or OR rules. The joint PDFs 
then have only two non-zero components,
$$
p(0,\dots,0) \ =\ \alpha,\quad\qquad
p(1,\dots,1) \ =\ 1-\alpha,\quad\qquad
\forall n~,
$$
all other $p(x_n,..,x_1)$ vanish and
$$
H[p_n]\ \equiv \ -\alpha\log(\alpha)\,-\,
(1-\alpha)\log(1-\alpha), 
\qquad\quad
\forall n~.
$$
The entropy density $h_\infty$ vanishes and
the excess entropy $E$ becomes $H[p_n]$; it 
vanishes for $\alpha\to0,1$, viz in the 
deterministic limit.

The excess entropy therefore fulfills
the concaveness criteria illustrated in
Fig.~\ref{complex_complex_plot}, vanishing
both in the absence of predictability (random
states) and for the case of strong predictability
(i.e.\ for deterministic systems). The excess entropy
does however not vanish in above example for
$0<\alpha<1$, when two predictable states are
superimposed statistically in an ensemble of
time series. Whether this behavior is compatible
with our intuitive notion of complexity is, to a
certain extent, a matter of taste.

\runinhead{Discussion}
The excess entropy is a nice tool for
time series analysis, satisfying several
basic criteria for complexity measures,
and there is a plethora of routes for
further developments, e.g.\ for systems
showing structured dynamical activity
both in the time as well as in the spatial
domain. The excess entropy is however exceedingly 
difficult to evaluate numerically and 
its scope of applications therefore limited 
to theoretical studies.

\subsection{Algorithmic and Generative Complexity}
\label{complex1_subsec_algorithmic_generative_complexity}
\index{algorithmic complexity|textbf}
\index{complexity!algorithmic|textbf}
\index{complexity!generative|textbf}

We have discussed so far descriptive approaches
using statistical methods for the construction 
of complexity measures. One may, on the other hand,
be interested in modelling the generative process.
The question is then: which is the simplest model
able to explain the observed data?

\runinhead{Individual Objects}
For the statistical analysis of a time series
we have been concerned with ensembles of
time series, as generated by the identical underlying
dynamical system, as well as with the limit of infinitely long
times. In this section we will be dealing with individual
objects composed of a finite number of $n$ symbols, like
$$
0000000000000000000000,
\qquad\quad
0010000011101001011001~.
$$
The question is then: which dynamical model can generate
the given string of symbols? One is interested, in particular,
in strings of bits and in computer codes capable of
reproducing them.

\runinhead{Turing Machine}
\index{Turing machine}
The reference computer codes in theoretical
informatics is the set of instructions
needed for a \qut{Turing machine} to carry
out a given computation. The exact definition
for a Turing machine is not of relevance here,
it is essentially a finite-state machine
working on a set of instructions called
code. The Turing machine plays a central
role in the theory of computability, e.g.\
when one is interested in examining how hard
it is to find the solution to a
given set of problems.

\runinhead{Algorithmic Complexity}
\index{algorithmic complexity}
\index{complexity!algorithmic}
The notion of algorithmic complexity tries
to find an answer to the question of how hard 
it is to reproduce a given time series in the
absence of prior knowledge.
\begin{quotation}
{\it Algorithmic Complexity.\enspace}
The \qut{algorithmic complexity} of a string of bits
is the length of the shortest program that prints
the given string of bits and then halts.
\end{quotation}
\index{Kolmogorov complexity}
\index{complexity!Kolmogorov}
The algorithmic complexity is also called
\qut{Kolmogorov complexity}. Note, that the 
involved computer or Turing machine is supposed
to start with a blank memory, viz with no
prior knowledge.

\runinhead{Algorithmic Complexity and Randomness}
Algorithmic complexity is a very powerful concept
for theoretical considerations in the context
of optimal computability. It has, however, two
drawbacks, being not computable and attributing
maximal complexity to random sequences.

A random number generator can only be approximated
by any finite state machine like the Turing machine
and would need an infinite code length to be perfect.
That is the  reason why real-world codes for random 
number generators are producing only 
\qut{pseudo random numbers}, with the degree of 
randomness to be tested by various statistical
measures. Algorithmic complexity therefore 
conflicts with the common postulate for complexity
measures to vanish for random state, compare
Fig.~\ref{complex_complex_plot}.

\runinhead{Deterministic Complexity}
\index{complexity!deterministic}
There is a vast line of research trying 
to understand the generative mechanism
of complex behavior not algorithmically
but from the perspective
of dynamical system theory, in particular
for deterministic systems. The question is then:
in the absence of noise, which are the features
needed to produce interesting and complex
trajectories?

Of interest are in this context the sensitivity 
to initial condition for systems having
a transition between chaotic and regular
states in phase space,
see Chap.~\ref{chap_networks2},
the effect of bifurcations and non-trivial
attractors like strange attractors,
see Chap.~\ref{chap_chaos1}, and the
consequences of feedback and tendencies
toward synchronization,
see Chap.~\ref{chap_synchro1}.
This line of research is embedded in the
general quest of understanding the properties
and the generative causes of complex
and adaptive dynamical systems.

\runinhead{Complexity and Emergence}
\index{complexity!and emergence}
Intuitively, we attribute a high 
degree of complexity to ever
changing structure emerging from
possibly simple underlying rules,
an example being the forest fires burning
their way through the forest along
self-organized fire fronts,
compare Fig.~\ref{automata_fig_fires}
for an illustration. This link between
complexity and \qut{emergence}
is, however, not easy to mathematize,
as no precise measure for emergence
has been proposed to date.


\section*{Exercises}
\addcontentsline{toc}{section}{Exercises} \markright{Exercises}
\begin{list}{}
\item \hspace*{-15pt}{\sc The Law of Large Numbers} \\
Generalize the derivation for the law of large numbers given in
Sect.~\ref{complex1_subsec_law_large_numbers} for the
case of $i=1,\dots,N$ independent discrete stochastic processes
$p^{(i)}_k$, described by their respective generating functionals
$G_i(x)=\sum_k p^{(i)}_k x^k$.
\item \hspace*{-15pt}{\sc Symbolization of Financial Data} \\
Generalize the symbolization procedure defined for the
joint probabilities $p_{\pm\pm}$ defined by
Eq.\ (\ref{complex1_joint_PDF_symbol}) to
joint probabilities $p_{\pm\pm\pm}$. E.g.\ $p_{+++}$
would measure the probability of three consecutive
increases. Download from the Internet the historical data 
for your favorite financial asset, like the Dow Jones or 
the Nasdaq stock indices, and analyze it with this symbolization
procedure. Discuss, whether it would be possible, as a matter
of principle, to develop in this way a money-making scheme.
\item \hspace*{-15pt}{\sc The OR Time Series with Noise} \\
Consider the time series generated by a logical OR, akin
to Eq.\ (\ref{complex1_time_series_XOR}). Evaluate the 
probability $p(1)$ for finding a 1, with and without averaging 
over initial conditions, both without and in presence of
noise. Discuss the result.
\item \hspace*{-15pt}{\sc Maximal Entropy Distribution Function} \\
Determine the probability distribution function $p(x)$, having
a given mean $\mu$ and a given variance $\sigma^2$, compare
Eq.\ (\ref{complex1_PDF_maximal_entropy_fixed_mu_sigma_condition}),
which maximizes the Shannon entropy.
\item \hspace*{-15pt}{\sc Two-Channel Markov Process}\\
Consider, in analogy to Eq.\ (\ref{complex1_two_channels})
the two-channel Markov process $\{\sigma_t,\tau_t\}$,
$$
\sigma_{t+1}\ =\ AND(\sigma_t,\tau_t),
\qquad
\tau_{t+1}\ =\ \left\{
\begin{array}{rcl}
OR(\sigma_t,\tau_t) &\quad& \mathrm{probability}\ 1-\alpha \\
\neg OR(\sigma_t,\tau_t) &\quad& \mathrm{probability}\ \alpha
\end{array}
\right.~.
$$
Evaluate the joint and marginal distribution
functions, the respective entropies and the
resulting mutual information. Discuss the
result as a function of noise strength $\alpha$.

\item \hspace*{-15pt}{\sc Kullback-Leibler Divergence}\\
Try to approximate an exponential distribution function by
a scale-invariant PDF, considering the Kullback-Leibler divergence
$K[p;q]$, Eq.\ (\ref{complex1_def_Kullback_Leibler_divergence}),
for the two normalized PDFs
$$
p(x)\ =\ \mathrm{e}^{-(x-1)},\qquad
q(x)\ =\ {\gamma-1\over x^\gamma},\qquad
x,\,\gamma\,>\,1~.
$$
Which exponent $\gamma$ minimizes $K[p;q]$? 
How many times do the graphs for $p(x)$ 
and $q(x)$ cross?

\item \hspace*{-15pt}{\sc Chi-Squared Test}\\
The quantity
\begin{equation}
\chi^2[p;q]\ =\ \sum_{i=1}^N {(p_i-q_i)^2\over p_i}
\label{complex1_chi_squared}
\end{equation}
measures the similarity of two normalized 
probability distribution functions $p_i$ and 
$q_i$. Show, that the Kullback-Leibler divergence
$K[p;q]$, Eq.\ (\ref{complex1_def_Kullback_Leibler_divergence}),
reduces to $\chi^2[p;q]/2$ if the two PDFs
are quite similar.

\item \hspace*{-15pt}{\sc Excess Entropy}\\
Use the representation
$$
E\ =\ \lim_{n\to\infty} E_n,\qquad\quad
E_n\ \approx\ H[p_n]\,-\,n\big(H[p_{n+1}]-H[p_{n}]\big)
$$
to prove that $E\ge0$, compare Eqs.\ (\ref{complex1_def_excess_entropy})
and (\ref{complex1_h_infty_h_n}),
as long as $H[p_n]$ is concave as a function of $n$.

\index{Tsallis entropy}
\index{entropy!Tsallis}
\item \hspace*{-15pt}{\sc Tsallis Entropy}\\
The \qut{Tsallis Entropy}
$$
H_q[p] \ =\ {1\over 1-q}\sum_k\left[ \big(p_k\big)^q-p_k\right],
\qquad\quad 0<q\le 1
$$
of a probability distribution function $p$ is a popular
non-extensive generalization of the Shannon entropy $H[p]$.
Prove that 
$$
\lim_{q\to1} H_q[p] \ =\ H[p]~,
\qquad\quad
H_q[p]\ \ge\ 0,
$$
and the non-extensiveness
$$
H_q[p]\ =\ H_q[p_X]+ H_q[p_Y]+(1-q)\,H_q[p_X]\, H_q[p_Y],
\qquad\quad
p = p_Xp_Y
$$
for two statistically independent systems $X$ and $Y$. 
For which distribution function $p$ is $H_q[p]$ maximal?
\end{list}


\def\refer#1#2#3#4#5#6{\item{\frenchspacing\sc#1}\hspace{4pt}
                       #2\hspace{8pt}#3 {\it \frenchspacing#4} {\bf#5}, #6.}
\def\bookref#1#2#3#4{\item{\frenchspacing\sc#1}\hspace{4pt}
                     #2\hspace{8pt}{\it#3}  #4.}

\addcontentsline{toc}{section}{Further Reading} 
\section*{Further Reading}
\markboth{\thechapter\enspace Complexity and Information Theory}{Further Reading}

We recommend for further readings 
introductions to information theory (Cover \& Thomas, 2006),
to  Bayesian statistics (Bolstad, 2004),
to complex system theory in general (Boccara, 2003),
and to algorithmic complexity (Li \& Vitanyi, 1997)

For further studies we recommend several
review articles, on evolutionary development of
complexity in organisms (Adami, 2002),
on complexity and predictability
(Boetta, Cencini, Falcioni \& Vulpiani, 2003),
a critical assessement of various complexity 
measures (Olbrich {\it et al.}, 2008) and
a thoughtful discussion on various approaches
to the notion of complexity (Manson, 2001).

For some further, somewhat more specialized
topics, we recommend Binder (2008) for a
perspective on the interplay between dynamical
frustration and complexity,
Binder (2009) for the question of decidability 
in complex systems, and Tononi \& Edelman (1998)
on possible interrelations between consciousness and
complexity.

{\baselineskip=15pt
\begin{list}{}{\leftmargin=2em \itemindent=-\leftmargin%
\itemsep=3pt \parsep=0pt \small}
\refer{Adami, C.}{2002}{What is complexity?}
{BioEssays}{24}{1085--1094}
\refer{Binder, P.-M.}{2008}
{Frustration in Complexity}
{Science}{320}{322--323}
\refer{Binder, P.-M.}{2009}
{The edge of reductionism}
{Nature}{459}{332--334}
\bookref{Boccara, N.}{2003}{Modeling Complex Systems.}{Springer,
Berlin}
\refer{Boetta, G., Cencini, M., Falcioni, M., Vulpiani, A.}
{2002}{Predictability: a way to characterize complexity.}
{Physics Reports}{356}{367--474}
\bookref{Bolstad, W.M.}{2004}
{Introduction to Bayesian statistics.}{Wiley-IEEE}
\bookref{Cover, T.M., Thomas, J.A.}{2006}
{Elements of information theory.}{Wiley-Interscience}
\bookref{Li, M., Vitanyi, P.M.B.}{1997}
{An introduction to Kolmogorov complexity and its applications.}
{1997}{Springer}
\refer{Manson, S.M.}{2001} {Simplifying complexity: 
a review of complexity theory.}{Geoforum}{32}{405--414}
\refer{Olbrich, E., Bertschinger, N., Ay, N., Jost, J.}{2008}
{How should complexity scale with system size?}
{The European Physical Journal B}{63}{407--415}
\refer{Tononi, G., Edelman, G.M.}{1998}
{Consciousness and complexity} {Science}{282}{1846}
%

%
\end{list}
\par}


 

\vspace{-20ex}
\chapter{Random Boolean Networks}
\label{chap_networks2}

\abstract
{Complex system theory deals with dynamical systems containing a very
large number of variables. The resulting dynamical behavior can be
arbitrary complex and sophisticated. It is therefore important to
have well controlled benchmarks, dynamical systems which can be
investigated and understood in a controlled way for large numbers of
variables.\newline
\indent Networks of interacting binary variables, i.e. boolean
networks, constitute such canonical complex dynamical system. They
allow the formulation and investigation of important
concepts like phase transition in the
resulting dynamical state. They are also recognized to be the
starting points for the modeling of gene expression and protein
regulation networks; the fundamental networks at the
basis of all life.}



\section{Introduction}

\runinhead{Boolean Networks} \index{boolean network|textbf}
\index{network!boolean|textbf} In this chapter, we describe the
dynamics of a set of $N$ binary variables.
\begin{quotation}
{\it Boolean Variables.\enspace}
\index{boolean!variable}
\index{variable!boolean}
A boolean or binary variable has two possible values,
typically 0 and 1.
\end{quotation}
The actual values chosen for the binary variable are
irrelevant{;} $\pm1$ is an alternative popular choice.
These elements interact with each other according to some given
interaction rules denoted {as} coupling functions.
\begin{quotation}
{\it Boolean Coupling Functions.\enspace}
\index{boolean!coupling function}A boolean function $\ \ \{0,1\}^K
\to\{0,1\}\ \ $ maps $K$ boolean variables onto a single one.
\end{quotation}
\index{dynamics!discrete time}The dynamics of the system is
considered to be discrete, $t=0,1,2,{\ldots}$. The
value of the variables at the next time step are determined by the
choice of boolean coupling functions.
%

\begin{figure}[t]
\centering
\includegraphics{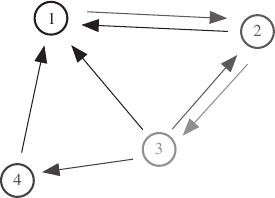}
\caption{Illustration of a boolean network with $N=4$ sites.
$\sigma_1(t+1)$ is determined by $\sigma_2(t)$, $\sigma_3(t)$ and
$\sigma_4(t)$ ($K=3$). The controlling elements of $\sigma_2$ are
$\sigma_1$ and $\sigma_3$ ($K=2$). The connectivity of $\sigma_3$
and $\sigma_4$ is $K=1$ }
\label{networks2_fig_booleanNetwork}
\vspace*{-10pt}
\end{figure}

\begin{quotation}
{\it {{The Boolean Network}}.\enspace}
\index{boolean network}The set of boolean coupling functions
interconnecting the $N$ boolean variables can be represented
graphically by a directed network,  the boolean
network.
\end{quotation}
In Fig.~\ref{networks2_fig_booleanNetwork} a small
boolean network is illustrated. Boolean networks at
first sight seem to be quite esoteric, devoid of the
practical significance for real-world phenomena.
Why are they then studied so intensively?

\enlargethispage*{15pt}

\runinhead{Cell Differentiation in Terms of Stable Attractors}
\index{cell differentiation}\index{gene expression network}
\index{network!gene expression}The field of boolean networks
{was given} the first big boost by the seminal study
of Kauffman in the late {1960s}. Kauffman casted
the problem of gene expression in terms of a gene regulation
network and introduced the so-called $N$-$K$ model in this
context. All cells of an animal contain the same genes and cell
differentiation, i.e.\ the fact that a skin cell differs from a
muscle cell, is due to differences in the gene activities in the
respective cells. Kauffman proposed that different stable
attractors, viz cycles, in his random boolean gene expression
network correspond to different cells in the
{bodies} of animals.

The notion is then that cell types correspond to different dynamical
states of a complex system, {i.e.} the gene expression
network, viz that gene regulation networks are the underpinnings of
life. This proposal by Kauffman has received strong support from
experimental studies in the last years. In Sect. 3.5.2 we will
discuss the case of the yeast cell division cycle.

\runinhead{Boolean Networks are Everywhere} Kauffman's original work
on gene expression networks was soon generalized to a wide spectrum
of applications, such as, to give a few examples, the modeling of
neural networks by random boolean networks and of the
\qut{punctuated equilibrium} in long-term evolution; a
concept that we will discuss in Chap.~\ref{chap_evolution1}.

\enlargethispage*{-12pt}

Dynamical systems theory (see Chap.~\ref{chap_chaos1}) deals
with dynamical systems containing a relatively small number of
variables. General dynamical systems with large numbers of
variables are very difficult to analyze and
control. Random boolean networks can hence be
considered, in a certain sense, as being of
prototypical importance in this field, as they provide well
defined classes of dynamical systems for which the thermodynamical
limit $N\to\infty$ can be taken. They show chaotic as well as
regular behavior, despite their apparent simplicity, and many
other typical phenomena of dynamical systems. In the thermodynamic
limit there can be phase transitions between chaotic and regular
regimes. These are the issues studied in this chapter.


\runinhead{{$\textbf{\textit{N}}$--$\textbf{\textit{K}}$} Networks}
\index{Kauffman network}There are several types of random boolean
networks. The most simple realization is the $N${--}$K$
model. It is made up of $N$ boolean variables, each variable
interacting exactly with $K$ other randomly chosen variables. The
respective coupling functions are also chosen randomly from the set
of all possible boolean functions mapping $K$ boolean inputs onto
one boolean output.

There is no known realization of $N${--}$K$ models in
nature. All real physical or biological problems have very
specific couplings determined by the structure and the physical
and biological interactions of the system considered. The topology
of the couplings is, however, often very complex and, in many
instances, completely unknown. It is then often a good starting
point to model the real-world system by a generic model, like the
{$N$--$K$} model.

\runinhead{Binary Variables}
\index{variable!boolean}Modeling real-world systems by a collection
of interacting binary variables is often a simplification, as
real-world variables are often continuous. For the case of the gene
expression network, one just keeps two possible states for every
single gene: active or inactive.

Thresholds, viz parameter regimes at which the dynamical behavior
changes qualitatively, are wide-spread in biological systems.
Examples are neurons, which fire or do not fire depending on the
total strength of presynaptic activity. \nobreak{Similar} thresholds
occur in metabolic networks in the form of activation
\nobreak{potentials} for the {chemical reactions involved}. Modeling real-world systems
based on threshold dynamics with binary variables is, then, a viable
first step towards an\break understanding.

\section{Random Variables and Networks}

Boolean networks have a rich variety of possible
concrete model realizations and we will discuss
in the following the most important ones.
\subsection{Boolean Variables and Graph Topologies}
\index{boolean!variable|textbf}
\index{variable!boolean|textbf}

\runinhead{Boolean Variables and State Space}
We denote by
$$
\sigma_i \ \in \ \{0,1\}, \qquad i = 1, 2, \ldots , N
$$
the $N$ binary variables and by $\Sigma_t$
the state of the system at time $t$,
\index{boolean network!state space}
\index{state space!boolean network}
\begin{equation}
\Sigma_t\ =\ \{\sigma_1(t),\sigma_2(t),\ldots,\sigma_N(t)\}~.
\end{equation}
$\Sigma_t$ can be thought of as a vector pointing to one of the
$\Omega=2^N$ edges of {an} \hbox{$N$-dimensional}
hypercube, where $\Omega$ is the number of possible configurations.
For numerical implementations and simulations it is useful to
consider $\Sigma_t$ as the binary representation of an
 integer number $0 \leq \Sigma_t < 2^N$.

\runinhead{Time Dependence}
Time is assumed to be discrete,
$$
\sigma_i\ =\ \sigma_i(t),\qquad\quad t\ =\ 1,\ 2,\
{\ldots}
$$
The value of a given boolean element $\sigma_i$ at the
next time step is determined by the values of $K$
controlling variables.
\begin{quotation}
{\it Controlling Elements.\enspace}
\index{boolean network!controlling elements}The controlling
elements $\sigma_{j_1(i)}$, $\sigma_{j_2(i)}$, $\ldots$,
$\sigma_{j_{K_i}(i)}$ of a boolean variable $\sigma_i$ determine its
time evolution by
\end{quotation}
\begin{equation}
\sigma_i(t+1) \  =\  f_i(\sigma_{j_1(i)}(t),\sigma_{j_2(i)}(t),
\ldots ,\sigma_{j_{K_i}(i)}(t))~. \label{boolean_m1}
\end{equation}
Here $f_i$ is a boolean function associated with $\sigma_i$. The
set of controlling elements might include $\sigma_i$ itself.
Some exemplary boolean functions are given in
Table \ref{boolean_f1}.


\begin{table}[b]
\vspace*{-7pt} \caption{Examples of boolean functions of three
arguments. (\textbf{a}) A particular random function.  (\textbf{b})
A canalizing function of the first argument. When $\sigma_1=0$,
    the function value is 1. If $\sigma_1=1$, then
    the output can be either 0 or 1.
(\textbf{c}) An additive function. The output is 1 (active) if at
least two inputs
    are active.
(\textbf{d}) The generalized XOR, which is true when the number of\break
1-bits is odd
         }
\label{boolean_f1}
\center
\begin{tabular*}{17.4pc}{@{}l@{\quad}l@{\quad}l@{\quad}l@{\quad}l@{\quad}l@{\quad}l@{}}
\hline\noalign{\smallskip}
$\sigma_{1}$ & $\sigma_{2}$ & $\sigma_{3}$ & \multicolumn{4}{c}{$f(\sigma_{1},\sigma_{2},\sigma_{3})$}\\[2pt]
\cline{4-7}\\[-6pt]
    &&&Random & Canalizing & Additive & Gen.~XOR\\
\noalign{\smallskip}\svhline\noalign{\smallskip}
0 & 0 & 0 & 0 & 1 & 0 & 0\\
0 & 0 & 1 & 1 & 1 & 0 & 1\\
0 & 1 & 0 & 1 & 1 & 0 & 1\\
0 & 1 & 1 & 0 & 1 & 1 & 0\\
1 & 0 & 0 & 1 & 0 & 0 & 1\\
1 & 0 & 1 & 0 & 1 & 1 & 0\\
1 & 1 & 0 & 1 & 0 & 1 & 0\\
1 & 1 & 1 & 1 & 0 & 1 & 1\\
\noalign{\smallskip}\hline\noalign{\smallskip}
\end{tabular*}
\end{table}

\enlargethispage*{12pt}
\runinhead{Model Definition} For a complete
definition of the model we then need to specify several
para\-meters:
\begin{itemize}
\index{boolean network!connectivity}
\item[--]{The Connectivity}:
The first step is to select the
connectivity $K_i$ of each element, i.e.\ the number
of its controlling elements. With
\[
\langle K \rangle\  =\  \frac{1}{N}\sum^N_{i=1}K_i
\]
\index{mean!connectivity}the average connectivity is defined. Here
we will consider mostly the case in which the connectivity is the
same for all nodes: $K_i = K$, $i = 1, 2,\ldots,
N$.
\index{boolean network!linkage}
\index{linkage!boolean network}
\item[--]{The Linkages}:
The second step is to select the specific set of controlling
elements
$\big\{\sigma_{j_1(i)}$, $\sigma_{j_2(i)}$, $\ldots$,
$\sigma_{j_{K_i}(i)} \big\}$
on which the element $\sigma_i$ depends. See
Fig.~\ref{networks2_fig_booleanNetwork}
for an illustration.
\index{boolean network!evolution rule}
\item[--]{The Evolution Rule}:
The third step is to choose the
boolean function $f_i$ determining the value of
\hbox{$\sigma_i(t+1)$} from the values of the linkages
$\big\{\sigma_{j_1(i)}(t)$, $\sigma_{j_2(i)}(t)$,
$\dots$, $\sigma_{j_{K_i}(i)}(t) \big\}$.
\end{itemize}

\runinhead{{The Geometry} of the Network}
\index{boolean network!geometry}The way the linkages are assigned
determines the topology of the network and networks can have highly
diverse topologies, see Chap.~\ref{chap_networks1}.
It is custom to consider two special cases:
%

\begin{figure}[t]
\centering
\includegraphics{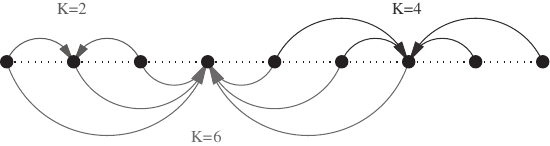}
\caption{Translational invariant linkages
  for a completely ordered one-dimensional lattice with
  connectivities $K=2,\ 4,\ 6$
  }
\label{boolean_booleanLinks}
\end{figure}

\begin{quotation}
{\it Lattice Assignment.\enspace} \index{boolean
network!geometry!lattice assignment} The boolean variables
$\sigma_i$ are assigned to the nodes of a regular lattice. The $K$
controlling elements $\big\{\sigma_{j_1(i)}$, $\sigma_{j_2(i)}$,
$\ldots$, $\sigma_{j_K(i)}\big\}$ are then chosen in a regular,
translational invariant manner, see Fig.~\ref{boolean_booleanLinks}
for an illustration.
\end{quotation}
\vspace{-12pt}
\begin{quotation}
{\it Uniform Assignment.\enspace} \index{boolean
network!geometry!uniform assignment} In {a} uniform
assignment the set of controlling elements are randomly drawn from
all $N$ sites of the network. This is the case for the
{$N$--$K$} model, also called {the {\em Kauffman net}}. In terms of graph theory
one {also speaks} of {an}
Erd\"os--R\'enyi random graph.
\end{quotation}
All intermediate cases are possible. Small-world networks, to give
an example, with regular short-distance links and random
long-distance links are popular models in network theory, as
discussed extensively in Chap.~\ref{chap_networks1}.

\vspace{-6pt}

\subsection{Coupling Functions}
\label{networks2_subsect_coupling_functions}
\index{boolean network!coupling functions|textbf}
\index{coupling functions|textbf}

\runinhead{Number of Coupling Functions}
The coupling function
$$
f_i:\qquad
\left\{\sigma_{j_1(i)},\ \dots,\ \sigma_{j_K(i)}\right\}\ \to\ \sigma_i
$$
has $2^K$ different arguments. To each argument value one can assign
either 0 or 1. Thus there are a total of
\begin{equation}
N_f\ =\  2^{\left(2^K\right)}\ =\  2^{2^K}\ =\
\left\{
\begin{array}{rcl}
4 &\quad& K=1 \\
16 &\quad& K=2 \\
256 &\quad& K=3
\end{array}
\right.
\end{equation}
possible coupling functions. In Table \ref{boolean_f1} we present
several examples for the case $K=3$, out of the $2^{2^3} = 256$
distinct $K=3$ boolean functions.

\runinhead{Types of Coupling Ensembles}
\index{coupling ensemble}There are a range of different possible
choices for the probability distribution of coupling functions.
{The following are}  some examples:
\begin{itemize}
\item[--] {Uniform Distribution}:
\index{coupling ensemble!uniform distribution}
{As introduced} originally by Kauffman, the
uniform distribution specifies all possible coupling functions to
occur with the same probability $1/N_f$.
\item[--] {Magnetization Bias\footnote{Magnetic
moments {often have} only two possible
directions (up or down in the language of spin-1/2 particles). A
compound is hence magnetic when more moments point into one of the
two possible directions, viz if the two directions are populated
unequally.}}: \index{coupling ensemble!magnetization bias} The
probability of a coupling function to occur is proportional to $p$
if the outcome is $0$ and proportional to $1-p$ if the outcome is
$1$.
\item[--] {Forcing Functions}:
\index{coupling ensemble!forcing functions}{Forcing
functions are also} called \qut{canalizing function}. The function
value is determined when one of its arguments, say
$m\in\{1,\ldots,K\}$, is given a specific value, say $\sigma_m=0$
(compare Table \ref{boolean_f1}). The function value is not
specified if the forcing argument has {another}
value, here when $\sigma_m=1$.
\item[--] {Additive Functions}:
\index{coupling ensemble!additive functions}In order to simulate
the additive properties of inter-neural synaptic activities one can
choose
$$
\sigma_i(t+1) \ =\ \Theta(f_i(t)),
\qquad
f_i(t)\ =\ h +\sum_{j=1}^N c_{ij}\,\sigma_j(t),
\qquad c_{ij}\in\{0,1\}~,
$$
where $\Theta(x)$ is the Heaviside step function and $h$ a bias. The
value of $\sigma_i(t+1)$ depends only on a weighted sum of its
controlling elements at time $t$.
\end{itemize}
\runinhead{Classification of Coupling Functions} \index{coupling
ensemble!classification}For small numbers of connectivity $K$ one
can completely classify all possible coupling functions:
\begin{itemize}
\item[--] $K=0$\\
  There are {only} two constant functions, $f=1$ and $f=0$.
\label{boolean_boolean_f2}
\item[--]$K=1$\\
\parbox{0.55\textwidth}{
  Apart from the two constant functions, which one may denote
  together by $\mathcal{A}$, there are the identity $1$ and
  the negation $\neg \sigma$, which one can lump together into
  a class $\mathcal B$.
                   } \hspace{4ex}
\parbox{0.5\textwidth}{
\begin{tabular}{@{}l@{\quad}ll@{\qquad}ll}
\hline\noalign{\smallskip} $\sigma$ & \multicolumn{2}{p{35pt}}{Class
$\mathcal{A}$} &
\multicolumn{2}{l}{Class $\mathcal{B}$}\\
\noalign{\smallskip}\svhline\noalign{\smallskip}
0 & 0 & 1 & 0 & 1\\
1 & 0 & 1 & 1 & 0\\
\noalign{\smallskip}\hline\noalign{\smallskip}
\end{tabular}
                    }
\item[--]$K=2$\\
  \label{networks2_K_2_boolean_functions}There are four classes of functions $f(\sigma_1,\sigma_2)$,
  with each class being invariant under the interchange
  $0 \leftrightarrow 1$ in either {the} arguments or {the} value of
  $f$: ${\mathcal A}$ (constant functions),
  ${\mathcal B}_1$ (fully canalizing functions for which
  one of the arguments determines the output deterministically),
  ${\mathcal B}_2$ (normal canalizing functions),
  ${\mathcal C}$ (non-canalizing functions, sometimes
  also denoted \qut{reversible functions}).
  Compare Table \ref{boolean_boolean_f3}.

\end{itemize}

\begin{table}[b]
\centering
\caption{{The} 16 boolean functions for $K=2$. For the
definition of
   the various classes see p.~\pageref{networks2_K_2_boolean_functions}
   and Aldana et al. (2003)}
\label{boolean_boolean_f3}
\begin{tabular}{@{}l@{\quad}l@{\qquad}l@{\quad}l@{\qquad}@{\quad}l@{\quad}l@{\quad}l@{\quad}l@{\quad}@{\quad}l@{\quad}l@{\quad}l@{\quad}l@{\quad}l@{\quad}l@{\quad}l@{\quad}l@{\quad}@{\quad}l@{\quad}l@{\quad}@{}}
\hline\noalign{\smallskip}
$\sigma_{1}$ & $\sigma_{2}$ &
\multicolumn{2}{l}{Class ${\mathcal A}$} &
\multicolumn{4}{l}{Class ${\mathcal B}_1$} &
\multicolumn{8}{l}{Class ${\mathcal B}_2$} &
\multicolumn{2}{l}{Class ${\mathcal C}$} \\
\noalign{\smallskip}\svhline\noalign{\smallskip}
0&0&1&0&0&1&0&1&1&0&0&0&0&1&1&1&1&0\\
0&1&1&0&0&1&1&0&0&1&0&0&1&0&1&1&0&1\\
1&0&1&0&1&0&0&1&0&0&1&0&1&1&0&1&0&1\\
1&1&1&0&1&0&1&0&0&0&0&1&1&1&1&0&1&0\\
\noalign{\smallskip}\hline\noalign{\smallskip}
\end{tabular}\vspace*{-5pt}
\end{table}

\vspace{-12pt}

\subsection{Dynamics}
\index{boolean network!dynamics|textbf}
\index{dynamics!boolean network|textbf}

\runinhead{Model Realizations}
\index{boolean network!model realizations}A given set of linkages
and boolean functions $\{f_i\}$ defines what one calls a {\em
realization} of the model. The dynamics then follows from
Eq.~(\ref{boolean_m1}). For the updating of all elements during one
time step one has several choices:
\begin{itemize}
\item[--] {Synchronous Update:}{\enspace}\index{synchronous updating}\index{updating!synchronous}All variables $\sigma_i(t)$ are updated simultaneously.

\item[--] {Serial Update (or asynchronous update):}{\enspace}\index{serial updating}\index{updating!serial}Only one variable is
updated at every step. This variable may be picked at random or by
some predefined ordering scheme.
\end{itemize}
The choice of updating does not affect thermodynamic properties,
like the phase diagram discussed in
Sect.~\ref{networks2_Mean_field_phase_diagram}. The occurrence and
the properties of cycles and attractors, as discussed in
Sect.~\ref{networks2_sec_cycles_attractors}, {however, crucially depends}  on the form of
\nobreak {update}.

\runinhead{Selection of the Model Realization}
There are several alternatives {for
choosing} the model realization during numerical simulations.

\vskip2pt

\begin{itemize}
\item[--] {The Quenched Model}\footnote{
\index{quenched!boolean network}\index{boolean network!quenched
model}An alloy made up of two or more substances is said to
    be \qut{quenched} when it is cooled so quickly that it remains
    stuck in a specific atomic configuration, which does not
    change anymore with time.}:
One specific realization of coupling functions is selected at the
beginning and kept throughout all time.
\item[--] {{The} Annealed Model}\footnote{
\index{annealed!boolean network}\index{boolean network!annealed
model}A compound is said to be \qut{annealed} when it {has
    been}
    kept long enough at elevated temperatures such that the
    thermodynamic stable configuration {has been} achieved.}:
    A new realization is randomly selected after each time step.
    Then either the linkages or the coupling functions or both change
    with every update, depending on the choice of the algorithm.
\item[--] {{The} Genetic Algorithm}:
\index{genetic algorithm}\index{algorithm!genetic}If the {network} is thought to approach a predefined goal, one
  may employ a genetic algorithm in which the system slowly modifies
  its realization with {passing time}.
\end{itemize}
Real-world systems are normally modeled by quenched systems with
synchronous updating. All interactions are then fixed for all times.

\vskip2pt

\runinhead{Cycles and Attractors}
\index{cycle}\index{attractor}\index{attractor!cyclic} Boolean
dynamics correspond to a trajectory within a finite state space of
size $\Omega =2^N$. Any trajectory generated by a dynamical system
with unmutable dynamical update rules, as for the quenched model,
will eventually lead to a cyclical behavior. No trajectory can
generate more than $\Omega$ distinct states in a row. Once a state
is revisited,
\vskip3pt
$$
\Sigma_t\ =\ \Sigma_{t-T}, \qquad\quad T<\Omega~,
$$

\noindent part of the original trajectory is retraced and cyclic behavior
follows. The resulting cycle acts as an attractor for a
set of initial conditions.

Cycles of length {1} are fixpoint attractors. The
fixpoint condition $\sigma_i(t+1)=\sigma_i(t)$ ($i=1,\ldots,N$) is
independent of the updating rules, viz synchronous vs.\
asynchronous. The order of updating the individual $\sigma_i$ is
irrelevant when none of them changes.

\begin{figure}[t]
\centering
\includegraphics{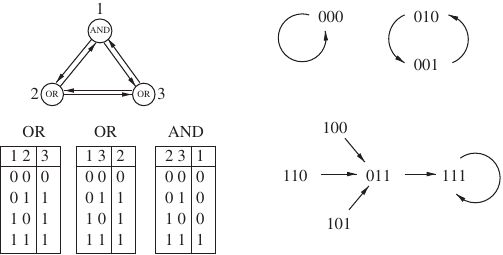}
\caption{A boolean network with $N=3$ sites and
         connectivities $K_i\equiv 2$.
\textit{Left}: Definition of the network linkage and coupling
      functions.
\textit{Right}: The complete network dynamics (from Luque and Sole,
2000) } \label{networks2_fig_booleanNetwork_example}
\vspace*{-14pt}
\end{figure}

\runinhead{An Example}
In Fig.~\ref{networks2_fig_booleanNetwork_example} a network with
$N=3$ and $K=2$ is fully defined. The time evolution of the $2^3=8$
states $\Sigma_t$ is given for synchronous updating. One can observe
one cycle of length 2 and two cycles of length 1 (fixpoints).

\section{The Dynamics of Boolean Networks}

We will now examine how we can characterize the dynamical
state of boolean networks in general and of N-K nets
in particular. Two concepts will turn out to
be of central importance, the relation of robustness 
to the flow of information and the characterization of
the overall dynamical state, which we will find to
be either frozen, critical or chaotic.

\subsection{The Flow of Information Through the Network}
\label{boolean_information_flow}
\index{flow!of information}



\runinhead{The Response to Changes} 
\index{boolean network!response to changes}
For random models the 
value of any given variable $\sigma_i$, or its change 
with time, is, per se, meaningless. Of fundamental 
importance, however, for quenched models is its response to
changes. We may either change the initial conditions, or some
specific coupling function, and examine its effect on the time
evolution of the variable considered.

\runinhead{Robustness}
Biological systems need to be robust. A gene 
regulation network, to give an example, for which 
even small damage routinely results in the death of 
the cell, will be at an evolutionary disadvantage 
with respect to a more robust gene expression set-up. 
Here we will examine the sensitivity of the dynamics 
with regard to the initial conditions. A system is robust
if two similar initial conditions lead to similar long-time
behavior.

\runinhead{The Hamming Distance and the Divergence of Orbits}
We consider two different initial states,\vspace*{4pt}
\[
\Sigma_0 = \{\sigma_1(0),\sigma_2(0),\ldots,\sigma_N(0)\}, \qquad
\tilde{\Sigma}_0 =
\{\tilde{\sigma}_1(0),\tilde{\sigma}_2(0),\ldots,
\tilde{\sigma}_N(0)\}~.\vspace*{4pt}
\]
Typically we are interested in the case when
$\Sigma_0$ and $\tilde\Sigma_0$ are close, viz when
they differ in the values of only a few elements. A 
suitable measure for the distance is the \qut{Hamming distance} 
$D(t)\in[0,N]$, \index{distance!Hamming}\index{Hamming distance}
\begin{equation} D(t)\ =\ \sum_{i \ = 1}^N \,
\Bigl(\sigma_i(t) -\tilde{\sigma}_i(t)\Bigr)^2~,
\label{boolean_dis}
\end{equation}
which is just the sum of elements {that} differ in $\Sigma_0$ and
$\tilde\Sigma_0$. As an example we consider\vspace*{4pt}
$$
\Sigma_1 \ =\ \{1,0,0,1\},\qquad\quad \Sigma_2 \ =\
\{0,1,1,0\},\qquad\quad \Sigma_3 \ =\ \{1,0,1,1\}~.\vspace*{4pt}
$$
We have 4 for the Hamming distance $\Sigma_1$-$\Sigma_2$ and 1 for
the Hamming distance $\Sigma_1$-$\Sigma_3$. If the system is robust,
two close-by initial conditions will never move far apart with time
passing{with passing time}, in terms of the Hamming distance.

\runinhead{The Normalized Overlap} The normalized overlap
$a(t)\in[0,1]$ between two configurations is defined as
\index{normalized overlap}
\begin{eqnarray} \nonumber
a(t)& =& 1 - {D(t)\over N}
\ =\ 1 - {1\over N}\sum_{i=1}^N \Big( \sigma_i^2(t) -
2\sigma_i(t)\tilde\sigma_i(t) + \tilde\sigma_i^2(t) \Big) \\[6pt]
& \approx &  {2\over N} \sum_{i=1}^N\, \sigma_i(t)\tilde\sigma_i(t)~,
\label{boolean_normalized_overlap}
\end{eqnarray}
where we have assumed the absence of any magnetization bias,
namely
$${1\over N}\sum_i\sigma_i^2\ \approx\  {1\over2}\
\approx\ {1\over N}\sum_i\tilde\sigma_i^2~,
$$
\looseness1 in the last step. The normalized overlap
Eq.~(\ref{boolean_normalized_overlap}) is then like a normalized
scalar product between $\Sigma$ and $\tilde\Sigma$. Two arbitrary
states have, on the average, a Hamming distance of $N/2$ and a
normalized overlap $a=1-D/N$ of 1/2.

\runinhead{Information Loss/Retention for Long Time Scales}
The difference between two initial states $\Sigma$ and
$\tilde\Sigma$ can also be interpreted as an information 
for the system. One then has than two possible behaviors:
\begin{itemize}
\index{information!loss}
\item[--]{Loss of Information:}\ \ \
                $\lim_{t\to\infty} a(t)\to 1$ \newline
  $a(t)\to1$ implies that two states are identical, or
  that they differ only by a finite number of elements,
  in the thermodynamic limit. This can happen when
  two states are attracted by the same cycle.
  All information about the starting states is~lost.
\index{information!retention}
\item[--] {Information Retention:}\ \ \
                $\lim_{t\to\infty} a(t)= a^*<1$ \newline
 The system ``remembers'' that the two configurations were
 initially different, with the difference measured by
 the respective Hamming distance.
\end{itemize}
The system is very robust when information is routinely lost.
Robustness depends on the value of $a^*$ when information is kept.
If $a^*>0$ then two trajectories retain a certain similarity for all
time scales.\vspace*{6pt}

\runinhead{Percolation of Information for Short Time {Scales}}
\index{percolation!information}\index{boolean network!percolation
of information} Above we considered how information present in
initial states evolves for very long times. Alternatively one may
ask, and this a typical question {in} dynamical
system theory, how information is processed for short times. We
write
\begin{equation}
D(t)\ \approx\  D(0)\,{\rm e}^{\lambda t}~,
\label{networks2_short_time_Lyaponov}
\end{equation}
\noindent where $0<D(0)\ll N$ is the initial Hamming distance and
where $\lambda$ is called the \qut{Lyapunov exponent}, which we
discussed in somewhat more detail in Chap.~\ref{chap_chaos1}.

The question is then whether two initially close trajectories, also
called \qut{orbits} within dynamical systems theory,
converge or diverge initially. One may generally distinguish between
three different types of behaviors or phases:

\begin{itemize}
\index{phase!chaotic}
\item[--] The Chaotic Phase:\ \ \ $\lambda>0$\\
The Hamming distance grows exponentially, {i.e.}
information is transferred to an exponential large number of
elements. Two initially close orbits {soon
become} very different. This behavior is found for large
connectivities $K$ and is not suitable for real-world biological
systems.
\index{phase!frozen}
\item[--] The Frozen Phase:\ \ \ $\lambda<0$\\
Two close trajectories typically converge, as they are attracted by
the same attractor. This behavior arises for small connectivities
$K$. The system is locally robust.\\[-10pt]
\index{phase!critical}\index{critical!phase}
\item[--] The Critical Phase:\ \ \ $\lambda=0$\\
An exponential time dependence, when present,
dominates all other contributions.
There is no exponential time dependence
when the Lyapunov exponent vanishes
and the Hamming distance then
typically depends algebraically on
time, $D(t)\propto t^\gamma$.
\end{itemize}

\vskip3pt
All three phases can be found in the {$N$--$K$}
model when $N\to\infty$. We will now study the
{$N$--$K$} model and determine its phase diagram.

\subsection{The Mean-Field Phase Diagram}
\label{networks2_Mean_field_phase_diagram}

A mean-field theory, also denoted \qut{molecular-field theory}
is a simple treatment of a microscopic model by
averaging the influence of many components,
lumping them together into a single mean-
or molecular-field. Mean-field theories are
ubiquitous and embedded into
the overall framework of the
\qut{Landau Theory of Phase Transitions},
which we are going to
discuss in Sect.~\ref{automata_Landau_theory}.

\runinhead{Mean-Field Theory}
\index{boolean network!mean-field theory}
\index{mean-field theory!boolean network}
We consider two initial states
$$
\Sigma_0,\qquad\quad \tilde{\Sigma}_0,\qquad\quad
D(0)\ =\ \sum_{i \ = 1}^N \,
\Bigl(\sigma_i -\tilde{\sigma}_i\Bigr)^2~.
$$
We remember that the Hamming distance $D(t)$ measures
the number of elements differing in $\Sigma_t$
and $\tilde\Sigma_t$.


\begin{figure}[t]
\centering
\includegraphics{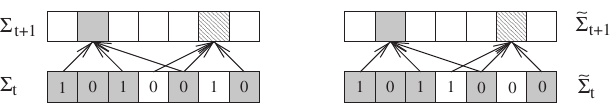}
\caption{{The time} evolution of the
overlap between two states $\Sigma_t$ and $\tilde
\Sigma_t${.} The vertices (given by the {\it{squares}}) can
have values 0 or 1. Vertices with the same value in both states
$\Sigma_t$ and $\tilde \Sigma_t$ are highlighted by a {\textit{gray}}
background. The values of vertices at the {next
time} step, $t+1$, can only differ if the corresponding arguments
are different. Therefore, the vertex with {\textit{gray}} background at time
$t+1$ must be identical in both states. The vertex with
{the} {\textit{striped}} background can have different values in
both states at time, $t+1$, with a probability $2\,p\,(1-p)$, where
$p/(1-p)$ are the probabilities of having vertices with 0/1,
respectively} \label{boolean_overlap}
\end{figure}

For the {$N$--$K$} model, every boolean coupling
function $f_i$ is as likely to occur and every variable is, on the
average, a controlling element for $K$ other variables.
{Therefore, the variables differing in $\Sigma_t$ and
$\tilde\Sigma_t$} affect on the average $KD(t)$ coupling functions,
see Fig.~\ref{boolean_overlap} for an illustration. Every coupling
function changes with probability {half of} its value,
in the absence of a magnetization bias. The number of elements
different in $\Sigma_{t+1}$ and $\tilde\Sigma_{t+1}$ , viz the
Hamming distance $D(t+1)$ will then~be
\begin{equation}
D(t+1)\ =\ {K\over 2} \, D(t),\qquad\quad
D(t)\ =\  \left({K\over 2}\right)^t D(0)\ =\
D(0)\, {\rm e}^{t \ln(K/2)}~.
\label{boolean_expgrow}
\end{equation}
The connectivity $K$ then determines the phase of the
{$N$--$K$} network:
\begin{itemize}
\index{phase!chaotic}
\item[--] {Chaotic}\ \ \ $K>2$\\
     Two initially close orbits diverge, the number of
     different elements, i.e.\ the relative Hamming
     distance grows exponentially with time $t$.

\index{phase!frozen}
\item[--] {Frozen}\ \ \ ($K<2$)\\
     The two orbits approach each other exponentially. All
     initial information contained $D(0)$ is lost.

\index{phase!critical}
\index{critical!phase}
\item[--]  {Critical}\ \ \ ($K_c=2$)\\
     The evolution of $\Sigma_t$ relative to $\tilde\Sigma_t$
     is driven by fluctuations. The power laws typical
     for critical regimes cannot be deduced within mean-field
     theory, which discards fluctuations.
\end{itemize}
The mean-field theory takes only average quantities into account.
The evolution law $D(t+1) = (K/2) D(t)$ holds only on the average.
Fluctuations, viz the deviation of the evolution from the mean-field
prediction, are however of importance only close to a phase
transition, i.e.\ close to the critical point $K=2$.

The mean-field approximation generally works well
for lattice physical systems in high spatial dimensions
and fails in low dimensions, compare Chap.~\ref{chap_chaos1}. The
Kauffman network has no dimension per se, but the connectivity $K$
plays an analogous role.

\runinhead{Phase Transitions in Dynamical Systems and the Brain}
\index{dynamical system!phase transition}\index{phase
transition!dynamical system}The notion of a \qut{phase transition}
originally comes from physics, where it denotes
the transition between two or more different physical
phases, like ice, water and gas, see Chap.~\ref{chap_chaos1},
which are well characterized by their respective order parameters.

The term phase transition therefore {classically denotes} a transition between two
stationary states. The phase transition discussed here involves the
characterization of the overall behavior of a dynamical system. They
are well defined phase transitions in the sense that $1-a^*$ plays
the role of an order parameter{;} its value uniquely
characterizes the frozen {phase} and the chaotic phase in
the thermodynamic limit.

\looseness1 An interesting, completely open and unresolved question is then,
whether dynamical phase transitions play a role in the most complex
dynamical system known, the mammalian brain. It is tempting to
speculate that the phenomena of consciousness {may}
result from a dynamical state characterized by a yet unknown order
parameter. {Were this} true, then this
phenomena would be \qut{emergent} in the strict physical sense, as
order parameters are rigorously defined only in the thermodynamic
limit.

Let us stress, however, that these considerations are very
speculative at this point. In Chap.~\ref{chap_cogSys1}, we will
discuss a somewhat more down-to-earth approach to cognitive systems
theory in general and to aspects of the brain dynamics in
particular.



\subsection{{The} Bifurcation Phase Diagram}
\label{boolean_bifurcation_phase_diagram}
\index{phase diagram!bifurcation}

In deriving {Eq.~}(\ref{boolean_expgrow}) we
assumed that the coupling functions $f_i$ of the
system \nobreak acquire the values $0$ and $1$ with the same
probability $p = 1/2$. We generalize this approach and consider the
case of a magnetic bias in which the coupling functions are
$$
f_i \ =\ \left\{
\begin{array}{rl}
0,& \quad \mbox{with probability\ $p$}\\
1,& \quad \mbox{with probability\ $1-p$}
\end{array}~.
\right.
$$
For a given value of the bias $p$ and connectivity $K$,
there are critical values
$$
K_c(p),\qquad\qquad p_c(K)~,
$$
such that for $K<K_c$ ($K>K_c$) {the} system is in the
frozen phase (chaotic phase). When we consider a fixed connectivity
and vary $p$, then $p_c(K)$ separates the system into a chaotic
{phase} and a frozen phase.

\runinhead{{The} Time Evolution of the Overlap}
\index{normalized overlap!dynamics}
We note that the overlap $a(t)=1-D(t)/N$
between two states $\Sigma_t$ and
$\tilde\Sigma_t$ at time $t$ is the probability
that two vertices have the same value both in
$\Sigma_t$ and in $\tilde\Sigma_t$.
The probability that all arguments of the function $f_i$
will be the same for both configurations is then
\begin{equation}
\rho_K\ =\ \big[\,a(t)\,\big]^K~.
\label{boolean_qtemp}
\end{equation}
As illustrated by Fig.~\ref{boolean_overlap}, the values at the next
time step differ with a probability $2p(1-p)$, but only if the
arguments of the coupling functions are non-different. Together with
the probability that at least one controlling element has different
values in $\Sigma_t$ and $\tilde\Sigma_t$, $1-\rho_K$,
this gives the probability, $(1-\rho_K) 2p(1-p)$,
of values being different in the next time step. We
then have
\begin{equation}
a(t+1)\ =\ 1 - (1-\rho_K)\,2p(1-p)
\ =\ 1 - {1-[a(t)]^K\over K_c}~,
\label{boolean_overp}
\end{equation}
where $K_c$ is given in terms of $p$ as
\begin{equation}
K_c\ =\ {1\over 2p(1- p)},
\qquad\quad p_c^{1,2}\ =\ {1\over 2}\pm \sqrt{{1\over4}-{1\over 2K}}~.
\label{boolean_K_c}
\end{equation}
The fixpoint $a^*$ of Eq.~(\ref{boolean_overp}) obeys
\index{fixpoint!flow of information}
\begin{equation}
a^*\ =\ 1 -{1-[a^*]^K\over K_c}~.
\label{boolean_over_star}
\end{equation}
%

\begin{figure}[t]
\centering
\includegraphics{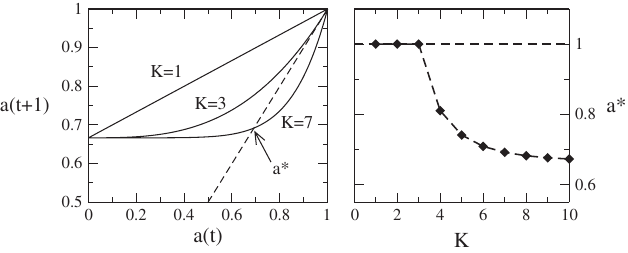}
\caption{Solution of the self-consistency condition
$a^*=1-\left[1-(a^*)^K\right]/K_c$, see
Eq.~(\protect{\ref{boolean_over_star}}).  \textit{Left}: Graphical
solution equating both sides.  \textit{Right:} Numerical result for
$a^*$ for $K_c = 3$.
       The fixpoint $a^*=1$ becomes unstable
       for $K>K_c=3$}
\label{boolean_mapping}
\end{figure}

\index{normalized overlap!self-consistency condition}
\index{self-consistency condition!normalized overlap}This
self-consistency condition for the normalized overlap can be solved
graphically or numerically by simple iterations, see
Fig.~\ref{boolean_mapping}.

\runinhead{Stability Analysis}
\index{flow!of information!stability}
The trivial fixpoint
$$
a^* \ =\ 1
$$
{always constitutes} a solution of
Eq.~(\ref{boolean_over_star}). We examine its stability under the
time evolution Eq.~(\ref{boolean_overp}) by considering a small
deviation $\delta a_t>0$ from the fixpoint solution,
$a_t=a^*-\delta a_t$:
\begin{equation}
1-\delta a_{t+1} \ =\ 1 - {1-[1-\delta a_t]^K\over K_c},
\qquad\quad
\delta a_{t+1} \ \approx\  {K\,\delta a_t\over K_c}~.
\label{boolean_bifurkation_stability}
\end{equation}
The trivial fixpoint $a^*=1$ therefore becomes unstable for
$K/K_c>1$, viz when $K>K_c=\big(2p(1-p)\big)^{-1}$.

\runinhead{Bifurcation}
Equation (\ref{boolean_over_star}) has two solutions for $K>K_c$, a
stable fixpoint $a^*<1$ and the unstable solution  $a^*=1$. One
speaks of a bifurcation, which is shown in
Fig.~\ref{boolean_mapping}. We note that
$$
K_c\Big|_{p=1/2} \ =\ 2~,
$$
in agreement with our previous mean-field result{, Eq.~}
(\ref{boolean_expgrow}), and that
$$
\lim_{K\to\infty} a^*\ =\
\lim_{K\to\infty}\left(
1 -{1-[a^*]^K\over K_c}\right) \ =\
1- {1\over K_c} \ =\ 1- 2p(1-p) ~,
$$
since $a^*<1$ for $K>K_c$, compare Fig.~\ref{boolean_mapping}.
Notice that $a^*=1/2$ for $p=1/2$ corresponds to the average
normalized overlap for two completely unrelated states in the
absence of {the} magnetization bias, $p=1/2$. Two
initial similar states then become completely uncorrelated for
$t\to\infty$ in the limit of infinite connectivity $K$.

\begin{figure}[t]
\centering
\includegraphics{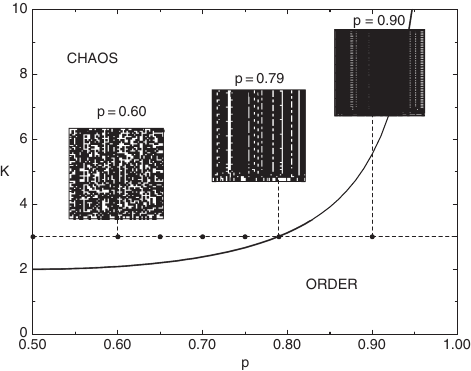}
\caption{Phase diagram for the {$N$--$K$} model.
   The \textit{curve} separating the chaotic {phase} from the ordered (frozen)
   phase is $K_c=[2p(1-p)]^{-1}$.
   The \textit{insets} are simulations for $N=50$ networks with $K=3$
   and $p=0.60$ (chaotic phase), $p=0.79$ (on the critical line)
   and $p=0.90$ (frozen phase). The site index runs {horizontally},
   the time {vertically}. Notice the fluctuations for $p=0.79$
   (from Luque and Sole, 2000)
 }
\label{boolean_critical_line}
\index{phase diagram!N-K model}
\end{figure}

\begin{figure}[t]
\centering
\includegraphics{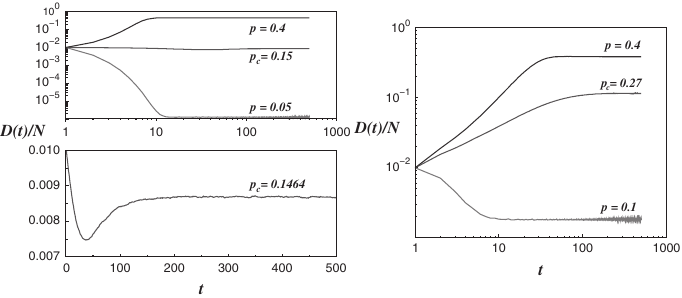}
\caption{Normalized Hamming distance $D(t)/N$ for a
         Kauffman net (\textit{left}) and a square lattice (\textit{right})
         with $N=10\,000$ variables, connectivity $K=4$
         and $D(0)=100$, viz $D(0)/N=0.01$.
\textit{Left}: (\textit{top}) Frozen phase ($p=0.05$), critical
($p_c \simeq 0.1464$)
            and chaotic ($p=0.4$) {phases}, plotted with a logarithmic
            scale{;}
   (\textit{bottom}) Hamming distance for the critical phase ($p=p_c$)
            but in a non-logarithmic graph.
\textit{Right}: Frozen phase ($p=0.1$), critical ($p_c\simeq 0.27$)
and chaotic
      ($p=0.4$) {phases}, plotted with a logarithmic scale. Note that
      $a^*=\lim_{t\to\infty}(1-D(t)/N)<1$ in the frozen state
      of the lattice system, compare Fig.~\ref{boolean_mapping}
      (from Aldana et al., 2003)}
\label{boolean_simulations_hamming}
\end{figure}

\runinhead{Rigidity of the Kauffman Net}
\index{Kauffman network!rigidity}We can connect the results for the
phase diagram of the {$N$--$K$} network
illustrated in Fig.~\ref{boolean_critical_line} with our discussion
on robustness, see Sect.~\ref{boolean_information_flow}.
\begin{itemize}

\item[--] {{The Chaotic} Phase:}\ \ \  $K>K_c$ \newline
\index{Kauffman network!chaotic phase}\index{phase!chaotic}The
infinite time normalized overlap $a^*$ is less than
{1} even when two trajectories $\Sigma_t$ and
$\tilde\Sigma_t$ {start} out very close to each
other. $a^*$, {however, always
remains} above the value expected for two completely unrelated
states. This is so as the two orbits {enter two different attractors
consecutively}, after which the Hamming distance remains constant,
modulo small-scale fluctuations {that} do not
contribute in the thermodynamic limit $N\to\infty$.

\item[--] {{The} Frozen Phase:}\ \ \  $K<K_c$ \newline
\index{Kauffman network!frozen phase}\index{phase!frozen}The
infinite time overlap $a^*$ is exactly one. All trajectories
approach essentially the same configuration independently of the
starting point, apart from fluctuations {that}
vanish in the thermodynamic limit. The system is said to
{\qut{order}}.
\end{itemize}
\runinhead{Lattice Versus Random Networks}
\index{phase!lattice vs.\ random boolean network}The complete loss
of information in the ordered phase observed for the Kauffman net
does not occur for lattice networks, for which $a^*<1$ for any
$K>0$. This behavior of lattice systems is born out by the results
of numerical simulations presented in
Fig.~\ref{boolean_simulations_hamming}. The finite range of the
linkages in lattice systems allows them to store information about
the initial data in spatially finite proportions of the system,
specific to the initial state. For the Kauffman graph every region
of the network is equally close to any other and local storage of
information is impossible.

\runinhead{Percolation Transition in Lattice Networks}
\index{percolation!transition!lattice}For lattice boolean networks
the frozen and chaotic phases cannot be distinguished
{by} examining the value of the long-term normalized
overlap $a^*$, as it is always smaller than unity. The lattice
topology, {however, allows} for a
connection with percolation theory. One considers a finite system,
e.g.\ a $100\times100$ square lattice, and two states $\Sigma_0$ and
$\tilde\Sigma_0$ {that} differ only along one edge.
If the damage, viz the difference in between $\Sigma_t$ and
$\tilde\Sigma_t$ spreads for long times to the opposite edge, then
the system is said to be percolating and in the chaotic phase. If
the damage never reaches the opposite edge, then the system is in
the frozen phase. Numerical simulations indicate, e.g.\ a critical
$p_c\simeq0.298$ for the two-dimensional square lattice with
connectivity $K=4$, compare Fig.~\ref{boolean_simulations_hamming}.

\runinhead{Numerical Simulations}
The results of the mean-field solution for the Kauffman net are
confirmed by numerical solutions of finite-size networks. In
Fig.~\ref{boolean_simulations_hamming} the normalized Hamming
distance, $D(t)/N$, is plotted for both Kauffman graphs and a
two-dimensional squared lattice, both containing $N=10\,000$
elements and connectivity $K=4$.

For both cases results are shown for parameters corresponding to the
frozen {phase} and to the chaotic phase, in addition to a
parameter close to the critical line. Note that $1-a^*=D(t)/N\to0$
in the frozen phase for the random Kauffman network, but not for the
lattice system.

\enlargethispage*{12pt}

\vspace{-12pt}
\subsection{Scale-Free Boolean Networks}
\index{boolean network!scale-free|textbf} 
\index{scale-free!boolean
network|textbf}

The Kauffman model is a reference model which 
can be generalized in various ways, e.g.\
by considering small-world or scale-free networks.

\runinhead{Scale-Free Connectivity Distributions}
\index{scale-free!distribution} 
Scale-free connectivity distributions
\begin{equation}
P(K)\ =\ \frac{1}{\zeta(\gamma)} K^{-\gamma},
\qquad\quad
\zeta(\gamma)\ =\ \sum_{K=1}^\infty K^{-\gamma},
\qquad\quad \gamma>1
\label{boolean_distribution}
\end{equation}
abound in real-world networks, as discussed in Chap.~\ref{chap_networks1}.
Here $P(K)$ denotes the probability to draw a coupling function
$f_i(\cdot)$ having $Z$ arguments. The distribution
Eq.~(\ref{boolean_distribution}) is normalizable for $\gamma > 1$.

The average connectivity $\langle K\rangle$ is
\begin{equation}
\langle K\rangle\ =\
\sum_{K=1}^\infty KP(K) \ =\
\left\{
\begin{array}{cll}
\infty &\ \ \mbox{if}& 1< \gamma \leq 2 \\
 & & \\
\frac{\zeta(\gamma-1)}{\zeta(\gamma)} <\infty
&\ \ \mbox{if}&  \gamma > 2
\end{array}
\right.~,
\label{boolean_firstmoment}
\end{equation}
where $\zeta(\gamma)$ is the Riemann {zeta} function.

\runinhead{Annealed Approximation}
\index{annealed!approximation}
We consider again two states $\Sigma_t$ and
$\tilde\Sigma_t$ and the normalized overlap
$$a(t)\ =\ 1-D(t)/N~,
$$
which is identical to the probability that two vertices in $\Sigma$
and $\tilde\Sigma$ have the same value. In
Sect.~\ref{boolean_bifurcation_phase_diagram} we derived, for a
magnetization bias $p$,
\begin{equation}
a(t+1)\  =\ 1- \left(1-\rho_K\right)2p(1-p)
\label{boolean_scalefree_a_plus_1}
\end{equation}
for the time-evolution of $a(t)$, where
\begin{equation}
\rho_K \ =\ \left[a(t)\right]^K \quad \to\quad
 \sum_{K=1}^\infty \left[a(t)\right]^K P(K)
\label{networks2_scaleFree_rho_K}
\end{equation}
is the average probability that the $K=1,2,\dots$ controlling
elements of the coupling function $f_i()$ are all identical. In
Eq.~(\ref{networks2_scaleFree_rho_K}) we have generalized
Eq.~(\ref{boolean_qtemp}) to a non-constant connectivity
distribution $P(K)$. We then find
\begin{equation}
a(t+1)\ = \ 1-2p(1-p)\left\{1-
\sum_{K=1}^\infty a^K(t)\, P(K)\right\}
\ \equiv\ F(a(t))~,
\label{boolean_scaleFree_self_consistency}
\end{equation}
compare Eq.~(\ref{boolean_overp}). Effectively we have used here an
annealed model, due to the statistical averaging in
Eq.~(\ref{networks2_scaleFree_rho_K}).

\runinhead{Fixpoints Within the Annealed Approximation}
\index{annealed!approximation!fixpoint}In the limit
$t\rightarrow\infty$, Eq.~(\ref{boolean_scaleFree_self_consistency})
becomes the self-consistency equation \index{self-consistency
condition!scale-free boolean net}
$$
a^*\ =\ F(a^*)~,
$$
for the fixpoint $a^*$, where $F(a)$ is defined as the
right-hand-side of Eq.~(\ref{boolean_scaleFree_self_consistency}).
Again, $a^*=1$ is always a  fixpoint of
Eq.~(\ref{boolean_scaleFree_self_consistency}), since $\sum_KP(K)=1$
per definition.

\runinhead{Stability of the Trivial Fixpoint}
\index{stability!trivial fixpoint}We repeat the stability analysis
of the trivial fixpoint $a^*=1$ of
Sect.~\ref{boolean_bifurcation_phase_diagram} and assume a small
deviation $\delta a>0$ from $a^*$:
$$
a^*-\delta a\ =\ F(a^*-\delta a) \ =\
F(a^*) - F'(a^*)\delta a,\qquad\quad
\delta a \ =\ F'(a^*)\delta a~.
$$
The fixpoint $a^*$ becomes unstable if
$F'(a^*)>1$. We find for $a^*=1$
\begin{eqnarray}
1\ =\
\lim_{a\rightarrow1^-}\frac{\mathrm{d}F(a)}{\mathrm{d}a}&=&2p(1-p)
\sum_{k=1}^\infty KP(K) \nonumber \\
&=& 2p(1-p)\,\langle K\rangle~.
\label{boolean_F-slope}
\end{eqnarray}
For $\lim_{a\rightarrow1^-}\mathrm{d}F(a)/{\mathrm{d}a}<1$ the
fixpoint $a^*=1$ is stable, otherwise it is unstable. The phase
transition is then given by
\begin{equation}
2p(1-p)\langle K\rangle\ =\ 1~.
\label{boolean_phasetransition}
\end{equation}
For the classical $N$--$K$ model all elements have the same
connectivity, $K_i=\langle K \rangle = K$, and
Eq.~(\ref{boolean_phasetransition}) reduces to
Eq.~(\ref{boolean_bifurkation_stability}).

\begin{figure}[t]
\centering
\includegraphics{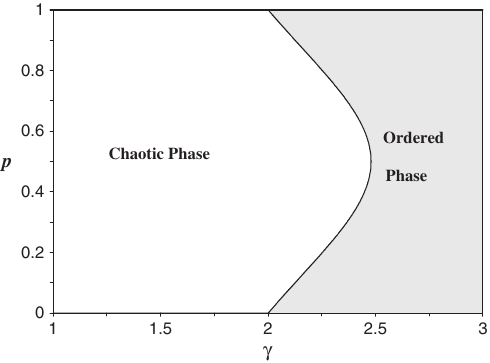}
\caption{Phase diagram for a scale-free boolean network with
connectivity distribution $\propto K^{-\gamma}$. The average
connectivity diverges for $\gamma<2$ and the network is chaotic for
all $p$ (from Aldana and Cluzel, 2003)}
\label{boolean_phaseDiagram_scalefree} \index{phase
diagram!scale-free model}
\end{figure}

\runinhead{{The} Frozen and Chaotic Phases for the Scale-Free Model}
\index{phase!frozen!scale-free model}
\index{scale-free!model!phases}For $1<\gamma\leq2$ the average
connectivity is infinite, see Eq.~(\ref{boolean_firstmoment}).
$F'(1)=2p(1-p)$ $\langle K\rangle$ is then always larger than unity and
$a^*=1$ unstable, as illustrated in
Fig.~\ref{boolean_phaseDiagram_scalefree}. Equation
(\ref{boolean_scaleFree_self_consistency}) {then
has} a stable fixpoint $a^*\neq1${;}  the system is in
the chaotic phase for all $p\in ]0,1[$.

For $\gamma>2$ the first moment of the connectivity distribution
$P(K)$ is finite and the phase diagram is identical to that of the
$N$--$K$ model shown in Fig.~\ref{boolean_critical_line}, with $K$
replaced by $\zeta(\gamma_c-1)/\zeta(\gamma_c)$. The phase diagram
in $\gamma$--$p$ space is presented in
Fig.~\ref{boolean_phaseDiagram_scalefree}. One finds that
$\gamma_c\in[2,2.5]$ for any value of $p$. There is no chaotic
scale-free network for $\gamma>2.5$. It is interesting to note that
$\gamma\in[2,3]$ for many real-world scale-free networks.

%
%

\section{Cycles and Attractors}
\label{networks2_sec_cycles_attractors}

We have emphasized so far the general properties of
boolean networks, such as the phase diagram. We now
turn to a more detailed inspection of the dynamics,
particulary regarding the structure of the attractors.

\subsection{Quenched Boolean Dynamics}
\index{quenched!dynamics|textbf}
\index{dynamics!boolean network!quenched|textbf}

\begin{figure}[t]
\centering
\includegraphics{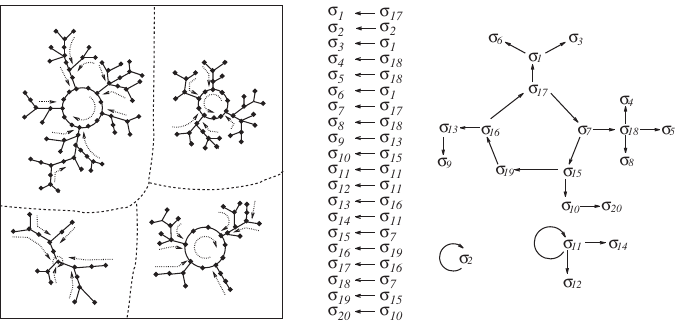}
\caption{{Cycles} and linkages.
\textit{Left}: Sketch of the state space where
every \textit{bold point} stands for a state 
$\Sigma_t=\{\sigma_1,\ldots ,\sigma_N\}$. 
The state space decomposes into distinct attractor
basins for each {cycle attractor} or fixpoint
attractor.  \textit{Right}: Linkage loops for an $N=20$ model with
$K=1$. The controlling elements are listed in the center column.
Each \textit{arrow} points from the controlling element toward the direct
descendant. There are three modules of uncoupled variables (from
Aldana et al., 2003)}
\label{network2_linkages_attractors_cycles}
\vspace*{-12pt}
\end{figure}

\runinhead{Self-{Retracting} Orbits}
\index{orbit!self-retracting}\index{self-retracting!orbit}
{From now on we consider} quenched
systems for which the coupling functions $f_i(\sigma_{i_1},\ldots
,\sigma_{i_K})$ are fixed for all times. Any orbit eventually
{partly retraces itself}, since the
state space $\Omega = 2^N$ is finite. The long-term trajectory is
therefore cyclic.
\begin{quotation}
{\it Attractors.\enspace}
\index{attractor!boolean network}An attractor $A_0$ of a discrete
dynamical system is a region $\{\Sigma_t\}\subset\Omega$ in phase
space that maps completely onto itself under the
time evolution $A_{t+1}= A_t\equiv A_0$.
\end{quotation}
Attractors are typically cycles
\index{attractor!cyclic}
\index{cycle!attractor}
$$
\Sigma^{(1)}\quad \to\quad \Sigma^{(2)}\quad\to\quad \ldots \quad
\to\quad \Sigma^{(1)}~,
$$
see Figs.~\ref{networks2_fig_booleanNetwork_example} and
\ref{network2_linkages_attractors_cycles} for some examples. Fixed
points are cycles of length 1.

\begin{quotation}
{\it The Attraction Basin.\enspace}
\index{attractor!basin}\index{basin of attraction!cycle}
The attraction basin $B$ of an attractor $A_0$ 
is the set $\{\Sigma_t\}\subset\Omega$ for which 
there is a time $T<\infty$ such that $\Sigma_T\in A_0$.
\end{quotation}
The probability to end up in a given cycle is 
directly proportional, for randomly drawn initial 
conditions, to the size of its basin of attraction. 
The three-site network illustrated in 
Fig.~\ref{networks2_fig_booleanNetwork_example}
is dominated by the fixpoint $\{1,1,1\}$, which 
is reached with probability 5/8 for random 
initial starting states.

\enlargethispage{12pt}

\runinhead{Attractors are Everywhere}
Attractors and fixpoints are generic features of dynamical systems
and are very important for their characterization, as they
dominate the time evolution in state space within their respective
basins of attraction. Random boolean networks allow for very
detailed studies of the structure of attractors and of the
connection to network topology. Of special interest in this context
is how various properties of the attractors, like the cycle length
and the size of the attractor basins, relate to the thermodynamic
differences between the frozen phase and the chaotic phase. These
are the issues that we shall now discuss.

%

\enlargethispage*{12pt}

\runinhead{Linkage Loops, Ancestors and Descendants}
\index{loop!linkage} Every variable $\sigma_i$ can appear as an
argument in the coupling functions for other elements;
it is said to act as a controlling element. The collections of all
such linkages can be represented graphically by a directed graph, as
illustrated in
Figs.~\ref{networks2_fig_booleanNetwork},
\ref{networks2_fig_booleanNetwork_example} and
\ref{network2_linkages_attractors_cycles}, with the vertices
representing the individual binary variables. Any given element
$\sigma_i$ can then influence a large number of different states
during the continued time evolution.

\begin{quotation}
{\it Ancestors and Descendants.\enspace}
\index{ancestor!boolean dynamics} \index{boolean dynamics!ancestor}
\index{descendant!boolean dynamics}
\index{boolean dynamics!{descendant}}The
elements a vertex affects consecutively via the coupling functions
are called its {descendants}. Going backwards
in time one find ancestors for each element.
\end{quotation}
In the 20-site network illustrated in
Fig.~\ref{network2_linkages_attractors_cycles} the
{descendants} of $\sigma_{11}$ are
$\sigma_{11}$,  $\sigma_{12}$ and $\sigma_{14}$.

When an element is its own descendant (and
ancestor) it is said to be part of a \qut{linkage loop}.
\index{linkage!loop}Different linkage loops can overlap, as
is the case for the linkage loops
$$
\sigma_1\to\sigma_2\to\sigma_3\to\sigma_4\to\sigma_1,\qquad\quad
\sigma_1\to\sigma_2\to\sigma_3\to\sigma_1
$$
shown in Fig.~\ref{networks2_fig_booleanNetwork}. Linkage loops are
disjoint for $K=1$, compare
Fig.~\ref{network2_linkages_attractors_cycles}.

\runinhead{Modules and Time Evolution}
\index{boolean network!time evolution}
The set of ancestors and descendants determines 
the overall dynamical \nobreak{dependencies.}%
\begin{quotation}
{\it Module.\enspace}
\index{boolean network!linkage module}
\index{module!boolean network} 
The collection of all ancestors and
descendants of a given element $\sigma_i$ is
called the module (or component) 
to which $\sigma_i$ belongs.
\end{quotation}
If we go through all variables $\sigma_i$, $i=1,\ldots ,N$ we find
all modules, with every element belonging to one and only one
specific module. Otherwise stated, disjoint modules correspond to
disjoint subgraphs, the set of all modules constitute the full
linkage graph. The time evolution is block-diagonal in terms of
modules; $\sigma_i(t)$ is independent of all variables
not belonging to its own module, for all times $t$.

In lattice networks the clustering coefficient
(see Chap.~\ref{chap_networks1}) is large and
closed linkage loops occur frequently. For big lattice
systems with a small mean linkage $K$ we expect far away spatial
regions to evolve independently, due the lack of long-range
connections.

\runinhead{Relevant Nodes and Dynamic Core}
Taking a look at dynamics of the 20-site model
illustrated in
Fig.~\ref{network2_linkages_attractors_cycles},
we notice that, e.g., the elements $\sigma_{12}$
and $\sigma_{14}$ just follow the dynamics of
$\sigma_{11}$, they are \qut{enslaved} by $\sigma_{11}$.
These two elements do not control
any other element and one could just delete them
from the system wihout qualitative changes to
the overall dynamics.
\begin{quotation}
{\it Relevant Nodes.\enspace}
\index{boolean network!relevant node}
A node is termed relevant if its state is
not constant and if it controls at least one other
relevant element (eventually itself).
\end{quotation}
\index{boolean network!dynamic core}
An element is constant if it evolves, indepedently
of the initial conditions, always to the same state
and not constant otherwise.
The set of relevant nodes, the dynamic core, controls
the overall dynamics. The dynamics of all other nodes can 
be disregarded without changing the attractor structure. The 
node $\sigma_{13}$ of the 20-site network illustrated in
Fig.~\ref{network2_linkages_attractors_cycles}
is relevant if the boolean function connecting it to
itself is either the identity or the negation
(see p.~\pageref{boolean_boolean_f2}). 

The concept of a dynamic core is of great importance for
practical applications. Gene expression networks may
be composed of thousands of nodes, but contain generally
a relatively small dynamic core controlling the overall
network dynamics. This is the case, e.g., for the gene
regulation network controlling the yeast cell cycle
discussed in Sect.~\ref{networks2_yeast_cell_cycle}.

\runinhead{Lattice Nets versus Kauffman Nets}\
For lattice systems the linkages are short-ranged 
and whenever a given element $\sigma_j$ acts
as a controlling element for another element $\sigma_i$ there is a
high probability that the reverse is also true, viz that $\sigma_i$
is an argument of $f_j$.

The linkages are generally non-reciprocal for the Kauffman
net; the probability for reciprocality is just $K/N$
and vanishes in the thermodynamic limit for finite $K$. The number
of disjoint modules in a random network therefore grows more slowly
than the system size. For lattice systems, {on the other hand, the number of
modules is} proportional to the size of the system. The
{differences} between lattice and Kauffman
networks translate to different cycle structures, as every periodic
orbit for the full system is constructed out of the individual
attractors of all modules present in the network considered.

\subsection[The ${\textit{K = 1}}$ Kauffman Network]
{The $\textbf{\textit{K = 1}}$ Kauffman Network}
\label{networks_2_K_1_Kauffman}
\index{Kauffman network!K=1|textbf}

We start our discussion of the cycle structure of Kauffman nets with
the case $K=1$, which can be solved exactly. The maximal length for
a linkage loop $l_{\max}$ is on the average of the order of
\begin{equation}
l_{\max}\ \sim \ N^{ 1 / 2} ~. \label{boolean_l_max}
\end{equation}
The linkage loops determine the cycle structure together with
the choice of the coupling ensemble. As an example we discuss
the case of an $N=3$ linkage loop.

\runinhead{The Three-site Linkage Loop with Identities}
\index{linkage!loop!K=1 network}For $K=1$ there are only two
non-constant coupling functions, i.e.\ the identity $I$
and the negation $\neg$, see p.~\pageref{boolean_boolean_f2}. We
start by considering the case of all the coupling functions being
the identity:
$$
ABC\ \to\ CAB\ \to\ BCA\ \to\ ABC\ \to\ \dots~,
$$
where we have denoted {by}  $A,B,C$ the values of the
binary variables $\sigma_i$, $i=1,2,3$. There are two cycles of
length {1}, in which all elements are identical. When
the three elements are not identical, the cycle length is
{3}. The complete dynamics is then:
$$
\begin{array}{rcl}
000 &\to & 000\\
111 &\to & 111\\
\end{array}
\qquad\qquad\qquad\qquad
\begin{array}{rcccccl}
100 &\to & 010 &\to& 001&\to& 100\\
011 &\to & 101 &\to& 110&\to& 011\\
\end{array}
$$

\runinhead{Three-Site Linkage Loops with Negations}
Let us consider now the case that all three 
coupling functions are negations:
$$
ABC\ \to\ \bar C\bar A\bar B\ \to\ BCA\ \to\ 
\bar A\bar B\bar C\ \to\ \dots
\qquad\quad
\bar A=\neg A,\ \mbox{etc.}~.
$$
The cycle length is 2 if all elements are identical
$$
\begin{array}{rcccl}
000 &\to & 111 &\to& 000
\end{array}
$$
and of length 6 if they are not.
$$
\begin{array}{rcccccccccccl}
100 &\to & 101 &\to& 001&\to& 011 &\to & 010 &\to& 110&\to& 100~.
\end{array}
$$
The complete state space $\Omega=2^3=8$ decomposes into two cycles,
one of length 6 and one of length 2.

\runinhead{Three-Site Linkage Loops with a Constant Function}
Let us see what happens if any of the coupling functions
{are} a constant function. For illustration
{purposes} we consider the case of two constant functions
0 and 1 and the identity:
\begin{equation}
ABC\ \to\ 0A1\ \to\ 001\ \to\ 001~.
\label{networks2_lingage_loop_3_constant}
\end{equation}

\enlargethispage{5pt}

Generally it holds that the cycle length is 1 if any of the
coupling functions is an identity and that there is then only a
single fixpoint attractor. Equation
(\ref{networks2_lingage_loop_3_constant}) holds for all
$A,B,C\in\{0,1\}$; the basin of attraction for 001 is therefore
the whole state space, and 001 is a global attractor.

The Kauffman net can contain very large linkage loops for $K=1$, see
Eq.~(\ref{boolean_l_max}), but then the probability that a given
linkage loop contains at least one constant function is also very
high. The average cycle length therefore remains short for the $K=1$
Kauffman net.

\runinhead{Loops and Attractors}
The attractors are made up of the set of
linkage loops. As an example we consider
a 5-site network with two linkage loops,
$$
A\ \to^{\hspace{-1.5ex}I\hspace{1ex}}\ B\ 
   \to^{\hspace{-1.5ex}I\hspace{1ex}}\ C\ 
   \to^{\hspace{-1.5ex}I\hspace{1ex}}\ A,
\qquad\quad
D\ \to^{\hspace{-1.5ex}I\hspace{1ex}}\ E\ 
   \to^{\hspace{-1.5ex}I\hspace{1ex}}\ D ~,
$$
with all coupling functions being the identity $I$.
The states 
$$
00000, \qquad\quad
00011, \qquad\quad
11100, \qquad\quad
11111  \qquad\quad
$$
are fixpoints in phase space $\Sigma=ABCDE$.
Examples of cyclic attrators of length 3 and 6 are 
$$
10000\ \to\ 01000\ \to\ 00100\ \to\ 10000
$$
and
$$
10010\ \to\ 01001\ \to\ 00110\ \to\ 10001\ 
\to\ 01010\ \to\  00101\ \to\ 10010~.
$$
In general, the length of an attractor is given
by the least common multiple of the periods of the
constituent loops. This relation holds for $K=1$
Boolean networks, for general $K$ the attractors
are composed of the cycles of the constituent
set of modules.

\runinhead{Critical $K=1$ Boolean networks}
When the coupling ensemble is selected uniformly, 
compare Sect.~\ref{networks2_subsect_coupling_functions},
the $K=1$ network is in the frozen state.
If we do however restrict our coupling ensemble
to the identity I and to the negation $\neg$,
the value of one node is just copied or inverted
to exactly one other node. There is no loss of
information anymore, when disregarding the two
constant $K=1$ coupling functions
(see p.~\pageref{boolean_boolean_f2}). The information
is not multiplied either, being transmitted to exactly one
and not more nodes. The network is hence critical,
as pointed out in Sect.~\ref{boolean_information_flow}.

\subsection[The ${\textit{K = 2}}$ Kauffman Network]
{The $\textbf{\textit{K = 2}}$ Kauffman Network}
\index{Kauffman network!K=2|textbf}

The $K=2$ Kauffman net is critical, as discussed in
Sects.~\ref{boolean_information_flow} and
\ref{networks2_Mean_field_phase_diagram}. When physical systems
undergo a (second-order) phase transition, power laws are
expected right at the point of transition for many
response functions; see the discussion in
Chap.~\ref{chap_chaos1}. It is therefore natural
to expect the same for critical dynamical
systems, such as a random boolean network.

This expectation was indeed initially born out of a series
of mostly numerical investigations, which indicated that
both the typical cycle lengths, as well as the mean number
of different attractors, would grow algebraically with $N$,
namely like $\sqrt{N}$. It was therefore tempting to relate
many of the power laws seen in natural organisms to the
behavior of critical random boolean networks.

\runinhead{Undersampling of the State Space}
\index{state space!undersampling}\looseness1The problem to determine
the number and the length of cycles is, however, numerically very
difficult. In order to extract power laws one has to simulate
systems with large $N$. The state space $\Omega=2^N$,
however, grows exponentially, so that an
exhaustive enumeration of all cycles is impossible. One has
therefore to resort to a weighted sampling of the state
space for any given network realization and to extrapolate from the
small fraction of states sampled to the full state space. This
method yielded the $\sqrt{N}$ dependence referred
to above.

The weighted sampling is, however, not without
problems; it might in principle undersample the state
space. The number of cycles found in the average state
space might not be representative for the overall number of cycles,
as there might be small fractions of state space with very high
number of attractors dominating the total number of
attractors.

This is indeed the case.  One can prove rigorously that the number
of attractors grows faster than any power for the $K=2$ Kauffman
net. One might still argue, however, that for biological
applications the result for the \qut{average state space} is
relevant, as biological systems are not too big
{anyway}. The hormone regulation network of mammals
contains of the order of 100 elements, the gene regulation network
of the order of 20\,000 elements.

%

\subsection[The ${\textit{K = N}}$ Kauffman Network]{The $\textbf{\textit{K = N }}$ Kauffman Network}
\index{Kauffman network!K=N|textbf}

Mean-field theory holds for the fully connected network
$K=N$ and we can evaluate the average number and length
of cycles using probability arguments.

\runinhead{The Random Walk Through Configuration Space}
\index{random!walk!configuration space}
We consider an orbit starting from an arbitrary configuration
$\Sigma_0$ at time $t = 0 $. The time evolution
generates a series of states
$$
\Sigma_0,\ \Sigma_1,\ \Sigma_2,\ \ldots
$$
through the configuration space of size $\Omega=2^N$. We consider
all $\Sigma_t$ to be uncorrelated, viz we consider a random walk.
This assumption holds due to the large connectivity
$K=N$.\enlargethispage*{12pt}

\begin{figure}[!t]
\centering
\includegraphics{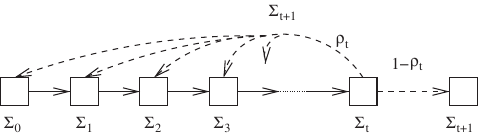}
\caption{{A} random walk in
configuration space. The relative probability of closing the loop at
time $t$, $\rho_t=(t+1)/\Omega$, is the probability that
$\Sigma_{t+1}\equiv\Sigma_{t'}$, with a certain
 $t'\in[0,t]$ }
\label{random_walk}
\end{figure}

\runinhead{Closing the Random Walk}
\index{random!walk!closed}The walk through configuration space
continues until we hit a previously visited point, 
see Fig.~\ref{random_walk}. We
define by
\begin{itemize}

\item[--] $q_t$:  the probability that the trajectory remains unclosed
after $t$ steps;
\smallskip

\item[--] $P_{t}$: the probability of terminating the excursion 
exactly at time $t$.
\end{itemize}
If the trajectory is still open at time $t$, we have already visited
$t + 1 $ different sites (including the sites $\Sigma_0$ and
$\Sigma_t$). Therefore, there are $t+1$ ways of terminating the
walk at the next time step. The relative probability of termination
is then \hbox{$\rho_t=(t+1)/\Omega$} and the overall probability
$P_{t+1}$ to terminate the random walk at time $t+1$ is
%

\vspace*{3pt}
$$
P_{t+1} \ =\ \rho_t\,q_t\ =\ {t+1\over\Omega}\,q_t~.
$$\vspace*{1pt}

The probability of still having an open
trajectory after $t+1$ steps is
\vspace*{3pt}
\[
q_{ t + 1}\ =\ q_t (1 - \rho_t)\ =\
q_t \left( 1 - \frac{t+1}{\Omega} \right)
\ =\ q_0\prod_{i=1}^{t+1}
\left(1-\frac{i}{\Omega}\right),
\qquad\quad q_0 = 1~.
\]
The phase space $\Omega=2^N$ diverges in the thermodynamic
limit $N\to\infty$ and the approximation
\vspace*{3pt}
\begin{equation}
q_t\ =\ \prod_{i=1}^t \left(1-\frac{i}{\Omega}\right)
\ \approx\   \prod_{i=1}^t \,e^{-i/\Omega}
\ =\   e^{-\sum_i i/\Omega}
\ =\   e^{-t(t+1)/(2\Omega)}
\label{boolean_q_t_larger_Omega}
\end{equation}
\vspace*{3pt}
becomes exact in this limit. For large times $t$ we have
$t(t+1)/(2\Omega)\approx t^2/(2\Omega)$ in
Eq.~(\ref{boolean_q_t_larger_Omega}). The probability
$$
\sum_{t=1}^\Omega\,P_t \ \simeq\
\int_0^\infty dt \,{t\over\Omega}\, e^{-t^2/(2\Omega)}
\ =\ 1
$$
for the random walk to close at all is unity.

\runinhead{Cycle Length Distribution}
\index{cycle!length distribution}\index{distribution!cycle length}
The probability $\langle N_c (L) \rangle$ that the system contains a
cycle of length $L$ is
\begin{equation}
\langle N_c (L) \rangle\ =\ {q_{t=L}\over \Omega}\,{\Omega\over L}
\ =\
{\exp[-L^2 / (2 \Omega)]\over L}~,
\label{boolean_N_c}
\end{equation}
where we used Eq.~(\ref{boolean_q_t_larger_Omega}). $\langle \cdots
\rangle$ denotes an ensemble average over realizations. In deriving
Eq.~(\ref{boolean_N_c}) we used the following
considerations:
\begin{enumerate}\leftskip7pt
\item[(i)\phantom{ii}] The probability that $\Sigma_{t+1}$ is identical to
$\Sigma_0$ is $1/\Omega$.

\item[(ii)\phantom{i}] There are $\Omega$ possible starting points (factor $\Omega$).

\item[(iii)]  Factor $1/L$ corrects for the {overcounting} of cycles when
      considering the $L$ possible starting sites of the $L$-cycle.
\end{enumerate}

\runinhead{Average Number of Cycles} \index{mean!number of cycles}
\index{cycle!average number} We are interested in the mean number
$\bar N_c$ of cycles,
\begin{equation}
\bar N_c \ =\
\sum_{L=1}^N\, \langle N_c(L) \rangle
\ \simeq \  \int_{1}^\infty dL\, \langle N_c(L) \rangle ~.
\label{networks2_bar_N_c}
\end{equation}
When going from the sum $\sum_L$ to the integral $\int \mathrm{d}L$
in Eq.~(\ref{networks2_bar_N_c}) we neglected
terms of order unity. We find
$$
\bar N_c\ =\ \int_{1}^\infty \mathrm{d}L~{\exp[-L^2/(2\Omega)]\over
L} \ =\ \underbrace{ \int_{1/\sqrt{2\Omega}}^1 \mathrm{d}u\,
{e^{-u^2}\over u}
           }_{\equiv\ I_1}
\,+\, \underbrace{ \int_{1}^\infty \mathrm{d}u\, {e^{-u^2}\over u}
           }_{\equiv\ I_2}
~,
$$
where we rescaled the variable by $u=L/\sqrt{2\Omega}$.
For the separation
$\int_{1/\sqrt{2\Omega}}^\infty =
 \int_{1/\sqrt{2\Omega}}^c+\int_c^\infty$
of the integral {above we used} $c=1$
for simplicity{;} any other finite value for $c$ would
do also the job.

The second integral, $I_2$, does not diverge as $\Omega \rightarrow
\infty$. For $I_1$ we have \vspace{-6pt}
\begin{eqnarray}
\nonumber I_1 &=& \int_{1/\sqrt{2\Omega}}^1
\mathrm{d}u\,{e^{-u^2}\over u} \ =\  \int_{1/\sqrt{2\Omega}}^1
\mathrm{d}u\, \frac{1}{u}\,
\Big(1-u^2+\frac{1}{2} u^4 + \ldots \Big) \\
&\approx& \ln (\sqrt{2\Omega})~,
\end{eqnarray}
\vspace{-6pt}%
\noindent since all further terms $\propto\int_{1/\sqrt{2\Omega}}^1
\mathrm{d}u\,u^{n-1}<\infty$ for $n=2,4,\ldots$ and
$\Omega\to\infty$. The aver-\vspace*{3pt} age number of cycles is
then
\begin{equation}
\bar N_c \ =\ \ln (\sqrt{2^N})\,+\,O(1)\ =\ {N\ln 2\over 2}\, +\,O(1)
\label {boolean_N_ctot}
\end{equation}
for the $N=K$ Kauffman net in thermodynamic limit $N\to\infty$.

\runinhead{Mean Cycle Length}
\index{mean!cycle length}
\index{cycle!average length}
The average length $\bar L$ of a random cycle is
\begin{eqnarray} \nonumber
\bar L & =& {1\over\bar N_c}
\sum_{L=1}^\infty\, L\,\langle N_c(L) \rangle \ \approx\
{1\over\bar N_c}
\int_{1}^\infty \mathrm{d}L\,L~{\exp[-L^2/(2\Omega)]\over L} \\[6pt]
& =& {1\over\bar N_c} \int_{1}^\infty \mathrm{d}L~e^{-L^2/(2\Omega)}
\ =\ {\sqrt{2\Omega}\over\bar N_c} \int_{1/\sqrt{2\Omega}}^\infty
\mathrm{d}u~e^{-u^2} \label{eq_mean_cycle_length}
\end{eqnarray}

\noindent after rescaling with $u=L/\sqrt{2\Omega}$ and using
Eq.~(\ref{boolean_N_c}). The last integral on the
\hbox{right-hand-side} of Eq.~(\ref{eq_mean_cycle_length}) converges
for $\Omega\to\infty$ and the mean cycle length $\bar L$
{consequently scales as}
\begin{equation}
\bar L \ \sim \ \Omega^{1/2}/N \ =\ 2^{N/2}/N
\label{networks2_bar_L_result}
\end{equation}
for the $K=N$ Kauffman net, when using
Eq.~(\ref{networks2_bar_N_c}), $\bar N_c\sim N$.

\vspace{-9pt}
\section{Applications}

\subsection{Living at the Edge of Chaos}
\label{networks2_live_edge_chaos}

\runinhead{Gene Expression Networks and Cell Differentiation}
\index{network!gene expression} \index{cell
differentiation!{$N$--$K$}
network}\looseness-1Kauffman introduced the $N$--$K$ model in the
late {1960s} for the purpose of modeling the
dynamics and time evolution of networks of interacting genes,
{i.e.} the gene expression network. In this model an
active gene might influence the expression of any other gene, e.g.\
when the protein transcripted from the first gene influences the
expression of the second gene.

The gene expression network of real-world cells is not random. The
web of linkages and connectivities among the genes in a living
organism is, however, very intricate, and to model the gene--gene
interactions as randomly linked is a good zero-th order
approximation. One might then expect to gain a generic insight into
the properties of gene expression {networks;}
insights {that} are independent of the particular
set of linkages and connectivities realized in any particular living
cell.

\runinhead{Dynamical Cell Differentiation} \index{cell
differentiation!dynamical}Whether random or not, the gene expression
network needs to result in a stable dynamics in order for the cell
to keep functioning. Humans have only a few hundreds of different
cell types in their bodies. Considering the fact that every single
cell contains the identical complete genetic
material, in 1969 Kauffman proposed an, at that time
revolutionary, suggestion that every cell type corresponds to a
distinct dynamical state of the gene expression network. It is
natural to assume that these states correspond to attractors, viz in
general to cycles. The average length $\bar L$ of a cycle in a
$N$--$K$ Kauffman net is
$$ \bar L\ \sim 2^{\alpha N}
$$
in the chaotic phase, e.g.\ for $N=K$ where $\alpha=1/2$, see
Eq.~(\ref{networks2_bar_L_result}), The mean cycle length $\bar L$
is exponentially large; consider that $N\approx 20\,000$ for the
human
genome. A~single cell would take the
universe's lifetime to complete a single cycle, {which is} an
unlikely setting. It then follows that gene expression networks of
living organisms cannot be operational in the chaotic phase.

\enlargethispage{12pt}

\runinhead{Living at the Edge of Chaos} \index{life!edge of
chaos}\index{chaos!life at the edge of}If the gene expression
network cannot operate in the chaotic phase there are but two
possibilities left: the frozen phase or the critical point. The
average cycle length is short in the frozen phase, see
Sect.~\ref{networks_2_K_1_Kauffman}{,} and the dynamics
stable. The system is consequently very resistant to
{damage} of the linkages.

\looseness-1But what about Darwinian evolution? Is too much
stability good for the adaptability of cells in a changing
environment? Kauffman suggested that gene expression networks
operate {\em at the edge of chaos}, an expression {that has become legendary}. By this he
meant that networks close to criticality {may}
benefit from the stability properties of the close-by frozen phase
and {at the same time exhibit}
enough sensitivity to changes in the network structure so that
Darwinian adaption remains possible.

But how {can} a system reach criticality by itself?
For the $N$--$K$ network there is no extended critical phase, only a
single critical point $K=2$. In Chap.~\ref{chap_automata1}
we will discuss mechanisms that allow certain adaptive
systems to evolve their own internal parameters autonomously
in such a way that they approach the critical point.
This phenomenon is called \qut{self-organized criticality}.

One could then assume that Darwinian evolution trims the gene
expression networks towards criticality: Cells in the chaotic phase
are unstable and {die;} cells deep in the frozen
phase cannot adapt to environmental changes and are selected out in
the course of time.

\begin{figure}[t]
\centering
\includegraphics{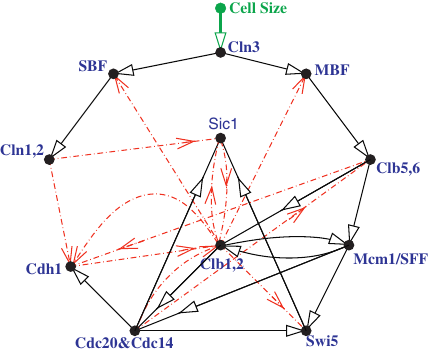}
\caption{The $N=11$ core network responsible for the yeast cell
cycle. {\textit{Acronyms}} denote protein names, 
{\textit{solid arrows}} excitatory connections and 
{\textit{dashed arrows}} inhibitory connections. Cln3 is
inactive in the resting state $G_1$ and becomes active 
when the cell reaches a certain size ({\textit{top}}), 
initiating the cell division process
(compare Li et al., 2004)
        }
\label{networks2_fig_yeast_11_net}
\end{figure}

\subsection{The Yeast Cell Cycle}
\label{networks2_yeast_cell_cycle}
\index{yeast cell cycle|textbf}
\index{cell!yeast cycle|textbf}
\index{cycle!yeast}

\runinhead{The Cell Division Process} \index{cell!division}Cells
have two tasks: to survive and to multiply. When a living 
cell grows too big, a cell division process starts. The cell
cycle has been studied intensively for the budding yeast. 
In the course of the division process the cell goes through a
distinct set of states
$$
G_1\ \to\ S \ \to\ G_2 \ \to\ M\ \to\ G_1~,
$$
with $G_1$ being the \qut{ground state} in physics slang, viz the
normal cell state and the chromosome division takes place during the
$M$ phase. These states are characterized by distinct gene
activities, i.e.\ by the kinds of proteins active in the cell. All
eukaryote cells have similar cell division cycles.

\runinhead{The Yeast Gene Expression Network} \index{network!gene
expression!yeast}From the $\approx800$ genes involved only
\hbox{11--13} core genes are actually regulating the part of the
gene expression network responsible for the division process{;} all
other genes are more or less just descendants of the core genes. The
cell dynamics contains certain checkpoints,
where the cell \nobreak division process {can} be stopped if
something {were to go} wrong. When eliminating the checkpoints a
core network with only 11 elements remains. This network is shown in
Fig.~\ref{networks2_fig_yeast_11_net}.

\begin{figure}[t]
\centering
\includegraphics{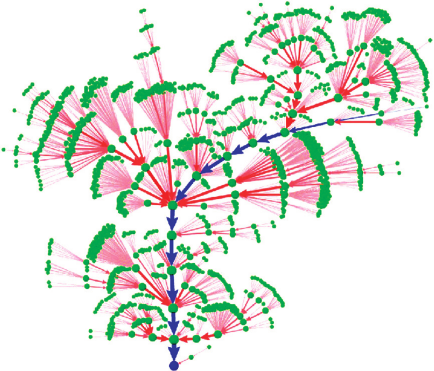}
\caption{The yeast cell cycle as an attractor trajectory of the gene
expression network. Shown are the 1764 states (\textit{green dots},
out of the $2^{11}=2048$ states in phase space $\Omega$) making up
the basin of attraction of the {biologically}
stable $G_1$ state (at the {\textit{bottom}}). After starting with the excited
$G_1$ normal state ({the} first state in {the}
biological pathway represented by {\it blue arrows}), compare
Fig.~\ref{networks2_fig_yeast_11_net}, the boolean dynamics runs
through the {known} intermediate states (blue arrows)
until the $G_1$ states attractor is again reached, representing the
two daughter cells (from Li et al., 2004)
        }
\label{networks2_fig_yeast_cycle}
\end{figure}

\runinhead{Boolean Dynamics}
The full dynamical dependencies are not yet
known for the yeast gene expression
network. The simplest model is to assume
\begin{equation}
\sigma_i(t)\ =\
\left\{
\begin{array}{rcl}
1 &\ \mbox{if}\ & a_i(t)>0\\
0 &\ \mbox{if}\ & a_i(t)\le0
\end{array}
\right.,
\qquad\quad
a_i(t)=\sum_j w_{ij}\sigma_j(t)~,
\label{networks2_yeast_dynamics}
\end{equation}
i.e.\ a boolean dynamics\footnote{Genes are
boolean variables in the sense that they are
either expressed or not. The quantitative 
amount of proteins produced by a given active
gene is regulated via a separate mechanism
involving microRNA, small RNA snippets.}
for the binary
variables $\sigma_i(t)=0,1$ representing
the activation/deactivation of protein $i$,
with couplings $w_{ij}=\pm\,1$ for an
excitatory/inhibitory functional relation.

\runinhead{Fixpoints}
The 11-site network has 7 attractors, all cycles of length
{1}, viz fixpoints. The dominating fixpoint has an
attractor basin of 1764 states, representing about 72\% of the state
space $\Omega=2^{11}=2048$. {Remarkably}, the
protein activation pattern of the dominant fixpoint corresponds
exactly to that of the experimentally determined $G_1$ ground state
of the living yeast cell.

\runinhead{The Cell Division Cycle}
\index{yeast cell cycle}
\index{cell!yeast cycle}
In the $G_1$ ground state the protein
Cln3 is inactive. When the cell reaches
a certain size it becomes expressed,
i.e.\ it becomes active. For the
network model one then just starts the
dynamics by setting
$$
\sigma_{Cln3}\ \to\ 1,
\qquad\quad \mbox{at}\quad t=0
$$
in the $G_1$ state. The {ensuing
simple boolean dynamics, induced by
Eq.~(\ref{networks2_yeast_dynamics})}, is depicted in
Fig.~\ref{networks2_fig_yeast_cycle}.

The remarkable result is that the system follows an attractor
pathway {that} runs through all experimentally known
intermediate cell states, reaching the ground state $G_1$ in 12
steps.

\begin{figure}[t]
\centering
\includegraphics{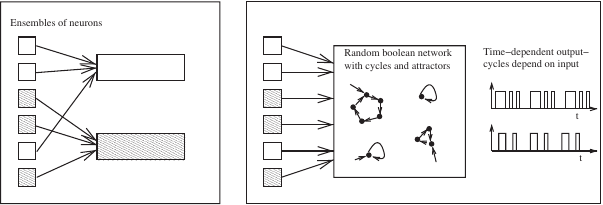}
 \caption{Illustration of ensemble (\textbf{a}) and time
(\textbf{b}) encoding. \textit{Left}: All receptor neurons
corresponding to the
      same class of input signals are combined,
      as {occurs} in the nose for different odors.
\textit{Right}: The primary input signals are mixed together
       by a random neural network close to criticality and
       the relative weights are time encoded by the
       output signal
        }
\label{networks2_fig_time_encoding}
\end{figure}

\runinhead{Comparison with Random Networks}
The properties of the boolean network depicted in
Fig.~\ref{networks2_fig_yeast_11_net} can be compared with those of
a random boolean network. A random network of the same size and
average connectivity would have more attractors with correspondingly
smaller basins of attraction. Living cells clearly need a robust
protein network to survive in harsh environments.

Nevertheless, the yeast protein network shows more or less the same
susceptibility to damage as a random network. The core yeast protein
network has an average connectivity of $\langle K\rangle
=27/11\simeq2.46$. The core network has only $N=11$ sites, a number
far too small to allow comparison with the properties of $N$--$K$
networks in the thermodynamic limit $N\to\infty$. Nevertheless, an
average connectivity of 2.46 is remarkably close to $K=2$,
{i.e.} the critical connectivity for $N$--$K$ networks.

\runinhead{Life as an Adaptive Network} \index{adaptive
system!life}Living beings are complex and adaptive dynamical
systems; a subject that we will further
dwell on in Chap.~\ref{chap_evolution1}.
The here discussed preliminary results on the
yeast gene expression network indicate that this statement is not
just an abstract notion. Adaptive regulative networks constitute the
core of all living.

\subsection{Application to Neural Networks}
\runinhead{Time Encoding by Random Neural Networks}
   \index{time!encoding}
There is some debate in neuroscience whether, and to which
{extent}, time encoding is used in neural
processing.
\begin{itemize}
\item[--] {Ensemble Encoding}:
   \index{ensemble!encoding}Ensemble encoding is present when the activity of
   a sensory input is transmitted via the firing
   of certain ensembles of {neurons}. Every sensory
   input, e.g.\ every different smell sensed by the
   nose, has its respective neural ensemble.
\item[--] {Time Encoding}:
   Time encoding is present if the same neurons
   transmit more than one {piece} of sensory information
   by changing their respective firing patterns.
\end{itemize}
Cyclic attractors in a dynamical ensemble are an obvious tool to
generate time encoded information. For random boolean networks as
well as for random neural networks appropriate initial conditions,
corresponding to certain activity patterns of the primary sensory
organs, will settle into a cycle, as discussed in
Sect.~\ref{networks2_sec_cycles_attractors}. The random network may
then be used to encode initial firing patterns by the time sequence
of neural activities resulting from the firing patterns of the
corresponding limiting cycle, see
Fig.~\ref{networks2_fig_time_encoding}.

\runinhead{Critical Sensory Processing}
\index{critical!sensory processing}The processing of incoming
information is qualitatively different in the various phases of the
$N$--$K$ model, as discussed in
Sect.~\ref{boolean_information_flow}.

The chaotic phase is unsuitable for information processing, any
input results in an unbounded response and saturation. The response
in the frozen phase is strictly proportional to the input and
{is} therefore well behaved, but also relatively
uninteresting. The critical state, on the other hand, has the
possibility of nonlinear signal amplification.

\begin{figure}[t]
\centering
\centering
\includegraphics{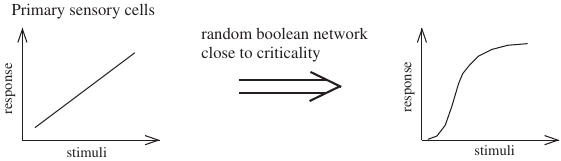}
\caption{The primary response of sensory receptors
         can be enhanced by many orders of magnitude using the
         non-linear amplification properties of a random neural
         network close to criticality%
        }
\label{networks2_fig_critical_processing}
\end{figure}

Sensory organs in animals can routinely process physical stimuli,
such as light, sound, pressure or odorant concentrations, which vary
by many orders of magnitude in intensity. The primary sensory cells,
e.g.\ the light receptors in the retina, {have, however} a linear sensibility to the intensity of the
incident light, with a relatively small dynamical range. It is
therefore conceivable that the huge \hbox{dynamical} range of sensory
information processing of animals is a collective effect, as it
occurs in a random neural network close to criticality. This
mechanism, which is plausible from the view of possible genetic
encoding mechanisms, is illustrated \break in
Fig.~\ref{networks2_fig_critical_processing}.

\section*{Exercises}
\addcontentsline{toc}{section}{Exercises}
\markright{Exercises}
{\sc $K=1$ Kauffman Net}
\begin{list}{}
%
\item Analyze some $K=1$ Kauffman nets with $N=3$ and a cyclic linkage
tree: $\sigma_1=f_1(\sigma_2)$, $\sigma_2=f_2(\sigma_3)$,
$\sigma_3=f_3(\sigma_1)$. Consider: \newline
(i) $f_1=f_2=f_3=$ identity, \newline
(ii) $f_1=f_2=f_3=$ negation and \newline
(iii) $f_1=f_2=$ negation, $f3=$ identity.\newline
Construct all cycles and their attraction basin.
\end{list}
\hspace*{-12pt}{\sc $N=4$ Kauffman Net}
\begin{list}{}
\item Consider {the} $N=4$ graph illustrated in
Fig.~\ref{networks2_fig_booleanNetwork}. Assume all coupling
functions to be generalized XOR-functions (1/0 if the number of
input-1's is odd/even). Find all cycles.
\end{list}
\hspace*{-12pt}{\sc Synchronous vs.\ Asynchronous Updating}
\begin{list}{}
\item  Consider the dynamics of the three-site network 
illustrated in Fig.~\ref{networks2_fig_booleanNetwork_example} 
under sequential asynchronous updating. At every time step 
first update $\sigma_1$ then $\sigma_2$ and then $\sigma_3$.
Determine the full network dynamics, find all cycles and fixpoints
and compare with the results for synchronous updating shown in
Fig.~\ref{networks2_fig_booleanNetwork_example}.
\end{list}
\hspace*{-12pt}{\sc Loops and Attractors}
\begin{list}{}
\item Consider, as in Sect.~\ref{networks_2_K_1_Kauffman},
      a $K=1$ network with two linkage loops,
$$
A\ \to^{\hspace{-1.5ex}I\hspace{1ex}}\ B\ 
   \to^{\hspace{-1.8ex}\neg\hspace{1ex}}\ C\ 
   \to^{\hspace{-1.5ex}I\hspace{1ex}}\ A,
\qquad\quad
D\ \to^{\hspace{-1.8ex}\neg\hspace{1ex}}\ E\ 
   \to^{\hspace{-1.8ex}\neg\hspace{1ex}}\ D~,
$$
with $I$ denoting the identity coupling and $\neg$
the negation, compare p.~\pageref{boolean_boolean_f2}.
Find all attractors by considering first the dynamics
of the individual linkage loops. Is there any state 
in phase space which is not part of any cycle?
\end{list}
\hspace*{-12pt}{\sc Relevant Nodes and Dynamic Core}
\begin{list}{}
\item How many constant nodes does the network shown in
Fig.~\ref{networks2_fig_booleanNetwork_example}
have? Replace then the AND function
with XOR and calculated the complete dynamics. How
many relevant nodes are there now?
\end{list}
\hspace*{-12pt}{\sc The Huepe and Aldana Network}
\index{Huepe--Aldana Network}
\begin{list}{}
\item Solve the boolean neural network with uniform
coupling functions and noise,
\vskip3pt
$$
\sigma_i(t+1)\ =\ \left\{
\begin{array}{lcc}
\phantom{-} \mbox{sign}\left(\sum_{j=1}^K\sigma_{i_j}(t)\right) &
\mbox{with probability} & 1-\eta , \\
& & \\
-\mbox{sign}\left(\sum_{j=1}^K\sigma_{i_j}(t)\right) & \mbox{with
probability} & \eta ,
\end{array}
\right.
$$
\vskip3pt
via mean-field theory, where $\sigma_i=\pm1$, by
considering the order parameter
\vskip3pt
$$
\Psi\ =\ \lim_{T\to\infty}{1\over T}\int_0^T |s(t)|\,\mathrm{d}t,
\qquad\quad s(t)\ =\ \lim_{N\to\infty} {1\over
N}\sum_{i=1}^N\,\sigma_i(t)~.
$$
\vskip3pt
See Huepe and Aldana-Gonz\'alez~(2002) and additional
hints in the solutions section.
\end{list}
\hspace*{-12pt}{\sc Bond Percolation}
\begin{list}{}
\item Consider a finite $L\times L$ two-dimensional square lattice. 
Write a code that generates a graph by adding with
probability $p\in[0,1]$ nearest-neighbor edges. Try to develop an
algorithm searching for a non-interrupted path of bonds from one
edge to the opposite edge; you might consult web resources. 
Try to determine the critical $p_c$, for $p>p_c$, a percolating 
path should be present with probability 1 for very large
systems $L$.
\end{list}

\def\refer#1#2#3#4#5#6{\item{\frenchspacing\sc#1}\hspace{4pt}
                       #2\hspace{8pt}#3 {\it \frenchspacing#4} {\bf#5}, #6.}
\def\bookref#1#2#3#4{\item{\frenchspacing\sc#1}\hspace{4pt}
                       #2\hspace{8pt}{\it#3}  #4.}

\addcontentsline{toc}{section}{Further Reading} 
\section*{Further Reading}
\markboth{\thechapter\enspace Random Boolean Networks}{Further Reading}

\looseness1 The interested reader may want to take a look at
Kauffman's (1969) seminal work on random boolean networks, or to
study his book (Kauffman, 1993). For reviews on boolean networks
please consult Aldana, Coppersmith and Kadanoff (2003)
and the corresponding chapter by B.~Drossel in Schuster (2008).

Examples of additional applications of boolean network theory
regarding the modeling of neural networks (Wang et al., 1990) 
and of evolution (Bornholdt and Sneppen,
1998) are also recommended. Some further interesting
original literature concerns the connection of
Kauffman nets with percolation theory (Lam, 1988), as well as the
exact solution of the Kauffman net with connectivity one (Flyvbjerg
and Kjaer, 1988), numerical studies of the Kauffman net (Flyvbjerg,
1989; Kauffman, 1969, 1990; Bastolla and Parisi, 1998), as well as
the modeling of the yeast reproduction cycle by boolean networks (Li
et al., 2004).

Some of the new developments concern the stability of the Kauffman
net (Bilke and Sjunnesson, 2001) and the number of attractors
(Samuelsson and Troein, 2003) and applications to time encoding by
the cyclic attractors (Huerta and Rabinovich, 2004) and nonlinear
signal amplification close to criticality (Kinouchi and Copelli,
2006).

{\baselineskip=15pt
\begin{list}{}{\leftmargin=2em \itemindent=-\leftmargin%
\itemsep=3pt \parsep=0pt \small}

\refer{Aldana-Gonzalez, M., Cluzel, P.}{2003}
   {A natural class of robust networks.}
{Proceedings of the National Academy of Sciences}{100}{8710--8714}

\bookref{Aldana-Gonzalez, M., Coppersmith, S., Kadanoff, L.P.}
{2003}{\rm{Boolean dynamics with random couplings.}}{In Kaplan, E.,
Marsden, J.E., Sreenivasan, K.R. (eds.) {\it Perspectives and
Problems in Nonlinear Science.
  A Celebratory Volume in Honor of Lawrence Sirovich}, pp. 23--89.
   Springer Applied Mathematical Sciences Series, Berlin}

\refer{Bastolla, U., Parisi, G.}{1998}{Relevant elements,
  magnetization and dynamical properties in Kauffman networks: A numerical
    study.}{Physica D}{115}{203--218}

\refer{Bilke, S., Sjunnesson, F.}{2001}{Stability of the Kauffman
model.} {Physical Review E}{65}{016129}

\refer{Bornholdt, S., Sneppen, K.}{1998}{Neutral mutations and
punctuated equilibrium in evolving genetic networks.}
 {Physical Review Letters}{81}{236--239}

\refer{Flyvbjerg, H.}{1989}{Recent results for random networks of
automata.}{Acta Physica Polonica B}{20}{321--349}

\refer{Flyvbjerg, H., Kjaer, N.J.}{1988}{Exact solution of Kauffman
model with connectivity one.}{Journal of Physics A: Mathematical and
General}{21}{1695--1718}

\refer{Huepe, C., Aldana-Gonz\'alez, M.}{2002}{Dynamical phase
transition in a neural network model with noise: An exact solution.}
{Journal of Statistical Physics}{108}{527--540}

\refer{Huerta, R., Rabinovich, M.}{2004}{Reproducible
       sequence generation in random neural ensembles.}
       {Physical Review Letters}{93}{238104}

\refer{Kauffman, S.~A.}{1969}{Metabolic stability and epigenesis in randomly
  constructed nets.} {Journal of\,\, Theoretical Biology}{22}{437--467}

\refer{Kauffman, S.A.}{1990}{Requirements for evolvability in complex
  systems -- orderly dynamics and frozen components.}{Physica D}{42}{135--152}

\bookref{Kauffman, S.A.}{1993}{The Origins of Order:
Self-Organization and
  Selection in Evolution.}{Oxford University Press, New York}

\refer{Kinouchi, O., Copelli, M.}{2006}{Optimal dynamical range of
excitable networks at criticality.} {Nature Physics}{2}{348--352}

\refer{Lam, P.M.}{1988}{A percolation approach to the Kauffman
model.} {Journal of Statistical Physics}{50}{1263--1269}

\refer{Li, F., Long, T., Lu, Y., Ouyang, Q., Tang, C.}{2004}
   {The yeast cell-cycle network is robustly designed.}
   {Proceedings of the National Academy Science}{101}{4781--4786}

\refer{Luque, B., Sole, R.V.}{2000}
  {Lyapunov exponents in random boolean networks.}
  {Physica A}{284}{33--45}

\refer{Samuelsson, B., Troein, C.}{2003}{Superpolynomial growth in
the number of attractors in Kauffman networks.}
       {Physical Review Letters}{90}{098701}

\bookref{Schuster, H.G. (Editor)}{2008}{Reviews of Nonlinear
Dynamics and Complexity: Volume 1.}{Wiley-VCH, New York}

\refer{Somogyi, R., Sniegoski, C.A.}{1996}{Modeling the complexity
  of genetic networks: Understanding multigenetic and pleiotropic regulation.}
  {Complexity}{1}{45--63}

\refer{Wang, L., Pichler, E.E., Ross, J.}{1990}{Oscillations and
chaos in neural networks -- an exactly solvable model.}
{Proceedings of the National Academy of Sciences of the United States of America}
{87}{9467--9471}

\end{list}
\par}


\vspace{-20ex}
\chapter{Cellular Automata and Self-Organized Criticality}
\label{chap_automata1}


\abstract{The notion of \qut{phase transition} is a key concept in
the theory of complex systems. We encountered an
important class of phase transitions in
Chap.~\ref{chap_networks2}, viz transitions in
the overall dynamical state induced by changing the average
connectivity in networks of randomly interacting boolean
variables.\newline \indent  The concept of phase transition
originates from physics. At its basis lies the \qut{Landau theory of
phase transition}, which we will discuss in this chapter. Right at
the point of transition between one phase and another, systems
behave in a very special fashion; they are said to be
\qut{critical}. Criticality is reached normally when tuning an
external parameter, such as the temperature for many physical phase
transitions or the average connectivity for the case of random
boolean networks.\newline \indent The central question discussed in
this chapter is whether \qut{self-organized criticality} is possible
in complex adaptive systems, i.e.\ whether a system can adapt its
own parameters in a way to move towards criticality on its own, as a
consequence of a suitable adaptive dynamics. The possibility of
self-organized criticality is a very intriguing outlook.
In this context, we discussed in Chap.~\ref{chap_networks2},
the notion of \qut{life at the edge of
chaos}, viz the hypothesis that the dynamical state of living beings
{may} be close to self-organized
criticality.\newline \indent We will introduce and discuss
\qut{cellular automata} in this chapter, an important and popular
class of standardized dynamical systems. Cellular automata allow a
very intuitive construction of models, such as the famous
\qut{sandpile model}, showing the {phenomenon}
of self-organized criticality. The chapter then concludes with a
discussion of whether self-organized criticality occurs in the most
adaptive dynamical system of all, namely in the context of long-term
evolution.}

\section{The Landau Theory of Phase Transitions}
\label{automata_Landau_theory}
\index{phase transition!Landau theory|textbf}
\index{Landau theory|textbf}

One may describe the physics of thermodynamic phases
either microscopically with the tools of statistical
physics, or by considering the general properties
close to a phase transition. The Landau theory of
phase transitions does the latter, providing a
general framework valid irrespectively of the
microscopic details of the material.

\runinhead{Second-Order Phase Transitions} \index{phase
transition!second-order} Phase transitions occur in many physical
systems when the number of components diverges, viz
\qut{macroscopic} systems. Every phase has characteristic
properties. The key property, which distinguishes one phase from
another, is denoted {the} \qut{order parameter}.
Mathematically one can classify the type of ordering according to
the symmetry {of} the ordering breaks.
%

\begin{figure}[t]
\centerline{\includegraphics{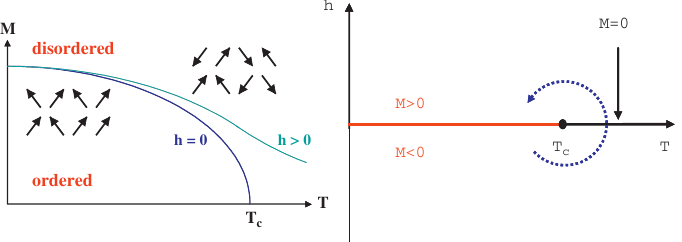}}
\caption{Phase diagram of a magnet in an external magnetic field
$h$. \textit{Left}: The order parameter $M$ (magnetization) as a
function of temperature across the phase transition. The {\it
arrows} illustrate typical arrangements of the local moments. In the
ordered phase there is a net magnetic moment (magnetization). For
$h=0$/$h>0$ the transition disorder--order is a sharp
transition/crossover. \textit{Right}: The $T-h$ phase diagram. A
sharp transition occurs only for vanishing external field $h$}
\label{automata_phaseTransGen}\vspace*{6pt}
\end{figure}

\begin{quotation}
{\it {The} Order Parameter.\enspace}
\index{order parameter} In a continuous or \qut{second-order} phase
transition the high-temperature phase has a higher symmetry than the
low-temperature phase and the degree of symmetry breaking can be
characterized by an order parameter $\phi$.
\end{quotation}
Note that all matter is disordered at high enough temperatures and
ordered phases occur at low to moderate temperatures in physical
systems.

\runinhead{Ferromagnetism in Iron}
\index{ferromagnetism} The classical example for a phase transition
is that of a magnet like iron. Above the Curie temperature of
$T_c=1043^\circ{\rm K}$ \index{temperature!Curie} the elementary
magnets are disordered, see Fig.~\ref{automata_phaseTransGen} for an
illustration. They fluctuate strongly and point in random
directions. The net magnetic moment vanishes. Below the Curie
temperature the moments point on the average {to} a
certain direction creating such a macroscopic magnetic field. Since
magnetic fields are generated by circulating currents and since an
electric current depends on time, one speaks of a breaking of
\qut{time-reversal symmetry} in the magnetic state of a ferromagnet
like iron. Some further examples of order parameters characterizing
phase transitions in physical systems are listed in Table~\ref{automata1_order_parameters}.

\runinhead{Free Energy} \index{free energy} A statistical mechanical
system {takes the configuration with the lowest energy at
zero temperature}. A physical system at finite temperatures $T>0$
does not minimize its energy but a quantity called {the}
{\em free energy F}, which differs from the energy by a term
proportional to the entropy and to the temperature.\footnote{Details
can be found in any book on thermodynamics and phase transitions,
{e.g.} Callen (1985), they are, however, not
necessary for an understanding of the following discussions.}

\index{temperature!transition} Close to the transition temperature
$T_c$ the order parameter $\phi$ is small and one assumes within the
Landau--Ginsburg model that the free energy density $f=F/V$,
\index{Landau--Ginsburg model}
$$
f \ =\ f(T,\phi,h)~,
$$
can be expanded for a small order parameter $\phi$
and a small external field $h$:
\begin{equation}
f(T,\phi,h)\ =\ f_{0}(T,h)\,-\,h\,\phi\,+\,a\,\phi^{2}
\,+\,b\,\phi^{4}\,+\,\ldots
\label{automata_eq_f_T_h}
\end{equation}
where the parameters $a=a(T)$ and $b=b(T)$ are functions of the
temperature $T$ and of an external field $h$, e.g.\ a magnetic field
for the case of magnetic systems. Note the linear coupling of the
external field $h$ to the order parameter in lowest order and that
$b>0$ (stability for large $\phi$), compare
Fig.~\ref{automata_fig_Landau_Ginzburg}.

\begin{table}[t]
\centering
\caption{Examples of important types of phase transitions in
physical systems. When the transition is continuous/discontinuous
one speaks of a second-/first-order phase transition. Note that most
order parameters are non-intuitive. {The}
superconducting state, notable for its ability to carry electrical
current {without dispersion}, breaks what
one calls the $U(1)$-gauge invariance of the normal
(non-superconducting) metallic state }
\label{automata1_order_parameters} \index{superconductivity}
\index{magnetism} \index{ferroelectricity} \index{Bose--Einstein
condensation} \index{liquid--gas transition}
\begin{tabular*}{23pc}{@{\extracolsep{\fill}}lll}
\hline\noalign{\smallskip}
Transition & Type & Order parameter $\phi$ \\
\noalign{\smallskip}\svhline\noalign{\smallskip}
Superconductivity & Second-order & U(1)-gauge \\
Magnetism  & Mostly second-order & Magnetization \\
Ferroelectricum  & Mostly second-order & Polarization \\
Bose--Einstein  & Second-order & Amplitude of $k=0$ state \\
Liquid--gas & First-order & Density \\
\noalign{\smallskip}\hline\noalign{\smallskip}
\end{tabular*}
\vspace*{-6pt}
\end{table}

\runinhead{Spontaneous Symmetry Breaking}
\index{spontaneous symmetry breaking}
All odd terms $\sim \phi^{2n+1}$ vanish
in the expansion (\ref{automata_eq_f_T_h}).
The reason is simple. The expression (\ref{automata_eq_f_T_h})
is valid for all temperatures close to $T_c$
and the disordered high-temperature
state is invariant under the symmetry operation
$$
f(T,\phi,h) \ =\ f(T,-\phi,-h),
\qquad \phi\ \leftrightarrow\ -\phi,
\qquad h\ \leftrightarrow\ -h~.
$$
This relation must therefore hold also for the exact
Landau--Ginsburg functional. When the temperature is lowered the
order parameter $\phi$ will acquire a finite expectation value. One
speaks of a \qut{spontaneous} breaking of the symmetry inherent to
the system.

\runinhead{{The} Variational Approach}
The Landau--Ginsburg functional (\ref{automata_eq_f_T_h}) expresses
the value {that} the free-energy would have for all
possible values {of} $\phi$. The true physical state,
which one calls the \qut{thermodynamical stable state}, is obtained
by finding the minimal $f(T,\phi,h)$ for all possible values of
$\phi$:

\begin{figure}[t]
\centerline{\includegraphics{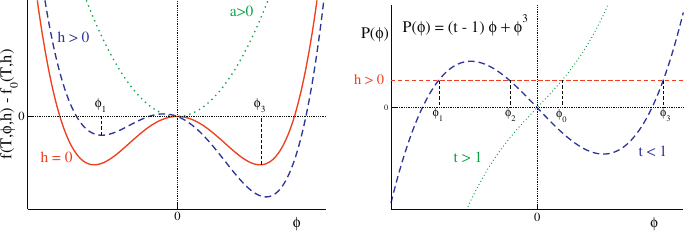}}
\caption{\textit{Left}: The functional dependence of the
Landau--Ginzburg free energy
$f(T,\phi,h)-f_0(T,h)=-h\,\phi+a\,\phi^2+b\,\phi^4$, with
$a=(t-1)/2$. Plotted is {the} free energy for $a<0$ and
$h>0$ (\textit{dashed line}) and $h=0$ ({\it full line}) and for $a>0$
(\textit{dotted line}). \textit{Right}: Graphical solution of
Eq.~(\ref{automata_h_phi_eq}) for a non-vanishing field
$h\ne0${;} $\phi_0$ is the order parameter in the
disordered phase ($t>1$, \textit{dotted line}), $\phi_1,\ \phi_3$
the stable solutions in the order phase ($t<1$, \textit{dashed
line}) and $\phi_2$ the unstable solution, compare the
{\textit{left-hand}} side
{illustration}}
\label{automata_fig_Landau_Ginzburg}
\vspace*{-18pt}
\end{figure}

\begin{eqnarray} \nonumber
\delta f& =& \left(-h+2\,a\,\phi+4\,b\,\phi^{3}\right)\,\delta\phi\ =\ 0, \\
0 & = & -h\,+\,2\,a\,\phi\,\,+4\,b\,\phi^{3}~,
\label{automata_delta_f}
\end{eqnarray}
where $\delta f$ and $\delta\phi$ denote small variations of the
free energy and of the order parameter, respectively. This solution
corresponds to a minimum in the free energy~if
\begin{equation} \label{automata_eq9.26}
\delta^{2}f\ > \ 0,
\qquad\quad
\delta^{2}f\ =\ \left(2\,a+12\,b\,\phi^{2}\right)\,(\delta \phi)^{2}~.
\end{equation}
One also says that the solution is \qut{locally stable}, since any
change in $\phi$ from its optimal value would raise the free energy.

\runinhead{Solutions for ${h = 0}$} We consider first the case with
no external field, $h=0$. The solution of
Eq.~(\ref{automata_delta_f}) is then
\begin{equation}
\label{automata_sol_phi_h_0}
\phi\ =\ \left \{
\begin{array}{rcl}
0 &\ \mbox{\ for}\ & a>0\\
\pm \sqrt{-a/(2\,b)} &\ \mbox{\ for}\ & a<0
\end{array}
\right. ~.
\end{equation}
The trivial solution $\phi=0$ is stable,
\begin{equation} \label{automata_eq9.28}
\left(\delta^{2}f\right)_{\phi=0}\ =\ 2\,a\,(\delta \phi)^{2}~,
\end{equation}
if  $a>0$. The\enlargethispage*{12pt} nontrivial solutions $ \phi=\pm \sqrt{-a/(2\,b)}$ of
{Eq.~}(\ref{automata_sol_phi_h_0}) are stable,
\begin{equation} \label{automata_eq9.29}
\left(\delta^{2}f\right)_{\phi \neq 0}\ =\ -4\,a\,(\delta \phi)^{2}~,
\end{equation}
for $a<0$. Graphically this is immediately evident, see
Fig.~\ref{automata_fig_Landau_Ginzburg}. For $a>0$ there is a single
global minimum at $\phi=0$, for $a<0$ we have two symmetric minima.

\runinhead{Continuous Phase Transition}
\index{phase transition!continuous} We therefore find that the
Ginsburg--Landau functional (\ref{automata_eq_f_T_h}) describes
continuous phase transitions when $a=a(T)$ changes sign at the
critical temperature $T_c$. Expanding $a(T)$ for small $T-T_c$ we
have
$$
a(T)\ \sim\ T-T_{c},
\qquad a\ =\ a_0\,(t-1),
\qquad t=T/T_{c},
\qquad a_0>0~,
$$
where we have used $a(T_c)=0$. For $T<T_c$ (ordered phase) the
solution {Eq.~}(\ref{automata_sol_phi_h_0}) then takes
the form
\begin{equation} \label{automata_eq9.30}
\phi\ =\ \pm \sqrt{\frac{a_0}{2\,b}\,(1-t)},
\qquad\quad t<1,
\qquad\quad T<T_c~.
\end{equation}

\runinhead{Simplification by Rescaling}
We can always rescale the order parameter $\phi$, the external
field $h$ and the free energy density $f$ such that
$a_0=1/2$ and $b=1/4$. We then have
$$
a={t-1\over 2},
\qquad\quad
f(T,\phi,h)-f_0(T,h) \ =\  -h\,\phi\,+\,\frac{t-1}{2}\,\phi^2
\,+\,\frac{1}{4}\,\phi^4
$$
and
\begin{equation} \label{automata_eq9.30b}
\phi\  =\ \pm \sqrt{1-t},
\qquad\quad
t\ =\ T/T_c
\end{equation}
for the non-trivial solution
{Eq.~}(\ref{automata_eq9.30}).

\runinhead{Solutions for $\textbf{\rm h}\ne{\bf 0}$}
The solutions of {Eq.~}(\ref{automata_delta_f}) are
determined in rescaled form by
\begin{equation}
h\ =\ (t-1)\,\phi\,+\,\phi^{3}
\ \equiv \ P(\phi)~,
\label{automata_h_phi_eq}
\end{equation}
see Fig.~\ref{automata_fig_Landau_Ginzburg}. 
In general one finds three solutions $\phi_1 <\phi_2
<\phi_3$. One can show (see the Exercises)
that the intermediate solution is always locally instable and that
$\phi_3$ ($\phi_1$) is globally stable for $h>0$ ($h<0$).

\begin{figure}[t]
\centerline{\includegraphics{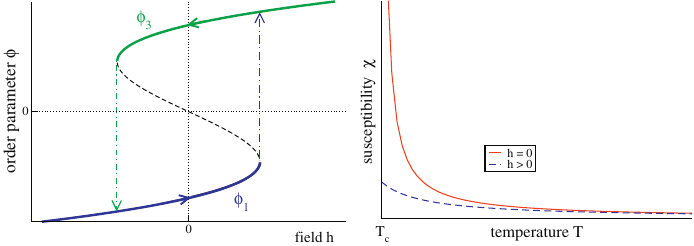}}
\caption{\textit{Left}: Discontinuous phase transition and
hysteresis in the Landau model. Plotted is the solution
$\phi=\phi(h)$ of $h=(t-1)\phi+\phi^3$ in the ordered phase ($t<1$)
when changing the field $h$. \textit{Right}: The susceptibility
$\chi=\partial\phi/\partial h$ for $h=0$ (\textit{solid line}) and
$h>0$ (\textit{dotted line}). The susceptibility divergence in the
absence of an external field ($h=0$), compare
Eq.~(\ref{automata_chi})
        }
\label{automata_fig_hysteresis}
\end{figure}

\runinhead{First-Order Phase Transition}
\index{phase transition!first-order} We note, see
Fig.~\ref{automata_fig_Landau_Ginzburg}, that the 
solution $\phi_3$ for $h>0$ remains locally stable 
when we vary the external field slowly (adiabatically)
$$
(h>0)\ \ \to\ \ (h=0)\ \ \to\ \ (h<0)
$$
in the ordered state $T<T_c$. At a certain critical 
field, see Fig.~\ref{automata_fig_hysteresis}, the 
order parameter changes sign abruptly, jumping from 
the branch corresponding to $\phi_3>0$ to the branch 
$\phi_1<0$. One speaks of hysteresis, a phenomenon typical
for first-order phase transitions.

\runinhead{Susceptibility}
When the system is disordered and approaches the phase transition
from above, it has an {increased} sensitivity
towards ordering under the influence of an external field~$h$.

\vspace{-4pt}
\begin{quotation}
{\it Susceptibility.\enspace}
\index{susceptibility}
The susceptibility $\chi$ of a system denotes its response
to an external field:
\begin{equation} \label{automata_eq9.22}
\chi \ =\ \left(\frac{\partial \phi}{\partial h}\right)_{T}~,
\end{equation}
where the subscript $T$ indicates that the temperature is kept
constant. The susceptibility measures the relative amount of
{the} induced order $\phi=\phi(h)$.
\end{quotation}

\vspace{-3pt}

\runinhead{Diverging Response}
Taking the derivative with respect to the external field $h$ in
Eq.~(\ref{automata_h_phi_eq}), $h =(t-1)\,\phi\,+\,\phi^{3}$, we
find for the disordered phase $T>T_c$,
\begin{equation}
1\ =\ \Big[(t-1)\,+\,3\,\phi^2 \Big]\,
{\partial\phi\over \partial h},
\qquad\quad
\chi(T)\Big|_{h\to0} \ =\ {1\over t-1} \ =\ {T_c\over T-T_c}~,
\label{automata_chi}
\end{equation}
since $\phi(h=0)=0$ for $T>T_c$. The susceptibility diverges at the
phase transition for $h=0$, see Fig.~\ref{automata_fig_hysteresis}.
This divergence is a typical precursor of ordering for a
second-order phase transition. Exactly at $T_c$, viz at criticality,
the response of the {system is,
strictly} speaking, infinite.

A non-vanishing external field $h\ne0$ induces a finite amount of
ordering $\phi\ne0$ at all {temperatures} and
the phase transition is masked, compare
Fig.~\ref{automata_phaseTransGen}. In this case, the susceptibility
is a smooth function of {the} temperature, see
Eq.~(\ref{automata_chi}) and Fig.~\ref{automata_fig_hysteresis}.

\section{Criticality in Dynamical Systems}
\label{section_criticality_dynamical_systems}
\index{dynamical system!criticality|textbf}
\index{criticality!dynamical system|textbf}

\runinhead{Length Scales}
Any physical or complex system {normally has}
well defined time and space scales. As an example we take a look at
the Schr\"odinger equation for the hydrogen atom,
\index{Schr\"odinger equation} \index{hydrogen atom}
$$
i\hbar {\partial \Psi(t,\textbf{\textit{ r}})\over \partial t} \ =\ H\,
\Psi(t,\textbf{\textit{ r}}), \qquad \quad H \ =\ -{\hbar^2 \Delta\over 2m} \,-\,
{Ze^2\over |\textbf{\textit{ r}} |}~,\vspace{-2pt}
$$
where\vspace{-2pt}
$$
\Delta\ =\ {\partial^2\over \partial x^2}+
         {\partial^2\over \partial y^2}+
         {\partial^2\over \partial z^2}\vspace{-2pt}
$$
is the Laplace operator. \index{Laplace operator} We do not need to
know the physical significance of the parameters to realize that we
can rewrite the differential operator $H$, called the \hbox{\qut{Hamilton}
operator}, as \index{Hamilton operator}\vspace{-2pt}
$$
H \ =\ -E_R\left(a_0^2\Delta \,+\, {2a_0\over |\textbf{\textit{r}} |}
           \right),
\qquad \quad E_R={mZ^2e^4\over 2\hbar^2}, \qquad \quad
a_0={\hbar^2\over mZe^2}~.\vspace{-2pt}
$$
The length scale $a_0=0.53\,\AA/Z$ is called the \qut{Bohr radius}
and the energy scale $E_R=13.6\ {\rm eV}$ the \qut{Rydberg energy},
which corresponds to a frequency scale of $E_R/\hbar =
3.39\cdot10^{15}\,{\rm Hz}$. The energy scale $E_R$ determines the
ground state energy and the characteristic excitation energies. The
length scale $a_0$ determines the mean radius of the ground state
wavefunction and all other radius-dependent properties.

\index{length scale} Similar length scales can be defined for
essentially all dynamical systems defined by a set of differential
equations. The damped harmonic oscillator and the diffusion
equations, e.g.\ are given by \index{harmonic oscillator!damped}
\begin{equation}
\ddot x(t)-\gamma\dot x(t)+\omega^2 x(t) = 0, \qquad\quad {\partial
\rho(t,\textbf{\textit{ r}})\over \partial t} \ =\ D\Delta \rho(t,\textbf{\textit{ r}})~.
\label{automata_damped_diff_eq}
\end{equation}
The parameters $1/\gamma$ and $1/\omega$, respectively, determine
the time scales for relaxation and oscillation, {and} $D$
is the diffusion constant.

\runinhead{Correlation Function}
\index{correlation!spatial} A suitable quantity to measure and
discuss the properties of the solutions of dynamical systems like
the ones defined by Eq.~(\ref{automata_damped_diff_eq}) is the
equal-time correlation function $S(r)$, which is the expectation
value \index{correlation function!equal-time}
\begin{equation}
S(r)\ =\
\langle\, \rho(t_0,\textbf{\textit{ x}})\,\rho(t_0,\textbf{\textit{ y}})\,\rangle,
\qquad\quad r=|\textbf{\textit{ x}}-\textbf{\textit{ y}} |~.
\label{automata_corr_funct}
\end{equation}
Here $\rho(t_0,\textbf{\textit{ x}})$ denotes the particle density, for the case
of the diffusion equation or when considering a statistical
mechanical system of interacting particles. The exact expression for
$\rho(t_0,\textbf{\textit{ x}})$ {in general depends}
on the type of dynamical system considered{;} for the
Schr\"odinger equation $\rho(t,\textbf{\textit{ x}})= \Psi^*(t,\textbf{\textit{ x}}) \Psi(t,\textbf{\textit{
x}})$, i.e.\ the probability to find the particle at time $t$ at the
point  $\textbf{\textit{ x}}$.

The equal-time correlation function then measures the
probability to find a particle at
position $\textbf{\textit{ x}}$ when there
is one at $\textbf{\textit{ y}}$.
$S(r)$ is directly
measurable in scattering experiments and therefore
a key quantity for the characterization of
a physical system. Often one is interested in
the deviation of the correlation from the
average behaviour. In this case one considers
$\langle\, \rho(\textbf{\textit{x}})\,
           \rho(\textbf{\textit{y}})\,\rangle
-\langle\, \rho(\textbf{\textit{x}})\,\rangle
 \langle\, \rho(\textbf{\textit{y}})\,\rangle
$
for the correlation function $S(r)$.

\runinhead{Correlation Length}
\index{correlation!length}
Of interest is the behavior of the equal-time
correlation function $S(r)$ for large
distances $r\to\infty$.
In general we have two possibilities:
\begin{equation}
S(r)\Big|_{r\to\infty}\quad \sim\quad
\left\{
\begin{array}{cl}
e^{-r/\xi} &\ \mbox{non-critical} \\
1/r^{d-2+\eta} &\ \mbox{critical}
\end{array}
\right. ~.
\label{automata_corr_length}
\end{equation}
\index{correlation function!critical}In any ``normal''
(non-critical) system, correlations over arbitrary large distances
cannot be built up, and the correlation function decays
exponentially with the \qut{\hbox{correlation} length} $\xi$. The notation
${d-2+\eta}>0$ for the decay exponent of the critical system is a
convention from statistical physics, where $d=1,2,3,\ldots $ is the
dimensionality of the system. \index{exponent!critical}

\runinhead{Scale-Invariance and Self-Similarity}
\index{self-similar!correlation function} If a control parameter,
often the temperature, of a physical system is tuned such that it
sits exactly at the point of a phase transition, the system is said
to be critical. At this point there are no characteristic length
scales.
\begin{quotation}
{\it Scale Invariance.\enspace}
\index{criticality!scale invariance}
\index{correlation function!scale invariance}
If a measurable quantity, like the
correlation function, decays like a power
of the distance  $\sim \left(1/r\right)^\delta$,
with a critical exponent $\delta$, the system
is said to be critical or scale-invariant.
\end{quotation}
\index{law!power} \index{scale invariance!power law} Power laws have
no scale; they are self-similar,
$$
S(r)\ =\ c_0\left({r_0\over r}\right)^\delta
      \ \equiv \ c_1\left({r_1\over r}\right)^\delta,
\qquad\quad
c_0\,r_0^\delta = c_1\,r_1^\delta~,
$$
for arbitrary distances $r_0$ and $r_1$.

\runinhead{Universality at the Critical Point}
\index{universality!critical systems}
\index{criticality!universality} The equal-time correlation function
$S(r)$ is scale-invariant at criticality, compare
Eq.~(\ref{automata_corr_length}). This is a surprising statement,
since we have seen before that the differential equations
determining the dynamical system have well defined time and length
scales. How then {does the solution of a dynamical system become}
effectively independent of the parameters entering its governing
equations?

\begin{figure}[t]
\centerline{\includegraphics{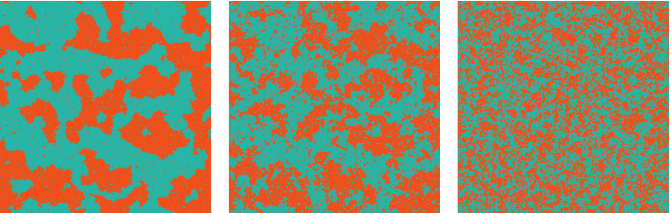}}
\caption{Simulation of the 2{D}-Ising model
$H=\sum_{<i,j>} \sigma_i\sigma_j$, $<i,j>$ nearest neighbors on a
square lattice. Two magnetization orientations $\sigma_i=\pm 1$
correspond to the {\it dark/light dots}. For $T<T_c$ (\textit{left},
ordered), $T\approx T_c$ ({\it middle}, critical) and $T>T_c$
(\textit{right}, disordered). Note the occurrence of fluctuations at
all length scales at criticality (self-similarity) }
\label{automata_pic_Ising} \index{Ising model} \index{model!Ising}
\end{figure}

Scale invariance implies that fluctuations {occur over all length scales}, albeit with varying
probabilities. This can be seen by observing snapshots of
statistical \hbox{mechanical} simulations of simple models, compare
Fig.~\ref{automata_pic_Ising}. The scale invariance of the
correlation function at criticality is a central result of the
theory of phase transitions and statistical physics. The properties
of systems close to a phase transition are not determined by the
exact values of their parameters, but by the structure of the
governing equations and their symmetries. This circumstance is
denoted \qut{universality} and constitutes one of the reasons
{for classifying phase
transitions} according to the symmetry of their order parameters,
see Table~\ref{automata1_order_parameters}.

\runinhead{Autocorrelation Function}
\index{autocorrelation function} \index{correlation
function!autocorrelation} The equal-time correlation function $S(r)$
measures real-space correlations. The corresponding quantity in
{the} time domain is the autocorrelation
function\vspace{3pt}
\begin{equation}
\Gamma(t)\ =\ { \langle A(t+t_0) A(t_0)\rangle - \langle A\rangle^2
\over \langle A^2\rangle -\langle A\rangle^2 }~,
\label{automata_auto_corr_funct}\vspace{3pt}
\end{equation}
which can be defined for any time-dependent measurable quantity $A$,
e.g.\ $A(t)=\rho(t,\vec r)$. Note that the autocorrelations are
defined relative to $\langle A\rangle^2$, viz the mean
(time-independent) fluctuations. The denominator in
Eq.~(\ref{automata_auto_corr_funct}) is a normalization convention,
namely $\Gamma(0)\equiv1$.


In the non-critical regime, viz the diffusive regime, no long-term
memory is present in the system and all information about the
initial state is lost exponentially,\vspace{3pt}
\begin{equation}
\Gamma(t)\ \sim\ e^{-t/\tau},
\qquad\quad t\to\infty~.
\label{automata_eq_relaxation_time}
\end{equation}
$\tau$ is called the relaxation time.\index{time!relaxation}
\index{relaxation time}
The relaxation  or autocorrelation time $\tau$ is the time scale of
diffusion processes.

\runinhead{Dynamical Critical Exponent}
\index{exponent!dynamical} The relaxation time entering
Eq.~(\ref{automata_eq_relaxation_time}) diverges at criticality, as
does the real-space correlation length $\xi$ entering
Eq.~(\ref{automata_corr_length}). One can then define an appropriate
exponent $z$, dubbed {the} \qut{dynamical critical
exponent} $z$, in order to relate the two power laws for $\tau$ and
$\xi$ via
$$
\tau \ \sim\ \xi^z,
\qquad \quad\mbox{for}
\qquad  \xi=|T-T_c|^{-\nu}\to\infty~.
$$
The autocorrelation time is divergent in
the critical state $T\to T_c$.

\runinhead{Self-Organized Criticality}
\index{self-organized criticality} We have seen that phase
transitions can be characterized by a set of exponents describing
the respective power laws of various quantities like the correlation
function or the autocorrelation function. The phase transition
occurs generally at a single point, viz $T=T_c$ for a
thermodynamical system. At the phase transition the system becomes
effectively independent of the details of its governing equations,
being determined by symmetries.

It then comes as a surprise that there should exist complex
dynamical systems {that} attain a critical state for
a finite range of parameters. This possibility, denoted
\qut{self-organized criticality} and the central subject of this
chapter, is to some {extent} counter intuitive. We
can regard the parameters entering the evolution equation as given
externally. Self-organized criticality then signifies that the
system effectively adapts to changes in the external parameters,
e.g. to changes in the given time and length scales, in such a way
that the stationary state becomes independent {of}
those changes.

\subsection[${\textit{1/f}}$ Noise]{$\textbf{\textit{1/f}}$ Noise}
\index{noise@1/f noise|textbf}
\index{Bak, Per!1/f noise|textbf}

So far we have discussed the occurrence of critical states in
classical thermodynamics and statistical physics. We now ask
ourselves for experimental evidence that criticality might play a
central role in certain time-dependent phenomena.\vspace*{2pt}

\runinhead{$\textbf{\textit{1/f}}$ Noise} The power spectrum of the
noise generated by many real-world dynamical processes falls off
inversely with frequency $f$. This $1/f$ noise has been observed for
various biological activities, like the heart beat rhythms, for
functioning electrical devices or for meteorological data series.
Per Bak and coworkers have pointed out that the ubiquitous $1/f$
noise could be the result of a self-organized phenomenon. Within
this view one may describe the noise as being generated by a
continuum of weakly coupled damped oscillators representing the
environment.\vspace*{2pt}

\runinhead{Power Spectrum of a Single Damped Oscillator}
\index{power spectrum} A system with a single relaxation time
$\tau$, see Eq.~(\ref{automata_damped_diff_eq}), and eigenfrequency
$\omega_0$ has a Lorentzian power spectrum\vspace*{2pt}
$$
S(\omega,\tau)\ =\  Re \int_0^\infty dt\, e^{i\omega t}
e^{-i\omega_0 t-t/\tau} \ =\ Re{-1\over i(\omega-\omega_0)-1/\tau} \
=\ {\tau\over 1 + \tau^2 (\omega-\omega_0)^2}~.\vspace*{2pt}
$$
For large frequencies $\omega\gg 1/\tau$ the power spectrum falls
off like $1/\omega^2$. Being interested in the large-$f$ behavior we
will neglect $\omega_0$ in the following.\vspace*{2pt}

\runinhead{Distribution of Oscillators}
The combined power or frequency spectrum of a continuum of
oscillators is determined by the distribution $D(\tau)$ of
relaxation times $\tau$. For a critical system relaxation occurs
over all time scales, as discussed in
Sect.~\ref{section_criticality_dynamical_systems} and we
may assume a scale-invariant distribution \index{relaxation
time!scale-invariant distribution}
\begin{equation}
D(\tau) \ \approx\  {1\over \tau^\alpha}
\label{automata_D_tau_critical}
\end{equation}
for the relaxation times $\tau$.
This distribution of relaxation times
yields a frequency spectrum
\begin{eqnarray} \nonumber
S(\omega) & =& \int d\tau\ D(\tau){\tau\over 1 + (\tau \omega)^2}
\ \sim\ \int d\tau\ {\tau^{1-\alpha} \over 1 + (\tau \omega)^2}
\\[6pt]
& =& {1\over \omega\, \omega^{1-\alpha}}
\int d(\omega\tau)\ {(\omega\tau)^{1-\alpha} \over 1 + (\tau \omega)^2}
\ \sim\ \omega^{\alpha-2}~.
\label{automata_1_over_f_noise}
\end{eqnarray}
For $\alpha=1$ we obtain $1/\omega$, the typical behavior of
$1/f$ noise.

The question is then how assumption (\ref{automata_D_tau_critical})
can be justified. The wide-spread appearance of $1/f$ noise can only
happen when scale-invariant distribution of relaxation times
{are} ubiquitous, viz if they {were} self-organized. The $1/f$ noise {therefore constitutes} an interesting motivation for the
search of possible mechanisms leading to self-organized criticality.

\section{Cellular Automata}
\label{automata1_sec_cellular_automata}
\index{cellular automata|textbf}

Cellular automata are finite state lattice systems with discrete
local update rules.
\begin{equation}
z_i\ \to f_i(z_i,z_{i+\delta},\ldots), \qquad\quad
z_i\in[0,1,\ldots, n]~, \label{automata_def}
\end{equation}
where $i+\delta$ denote neighboring sites of site $i$. Each site or
``cell'' of the lattice follows a prescribed rule evolving in
discrete time steps.  At each step the new value for a cell depends
only on the current state of itself and on the state of its
neighbors.

Cellular automata differ from the dynamical networks we studied in
Chap.~\ref{chap_networks2}, in two aspects:
\begin{enumerate}\leftskip5pt
\item[(i)\phantom{i}] The update functions are all
           identical: $f_i()\equiv f()$, viz
           they are \hbox{translational} \hbox{invariant}.

\item[(ii)] The number $n$ of states per cell is usually larger
            than 2 (boolean case).
\end{enumerate}
Cellular automata can give rise to extremely complex behavior
despite their deceptively simple dynamical structure. We note
that cellular automata are always updated synchronously and
never sequentially or randomly. The state of all cells is
updated simultaneously.

\runinhead{Number of Update Rules}
\index{cellular automata!updating rules!number}
The number of possible update rules is huge.
Take, e.g.\ a two-dimensional model (square lattice),
where each cell can take
only one of two possible states,
$$
z_i=0,\quad (\mbox{dead}), \qquad\qquad z_i=1,\quad (\mbox{alive})~.
$$
We consider, for simplicity, rules for which the evolution of a
given cell to the next time step depends on the current state of the
cell and on the values of each of its {eight} nearest
neighbors. In this case there are
$$
2^9=512\ \ \mbox{configurations},
\qquad\qquad
2^{512}=1.3\times 10^{154}\ \ \mbox{possible rules}~,
$$
since any one of the 512 {configurations}
can be mapped independently to \qut{live} or \qut{dead}. For
comparison note that the universe is only of the order of $3\times
10^{17}$ seconds old.

\runinhead{Totalistic Update Rules} \index{cellular
automata!updating rules!totalistic} It clearly does not make sense
to explore systematically the consequences of arbitrary updating
rules. One simplification is to consider a mean-field approximation
{that} results in a subset of rules called
\qut{totalistic}. For mean-field rules the new state of a cell
depends only on the total number of living neighbors and on its own
state. The {eight-cell} neighborhood
has\vspace{6pt}
\smallskip

\centerline{
9\ \ \mbox{possible total occupancy states of neighboring sites},
           }
\smallskip

\centerline{
$ 2\cdot9=18$\ \ configurations,\hspace{4ex}
$2^{18}=262,144$\ \ totalistic rules~.
           }
\smallskip

\vspace{6pt}

\noindent {This is a} large number, but it is
exponentially smaller than the number of all possible update rules
for the same neighborhood.


\subsection{Conway's Game of Life}
\index{game of life|textbf}
\index{Conway's game of life|textbf}

The \qut{game of life} takes its name because it attempts to
simulate the reproductive cycle of a species. It is formulated on a
square lattice and the update rule involves the {eight-cell} neighborhood. A new offspring needs exactly three
parents in its neighborhood. A living cell dies of loneliness if it
has less than two live neighbors, and of overcrowding if it has more
than three live neighbors. A living cell feels comfortable with two
or three live neighbors{;} in this case it survives. The
complete set of updating rules is listed in Table
\ref{automata1_table_rules_game_life}.

\begin{figure}[t]
\centerline{\includegraphics{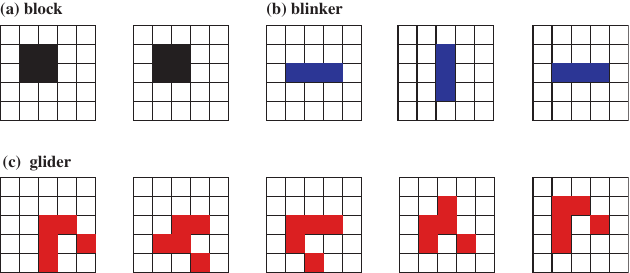}}
\caption{Time evolution of some living configurations for the game
of life, see Table \ref{automata1_table_rules_game_life}.
(\textbf{a}) The \qut{block}; it quietly survives. (\textbf{b}) The
\qut{blinker}; it oscillates with period 2. (\textbf{c}) The \qut{glider}; it
shifts by ($-$1, 1) after four time steps} \label{automata_fig_life0}
\end{figure}

\runinhead{Living Isolated Sets}
The time evolution of an initial set of a cluster of living cells
can show extremely varied types of behavior.
{Fixpoints} of the updating rules, such as a
square
$$
\Big\{(0,0),(1,0),(0,1),(1,1)\Big\}
$$
\index{game of life!block}of four neighboring live cells, survive
unaltered. There are many configurations of living cells which
oscillate, such as three live cells in a row or column,
$$
\Big\{(-1,0),(0,0),(1,0)\Big\},
\qquad\quad
\Big\{(0,-1),(0,0),(0,1)\Big\}~.
$$
\index{game of life!blinker}It constitutes a fixpoint of $f(f(.))$,
alternating between a vertical and a horizontal bar. The
configuration
$$
\Big\{(0,0),(0,1),(0,2),(1,2),(2,1)\Big\}
$$
\index{game of life!glider}is dubbed \qut{glider}, since it returns
to its initial shape after four time steps but {is} displaced by
$(-1,1)$, see Fig.~\ref{automata_fig_life0}. It constitutes a
fixpoint of $f(f(f(f(.))))$ times the translation by $(-1,1)$. The
glider continues to propagate until it encounters a cluster of other
living cells.

\begin{table}[b]
\vspace*{-12pt} \centering \caption{Updating rules for the game of
life{;} $z_i=0,1$ {corresponds}
to empty and living cells. An \qut{x} as an entry denotes what is
going to happen for the respective number of living neighbors}
\label{automata1_table_rules_game_life}
\begin{tabular*}{18pc}{@{\extracolsep{20pt}}llllllll}
\hline\noalign{\smallskip}
$z_i(t)$         &  $z_i(t+1)$          & \multicolumn{5}{l@{}}{Number of living neighbors} \\
\cline{3-7}

         &            &  \\
 &  &\ 0 & 1 & 2 & 3 & 4..8 \\
\noalign{\smallskip}\svhline\noalign{\smallskip}
       0 & 1          &    &   &   & x &   \\
         & 0          &\ x & x & x &   & x \\ [3pt]
       1 & 1          &    &   & x & x &   \\
         & 0          &\ x & x &   &   & x \\
         \noalign{\smallskip}\hline\noalign{\smallskip}
\end{tabular*}
{}\vspace*{-12pt}
\end{table}

\enlargethispage*{12pt}

\runinhead{The Game of Life as a Universal Computer}
 \index{game of life!universal computing} \looseness-1 It is interesting to
investigate, from an engineering point of view, all possible
interactions between initially distinct sets of living cells in the
game of life. {In this context
one finds} that it is possible to employ gliders for the propagation
of information over arbitrary distances. One can prove that
arbitrary calculations can be performed by the game of life, when
identifying the gliders with bits. Suitable and complicated initial
configurations are necessary for this purpose, in addition to
dedicated living subconfigurations  performing logical computations,
in analogy to electronic gates, when hit by one or more \nobreak gliders.

\subsection{{The} Forest Fire Model}
\index{model!forest fire|textbf}
\index{forest fire model|textbf}
The forest fires automaton is a very simplified model
of real-world forest fires. It is formulated on
a square lattice with three possible states per cell,
$$
z_i=0,\quad (\mbox{empty}), \qquad\qquad z_i=1,\quad (\mbox{tree}),
\qquad\qquad z_i=2,\quad (\mbox{fire})~.
$$
A tree {sapling} can grow on every empty cell
with probability $p<1$. There is no need for nearby parent trees, as
sperms are carried by wind over wide distances. Trees
{do not} die in this model, but they catch fire from
any burning nearest neighbor tree. The rules are:

\begin{table}[h]
\centering
\begin{tabular*}{18pc}{@{\extracolsep{\fill}}llll}
\hline\noalign{\smallskip}
$z_i(t)$ & $z_i(t+1)$ \ &&  Condition \\
\noalign{\smallskip}\svhline\noalign{\smallskip}
Empty & Tree  &\ & With probability $p<1$ \\
Tree  & Tree  &\ & No fire close by \\
Tree  & Fire  &\ & At least one fire close by \\
Fire  & Empty &\ & Always \\
\noalign{\smallskip}\hline\noalign{\smallskip}
\end{tabular*}
{}
\end{table}
\enlargethispage*{12pt}

The forest fire automaton differs from typical rules, such as
Conway's game of life, {because it has} a
stochastic component. In order to have an interesting dynamics one
needs to adjust the growth rate $p$ as a function of system size, so
as to keep the fire {burning
continuously}. The fires burn down the whole forest when trees grow
too fast. When the growth rate is too low, on the other hand, the
fires, {being surrounded by ashes, may die out completely}.

\begin{figure}[t]
\centerline{\includegraphics{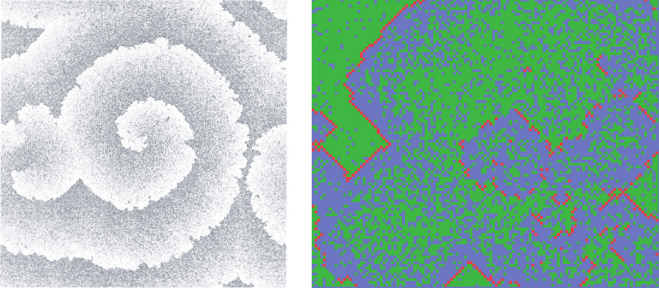}}
\caption{Simulations of the forest fire model. \textit{Left}: Fires
burn in characteristic spirals for a growth probability $p=0.005$
and no lightning, $f=0$ (from Clar et~al. 1996). \textit{Right}: A
snapshot of the forest fire model with a growth probability $p=0.06$
and a lightning probability $f=0.0001$. Note the characteristic fire
fronts with trees in front and ashes behind }
\label{automata_fig_fires}\vspace*{-6pt}
\end{figure}

When adjusting the growth rate properly one reaches a steady state,
the system having fire fronts continually sweeping through the
forest, as  is observed for real-world forest fires; { this is
illustrated} in Fig.~\ref{automata_fig_fires}. In large systems
stable spiral structures form and set up a steady rotation.

\runinhead{Criticality and Lightning}
\index{forest fire model!lightning} The forest fire model, as
defined above, is not critical, since the characteristic time scale
$1/p$ for the regrowth of trees governs the dynamics. This time
scale translates into a characteristic length scale $1/p$, which can
be observed in Fig.~\ref{automata_fig_fires}, via the propagation
rule for the fire.

Self-organized criticality can, however, be induced in the forest
fire model when introducing an additional rule, namely that a tree
might ignite spontaneously with a small probability $f$, when struck
by {} lightning, {causing} also small
patches of forest to burn. We will not discuss this mechanism in
detail here, treating instead in the next section the occurrence of
self-organized criticality in the sandpile model on a firm
mathematical basis.

\vspace*{-6pt}
\section{The Sandpile Model and Self-Organized Criticality}
\index{sandpile model|textbf}

\runinhead{Self-Organized Criticality} \index{sandpile
model!self-organized criticality} We have learned in
Chap.~\ref{chap_networks2} about
the concept \qut{life at the edge of chaos}. Namely, that certain
dynamical and organizational aspects of living organisms may be
critical. Normal physical and dynamical systems, however, show
criticality only for selected parameters, e.g.\ $T=T_c$, see
Sect.~\ref{automata_Landau_theory}. For criticality to be
biologically relevant, the system must evolve into a critical state
starting from a wide range of initial states -- one speaks of
\qut{self-organized criticality}.

\runinhead{The Sandpile Model}
\index{sandpile model!updating rule} \index{Bak, Per!sandpile model}
Per Bak and coworkers introduced a simple cellular automaton
{that} mimics the properties of sandpiles,
{i.e.} the BTW model. Every cell is characterized by a
force
$$
z_i\ =\ z(x,y)\ =\ 0,\ 1,\ 2,\ \ldots, \qquad \quad x,y=1,\ldots, L
$$
on a finite $L\times L$ lattice. There is no one-to-one
correspondence of the sandpile model to real-world
sandpiles. Loosely speaking one may identify the
force $z_i$ with the slope of real-world sandpiles.
But this analogy is not rigorous, as the slope of a
real-world sandpile is a continuous variable.
The slopes belonging to two neighboring cells should
therefore be similar, whereas the values of
$z_i$ and $z_j$ on two neighboring cells can differ by
an arbitrary amount within the sandpile model.

\index{sand toppling}
The sand begins to topple when the slope gets too big:
$$
z_j\ \to \  z_j\, -\, \Delta_{ij},
\qquad \quad\mbox{if}
\qquad z_j>K~,
$$
where $K$ is the threshold slope and
with the toppling matrix
\begin{equation}
\Delta_{i,j} \ =\ \left\{
\begin{array}{rll}
 4  &   i=j  & \\
-1  & i, j & {\rm \ \ nearest\ neighbors} \\
 0  &      &  {\rm \ \ otherwise}
\end{array}\right. ~.
\label{automata_sand_pile_model}
\end{equation}
This update rule is valid for the {four}-cell
neighborhood $\{(0,\pm1),(\pm1,0)\}$. The threshold $K$ is
arbitrary, a shift in $K$ simply shifts {}$z_i$. It is
{customary} to consider $K=3$. Any initial random
configuration will then relax into a steady-state final
configuration (called {the} stable state) with
$$
z_i\ =\ 0,\ 1,\ 2,\ 3, \qquad \quad\mbox{(stable state)}~.
$$

\runinhead{Open Boundary Conditions}
\index{open boundary conditions} \index{sandpile model!boundary
conditions} The update rule
{Eq.~}(\ref{automata_sand_pile_model}) is conserving:
\begin{quotation}
{\it Conserving Quantities.\enspace}
\index{dynamics!conserving} \index{sandpile model!local conservation
of sand}
 If there is a quantity {that} is not
changed by the update rule it is said to be conserving.
\end{quotation}
The sandpile model is locally conserving. The total height $\sum_j
z_j$ is constant due to $\sum_j \Delta_{i,j} =0$.
Globally{, however,} it is not conserving, as one
uses open boundary conditions for which excess sand is lost at the
boundary. When a site at the boundary topples, some sand is lost
there and the total $\sum_j z_j$ is reduced by one.

\index{sandpile model!real-world sandpile} However, here
{we have} only a vague relation of the BTW model to
real-world sandpiles. The conserving nature of the sandpile model
mimics the fact that sand grains cannot be lost in real-world
sandpiles. This interpretation {,
however, contrasts} with the previously assumed correspondence of
{} $z_i$ with the slope of real-world sandpiles.

\begin{figure}[t]
\centerline{\includegraphics{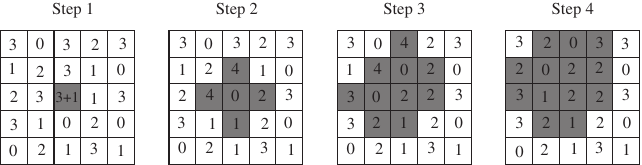}}
\caption{The progress of an avalanche, with duration $t=3$ and size
$s=13$, for a sandpile configuration on a {$5 \times 5$} lattice
with $K=3$. The height of the sand in each cell is indicated by the
numbers. The {\it shaded region} is where the avalanche has
progressed. The avalanche stops after\break step 3
\label{automata_avalanche}}
\end{figure}

\runinhead{Avalanches}
\index{avalanche!sandpile} When starting from a random initial state
with $z_i\ll K$ the system settles in a stable configuration when
adding \qut{grains of sand} for a while. When a
{grain of sand} is added to a site with $z_i=K$
$$
z_i\ \to \ z_i+1,
\qquad\quad z_i=K~,
$$
a toppling event is induced, which may in turn lead to a whole
series of topplings. The resulting avalanche is characterized by its
duration $t$ and the size $s$ of affected sites. It continues until
a new stable configuration is reached. In
Fig.~\ref{automata_avalanche} a small avalanche is
shown.\vspace*{3pt}

\runinhead{Distribution of Avalanches}
\index{avalanche!distribution!size}
\index{avalanche!distribution!length} We define with $D(s)$ and
$D(t)$ the distributions of the size and of the duration of
avalanches. One finds that they are scale-free,\vspace*{3pt}
\begin{equation}
D(s)\ \sim\ s^{-\alpha_s}, \qquad\quad D(t)\ \sim\ t^{-\alpha_t}~,
\label{automata_D_s_t}\vspace*{3pt}
\end{equation}
as we will discuss in the next section. Equation
(\ref{automata_D_s_t}) expresses the essence of self-organized
criticality. We expect these scale-free relations to be valid for a
wide range of cellular automata with conserving dynamics,
independent of the special values of the parameters entering the
respective update functions. Numerical simulations and analytic
approximations {for $d =2$ dimensions yield}\vspace*{3pt}
$$
\alpha_s\ \approx\ {5\over 4}, \qquad\quad \alpha_t\ \approx\
{3\over 4}~.\vspace*{3pt}
$$
\runinhead{Conserving Dynamics and Self-Organized Criticality}
\index{self-organized criticality!conserving dynamics} We note that
the toppling events of an avalanche are (locally) conserving.
Avalanches of arbitrary large sizes must therefore occur, as sand
can be lost only at the boundary of the system. One can indeed prove
that Eqs.~(\ref{automata_D_s_t}) are valid only for locally
conserving models. Self-organized criticality breaks down as soon as
there is a small but non-vanishing probability to {lose} sand
somewhere inside the system.\vspace*{3pt}

\runinhead{Features of the Critical State} The empty board, when all
cells are initially empty, $z_i\equiv0$, is not critical. The system
remains in the frozen phase when adding sand; compare
Chap.~\ref{chap_networks2}, as long as most
$z_i<K$. Adding one sand corn after the other
the critical state is slowly approached. There is
no way to avoid the critical state.

Once the critical state is achieved the system remains critical.
This critical state is paradoxically also the  point at which the
system is dynamically most unstable. It has an unlimited
susceptibility to an external driving (adding a
{grain of sand}), using the terminology of
Sect.~\ref{automata_Landau_theory}, as a single added
{grain of sand} can trip avalanches of arbitrary
size.

It needs to be noted that the dynamics of the sandpile model is
deterministic, once the {grain of sand} has been
added, and that the disparate fluctuations in terms of induced
avalanches are features of the critical state per se and not due to
any hidden stochasticity, as discussed in
Chap.~\ref{chap_chaos1}, or due to any hidden deterministic chaos.

\section{Random Branching Theory}
\label{automata_random_branching_theory}
\index{random branching!theory|textbf}

Branching theory deals with the growth of networks via branching.
Networks \hbox{generated} by branching processes are
loopless; they typically arise in theories of
evolutionary processes. 

\subsection[Branching Theory of Self-Organized Criticality]
           {Branching Theory of Self-Organized Criticality}
\index{self-organized criticality!branching theory}

Avalanches have an intrinsic relation to branching 
processes: at every time step the avalanche can either
continue or stop. Random branching theory is hence a
suitable method for studying self-organized criticality.

\runinhead{Branching in Sandpiles}
\index{random branching!sandpile}
A typical update during an avalanche is of the form
$$
\begin{array}{rcccl}
\mbox{time 0:} &\quad& z_i\to z_i-4 &\quad& z_j\to z_j + 1~, \\
\mbox{time 1:} &\quad& z_i\to z_i+1 &\quad& z_j\to z_j - 4~,
\end{array}
$$
when two neighboring cells $i$ and $j$ {initially have} $z_i=K+1$ and $z_j=K$. This implies that
an avalanche typically intersects with itself. Consider, however, a
general \hbox{$d$-dimensional} lattice with $K=2d-1$. The
self-interaction of the avalanche becomes unimportant in the limit
$1/d\to 0$ and the avalanche can be mapped rigorously to a random
branching process. Note that we encountered an analogous situation
in the context of high-dimensional or random graphs, discussed in
Chap.~\ref{chap_networks1}, which are also loopless
in the thermodynamic limit.

\runinhead{Binary Random Branching} In $d\to\infty$ the notion of
neighbors {loses} meaning, avalanches then have no
spatial structure. Every toppling event affects $2d$ neighbors, on a
$d$-dimensional hypercubic lattice. However, only the cumulative
probability of toppling of the affected cells is relevant, due to
the absence of geometric constraints in the limit $d\to\infty$. All
{that} is important then is the question whether an
avalanche continues, increasing its size continuously, or whether it
stops.

\begin{figure}[t]
\centerline{\includegraphics{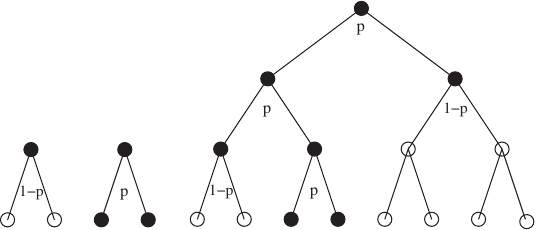}}
\caption{Branching processes. \textit{Left}: The two possible
processes of order $n=1$. \textit{Right}: A generic process of order
$n=3$ with an avalanche of size $s=7$ }
\label{automata_branching_1}\vspace*{-6pt}
\end{figure}

We can therefore consider the case of binary branching, viz that a
toppling event creates two new active sites.
\begin{quotation}
{\it Binary Branching.\enspace}\index{random branching!binary}An active site of an avalanche topples with the probability $p$ and
creates {two} new active sites.
\end{quotation}
For $p<1/2$ the number of new active sites decreases on the average
and the avalanche dies out. $p_c=1/2$ is the critical state with (on
the average) conserving dynamics. See
Fig.~\ref{automata_branching_1} for some examples of
{branching} processes.

\begin{figure}[t]
\centerline{\includegraphics{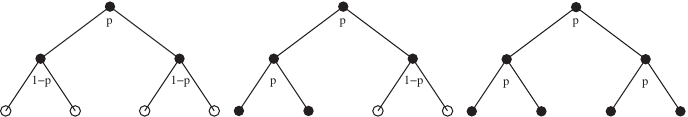}}
\caption{Branching processes of order $n=2$ with avalanches of sizes
$s=3,5,7$ (\textit{left}, \textit{middle}, \textit{right}) and
boundaries $\sigma=0,2,4$} \label{automata_branching_2}
\end{figure}

\runinhead{Distribution of Avalanche Sizes}
\index{avalanche!distribution!size}
The properties of avalanches are determined by
the probability distribution,
$$
P_n(s,p), \qquad\quad
\sum_{s=1}^\infty P_n(s,p)=1~,
$$
describing the probability to find an avalanche of size $s$ in a
branching process of order $n$. Here $s$ is the (odd) number of
sites inside the avalanche, see Figs.~\ref{automata_branching_1} and
\ref{automata_branching_2} for some examples.

\runinhead{Generating Function Formalism} \index{probability
generating function} {In Chap.~\ref{chap_networks2}, we introduced} the
generating\break \hbox{functions} for probability distribution. This
formalism is very useful when one has to deal with independent
stochastic processes, as the joint probability of two independent
stochastic processes is equivalent to the simple multiplication of
the corresponding generating functions.

We define via
\begin{equation}
f_n(x,p) \ =\ \sum_s P_n(s,p)\, x^s, \qquad\quad f_n(1,p)  = \sum_s
P_n(s,p)  = 1 \label{automata_gen_function}
\end{equation}
the generating functional $f_n(x,p)$ for the probability
distribution $P_n(s,p)$. We note that\vspace*{3pt}
\begin{equation}
P_n(s,p) \ =\ \frac 1 {s!} \,{\partial^s f_n(x,p)\over \partial
x^s}\Big|_{x=0}, \qquad\quad n,\,p\ \ \mbox{fixed}~.
\label{automata_P_form_gen_function}\vspace*{3pt}
\end{equation}
\runinhead{Small Avalanches} \index{avalanche!small} For small $s$
and large $n$ one can evaluate the probability for small avalanches
to occur by hand and one finds for the corresponding generating
functionals:
$$
P_n(1,p)=1-p,\qquad P_n(3,p)=p(1-p)^2,\qquad P_n(5,p)=2p^2(1-p)^3~,
$$
compare Figs.~\ref{automata_branching_1} and
\ref{automata_branching_2}. Note that $P_n(1,p)$ is the probability
to find an avalanche of just one site.

\runinhead{{The} Recursion Relation} For generic $n$ the recursion
relation
\begin{equation}
f_{n+1}(x,p)\ =\ x\,(1-p) \, +\, x\,p\,f_n^2(x,p)
\label{automata_recursion_rel}
\end{equation}
is valid. To see why, one considers building the branching network
backwards, adding a site at the top:
\begin{itemize}
\item[--]{With the probability $(1-p)$}\\ one adds a single-site
avalanche described by the generating functional $x$.
\item[--]{With the probability $p$}\\ one adds a site,
described by the generating functional $x$, which
generated two active sites, described each by the generating
functional $f_n(x,p)$.
\end{itemize}

\index{random branching!decomposition}
In the terminology of branching theory, one also speaks
of a decomposition of the branching process after its
first generation, a standard procedure.

\runinhead{{The} Self-Consistency Condition}
\index{self-consistency condition!avalanche size distribution}
For large $n$ and finite $x$ the generating functionals
$f_n(x,p)$ and $f_{n+1}(x,p)$ become identical,
leading to the self-consistency condition
\begin{equation}
f_{n}(x,p)\ =\ f_{n+1}(x,p)\ =\ x\,(1-p) \, +\, x\,p\,f_n^2(x,p)~,
\label{automata_self_consistency}
\end{equation}
%

\noindent with the solution\enlargethispage*{12pt}
\begin{equation}
f(x,p)\ \equiv\ f_n(x,p)\ =\
{1-\sqrt{1-4x^2p(1-p)}\over 2xp}
\label{automata_solution_f}
\end{equation}
for the generating functional $f(x,p)$. The
normalization condition
$$
f(1,p)\ =\ {1-\sqrt{1-4^2p(1-p)}\over 2p}
      \ =\ {1-\sqrt{(1-2p)^2}\over 2p} \ =\ 1
$$
is fulfilled for $p\in[0,1/2]$. For $p>1/2$ the
last step in above equation would not be correct.

\runinhead{The Subcritical Solution}
\index{avalanche!subcritical} Expanding
Eq.~(\ref{automata_solution_f}) in powers of $x^2$ we find terms
like
$$
{1\over p}\,
\Big[4p(1-p)\Big]^k\, {\left(x^2\right)^k\over x} \ =\
{1\over p}\,
\Big[4p(1-p)\Big]^k\, x^{2k-1}~.
$$
Comparing {this} with the definition of the generating
functional {Eq.~}(\ref{automata_gen_function}) we note
that $s=2k-1$, $k=(s+1)/2$ and that
\begin{equation}
P(s,p) \ \sim\ {1\over p}\,
\sqrt{4p(1-p)}\,\Big[4p(1-p)\Big]^{s/2}
 \ \sim\ e^{-s/s_c(p)}~,
\label{automata_sub_critical_sol}
\end{equation}


\noindent where we have used the relation
$$
a^{s/2}\ =\ e^{\ln(a^{s/2})}\ =\ e^{-s(\ln a)/(-2)},
\qquad\quad a=4p(1-p)~,
$$
and where we have
defined the avalanche correlation size
$$
s_c(p)\ =\ {-2\over \ln[4p(1-p)]},
\qquad\quad
\lim_{p\to 1/2}s_c(p)\ \to\ \infty~.
$$
For $p<1/2$ the size correlation length $s_c(p)$ is finite and the
avalanche is consequently not scale-free, see
Sect.~\ref{section_criticality_dynamical_systems}. The
characteristic size of an avalanche $s_c(p)$ diverges for $p\to
p_c=1/2$. Note that $s_c(p)>0$ for $p\in]0,1[$.

\runinhead{The Critical Solution}
\index{avalanche!critical} \index{critical!avalanche} We now
consider the critical case with
$$
p=1/2, \qquad \quad 4 p (1-p)\ =\ 1,
\qquad\quad f(x,p)\ =\ {1-\sqrt{1-x^2}\over x}~.
$$
The expansion of $\sqrt{1-x^2}$ with respect to $x$ is
$$
\sqrt{1-x^2}\ =\ \sum_{k=0}^\infty
{ {1\over 2}
\left({1\over 2}-1\right)
\left({1\over 2}-2\right)
\cdot\cdot\cdot
\left({1\over 2}-k+1\right)
\over k! }\, \Big(-x^2\Big)^k
$$
in Eq.~(\ref{automata_solution_f}) and therefore
$$
P_c(k)\,\equiv\,
P(s=2k-1,p=1/2) \,= \,
{ {1\over 2}
\left({1\over 2}-1\right)
\left({1\over 2}-2\right)
\cdot\cdot\cdot
\left({1\over 2}-k+1\right)
\over k! }\,(-1)^{k}~.
$$
This expression is still unhandy. We are, however, only interested
in the asymptotic behavior for large avalanche sizes $s$. For this
purpose we consider the recursive relation\
$$
P_c(k+1) \ = \
{1/2-k\over k+1}(-1)P_c(k)  \ =\
{1-1/(2k)\over 1+1/k}P_c(k)\vspace{3pt}
$$
in the limit of large $k=(s+1)/2$, where
$1/(1+1/k)\approx 1-1/k$,
$$
P_c(k+1) \ \approx \
\Big[1-1/(2k)\Big]\,\Big[1-1/k\Big]\,P_c(k) \ \approx\
\Big[1-3/(2k)\Big]\,P_c(k)~.
$$
This asymptotic relation leads to\vspace{3pt}
$$
{P_c(k+1) - P_c(k)\over 1} \ =\ {-3\over 2k}\,P_c(k),
\qquad \quad
{\partial P_c(k)\over\partial k} \ =\
{-3\over 2k}\,P_c(k)~,\vspace{3pt}
$$
with the solution
\begin{equation}
P_c(k)\ \sim\ k^{-3/2},
\qquad \quad
D(s)\ =\ P_c(s)\ \sim\ s^{-3/2},
\qquad \quad
\alpha_s\ =\ {3\over2}~,
\label{automata_P_c_scaling}
\end{equation}
for large $k,s$, since $s=2k-1$.

\runinhead{Distribution of Relaxation Times}
\index{relaxation time!distribution}
The distribution of the duration $n$ of avalanches
can be evaluated in a similar fashion. For
this purpose one considers the
probability distribution function
$$
Q_n(\sigma,p)
$$
for an avalanche of duration $n$ to have $\sigma$ cells at the
boundary, see Fig.~\ref{automata_branching_2}.

One can then derive a recursion relation analogous to
Eq.~(\ref{automata_recursion_rel}) for the corresponding generating
functional and solve it self-consistently. We leave this as an
exercise for the reader.

The distribution of avalanche durations
is then given by considering
$Q_n~=~Q_n\break (\sigma=0,p=1/2)$, i.e.\ the probability
that the avalanche stops after $n$ steps.
One finds
\begin{equation}
Q_n\ \sim\ n^{-2},
\qquad \quad
D(t)\ \sim\ t^{-2},
\qquad \quad
\alpha_t\ =\ {2}~.
\label{automata_Q_scaling}
\end{equation}

\runinhead{Tuned or Self-Organized Criticality?}
\index{self-organized criticality!vs.\ tuned criticality} 
The random branching model discussed in this section had 
only one free parameter, the probability $p$. This model 
is critical only for $p\to p_c=1/2$, giving rise to the 
impression that one has to fine tune the parameters in 
order to obtain criticality, just like in
ordinary phase transitions.

This, however, is not the case. As an
example we could generalize the sandpile model to continuous forces
$z_i\in[0,\infty]$ and to the update rules
$$
z_i\ \to \  z_i\, -\, \Delta_{ij},
\qquad \quad\mbox{if}
\qquad z_i>K~,
$$
and\vspace{3pt}
\begin{equation}
\Delta_{i,j} \ =\ \left\{
\begin{array}{rll}
 K  &   i=j  & \\
-c\,K/4  & i, j & \mbox{\ \ nearest\ neighbors} \\
-(1-c)\,K/8  & i, j & \mbox{\ \ next-nearest\ neighbors} \\
 0  &      &  {\rm \ \ otherwise}
\end{array}\right.
\label{automata_generalized_sand_pile_model}\vspace{3pt}
\end{equation}
for a square-lattice with four nearest neighbors and
eight next-nearest neighbors (Manhattan distance). The
update rules are conserving,
$$
\sum_j \Delta_{ij}\ =\ 0,
\qquad \quad
\forall c\in[0,1]~.
$$
For $c=1$ this model corresponds to the continuous field
generalization of the BTW model. The model defined by
Eqs.~(\ref{automata_generalized_sand_pile_model}), 
which has not yet been studied in the literature, 
might be expected to map in the limit $d\to\infty$ to
an appropriate random branching model with $p=p_c=1/2$ 
and to be critical for all values of the parameters 
$K$ and $c$, due to its conserving dynamics.

\begin{figure}[t]
\centerline{
\includegraphics[width=0.45\hsize]{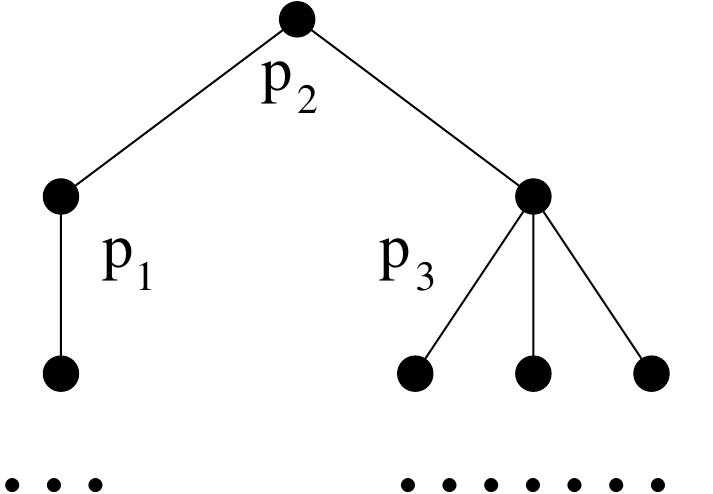}
\hspace{3ex}
\includegraphics[width=0.40\hsize]{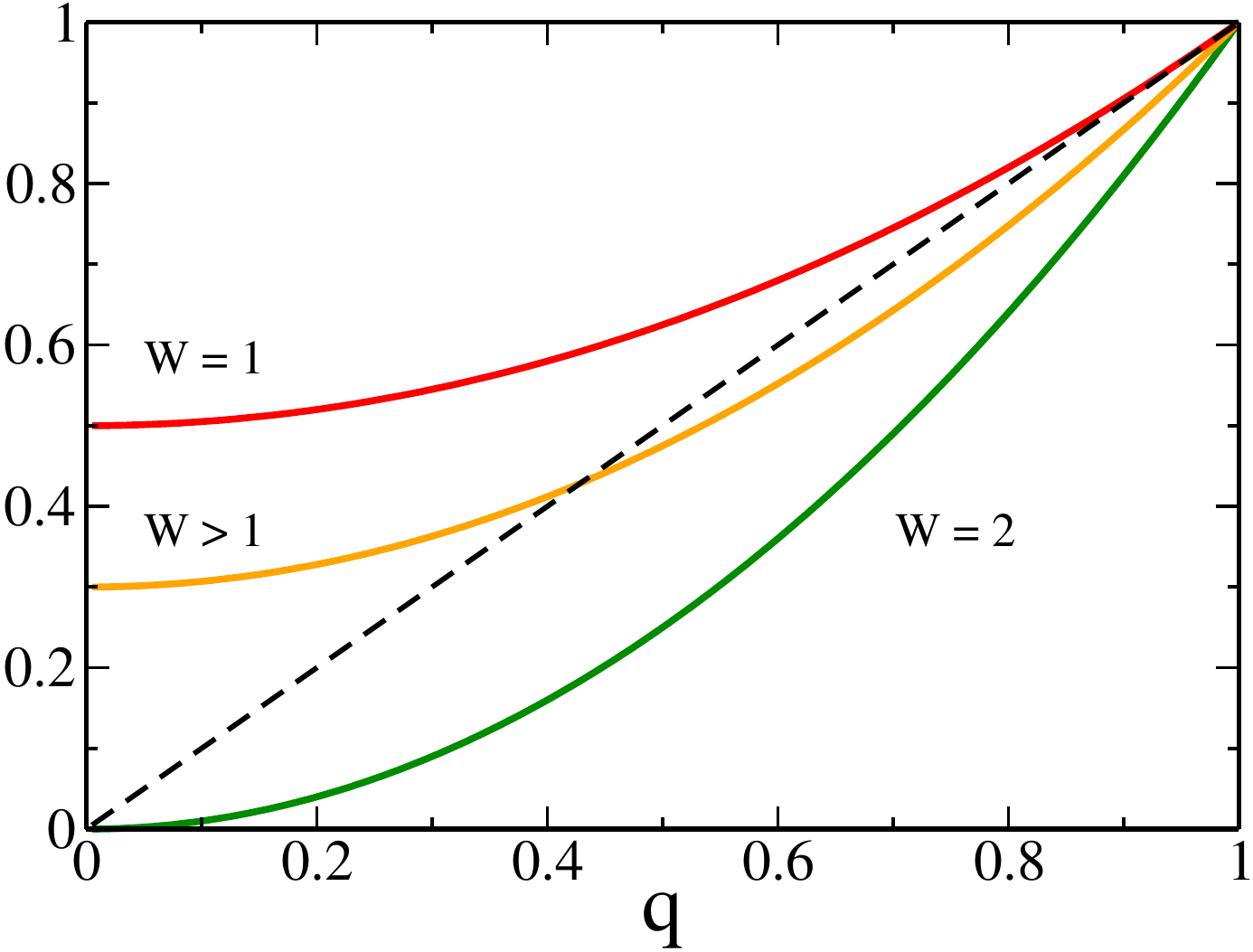}
           }
\caption{Galton-Watson processes.
\textit{Left}: Example of a reproduction tree, $p_m$ being
the probabilities of having $m=0,\ 1,\ \dots$ offsprings.
\textit{Right}: Graphical solution for the fixpoint
equation (\ref{automata1_Galton_Watson_example}), for
various average numbers of offsprings $W$.
        }
\label{automata_Galton_Watson}
\end{figure}

\subsection[Galton-Watson Processes]
           {Galton-Watson Processes}

Galton-Watson processes are generalizations 
of the binary branching processes considered so far,
with interesting applications in evolution theory 
and some everyday experiences.

\runinhead{The History of Family Names}
Family names are handed down traditionally
from father to son. Family names regularly die out,
leading over the course of time to a substantial
reduction of the pool of family names. This
effect is especially pronounced in countries
looking back on millenia of cultural continuity,
like China, where 22\% of the population are sharing 
only three family names. 

The evolution of family names is described by a
Galton-Watson process and a key quantity of
interest is the extinction probabilty, viz the
probability that the last person bearing
a given family name dies without descendants.

\runinhead{The Galton-Watson Process}
\index{Galton-Watson process}
The basic reproduction statistics determines
the evolution of family names, see
Fig.~\ref{automata_Galton_Watson}.

We denote with $p_m$ the probability that an
individual has $m$ offsprings and with $G_0(x)=\sum_m p_m x^m$
its generating function. Defining with $p_m^{(n)}$ the
probability of finding a total of $m$ descendants in
the $n$-th generation, we find the recursion relation
$$
G^{(n+1)}(x) \ =\ \sum_m p_m^{(n)}\left[G_0(x)\right]^m
\ =\  G^{(n)}(G_0(x)),
\qquad\quad
G^{(n)}(x) \ =\ \sum_m p_m^{(n)} x^m
$$
for the respective generating function. 
Using the intial condition $G^{(0)}(x)=x$ 
we may rewrite this recursion relation as
\begin{equation}
G^{(n)}(x) \ =\ G_0(G_0(\dots G_0(x)\dots)) \ =\
G_0\left(G^{(n-1)}(x)\right)~.
\label{automata1_Galton_Watson_self_consistency}
\end{equation}
This recursion relation is the basis for
all further considerations; we consider here
the extinction probability $q$.

\runinhead{Extinction Probability}
The reproduction process dies out when 
there is a generation with zero members.
The probability of having zero persons bearing the
given family name in the $n$-th generation is
\begin{equation}
q\ =\ p_0^{(n)}\ =\ G^{(n)}(0) \ =\ G_0\left(G^{(n-1)}(0)\right)
\ =\ G_0(q)~,
\label{automata1_Galton_Watson_q}
\end{equation}
where we have used the recursion relation
Eq.~(\ref{automata1_Galton_Watson_self_consistency})
and the stationary condition
$G^{(n)}(0)\approx G^{(n-1)}(0)$. The
extinction probability $q$ is hence given by the fixpoint
$q=G_0(q)$ of the generating functional
$G_0(x)$ of the reproduction probability.

\runinhead{Binary Branching as a Galton-Watson Process}
As an example we consider the case
that
$$
G_0(x) \ =\ 1-{W\over 2} + {W\over 2} x^2,
\qquad\quad
G_0^{\,\prime}(1) \ =\ W~,
$$
viz that people may not have but either zero or two
sons, with probabilities $1-W/2$ and $W/2<1$
respectively. The expected number of offsprings
$W$ is also called the fitness in evolution theory, 
see Chap.~\ref{chap_evolution1}. This setting
corresponds to the case of binary branching,
see Fig.~\ref{automata_branching_1}, with
$W/2$ being the branching probability, describing 
the reproductive dynamics of unicellular
bacteria.

The self-consistency condition (\ref{automata1_Galton_Watson_q}) 
for the extinction probability $q\ =\ q(W) $ then reads
\begin{equation}
q \ =\ 1-{W\over 2} + {W\over 2} q^2,
\qquad\quad
q(W) \ =\ {1\over W} \pm \sqrt{
{1\over W^2} - {(2-W)^2\over W^2}}~,
\label{automata1_Galton_Watson_example}
\end{equation}
with the smaller root being here of relevance.
The extinction probability vanishes
for a reproduction rate of two,
$$
q(W) \ =\ \left\{
\begin{array}{ccc}
0 &\quad& W=2 \\
q\in\,]0,1[ &\quad& 1<W<2 \\
1 &\quad & W \le 1
\end{array}
          \right.
$$
and is unity for a fitness below one,
compare Fig.~\ref{automata_Galton_Watson}.

\section{Application to Long-Term Evolution}
\label{section_Bak_Sneppen}
\index{evolution!long-term|textbf}

An application of the techniques developed in this
chapter can be used to study a model for the
evolution of species proposed by Bak and Sneppen.

\begin{figure}[t]
\centerline{\includegraphics{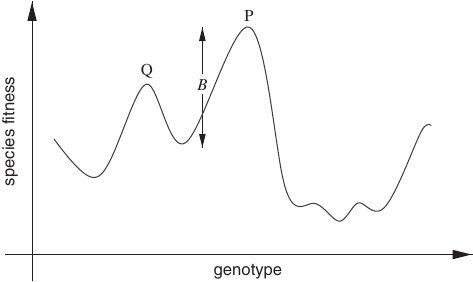}}
\caption{A one-dimensional fitness landscape.
  A species evolving from an adaptive peak $P$ to
  a new adaptive peak $Q$ needs to overcome the fitness barrier $B$}
\label{automata_fitnessBarrier}
\end{figure}

\runinhead{Fitness Landscapes}
Evolution deals with the adaption of species and their fitness
relative to the ecosystem they {live} in.
\begin{quotation}
{\it Fitness Landscapes.\enspace}
\index{fitness landscape} The function {that}
determines the chances of survival of a species, its fitness, is
called the fitness landscape.
\end{quotation}
In Fig.~\ref{automata_fitnessBarrier} a simple fitness landscape, in
which there is only one dimension in the genotype (or
phenotype)\footnote{\looseness1The term \qut{genotype} denotes the
ensemble of genes. The actual form of an organism, the
\qut{phenotype}, is determined by the genotype plus environmental
factors, like food supply during growth.} space, is illustrated.

\index{fitness!maximum} \index{fitness!barrier} The population will
spend most of its time in a local fitness maximum, whenever the
mutation rate is low with respect to the selection rate, since there
are fitness barriers, see Fig.~\ref{automata_fitnessBarrier},
between adjacent local fitness maxima. Mutations are random
processes and the evolution from one local fitness maximum to the
next can then happen only through a stochastic escape, a process we
discussed in Chap.~\ref{chap_chaos1}.

\runinhead{Coevolution} \index{coevolution} It is important to keep
in mind for the following discussion that an ecosystem, and with it
the respective fitness landscapes, is not static on long time
scales. The ecosystem is the result of the combined action of
geophysical factors, such as the average rainfall and temperature,
and biological influences, viz the properties and actions of the
other constituting species. The evolutionary progress of one species
will therefore, {in general, trigger}
adaption processes in other species appertaining to the same
ecosystem, a process denoted \qut{coevolution}.

\runinhead{Evolutionary Time Scales}
\index{evolution!time scales} \index{evolution!fitness barrier} In
the model of Bak and Sneppen there are no explicit fitness
landscapes like the one illustrated in
Fig.~\ref{automata_fitnessBarrier}. Instead the model attempts to
mimic the effects of fitness landscapes, viz the influence of all
the other species making up the ecosystem, by a single number, the
\qut{fitness barrier}. The time needed for a stochastic escape from
one local fitness optimum increases exponentially with
{the} barrier height. We may therefore assume that the
average time $t$ it takes to mutate across a fitness barrier of
height $B$ scales as
\begin{equation}
t\ =\ t_0\, e^{B/T}~,
\label{automata_times}
\end{equation}
where $t_0$ and $T$ are constants. The value of $t_0$ merely sets
the time scale and is not important. The parameter $T$ depends on
the mutation rate, and the assumption that mutation is low implies
that $T$ is small compared with the typical barrier heights $B$ in
the landscape. In this case the time scales $t$ for crossing
slightly different barriers are distributed over many orders of
magnitude and only the lowest barrier is\break
relevant.\vspace*{3pt}

\runinhead{The Bak and Sneppen Model}
\index{Bak--Sneppen model} \index{model!Bak--Sneppen} The Bak and
Sneppen model is a phenomenological model for the evolution of
barrier heights. The number $N$ of species is fixed and each species
has a respective barrier\vspace{3pt}
$$
B_i\ =\ B_i(t)\,\in\,[0,1],\ \qquad\quad t=0,1,2,\ldots\vspace{3pt}
$$
for its further evolution. The initial
$B_i(0)$ are drawn randomly from $[0,1]$.
The model then consists of the repetition of two steps:
\begin{enumerate}\leftskip5pt
\item [(1)]     {The times for a stochastic escape are exponentially distributed,
      see Eq.~(\ref{automata_times}). It is therefore reasonable to
      assume that the species with the lowest barrier $B_i$ mutates
      and escapes first. After escaping, it will adapt quickly to a new
      local fitness maximum. At this point it will then have a
      new barrier for mutation, which is assumed to be uniformly
      distributed in $[0,1]$.}
\item [(2)]{ The fitness function for a species $i$ is given by the
      ecological environment it lives in, which is
      made up of all the other species.
      When any given species mutates it therefore
      influences the fitness landscape
      for a certain number of other species. Within
      the Bak and Sneppen model this translates into
      assigning new random barriers $B_j$ for $K-1$
      neighbors of the mutating species $i$.}
\end{enumerate}
The Bak and Sneppen model therefore tries to capture two essential
ingredients of long-term evolution: The exponential distribution of
successful mutations and the interaction of species via the change
of the overall ecosystem, when one constituting species
evolves.\vspace*{3pt}

\runinhead{The Random Neighbor Model} \index{model!random
neighbors} \index{random!neighbor model} The topology of the
interaction between species in the Bak--Sneppen model is unclear. It
might be chosen {as} two-dimensional, if the species are thought to
live geographically separated, or one-dimensional in a toy model. In
reality the topology is complex and can be assumed to be, in first
approximation, random, resulting in the soluble random neighbor
model.\vspace*{3pt}

\begin{figure}[t]
\vspace{6pt}
\centerline{\includegraphics{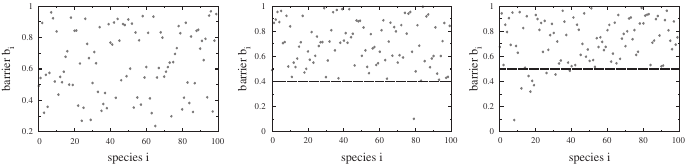}}
\caption{The barrier values ({\it dots}) for a 100 species one-dimensional
Bak--Sneppen model after 50, 200 and 1600 steps of a simulation. The
{\it horizontal line} in each frame represents the approximate position of
the upper edge of the \qut{gap}. A few species have barriers below
this level, indicating that they were involved in an avalanche at
the moment when the snapshot of the system\break was taken}
\label{automata_snappen}
\end{figure}

\runinhead{Evolution of Barrier Distribution}
\index{evolution!barrier distribution} \index{distribution!fitness
barrier} Let us discuss qualitatively the redistribution of barrier
heights under the dynamics, the sequential repetition of step (1)
and (2) above, see Fig.~\ref{automata_snappen}. The initial barrier
heights are uniformly distributed over the interval $[0,1]$ and the
lowest barrier, removed in step (1), is small. The new heights
reassigned in steps (1) and (2) will therefore lead, on the average,
to an increase of the average barrier height with {passing time}.

With increasing average barrier height {the characteristic
lowest barrier is also raised} and eventually a steady state will be
reached, just as in the sandpile model discussed previously. It
turns out that the characteristic value for the lowest barrier is
about $1/K$ at equilibrium  in the mean-field approximation and that
the steady state is critical.

\runinhead{Molecular Field Theory}
\index{mean-field theory!Bak--Sneppen model} In order to solve the
Bak--Sneppen model, we define the barrier distribution function,
$$
p(x,t)~,
$$
viz the probability to find a barrier of hight $x\in[0,1]$ at time
step $t=1,2,\dots$. In addition, we define with $Q(x)$ the
probability to find a barrier above $x$:
\begin{equation}
Q(x)\ =\ \int_x^1 \mathrm{d}x'\,p(x'), \qquad \quad Q(0)=1, \qquad
Q(1)=0~. \label{automata_Q_x}
\end{equation}
The dynamics is governed by the size of the
smallest barrier. The distribution function
$p_1(x)$ for the lowest barrier is
\begin{equation}
p_1(x)\ =\ N\,p(x)\,Q^{N-1}(x)~,
\label{automata_p_1}
\end{equation}
given by the probability $p(x)$ for one barrier (out of the $N$
barriers) to have the barrier height $x$, while all the other $N-1$
barriers are larger. $p_1(x)$ is normalized,
$$
\int_0^1 \mathrm{d}x\, p_1(x) \ =\ (-N)\int_0^1 \mathrm{d}x\,
Q^{N-1}(x){\partial Q(x)\over \partial x} \ =\
-Q^N(x)\Big|_{x=0}^{x=1} \ =\ 1~,
$$
where we used $p(x) = - Q'(x)$, $Q(0)=1$ and $Q(1)=0$, see
Eq.~(\ref{automata_Q_x}).

\runinhead{Time Evolution of Barrier Distribution}
\index{dynamics!Bak--Sneppen model} The time evolution for the
barrier distribution consists in taking away one (out of $N$)
{barrier}, the lowest, via
$$
p(x,t)\,-\,{1\over N}\,p_1(x,t)~,
$$
and by removing randomly $K-1$ barriers from the remaining $N-1$
barriers, and adding $K$ random barriers:\vspace*{-4pt}
\begin{eqnarray}
\label{automata_p_evolution}
p(x,t+1)& =& p(x,t)\,-\,{1\over N}\,p_1(x,t) \\
&-&{K-1\over N-1}\left(
p(x,t)\,-\,{1\over N}\,p_1(x,t)
                   \right)
\,+\,{K\over N}~.
\nonumber
\end{eqnarray}
We note that $p(x,t+1)$ is normalized whenever $p(x,t)$
and $p_1(x,t)$ were normalized correctly:
\begin{eqnarray*}
\int_0^1 \mathrm{d}x\ p(x,t+1)& =& 1-{1\over N} - {K-1\over
N-1}\left(1-{1\over N}\right)
           +{K\over N}\\
& = & \left(1-{K-1\over N-1}\right){N-1\over N} + {K\over N}
\ = \ {N-K\over N} + {K\over N} \ \equiv\ 1~ .
\end{eqnarray*}

\runinhead{Stationary Distribution}
\index{distribution!fitness barrier!stationary} \index{stationary
distribution!Bak--Sneppen model} After many iterations of
{Eq.~}(\ref{automata_p_evolution}) the barrier
distribution will approach a stationary solution
$p(x,t+1)=p(x,t)\equiv p(x)$, as can be observed from the numerical
simulation shown in Fig.~\ref{automata_snappen}. The stationary
distribution corresponds to the fixpoint condition\vspace{4pt}
$$
0 \ = \ p_1(x){1\over N}\left({K-1\over N-1} -1 \right)
\,-\, p(x)\,{K-1\over N-1}\, +\, {K\over N}\vspace{4pt}
$$
of Eq.~(\ref{automata_p_evolution}). Using the expression
$p_1=Np\,Q^{N-1}$, see Eq.~(\ref{automata_p_1}), for $p_1(x)$ we
then have\vspace{4pt}
$$
0 \ = \ Np(x)\,Q^{N-1}(x)(K-N)
\,-\, p(x)\,(K-1)N\, +\, K(N-1)~.\vspace{4pt}
$$
Using $p(x)=-{\partial Q(x)\over \partial x}$ we obtain
\begin{eqnarray*}
0& =& N(N-K){\partial Q(x)\over \partial x}Q^{N-1}
\,+\, (K-1)N{\partial Q(x)\over \partial x}
\,+\, K(N-1)\\[6pt]
0& =& N(N-K)\,Q^{N-1}\,dQ \,+\, (K-1)N\,dQ \,+\,
K(N-1)\,\mathrm{d}x~.
\end{eqnarray*}
We can integrate this last expression with respect to $x$,\vspace{4pt}
\begin{equation}
0\ =\ (N-K)\,Q^{N}(x)
\,+\, (K-1)N\,Q(x)
\,+\, K(N-1)\,(x-1)~,
\label{automata_BS_NK}
\end{equation}
where we took care of the boundary condition $Q(1)=0$, $Q(0)=1$.

\runinhead{Solution in the Thermodynamic Limit}
\index{distribution!fitness barrier!thermodynamic limit} The
polynomial Eq.~(\ref{automata_BS_NK}) simplifies in the
thermodynamic limit, with $N\to\infty$ and $K/N\to0$, to\vspace{4pt}
\begin{equation}
0\ =\ Q^{N}(x) \,+\, (K-1)\,Q(x)
\,-\, K\,(1-x)~.
\label{automata_BS_K}\vspace{4pt}
\end{equation}
We note that $Q(x)\in[0,1]$ and that $Q(0)=1$, $Q(1)=0$. There must
{therefore be} some $x\in]0,1[$ for which
$0<Q(x)<1$. Then\vspace{4pt}
\begin{equation}
Q^N(x)\to 0,
\qquad\quad
Q(x)\ \approx\ {K\over K-1}\,(1-x)~.
\label{automata_BS_Q_small}
\end{equation}
%

\begin{figure}[t]
\centerline{\includegraphics{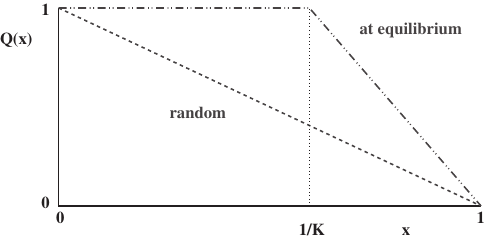}}
\caption{The distribution $Q(x)$ to find a fitness barrier
         larger than $x\in[0,1]$ for the Bak and Sneppen model,
        for the case of random barrier distribution (\textit{dashed line})
        and the stationary distribution (\textit{dashed-dotted line}),
        compare Eq.~(\ref{automata1_Q_equilibrium})
        }
\label{automata_Bak_Sneppen_Q}
\end{figure}
\noindent Equation (\ref{automata_BS_Q_small}) remains valid as long as $Q<1$,
or $x>x_c$:
$$
1\ =\ {K\over K-1}(1-x_c), \qquad\quad  x_c\ =\ {1\over K}~.
$$
We then have in the limit $N\to\infty$
\begin{equation}
\lim_{N\to\infty}Q(x)\ =\ \left\{
\begin{array}{ccl}
1 &\ \mbox{for}\ & x<1/K \\
(1-x)K/(K-1) &\ \mbox{for}\ & x>1/K
\end{array}
\right.~ ,
\label{automata1_Q_equilibrium}
\end{equation}
compare Fig.~\ref{automata_Bak_Sneppen_Q}, and, using $p(x) =
-{\partial Q(x)/ \partial x}$,
\begin{equation}
\lim_{N\to\infty}p(x)\ =\ \left\{
\begin{array}{ccl}
0 &\ \mbox{for}\ & x<1/K \\
K/(K-1) &\ \mbox{for}\ & x>1/K
\end{array}
\right.~.
\label{automata1_p_equilibrium}
\end{equation}
This result compares qualitatively well with the numerical results
presented in Fig.~\ref{automata_snappen}. Note, however, that the
mean-field solution Eq.~(\ref{automata1_p_equilibrium}) does not
predict the exact critical barrier height, which is somewhat larger
for $K=2$ and {a} one-dimensional arrangement of
neighbors, as in Fig.~\ref{automata_snappen}.

\runinhead{$\textbf{{{1}/N}}$ Corrections}
Equation (\ref{automata1_p_equilibrium}) cannot be rigorously true for
$N<\infty$, since there is a finite probability for barriers with
$B_i<1/K$ to reappear at every step. One can expand the solution of
the self-consistency Eq.~(\ref{automata_BS_NK}) in powers of $1/N$.
One finds
\begin{equation}
p(x) \ \simeq\ \left\{
\begin{array}{rcl}
K/N &\ \mbox{for}\ & x<1/K\\
K/(K-1) &\ \mbox{for}\ & x>1/K
\end{array}
\right. ~.
\label{automata_BS_solution}
\end{equation}
We leave the derivation as an exercise {for} the
reader.

\enlargethispage{-12pt}

\runinhead{Distribution of {the} Lowest Barrier}
If the barrier distribution is zero below
the self-organized threshold $ x_c =1/K$
and constant above, then
the lowest barrier must be below $x_c$
with equal probability:\vspace{3pt}
\begin{equation}
p_1(x) \quad \to\quad \left\{
\begin{array}{rcl}
K &\ \mbox{for}\ & x<1/K\\
0 &\ \mbox{for}\ & x>1/K
\end{array}
\right. ,
\qquad\quad
\int_0^1 dx\,p_1(x)=1~.
\label{automata_p_1_solution}\vspace{3pt}
\end{equation}
Equations (\ref{automata_p_1_solution}) and (\ref{automata_p_1}) are
consistent with {Eq.~}(\ref{automata_BS_solution}) for
$x<1/K$.

\begin{figure}[t]
\centerline{\includegraphics{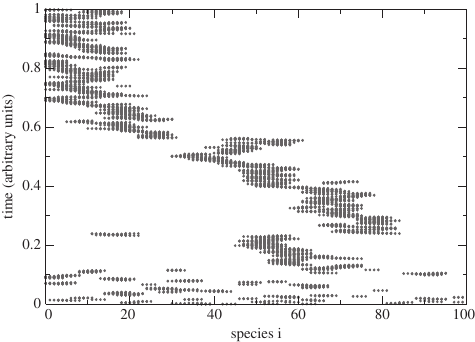}}
\caption{A time series of evolutionary activity in a simulation
         of the one-dimensional Bak--Sneppen model with $K=2$
         showing coevolutionary avalanches interrupting the
         punctuated equilibrium.  Each {\it dot} represents the action
         of choosing a new barrier value for one species}
\label{automata_snappen_avalanches}
\end{figure}

\runinhead{Coevolution and Avalanches} \index{coevolution!avalanche}
\index{avalanche!coevolution} When the species with the lowest
barrier mutates we assign new random barrier heights to it and to
its $K-1$ neighbors. This causes an avalanche of evolutionary
adaptations whenever one of the new barriers
becomes the new lowest fitness barrier. One calls this
phenomenon \qut{coevolution} since the
evolution of one species drives the adaption of other species
belonging to the same ecosystem. We will discuss this and other
aspects of evolution in more detail in Chap.~\ref{chap_evolution1}. In
Fig.~\ref{automata_snappen_avalanches} this process is illustrated
for the one-dimensional model. The avalanches in the system are
clearly visible and well separated in time. In between the
individual avalanches the barrier distribution does not change
appreciably; one speaks of a \qut{punctuated
equilibrium}. \index{punctuated equilibrium}

\runinhead{Critical Coevolutionary Avalanches}
\index{critical!coevolutionary avalanches} {In
Sect.~\ref{automata_random_branching_theory} we discussed} the
connection between avalanches and random branching. The branching
process is critical when it goes on with a probability of 1/2. To
see whether the coevolutionary avalanches within the Bak and Sneppen
model are critical we calculate the probability $p_{\mathrm{bran}}$
that at least one of the $K$ new, randomly selected, fitness
barriers will be the new lowest barrier.

With probability $x$ one of the new random
{barriers} is in $[0,x]$ and below the actual
lowest barrier, which is distributed with $p_1(x)$, see
Eq.~(\ref{automata_p_1_solution}). We then have
$$
p_{\mathrm{bran}}\ =\ K\,\int_0^{1} p_1(x)\,x\,\mathrm{d}x\ =\
 K\,\int_0^{1/K} K\,x\,\mathrm{d}x\ =\
{K^2\over 2}\,x^2\Big|_0^{1/K}\ \equiv\ {1\over 2}~,
$$
viz the avalanches are critical. The distribution of the size $s$ of
the coevolutionary avalanches is then
$$
D(s)\ \sim\ \left({1\over s}\right)^{3/2}~,
$$
as evaluated within the random branching approximation, see
Eq.~(\ref{automata_P_c_scaling}), and independent of $K$. The size
of a coevolutionary avalanche can be
{arbitrarily} large and involve, in extremis, a
finite fraction of the ecosystem, compare
Fig.~\ref{automata_snappen_avalanches}.

\runinhead{Features of the Critical State}
The sandpile model evolves into a critical state under the influence
of an external driving, when adding one {grain of
sand} after another. The critical state is characterized by a
distribution of slopes (or heights) $z_i$, one of its
characteristics being a discontinuity{;} there is a
finite fraction of slopes with $z_i=Z-1$, but no slope with $z_i=Z$,
apart from some of the sites participating in an avalanche.

In the Bak and Sneppen model the same process occurs, but without
external drivings. At criticality the barrier  distribution
$p(x)=\partial Q(x)/\partial x$ has a discontinuity at $x_c=1/K$,
see Fig.~\ref{automata_Bak_Sneppen_Q}. One could say, cum grano
salis, that the system has developed an \qut{internal phase
transition}, namely a transition in the barrier distribution $p(x)$,
an internal variable. This emergent state for $p(x)$ is a many-body
or collective effect, since it results from the mutual reciprocal
interactions of the species participating in the formation of the
ecosystem.


\vspace*{-6pt}

\section*{Exercises}
\addcontentsline{toc}{section}{Exercises}
\markboth{\thechapter\enspace Cellular Automata and 
Self-Organized Criticality}{Exercises}
\begin{list}{}
\item \hspace*{-17pt}{\sc Solutions of the Landau--Ginzburg Functional} \\
Determine the order parameter for $h\ne0$ via
Eq.~(\ref{automata_h_phi_eq}) and
Fig.~\ref{automata_fig_Landau_Ginzburg}. Discuss the local stability
condition {Eq.~}(\ref{automata_eq9.26}) for the three
possible solutions and their global stability. Note that $F=f\,V$,
where $F$ is the free energy, $f$ the free energy density and $V$
the volume.
\item \hspace*{-17pt}{\sc Entropy and Specific Heat Within the Landau Model} \\
Determine the entropy $S(T)={\partial F\over \partial T}$ and the
specific heat $c_V=T{\partial S\over \partial T}$ within the
Landau--Ginzburg theory Eq.~(\ref{automata_eq_f_T_h})
for phase transitions.
\item\hspace*{-17pt} {\sc The Game of Life} \\
Consider the evolution of the following states, see
Fig.~\ref{automata_fig_life0}, under the rules for 
Conway's game of life:

\{(0,0),(1,0),(0,1),(1,1)\}

\{(0,-1),(0,0),(0,1)\}

\{(0,0),(0,1),(1,0),($-$1,0),(0,$-$1)\}

\{(0,0),(0,1),(0,2),(1,2),(2,1)\}

The predictions can be checked with Java-applets you 
may easily find in the Internet.
\item \hspace*{-17pt}{\sc The Game of Life on a Small-World Network} \\
Write a program to simulate the game of life on a 2D lattice.
Consider this lattice as a network with every site having edges to
its eight neighbors. Rewire the network such that (a) the local
connectivities $z_i\equiv 8$ are retained for every site and (b) a
small-world network is obtained. This can be achieved by cutting two
arbitrary links with probability $p$ and rewiring the four resulting
stubs randomly.

Define an appropriate dynamical order parameter and
characterize the changes as a function of the rewiring
probability. Compare Chap.~\ref{chap_networks1} and
Chap.~\ref{chap_chaos1}.
\item \hspace*{-17pt}{\sc The Forest Fire Model} \\
Develop a mean-field theory for the forest fire model by introducing
appropriate probabilities to find cells with trees, fires and ashes.
Find the critical number of nearest neighbors $Z$ for fires to
continue burning.
\item\hspace*{-17pt} {\sc The Realistic Sandpile Model} \\
Propose a cellular automata model that simulates
the physics of real-world sandpiles somewhat more realistically than
the BTW model. The cell values $z(x,y)$ should correspond to the
local height of the sand. Write a program to simulate
the model.
\item \hspace*{-17pt}{\sc The Random Branching Model} \\
Derive the distribution of avalanche durations
Eq.~(\ref{automata_Q_scaling}) in analogy to the steps
explained in Sect.~\ref{automata_random_branching_theory},
by considering a recursion relation for the integrated
duration probability $\tilde Q_n=\sum_{n'=0}^n Q_n(0,p)$, viz
for the probability that an avalanche last maximally $n$
time steps.
\item \hspace*{-17pt}{\sc The Galton-Watson Process} \\
Use the fixpoint condition, Eq.~(\ref{automata1_Galton_Watson_q})
and show that the extinction probability is unity if the
average reproduction rate is smaller than one.

\item \hspace*{-17pt}{\sc The Bak and Sneppen Model} \\
Write a program to simulate the Bak and Sneppen model
in Sect.~\ref{section_Bak_Sneppen} and compare
it with the molecular field solution
Eq.~(\ref{automata_BS_NK}).
\end{list}


\def\refer#1#2#3#4#5#6{\item{\frenchspacing\sc#1}\hspace{4pt}
                       #2\hspace{8pt}#3 {\it\frenchspacing#4} {\bf#5}, #6.}
\def\bookref#1#2#3#4{\item{\frenchspacing\sc#1}\hspace{4pt}
                     #2\hspace{8pt}{\it#3}  #4.}

\enlargethispage{-12pt}

\addcontentsline{toc}{section}{Further Reading} 
\section*{Further Reading}

\markboth{\thechapter\enspace Cellular Automata and Self-Organized Criticality}{Further Reading}

Introductory texts to cellular automata  and to the game of life are
 Wolfram (1986), Creutz (1997) and Berlekamp et~al. (1982). For a review of the forest fire and several
related models, see Clar et~al. (1996); for a review
of sandpiles{, see} Creutz (2004), and for a general
review of self-organized criticality, { see} Paczuski and
Bak (1999). Exemplary textbooks on statistical physics and phase
transitions {have been written by} Callen (1985) and
Goldenfeld (1992).

Some general features of $1/f$ noise are discussed by Press
(1978){;} its possible relation to self-organized
criticality has been postulated by Bak et~al. (1987).
The formulation of the Bak and Sneppen (1993) model for long-term
\hbox{coevolutionary} processes and its mean-field solution are
discussed by Flyvbjerg et~al. (1993).

The interested reader {may also glance at} some original research literature, such as
a numerical study of the sandpile model (Priezzhev et~al. 1996) and the application of random branching theory to
the sandpile model (Zapperi et~al. 1995). The
connection of self-organized criticality to local conservation rules
is worked out by Tsuchiya and Katori (2000), and the forest fire
model with lightning {is} introduced by Drossel and
Schwabl (1992).

{\baselineskip=15pt
\begin{list}{}{\leftmargin=2em \itemindent=-\leftmargin%
\itemsep=3pt \parsep=0pt \small}
\refer{Bak, P., Sneppen, K.}{1993}{Punctuated equilibrium and
  criticality in a simple model of
  evolution.}{Physical Review Letters}{71}{4083--4086}
\refer{Bak, P., Tang, C., Wiesenfeld, K.}{1987}
{Self-organized criticality: An explanation
of $1/f$ noise.}{Physical Review Letters}{59}{381--384}
\bookref{Berlekamp, E., Conway, J.,  Guy, R.}{1982}{Winning Ways for
Your Mathematical Plays, Vol. 2.}{Academic Press, New York}
\bookref{Callen, H.B.}{1985}{Thermodynamics and Introduction to
Thermostatistics.}{Wiley, New York}
\refer{Clar, S., Drossel, B., Schwabl, F.}{1996}{Forest fires and
other examples of self-organized criticality.} {Journal of Physics:
Condensed Matter}{8}{6803--6824}
\bookref{Creutz, M.}{1997}{\rm Cellular automata and self-organized
criticality.} {In G. Bhanot, S. Chen and P.~Seiden (eds). {\it Some
New Directions in Science on Computers},  pp. 147--169, World
Scientific, Singapore}
\refer{Creutz, M.}{2004}{Playing with sandpiles.}
{Physica A}{340}{521--526}
\refer{Drossel, B., Schwabl, F.}{1992}
{Self-organized critical forest-fire model.}
{Physical Review Letters}{69}{1629--1632}
\refer{Flyvbjerg, H., Sneppen, K., Bak, P.}{1993}{Mean field
theory for a simple model of evolution.} {Physical Review
Letters}{71}{4087--4090}
\bookref{Goldenfeld, N.}{1992}{Lectures on Phase Transitions and
the Renormalization Group.}{Perseus Publishing, Reading, MA}
\bookref{Newman, M.E.J., Palmer, R.G.}{2002}{Models of Extinction.}
{Oxford University Press,\break New~York}
\bookref{Paczuski, M., Bak. P.}{1999}
{\rm Self organization of complex systems.}
{In: {\it Proceedings of 12th Chris Engelbrecht Summer School};
also available as \url{http://www.arxiv.org/abs/cond-mat/9906077}}
\refer{Press, W.H.}{1978}{Flicker noises in astronomy and elsewhere.}
{Comments on Modern Physics, Part C}{7}{103--119}
\refer{Priezzhev, V.B., Ktitarev, D.V., Ivashkevich,
E.V.}{1996}{Formation of avalanches and critical exponents in an
abelian sandpile model.} {Physical Review Letters}{76}{2093--2096}
\refer{Tsuchiya, T., Katori, M.}{2000}{Proof of breaking of
self-organized criticality in a nonconservative abelian sandpile
model.} {Physical Review Letters}{61}{1183--1186}
\bookref{Wolfram, S., editor}{1986}{Theory and Applications of Cellular
Automata.}{World Scientific, Singapore}
\refer{Zapperi, S., Lauritsen, K.B., Stanley,
H.E.}{1995}{Self-organized branching processes: Mean-field theory
for avalanches.} {Physical Review Letters}{75}{4071--4074}
\end{list}
\par}
 
\addtocontents{toc}{\protect\enlargethispage*{12pt}}

\chapter{Darwinian Evolution, Hypercycles and Game Theory}
\label{chap_evolution1}


\abstract {Adaptation and evolution are quasi synonymous in popular
language and Darwinian evolution is a prime application of complex
adaptive system theory. We will see that adaptation does not
happen automatically and discuss the concept of \qut{error
catastrophe} as a possible root for the downfall of a species.
Venturing briefly into the mysteries surrounding the origin of life,
we will investigate the possible advent of a \qut{quasispecies} in
terms of mutually supporting hypercycles. The basic theory of 
evolution is furthermore closely related to game theory, the
mathematical theory of~interacting agents, viz of rationally acting
economic persons.\newline\indent We will learn in this chapter, on
the one hand, that every complex dynamical system has its distinct
characteristics to be considered. In the case of Darwinian evolution
these are concepts like fitness, selection and mutation. General
notions from complex system theory are, on the other hand, important
for a thorough understanding. An example is the phenomenon of
stochastic escape discussed in Chap.~\ref{chap_chaos1}, which is
operative in the realm of Darwinian evolution.}

\section{Introduction}
\label{evolution_introduction}

\runinhead{Microevolution}
\index{microevolution}\index{evolution!microevolution}The ecosystem
of the earth is a complex and adaptive system. It formed via
\hbox{Darwinian} evolution through species differentiation and
adaptation to a changing environment. A set
of inheritable traits, the genome, is passed from parent to
offspring and the reproduction
success is determined by the outcome of random mutations and natural
selection -- a process denoted \qut{microevolution}\footnote{Note
that the term \qut{macroevolution}, coined to describe the evolution
at the level of organisms, is nowadays somewhat obsolete.}
\begin{quotation}
{\it Asexual Reproduction.\enspace}
\index{asexual reproduction}
\index{reproduction!asexual}
One speaks of asexual reproduction when an individual
has a single parent.\clearpage
\end{quotation}\vspace*{-6pt}\enlargethispage*{10pt}
\vspace*{-12pt}Here we consider mostly models for asexual reproduction,
though most concepts can be easily generalized to the case
of sexual reproduction.

\begin{table}[b!]
\vspace*{-12pt}\centering \caption{Genome size $N$ and the
spontaneous mutation rates $\mu$, compare
Eq.~(\ref{evolution_trans_1}), per base for two RNA-based bacteria
and DNA-based eucaryotes. From Jain and Krug (2006) and Drake et al.
(1998)
        }
\label{evolution1_table_rates}
\index{genome!size}
{\begin{tabular*}{300pt}{@{\extracolsep{\fill}}llll@{}}
\hline\noalign{\smallskip}
Organism  & Genome size $ \ $ & Rate per base $ \ $  & Rate per genome \\
\noalign{\smallskip}\svhline\noalign{\smallskip}
Bacteriophage $Q\beta$ & 4.5 $\times 10^{3}$ & 1.4 $\times 10^{-3}$ & 6.5 \\
Bacteriophage $\lambda$ & 4.9 $\times 10^{4}$ & 7.7 $\times 10^{-8}$ &
0.0038 \\
\emph{E. Coli} & 4.6 $\times 10^{6}$ & 5.4 $\times 10^{-10}$ & 0.0025 \\
\emph{C. Elegans} & 8.0 $\times 10^{7}$ & 2.3 $\times 10^{-10}$ & 0.018 \\
Mouse  & 2.7 $\times 10^{9}$ & 1.8 $\times 10^{-10}$ & 0.49 \\
Human & 3.2 $\times 10^{9}$ & 5.0 $\times 10^{-11}$ & 0.16 \\
\noalign{\smallskip}\hline\noalign{\smallskip}
\end{tabular*}}
{}
\end{table}

\vspace*{6pt}

\runinhead{Basic Terminology}
Let us introduce some basic variables
needed to formulate the approach.
\begin{itemize}
\item[--] {Population $M$: The number of individuals.\\
\index{population}We assume here that $M$ does not change
  with time, modeling the competition for
  a limited supply of resources.}
\item[--] {Genome $N$: Size of the genome.\\
\index{genome}We encode the inheritable traits by a set of $N$ binary
  variables,
$$
{\bf s} = (s_1,s_2,\ldots ,s_N), \qquad\quad s_i =\ \pm 1 ~.
$$
$N$ is considered fixed.}
\item[--] {Generations \\
\index{evolution!generation}\index{population!generation}We consider time sequences of non-overlapping generations,
  like in a wheat field. The population present at time
  $t$ is replaced by their {offspring} at
  generation $t+1$.}
\end{itemize}
In Table \ref{evolution1_table_rates} some typical values for the
size $N$ of the genome are listed. Note the three orders of
magnitude between simple eucaryotic life forms and the human genome.

\vspace*{6pt}
\runinhead{State of the Population} \index{state space!population}
The state of the population at time $t$ can be described by
specifying the genomes of all the individuals,
$$
\{{\bf s}^\alpha(t)\},\qquad\quad \alpha=1\dots M,
\qquad\quad {\bf s}=(s_1,\dots,s_N)~.
$$
We define by
\begin{equation}
X_{\bf s}(t),\qquad\quad \sum_{\bf s}X_{\bf s}(t)= M~,
\label{evolution_nu_s}
\end{equation}
the number of individuals with genome ${\bf s}$ for each of the
$2^N$ points ${\bf s}$ in the genome space. Typically, most of these
occupation numbers vanish{;} biological populations are
extremely sparse in genome space.

\vspace*{6pt}

\runinhead{Combinatorial Genetics of Alleles}
\index{genetics!combinatorial}\index{alleles}Classical genetics
focuses on the presence (or absence) of a few characteristic traits.
These traits are determined by specific sites, denoted \qut{loci},
in the genome. The genetic realizations of these specific loci are
called \qut{alleles}. Popular examples are alleles for blue, brown
and green eyes.

Combinatorial genetics deals with the frequency change of the
appearance of a given allele resulting from environmental changes
during the evolutionary process. Most visible evolutionary changes
are due to a remixing of alleles, as mutation induced changes in the
genome are relatively rare{;} compare the mutation rates listed in
Table \ref{evolution1_table_rates}.\vspace*{3pt}

\runinhead{Beanbag Genetics Without Epistatic Interactions}
\index{genetics!{beanbag}}\index{beanbag genetics}\index{epistatic
interactions}One calls \qut{epistasis} the fact that the effect of
the presence of a given allele in a given locus may depend on which
alleles are present in some other loci. Classical genetics neglects
epistatic interactions. The resulting picture is often called
\qut{beanbag genetics}, as if the genome were nothing but a bag
carrying the different alleles within itself.\vspace*{3pt}

\runinhead{Genotype and Phenotype} We note that the physical
appearance of an organism is not determined exclusively by gene
expression. One distinguishes between the genotype and the
phenotype.
\begin{itemize}
\item[--] {{The} Genotype}:
\index{genotype}\index{genome!genotype}The genotype of an organism is the class to which
that organism belongs as determined by the DNA that
was passed to the organism by its parents at the
organism's conception.
\item[--] {{The} Phenotype}:
\index{phenotype}\index{genome!phenotype}The phenotype of an organism is the class to which that organism
belongs as determined by the physical
and behavioral characteristics of the organism,
for example its size and shape,
its metabolic activities and its pattern of movement.
\end{itemize}
Selection acts, strictly speaking, only upon phenotypes, but only
the genotype is bequeathed. The variations in phenotypes then act as
a source of noise for the selection process.\vspace*{3pt}

\runinhead{Speciation}
\index{speciation}\index{evolution!speciation}One denotes
{by} \qut{speciation} the process leading to the
differentiation of an initial species into two distinct species.
Speciation occurs due to {adaptation} to
different ecological niches, often in distinct geographical
environments. We will not treat the various theories proposed for
speciation here.

\vspace*{3pt}
\section{Mutations and Fitness in a Static Environment}
\label{evolution_mutations_fitness}

\runinhead{Constant Environment} \index{environment!constant}We
consider here the environment to be {static;} an
assumption {that is} justified for the case of short-term
evolution. This assumption clearly breaks down for long time scales,
as already discussed in Chap.~\ref{chap_automata1}
since the evolutionary change of one species might
lead to repercussions all over the ecosystem to which it
appertains.

\runinhead{Independent Individuals} An important issue in the theory
of evolution is the emergence of specific kinds of social behavior.
Social behavior can only arise if the individuals of the same
population interact. We discuss some of these issues in
Sect.~\ref{evolution_coevolution} in the context of game theory.
Until then we assume non-interacting individuals, which implies that
the fitness of a given genetic trait is independent of the frequency
of this and of other alleles, apart from the overall competition for
resources.

\runinhead{Constant Mutation Rates} \index{mutation!rate} We
furthermore assume that the mutation rates are
\begin{itemize}
\item[--] constant over time,
\item[--] independent of the locus in the genome, and
\item[--]not subject to genetic control.
\end{itemize}
Any other assumption would require {a} detailed
microbiological {modeling;} a subject
beyond our scope.

\runinhead{Stochastic Evolution} \index{dynamics!evolution}
\index{stochastic!evolution} \index{evolution!stochastic} The
evolutionary process can then be modeled as a three-stage
sto\-chas\-tic process:
\begin{enumerate}
\item {Reproduction}:
\index{population!reproduction}The individual $\alpha$ at generation $t$ is the
  offspring of an individual $\alpha'$
  living at generation $t-1$. Reproduction
  is thus represented as a stochastic map
\begin{equation}
\alpha\quad \longrightarrow\quad \alpha'=G_t(\alpha)~,
\end{equation}
where $G_t(\alpha)$ is the parent of the individual $\alpha$,
and is chosen at random among the $M$ individuals living at
generation $t-1$.
\item {Mutation}:
\index{genome!mutation}\index{evolution!mutation}The genomes of the {offspring} differ from the respective
 genomes of their parents through random changes.
\item {Selection}:
\index{evolution!selection}The number of surviving {offspring} of each individual
   depends on its genome{;} it is proportional to
   its \qut{fitness}, which is a functional of the genome.
\end{enumerate}
\runinhead{Point Mutations and Mutation Rate}
\index{point mutation}\index{mutation!rate}Here we consider mostly independent point mutations,
namely that every element of the genome
is modified independently of the other elements,
\begin{equation}
s_i^\alpha(t)\ =\ -s_i^{G_t(\alpha)}(t-1)\qquad
\hbox{with probability }\ \  \mu~,
\label{evolution_trans_1}
\end{equation}
where the parameter $\mu\in [0,1/2]$ is the microscopic
\qut{mutation rate}. In real organisms, more complex phenomena take
place, like global rearrangements of the genome, copies of some part
of the genome, displacements of blocks of elements from one location
to another, and so on. The values for the real-world mutation rates
$\mu$ for various species listed in Table
\ref{evolution1_table_rates} are therefore to be considered
{as} effective mutation rates.

\runinhead{Fitness and Fitness Landscape}
\index{fitness landscape}The fitness $W({\bf s})$, also called
\qut{Wrightian fitness}, of a genotype trait $\bf s$ is proportional
to the average number of offspring an individual possessing the
trait $\bf s$ has. It is strictly positive and can therefore be
written as
\begin{equation}
W({\bf s})\ =\ e^{k F({\bf s})} \ \ \propto\ \ \hbox{average number
of {offspring} of $\bf s$.}
\label{evolution_fit_1}
\end{equation}
Selection acts in first place upon phenotypes, but we neglect here
the difference, considering the variations in phenotypes as a source
of noise, as discussed above. The parameters in
Eq.~(\ref{evolution_fit_1}) are denoted:
\begin{itemize}
\item[--] {$W({\bf s})$}: Wrightian fitness,
\index{Wrightian fitness}\index{fitness!Wrightian}

\item[--] {$F({\bf s})$}: fitness landscape,
\item[--] {$k$}: inverse selection temperature,\footnote{The
probability to find a state with energy $E$ in
a thermodynamic system with temperature $T$ is proportional
to the Boltzmann factor $\exp(-\beta\,E)$. The inverse temperature
is $\beta=1/(k_B T)$, with $k_B$ being the Boltzmann constant.} and
\index{temperature!inverse of selection}

\item[--] {$w({\bf s})$}: Malthusian fitness, when
rewriting Eq.~(\ref{evolution_fit_1}) as $\ \displaystyle W({\bf
s})\ =\ e^{w({\bf s})\Delta t}$,\newline where $\Delta t$ is the
generation time. \index{Malthusian fitness}\index{fitness!Malthusian}
\end{itemize}
We will work here with discrete time, viz with non-overlapping
generations, and make use only of the Wrightian fitness
$W({\bf s})$.

\runinhead{Fitness of Individuals Versus Fitness of Species}
\index{fitness!individual vs.\ species}\index{species!fitness}We
remark that this notion of fitness is a concept defined at the level
of individuals in a homogeneous population. The resulting fitness of
a species or of a group of species needs to be explicitly evaluated
and is model-dependent.

\runinhead{Fitness Ratios}
\index{fitness!ratio}
The assumption of a constant population size
makes the reproductive success a {\em relative\/}
notion. Only the ratios
\vspace*{6pt}
\begin{equation}
{W({\bf s}_1)\over W({\bf s}_2)} \ =\
{e^{kF({\bf s}_1)}\over e^{kF({\bf s}_2)} } \ =\
e^{k[F({\bf s}_1)-F(({\bf s}_2)]}
\label{evolution_W_exp}
\vspace*{6pt}
\end{equation}
are important. It follows that the quantity $W({\bf s})$ is defined
up to a proportionality constant and, accordingly, the fitness
landscape $F({\bf s})$ only up to an additive constant,
{} much like the energy in physics.

\begin{figure}[t]
\centering
\includegraphics{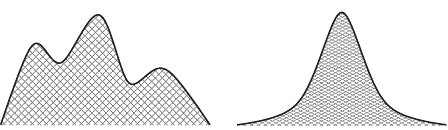}
\caption{{(Smooth)}
one-dimensional model fitness
 landscapes $F({\bf s})$. Real-world fitness landscapes{, however, contain} discontinuities. \textit{Left}: A
fitness landscape with peaks and valleys, metaphorically also called
a \qut{rugged landscape}.
 \textit{Right}: A fitness landscape containing a single smooth peak,
as
 described by Eq.~(\ref{evolution_fuji_F})}
\label{evolution1_fitnessLandscapes}
\end{figure}

\runinhead{{The} Fitness Landscape}
The graphical representation of the fitness function $F({\bf s})$ is
not really possible for real-world fitness functions, due to the
high {dimensional} $2^N$ of the genome
space. It is nevertheless customary to draw {a} fitness
landscape, like the one shown in
Fig.~\ref{evolution1_fitnessLandscapes}. {However, one must bear in mind that} these
illustrations are not to be taken at face value, apart from model
considerations.

\runinhead{{The} Fundamental Theorem of Natural
Selection} \index{theorem!fundamental of natural selection}
\index{evolution!fundamental theorem} The so-called fundamental
theorem of natural selection, first stated by Fisher in 1930, deals
with adaptation in the absence of mutations and in the thermodynamic
limit $M\to\infty$. An infinite population size allows one
to neglect fluctuations.

The theorem states that the average fitness of the population cannot
decrease in time under these circumstances, and that the average
fitness becomes stationary only when all individuals in the
population have the maximal reproductive fitness.

The proof is straightforward. We define by
\begin{equation}
\langle W\rangle_t\ \equiv\
{1\over M}\sum_{\alpha=1}^M
W\left({\bf s}^\alpha(t)\right)
\ =\ {1\over M}
\sum_{\bf s}W\left({\bf s}\right)X_{\bf s}(t)~,
\label{evolution_W_new}
\end{equation}
the average fitness of the population. Note that the $\sum_{\bf s}$
in {Eq.~}(\ref{evolution_W_new}) contains $2^N$ terms.
The evolution equations are given in the absence of mutations by
\begin{equation}
X_{\bf s}(t+1)\ =\ \frac{W({\bf s})}{\langle W\rangle_t}
                     \,X_{\bf s}(t)~,
\label{evolution_no_mut}
\end{equation}
where ${W({\bf s})}/{\langle W\rangle_t}$ is the relative
reproductive success. The overall population size remains
constant,
\begin{equation}
\sum_{\bf s}
X_{\bf s}(t+1)\ =\ \frac{1}{\langle W\rangle_t}\sum_{\bf s}
X_{\bf s}(t)W({\bf s})\ =\ M~,
\end{equation}
where we have used Eq.~(\ref{evolution_W_new}) for $\langle
W\rangle_t$. Then
\begin{eqnarray}\nonumber
\langle W\rangle_{t+1}& =& {1\over M}
\sum_{\bf s}W\left({\bf s}\right)X_{\bf s}(t+1)
\ =\
{ {1\over M}\sum_{\bf s} W^2({\bf s})X_{\bf s}(t)\over
  {1\over M}\sum_{{\bf s}'} W({\bf s}')X_{{\bf s}'}(t)} \\
& =& {\langle W^2\rangle_t\over \langle W\rangle_t}
\ \geq\  \langle W\rangle_t~,
\end{eqnarray}
since
$\langle W^2\rangle_t -\langle W\rangle_t^2
=\langle \Delta W^2\rangle_t \ge0$.
The steady state
$$
\langle W\rangle_{t+1}\ =\ \langle W\rangle_{t},
\qquad\quad
\langle W^2\rangle_t \ =\ \langle W\rangle_t^2~,
$$
is only possible when all individuals $1\ldots M$ in the population
have the same fitness, viz the same genotype.

\section{Deterministic Evolution}
\label{evolution_deterministic}
\index{deterministic evolution|textbf}

Mutations are random events and the evolution process is therefore a
stochastic process. But stochastic fluctuations become irrelevant in
the limit of infinite population size $M\to\infty${;}
they average out. In this limit the equations governing evolution
become deterministic and only the average transition rates are
relevant. One can then study in detail the condition necessary for
adaptation to occur for various mutation rates.

\subsection{Evolution Equations}
\index{equation!deterministic evolution|textbf}
\index{evolution equations|textbf}

\runinhead{{The} Mutation Matrix} \index{mutation!matrix}
\index{matrix!mutation} The mutation matrix\vspace{-2pt}
\begin{equation}
Q_\mu({\bf s}'\to {\bf s}), \qquad\quad \sum_{\bf s} Q_\mu({\bf
s}'\to {\bf s}) = 1 \label{evolution_Q_matrix}\vspace{-2pt}
\end{equation}
denotes the probabilities of obtaining a genotype ${\bf s}$ when
attempting to reproduce an individual with genotype ${\bf s}'$. The
mutation rates $Q_\mu({\bf s}'\to {\bf s})$ may depend on a
parameter $\mu$ determining the overall mutation rate. The mutation
matrix includes the absence of any mutation, viz the transition
$Q_\mu({\bf s}'\to {\bf s}')$. It is normalized.

\runinhead{Deterministic Evolution with Mutations} We generalize
Eq.~(\ref{evolution_no_mut}), {which is} valid in the
absence of mutations, by including the effect of mutations via the
mutation matrix $Q_\mu({\bf s}'\to {\bf s})$:\vspace{-2pt}
$$
{X_{\bf s}(t+1)/ M}\ =\ \left(\sum_{{\bf s}'}X_{{\bf s}'}(t)W({\bf
s}') Q_\mu({\bf s}'\to {\bf s})\right) \Bigg/ \left(\sum_{{\bf
s}'}W_{{\bf s}'}X_{{\bf s}'}(t)\right)~,\vspace{-2pt}
$$
or\vspace{-2pt}
\begin{equation}
x_{\bf s}(t+1)\ =\ {\sum_{{\bf s}'}x_{{\bf s}'}(t)W({\bf s}')
Q_\mu({\bf s}'\to {\bf s}) \over \langle W\rangle_t}, \qquad\quad
\langle W\rangle_t = \sum_{{\bf s}'}W_{{\bf s}'}x_{{\bf s}'}(t)~,
\label{evolution_x_t+1_Q}\vspace{-2pt}
\end{equation}
where we have introduced the  normalized population variables\vspace{-2pt}
\begin{equation}
x_{\bf s}(t)\ =\ {X_{\bf s}(t)\over M}
,\qquad\quad \sum_{\bf s}x_{\bf s}(t)= 1~.
\label{evolution_x_s}\vspace{-2pt}
\end{equation}
The evolution dynamics Eq.~(\ref{evolution_x_t+1_Q}) retains the
overall size $\sum_{\bf s}X_{\bf s}(t)$ of the population, due to
the normalization of the mutation matrix $Q_\mu({\bf s}'\to{\bf
s})$, Eq.~(\ref{evolution_Q_matrix}).

\runinhead{{The} Hamming Distance}
\index{Hamming distance}
\index{distance!Hamming}
The Hamming distance\vspace{-2pt}
\begin{equation}
d_{\rm H}({\bf s},{\bf s}')\ =\ \sum_{i=1}^N {(s_i-s'_i)^2\over 4}
\ =\ {N\over 2}-{1\over2} \sum_{i=1}^N s_i s'_i
\label{evolution1_hamming_distance}\vspace{-2pt}
\end{equation}
measures the number of units that are
different in two genome configurations
$\bf s$ and $\bf s'$, e.g.\ before and after
the effect of a mutation event.

\runinhead{The Mutation Matrix for Point Mutations}
\index{point mutation}\index{mutation!point}We consider the
{}simplest mutation pattern, viz the case of fixed
genome length $N$ and random transcription errors afflicting only
individual loci. For this case, namely point mutations, the overall
mutation probability
\begin{equation}
Q_\mu({\bf s}'\to {\bf s})\ =\ \mu^{d_{\rm H}}(1-\mu)^{N-d_{\rm H}}
\label{evolution_eq_point}
\end{equation}
is the product of the independent mutation probabilities for all
loci $i=1,\ldots ,N$, with $d_H$ denoting the Hamming distance
$d_{\rm H}({\bf s},{\bf s}')$
given by Eq.~(\ref{evolution1_hamming_distance}) and
$\mu$ the mutation rate $\mu$ defined in Eq.~(\ref{evolution_trans_1}).
One has
$$
\sum_{\bf s}Q_\mu({\bf s}'\to{\bf s}) \ =\
\sum_{d_H} {N \choose d_H} (1-\mu)^{N-d_N} \mu^{d_N} \ =\
(1-\mu+\mu)^N \ \equiv\ 1
$$
and the mutation matrix defined by Eq.~(\ref{evolution_eq_point})
is consequently normalized. We rewrite the mutation matrix as
\begin{equation}
Q_\mu({\bf s}'\to {\bf s})\ =\
\ \propto\  \exp\Big([\log(\mu)-\log(1-\mu)]d_H\Big)
\ \propto\  \exp\left(\beta\sum_i s_is_i'\right)~,
\label{evolution_eq_point_exp}
\end{equation}
where we denoted by $\beta$ an effective inverse temperature,
defined by
\index{temperature!inverse of selection}
\begin{equation}
\beta\ =\ {1\over 2}\log\left(1-\mu\over \mu\right)~.
\label{evolution_beta}
\end{equation}
The relation of the evolution equation
(\ref{evolution_eq_point_exp}) to the partition function of a
thermodynamical system, hinted at by the terminology
\qut{inverse temperature} will become evident
below.

\runinhead{Evolution Equations for Point Mutations}
\index{evolution equations!point mutations}Using the exponential
representation $W({\bf s})=\exp[kF({\bf s})]$, see
Eq.~(\ref{evolution_fit_1}), of the fitness $W({\bf s})$ and
Eq.~(\ref{evolution_eq_point_exp}) for the mutation matrix, we can write
the evolution Eq.~(\ref{evolution_x_s})~via
\begin{equation}
x_{\bf s}(t+1)\ =\ {1\over \langle W\rangle_t}
\sum_{{\bf s}'} x_{{\bf s}'}(t)
\exp\left(\beta\sum_is_i s'_i+kF({\bf s}')\right)
\label{evolution_map}
\end{equation}
in a form that is suggestive of a statistical mechanics
analogy.

\runinhead{Evolution Equations in Linear Form}
\index{evolution equations!linear}The evolution
Eq.~(\ref{evolution_map}) is non-linear in the dynamical variables
$x_{\bf s}(t)$, due to the normalization factor $1/\langle
W\rangle_t$. A suitable change of variables does, however, allow
{the
evolution equation to be cast into a linear form}.

{For this purpose we
introduce}  the unnormalized variables $y_{\bf s}(t)$ via
\begin{equation}
x_{\bf s}(t)\ =\
\frac{y_{\bf s}(t)}{\sum_{{\bf s}'}y_{{\bf s}'}(t)},
\qquad \quad
\langle W\rangle_t=\sum_{\bf s}W({\bf s})x_{\bf s}(t)
= {\sum_{\bf s}W({\bf s})y_{\bf s}(t) \over
\sum_{{\bf s}'}y_{{\bf s}'}(t)}~.
\label{evolution_1_def_y}
\end{equation}
Note that {} $y_{\bf s}(t)$ are determined by
Eq.~(\ref{evolution_1_def_y}) implicitly and that the normalization
$\sum_{{\bf s}'}y_{{\bf s}'}(t)$ can be chosen freely for every
generation $t=1,2,3,\ldots  $. The evolution
Eq.~(\ref{evolution_map}) then becomes
\begin{equation}
y_{\bf s}(t+1)\ =\ Z_t\,\sum_{{\bf s}'} y_{{\bf s}'}(t)
\exp\left(\beta\sum_is_is'_i+kF({\bf s}')\right)~,
\label{evolution_map2}
\end{equation}
where
$$
Z_t \ =\ {\sum_{{\bf s}'}y_{{\bf s}'}(t+1) \over
\sum_{\bf s}W({\bf s})y_{\bf s}(t) }~.
$$
Choosing a different normalization for  $y_{{\bf
s}}(t)$ and for  $y_{{\bf s}}(t+1)$ we may achieve
$Z_t\equiv1$. Equation (\ref{evolution_map2}) is then linear in
{} $y_{\bf s}(t)$.

\runinhead{Statistical Mechanics of the Ising Model}
In the following we will make use of analogies to notations commonly
used in statistical mechanics. Readers who are unfamiliar with the
mathematics of the one-dimensional Ising model may skip the
mathematical details and concentrate on the interpretation of the
results.

We write the linear evolution Eq.~(\ref{evolution_map2}) as
\begin{equation}
y_{\bf s}(t+1)\ =\ \sum_{{\bf s}'}
e^{\beta H[{\bf s},{\bf s}']}\, y_{{\bf s}'}(t),\qquad\quad
y_{{\bf s}(t+1)}\ =\ \sum_{{\bf s}(t)}
e^{\beta H[{\bf s}(t+1),{\bf s}(t)]}\, y_{{\bf s}(t)} ~,
\label{evolution_map3}
\end{equation}
where we denote by $H[{\bf s},{\bf s}']$
an effective Hamiltonian\footnote{The energy of a
state depends in classical mechanics on the values of
the available degrees of freedom, like
the position and the velocity of a particle.
This function is denoted Hamiltonian. In
Eq.~(\ref{evolution_eq_H_s_s_prime}) the Hamiltonian
is a function of the binary variables ${\bf s}$ and ${\bf s}'$.}
\begin{equation}
\beta H[{\bf s},{\bf s}']\ =\
\beta\sum_{i}s_i s_i'\,+\,k F({\bf s}')~,
\label{evolution_eq_H_s_s_prime}
\end{equation}
and where we renamed the variables ${\bf s}$ by ${\bf s}(t+1)$ and
${\bf s}'$ by ${\bf s}(t)$. Equation (\ref{evolution_map3}) can be
solved iteratively,
\begin{equation}
y_{{\bf s}(t+1)}\ =\ \sum_{{\bf s}(t),\dots,{\bf s}(0)}
e^{\beta H[{\bf s}(t+1),{\bf s}(t)]}\,\cdot\cdot\cdot\,
e^{\beta H[{\bf s}(1),{\bf s}(0)]}\,
y_{{\bf s}(0)}~,
\label{evolution_map4}
\end{equation}
with the two-dimensional Ising-type Hamiltonian\footnote{Any
system of binary variables is equivalent to a system of interacting
Ising spins, which retains only the classical contribution to the
energy of interacting quantum mechanical spins (the magnetic
moments).}
\index{Ising model!deterministic evolution}
\begin{equation}
\beta H\ =\ \beta\sum_{i,t}s_i(t+1) s_i(t)\,+\,k\sum_t F({\bf s}(t))~.
\label{evolution_ising_hamil}
\end{equation}

\runinhead{A Short Detour: The Bra-ket Notation}
\index{bra-ket notation}
The evolution equation (\ref{evolution_map4}) can be carried
out in a straight-forward manner. For readers interested in
the cross-correlations to the quantum mechanics of transfer matrices
we make here a small detour into the Bra-ket notation,
which may otherwise be skipped.

One denotes with the \qut{bra} $\langle y|$ and
with the \qut{ket} $|y\rangle$
the respective row and column vectors
$$
\langle y|\ \hat{=}\ (y_1^*,y_2^*,\dots,y_{2^N}^*), \qquad
|y\rangle \ \hat{=}\
\left(
\begin{array}{c}
y_1\\ \vdots \\ y_{2^N}
\end{array}
\right)~,
\qquad y_j\ \hat{=}\ y_{\bf s}
$$
of a vector ${\bf y}$, where $y_j^*$ is the conjugate
complex of $y_j$. Our variables are, however, all real
and $y_j^*\equiv y_j$. The scalar product ${\bf x}\cdot{\bf y}$ of
two vectors is then
$$
{\bf x}\cdot{\bf y} \ \equiv\  \sum_j\, x_j^*\, y_j \ =\
\langle x|y\rangle~.
$$
The expectation value $\langle A\rangle_y$ is given in bra-ket
notation as
$$
\langle A\rangle_y \ =\ \sum_{i,j} \,y_i^*\,A_{ij}\, y_j \ =\ \langle y|A|y\rangle~,
$$
where $A_{ij}$ are the elements of the matrix $A$.
In this notation we may rewrite the evolution equation
(\ref{evolution_map4}) as
\begin{equation}
y_{{\bf s}(t+1)}\ =\ \langle {\bf s}(t+1)|e^{\beta H}|y(0)\rangle~,
\label{evolution1_eq_y_y_bra_ket}
\end{equation}
with $y_{\bf s}(t)=\langle{\bf s}|y(t)\rangle$.
We are interested in the asymptotic limit $t\to\infty$ of the
population state $|y(t)>$.

\vspace*{-6pt}

\subsection{Beanbag Genetics -- Evolutions Without Epistasis}
\label{evolution1_beanbag}\index{beanbag genetics|textbf}
\index{evolution!without epistasis}

\runinhead{{The} Fujiyama Landscape}
\index{Fujiyama landscape}
\index{fitness landscape!Fujiyama}
The fitness function
\begin{equation}
F({\bf s})\ =\ \sum_{i=1}^{N}\,h_is_i,
\qquad\quad
W({\bf s})\ =\ \prod_{i=1}^{N}\,e^{kh_is_i}~,
\label{evolution_fuji_F}
\end{equation}
is denoted the \qut{Fujiyama landscape} since it
corresponds to a single smooth peak as illustrated in
Fig.~\ref{evolution1_fitnessLandscapes}. To see why, we consider the
case $h_i>0$ and rewrite Eq.~(\ref{evolution_fuji_F})~as
$$
F({\bf s}) \ =\ {\bf s}_0\cdot{\bf s}, \qquad\quad {\bf s}_0\ =\
(h_1,h_2,\ldots ,h_{N})~.
$$
The fitness of a given genome ${\bf s}$ is directly proportional
to the scalar product with the master sequence ${\bf s}_0$,
with a well defined gradient pointing towards the master sequence.

\runinhead{{The} Fujiyama Hamiltonian}
No epistatic interactions are present in the smooth peak landscape
Eq.~(\ref{evolution_fuji_F}). In terms of the corresponding
Hamiltonian, see\break Eq.~(\ref{evolution_ising_hamil}), this fact
expresses itself as
\begin{equation}
\beta H\ =\ \beta \sum_{i=1}^N H_i, \qquad\quad
H_i \ =\
\sum_t s_i(t+1) s_i(t)\,+\,{kh_i\over\beta}\sum_t s_i(t)~.
\label{evolution_fuji_hamil}
\end{equation}
Every locus $i$ corresponds exactly to {the}
one-dimensional $t=1,2,\dots $ Ising-model $\beta H_i$ in an
effective uniform magnetic field $kh_i/\beta$.

\runinhead{{The} Transfer Matrix}
\index{matrix!transfer}\index{Ising model!transfer matrix}The
Hamiltonian {Eq.~}(\ref{evolution_fuji_hamil}) does not
contain interactions between different loci of the genome; we can
just consider a single Hamiltonian $H_i$ and find for the iterative
solution {Eq.~}(\ref{evolution_map4})
\begin{equation}
\langle y_i(t+1)|e^{\beta H_i}|y_i(0)\rangle\ =\
\langle y_i(t+1)|\left(\prod_{t'=0}^tT_{t'}\right)|y_i(0)\rangle~,
\label{evolution_fuji_prod}
\end{equation}
with the $2\times2$ transfer matrix
$T_t=e^{\beta H_i[s_i(t+1),s_i(t)]}$
given by
\begin{equation}
\left(T_t\right)_{s,s'} \ =\ <s|T_t|s'>,
\qquad\quad
T_t \ =\ \left(
\begin{array}{cc}
e^{\beta+kh_i} & e^{-\beta} \\
e^{-\beta} & e^{\beta-kh_i}
\end{array}
\right)~,
\label{evolution_fuji_trans_mat}
\end{equation}
where we have used $s,s'=\pm1$ and the
symmetrized form
$$
\beta H_i\ =\
\beta \sum_t s_i(t+1) s_i(t)\,+\,{kh_i\over 2}\sum_t
\Big[s_i(t+1)+s_i(t)\Big]~.
$$
of the one-dimensional Ising model.

\runinhead{Eigenvalues of the Transfer Matrix}
\index{transfer matrix!1D Ising model}
We consider
$$
h_i \equiv 1
$$
and evaluate the eigenvalues $\omega$ of $T_t$:
$$
\omega^2\, -\, 2\omega\,e^\beta\,\cosh(k)
\,+\, e^{2\beta}\, -\, e^{-2\beta}\ =\ 0~.
$$
The solutions are
$$
\omega_{1,2}\ =\
e^\beta\,\cosh(k)\, \pm\,
\sqrt{e^{2\beta}\cosh^2(k)-e^{2\beta}+e^{-2\beta}}~.
$$

\noindent The larger eigenvalue $\omega_1$ thus has the form
\begin{equation}
\omega_{1}\ =\
e^\beta\,\cosh(k)\, +\,
 \sqrt{e^{2\beta}\sinh^2(k)+e^{-2\beta}} ~.
\label{evolution_fuji_omega}
\end{equation}
\runinhead{Eigenvectors of {the} Transfer Matrix}
\index{transfer matrix!1D Ising model}
For $\omega_1>\omega_2$ the
eigenvector $|\omega_{1}\rangle$
corresponding to the larger eigenvalue $\omega_1$
dominates in the $t\to\infty$ limit and its
components determine the genome distribution.
It is determined by
$$
\left(\begin{array}{c}
\langle+|\omega_1\rangle\\
\langle-|\omega_1\rangle
\end{array} \right)
\ =\
\left(\begin{array}{c}
A_+\\ A_-
\end{array} \right),
\qquad\quad
\left(e^{\beta+k}-\omega_1\right)A_+\,+\,e^{-\beta}A_- = 0~,
$$
where
$$
\omega_1\,-\, e^{\beta+k} \ =\
\sqrt{e^{2\beta}\sinh^2(k)+e^{-2\beta}}
\,-\,
e^\beta\,\sinh(k)~.
$$
This yields
\begin{equation}
\left(\begin{array}{c}
A_+\\ A_-
\end{array} \right) \ =\
{1\over \sqrt N_\omega}
\left(\begin{array}{c}
e^{-\beta}
\\
\sqrt{e^{2\beta}\sinh^2(k)+e^{-2\beta}}
-e^\beta\sinh(k)
\end{array} \right),
\label{evolution_eigen_A}
\end{equation}
with the normalization
\begin{eqnarray*}
N_\omega& =& A_+^2+A_-^2 \ =\
e^{-2\beta} + e^{2\beta}\sinh^2(k)
\\ & &
+\left({e^{2\beta}\sinh^2(k)+e^{-2\beta}} \right)
+2e^{\beta}\sinh(k)
\sqrt{e^{2\beta}\sinh^2(k)+e^{-2\beta}}
\\ &=&
2e^{-2\beta}+ e^{2\beta}\sinh^2(k)
-2e^{\beta}\sinh(k) \sqrt{e^{2\beta}\sinh^2(k)+e^{-2\beta}}~.
\end{eqnarray*}

\runinhead{{The} Order Parameter} \index{order
parameter!Fujiyama landscape}\index{Fujiyama
landscape!adaptation}\index{adaptation!Fujiyama landscape}The
one-dimensional Ising model does not have phase transitions. Thus we
reach the conclusion that evolution in the Fujiyama landscape takes
place in a single phase, where there is always some degree of
adaptation. One can evaluate the amount of adaptation by introducing
the order parameter\footnote{The concept of order parameters in the
theory of phase transition is discussed in Chap.~\ref{chap_automata1}.}
\begin{equation}
m \ =\ \lim_{t\to\infty}\langle s(t)\rangle
  \ =\  A_+\,-\,A_- ~,
\label{evolution_ising_order_par_def}
\end{equation}
which corresponds to the uniform magnetization in the
Ising model analogy. One obtains
\begin{equation}
m \ =\ {1\over N_\omega}\Big[
e^{-\beta}
-\sqrt{e^{2\beta}\sinh^2(k)+e^{-2\beta}}
+e^{\beta}\sinh(k)
\Big]~.
\label{evolution_ising_order_par_1}
\end{equation}
In order to interpret this result for the amount $m$ of adaptation
in the smooth Fujiyama landscape we recall that (see
Eqs.~(\ref{evolution_beta}) and (\ref{evolution_fit_1}))
$$
\beta\ =\ {1\over 2}\log\left(1-\mu\over \mu\right),
\qquad\quad
W({\bf s})\ =\ e^{k F({\bf s})}~,
$$
where $\mu$ is the mutation rate for point mutations. Thus we see
that there is some degree of adaptation whenever the fitness
landscape does not vanish ($k>0$). Note that
$\mu\to1/2$, $\beta\to0$ corresponds to a diverging temperature
in the Ising model analogy (\ref{evolution_fuji_hamil}),
but with an diverging effective magnetic field $k h_i/\beta$.

\subsection{Epistatic Interactions and the Error Catastrophe}
\label{evolution1_epistatic}
\index{error catastrophe|textbf}
\index{epistatic interactions|textbf}

The result of the previous Sect. \ref{evolution1_beanbag},
{i.e.} the occurrence of adaptation in a smooth fitness
landscape for any non-trivial model parameter, is due to the absence
of epistatic interactions in the smooth fitness landscape. Epistatic
interactions introduce a phase transition to a non-adapting regime
once the mutation rate becomes too high.

\runinhead{The Sharp Peak Landscape} \index{fitness landscape!sharp
peak}\index{sharp peak landscape} One possibility to study this
phenomenon is the limiting case of very strong epistatic
interactions; in this case, a single element of the
genotype does not give any information on the value of the fitness.
This fitness is defined by the equation
\begin{equation}
W({\bf s})\ =\ \cases{1 &if \ ${\bf s}={\bf s}_0$ \cr
1-\sigma &otherwise }~.
\label{evolution_tower}
\end{equation}
It is also denoted a fitness landscape with a \qut{tower}. In this
case, all genome sequences have the same fitness, which is lower
than the one of the master sequence ${\bf s}_0$. The corresponding
landscape $F({\bf s})$, defined by $W({\bf s})=e^{kF({\bf s})}$ is
then equally \nobreak discontinuous. This landscape has no gradient
pointing towards the master sequence of maximal fitness.

\runinhead{Relative Notation}
We define by $x_k$ the fraction of the
population whose genotype
has a Hamming distance $k$
from the preferred genotype,
\begin{equation}
x_k(t)\ =\
\frac{1}{M}\sum_{\bf s}\delta_{d_{\rm H}({\bf s},{\bf s}_0),k}
\, X_{\bf s}(t)~.
\end{equation}
The evolution equations can be formulated entirely in terms of these
$x_k${;} they correspond to the fraction of the
population being $k$ point mutations away from the master sequence.

\enlargethispage{14pt}

\runinhead{Infinite Genome Limit}
We take the $N\to\infty$ limit and scale the mutation rate, see
Eq.~(\ref{evolution_trans_1}),
\begin{equation}
\mu\ =\ u/ N~,
\label{evolution1_mu_u_N}
\end{equation}
for point mutations
such that the average number of mutations
$$
u\ =\ N\mu
\vspace*{-3pt}
$$
occurring at every step remains finite.


\runinhead{{The} Absence of Back Mutations}
We consider starting from the optimal genome ${\bf s}_0$ and
consider the effect of mutations. Any successful mutation increases
the distance $k$ from the optimal genome ${\bf s}_0$. Assuming $u\ll
1$ in Eq.~(\ref{evolution1_mu_u_N}) implies that
\begin{itemize}
\item[--] multiple mutations do  not appear, and that
\item[--] one can neglect back mutations that reduce the
      value of $k$, since they have a relative probability proportional to
$$
{k\over N-k}\ \ll\  1~.
$$
\end{itemize}
\vspace*{-5pt}
\runinhead{{The} Linear Chain Model}
\index{sharp peak landscape!linear chain model}The model so
defined {consequently has
the structure} of a linear chain. $k=0$ being the starting point of
the chain.

\begin{figure}[t]
\centering
\includegraphics{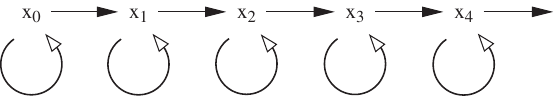}
\caption{{The} linear chain model for
the tower landscape, Eq.~(\ref{evolution_tower}), with $k$ denoting
the number of point mutations necessary to reach the optimal genome.
The population fraction $x_{k+1}(t+1)$ is only influenced by the
value of $x_k$ and its own value at time $t$}
\label{evolution1_linear_chain}
\vspace*{-6pt}
\end{figure}

We have two parameters: $u$, which measures the mutation rate and
$\sigma$, {which} measures the strength of the
selection. Remembering that the fitness $W({\bf s})$ is proportional
to the number of {offspring}, see
Eq.~(\ref{evolution_tower}), we then find
\begin{eqnarray}
\label{evolution1_linear_chain_model_0}
x_0(t+1)&=& {1\over\langle W\rangle}\Big[x_0(t)\left(1-u\right)\Big]~, \\
\label{evolution1_linear_chain_model_1}
x_1(t+1)&=& {1\over\langle W\rangle}\Big[
u x_0(t)+\left(1-u\right)\left(1-\sigma\right)x_1(t) \Big]~;\\
\label{evolution1_linear_chain_model_k} x_k(t+1)&=& {1\over\langle
W\rangle} \Big[u x_{k-1}(t)+(1-u)x_k(t)\Big](1-\sigma)~, \qquad k>1,
\end{eqnarray}
where $\langle W\rangle$ is the average fitness. These equations
describe a linear chain model as illustrated in
Fig.~\ref{evolution1_linear_chain}. The population of individuals
with the optimal genome $x_0$ constantly loses members due to mutations.
But it also has a higher number of offspring than all
other populations due to its larger fitness.

\runinhead{Stationary Solution}
\index{sharp peak landscape!stationary solution}\index{stationary
solution!sharp peak landscape} The average fitness of the population
is given by
\begin{equation}
\langle W\rangle\ =\ x_0\,+\, (1-\sigma)(1-x_0)
\ =\ 1-\sigma(1-x_0)~.
\label{evolution1_w_bar}
\end{equation}

We look for the stationary distribution $\{x_k^*\}$.
The equation for $x_0^*$
does not involve the $x_k^*$ with $k>0$:
$$
x_0^*\ =\ {x_0^*(1-u)\over 1-\sigma(1-x_0^*)},
\quad\qquad
1-\sigma(1-x_0^*)\ =\ 1-u~.
$$
The solution is
\begin{equation}
x_0^*\ =\
\cases{1-u/\sigma  &if $u<\sigma$\cr
0  &if $u\geq\sigma$}~,
\label{evolution_sol_x_0_star}
\end{equation}
due to the normalization condition $x_0^*\le1$.
For $u>\sigma$ the model becomes ill defined.
The stationary solutions for the $x_k^*$ are for $k=1$
$$
x_1^*\ =\ {u\over 1-\sigma(1-x_0^*)-(1-u)(1-\sigma)}\, x_0^*~,
$$
which follows directly from Eqs.\ (\ref{evolution1_linear_chain_model_1})
and (\ref{evolution1_w_bar}), and for $k>1$
\begin{equation}
x_k^*\ =\ {(1-\sigma)u\over 1-\sigma(1-x_0^*)-(1-u)(1-\sigma)}\, x_{k-1}^*~,
\label{evolution_sol_x_gen_star}
\vspace*{5pt}
\end{equation}
which follows from Eqs.\ (\ref{evolution1_linear_chain_model_k}) and
(\ref{evolution1_w_bar}).

\begin{figure}[t]
\centering
\includegraphics{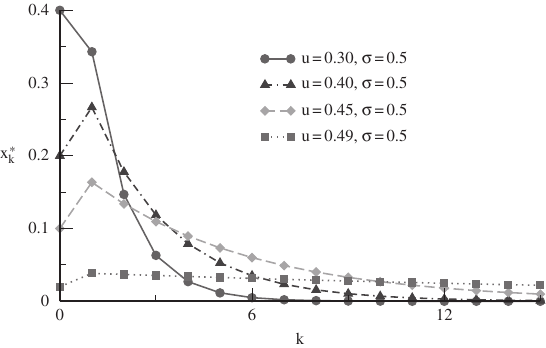}
\caption{Quasispecies formation within the sharp peak fitness
landscape, Eq.~ (\ref{evolution_tower}). The stationary population
densities $x^*_k$, see Eq.~(\ref{evolution_sol_x_gen_star}), are
peaked around the genome with maximal fitness, $k=0$. The population
tends to spread out in genome space when the overall mutation rate
$u$ approaches the critical point $u\to\sigma$}
\label{evolution1_sharpPeak_QS}
\end{figure}

\runinhead{Phase Transition and {the} Order Parameter}
\index{phase transition!sharp peak landscape} We can thus
distinguish two regimes determined by the magnitude of the mutation
rate $\mu=u/N$ relative to the fitness parameter $\sigma$, with
$$
u\ =\ \sigma
$$
being the transition point. In physics language the epistatic
interaction corresponds to many-body interactions and the occurrence
of a phase transition in the sharp peak model is due to the
many-body interactions which were absent in the smooth fitness
landscape model considered in
Sect.~\ref{evolution1_beanbag}.

\runinhead{The Adaptive Regime and Quasispecies}
\index{evolution!adaptive regime}
\index{adaptive regime}
\index{regime!adaptive}
\index{regime!adaptive}
In the regime of small mutation rates, $u<\sigma$,
one has $x^*_0>0$ and in fact
the whole population lies a finite distance away
from the preferred genotype. To see why, we note
that
$$
\sigma(1-x_0^*)\ =\ \sigma(1-1+u/\sigma)\ =\ u
$$
and take a look at Eq.~(\ref{evolution_sol_x_gen_star}):
$$
{(1-\sigma)u\over 1-u-(1-u)(1-\sigma)} \ =\
\left({1-\sigma\over 1-u}\right)
\left({u\over \sigma}\right) \ \le\ 1,
\qquad\quad \mbox{for}\ \ u\,<\,\sigma~.
$$
The $x_k^*$ therefore form a geometric series,
$$
x_k^*\ \sim\ \left({1-\sigma\over 1-u} {u\over \sigma}\right)^k~,
$$
which is summable when $u<\sigma$. In this adaptive regime the
population forms what Manfred Eigen denoted a \qut{quasispecies}, see Fig.~\ref{evolution1_sharpPeak_QS}.
\begin{quotation}
{\it Quasispecies.\enspace}
\index{evolution!quasispecies}\index{quasispecies}
\index{species!quasispecies}A quasispecies is a population of
genetically close but not identical individuals.
\end{quotation}
\runinhead{{The} Wandering Regime and {The} Error Threshold}
\index{evolution!wandering regime}\index{wandering regime}
\index{regime!wandering}\index{error threshold} In the regime of a
large mutation rate, $u>\sigma$, we have $x_k^*=0$, $\forall k$. In
this case, a closer look at the finite genome situation shows that
the population is distributed in an essentially uniform way over the
whole genotype space. The infinite genome limit {therefore becomes} inconsistent, since the whole
population lies an infinite number of mutations away from the
preferred genotype. In this {\em wandering regime\/} the effects of
finite population size are prominent.

\begin{quotation}
{\it Error Catastrophe.\enspace}
\index{error catastrophe}The transition from the adaptive
(quasispecies) regime to the wandering regime is denoted the
\qut{error threshold} or \qut{error \nobreak catastrophe}.
\end{quotation}
The notion of error catastrophe is a quite generic feature of
quasispecies theory, independent of the exact nature of the fitness
landscape containing epistatic interactions. A quasispecies
{can no longer adapt}, once its
mutation rate becomes too large. In the real world the error
catastrophe implies extinction.

\section{Finite Populations and Stochastic Escape}
\label{evolution_finite populations}

\runinhead{Punctuated Equilibrium}
\index{punctuated equilibrium}Evolution is not a steady process,
there are regimes of rapid increase of the fitness and phases of
relative stasis. This kind of overall dynamical behavior is denoted
{the} \qut{punctuated equilibrium}.

In this context, adaptation can result  either from
local optimization of the fitness of a single species
or via coevolutionary avalanches, as
discussed in Chap.~\ref{chap_automata1}.
\begin{quotation}
{\it {The} Neutral Regime.\enspace}
\index{neutral regime}
\index{regime!neutral}
\index{evolution!neutral regime}
The stage where evolution is essentially driven by random
mutations is called the neutral (or wandering) regime.
\end{quotation}
The quasispecies model is inconsistent in the neutral regime. In
fact, the population spreads out in genome space in the neutral
regime and the infinite population limit is {no longer reachable}. In this situation, the fluctuations
of the reproductive process in a finite population have to be taken
into account.

\vspace{3pt}

\runinhead{Deterministic Versus Stochastic Evolution}
\index{deterministic evolution!vs.\ stochastic evolution}Evolution
is driven by stochastic processes, since mutations are random
events. {Nevertheless, randomness averages out} and the
evolution process becomes deterministic in the thermodynamic limit,
as discussed in Sect.~\ref{evolution_deterministic}, when the number
$M$ of individuals diverges, $M\to\infty$.

Evolutionary processes in populations with a finite number
of individuals differ from deterministic evolution
quantitatively and sometimes also qualitatively, the
later being our focus of interest here.
\begin{quotation}
{\it Stochastic Escape.\enspace}
\index{stochastic escape!evolution}Random mutations in a finite
population might lead to a decrease in the fitness and to a loss of
the local maximum in the fitness landscape with a resulting
dispersion of the quasispecies.
\end{quotation}
\vskip-\lastskip\pagebreak

\noindent We have given a general account of the theory of
stochastic escape in Chap.~\ref{chap_chaos1}. Here we will discuss
in some detail under which circumstances this phenomenon is
important in evolutionary processes of small populations.

\subsection{Strong Selective Pressure and~Adaptive~Climbing}
\label{evolution_subsec_adaptive_climbing}
\index{dynamics!adaptive climbing|textbf}

\runinhead{Adaptive Walks}
\index{adaptive walk}\index{walk!adaptive}
We consider a  coarse-grained description of population dynamics
for finite populations. We assume that
\begin{description}
\item[(a)] {the population is finite,}
\item[(b)] {the selective pressure is very strong, {and}}
\item[(c)] {the mutation rate is small.}
\end{description}
It follows from (b) that one can represent the population by a
single point in genome space{;} the genomes of all
individuals are taken to be equal. The evolutionary dynamics is then
{the following}:

\begin{description}
\item[(A)] {At each time step, only one genome element of
some individual in the population mutates.}

\item[(B)] {If, because of this mutation,
one obtains a genotype with higher fitness, the new genotype spreads
rapidly throughout the entire population, which then moves
{altogether} to the new position in genome
space.}

\item[(C)] {If the
fitness of the new genotype is lower, the mutation is
rejected and the population remains at the
old position.}
\end{description}

Physicists would call this type of dynamics a Monte Carlo process at
zero temperature. As {} is well known, this algorithm
does not lead to a global optimum, but to a ``typical" local
optimum. {Step} {\bf (C)} holds only for the
infinite population limit. We will relax this condition further
below.

\begin{figure}[t]
\centering
\includegraphics{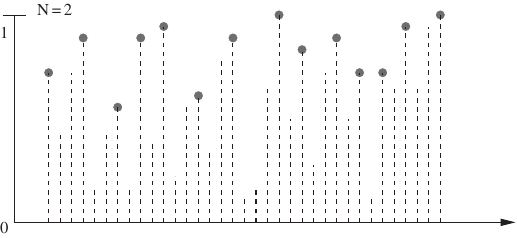}
\caption{{Local} fitness optima in a
one-dimensional random fitness
{distribution;} the number of neighbors is
two. This simplified picture does not corresponds directly to the
$N=2$ random energy model, for which there are just $2^2=4$ states
in genome space. It shows, however, that random distributions may
exhibit an enormous number of local optima (\textit{filled
circles}), which are characterized by lower fitness values both on
the left-hand side as well as on the right-hand side}
\label{evolution1_random_optima}
\end{figure}

\runinhead{The Random Energy Model}
\index{model!random energy}
\index{evolution!random energy model}
It is thus important to investigate the statistical
properties of the local optima, which depend
on the properties of the fitness landscape. A suitable
approach is to assume a random distribution of the fitness.
\begin{quotation}
{\it The Random Energy Model.\enspace}
The fitness landscape $F({\bf s})$ is uniformly distributed
between 0 and 1.
\end{quotation}
The random energy model is illustrated in
Fig.~\ref{evolution1_random_optima}. It captures, as we
will see further below two ingredients expected for
real-world fitness landscapes, namely a large number
of local fitness optima close to the global fitness maximum.

\runinhead{Local Optima in the Random Energy Model}
\index{local optima}
Let us denote by $N$ the number of genome elements.
The probability that a point with fitness $F({\bf s})$
is a local optimum is simply given by
$$
F^N \ =\  F^N({\bf s})~,
$$

\pagebreak

\noindent  since we have to impose that the $N$ nearest neighbors
$$
(s_1,\ldots,-s_i,\ldots,s_N), \qquad (i=1,\ldots,N),\qquad {\bf s}\
=\ (s_1,\ldots,s_N)~,
$$
of the point have fitness less than $F$. The probability
that a point in genome space is a local optimum is given by
\begin{equation}
{\rm P}\left\{\hbox{local optimum}\right\}\ =\ \int_0^1F^N
\mathrm{d}F\ =\ {1\over N+1}~,
\end{equation}
since the fitness $F$ is equally distributed in $[0,1]$. There are
therefore many local optima, namely $2^N/(N+1)$. A schematic picture
of the large number of local optima in a random distribution is
given {in} Fig.~\ref{evolution1_random_optima}.

\runinhead{Average Fitness at a Local Optimum}
\index{fitness!average}
The typical fitness of a local optimum is
\begin{equation}
F_{typ} \ =\ {1\over 1/(N+1)}\int_0^1F\,F^N \mathrm{d}F\ =\
{N+1\over N+2} \ =\ {1+1/N\over 1+2/N} \ \approx\ 1-1/N ~,
\label{evolution_F_typ}
\end{equation}
viz very close the global optimum of $1$, when the genome length $N$
is large. At every successful step the distance from the top is
divided, on average, by a factor {of} 2.

\runinhead{Successful Mutations} \index{mutation!adaptive climbing}
We now consider the adaptation process. Any mutation results in a
randomly distributed fitness of the offspring. A mutation is
successful whenever the fitness of the offspring is bigger than the
fitness of its parent. The typical fitness attained after $\ell$
successful steps is then of the order of
$$
1- {1\over 2^{\ell+1}}~,
$$
when starting ($l=0$) from an average initial fitness of 1/2. It
follows that the typical number of successful mutations after which
an optimum is attained {is}
\begin{equation}
F_{typ} \ =\
1-1/N\ =\ 1- {1\over 2^{\ell_{typ}+1}},
\qquad\quad
\ell_{typ}+1\ =\ {\log N\over\log 2}~,
\label{evolution_mut_successful}
\end{equation}
{i.e.} it is relatively small.

\begin{figure}[t]
\centering
\includegraphics{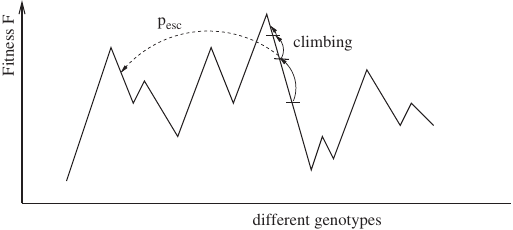}
\caption{{Climbing}
process and stochastic escape. The higher the fitness,
         the more difficult it becomes to climb further.
         With an escape probability $p_{\mathrm{esc}}$ the population
         jumps somewhere else and escapes a local optimum}
\label{evolution1_fitness_climbing}
\end{figure}

\runinhead{{The Time Needed} for One Successful Mutation}
\index{time!successful mutation}\index{mutation!time scale} Even
though the number of successful mutations
{Eq.~}(\ref{evolution_mut_successful}) needed to arrive
at the local optimum is small, the time to climb to the local peak
can be very long{;} see
Fig.~\ref{evolution1_fitness_climbing} for an illustration of the
climbing process.

We define by
$$
t_F \ =\ \sum_{n} \, n\, P_n,
\qquad n:\ \mbox{number of generations}
$$
the average number of generations {necessary} for
the population with fitness $F$ to achieve one successful mutation,
with $P_n$ being the probability that it takes exactly $n$
generations. We obtain:
\begin{eqnarray}
t_F& =& 1\,(1-F)\,+\, 2\,(1-F)F
      \,+\, 3\,(1-F)F^2 \,+\, 4\,(1-F)F^3 \,+\, \cdots
\nonumber\\[3pt]
& =& {1-F\over F}\sum_{n=0}^\infty n\, F^n
\ =\ {1-F\over F}\left(F{\partial\over\partial F}\sum_{n=0}^\infty F^n\right)
\ =\ (1-F){\partial\over\partial F}{1\over 1-F}
\nonumber\\[3pt]
& = & {1\over 1-F}~.
\label{evolution_t_F}
\end{eqnarray}
The average number of generations {necessary} to
further increase the fitness by a successful mutation diverges close
to the global optimum $F\to1$.

\vspace*{5pt} \runinhead{{The} Total Climbing Time}
\index{time!adaptive climbing}\index{adaptation!time scale}Every
successful mutation decreases the distance $1-F$ to the top by 1/2
and therefore increases the factor $1/(1-F)$ on the average by
{2}. The typical number $\ell_{\mathrm{typ}}$, see
Eq.~(\ref{evolution_mut_successful}), of successful mutations needed
to arrive at a local optimum determines, via
Eq.~(\ref{evolution_t_F}), the expected total number of generations
$T_{\mathrm{opt}}$ to arrive at the local optimum. It is therefore
on the average
\begin{eqnarray}
T_{\mathrm{opt}}& =& 1\,t_F\,+\, 2\,t_F \,+\, 2^2\,t_F \,+\,\dots
 \,+\, 2^{\ell_{\mathrm{typ}}}t_F
\nonumber\\[3pt]
       &=& t_F\,{1-2^{\ell_{\mathrm{typ}}+1} \over 1-2}
\ \approx\ t_F\, 2^{\ell_{\mathrm{typ}}+1} \ =\ t_F\,
e^{(\ell_{\mathrm{typ}}+1)\log 2}
\nonumber\\[3pt]
&\approx & t_F\,e^{\log N}
\ =\  {N\over 1-F} \ \approx\ 2\,N~,
\label{evolution_t_opt}
\end{eqnarray}
where we have used Eq.~(\ref{evolution_mut_successful}) and
$F\approx 1/2$ for a typical starting fitness.
The time needed to
climb to a local maximum in the random fitness landscape is
therefore proportional to the length of the genome.

\enlargethispage{8pt}

\vspace*{-6pt}

\subsection{Adaptive Climbing Versus Stochastic Escape}
\index{adaptive climbing!vs.\ stochastic escape|textbf}
\index{stochastic escape!vs.\ adaptive climbing|textbf}

In {Sect.~}\ref{evolution_subsec_adaptive_climbing} the average
properties of adaptive climbing have been evaluated. We
{now take} the fluctuations in the reproductive
process into account and compare the typical time scales for a
stochastic escape with {those} for adaptive climbing.

\runinhead{Escape Probability}
\index{stochastic escape!probability} When a favorable mutation
appears it spreads instantaneously into the whole population, under
the condition of strong selection limit, as assumed in our model.

We consider a population situated at a local optimum or very close
to a local optimum. Every point mutation then leads to a lower
fitness and the probability $p_{\mathrm{esc}}$ for stochastic escape
is
$$
p_{\mathrm{esc}}\ \approx\ u^M~,
$$
where $M$ is the number of individuals in the population and
$u\in[0,1]$ the mutation rate per genome, per individual and per
generation, compare Eq.~(\ref{evolution1_mu_u_N}). The escape can
only happen when a mutation occurs in every member of the population
within the same generation (see also
Fig.~\ref{evolution1_fitness_climbing}). If a single individual does
not mutate it retains its higher fitness of the present local
optimum and all other mutations are discarded within the model,
assuming a strong selective pressure.

\runinhead{Stochastic Escape and Stasis}
{We now consider a population climbing towards a
local optimum}. The probability that the fitness of a given
individual increases is $(1-F)u${. It} needs to
mutate with a probability $u$ and to achieve a higher fitness, when
mutating, with probability $1-F$. We denote {by}
$$
a\ =\ 1-(1-F)u
$$
the probability that the fitness of an individual does not increase
with respect to the current fitness $F$ of the population. The
probability $q_{\mathrm{bet}}$ that at least one better genotype is
found is then given by
$$
q_{\mathrm{bet}}\ =\ 1-a^M~.
$$
Considering a population close to a local optimum,
a situation typical for real-world ecosystems,
we can then distinguish between two evolutionary regimes:
\begin{itemize}
\item[--] {Adaptive Walk}:
   \index{adaptive walk}\index{walk!adaptive}
   The escape probability $p_{\mathrm{esc}}$ is much smaller
   than the probability to increase the fitness,
   $q_{\mathrm{bet}}\gg p_{\mathrm{esc}}$. The population continuously
   increases its fitness via small mutations.
\item[--] {{The} Wandering Regime}:
  \index{wandering regime}\index{regime!wandering}
  Close to a local optimum the adaptive dynamics slows down
  and the probability of stochastic escape $p_{\mathrm{esc}}$
  becomes comparable to that of an adaptive process,
  $p_{\mathrm{esc}}\approx q_{\mathrm{bet}}$. The population wanders around
  in genome space, starting a new adaptive walk after
  every successful escape.
\end{itemize}

\runinhead{Typical Escape Fitness} \index{stochastic escape!typical
fitness}During the adaptive walk regime the fitness $F$ increases
steadily, until it reaches a certain typical fitness
$F_{\mathrm{esc}}$ for which the probability of stochastic escape
becomes substantial, i.e.\ when $p_{\mathrm{esc}}\approx
q_{\mathrm{bet}}$ and\vspace*{3pt}
$$
p_{\mathrm{esc}}\ =\ u^M\ =\
1-\left[1-(1-F_{\mathrm{esc}})u\right]^M \ =\
q_{\mathrm{bet}}\vspace*{3pt}
$$
holds. As $(1-F_{\mathrm{esc}})$ is then small we can expand {the}
above expression in $(1-F_{\mathrm{esc}})$,\vspace*{3pt}
$$
u^M \ \approx\ 1-\left[1-M(1-F_{\mathrm{esc}})u\right] \ =\
M(1-F_{\mathrm{esc}})u~,\vspace*{3pt}
$$

\noindent obtaining\vspace*{3pt}
\begin{equation}
1-F_{\mathrm{esc}} \ =\ {u^{M-1}/M}~.
\label{evolution_F_esc}\vspace*{3pt}
\end{equation}
The fitness $F_{\mathrm{esc}}$ necessary for the stochastic escape
to become relevant is exponentially close to the global optimum
$F=1$ for large populations $M$.\vspace*{3pt}

\runinhead{The Relevance of Stochastic Escape}
\index{stochastic escape!relevance for evolution}The stochastic
escape occurs when a local optimum is reached, or when we are close
to a local optimum. We may estimate the importance of the escape
process relative to that of the adaptive walk by comparing the
typical fitness $F_{\mathrm{typ}}$ of a local optimum achieved by a
typical climbing process with the typical fitness $F_{\mathrm{esc}}$
needed for the escape process to become important:
$$
F_{\mathrm{typ}}\ =\ 1-{1\over N}\ \equiv\ F_{\mathrm{esc}} \ =\
1-{u^{M-1}\over M}, \quad\qquad {1\over N}\ =\ {u^{M-1}\over M}~,
$$
where we have used
{Eq.~}(\ref{evolution_F_typ}) for $F_{\mathrm{typ}}$.
The last expression is now independent of the details of
the fitness landscape, containing only the measurable
parameters $N,\ M$ and $u$. This condition can be fulfilled
only when the number of individuals $M$ is much smaller than
the genome length $N$, as $u<1$. The phenomenon of stochastic
escape occurs only for very small populations.

\enlargethispage{12pt}

\vspace*{2pt}
\section{Prebiotic Evolution}
\label{evolution_prebiotic}
\index{prebiotic evolution|textbf}
\index{evolution!prebiotic|textbf}

\index{life!origin}Prebiotic evolution deals with the question of
the origin of life. Is it possible to define chemical autocatalytic
networks in the primordial soup having properties akin to those of
the metabolistic reaction networks going on
continuously~in~every~living~cell?

\subsection{Quasispecies Theory}
\index{quasispecies|textbf}

\index{Eigen, Manfred!quasispecies theory}The quasispecies theory
was introduced by Manfred Eigen to describe the evolution of a
system of information carrying macromolecules through a set of
equations for chemical kinetics, \index{equation!chemical kinetic}
\index{dynamics!macromolecules}\index{chemical reactions}
\begin{equation}
{\mathrm{d} \over \mathrm{d} t} x_i\ =\ \dot{x}_i\ =\ W_{ii}x_i\,+\,
\sum_{j\ne i} W_{ij}x_j\,-\, x_i\phi(t)~, 
\label{evolution_eigen_x}
\end{equation}
where the $x_i$ denote the concentrations of $i=1\ldots N$
molecules. $W_{ii}$ is the (autocatalytic) self-replication rate and
the off-diagonal terms $W_{i,j}$ ($i\ne j$) the respective mutation
rates.

\runinhead{Mass Conservation}
\index{mass conservation}We can choose the flux $-x\phi(t)$ in
Eigen's equations (\ref{evolution_eigen_x}) for prebiotic evolution
such that the total concentration $C$, viz the total mass
$$
C\ =\ \sum_i x_i
$$
is conserved for long times. Summing Eq.~(\ref{evolution_eigen_x})
over $i$ we obtain
\begin{equation}
\dot C\ =\ \sum_{ij} W_{ij} x_j\,-\, C\,\phi,
\qquad\quad
\phi(t)\ =\ \sum_{ij}W_{ij}x_j(t)~,
\label{evolution_eigen_c}
\end{equation}
for a suitable choice for the field $\phi(t)$, leading to
\begin{equation}
\dot C\ =\ \phi\,(1-C), \qquad\quad {\mathrm{d}\over \mathrm{d}t}
(C-1)\ =\ -\phi\,(C-1)~. \label{evolution_mass_conservation}
\end{equation}
The total concentration $C(t)$ will therefore approach 1 for
$t\to\infty$ for $\phi>0$, which we assume {to be the case here}, implying total mass conservation. In
this case the autocatalytic rates $W_{ii}$ dominate with respect to
the transmolecular mutation rates $W_{ij}$ ($i\ne j$).

\runinhead{Quasispecies}
\index{quasispecies!prebiotic}
We can write the evolution equation
(\ref{evolution_eigen_x}) in matrix form
\begin{equation}
{\mathrm{d}\over \mathrm{d}t}\textbf{\textit{ x}}(t)\ =\ \left( W-
1\phi\right)\textbf{\textit{ x}}(t), \qquad\quad \textbf{\textit{ x}}=\left(
\begin{array}{c}
x_1\\ x_1 \\ \cdot\cdot\cdot \\ x_N
\end{array}
\right)~,
\label{evolution_eigen_vec}
\end{equation}
where  $W$ is the matrix $\{W_{ij}\}$. We
assume here for simplicity a
symmetric mutation matrix $W_{ij}=W_{ji}$.
The solutions of the linear differential equation
(\ref{evolution_eigen_vec}) are then given
in terms of the eigenvectors $\vec{e}_\lambda$
of $W$:
$$
W\textbf{\textit{e}}_\lambda\ =\ \lambda\,\textbf{\textit{e}}_\lambda,
\qquad\quad
\textbf{\textit{ x}}(t)\ =\ \sum_\lambda a_\lambda(t) \textbf{\textit{ e}}_\lambda,
\qquad\quad
\dot a_\lambda\ =\ \left[\lambda-\phi(t)\right]a_\lambda~.
$$
The eigenvector $\textbf{\textit{ e}}_{\lambda_{\max}}$ with the
largest eigenvalue $\lambda_{\max}$ will dominate for $t\to\infty$,
due to the overall mass conservation
{Eq.~}(\ref{evolution_mass_conservation}). The flux will adapt to
the largest eigenvalue,\vspace*{3pt}
$$
\lim_{t\to\infty}\Big(\lambda_{\max}-\phi(t)\Big)\ \to\
0~,\vspace*{3pt}
$$
leading to the stationary condition $\dot x_i=0$ for the evolution
Eq.~(\ref{evolution_eigen_vec}) in the long time limit.

If $W$ is diagonal (no mutations) a single macromolecule will remain
in the primordial soup for $t\to\infty$. For small but finite
mutation rates $W_{ij}$ ($i\ne j$), a quasispecies will emerge, made
up of different but closely related macromolecules.

\runinhead{The Error Catastrophe}
\index{error catastrophe!prebiotic evolution}{The mass conservation equation}
(\ref{evolution_mass_conservation}) cannot be retained when the
mutation rates become too big, viz when the eigenvectors $\vec
e_\lambda$ become extended. In this case the flux $\phi(t)$
diverges, see Eq.~(\ref{evolution_eigen_c}), and the quasispecies
model {consequently becomes}
inconsistent. This is the telltale sign of the error catastrophe.

The quasispecies model {Eq.~}(\ref{evolution_eigen_x}) is
equivalent to the random energy model for microevolution studied in
Sect.~\ref{evolution_finite populations}, with the autocatalytic
rates $W_{ii}$ corresponding to the fitness of the $x_i$, which
corresponds to the states in genome space. The analysis carried
through in Sect.~\ref{evolution1_epistatic} for the occurrence of an
error threshold is therefore also valid for Eigen's prebiotic
evolutionary equations.

\subsection{Hypercycles and Autocatalytic Networks}
\label{evolution1_hypercycles}
\index{hypercycle|textbf}
\index{network!autocatalytic|textbf}

\runinhead{RNA World}
\index{prebiotic evolution!RNA world} \index{RNA world}
The macromolecular evolution equations (\ref{evolution_eigen_x}) do
not contain terms describing the catalysis of molecule $i$ by
molecule $j$. This process is, however, important both for the
prebiotic evolution, as stressed by Manfred Eigen, as well as for
the protein reaction network in living cells. \index{Eigen,
Manfred!hypercycle}

\begin{figure}[t]
\centering\includegraphics{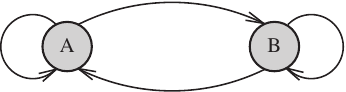}
\caption{The simplest hypercycle. A and B are self-replicating
molecules. A acts as a catalyst for B, i.e.\ the replication rate of
B increases with the concentration of A. Likewise the presence of B
favors the replication of A} \label{evolution1_hyper1}
\vspace*{10pt}
\end{figure}

\begin{figure}[t]
\centering
\includegraphics{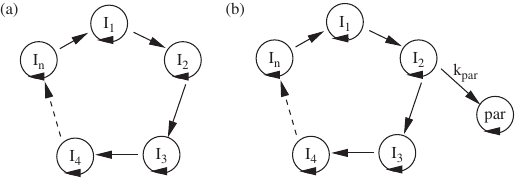}
\caption{Hypercycles of higher order. (\textbf{a}) A hypercycle of
order $n$ consists of $n$ cyclically coupled self-replicating
molecules $I_i$, and each molecule provides catalytic support for
the subsequent molecule in the cycle. (\textbf{b}) A hypercycle with
a single self-replicating parasitic molecule ``par'' coupled to it
via $k_{\mathrm{par}}$. The parasite gets catalytic support from
$I_2$ but does not give back catalytic support to the molecules in
the hypercycle} \label{evolution1_hyper2} \vspace*{4pt}
\end{figure}

\begin{quotation}
{\it Hypercycles.\enspace}
\index{hypercycle}Two or more molecules may form a stable catalytic
(hyper) cycle when the respective intermolecular catalytic rates are
large enough to mutually support their respective synthesis.
\end{quotation}\pagebreak
\noindent An illustration of some hypercycles is given
in Figs.~\ref{evolution1_hyper1} and
\ref{evolution1_hyper2}. The most likely chemical candidate for the
constituent molecules is RNA, functioning both
enzymatically and as a precursor of the genetic material.
One speaks also of an ``RNA world''.

\runinhead{Reaction Networks}
\index{network!reaction}
We disregard mutations in the following and
consider the catalytic reaction equations
\begin{eqnarray}
\dot x_i &=& x_i\left(
\lambda_i+\sum_j \kappa_{ij}x_j-\phi
              \right)
\label{evolution_reaction_network}  \\
\phi &=& \sum_k x_k \left(
\lambda_k+\sum_j \kappa_{kj}x_j
              \right)~,
\label{evolution_reaction_phi}
\end{eqnarray}
\index{growth rate!autocatalytic}where $x_i$ are
the respective concentrations, $\lambda_i$ the
autocatalytic growth rates and  $\kappa_{ij}$ the
transmolecular catalytic rates. The field $\phi$ has been chosen,
Eq.~(\ref{evolution_reaction_phi}), such that the total
concentration $C=\sum_i x_i$ remains constant
$$
\dot C\ =\ \sum_i \dot x_i\ =\
\sum_i x_i \left( \lambda_i+\sum_j \kappa_{ij}x_j
              \right)- C\,\phi \ =\ (1-C)\,\phi\ \to\ 0
$$
\index{self-consistency condition!mass conservation}for $C\to1$.

\begin{figure}[t]
\centering
\includegraphics{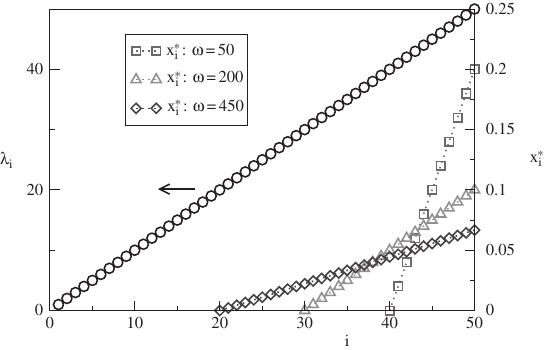}
\caption{The autocatalytic growth rates $\lambda_i$ (\textit{left
axis}), as in Eq.~(\ref{evolution_hom_net}) with $\alpha=1$, and the
stationary solution $x_i^*$ (\textit{right axis}) of the
concentrations, Eq.~(\ref{evolution1_hypocycle_x_*}), constituting a
prebiotic quasispecies, for various mean intercatalytic rates
$\kappa\equiv\omega$. The \textit{horizontal axis} $i=1,2,\ldots,50$ denotes the
respective molecules } \label{evolution1_fig_hypocycle_net}
\end{figure}

\runinhead{The Homogeneous Network}
We consider the case of homogeneous \qut{interactions}
$\kappa_{i\ne j}$
and uniformly distributed autocatalytic growth rates:
\begin{equation}
\kappa_{i\ne j}\ =\ \kappa, \qquad\quad
\kappa_{ii}\ =\ 0, \qquad\quad
\lambda_i\ =\ \alpha\,i~,
\label{evolution_hom_net}
\end{equation}
compare Fig.~\ref{evolution1_fig_hypocycle_net}, leading to
\begin{equation}
\dot x_i \ =\ x_i\left(\lambda_i+\kappa\sum_{j\ne i} x_j-\phi\right)
\ =\ x_i\,\Big(\lambda_i+\kappa-\kappa x_i-\phi\Big)~,
\label{evolution_react_net_hom}
\end{equation}
where we have used $\sum_i x_i=1$. The fixed points $x_i^*$ of
{Eq.~}(\ref{evolution_react_net_hom}) are
\begin{equation}
x_i^*\ =\ \left\{
\begin{array}{c}
(\lambda_i+\kappa-\phi)/\kappa \\ 0
\end{array}
          \right.
\qquad \qquad \lambda_i=\alpha,2\alpha,\ldots,N\alpha~,
\label{evolution_fixed_points}
\end{equation}
where the non-zero solution is valid for $\lambda_i-\kappa-\phi>0$.
The flux $\phi$ in Eq.~(\ref{evolution_fixed_points}) needs to obey
Eq.~(\ref{evolution_reaction_phi}), as {the}
self-consistency condition.

\runinhead{The Stationary Solution}
\index{stationary solution!prebiotic evolution}The case of
homogeneous interactions, Eq.~(\ref{evolution_hom_net}), can be
solved analytically. Dynamically, the $x_i(t)$ with the largest
growth rates $\lambda_i$ will dominate and obtain a non-zero
steady-state concentration $x_i^*$. We may therefore assume that
there {exists} an $N^*\in[1,N]$ such that
\begin{equation}
x_i^*\ =\ \left\{
\begin{array}{ccl}
(\lambda_i+\kappa-\phi)/\kappa
&\quad&  N^* \le i \le N \\
0 &\quad & 1\le i< N^*
\end{array}
          \right. ~,
\label{evolution1_hypocycle_x_*}
\end{equation}
compare Fig.~\ref{evolution1_fig_hypocycle_net}, where $N^*$ and
$\phi$ are determined by the normalization condition
\begin{eqnarray}
1& =& \sum_{i=N^*}^N x_i^*
\ =\ \sum_{i=N^*}^N {\lambda_i+\kappa-\phi\over \kappa}
\ =\ {\alpha\over\kappa} \sum_{i=N^*}^N i
+\left[{\kappa-\phi\over\kappa}\right]\Big(N+1-N^*\Big)
\nonumber \\
& =&
{\alpha\over2\kappa}\,\Big[ N(N+1)-N^*(N^*-1)
                    \Big]
+\left[{\kappa-\phi\over\kappa}\right]\Big(N+1-N^*\Big)
\label{evolution_cond_c_const}
\end{eqnarray}
and by the condition that $x_i^*=0$ for $i=N^*-1$:
\begin{equation}
0 \ =\
{\lambda_{N^*-1}+\kappa-\phi\over \kappa}
\ =\ {\alpha (N^*-1)\over \kappa} + {\kappa-\phi\over\kappa}~.
\label{evolution_cond_N_star_1}
\end{equation}
We eliminate $(\kappa-\phi)/\kappa$ {from}
Eqs.~(\ref{evolution_cond_c_const}) and
(\ref{evolution_cond_N_star_1}) for large $N$, $N^{*}$:
\begin{eqnarray*}
{2\kappa\over \alpha} & \simeq &
N^2-\left(N^*\right)^2 -2N^*\left(N-N^*\right) \\
& = & N^2 -2N^*N+\left(N^*\right)^2  \ =\
(N-N^*)^2~.
\end{eqnarray*}
The number of surviving species $N-N^*$ is therefore
\begin{equation}
N-N^* \ \simeq\ \sqrt{2\kappa\over\alpha}~,
\label{evolution_hyper_cycle}
\end{equation}
which is non-zero for a finite and positive
inter-molecular catalytic rate $\kappa$. A 
hypercycle of mutually supporting species 
(or molecules) has formed.
\index{hypercycle!prebiotic evolution}

\runinhead{The Origin of Life}
\index{origin of life}\index{life!origin} 
The scientific discussions concerning the origin 
of life are highly controversial to date and it 
is speculative whether hypercycles have anything to
do with it. Hypercycles describe closed systems of 
chemical reactions which have to come to a stillstand 
eventually, as a consequence of the continuous energy 
dissipation. In fact, a tellpoint sign of biological 
activities is the buildup of local structures, resulting 
in a local reduction of entropy, possible only at the 
expense of an overall increase of the environmental entropy.
Life, as we understand it today, is possible only as an open
system driven by a constant flux of energy.

Nevertheless it is interesting to point out that
Eq.~(\ref{evolution_hyper_cycle}) implies a clear 
division between molecules $i=N^*,\ \ldots ,\ N$ 
which can be considered to form a primordial 
\qut{life form} separated by molecules 
$i=1,\ldots ,N^*-1$ belonging to the \qut{environment}, 
since the concentrations of the latter are reduced to 
zero. This clear separation between participating and 
non-participating substances is a result of the 
non-linearity of the reaction equations
(\ref{evolution_reaction_network}). The linear
evolution equations (\ref{evolution_eigen_x})
would, on the other hand, result in a continuous density
distribution, as illustrated in Fig.~\ref{evolution1_sharpPeak_QS}
for the case of the sharp peak fitness landscape. One could then
conclude that life is possible only via cooperation, resulting from
non-linear evolution equations.

\section{Coevolution and Game Theory}
\label{evolution_coevolution}
\index{game theory|textbf}

The average number of offsprings, viz
the fitness, is the single relevant
reward function within Darwinian evolution.
There is hence a direct connection between
evolutionary processes and game theory,
which deals with interacting agents trying
to maximize a single reward function denoted
utility. Several types of games may be considered
in this context, namely games of interacting
species giving rise to coevolutionary phenomena
or games of interacting members of the same species,
pursuing distinct behavioral strategies.

\begin{figure}[t]
\centering
\includegraphics{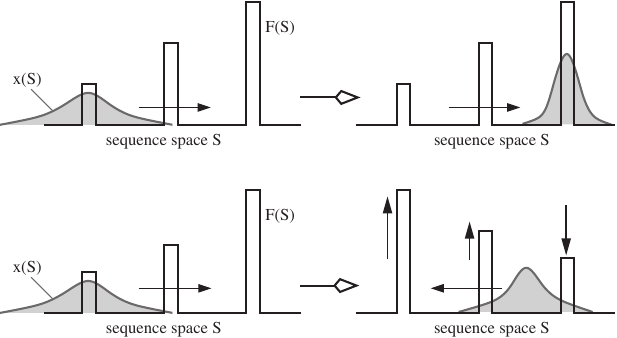}
\caption{\textit{Top}: Evolutionary process of a single (quasi)
species in a fixed fitness landscape (fixed ecosystem), here with
tower-like structures, see Eq.~(\ref{evolution_tower}).
\textit{Bottom}: A coevolutionary process might be regarded as
changing the respective fitness landscapes}
\label{evolution_fig_co_games} 
\end{figure}

\runinhead{Coevolution}
In the discussion so far we first considered 
the evolution of a single species and then 
in Sect.~\ref{evolution1_hypercycles}, the
stabilization of an \qut{ecosystem} made of 
a hypercycle of mutually supporting species.
\begin{quotation}
{\it Coevolution.\enspace}
\index{coevolution}
 When two or more species form an interdependent
ecosystem the evolutionary progress of part of the ecosystem will
generally induce coevolutionary changes also in the other species.
\end{quotation}
One can view the coevolutionary process also as a change in the
respective fitness landscapes, see
Fig.~\ref{evolution_fig_co_games}. A prominent example of phenomena
arising from coevolution is the \qut{red queen} phenomenon.
%

\begin{quotation}
{\it The Red Queen Phenomenon.\enspace}
\index{red queen phenomenon}\index{coevolution!red queen
phenomenon}When two or more species are interdependent then
\qut{It takes all the running, to stay in place}
{(from Lewis Carroll's children's book \qut{Through the Looking
Glass})}.
\end{quotation}
A well-known example of the red queen phenomenon is the \qut{arms
race} between predator and prey commonly observed in natural
ecosystems. \index{coevolution!arms race}

\runinhead{The Green World Hypothesis} \index{coevolution!green
world hypothesis} Plants abound in real-world ecosystems, geology
and\enlargethispage*{10pt} climate permitting, they are rich and
green. Naively one may expect that herbivores should proliferate
when food is plenty, keeping vegetation constantly down. This
doesn't seem to happen in the world and Hairston, Smith and
Slobodkin proposed that coevolution gives rise to a trophic cascade,
where predators keep the herbivores substantially below the support
level of the bioproductivity of the plants. This \qut{green world
hypothesis} arises natural in evolutionary models, but has been
difficult to verify in field studies.

\runinhead{Avalanches and Punctuated Equilibrium} In
Chap.~\ref{chap_automata1} we discussed the Bak
and Sneppen model of coevolution. It may explain
the occurrence of coevolutionary avalanches within a state of
punctuated equilibrium.
\begin{quotation}
{\it Punctuated Equilibrium.\enspace}
\index{punctuated equilibrium}
Most of the time the ecosystem is in equilibrium,
in the neutral phase. Due to rare stochastic processes
periods of rapid coevolutionary processes are induced.
\end{quotation}
The term punctuated equilibrium was proposed by Gould and
Eldredge in 1972 to describe a characteristic feature of the evolution
of simple traits observed in fossil records. In contrast to the
gradualistic view of evolutionary changes, these traits typically
show long periods of stasis interrupted by very rapid changes.

The random events leading to an increase in genome
optimization might be a rare mutation bringing one or
more individuals to a different peak in the fitness
landscape (microevolution) or a coevolutionary avalanche.

\runinhead{Strategies and Game Theory}
One is often interested, in contrast to the stochastic considerations
discussed so far, in the evolutionary processes giving rise
to very specific survival strategies. These questions
can be addressed within game theory, which deals with
strategically interacting agents in economics and beyond.
When an animal meets another animal it has to decide,
to give an example,
whether confrontation, cooperation or defection is the
best strategy. The basic elements of game theory are:
\begin{itemize}
\item[--] {Utility}:
\index{game theory!utility}Every participant, also called {an} agent,  plays for himself,
   trying to maximize its own utility.

\item[--]{Strategy}:
\index{game theory!strategy}Every participant follows a set of
     rules of what to do when encountering an opponent; the strategy.

\item[--] {Adaptive Games}:
\index{adaptation!game theory}In adaptive games the participants
    change their strategy
   in order to maximize future return. This change can be
   either deterministic or \nobreak stochastic.

\item[--] {Zero-Sum Games}:
\index{game theory!zero-sum}When the sum of utilities is constant,
   you can only win what the others lose.

\item[--] {Nash Equilibrium}: \index{game theory!Nash equilibrium}
\index{Nash equilibrium}Any strategy change by {a} participant leads
   to a reduction of his utility.
\end{itemize}

\vspace*{6pt}
\runinhead{Hawks and Doves} \index{game!Hawks and
Doves}\index{Hawks and Doves game}This simple evolutionary game
tries to model competition in terms of expected utilities between
aggressive behavior (by the \qut{hawk}) and peaceful (by the
\qut{dove}) demeanor. The rules are: \vspace*{-10pt}
\begin{table}[h]
\centering
{\begin{tabular*}{\textwidth}{@{\extracolsep{\fill}}llllp{140pt}@{}} 
\hline\noalign{\smallskip}
Dove meets Dove &\ & $A_{DD}=V/2$ &\ & They divide the territory. \\
Hawk meets Dove &\ & $A_{HD}=V$, $A_{DH}=0$ &\ &
The Hawk gets all the territory, the Dove retreats and gets nothing. \\
Hawk meets Hawk &\ & $A_{HH}=(V-C)/2$ & \ &
They fight, get injured, and win half the territory. \\
\noalign{\smallskip}\hline\noalign{\smallskip}
\end{tabular*}}
{}
\end{table}

\enlargethispage*{-10pt}

The expected returns, the utilities, can be cast in
matrix form,
$$
A \ =\
\left(
\begin{array}{cc}
A_{HH} &A_{HD} \\
A_{DH} & A_{DD}
\end{array}
\right) \ =\
\left(
\begin{array}{cc}
{1\over2}(V-C) & V \\
0 &  {V\over2}
\end{array}
\right) ~.
$$
$A$ is denoted the \qut{payoff} matrix. \index{payoff matrix}
\index{game theory!payoff matrix}\index{matrix!payoff}The question
is then, under which conditions it pays to be peaceful or
aggressive.

\runinhead{Adaptation by Evolution} 
 \index{adaptation!game theory} The introduction of reproductive
capabilities for the participants turns the hawks-and-doves game
into an evolutionary game. In this context one considers the
behavioral strategies to result from the expression of distinct
\nobreak alleles.

The average number of {offspring} of a player
is proportional to its fitness, which in turn is
assumed to be given by its expected utility,
\begin{equation}
\begin{array}{rcl}
\dot x_H & = & \Big( A_{HH} x_H + A_{HD}x_D-\phi(t)\Big)\,x_H \\
\dot x_D & = & \Big( A_{DH} x_H + A_{DD}x_D-\phi(t)\Big)\,x_D
\end{array}~,
\label{evolution_hawks_doves}
\end{equation}
where $x_D$ and $x_H$ are the density of doves and hawks,
respectively, and where the flux
$$
\phi(t)\ =\ x_HA_{HH}x_H + x_HA_{HD}x_D + x_DA_{DH}x_H + x_DA_{DD}x_D
$$
ensures an overall constant population, $x_H+x_D=1$.

\runinhead{{The} Steady State Solution}
\index{stationary solution!Hawks and Doves}We are interested in the
steady-state solution of\break Eq.~(\ref{evolution_hawks_doves}), with
$\dot x_D=0=\dot x_H$. Setting
$$
x_H\ = x, \qquad\quad x_D=1-x~,
$$
we find
$$
\phi(t)\ =\ {x^2\over2}(V-C) + Vx(1-x) + {V\over2}(1-x)^2
       \ =\ {V\over2} - {C\over 2}x^2
$$
and
\begin{eqnarray*}
\dot x& = & \left({V-C\over 2}x + V(1-x) -\phi(t)\right)\,x
\ =\  \left({V\over2}-{V\over2}x+{C\over 2}\left(x^2-x\right) \right)\,x
\\[6pt]
 & = & {C\over2}\,x\,\left(x^2-{C+V\over C}x+{V\over C}\right)
 \ = \ {C\over2}\,x\,\left(x-1\right)\left(x-V/C\right)\\[6pt]
& =& -{{\rm d}\over {\rm d}x}V(x)~,
\end{eqnarray*}
with
$$
V(x) \ =\ -{x^2\over 4}V \, +\, {x^3\over6}(V+C)\,-\,{x^4\over 8}C~.
$$
The steady state solution is given by\enlargethispage*{12pt}
$$
V'(x)\ =\ 0,
\qquad\quad x \ =\ V/C~,
$$
apart from the trivial solution $x=0$ (no hawks) and
$x=1$ (only hawks).
For $V>C$ there will be no doves left in the population,
but for $V<C$ there will be an equilibrium with
$x=V/C$ hawks and $1-V/C$ doves. A population consisting
exclusively of cooperating doves ($x=0$) is unstable
against the intrusion of hawks.

\runinhead{The Prisoner's Dilemma}
\index{Prisoner's dilemma}
\index{game!Prisoner's dilemma}
The payoff matrix of the prisoner's dilemma
is given by
\begin{equation}
A \ =\
\left(
\begin{array}{cc}
R & S \\
T & P
\end{array}
\right)
\qquad \quad
\begin{array}{c}
T>R>P>S \\
2R>S+T
\end{array}
\qquad \quad
\begin{array}{rcl}
{\rm cooperator} &\hat{=} & {\rm dove} \\
{\rm defector}   &\hat{=} & {\rm hawk}
\end{array}~.
\label{evolution_prisoners_dilemma}
\end{equation}
Here \qut{cooperation} between the two prisoners is implied and not
cooperation between a suspect and the police. The prisoners are best
off if both keep silent. The standard values are
$$
T=5,\qquad
R=3,\qquad
P=1,\qquad
S=0~.
$$
The maximal global utility $NR$ is obtained when everybody
cooperates, but in a situation where agents interact randomly, the
only stable Nash equilibrium is when everybody defects, with a
global utility $NP$:
\begin{eqnarray*}
\mbox{reward for cooperators} \ =\ R_c &=& \Big[ RN_c+S(N-N_c)\Big]/N~,\\
\mbox{reward for defectors} \ =\ R_d &=& \Big[TN_c+P(N-N_c)\Big]/N ~,
\end{eqnarray*}
where $N_c$ is the number of cooperators and $N$ the total
number of agents. The \hbox{difference} is
$$
R_c-R_d\ \sim\ (R-T)N_c\,+\,(S-P)(N-N_c)\ <\ 0~,
$$
as $R-T<0$ and $S-P<0$. The reward for cooperation is always
smaller than that for defecting.
\begin{figure}[t]
\centering
\includegraphics{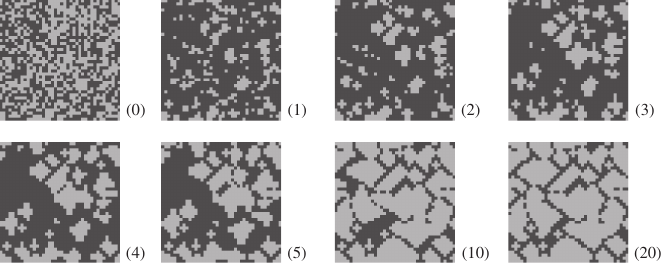}
\caption{Time series of the spatial distribution of cooperators
({\textit{gray}})
  and defectors (\textit{black}) on {a} lattice of size $N=40\times 40$. The time is
  given by the numbers of generations in {\it brackets}. Initial condition:
  Equal number of defectors and cooperators, randomly distributed.
  Parameters for the payoff matrix, $\{T;R;P;S\}=\{3.5;3.0;0.5;0.0\}$
(from Schweitzer et al., 2002)
  }
\label{evolution_lattix_pisoner}\vspace*{-6pt}
\end{figure}

\runinhead{Evolutionary Games on a Lattice}
\index{game theory!lattice}The adaptive dynamics of evolutionary
games can change completely when the individual agents are placed on
a regular lattice and when they adapt their strategies based on past
observations. A possible simple rule is the following:
\begin{itemize}
\item [--]At each generation (time step)
every agent evaluates its own payoff when
interacting with its four neighbors, as well
as the payoff of its neighbors.

\item[--]
The individual agent then compares his own payoff
one-by-one with the payoffs obtained
by his four neighbors.

\item[--]
The agent then switches his strategy (to cooperate or to defect) to
the strategy of {his} neighbor if the neighbor
{received} a higher payoff.
\end{itemize}
This simple rule can lead to complex real-space patterns of
defectors intruding in a background of cooperators, see
Fig.~\ref{evolution_lattix_pisoner}. The details depend on the value
chosen for the payoff matrix.\vspace*{4pt}

\runinhead{Nash Equilibria and Coevolutionary Avalanches}
\index{Nash equilibrium}\index{coevolution!avalanche} Coevolutionary
games on a lattice {eventually lead} to an
equilibrium state, which by definition has to be a Nash equilibrium.
If such a state is perturbed from the outside, a self-critical
coevolutionary avalanche may follow, in close relation to the
sandpile model discussed in Chap.~\ref{chap_automata1}.


\runinhead{Game Theory and Memory}
Standard game theory deals with an anonymous society
of agents, with agents having no memory of previous
encounters. Generalizing this standard setup it is
possible to empower the agents with a memory of their
own past strategies and achieved utilities. Considering
additionally individualized societies, this memory may
then include the names of the opponents encountered previously,
and this kind of games provides the basis for
studying the emergence of sophisticated survival strategies,
like altruism, via evolutionary processes.

\runinhead{Opinion Dynamics}
\index{opinion dynamics}
\index{dynamics!opinion}
Agents in classical game theory aim to maximize their
respective utilities. Many social interactions between 
interacting agents however do not need explicitly the 
concept of rewards or utilities in order to 
describe interesting phenomena.

Examples of reward-free games are opinion dynamics models.
In a simple model for continous opinion dynamics 
$i=1,\ \dots,\ N$ agents have continous opinions
$x_i=x_i(t)$. When two agents interact they change
their respective opinions according to
\begin{equation}
x_i(t+1)\ =\  \left\{
\begin{array}{ccl}
[ x_i(t)+x_j(t)]/2 &\qquad & |x_i(t)-x_j(t)| < \theta \\
x_i(t) &\qquad & |x_i(t)-x_j(t)| \ge \theta 
\end{array}
              \right.~,
\label{evolution1_opinion_dynamics}
\end{equation}
where $\theta$ is the confidence interval. Consensus
can be reached step by step only when the initial
opinions are not too contrarian. For large
confidence intervals $\theta$, relative
to the intial scatter of opinions, global 
consensus will be reached, clusters of
opinions emerge on the other side for a 
small confidence interval.


\section*{Exercises}
\addcontentsline{toc}{section}{Exercises}
%

{\sc {The One-Dimensional} Ising Model}
\begin{list}{}
\item
Solve the one-dimensional Ising model\vspace*{4pt}
$$
H \ =\ J\sum_i s_i s_{i+1}\,+\,B\sum_i s_i\vspace*{4pt}
$$
by the transfer matrix method presented in
Sect.~\ref{evolution1_beanbag} and calculate the free energy
$F(T,\!B)$, the magnetization $M(T,B)$ and the susceptibility
$\chi(\!T)=\break \lim_{B\to0}{\partial M(T,B)\over\partial B}$.
\end{list}
\hspace*{-12pt}{\sc Error Catastrophe}
\begin{list}{}
\item For the prebiotic quasispecies model Eq.~(\ref{evolution_eigen_vec})
consider tower-like autocatalytic reproduction rates $W_{jj}$ and
mutation rates $W_{ij}$ ($i\ne j$) of the form
\vspace*{5pt}
$$
W_{ii} \ =\ \left\{
\begin{array}{cc}
1 & i=1\\
1-\sigma & i>1
\end{array} \right. , \qquad\quad
W_{ij} \ =\ \left\{
\begin{array}{cl}
u_+ & i=j+1\\
u_- & i=j-1\\
0   & i\ne j \ \mbox{otherwise}
\end{array} \right. ~,
\vspace*{5pt}
$$
with $\sigma,u_\pm\in[0,1]$. Determine the error catastrophe for the
two cases $u_+=u_-\equiv u$ and $u_+=u$, $u_-=0$. Compare {it} to
the results for the tower landscape discussed in
Sect.~\ref{evolution1_epistatic}.\newline Hint: For the stationary
eigenvalue equation (\ref{evolution_eigen_vec}), with $\dot x_i=0$
($i=1,\ldots$), write $x_{j+1}$ as a function of $x_{j}$ and
$x_{j-1}$. This two-step recursion relation leads to a $2\times 2$
matrix. Consider the eigenvalues/vectors of this matrix, the initial
condition for $x_1$, and the normalization condition
$\sum_i x_i<\infty$ valid in the adapting\break regime.
\end{list}
\hspace*{-12pt}{\sc Models of Life}
\begin{list}{}
\item Go to the Internet, e.g.\ \url{http://cmol.nbi.dk/javaapp.php}, and
try {} a few JAVA applets simulating models of life.
Select a model of your {choice} and study the
literature given.
\end{list}
\hspace*{-12pt}{\sc Competition for Resources}
\begin{list}{}
\item The competition for scarce resources has been modelled in
      the quasispecies theory, see Eq.~(\ref{evolution_eigen_x}),
      by an overall constraint on population density. With
\begin{equation}
\dot x_i \ =\ W_{ii} x_i 
\qquad\quad W_{ii} \ =\  f r_i-d,
\qquad\quad \dot f \ =\ a-f \sum_i r_i x_i
\label{evolution1_competition_resources}
\end{equation}
     one models the competition for the resource $f$ explicitly,
     with $a$ ($fr_i$) being the regeneration rate of the resource 
     $f$ (species $i$) and $d$ the mortality rate. 
     Eq.\ (\ref{evolution1_competition_resources}) does not
     contain mutation terms $\sim W_{ij}$ describing a
     simple ecosystem.

     Which is the steady-state value of the total population
     density $C= \sum_i x_i$ and of the resource level $f$?
     Is the ecosystem stable?
\end{list}
\hspace*{-12pt}{\sc Hypercycles}
\begin{list}{}
\item
Consider the reaction {equations}
(\ref{evolution_reaction_network}) and
(\ref{evolution_reaction_phi}) for $N=2$ molecules and a homogeneous
network. Find the fixpoints and discuss their stability.
\end{list}
\hspace*{-12pt}{\sc {The} Prisoner's Dilemma on a Lattice}
\begin{list}{}
\item Consider the stability of intruders in the prisoner's dilemma
{Eq.~}(\ref{evolution_prisoners_dilemma}) on a square
lattice, as the one illustrated in
Fig.~\ref{evolution_lattix_pisoner}. Namely, the case of just one
and of two adjacent defectors/cooperators in a background of
cooperators/defectors.  Who {survives}?
\end{list}
\hspace*{-12pt}{\sc Nash Equilibrium}
\begin{list}{}
\item Examine the Nash equilibrium and its optimality for the
following two-player game:

Each player acts either cautiously or {riskily}. A
player acting cautiously always receives a low pay-off. A player
playing {riskily} gets a high pay-off if the other
player also takes a risk. Otherwise, the risk-taker obtains no
reward.
\end{list}


\def\refer#1#2#3#4#5#6{\item{\frenchspacing\sc#1}\hspace{4pt}
                       #2\hspace{8pt}#3 {\it\frenchspacing#4} {\bf#5}, #6.}
\def\bookref#1#2#3#4{\item{\frenchspacing\sc#1}\hspace{4pt}
                     #2\hspace{8pt}{\it#3}  #4.}

\addcontentsline{toc}{section}{Further Reading} 
\section*{Further Reading}

\markboth{\thechapter\enspace Statistical Modeling of Darwinian Evolution}{Further
Reading}

A comprehensive account of the earth's biosphere can be found in
Smil (2002); a review article on the statistical
approach to Darwinian evolution in Peliti (1997) and Drossel (2001).
Further general textbooks on evolution, game-theory
and hypercycles are Nowak (2006),
Kimura (1983), Eigen (1971),
Eigen and Schuster (1979) and Schuster (2001).
For a review article on evolution and speciation
see Drossel (2001), for an assessment of
punctuated equilibrium Gould and Eldredge (2000).

The relation between life and self-organization is further discussed
by Kauffman (1993), a review of the prebiotic RNA
world can be found in Orgel (1998) and critical discussions of
alternative scenarios for the origin of life in
Orgel (1998) and Pereto (2005).

The original formulation of the fundamental theorem of natural
selection was given by Fisher (1930). For the
reader interested in coevolutionary games we refer to Ebel and
Bornholdt (2002); for an interesting application of
game theory to world politics as an evolving complex system see
Cederman (1997) and for a field study on the green world
hypothesis Terborgh {et al.} (2006).

{\baselineskip=15pt
\begin{list}{}{\leftmargin=2em \itemindent=-\leftmargin%
\itemsep=3pt \parsep=0pt \small}
\bookref{Cederman, L.-E.}{1997}{Emergent Actors
         in World Politics.}
         {Princeton University Press\break Princeton,~N}
\refer{Drake, J.W., Charlesworth, B., Charlesworth, D.}{1998}
   {Rates of spontaneous mutation.}{Genetics}{148}{1667--1686}
\refer{Drossel, B.}{2001}{Biological evolution and statistical physics.} {Advances in Physics}{2} {\hbox{209--295}}
\refer{Ebel, H., Bornholdt, S.}{2002}{Coevolutionary games
       on networks.}{Physical Review E}{66}{056118}
\refer{Eigen, M.}{1971}{Self organization of matter and the evolution
  of biological macromolecules.}{Naturwissenschaften}{58}{465}
\bookref{Eigen, M., Schuster, P.}{1979}{The Hypercycle --
         A Principle of Natural Self-Organization.}
         {Springer, Berlin}
%

%
\bookref{Fisher, R.A.}{1930} {The Genetical Theory of Natural
     Selection.} {Dover, New York}
\bookref{Gould, S.J., Eldredge, N.}{2000}{\rm{Punctuated equilibrium
comes of age.}}
   {In H. Gee (ed), {\it Shaking the Tree: Readings from Nature in
   the History of Life}}
   {University Of Chicago Press Chicago, IL.}
\bookref{Jain, K., Krug, J.}{2006}{\rm{Adaptation in simple and complex
      fitness landscapes.}}{In Bastolla, U., Porto, M,
       Roman, H.E., Vendruscolo, M. (eds.) {\it Structural Approaches to Sequence Evolution:
       Molecules, Networks and Populations}} {AG Porto, Darmstadt}
\bookref{Kauffman, S.A.} {1993}{The Origins of Order.}
        {Oxford University Press New York}

\bookref{Kimura, M.}{1983}{The Neutral Theory of Molecular
     Evolution.} {Cambridge University Press Cambridge}

\bookref{Nowak, M.A.}{2006}
{Evolutionary Dynamics: Exploring the Equations of Life.}
{Harvard University Press Cambridge, MA}
\refer{{Orgel}, L.E}{1998}{The origin of
{life}: A review of facts and speculations.} {Trends
in Biochemical Sciences}{23}{491--495}
\bookref{Peliti, L.}{1997}
 {Introduction to the Statistical Theory of Darwinian Evolution.}
 {{ArXiv} preprint cond-mat/9712027}
\refer{Pereto, J.}{2005}{Controversies on the origin of life.}
{International Microbiology}{8}{23--31}
\bookref{Schuster, H.G.}{2001}{Complex Adaptive Systems
         -- An Introduction.}{Scator, Saarbr\"ucken}

\refer{Schweitzer, F., Behera, L., M\"uhlenbein, H.}{2002}
   {Evolution of cooperation in a spatial prisoner's dilemma.}
  {Advances in Complex Systems}{5}{269--299}

\bookref{Smil, V.}{2002}{The Earth's Biosphere:
          Evolution, Dynamics, and Change.}
      {MIT Press,\break \hbox{Cambridge, MA}}

\refer{Terborgh, J., Feeley, K., Silman, M., Nunez, P., Balukjian, B.}{2006}
      {Vegetation dynamics of predator-free land-bridge islands.}
      {Journal of Ecology}{94}{253--263}

\end{list}
\par}


\vspace{-20ex}
\chapter{Synchronization Phenomena}
\label{chap_synchro1}

\abstract{Here we consider the dynamics of complex systems 
constituted of interacting \hbox{local} computational units 
that have their own non-trivial dynamics. An example for a 
\hbox{local} dynamical system is the time evolution of an 
infectious disease in a certain city that is weakly influenced 
by an ongoing outbreak of the same disease in another city; or 
the case of a neuron in a state where it fires spontaneously 
under the influence of the afferent axon potentials.\newline
\indent A fundamental question is then whether the time evolutions
of these local units will remain dynamically independent of
each other or whether, at some point, they will start to change
their states all in the same rhythm. This is the notion of 
\qut{synchronization}, which we will study throughout this 
chapter, learning that the synchronization process may
be driven either by averaging dynamical variables
or through causal mutual influences.}

\section{Frequency Locking}
\label{syncho_frenquency_locking}

In this chapter we will be dealing mostly with autonomous
dynamical systems which may synchronize spontaneously. A
dynamical system may also be driven by outside influences,
being forced to follow the external signal synchronously.

\runinhead{The Driven Harmonic Oscillator} As
an example we consider the
driven harmonic oscillator 
\index{harmonic oscillator!driven}\vspace{3pt}
\begin{equation}
\ddot x\,+\,\gamma\, \dot x\, +\,\omega_0^2\,x\ =\ F\left(e^{i\omega
t}+c.c.\right), \qquad\quad \gamma \ >\ 0~.
\label{synchro_driven_oscillator}
\end{equation}
In the absence of external driving, $F\equiv0$, the solution is
\begin{equation}
x(t)\ \sim\ e^{\lambda t} , \qquad\quad \lambda_{\,\,\,\,\pm} =
-{\gamma\over2}\pm\sqrt{{\gamma^2\over4}-\omega_0^2}~,
\label{synchro_oscill_damped}\vspace{3pt}
\end{equation}
which is damped/critical/overdamped for
$\gamma<2\omega_0$, $\gamma=2\omega_0$
and $\gamma>2\omega_0$.
\index{critical!driven harmonic oscillator}

\runinhead{Frequency Locking} \index{frequency locking} In the long
time limit, $t\to\infty$, the dynamics of the system follows the
external driving, for all $F\ne 0$, due the damping $\gamma>0$. We
therefore consider the ansatz
\begin{equation}
x(t)\ =\ ae^{i\omega t}+c.c.,
\label{synchro_Ansatz_driven_oscillator}
\end{equation}
where the amplitude $a$ may contain an additional time-independent
phase. Using this ansatz for
{Eq.~}(\ref{synchro_driven_oscillator}) we obtain
\begin{eqnarray*}
F& =&
a\left(-\omega^2+i\omega\gamma+\omega_0^2\right) \\
& =&
-a\left(\omega^2-i\omega\gamma-\omega_0^2\right) \ =\
-a\left(\omega+i\lambda_+\right)
  \left(\omega+i\lambda_-\right)~,
\end{eqnarray*}
where the eigenfrequencies $\lambda_\pm$ are given by
Eq.~(\ref{synchro_oscill_damped}). The solution for the amplitude
$a$ can then be written in terms of {} $\lambda_\pm$
or alternatively as
\begin{equation}
a\ =\ {-F\over \left(\omega^2 -\omega_0^2\right)
-i\omega\gamma}~.
\label{synchro_oscillator_resonance}
\end{equation}
The response becomes divergent, viz $a\to\infty$, at resonance
$\omega=\omega_0$ and small damping $\gamma\to0$.\vspace*{3pt}

\runinhead{{The} General Solution} The driven, damped
harmonic oscillator
Eq.~(\ref{synchro_driven_oscillator}) is an
inhomogeneous linear differential equation and its general solution
is given by the superposition of the special solution
Eq.~(\ref{synchro_oscillator_resonance}) with the
general solution of the homogeneous system
Eq.~(\ref{synchro_oscill_damped}). The
latter dies out for $t\to\infty$ and the system
synchronizes with the external driving frequency $\omega$.
\index{synchronization!driven oscillator}

\section{Synchronization of Coupled Oscillators}
\label{syncho_coupled_oscillators}

Any set of local dynamical systems may synchronize,
whenever their dynamical behaviours are similary and
the mutual couplings substantial. We start by
discussing the simplest non-trivial set-up, viz
harmonically coupled harmonic oscillators.

\runinhead{Limiting Cycles} \index{limiting cycle} A free rotation
$$
\vec x(t)\ =\ r\,\Big(\cos(\omega t+\phi_0),\sin(\omega t+\phi_0)\Big),
\qquad \theta(t)\ =\ \omega t+\theta_0,
\qquad \dot\theta \ =\ \omega
$$
often occurs (in suitable coordinates) as limiting cycles of
dynamical systems, see Chap.~\ref{chap_chaos1}. One can then use the
phase variable $\theta(t)$ for an effective
description.\vspace*{4pt}

\runinhead{Coupled Dynamical Systems} \index{oscillator!coupled}We
consider a collection of individual dynamical systems $i=1,\dots,N$,
which have limiting cycles with natural frequencies $\omega_i$. The
coupled system then obeys
\begin{equation}
\dot\theta_i \ =\ \omega_i\,+\,\sum_{j=1}^N\,
\Gamma_{ij}(\theta_i,\theta_j),
\qquad\quad
i=1,\dots,N~,
\label{synchro_coupled_limit_cycles}
\end{equation}
where the $\Gamma_{ij}$ are suitable coupling
constants.\vspace*{4pt}

\runinhead{The Kuramoto Model} \index{Kuramoto model}
\index{model!Kuramoto} A particularly tractable 
choice for the coupling constants $\Gamma_{ij}$ 
has been proposed by Kuramoto:
\begin{equation}
\Gamma_{ij}(\theta_i,\theta_j)\ =\ {K\over N}\,
\sin(\theta_j-\theta_i)~,
\label{synchro_Kuramoto_model}
\end{equation}
where $K\ge0$ is the coupling strength and the factor
$1/N$ ensures that the model is well behaved in the
limit $N\to\infty$.

\runinhead{Two Coupled Oscillators} We consider first the case
$N=2$:
$$
\dot \theta_1\ =\ \omega_1 \,+\, {K\over 2}\, \sin(\theta_2-\theta_1),
\quad\qquad
\dot \theta_2\ =\ \omega_2 \,+\, {K\over 2}\, \sin(\theta_1-\theta_2)~,
$$
or
\begin{equation}
\Delta\dot\theta\ =\ \Delta\omega \,-\,
   K\, \sin(\Delta\theta),
\qquad
\Delta\theta = \theta_2-\theta_1,
\qquad
\Delta\omega = \omega_2-\omega_1~.
\label{synchro_2_CO}
\end{equation}
The system has a fixpoint $\Delta\theta^*$ for  which
\begin{equation}
{\mathrm{d}\over \mathrm{d}t}\Delta\theta^*\ =\ 0, \qquad\quad
   \sin(\Delta\theta^*) \ =\
{\Delta\omega \over K}
\label{synchro_2_CO_fix}
\end{equation}
and therefore
\begin{equation}
\Delta\theta^*\in[-\pi/2,\pi/2],
\qquad\quad
K\ >\ |\Delta\omega|~.
\label{synchro1_synchro_3_CO_fix}
\end{equation}
This condition is valid for attractive coupling
constants $K>0$. For repulsive $K<0$ anti-phase
states are stabilized.
\index{fixpoint!two coupled oscillators}
We analyze the stability of the
fixpoint using $\Delta\theta=\Delta\theta^*+\delta$ and
Eq.~(\ref{synchro_2_CO}). We obtain
$$
{\mathrm{d}\over \mathrm{d}t}\delta\ =\
-\left(K\cos\Delta\theta^*\right)\,\delta, \qquad\qquad \delta(t) \
=\ \delta_0\,e^{-K\cos\Delta\theta^* t}~.
$$
%

\begin{figure}[t]
\centering
\includegraphics{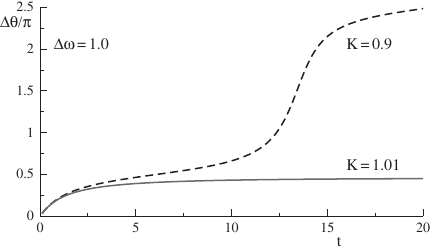}
\caption{The relative phase $\Delta\theta(t)$ of two coupled
oscillators, obeying Eq.~(\ref{synchro_2_CO}), with $\Delta\omega=1$
and a critical coupling strength $K_c=1$. For an undercritical
coupling strength $K=0.9$ the relative phase increases steadily, for
an overcritical coupling $K=1.01$ {it locks}}
\label{synchro1_twoCoupledOscillators}\vspace*{-12pt}
\end{figure}

The fixpoint is stable since $K>0$ and $\cos\Delta\theta^*>0$, due
to Eq.~(\ref{synchro1_synchro_3_CO_fix}). We therefore have a
bifurcation.
\begin{itemize}
\item[--] For
$ K<|\Delta\omega|$ there is no phase coherence
between the two oscillators, they are
drifting with respect to each other.

\item[--] For $ K>|\Delta\omega|$ there is phase locking and
the two oscillators rotate together with a constant
phase difference.
\end{itemize}
This situation is illustrated in
Fig.~\ref{synchro1_twoCoupledOscillators}.

\runinhead{Natural Frequency Distribution} \index{natural
frequencies} \index{distribution!natural frequencies} We now
consider the case of many coupled oscillators, $N\to\infty$. The
individual systems have different individual frequencies $\omega_i$
with a probability distribution
\begin{equation}
g(\omega)\ =\ g(-\omega), \qquad\qquad \int_{-\infty}^\infty
g(\omega)\,\mathrm{d}\omega \ =\ 1~. \label{synchro_g_minus_plus}
\end{equation}
We note that the choice of a zero average frequency
$$
\int_{-\infty}^\infty \omega\, g(\omega)\,\mathrm{d}\omega \ =\ 0
$$
implicit in {Eq.~}(\ref{synchro_g_minus_plus}) is
actually generally possible, as the dynamical {equations}
(\ref{synchro_coupled_limit_cycles}) and
(\ref{synchro_Kuramoto_model}) {are} invariant under a
global translation
$$
\omega\ \to\ \omega+\Omega,
\qquad\quad
\theta_i\ \to\ \theta_i+\Omega t~,
$$
with $\Omega$ being the initial non-zero
mean frequency.

\runinhead{{The} Order Parameter} \index{order
parameter!Kuramoto model} The complex order parameter
\begin{equation}
r\,e^{i\psi} \ =\ {1\over N}\sum_{j=1}^N\, e^{i\theta_j}
\label{synchro_complex_order_parameter}
\end{equation}
is a macroscopic quantity that can be interpreted as the
collective rhythm produced by the assembly of the interacting
oscillating systems. The radius $r(t)$ measures the degree
of phase coherence and $\psi(t)$ corresponds to the average
phase.

\runinhead{Molecular Field Representation} \index{mean-field
theory!Kuramoto model}We rewrite the order parameter
definition\break Eq.~(\ref{synchro_complex_order_parameter}) as
$$
r\, e^{i(\psi-\theta_i)} \ =\
{1\over N}\sum_{j=1}^N\, e^{i(\theta_j-\theta_i)},
\qquad\quad
r\sin(\psi-\theta_i) \ =\
{1\over N}\sum_{j=1}^N\, \sin(\theta_j-\theta_i)~,
$$
retaining the imaginary component of the first term. Inserting the
second expression into the governing {equation }
(\ref{synchro_coupled_limit_cycles}) we find
\begin{equation}
\dot \theta_i\ =\ \omega_i\,+\, {K\over N}
\sum_j\,\sin(\theta_j-\theta_i) \ =\
\omega_i\,+\,Kr\sin(\psi-\theta_i)~.
\label{synchro_dot_theta_r}
\end{equation}
The motion of every individual oscillator $i=1,\ldots,N$ is coupled
to the other oscillators only through the mean-field phase
$\psi${;} the coupling strength being proportional to
the mean-field amplitude $r$.

\begin{figure}[t]
\centering
\includegraphics{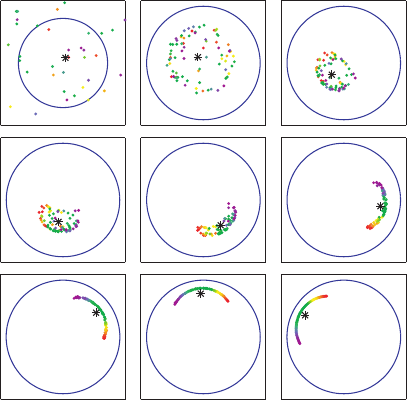}
\caption{Spontaneous synchronization in a network of limit cycle
oscillators with distributed individual frequencies. Color coding:
slowest (\textit{red})--fastest (\textit{violet}) natural frequency.
With respect to
{Eq.~}(\ref{synchro_coupled_limit_cycles}) an additional
distribution of individual radii $r_i(t)$ has been assumed, the
\textit{asterisk} denotes the mean field $re^{i\psi}=\sum_i r_i
e^{i\theta_i}/N$, compare
Eq.~(\ref{synchro_complex_order_parameter}), and the individual
radii $r_i(t)$ are slowly relaxing (from Strogatz, 2001)}
\label{synchro_fig_harmonicSyncro}
\end{figure}

The individual phases $\theta_i$ are drawn towards the
self-consistently determined mean phase $\psi$, as can be seen in
the numerical simulations presented in
Fig.~\ref{synchro_fig_harmonicSyncro}. Mean-field theory is exact
for the Kuramoto model. It is nevertheless non-trivial to solve, as
the self-consistency condition
(\ref{synchro_complex_order_parameter}) needs to be fulfilled.

\runinhead{{The} Rotating Frame of Reference} \index{rotating frame
of reference} \index{variable!rotating frame} The order parameter
$re^{i\psi}$ performs a free rotation in the thermodynamic
limit,\vspace*{3pt}
$$
r(t)\ \to\ r, \qquad\quad \psi(t)\ \to\ \Omega t, \qquad\quad N\
\to\ \infty~,\vspace*{3pt}
$$
and one can transform via\vspace*{3pt}
$$
\theta_i\ \to\ \theta_i+\psi\ =\ \theta_i+\Omega t, \qquad\quad
\dot\theta_i\ \to\ \theta_i+\Omega, \qquad\quad \omega_i\ \to\
\omega+\Omega\vspace*{4pt}
$$
to the rotating frame of reference. The governing
equation (\ref{synchro_dot_theta_r}) then becomes
\begin{equation}
\dot \theta_i\ =\ \
\omega_i\,-\,Kr\sin(\theta_i)~.
\label{synchro_dot_theta_rotating}
\end{equation}
This expression is identical to the one for the case of two coupled
oscillators, Eq.~(\ref{synchro_2_CO}), when substituting $Kr$ by
$K$. It then follows directly that $\omega_i=Kr$ constitutes a
special point.\vspace*{4pt}

\begin{figure}[t]
\centering
\includegraphics{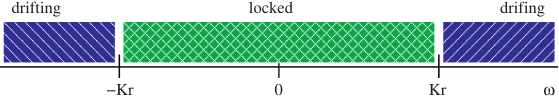}
\caption{The region of locked and drifting
         natural frequencies $\omega_i\to\omega$ within the
         Kuramoto model}
\label{synchro_fig_kuramoto_Kr}
\index{synchronization!Kuramoto model}
\end{figure}

\runinhead{Drifting and Locked Components} \index{Kuramoto
model!locked component} \index{Kuramoto model!drifting
component}Equation (\ref{synchro_dot_theta_rotating}) has a fixpoint
$\theta_i^*$ for which $\dot\theta_i^*=0$ and
\begin{equation}
Kr\sin(\theta_i^*)\ =\ \omega_i,
\qquad\quad |\omega_i|\, <\, K r,
\qquad\quad \theta_i^*\in[-{\pi\over2},{\pi\over 2}]~.
\label{synchro_fixed_point_theta_i}
\end{equation}
$\dot\theta_i^*=0$ in the rotating frame of reference means that the
participating limit cycles oscillate with the average frequency
$\psi${;} they are \qut{locked} to $\psi$, see
Figs.~\ref{synchro_fig_harmonicSyncro} and
\ref{synchro_fig_kuramoto_Kr}.

For $|\omega_i|> K r$ the participating limit cycle {\em drifts},
{i.e.} $\dot\theta_i$ never vanishes. They do, however,
slow down when they approach the locked oscillators, see
{Eq.~}(\ref{synchro_dot_theta_rotating}) and
Fig.~\ref{synchro1_twoCoupledOscillators}.

\runinhead{Stationary Frequency Distribution} \index{stationary
solution!Kuramoto model} \index{distribution!stationary frequencies}
We denote by
$$
\rho(\theta,\omega)\,\mathrm{d}\theta
$$
the fraction of drifting oscillators with natural frequency $\omega$
that lie between $\theta$ and $\theta+\mathrm{d}\theta$. It obeys
the continuity equation
$$
{\partial\rho\over\partial t}\,+\,
{\partial\over\partial\theta}\Big(\rho\,\dot\theta\Big) \ =\ 0~,
$$
where $\rho\dot\theta$ is the respective current density.
In the stationary case, $\dot\rho=0$, the
stationary frequency distribution
$\rho(\theta,\omega)$ needs to be inversely proportional
to the speed
$$
\dot\theta \ =\ \omega \,-\, Kr\sin(\theta)~.
$$
The oscillators pile up at slow places and thin out at fast
places on the circle. Hence
\begin{equation}
\rho(\theta,\omega) \ =\ {C\over |\omega - Kr\sin(\theta)|},
\qquad\quad \int_{-\pi}^{\pi}\rho(\theta,\omega) \,\mathrm{d}\theta\
=\ 1~, \label{synchro_rho_theta_omega}
\end{equation}
for $\omega>0$, where
$C$ is an appropriate normalization constant.

\runinhead{Formulation of the Self-Consistency Condition}
\index{self-consistency condition!Kuramoto model} We write the
self-consistency condition (\ref{synchro_complex_order_parameter})
as
\begin{equation}
\langle e^{i\theta}\rangle \ =\
\langle e^{i\theta}\rangle_{\rm locked} \,+\,
\langle e^{i\theta}\rangle_{\rm drifting} \ = \
r\,e^{i\psi}\ \equiv\ r~,
\label{synchro_brackets}
\end{equation}
where the brackets $\langle \cdot \rangle$ denote
population averages and where we have used the fact that
we can set the average phase $\psi$ to zero.

\runinhead{Locked Contribution} \index{Kuramoto model!locked
component} The locked contribution is
$$
\langle e^{i\theta}\rangle_{\rm locked} \ =\
\int_{-Kr}^{Kr}\,e^{i\theta^*(\omega)}g(\omega)\,\mathrm{d}\omega \
=\ \int_{-Kr}^{Kr}\,\cos\left((\theta^*(\omega)\right)
\,g(\omega)\,\mathrm{d}\omega~,
$$
where we have assumed $g(\omega)=g(-\omega)$ for the
distribution $g(\omega)$ of the natural frequencies within the
rotating frame of reference. Using
Eq.~(\ref{synchro_fixed_point_theta_i}),
$$
\mathrm{d}\omega\ =\ Kr\cos\theta^*\,\mathrm{d}\theta^*~,
$$
for $\theta^*(\omega)$ we obtain
\begin{eqnarray}
\label{synchro_contribution_locked} \langle e^{i\theta}\rangle_{\rm
locked} & =& \int_{-\pi/2}^{\pi/2}\,\cos(\theta^*)\,
g(Kr\sin\theta^*)\,Kr\,\cos(\theta^*)\,\mathrm{d}\theta^*\\[6pt]
& =& Kr\int_{-\pi/2}^{\pi/2}\,\cos^2(\theta^*)\,
g(Kr\sin\theta^*)\,\mathrm{d}\theta^*~. \nonumber
\end{eqnarray}
\runinhead{{The} Drifting Contribution} \index{Kuramoto
model!drifting component} The drifting contribution
$$
\langle e^{i\theta}\rangle_{\rm drifting} \ =\
\int_{-\pi}^{\pi}\,\mathrm{d}\theta\, \int_{|\omega|>Kr}
\mathrm{d}\omega\, e^{i\theta}\rho(\theta,\omega)g(\omega) \ =\ 0
$$
to the order parameter actually vanishes. Physically this is clear:
oscillators {that} are not locked to the mean field
cannot contribute to the order parameter. Mathematically it follows
from $g(\omega)=g(-\omega)$,
$\rho(\theta+\pi,-\omega)=\rho(\theta,\omega)$ and
$e^{i(\theta+\pi)} =-e^{i\theta}$.

\begin{figure}[t]
\centering
\includegraphics{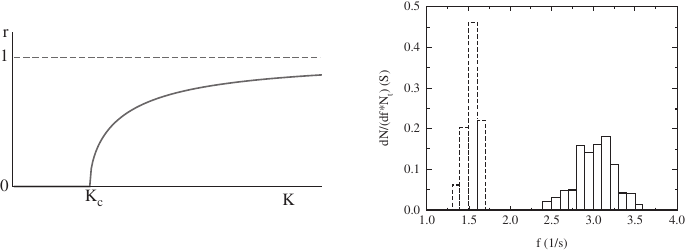}
\caption{\textit{Left}: The solution $r=\sqrt{1-K_c/K}$ for the
order parameter $r$ in the Kuramoto model. \textit{Right}:
Normalized distribution for the frequencies of clappings of one
chosen individual from 100 samplings (N\'eda et al.,
{2000a, b})
         }
\label{synchro_fig_kuramoto_r_K}
\end{figure}

\runinhead{Second-Order Phase Transition} \index{phase
transition!Kuramoto model}The population average $ \langle
e^{i\theta}\rangle$ of the order parameter
{Eq.~}(\ref{synchro_brackets}) is then just the locked
contribution {Eq.~}(\ref{synchro_contribution_locked})
\begin{equation}
r\ =\ \langle e^{i\theta}\rangle \ \equiv \ \langle
e^{i\theta}\rangle_{\rm locked} \ =\
Kr\int_{-\pi/2}^{\pi/2}\,\cos^2(\theta^*)\,
g(Kr\sin\theta^*)\,\mathrm{d}\theta^*~.
\label{synchro_final_self_consistency}
\end{equation}
For $K<K_c$ Eq.~(\ref{synchro_final_self_consistency}) has only the
trivial solution $r=0${;} for $K>K_c$ a finite order
parameter $r>0$ is stabilized, see
Fig.~\ref{synchro_fig_kuramoto_r_K}. We therefore have a
second-order phase transition, as discussed in
Chap.~\ref{chap_automata1}.

\runinhead{Critical Coupling} \index{critical!coupling}The critical
coupling strength $K_c$ can be obtained considering the limes
$r\to0+$ in Eq.~(\ref{synchro_final_self_consistency}):
\begin{equation}
1 \ =\
K_c\,g(0)\,\int_{-\pi/2}^{\pi/2}\cos^2\theta^*\,\mathrm{d}\theta^*
  \ =\ K_c\,g(0){\pi\over 2},
\qquad\quad
K_c \,=\,{2\over \pi g(0)}~.
\label{synchro_}
\end{equation}
The self-consistency condition
{Eq.~}(\ref{synchro_final_self_consistency}) can actually
be solved exactly with the result
\begin{equation}
r\ =\ \sqrt{1-{K_c\over K}},
\qquad\quad
K_c \ =\ {2\over \pi g(0)}~,
\label{synchro_Kuramoto_exact}
\end{equation}
as illustrated in Fig.~\ref{synchro_fig_kuramoto_r_K}.

\runinhead{{The} Physics of Rhythmic Applause}
\index{Kuramoto model!rhythmic applause}\index{rhythmic applause}
\index{synchronization!applause}A nice application of the Kuramoto
model is the synchronization of {the} clapping of an
audience after a performance, which happens when everybody claps at
a slow frequency and in tact. In this case the distribution of
\qut{natural clapping frequencies} is quite narrow and $K>K_c\propto
1/g(0)$.

When an individual wants to express especial satisfaction with the
performance he/she increases the clapping frequency by about a
factor of 2, as measured experimentally, in order to
increase the noise level, which just depends on the clapping
frequency. Measurements have shown, see
Fig.~\ref{synchro_fig_kuramoto_r_K}, that the distribution of
natural clapping frequencies is broader when the clapping is fast.
This leads to a drop in $g(0)$ and {then} $K<K_c\propto 1/g(0)$.
No synchronization is possible when the applause is intense.

\section{Synchronization with Time Delays}
\label{syncho_time_delay}
\index{synchronization!time delays|textbf}
\index{time delays!synchronization|textbf}

Synchronization phenomena need the exchange of
signals from one subsystem to another and this
information exchange typically needs a certain time.
These time delays become important when they
are comparable to the intrinsic time scales
of the individual subsystems. A short
introduction into the intricacies of
time-delayed dynamical systems
has been given in Sect.~\ref{sect_chaos_time_delays},
here we discuss the effect of time delays
on the synchronization process.

\runinhead{The Kuramoto Model with Time Delays}
\index{Kuramoto model!time delays}
\index{time delays!Kuramoto model}
We start with two limiting-cycle oscillators,
coupled via a time delay $T$:
$$
\dot \theta_1(t)\,=\,\omega_1 + {K\over 2}\, \sin[\theta_2(t-T)-\theta_1(t)],
\qquad
\dot \theta_2(t)\,=\,\omega_2 + {K\over 2}\, \sin[\theta_1(t-T)-\theta_2(t)]~.
$$
In the steady state,
\begin{equation}
\theta_1(t) \ =\ \omega\, t,
\qquad \qquad
\theta_2(t) \ =\ \omega\, t \,+\,\Delta\theta^* ~,
\label{synchro1_time_delay_ansatz_2}
\end{equation}
there is a synchronous oscillation with
a yet to be determined locking frequency $\omega$
and a phase slip $\Delta\theta^*$.
Using \ $\sin(\alpha+\beta)=\sin(\alpha)\cos(\beta)
                         +\cos(\alpha)\sin(\beta)$ \
we find
\begin{eqnarray}
\label{synchro_omega_1_time_delay}
\omega &=& \omega_1 + {K\over 2}\, \big[
-\sin(\omega T)\cos(\Delta\theta^*)
+\cos(\omega T)\sin(\Delta\theta^*)
                                   \big], \\
\omega &=& \omega_2 + {K\over 2}\, \big[
-\sin(\omega T)\cos(\Delta\theta^*)
-\cos(\omega T)\sin(\Delta\theta^*)
                                   \big]~.
\nonumber
\end{eqnarray}
Taking the difference we obtain
\begin{equation}
\Delta\omega\ =\ \omega_2-\omega_1 \ =\
K\sin(\Delta\theta^*)\cos(\omega T)~,
\label{synchro_2_CO_fix_time_delay}
\end{equation}
which generalizes Eq.~(\ref{synchro_2_CO_fix})
to the case of a finite time delay $T$.
Eqs.~(\ref{synchro_omega_1_time_delay})
and (\ref{synchro_2_CO_fix_time_delay}) then
determine together locking
frequency $\omega$ and the phase slip
$\Delta\theta^*$.

\begin{figure}[t]
\centering
\includegraphics{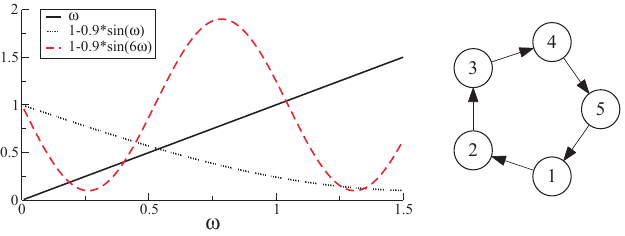}
\caption{\textit{Left}:
Graphical solution of the self-consistency condition
(\ref{synchro1_2_time_delay_omega_1}), given by the intersections
of the {\it solid line} with the {\it dashed lines}, for the locking
frequency $\omega$, and time delays $T=1$ (one solution) and
$T=6$ (three solutions in the inteval $\omega\in[0,1.5]$).
The coupling constant is $K=1.8$.
\textit{Right}: An example of a directed ring, containing five sites
} \label{synchro_fig_timeDelayOmega}
\end{figure}

\runinhead{Multiple Synchronization Frequencies}
For finite time delays $T$, there are generally
more than one solution for the synchronization
frequency $\omega$. For concreteness we consider
now the case
\begin{equation}
\omega_1\ =\ \omega_2\ \equiv\ 1,
\qquad\quad
\Delta\theta^* \ \equiv\ 0,
\qquad\quad
\omega \ =\  1 - {K\over 2} \sin(\omega T)~,
\label{synchro1_2_time_delay_omega_1}
\end{equation}
compare Eqs.~(\ref{synchro_2_CO_fix_time_delay})
and (\ref{synchro_omega_1_time_delay}).
This equation can be solved graphically,
see Fig.~\ref{synchro_fig_timeDelayOmega}.

For $T\to0$ the two oscillators are phase locked,
oscillating with the original natural frequency
$\omega=1$. A finite time delay then leads to
a change of the synchronization frequency and
eventually, for large enough time delay $T$ and
couplings $K$, to multiple solutions for the
locking frequency. These solutions
are stable for
\begin{equation}
K\cos(\omega T) \ > \ 0~;
\label{chaos1_time_delay_stability}
\end{equation}
we leave the derivation as an exercise to the
reader. The time delay such results in a qualitative
change in the structure of the phase space.

\runinhead{Rings of Delayed-Coupled Oscillators}
As an example of the possible complexity arising
from delayed couplings we consider a ring of
$N$ oscillators, as illustrated in
Fig.~\ref{synchro_fig_timeDelayOmega},
coupled unidirectionally,
\begin{equation}
\dot \theta_j \ = \ \omega_j \,+\,
K\sin[\theta_{j-1}(t-T)-\theta_j(t)],
\qquad\quad
j=1,..,N~.
\label{synchro1_time_delay_ring}
\end{equation}
The periodic boundary conditions imply that
$N+1\hat{=}1$ in Eq.~(\ref{synchro1_time_delay_ring}).
We specialize to the uniform case $\omega_j\equiv 1$.
The network is then invariant under rotations of
multiples of $2\pi/N$.

We consider plane-wave solutions\footnote{In the
complex plane $\psi_j(t)=e^{i\theta_j(t)}=e^{i(\omega t-kj)}$
corresponds to a plane wave on a periodic ring.
Eq.~(\ref{synchro1_time_delay_ring})
is then equivalent to the phase evolution of the
wavefunction $\psi_j(t)$. The system is invariant
under translations $j\to j+1$ and the discrete
momentum $k$ is therefore a good quantum number,
in the jargon of quantum mechanics.
The periodic boundary condition $\psi_{j+N}=\psi_j$
is satisfied for the momenta $k = 2\pi n_k/N$.
}
with frequency $\omega$ and momentum $k$,
\begin{equation}
\theta_j \ = \ \omega\,t \,-\, k\,j,
\qquad\quad
k = n_k{2\pi\over N},
\qquad\quad
n_k=0,..,N-1~,
\label{synchro1_time_delay_ring_plane_wave_sol}
\end{equation}
where $j=1,..,N$. For $N=2$ only in-phase
$k=0$ and anti-phase $k=\pi$ solutions
exist. The locking frequency $\omega$
is then determined by the self-consistency
condition
\begin{equation}
\omega \ =\ 1 \,+\, K\sin(k-\omega T)~.
\label{synchro1_time_delay_ring_omega_self}
\end{equation}
For a given momentum $k$ a set of solutions
is obtained. The resulting solutions
$\theta_j(t)$ are characterized by complex
spatio-temporal symmetries, oscillating
fully in phase only for vanishing momentum
$k\to0$.  Note however, that additional
unlocked solutions cannot be excluded
and may show up in numerical solutions.
It is important to remember in this context,
as discussed in Sect.~\ref{sect_chaos_time_delays},
that initial conditions in the entire
interval $t\in[-T,0]$ need to be provided.

\enlargethispage{-12pt}

\section{Synchronization via Aggregate Averaging}
\label{syncho_aggregate_averaging}
\index{synchronization!aggregate averaging|textbf}

The synchronization of the limiting cycle oscillators 
discussed in Sect.~\ref{syncho_coupled_oscillators} 
is mediated by the molecular field, which is an 
averaged quantity. Averaging plays a central role
in many synchronization processes and may act
both on a local basis and on a global level.
Alternatively, synchronization may be driven by
the casual influence of temporally well defined events,
a route to synchronization we will discuss in
Sect.~\ref{syncho_causal_signaling}.

\runinhead{Pairwise Averaging}
The coupling term of the Kuramoto model,
see Eq.~(\ref{synchro_Kuramoto_model}),
contains differences $\theta_i-\theta_j$
in the respective dynamical variables
$\theta_i$ and $\theta_j$. With an 
appropriate sign of the coupling constant,
this coupling results in a driving 
force towards the average,
$$
\theta_1 \ \to\ {\theta_1+\theta_2\over 2},
\qquad\quad
\theta_2 \ \to\ {\theta_1+\theta_2\over 2}~.
$$
This driving force competes with the differences
in the time-development of the individual
oscillators, which is present whenever their
natural frequencies $\omega_i$ and $\omega_j$
do not coincide. A detailed analysis is then
necessary, as carrried out 
in Sect.~\ref{syncho_coupled_oscillators},
in order to study this competion between 
the synchronizing effect of the coupling 
and the desynchronizing influence of
a non-trivial natural frequency distribution.

\runinhead{Aggregate Variables}
\index{variable!aggregate}
Generalizing above considerations we consider now
a set of dynamical variables $x_i$, with 
$\dot x_i=f_i(x_i)$ being the
evolution rule for the isolated units.
The geometry of the couplings is given by the
normalized weighted adjacency matrix 
$$
A_{ij},
\quad\qquad
\sum_j A_{ij}\ =\ 1~.
$$
The matrix elements
are $A_{ij}>0$ if the units $i$ and $j$ are 
coupled, and zero otherwise, compare
Chap.~\ref{chap_networks1}, with $A_{ij}$
representing the relative weight of the
link. We define now the aggregate variables
$\bar x_i=\bar x_i(t)$ by
\begin{equation}
\bar x_i \ =\ (1-\kappa_i) x_i\,+\, \kappa_i\sum_j A_{ij} x_j~,
\label{synchro1_aggregate_variable}
\end{equation}
where $\kappa_i\in[0,1]$ is the local coupling strength.
The aggregate variables $\bar x_i$ correspond to
a superposition of $x_i$ with the weighted
mean activtiy $\sum_j A_{ij} x_j$ of all its neighbors.

\runinhead{Coupling via Aggregate Averaging}
A quite general class of dynamical networks
can now be formulated in terms of aggregate
variables through
\begin{equation}
\dot x_i \ =\ f_i(\bar x_i), \qquad\quad
i=1,\ \dots,\ N~,
\label{synchro1_network_AV_def}
\end{equation}
with the $\bar x_i$ given by
Eq.~(\ref{synchro1_aggregate_variable}).
The $f_i$ describe the local dynamical
systems which could be, e.g., 
harmonic oscillators, relaxation oscillators 
or chaotic systems.

\runinhead{Expansion around the Synchronized State}
In order to expand Eq.~(\ref{synchro1_network_AV_def})
around the globally synchronized state we 
first rewrite the aggregate variables as
\begin{eqnarray}
\label{synchro1_AV_expansion}
\bar x_i & =& (1-\kappa_i) x_i\,+\, \kappa_i\sum_j A_{ij} (x_j-x_i+x_i) \\
         & =& x_i\Big(1-\kappa_i +\kappa_i\sum_j A_{ij} \Big) 
                    +   \kappa_i \sum_j A_{ij} (x_j-x_i) 
         \ =\ x_i \,+\, \kappa_i \sum_j A_{ij} (x_j-x_i)~,
\nonumber
\end{eqnarray}
where we have used
the normalization $\sum_j A_{ij}=1$.
The differences in activies $x_j-x_i$ are small
close to the synchronized state and we may
expand
\begin{equation}
f_i(\bar x_i) \ \approx\  f_i(x_i) 
\,+\, f_i'(x_i)\kappa_i\sum_j A_{ij} (x_j-x_i)~.
\label{synchro1_AA_f_i_expansion}
\end{equation}
Differential couplings $\sim(x_j-x_i)$
between the nodes of the network are hence
equivalent, close to synchronization, to 
the aggregate averaging of the local 
dynamics via the respective $\bar x_i$.

\runinhead{General Coupling Functions}
We may go one step further and define with
\begin{equation}
\dot x_i \ =\ f(x_i) \,+\, \sum_j g_{ij}(x_j-x_i)
\label{synchro1_AA_general_couplings}
\end{equation}
a general system of $i=1,\ \dots,\ N$ dynamical 
units interacting via the coupling functions
$g_{ij}(x_j-x_i)$. Close to the synchronized state
we may expand Eq.~(\ref{synchro1_AA_general_couplings}) as
\begin{equation}
\dot x_i \ \approx\ f(x_i) \,+\, \sum_j g_{ij}'(0)(x_j-x_i),
\qquad\quad
g_{ij}'(0)\ \hat{=}\ f_i'(x_i)\kappa_iA_{ij}~.
\label{synchro1_AA_general_couplings_expansion}
\end{equation}
The equivalence of $g_{ij}'(0)$ and 
$f_i'(x_i)\kappa_iA_{ij}$ is only local in time, 
with the later being time dependent, but this
equivalence is sufficient for a local stability 
analysis; the synchronized state of
the system with differential couplings,
Eq.~(\ref{synchro1_AA_general_couplings}), is
locally stable then and only then if the
corresponding system with aggregate couplings,
Eq.~(\ref{synchro1_network_AV_def}), is also stable
against perturbations.

\runinhead{Synchronization via Aggregated Averaging}
The equivalence of Eqs.~(\ref{synchro1_network_AV_def})
and (\ref{synchro1_AA_general_couplings}) tells us
that the driving forces leading to synchronization
are aggregated averaging processes of neighboring
dynamical variables. 

Till now we considered globally synchronized states. 
Synchronization processes are however in general 
quite intricate processes, we mention here two 
alternative possibilities. Above discussion concerning
aggregate averaging remains however valid, when 
generalized suitably, also for these more 
generic synchronized states. 
\begin{itemize}
\item[--] We saw, when discussing the Kuramoto 
      model in Sect.~\ref{syncho_coupled_oscillators},
      that generically not all nodes of a network
      participate in a synchronization process.
      For the Kuramoto model the oscillators with
      natural frequencies far away from the average
      do not become locked to the time development of
      the order parameter, see 
      Fig.~\ref{synchro_fig_kuramoto_Kr}, retaining
      drifting trajectories.
\item[--] Generically, synchronization takes the form
      of coherent time evolution with phase lags,
      we have seen an example when discussing
      two coupled oscillators in
      Sect.~\ref{syncho_coupled_oscillators}.
      The synchronized orbit is then
$$
      x_i(t) \ =\ x(t) + \Delta x_i,
      \qquad\quad
      \Delta x_i\ \hbox{const.}~,
$$
      viz the elements $i=1,\dots,N$ are all locked in.

\end{itemize}

\runinhead{Stability Analysis via the Second-Largest Lyapunov Exponent}
The stability of a globally synchronized state,
$x_i(t)= x(t)$ for $i=1,\, \dots,\ N$, can be 
determined by considering small perturbations,
viz
\begin{equation}
x_i(t) \ =\ x(t) \,+\, \delta_i\,c^t,
\qquad\quad
|c|^t=e^{\lambda t}~,
\label{synchro1_small_perturbation}
\end{equation}
where $\lambda$ is the Lyapunov exponent.
The eigenvectors $(\delta_1,\, \dots,\, \delta_N)$
of the perturbation are determined by the
equation of motion linearized around the synchronized
trajectory. There is one Lyapunov exponent for 
every eigenvector, $N$ in all:
$$
\lambda_1\ \ge\ \lambda_2\ \ge\ \lambda_3\ \ge\ \dots \ge \lambda_N~.
$$
In general the largest eigenvector $\lambda_1>0$ will 
correspond to the synchronized direction,
$$
\lambda_1,\qquad\quad 
\delta_i\ =\ \delta,
\qquad\quad
i=1,\, \dots,\, N~,
$$
corresponding to the dominant flow in phase space.
The second largest Lyapunov exponent determines hence
the stabilty of the synchronized orbit:
$$
(\lambda_2<0)
\qquad\quad
\Leftrightarrow
\qquad\quad
\hbox{stability}~,
$$
and vice versa.

\runinhead{Coupled Logistic Maps}
As an example we consider two coupled logistic maps,
see Fig.~\ref{fig_chaos_logistic},
\begin{equation}
x_i(t+1) \ =\ r\, \bar x_i(t)\,\big(1-\bar x_i(t)\big),
\qquad \quad
i=1,\, 2,
\qquad \quad
r\in[0,4]~,
\label{synchro1_coupled_logistic_maps}
\end{equation}
with
$$
\bar x_1= (1-\kappa) x_1 +\kappa x_2,
\qquad \quad
\bar x_2= (1-\kappa) x_2 +\kappa x_1 
$$
and $\kappa\in[0,1]$ being the coupling strength.
Using Eq.~(\ref{synchro1_small_perturbation})
as an Ansatz we obtain
$$
c
\left(
\begin{array}{c}
\delta_1 \\ \delta_2
\end{array}
\right) \ =\ r\big(1-2x(t)\big)
\left(
\begin{array}{cc}
(1-\kappa) & \kappa\\
\kappa & (1-\kappa)
\end{array}
\right) 
\left(
\begin{array}{c}
\delta_1 \\ \delta_2
\end{array}
\right)~,
$$
which determines $c$ as the eigenvalues
of the Jacobian of 
Eq.~(\ref{synchro1_coupled_logistic_maps}).
We have hence two local pairs of eigenvalues
and eigenvectors, namely
\begin{eqnarray*}
c_1 & =& r(1-2x)\phantom{(1-2\kappa)} 
\quad\qquad            (\delta_1,\delta_2)={1\over\sqrt 2}(1,1) \\
c_2 & =& r(1-2x)(1-2\kappa) 
\quad\qquad            (\delta_1,\delta_2)={1\over\sqrt 2}(1,-1) 
\label{synchro1_eigenvalues_coupled_logistic}
\end{eqnarray*}
corresponding to the respective local Lyapunov exponents,
$\lambda = \log|c|$,
\begin{equation}
\lambda_1 \ =\ \log|r(1-2x)|,
\qquad\quad
\lambda_2 \ =\ \log|r(1-2x)(1-2\kappa)|~.
\label{synchor1_coupled_logist_local_Lyapunov}
\end{equation}
As expected, $\lambda_1>\lambda_2$, since $\lambda_1$
corresponds to a perturbation along the synchronized
orbit.  The overall stability of the synchronized
trajectory can be examined by averaging above local
Lyapunov exponents over the full time development,
obtaining such the maximal Lyapunov exponent,
see Eq.~(\ref{chaos1_exponent_Lyapunov_maximal}).

\runinhead{Synchronization of Coupled Chaotic Maps}
The maximal Lyapunov exponent needs to be evaluated
numerically, but we can obtain an upper bound for 
the coupling strength $\kappa$ needed for stable 
synchronization by observing that
$|1-2x|\le 1$ and hence
$$
|c_2| \ \le\ r|1-2\kappa|~.
$$
The synchronized orbit is stable for $|c_2|<1$. 
Considering the case $\kappa\in[0,1/2]$
we find 
$$
1\ >\ r(1-2\kappa_s) \ \ge \ |c_2|,
\qquad\quad
\kappa_s \ >\ {r-1\over 2 r}
$$
for the upper bound for $\kappa_s$. The
logistic map is chaotic for 
$r>r_\infty\approx 3.57$ and above
result, being valid for all $r\in[0,4]$,
therefore proves that also chaotic 
coupled systems may synchronize.

For the maximal reproduction rate, $r=4$,
synchronization is guaranteed for 
$3/8<\kappa_s\le 1/2$. Note that $\bar x_1=\bar x_2$
for $\kappa=1/2$, synchronization through 
aggregate averaging is hence achieved in
one step for $\kappa=1/2$.

\section{Synchronization via Causal Signaling}
\label{syncho_causal_signaling}
\index{synchronization!causal signaling|textbf}

The synchronization of the limiting cycle oscillators 
discussed in Sect.~\ref{syncho_coupled_oscillators} 
is very slow, see Fig.~\ref{synchro_fig_harmonicSyncro}, 
as the information between the different oscillators 
is exchanged only indirectly via the molecular field, 
which is an averaged quantity. Synchronization may be 
sustantially faster, when the local dynamical units
influence each other with precisely timed signals, the
route to synchronization discussed here. 

Relaxational oscillators, like the van der Pol 
oscillator discussed in Chap.~\ref{chap_chaos1}
have a non-uniform cycle and the timing of
the stimulation of one element by another is 
important. This is a characteristic property of 
real-world neurons in particular and of many models 
of artificial neurons, like the so-called 
integrate-and-fire models. Relaxational oscillators
are hence well suited to study the phenomena of
synchronization via causal signaling.

\begin{figure}[t]
\centering
\includegraphics{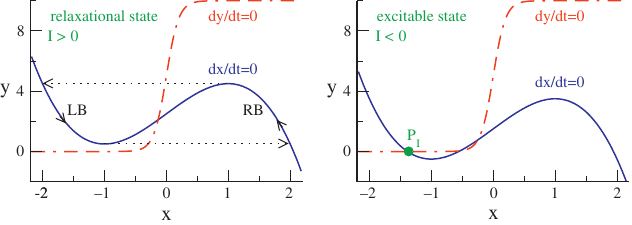}
\caption{The $\dot y=0$ (\textit{thick dashed-dotted lines}) and the
$\dot x=0$ (\textit{thick full lines}) isocline of the Terman--Wang
oscillator, Eq.~(\ref{synchro_Terman_Wang}), for $\alpha=5$,
$\beta=0.2$, $\epsilon=0.1$.  \textit{Left}: $I=0.5$ with the
limiting relaxational cycle for $\epsilon\ll1$ (\textit{thin dotted
line with arrows}). \textit{Right}: $I=-0.5$ with the stable
fixpoint: $P_I$} \label{synchro_fig_Terman_Wang_fg}
\end{figure}

\runinhead{Terman--Wang Oscillators} \index{Terman--Wang
oscillator}\index{oscillator!Terman--Wang} \index{relaxation
oscillator!Terman--Wang}There are many variants of relaxation
oscillators relevant for describing integrate-and-fire neurons,
starting from the classical Hodgkin--Huxley equations. Here we
discuss the particularly transparent dynamical system
introduced by Terman and Wang, namely
\begin{equation}
\begin{array}{rcl}
\dot x & =& f(x)-y+I \\[3pt]
\dot y & =& \epsilon\,\big(g(x)-y\big)
\end{array}
\qquad\quad
\begin{array}{rcl}
f(x) & =& 3x-x^3+2 \\[3pt]
g(x) & =& \alpha\,\big(1+\tanh(x/\beta)\big)
\end{array}~.
\label{synchro_Terman_Wang}
\end{equation}
Here $x$ corresponds in neural terms to the membrane
potential and
$I$ represents the external stimulation to the
neural oscillator.
The amount of dissipation is given by\vspace*{3pt}
$$
{\partial \dot x\over\partial x}+
{\partial \dot y\over\partial y} \ =\
3-3x^2-\epsilon \ =\ 3(1-x^2)-\epsilon~.\vspace*{3pt}
$$

\noindent For small $\epsilon\ll1$ the system takes up
energy for membrane potentials $|x|<1$ and
dissipates energy for $|x|>1$.\vspace*{3pt}

\runinhead{Fixpoints} \index{fixpoint!Terman--Wang oscillator}The
fixpoints are determined via
$$
\begin{array}{rcl}
\dot x&=&0\\[3pt]
\dot y&=&0
\end{array}
\qquad\quad
\begin{array}{rcl}
 y&=&f(x)+I\\[3pt]
 y&=&g(x)
\end{array}
\qquad\quad
$$

\noindent by the intersection of the two functions 
$f(x)+I$ and $g(x)$, see 
Fig.~\ref{synchro_fig_Terman_Wang_fg}. 
We find two parameter regimes:
\begin{itemize}
\item[--] For $I\ge0$ we have one unstable fixpoint $(x^*,y^*)$ with
$x^*\simeq0$.
\item[--] For $I<0$ and $|I|$ large enough we have two additional fixpoints
given by the crossing of the sigmoid
$\alpha\big(1+\tanh(x/\beta)\big)$ with the left branch (LB)
of the cubic $f(x)=3x-x^3+2$, with one fixpoint being stable.
\end{itemize}
The stable fixpoint $P_I$ is indicated in
Fig.~\ref{synchro_fig_Terman_Wang_fg}.\vspace*{3pt}

\looseness1\runinhead{{The} Relaxational Regime}
\index{relaxational regime}\index{regime!relaxational}For the case
$I>0$ the Terman--Wang oscillator relaxes in the long time limit to
a periodic solution, see Fig.~\ref{synchro_fig_Terman_Wang_fg},
which is very similar to the limiting relaxation oscillation of the
Van der Pol oscillator discussed in\break Chap.~\ref{chap_chaos1}.

\begin{figure}[t]
\centering
\includegraphics{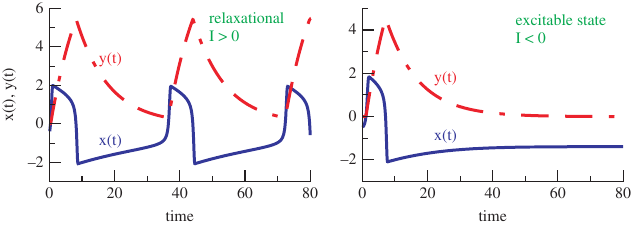}
\caption{Sample trajectories $y(t)$ (\textit{thick dashed-dotted
lines}) and $x(t)$ (\textit{thick full lines}) of
the Terman--Wang oscillator
Eq.~(\ref{synchro_Terman_Wang}) for $\alpha=5$,
$\beta=0.2$, $\epsilon=0.1$.
\textit{Left}: $I=0.5$ exhibiting spiking behavior,
having silent/active phases for negative/positive $x$.
\textit{Right}: $I=-0.5$, relaxing to the stable
fixpoint} \label{synchro_fig_Terman_Wang_trajectories}\vspace*{8pt}
\end{figure}

\runinhead{Silent and Active Phases} In its relaxational regime, the
periodic solution jumps very fast (for $\epsilon\ll1$) between
trajectories {that} approach closely the right
branch (RB) and the left branch (LB) of the $\dot x=0$ isocline. The
time development on the RB and the LB are, however, not symmetric,
see Figs.~\ref{synchro_fig_Terman_Wang_fg} and
\ref{synchro_fig_Terman_Wang_trajectories}, and we can distinguish
two regimes:

\begin{quotation}
 {{The} Silent Phase}.\enspace \index{Terman--Wang
oscillator!silent phase}\index{silent phase}\index{phase!silent}We
call the relaxational dynamics close to the LB ($x<0$) of the $\dot
x=0$ isocline the silent phase or the refractory period.
\end{quotation}
\begin{quotation}
{{The} Active Phase}.\enspace
\index{Terman--Wang oscillator!active phase}
\index{active phase}\index{phase!active}
We call the relaxational dynamics close to the RB ($x>0$) 
of the $\dot x=0$ isocline the active phase. 
\end{quotation}
%

\noindent The relative rate of the time development $\dot y$ in the
silent and active phases are determined by the parameter
$\alpha$, compare Eq.~(\ref{synchro_Terman_Wang}).

The active phase on the RB is far from the $\dot y=0$
isocline for $\alpha\gg1$, see
Fig.~\ref{synchro_fig_Terman_Wang_fg},
and the time development $\dot y$ is then fast. The silent phase on
the LB is, however, always close to the $\dot y=0$ isocline and the
system spends considerable time there.

\runinhead{The Spontaneously Spiking State and
the Separation of Time Scales} \index{Terman--Wang
oscillator!spiking state}\index{time scale separation}In its
relaxational phase, the Terman--Wang oscillator can therefore be
considered as a spontaneously spiking neuron, see
Fig.~\ref{synchro_fig_Terman_Wang_trajectories}, with the spike
corresponding to the active phase, which might be quite short
compared to the silent phase for\break $\alpha\gg1$.

The Terman--Wang differential equations
(\ref{synchro_Terman_Wang}) are examples of a standard technique
within dynamical system theory, the coupling of a slow variable,
$y$, to a fast variable, $x$, which results in a separation of time
scales. When the slow variable $y(t)$ relaxes below a certain
threshold, see Fig.~\ref{synchro_fig_Terman_Wang_trajectories}, the
fast variable $x(t)$ responds rapidly and resets the slow variable.
We will encounter further applications of this procedure in
Chap.~\ref{chap_cogSys1}.

\begin{figure}[t]
\centering
\includegraphics{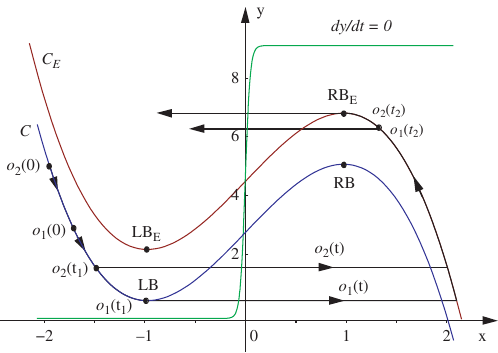}
\caption{Fast threshold modulation for two excitatory coupled
Terman--Wang oscillators,
{Eq.~}(\ref{synchro_Terman_Wang}) $o_1=o_1(t)$ and $o_2=o_2(t)$,
which start at time $0$. When $o_1$ jumps at $t=t_1$ the cubic $\dot
x=0$ isocline for $o_2$ is raised from $C$ to $C_E$. This induces
$o_2$ to jump as well. Note that the jumping from the right branches
($RB$ and $RB_E$) back to the left branches occurs in the reverse
order: $o_2$ jumps first (from Wang, 1999)}
\label{synchro_fig_Terman_Wang_2}
\end{figure}

\runinhead{The Excitable State}
\index{Terman-Wang oscillator!excitable state}The neuron has an
additional phase with a stable fixpoint $P_I$ on the LB (within the
silent region), for negative external stimulation (suppression)
$I<0$. The dormant state at the fixpoint $P_I$ is \qut{excitable}: A
positive external stimulation above a  small threshold will force a
transition into the active phase, with the neuron spiking
continuously.

\runinhead{Synchronization via Fast Threshold Modulation}
\index{relaxation oscillator!synchronization}
\index{synchronization!relaxation oscillator}\index{fast threshold
modulation} Limit cycle oscillators can synchronize, albeit slowly,
via the common molecular field, as discussed in
Sect.~\ref{syncho_coupled_oscillators}. A much faster
synchronization can be achieved via {\em fast threshold
synchronization} for a network of interacting relaxation
oscillators.

The idea is simple. Relaxational oscillators have 
distinct states during their cycle; we called them the
\qut{silent phase} and the \qut{active phase} for the
case of the\break Terman--Wang oscillator. We then 
assume that a neural oscillator in its (short) 
active phase changes the threshold $I$ of
the other neural oscillator in
Eq.~\ref{synchro_Terman_Wang} as
$$
I\ \to\ I\,+\,\Delta I,
\qquad\quad
\Delta I\,>\,0~,
$$
\noindent such that the second neural oscillator 
changes from an excitable state to the oscillating 
state. This process is illustrated graphically in 
Fig.~\ref{synchro_fig_Terman_Wang_2}; it corresponds
to a signal send from the first to the second dynamical
unit. In neural terms: when the first neuron fires, 
the second neuron follows suit.

\runinhead{Propagation of Activity}
We consider a simple model

\begin{figure}[t]
\centering
\includegraphics{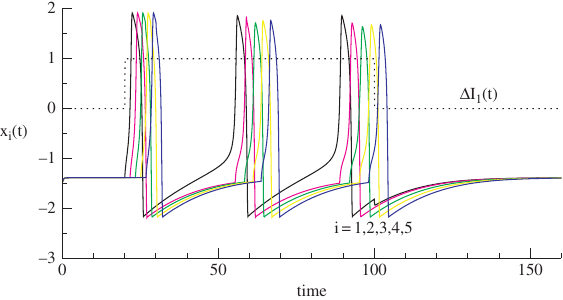}
\caption{Sample trajectories $x_i(t)$ (\textit{lines}) 
for a line of coupled Terman--Wang oscillators, an 
example of synchronization via causal signaling. 
The relaxational oscillators are in excitable states,
see Eq.~(\ref{synchro_Terman_Wang}), with
$\alpha=10$, $\beta=0.2$, $\epsilon=0.1$ and $I=-0.5$.
For $t\in[20,100]$ a driving current $\Delta I_1=1$ 
is added to the first oscillator. $x_1$ then starts 
to spike, driving the other oscillators one by one
via a fast threshold modulation.}
\label{synchro1_fig_fastThresholdModulation}
\end{figure}

\smallskip
\centerline{\Large \framebox{\ 1\ }\hspace{1.5ex} $\Rightarrow$
\hspace{1.5ex} \framebox{\ 2\ }\hspace{1.5ex} $\Rightarrow$
\hspace{1.5ex} \framebox{\ 3\ }\hspace{1.5ex} $\Rightarrow$
\hspace{1.5ex}$\dots$
           }
\smallskip

\noindent of $i=1,\ldots,N$ coupled oscillators $x_i(t)$, $y_i(t)$,
all being initially in the excitable state with $I_i\equiv -0.5$.
They are coupled via fast threshold modulation, specifically via
\begin{equation}
\Delta I_i(t)\ =\ \Theta(x_{i-1}(t))~,
\label{synchro1_Eq_fastThresholdModulation}
\end{equation}
where $\Theta(x)$ is the Heaviside step function. That is, we define
an oscillator $i$ to be in its active phase whenever $x_i>0$. The
resulting dynamics is shown in
Fig.~\ref{synchro1_fig_fastThresholdModulation}. The chain is driven
by setting the first oscillator of the chain into the spiking state
for a certain period of time. All other oscillators start to spike
consecutively in rapid sequence.

\section[Synchronization and Object Recognition in Neural Networks]
{Synchronization and Object Recognition\newline in Neural Networks}
\label{syncho_object_recognition}
\index{synchronization!object recognition|textbf}
\index{neural network!synchronization|textbf}

Synchronization phenomena can be observed in many realms 
of the living world. As an example we discuss here the
hypothesis of object definition via synchronous 
neural firing, a proposal by Singer and von der Malsburg
which is at the same time both fascinating and controversial.

\runinhead{Temporal Correlation Theory}
\index{temporal correlation theory}\index{correlation!temporal}
The neurons in the brain have time-dependent activities and can be
described by generalized relaxation oscillators, as outlined
in the previous Section. The \qut{temporal correlation theory}
assumes that not only the average activities of
individual neurons (the spiking rate) are important,
but also the relative phasing of the individual spikes.
Indeed, experimental evidence supports the notion of
object definition in the visual cortex via synchronized firing.
In this view neurons encoding the individual constituent
parts of an object, like the mouth and the eyes of a face,
fire in tact. Neurons being activated simultaneously by
other objects in the visual field, like a camera, would
fire independently.

\runinhead{The LEGION Network of Coupled Relaxation Oscillators}
\index{LEGION network}\index{network!LEGION} As an example of how
object definition via coupled relaxation oscillators can be achieved
we consider the LEGION (local excitatory globally inhibitory
oscillator network) network by Terman and Wang. Each oscillator $i$
is defined as
\begin{equation}
\begin{array}{rcl}
\dot x_i & =& f(x_i)-y_i+I_i+S_i+\rho \\
\dot y_i & =& \epsilon\,\big(g(x_i)-y_i\big)
\end{array}
\qquad\quad
\begin{array}{rcl}
f(x) & =& 3x-x^3+2 \\
g(x) & =& \alpha\,\big(1+\tanh(x/\beta)\big)
\end{array}
~.
\label{synchro_eq_LEGION}
\end{equation}
There are two terms in addition to the ones necessary
for the description of a single oscillator,
compare Eq.~(\ref{synchro_Terman_Wang}):

\begin{tabular}{rcl}
$\rho$ &: & a random-noise term and \\
$S_i$ &: & the interneural interaction.
\end{tabular}

\noindent
The interneural coupling in
Eq.~(\ref{synchro_eq_LEGION})
occurs exclusively via the
modulation of the threshold, the
three terms $I_i+S_i+\rho$
constitute an effective threshold.

\runinhead{Interneural Interaction}
The interneural interaction is given for the LEGION network by
\begin{equation}
S_i\ =\ \sum_{l\in N(i)} T_{il}\,\Theta(x_l-x_c)
\,-\, W_z\Theta(z-z_c)~,
\label{synchro_LEGION_S}
\end{equation}
where $\Theta(z)$ is the Heaviside step function. The parameters
have the following \hbox{meaning:}

\begin{tabular}{rcl}
$T_{il}>0$ &:& Interneural excitatory couplings. \\
$N(i)$     &:& Neighborhood of neuron $i$.\\
$x_c$      &:& Threshold determining the active phase.\\
$z$        &:& Variable for the global inhibitor.\\
$-W_z<0$   &:& Coupling to the global inhibitor $z$.\\
$z_c$      &:& Threshold for the global inhibitor.
\end{tabular}

\begin{figure}[t]
\centering
\includegraphics{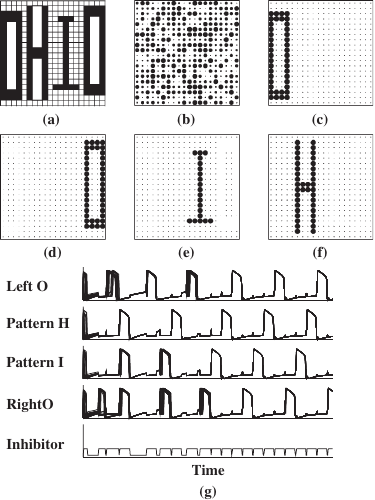}
\caption{(\textbf{a}) A pattern used to stimulate a $20\times 20$
LEGION network.  (\textbf{b}) Initial random activities of the
relaxation oscillators.  (\textbf{c}, \textbf{d}, \textbf{e},
\textbf{f}) Snapshots of the activities at different sequential
times.
(\textbf{g}) The corresponding time-dependent activities of selected
oscillators and of the global inhibitor (from Wang, 1999)}
\label{synchro_fig_LEGION_1}
\vspace*{-12pt}
\end{figure}

\runinhead{{Global Inhibition}}
\index{inhibition!global}Global inhibition is a quite generic
strategy for neural networks with selective gating capabilities. A
long-range or global inhibition term assures that only one or only a
few of the local computational units are active coinstantaneously.
In the context of the Terman--Wang LEGION network it is assumed to
have the dynamics
\begin{equation}
\dot z\ =\ \left(\sigma_z-z\right)\phi,
\qquad\qquad \phi\,>\,0~,
\label{synchro_LEGION_z}
\end{equation}
where the binary variable $\sigma_z$ is determined
by the following rule:

\begin{description}
\item[$\sigma_z=1$] if at least one oscillator is active.
\item[$\sigma_z=0$] if all oscillators are silent or in
the excitable state.
\end{description}

This rule is very non-biological, the LEGION network is just a proof
of {the} principle for object definition via fast
synchronization. When at least one oscillator is in its active phase
the global inhibitor is activated, $z\to 1$, and inhibition is
turned off whenever the network is completely inactive.

\runinhead{Simulation of the LEGION Network}
A simulation of a $20\times20$ LEGION network is presented in
Fig.~\ref{synchro_fig_LEGION_1}. We observe {the
following}:

\begin{itemize}
\item[--] The network is able to discriminate between different
      input objects.
\item[--] Objects are characterized by the coherent activity of the
      corresponding neurons, while neurons not belonging to the
      active object are in the excitable state.
\item[--] Individual input objects pop up randomly one after
      the other.
\end{itemize}

\runinhead{Working Principles of the LEGION Network}
\index{LEGION network!working principles}
The working principles of the LEGION network are
the following:

\begin{itemize}
\item[--] When the stimulus begins there will be a single
  oscillator $k$, which will jump first into the active phase,
  activating the global inhibitor{,
  Eq.~}(\ref{synchro_LEGION_z}),
  via $\sigma_z\to1$. The noise term $\sim\rho$ in
  Eq.~(\ref{synchro_eq_LEGION}) determines the
  first active unit randomly from~the set of all
  units receiving an input signal $\sim I_i$, whenever
  all input signals have the~same strength.
\item[--] The global inhibitor then suppresses the activity
      of all other oscillators, apart from the
      stimulated neighbors of $k$, which {also jump} into
      the active phase, having set the parameters
      such that
        $$
        I+T_{ik}-W_z\ >\ 0, \qquad\quad I:\ {\rm stimulus}
        $$
            is valid. The additional condition
        $$
        I-W_z\ <\ 0
        $$
            assures, that units receiving an input, but
            not being topologically connected to
            the cluster of active units, are suppressed.
            No two distinct objects can {then} be
            activated coinstantaneously.
\item[--] This process continues until all oscillators
            representing the stimulated pattern are active.
        As this process is very fast, all active oscillators
            fire nearly simultaneously, compare also
            Fig.~\ref{synchro1_fig_fastThresholdModulation}.
\item[--]  When all oscillators in a pattern oscillate in phase,
       they {}also jump back to the silent state simultaneously.
       At that point the global inhibitor is turned off:
         $\sigma_z\to0$ in Eq.~(\ref{synchro_LEGION_z})
         and the game starts {} again
         with a different pattern.
\end{itemize}

\runinhead{Discussion} Even though the network {nicely performs} its task of object recognition via coherent
oscillatory firing, there are a few aspects worth noting:

\begin{itemize}
\item[--] The functioning of the network depends on the
      global inhibitor {} triggered by the specific
      oscillator {that} jumps first. This might be
      difficult to realize in biological networks,
      like the visual cortex,
      which do not have well defined boundaries.
\item[--] The first active oscillator {sequentially recruits} all other
      oscillators belonging to its pattern. {}
This happens very fast via the mechanism of rapid threshold
modulation. The synchronization is therefore not a collective
process in which the input data is processed in
parallel{;} a property assumed to be important for
biological networks.
\item[--] The recognized pattern remains active {for exactly}
      one cycle and no longer.
\end{itemize}

\noindent
We notice, however, that the design of neural
networks capable of fast synchronization via a
collective process remains a challenge, since
collective processes have an inherent tendency
towards slowness, due to the need to
exchange information, e.g.\ via molecular fields.
Without reciprocal information exchange, a true collective
state, as an emergent property of the
constituent dynamical units, is not possible.



\section{Synchronization Phenomena in Epidemics}
\label{syncho_epidemics}
\label{synchronization!epidemics|textbf}

\begin{figure}[t]
\centering
\includegraphics{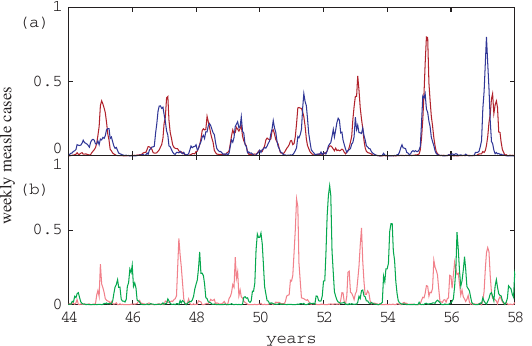}
\caption{Observation of the number of infected persons in a study on
illnesses.  (\textbf{a}) Weekly cases of measle cases in Birmingham
(\textit{red line}) and Newcastle (\textit{blue line}).
 (\textbf{b}) Weekly cases of measle cases in Cambridge
(\textit{green line}) and {in} Norwich (\textit{pink
line}) (from He, 2003)} \label{synchro_fig_epidemics_cities}\vspace*{-8pt}
\end{figure}

There are illnesses, like measles, {that} come and
go recurrently. Looking at the local statistics of measle outbreaks,
see Fig.~\ref{synchro_fig_epidemics_cities}, one can observe that
outbreaks occur in quite regular time intervals within a given city.
Interestingly though, these outbreaks can be either in phase
(synchronized) or out of phase between different cities.

The oscillations in the number of infected persons are 
definitely not harmonic, they share many characteristics 
with relaxation oscillations, which typically have
silent and active phases, compare
Sect.~\ref{syncho_causal_signaling}.

\runinhead{The SIRS Model}
\index{SIRS model}
\index{model!SIRS}
A standard approach to model the dynamics of
infectious diseases is the SIRS model. At
any time an individual can belong to one of
the three classes:
\smallskip

\begin{tabular}{rcp{0.90\hsize}}
S &:& susceptible,\\
I &:& infected, \\
R &:& recovered.
\end{tabular}
\smallskip

\noindent
The dynamics is governed by the following rules:
\begin{enumerate}\leftskip5pt

\item[(a)] Susceptibles pass to the infected state, with a
      certain probability, after coming
 into contact with one infected individual.
\item[(b)] Infected individuals pass to the recovered state after
      a fixed period of time $\tau_I$.
\item[(c)] Recovered individuals return to the susceptible state
      after a recovery time $\tau_R$,
      when immunity is lost,
      and the S$\to$I$\to$R$\to$ S cycle is complete.
\end{enumerate}
When $\tau_I\to\infty$ (lifelong immunity) the model reduces
to the SIR-model.

\runinhead{{The} Discrete Time Model}
\index{dynamics!discrete time}We consider a discrete time SIRS
model with $t=1,2,3,\ldots$ and $\tau_I=1$: The infected phase
{is} normally short and we can use it to set the unit
of time. The recovery time $\tau_R$ is then a multiple of
$\tau_I=1$.

\noindent
We define with
\smallskip

\begin{figure}[t]
\vspace*{12pt}\centering
\includegraphics{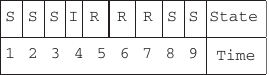}
\caption{Example of the course of an individual infection within the
SIRS model with an infection time $\tau_I=1$ and a recovery time
$\tau_R=3$. The number of individuals recovering at time $t$ is just
the sum of infected individuals at times $t-1$, $t-2$ and $t-3$,
compare Eq.~(\ref{synchro_s_t_x})} \label{synchro_fig_SIRS_example}
\end{figure}

\begin{tabular}{cp{0.90\hsize}}
$x_t$ & the fraction of infected individuals at time $t$, \\
$s_t$ & the percentage of susceptible individuals at time $t$,
\end{tabular}
\smallskip

\noindent
which obey
\begin{equation}
s_t \ =\ 1-x_t-\sum_{k=1}^{\tau_R} x_{t-k}
    \ =\ 1-    \sum_{k=0}^{\tau_R} x_{t-k}~,
\label{synchro_s_t_x}
\end{equation}

\noindent as the fraction of susceptible individuals is just {1}
minus the number of infected individuals minus the number of
individuals in the recovery state, compare
Fig.~\ref{synchro_fig_SIRS_example}.

\begin{figure}[t]
\centering
\includegraphics{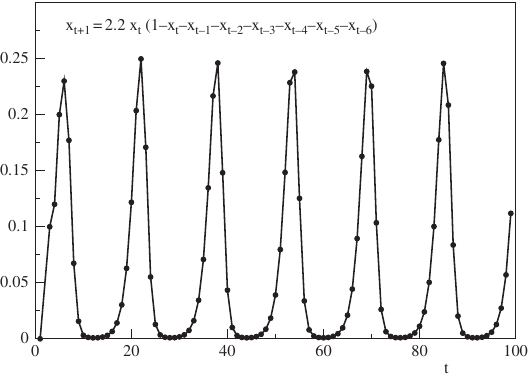}
\caption{Example of a solution to the SIRS model,
Eq.~(\ref{synchro_infection_transmission}), for $\tau_R=6$. The
number of infected individuals might drop to very low values during
the silent phase in between two outbreaks as most of the population
is first infected and then immunized during an outbreak}
\label{synchro_fig_SIRS_1}
\vspace*{-8pt}
\end{figure}

\runinhead{{The} Recursion Relation}
\index{SIRS model!recursion relation}We denote with $a$ the rate of
transmitting an infection when there is a contact between an
infected {individual} and a susceptible individual:
\begin{equation}
x_{t+1} \ =\ ax_ts_t \ =\ a x_t
\left(1-\sum_{k=0}^{\tau_R} x_{t-k}\right)~.
\label{synchro_infection_transmission}
\end{equation}
\runinhead{Relation to the Logistic Map}
\index{SIRS model!logistic map}\index{logistic map!SIRS model}
\index{SIRS model!logistic map}For $\tau_R=0$ the discrete time
SIRS model {}
(\ref{synchro_infection_transmission}) reduces to the logistic map
$$
x_{t+1} \ =\ a x_t\left(1-x_t\right)~,
$$
which we studied in Chap.~\ref{chap_chaos1}.
For $a<1$ it has only the trivial fixpoint
$x_t\equiv0$, the illness dies out. The non-trivial
steady state~is
$$
x^{(1)} \ =\ 1-{1\over a},
\qquad\quad {\rm for} \quad 1<a<3~.
$$
For $a=3$ there is a Hopf bifurcation and for $a>3$ the system
oscillates with a period of {2}. Equation
(\ref{synchro_infection_transmission}) has a similar behavior, but the
resulting oscillations may depend on the initial condition and
{for
$\tau_R\gg\tau_I\equiv1$ show} features characteristic of relaxation
oscillators, see Fig.~\ref{synchro_fig_SIRS_1}.

\runinhead{Two Coupled Epidemic Centers}
\index{SIRS model!coupled}
We consider now two epidemic centers
with variables
$$
s_t^{(1,2)}, \qquad\quad x_t^{(1,2)}~,
$$
denoting the fraction of susceptible/infected individuals
in the respective cities. Different dynamical couplings
are conceivable, via exchange or visits of susceptible or
infected individuals. We consider with
\begin{equation}
x_{t+1}^{(1)} \ =\ a\left(x_t^{(1)}+e\,x_t^{(2)}\right)s_t^{(1)},
\qquad\quad
x_{t+1}^{(2)} \ =\ a\left(x_t^{(2)}+e\,x_t^{(1)}\right)s_t^{(2)}
\label{synchro_SIRS_coupling}
\end{equation}
the visit of a small fraction $e$ of infected individuals to the
other center. Equation (\ref{synchro_SIRS_coupling}) determines the
time evolution of the epidemics together with
Eq.~(\ref{synchro_s_t_x}), generalized to both centers.
For $e=1$ there is no distinction between the two centers
anymore and their dynamics can be merged via
$x_t=x_t^{(1)}+x_t^{(2)}$ and
$s_t=s_t^{(1)}+s_t^{(2)}$ to the one of a single center.

\runinhead{In Phase Versus Out of Phase Synchronization}
\index{synchronization!in phase vs.\ out of phase}We have seen in
Sect.\ \ref{syncho_coupled_oscillators} that a strong coupling of
relaxation oscillators during their active phase leads in a quite
natural way to a fast synchronization. Here the active phase
corresponds to an outbreak of the illness and
Eq.~(\ref{synchro_SIRS_coupling}) indeed implements a coupling
equivalent to the fast threshold modulation discussed in
Sect.~\ref{syncho_causal_signaling}, since the coupling is
proportional to the fraction of infected individuals.

In Fig.~\ref{synchro_fig_SIRS_coupled} we present the results from a
numerical simulation of the coupled model, illustrating the typical
behavior. We see that the outbreaks of epidemics in the SIRS model
{indeed occur} in phase for a moderate to
large coupling constant $e$. For very small coupling $e$ between the
two centers of epidemics {on the other hand, the synchronization becomes}
antiphase, as {is sometimes observed}
in reality, see Fig.~\ref{synchro_fig_epidemics_cities}.

\runinhead{Time Scale Separation}
\index{time scale separation!SIRS model}The reason for the
occurrence of out of phase synchronization is the emergence of two
separate time scales in the limit $t_R\gg1$ and $e\ll1$. A small
seed $\sim e a x^{(1)} s^{(2)}$ of infections in the second city
needs substantial time to induce a full-scale outbreak, even via
exponential growth, when $e$ is too small. But in order to remain in
phase with the current outbreak in the first city the outbreak
occurring in the second city may not lag too far behind. When the
dynamics is symmetric under exchange $1\leftrightarrow2$ the system
then settles in antiphase cycles.

\vspace*{8pt}
\section*{Exercises}
\addcontentsline{toc}{section}{Exercises}
{\sc The Driven Harmonic Oscillator}
\begin{list}{}
\item Solve the driven harmonic oscillator,
Eq.~(\ref{synchro_driven_oscillator}), for all times $t$
and compare it with the long time solution $t\to\infty$,
Eqs.~(\ref{synchro_Ansatz_driven_oscillator}) and
(\ref{synchro_oscillator_resonance}).
\end{list}

\hspace*{-12pt}{\sc Self-Synchronization}
\begin{list}{}
\item Consider an oscillator with feedback,
$$
\dot\theta(t)\ =\ \omega_0\,+\,K\sin[\theta(t-T)-\theta(t)]~.
$$
Discuss the self-synchronization in analogy to
Sect.~\ref{syncho_time_delay}, the stability of
the steady-state solutions and the auto-locking
frequencies in the limit of strong self-coupling
$K\to\infty$.
\end{list}

\begin{figure}[t]
\centering
\includegraphics{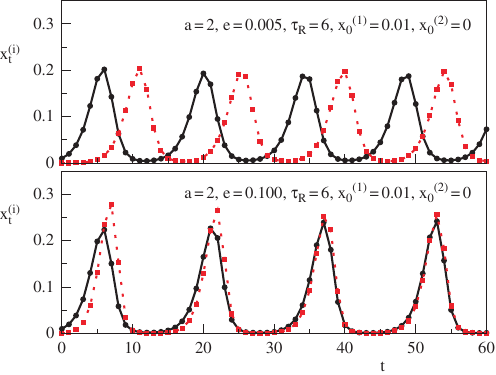}
\caption{Time evolution of the fraction of infected individuals
$x^{(1)}(t)$ and $x^{(2)}(t)$ within the SIRS model,
Eq.~(\ref{synchro_SIRS_coupling}), for two epidemic centers $i=1,2$
with recovery times $\tau_R=6$ and infection rates $a=2$, see
Eq.~(\ref{synchro_infection_transmission}). For a very weak coupling
$e=0.005$ (\textit{top}) the outbreaks occur out of phase, for a
moderate coupling $e=0.1$ (\textit{bottom}) in phase }
\label{synchro_fig_SIRS_coupled}
\end{figure}

\hspace*{-12pt}{\sc Synchronization of Chaotic Maps}
\index{Bernoulli shift map} 
\begin{list}{}
\item The Bernoulli shift map $f(x)=ax\ \hbox{mod}\ 1$ 
with $x\in[0,1]$ is chaotic for $a>1$. Consider with
\begin{equation}
\begin{array}{rcl}
x_1(t+1)& =& f\Big((1-\kappa)x_1(t)+\kappa x_2(t-T)\Big) \\
x_2(t+1)& =& f\Big((1-\kappa)x_2(t)+\kappa x_1(t-T)\Big)
\end{array}
\label{synchro_coupled_Bernoulli}
\end{equation}
two coupled chaotic maps, with $\kappa\in[0,1]$ being the
coupling strength and $T$ the time delay, compare
Eq.~(\ref{synchro1_network_AV_def}). Discuss the
stability of the synchronized states $x_1(t)=x_2(t)\equiv \bar x(t)$ 
for general time delays $T$. What drives the synchronization
process?
\end{list}

\hspace*{-12pt}{\sc The Terman--Wang Oscillator}
\begin{list}{}
\item Discuss the stability of the fixpoints of the Terman--Wang
oscillator, Eq.~(\ref{synchro_Terman_Wang}). Linearize the
differential equations around the fixpoint solution and consider
the limit $\beta \to  0$.
\end{list}

\hspace*{-12pt}{\sc The SIRS Model -- Analytical}
\begin{list}{}
\item Find the fixpoints $x_t\equiv x^*$ of the SIRS model{,
Eq.~}(\ref{synchro_infection_transmission}), for all $\tau_R$, as a
function of $a$ and study their stability for $\tau_R=0,1$.
\end{list}

\hspace*{-12pt}{\sc The SIRS Model -- Numerical}
\begin{list}{}
\item
Study the SIRS model{, Eq.~}
(\ref{synchro_infection_transmission}), numerically for various
parameters $a$ and $\tau_R=0,1,2,3$. Try to reproduce
Figs.~\ref{synchro_fig_SIRS_1} and
\ref{synchro_fig_SIRS_coupled}.
\end{list}


\def\refer#1#2#3#4#5#6{\item{\frenchspacing\sc#1}\hspace{4pt}
                       #2\hspace{8pt}#3 {\it\frenchspacing#4} {\bf#5}, #6.}
\def\bookref#1#2#3#4{\item{\frenchspacing\sc#1}\hspace{4pt}
                     #2\hspace{8pt}{\it#3}  #4.}

\addcontentsline{toc}{section}{Further Reading} 

\enlargethispage{-24pt}

\section*{Further Reading}

\markboth{\thechapter\enspace Synchronization Phenomena}{Further Reading}

A nice review of the Kuramoto model, together with historical
annotations, has been published by Strogatz (2000), 
for a textbook containing many examples of synchronization
see Pikovsky {\it et al.} (2003).
Some of the material discussed in this chapter requires 
a certain background in theoretical neuroscience, 
see e.g.\ Dayan and Abbott (2001).

We recommend that the interested reader takes a look 
at some of the original research literature, such as 
the exact solution of the Kuramoto (1984) model, 
the \hbox{Terman} and Wang (1995) relaxation oscillators, 
the concept of fast threshold synchronization (Somers and Kopell,
1993), the temporal correlation hypothesis for cortical networks
(von der Malsburg and Schneider, 1886), and its
experimental studies (Gray et al., 1989), the LEGION network (Terman
and Wang, 1995), the physics of synchronized clapping (N\'eda et
al., 2000a, b) and synchronization phenomena within
the SIRS model of epidemics (He and Stone, 2003). For
an introductory-type article on synchronization with
delays see (D'Huys et al, 2008).

{\baselineskip=15pt
\begin{list}{}{\leftmargin=2em \itemindent=-\leftmargin%
\itemsep=3pt \parsep=0pt \small}
\bookref{Dayan, P., Abbott, L.F.}{2001}{Theoretical Neuroscience:
Computational and Mathematical Modeling of Neural Systems.}{MIT
Press, Cambridge}

\refer{D$^\prime$Huys, O.,Vicente, R., Erneux, T., 
       Danckaert, J., Fischer, I.}{2008}
{Synchronization properties of network motifs:
         Influence of coupling delay and symmetry.}
  {Chaos}{18}{037116}
\refer{Gray, C.M., K\"onig, P., Engel, A.K., Singer, W.}{1989}
{Oscillatory responses in cat visual cortex exhibit incolumnar
synchronization which reflects global stimulus properties.}
{Nature}{338}{334--337}
\refer{He, D., Stone, L.}{2003}{Spatio-temporal synchronization of
recurrent epidemics.} {Proceedings of the Royal Society London
B}{270}{1519--1526}
\bookref{Kuramoto, Y.}{1984}{Chemical Oscillations,
Waves and Turbulence.}{Springer, Berlin}
\refer{N\'eda, Z., Ravasz, E., Vicsek, T., Brechet, Y., 
       Barab\'asi, A.L.}
       {2000a}{Physics of the rhythmic applause.}
{Physical Review E}{61}{6987--6992}
\refer{N\'eda, Z., Ravasz, E., Vicsek, T., Brechet, 
       Y., Barab\'asi, A.L.}
       {2000b}{The sound of many hands clapping.}
       {Nature}{403}{849--850}
\bookref{Pikovsky, A., Rosenblum, M., Kurths, J.}{2003}
{Synchronization: A Universal Concept in Nonlinear Sciences.}
{Cambridge University Press}
\refer{Somers, D., Kopell, N.}{1993}{Rapid synchronization through
fast threshold modulation.} {Biological Cybernetics}{68}{398--407}
\refer{Strogatz, S.H.}{2000}{From Kuramoto to Crawford: Exploring
the onset of synchronization in populations of coupled oscillators.}
{Physica D}{143}{1--20}
\refer{Strogatz, S.H.}{2001}{Exploring complex networks.}
{Nature}{410}{268--276}
\refer{Terman, D., Wang, D.L.}{1995}{Global competition and local
cooperation in a network of neural oscillators.} {Physica
D}{81}{148--176}
\refer{von der Malsburg, C., Schneider, W.}{1886}{A neural
cocktail-party processor.} {Biological Cybernetics}{54}{29--40}
\bookref{Wang, D.L.}{1999}{Relaxation oscillators and networks.} {In
Webster, J.G. (ed.) {\it Encyclopedia of Electrical and Electronic
Engineers}, pp.\ 396--405, Wiley, New York}
\end{list}
\par}
 

\vspace{-20ex}
\chapter{Elements of Cognitive {Systems} Theory}
\label{chap_cogSys1}

\vspace{-12pt}
\abstract{The brain is without doubt the most complex adaptive
system known to humanity, arguably also a complex system about
which we know very little.\\
\hspace*{12pt} Throughout this book we have considered and developed
general guiding principles for the understanding of complex networks
and their dynamical properties; principles and concepts transcending
the details of specific layouts realized in real-world complex
systems. We follow the same approach here, considering the brain as
just one example of what is called a cognitive system, a specific
instance of what one denotes, cum grano salis, a living dynamical
system.\\
\hspace*{12pt} In the first part we will treat general
layout considerations concerning dynamical organizational
principles, an example being the role of diffuse controlling and
homeostasis for stable long-term cognitive information processing.
Special emphasis will be given to the motivational problem -- how
the cognitive system decides what to do -- in terms of survival
parameters of the living dynamical system and the so-called
emotional diffusive control.\\
\hspace*{12pt} In the second part we will discuss two
specific generalized neural networks implementing various aspects of
these general principles: a dense and homogeneous
associative network (dHAN) for environmental data representation and
associative thought processes, and the simple recurrent network
(SRN) for concept extraction from universal prediction tasks.}

\enlargethispage*{12pt}

\section{Introduction}
\label{cogSys_introduction}
We start with a few basic considerations
concerning the general setting.

\begin{figure}[t]
\centerline{\includegraphics{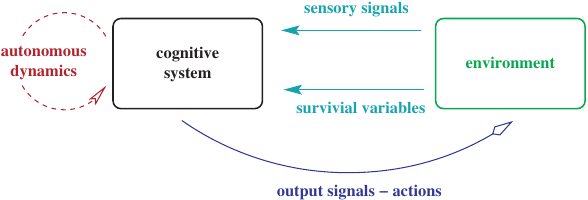}}
\caption{A cognitive system is placed in an environment
         (compare Sect.~\ref{cogSys_memory}) from
         which it receives two kinds of signals. The
         status of the survival parameters,
         which it needs to regulate
         (see Sect.~\ref{cogSys_benchmarks}),
         and the standard sensory input. The cognitive system
         generates output signals via its autonomous dynamics,
         which act back onto the outside world, viz
         the environment}
\label{cogSys_fig_CS_illust} 
\end{figure}

\runinhead{What is a Cognitive System?} A cognitive system may be
either biological, like the brain, or artificial. It is, in both
instances, a dynamical system embedded into an environment, with
which it mutually interacts.
\begin{quotation}
{\it Cognitive Systems.\enspace}
\index{cognitive system}
A cognitive system is a continuously active complex adaptive
system autonomously exploring and reacting to the environment with
the capability to \qut{survive}.
\end{quotation}
For a cognitive system, the only information source about the
outside is given, to be precise, by its sensory data input stream,
viz the changes in a subset of variables triggered by biophysical
processes in the sensory organs or sensory units. The cognitive
system does therefore not react directly to environmental events but
to the resulting changes in the sensory data input stream, compare
Fig.~\ref{cogSys_fig_CS_illust}.

\runinhead{Living Dynamical Systems} A cognitive system is an
instance of a living dynamical system, being dependent on a
functioning physical support unit, the body. The cognitive system is
terminated when its support unit ceases to work properly.
\begin{quotation}{\it Living Dynamical Systems.\enspace}
\index{dynamical system!living}\index{living dynamical system} A
dynamical system is said to \qut{live} in an abstract sense if it
needs to keep the ongoing dynamical activity in certain parameter
regimes.
\end{quotation}
As an example we consider a dynamical
variable $y(t)\ge0$,
part of the cognitive system, corresponding
to the current amount of pain or hunger.
This variable could be directly set by the
physical support unit, i.e.\ the body, of the
cognitive system, telling the dynamical system
about the status of its support unit.

The cognitive system can influence the value of $y(t)$ indirectly
via its motor output signals, activating its actuators, e.g.\ the
limbs. These actions will, in general, trigger changes in the
environment, like the uptake of food, which in turn will influence
the values of the respective survival variables. One could then
define the termination of the cognitive system when $y(t)$
surpasses a certain threshold $y_c$. The system
\qut{dies} when $y(t)>y_c$. These issues will be treated
in depth in Sect.~\ref{cogSys_benchmarks}.

\enlargethispage{12pt}

\runinhead{Cognition Versus Intelligence} \index{artificial
intelligence!vs.\ cognition} A cognitive system is not necessarily
intelligent, but it might be in principle. Cognitive system theory
presumes that artificial intelligence can be achieved only once
autonomous cognitive systems have been developed. This stance {is
somewhat in contrast} with the usual paradigm of artificial
intelligence (AI), which follows an all-in-one-step approach to
intelligent\break systems.

\runinhead{Universality} \index{universality!cognitive systems}
Simple biological cognitive systems are dominated by cognitive
capabilities and algorithms hard-wired by gene expression. These
features range from simple stimulus--response reactions to
sophisticated internal models for limb \hbox{dynamics}.

A priori information is clearly very useful for task solving in
particular and for cognitive systems in general. A main research
area in AI is therefore the development of efficient algorithms
making maximal use of a priori information about the environment. A
soccer-playing robot {normally does not}
acquire the ball dynamics from individual experience. Newton's law
is given to the robot by its programmer and hard-wired within its
code lines.

Cognitive system theory examines, on the other hand, universal
principles and algorithms necessary for the realization of an
autonomous cognitive system. This chapter will be devoted to the
discussion and
possible implementations of such
universal principles.

A cognitive system should therefore be able to operate in a wide
range of environmental conditions, performing tasks of different
kinds. A rudimentary cognitive system does not need to be efficient.
Performance boosting specialized algorithms can always be added
afterwards.

\runinhead{A Multitude of Possible Formulations} Fully
functional autonomous cognitive systems may possibly have very
different conceptual foundations. The number of consistent
approaches to cognitive system theory is not known, it may be
substantial. This is a key difference to other areas of research
treated in this book, like graph theory, and {is}
somewhat akin to ecology, as there are a multitude of fully
functional ecological systems.

It is, in any case, a central challenge to scientific
research to formulate and to examine
self-consistent building principles for
rudimentary but autonomous cognitive systems.
The venue treated in this chapter represents
a specific approach towards the formulation and
the understanding of the basic requirements
needed for the construction of a cognitive system.

\runinhead{Biologically Inspired Cognitive Systems} \index{cognitive
system!biologically inspired} Cognitive system theory has two
long-term targets: To understand the functioning of the human brain
and to develop an autonomous cognitive system. The realization of
both goals is still far away, but they may be combined to a certain
degree. The overall theory is however at an early stage and it is
presently unclear to which extent the first implemented artificial
cognitive systems will resemble our own cognitive organ, the brain.

\vspace*{-6pt}\enlargethispage*{18pt}
\section{Foundations of Cognitive {Systems} Theory}
\label{cogSys_foundations}

\subsection{Basic Requirements for the Dynamics}
\label{cogSys_requirements}

\runinhead{Homeostatic Principles} \index{homeostatic
principles!cognitive system} Several considerations suggest that
self-regulation via adaptive means, viz homeostatic principles, are
widespread in the domain of life in general and for biological
cognitive systems in particular.

\begin{itemize}
\item[--] There are concrete instances for neural algorithms,
      like the formation of topological neural maps, based on
      general, self-regulating feedback. An example is
      the topological map connecting the retina to the
      primary optical cortex.
\item[--] The number of genes responsible for the development of the brain
      is relatively low, perhaps a few thousands. The growth of
      about 100 billion neurons and of around $10^{15}$ synapses
      can only result in a functioning cognitive system if very
      general self-regulating and self-guiding algorithms are used.
\item[--] The strength and the number of neural pathways interconnecting
      different regions of the brain or connecting sensory organs
      to the brain may vary substantially during development or during
      lifetime, e.g.\ as a consequence of injuries. This implies, quite
      generally, that the sensibility of neurons to the average
      strength of incoming stimuli must be adaptive.
\end{itemize}
It is tempting to speak in this context of \qut{target-oriented
self-organization}, since mere ``blind'', viz basic
self-organizational processes might be insufficient tools for the
successful self-regulated development of the brain in a first step
and of the neural circuits in a second step.

\runinhead{Self-Sustained Dynamics} \index{dynamics!self-sustained}
Simple biological neural networks, e.g.\ the ones in most worms,
just perform stimulus--response tasks. Highly developed mammal
brains, on the other side, are not directly driven by external
stimuli. Sensory information influences the ongoing, self-sustained
neuronal dynamics, but the outcome cannot be predicted from the
outside viewpoint.

Indeed, the human brain is {on the whole}
occupied with itself and continuously active even in the sustained
absence of sensory stimuli. A central theme of cognitive
{systems} theory is therefore to formulate, test
and implement the principles {that} govern the
autonomous dynamics of a cognitive system.

\begin{figure}[t]
\centerline{\includegraphics{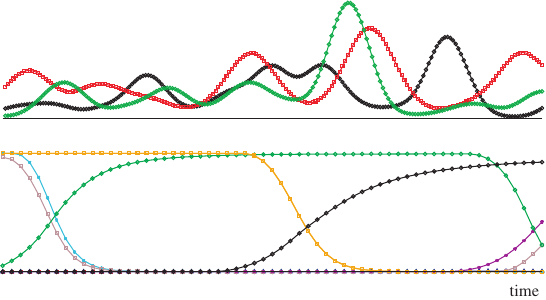}}
\caption{{Fluctuating} {\it
(top)} and transient state {\it (bottom)} dynamics}
\label{cogSys_fig_transStates}
\end{figure}

\runinhead{Transient State Versus Fluctuation Dynamics} There is a
plurality of approaches for the characterization of the time
development of a dynamical system. A key questions in this context
regards the repeated occurrence of well defined \nobreak dynamical
states, that is, of states allowing for a well defined
characterization of the current dynamical state of the cognitive
system, like the ones illustrated in
Fig.~\ref{cogSys_fig_transStates}.
\begin{quotation}{\it Transient States.\enspace} \index{transient state dynamics!cognitive
system} A transient state of a dynamical system corresponds to a
quasistationary plateau in the value of the variables.
\end{quotation}
Transient state dynamics can be defined mathematically in a rigorous
way. It is present in a dynamical system if the governing equations
of the system contain parameters {that} regulate the
length of the transient state, viz whenever it is possible, by
tuning theses parameters, to prolong the length of the plateaus
arbitrarily.

In the case of the human brain, several experiments
indicate the occurrence of spontaneously activated
transient neural activity patterns in the cortex,\footnote{See,
e.g., Abeles et al. (1995) and Kenet et al. (2003).}
on timescales corresponding to the cognitive
timescale\footnote{Humans can distinguish cognitively
about 10--12 objects per second.}
of about ${80}{-}{100}\,\mbox{ms}$. It is therefore
natural to assume that both fluctuating states
and those corresponding to transient activity
are characteristic for biological inspired
cognitive systems. In this chapter we will
especially emphasize the transient state dynamics
and discuss the functional roles of the transient
attractors generated by this kind of dynamics.

\runinhead{Competing Dynamics} \index{competing
dynamics} The brain is made up of many distinct regions
{that} are highly interconnected. The resulting
dynamics is thought to be partly competing.

\begin{quotation}{\it Competing Dynamics.\enspace}
\index{dynamics!competing} \index{cognitive system!competing
dynamics} A dynamical system made up of a collection of interacting
centers is said to show competing dynamics if active centers try to
suppress the activity level of the vast majority of competing
centers.
\end{quotation}

In neural network terminology, competing dynamics is also called a
{\em winners-take-all} setup. In the extreme case, when only a
single neuron is active at any given time, one speaks of a {\em
winner-take-all} situation.

\begin{quotation}{\it The Winning Coalition.\enspace} \index{winning
coalition}\index{cognitive system!winning coalition} In a
winners-take-all network the winners are normally formed by an
ensemble of mutually supportive centers, which one also denotes the
\qut{winning coalition}.
\end{quotation}

\noindent A winning coalition needs to be stable for a certain minimal period
of time, in order to be well characterized. Competing dynamics
therefore {frequently results} in
transient state dynamics.

Competing dynamics in terms of {dynamically}
forming winning coalitions is a possible principle for achieving the
target-oriented self-organization needed for a self-regulating
autonomously dynamical systems. We will treat this subject
 in detail in Sect.~\ref{cogSys_dhan}.

\runinhead{States-of-the-Mind and the Global Workspace}
\index{cognitive system!states-of-the-mind}\index{cognitive
system!global workspace}\index{global workspace} A highly developed
cognitive system is capable {of generating}
autonomously a very large number of different transient states,
which represent the \qut{states-of-the-mind}. This feature plays an
important role in present-day investigations of the neural
correlates of consciousness, which we shall now briefly mention
for completeness. We will not discuss the relation of
cognition and consciousness any further in this chapter.

Edelman and Tononi\footnote{See Edelman and Tononi (2000).}
 argued that these states-of-the-mind can be
characterized by \qut{critical reentrant events}, constituting
transient conscious states in the human brain. Several authors have
proposed the notion of a \qut{global workspace}. This workspace
would be the collection of neural ensembles contributing to global
brain dynamics. It could serve, {among other
things}, as an exchange platform for conscious experience and
working memory.\footnote{See Dehaene and Naccache (2003), and Baars
and Franklin (2003).} The constituting neural ensembles of the
global workspace have also been dubbed \qut{essential nodes}, i.e.\
ensembles of neurons responsible for the explicit representation of
particular aspects of visual scenes or other sensory
information.\footnote{See Crick and Koch (2003).}

\runinhead{Spiking Versus Non-Spiking Dynamics}
\index{dynamics!spiking vs.\ non-spiking}\index{spikes} Neurons emit
an axon potential called a spike, which lasts about a
millisecond. They then need to recover for about 10~ms, the
refractory period. Is it then important for a biologically inspired
cognitive system to use spiking dynamics? We note here in passing
that spiking dynamics can be generated by interacting relaxation
oscillators, as discussed in Chap.~\ref{chap_synchro1}.

The alternative would be to use a network of local computational
units having a continuously varying activity,
somewhat akin to the average spiking intensity of neural ensembles.
There are two important considerations in this context:

\begin{itemize}
\item[--] {At present, it does not seem} plausible that spiking dynamics
      is a condition \textit{sine qua non} for a cognitive system. It
      might be suitable for a biological system, but not
      a fundamental prerequisite.
\item[--] Typical spiking frequencies are in the range of 5--50 spikes
      per second. A typical cortical neuron receives input from
      about ten thousand other neurons, viz 50--500 spikes
      per millisecond. The input signal for typical neurons is
      therefore quasicontinuous.
\end{itemize}
{The exact} timing of neural spikes is clearly
important in many areas of the brain, e.g.\ for the processing of
acoustic data. Individual incoming spikes are also of relevance,
when they push the {postsynaptic} neuron above
the firing threshold. {However, the above considerations indicate} a
reduced importance of precise spike timing for the average
all-purpose neuron.

\runinhead{Continuous Versus Discrete Time Dynamics}
\index{dynamics!discrete time} Neural networks can be modeled either
by using a discrete time formulation $t=1,2,3,\ldots $ or by
employing continuous time $t\in[0,\infty]$.
\begin{quotation}{\it Synchronous and Asynchronous Updating.\enspace}
\index{asynchronous updating} \index{synchronous updating}
\index{updating!synchronous} \index{updating!asynchronous}
A dynamical system with discrete time is updated synchronously
(asynchronously) when all variables are
evaluated simultaneously (one after another).
\end{quotation}
\vspace*{-18pt}For a continuous time formulation there is no
difference between synchronous and asynchronous updating however, it
matters for a dynamical system with discrete time, as we discussed
in Chap.~\ref{chap_networks2}.

The dynamics of a cognitive system needs to be stable. 
This condition requires that the overall dynamical 
feature cannot depend, e.g., on the number of components 
or on the local numerical updating procedure. Continuous 
or quasi-continous time is therefore the only
viable option for real-world cognitive systems.

\runinhead{Continuous Dynamics and Online Learning}
{The above} considerations indicate that a
biologically inspired cognitive system should be continuously
active.
\begin{quotation}{\it Online Learning.\enspace} \index{online
learning}\index{learning!online} When a neural network type system
learns during its normal mode of operation one speaks of \qut{online
learning}. The case of \qut{offline learning} is given when learning
and performance are separated in time.
\end{quotation}
Learning is a key aspect of cognition and online learning is the
only possible learning paradigm for an autonomous cognitive system.
Consequently there can be no distinct training and performance
modes. We will come back to this issue in
Sect.~\ref{cogSys_subsec_online_learning}.

\subsection{Cognitive Information Processing Versus Diffusive Control}
\label{cogSys_vs}

A cognitive system is an (exceedingly) complex adaptive system
per excellence. As such it needs to be adaptive on several
levels.

Biological considerations suggest to use networks of local
computational units with primary variables
${\veci{x}_i}=(x_i^0,x_i^1,\ldots )$. Typically $x_i^0$ would
correspond to the average firing rate and the other $x_i^\alpha$
($\alpha=1,\ldots$) would characterize different dynamical
properties of the ensemble of neurons represented by the local
computational unit as well as the (incoming) synaptic weights.

The cognitive system, as a dynamical system, is governed
by a set of differential equations, such as
\begin{equation}
\dot{\veci x}_i \ =\ {\veci f}_i(\veci x_1,\ldots ,\veci x_N),
\qquad\qquad i=1,\ldots ,N~. \label{cogSys_time_evol_x}
\end{equation}
\runinhead{Primary and Secondary Variables} \index{cognitive
system!variables!primary {and} secondary} The functions
${\veci f}_i$ governing the time evolution {equation}
(\ref{cogSys_time_evol_x}) of the primary variables $\{\veci x_i\}$
generally depend on a collection of parameters $\{\vec \gamma_i\}$,
such as learning rates, firing thresholds, etc.:
\begin{equation}
{\veci f}_i(\veci x_1,\ldots ,\veci x_N) \ =\
  {\veci f}_i(\gamma_1,\gamma_2,\ldots |\veci x_1,\veci x_2,\ldots )~.
\label{cogSys_f_i_gamma}
\end{equation}
The time evolution of the system is fully determined by
Eq.~(\ref{cogSys_time_evol_x}) whenever the parameters $\gamma_j$
are unmutable, that is, genetically predetermined. Normally, however,
the cognitive system needs to adjust a fraction
of these parameters with time, viz
\begin{equation}
\dot{\gamma}_i \ =\ g_i(\gamma_1,\gamma_2,\ldots |
                  \veci x_1,\veci x_2,\ldots )~,
\label{cogSys_time_evol_gamma}
\end{equation}
In principle one could merge {} $\{\veci x_j\}$ and
{} $\{\gamma_i\}$ into one large set of dynamical
variables $\{y_l\}=\{\gamma_i|\veci x_j\}$. It is, however,
meaningful to keep them separated whenever their respective time
evolution differs qualitatively and quantitatively.
\begin{quotation}{\it Fast and Slow Variables.\enspace} 
\index{variable!fast and slow|textbf} When the average rate
changes of two variables $x=x(t)$ and $y=y(t)$ are typically very
different in magnitude, $|\dot x|\gg|\dot y|$, then one calls $x(t)$
the fast variable and $y(t)$ the slow variable.
\end{quotation}
The parameters $\{\gamma_j\}$ are, per definition,
slow variables. One can then also call them
\qut{secondary variables} as they follow the
long-term average of the primary variables $\{\veci x_i\}$.

\runinhead{Adiabatic Approximation}
\index{adiabatic approximation}
The fast variables $\{\veci x_i\}$ change rapidly
with respect to the time development of the slow
variables $\{\gamma_j\}$ in Eq.~(\ref{cogSys_time_evol_gamma}).
It is then often a good approximation to substitute the
$\veci x_i$ by suitable time-averages
$\langle \veci x_i\rangle_t$. In physics
jargon one speaks then of an
\qut{adiabatic approximation}.

\runinhead{Adaptive Parameters}
\index{cognitive system!adaptive parameters}
A cognitive system needs to self-adapt over a wide range
of structural organizations, as discussed in
Sect.~\ref{cogSys_requirements}. Many parameters relevant for the
sensibility to presynaptic activities, for short-term and long-term
learning, to give a few examples, need therefore to be adaptive, viz
time-dependent.
\begin{quotation}{\it Metalearning.\enspace}
\index{metalearning}\index{learning!meta} The
time evolution of the slow variables, the parameters, is called
\qut{metalearning} in the context of cognitive
systems theory.
\end{quotation}
\index{synaptic!strength} With (normal) learning we denote the
changes in the synaptic strength, i.e.\ the connections between
distinct local computational units. Learning (of memories) therefore
involves part of the primary variables.

\begin{figure}[t]
\centerline{\includegraphics{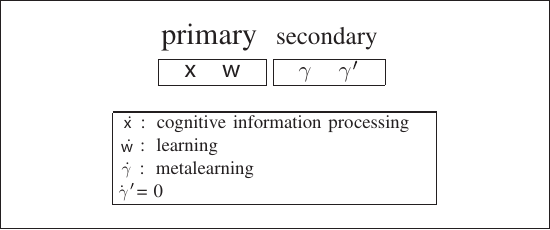}}
%
\caption{General classification scheme for the variables and the
parameters of a cognitive system. The variables can be categorized
as primary variables and as secondary variables (parameters). The
primary variables can be subdivided into the variables
characterizing the current state of the local computational units
$x$ and into generalized synaptic weights $w$. The \qut{parameters}
$\gamma$ are slow variables adjusted for homeostatic regulation. The
true unmutable (genetically predetermined) parameters are
{} $\gamma \,\,'$} \label{cogSys_fig_parameters}
\end{figure}

The other primary variables characterize the current state of a
local computational unit, such as the current average firing rate.
Their time evolution corresponds to the actual {\em cognitive
information processing}, see Fig.~\ref{cogSys_fig_parameters}.

\runinhead{Diffusive Control} Neuromodulators, like dopamine,
serotonin, noradrenaline and acetylcholine, serve in the brain as
messengers for the transmission of general information about the
internal status of the brain, and for overall system state control.
A release of a neuromodulator by the appropriate specialized neurons
does not influence individual target neurons, but extended cortical
areas.
\begin{quotation}{\it Diffusive Control.\enspace}
\index{diffusive control}\index{cognitive
system!diffusive control} A signal by a given part of a dynamical
system is called a \qut{diffusive control signal} if it tunes the
secondary variables in an extended region of the system.
\end{quotation}
A diffusive control signal\footnote{Note that
neuromodulators are typically released in the intercellular
medium from where they physically diffuse towards the
surrounding neurons.} does not influence the status of
individual computational units directly, i.e.\ their primary
variables. Diffusive control has a wide range of tasks. It plays an
important role in metalearning and reinforcement learning.

As an example of the utility of diffusive control signals we mention
the \qut{learning from mistakes} approach, see
Sect.~\ref{cogSys_memory}. Within this paradigm synaptic
plasticities are degraded after an unfavorable action
has been performed. For this purpose a diffusive
control signal is generated whenever a mistake has been made,
with the effect that all previously active synapses are weakened.

\subsection{Basic Layout Principles}
\label{cogSys_principles}\index{cognitive system!basic
principles|textbf} There is, at present, no fully developed theory
for real-world cognitive systems. Here we discuss some recent
proposals for a possible self-consistent set of requirements for
biologically inspired cognitive systems.

\begin{description}
\item[(A)] {} Absence of A Priori Knowledge About the Environment\\
\index{cognitive system!basic principles!a priori
knowledge}Preprogrammed information about the outside world is
normally a necessary ingredient for the performance of robotic
systems at least within the artificial intelligence paradigm.
{However, a
rudimentary system needs to perform dominantly on the base of
universal principles}.

\item[(B)] Locality of Information Processing\\
\index{cognitive system!basic principles!locality}Biologically
inspired models need to be scalable and adaptive to structural
modifications. This rules out steps in information processing
needing non-local information, as is the case for the standard
back-propagation algorithm, viz the minimization of a global error
function.

\item[(C)] {Modular Architecture}\\
Biological observations motivate a modular approach, with every
individual module being structurally homogeneous. An autonomous
cognitive system needs modules for various cognitive tasks and
diffusive control. Well defined interface specifications are then
needed for controlled intermodular information exchange. Homeostatic
principles are necessary for the determination of the intermodule
connections, in order to allow for scalability and adaptability to
structural modifications.

\item[(D)] {Metalearning via Diffusive Control}\\
\index{diffuse control!metalearning}Metalearning, i.e.\ the tuning
of control parameters for learning and sensitivity to internal and
external signals, occurs exclusively via diffusive control. The
control signal is generated by diffusive control units, which
analyze the overall status of the network and become active when
certain conditions are achieved.

\item[(E)] {Working Point Optimization}\\
\index{cognitive system!basic principles!working point}\index{working point optimization}The length of the stability
interval of the transient states relative to the length of
{the} transition time from one state-of-mind to the next
(the working point of the system) needs to be self-regulated by
homeostatic principles.

Learning influences the dynamical behavior of the cognitive system
in general and the time scales characterizing the transient state
dynamics in particular. Learning rules {therefore need} to be formulated in a way that autonomous
working point optimization is guaranteed.
\end{description}

\vspace{12pt}

\runinhead{The Central Challenge} The discovery and understanding of
universal principles, especially for cognitive information
processing, postulated in (A)--(F) is the key {to}
ultimately understanding the brain or {to} building an
artificial cognitive system. In Sect.~\ref{cogSys_prediction} we
will discuss an example for a universal principle, namely
environmental model building via universal prediction tasks.

\vspace*{2pt}

\runinhead{{The} Minimal Set of Genetic Knowledge}
\index{cognitive system!basic principles!a priori knowledge} No
cognitive system can be universal in a strict sense. Animals, to
give an example, do not need to learn that hunger and pain are
negative reward signals. This information is genetically
preprogrammed. Other experiences are not genetically fixed, e.g.\
some humans like the taste of coffee, others do not.

No cognitive system could be functioning with strictly zero a priori
knowledge, it would have no \qut{purpose}. A minimal set of goals is
necessary, as we will discuss further in depth in
Sect.~\ref{cogSys_motivation}. A minimal goal of fundamental
significance is to \qut{survive} in the sense that certain internal
variables need to be kept within certain parameter ranges. A
biological cognitive system needs to keep the pain and hunger
signals {that} it receives from its own body at low
average levels, otherwise its body would die. An artificial system
could be given corresponding tasks.

\vspace*{2pt}

\runinhead{Consistency of Local Information Processing with
Diffusive Control} We note that the locality principle (B) for
cognitive information processing is consistent with non-local
diffusive control (D). Diffusive control regulates the overall
status of the system, like attention focusing and sensibilities, but
it does not influence {the actual information processing directly}.

\vspace*{2pt}

\runinhead{Logical Reasoning Versus Cognitive Information
Processing} \index{artificial intelligence!logical reasoning} Very
intensive research on logical reasoning theories is carried out in
the context of AI. From (A) it follows that logical manipulation of
concepts is, however, not suitable as an exclusive framework for
universal cognitive systems. Abstract concepts cannot be formed
without substantial knowledge about the environment, but this
knowledge is acquired by an autonomous cognitive system only
step-by-step during its \qut{lifetime}.

\subsection{Learning and Memory Representations}
\label{cogSys_memory}

With \qut{learning} one denotes quite generally all 
modifications that influence the dynamical state and the
behavior. One distinguishes the learning of memories and actions.

\begin{quotation}{\it Memories.\enspace} 
\index{memory} \index{cognitive system!memory} By
memory one denotes the storage of a pattern found within the
incoming stream of sensory data, which presumably encodes
information about the environment.
\end{quotation}
The storage of information about its own actions, i.e.\ about the
output signals of a cognitive system is also covered by this
definition. Animals probably do not remember the output signal of 
the motor cortex directly, but rather the optical or acoustical
response of the environment as well as the feedback of its body via
appropriate sensory nerves embedded in the muscles.

\runinhead{The Outside World -- The Cognitive System as an Abstract
Identity} \index{cognitive system!environment}\index{environment} A
rather philosophical question is whether there is, from the
perspective of a cognitive system, a true outside world. The
alternative would be to postulate that only the internal
representations of the outside world, {i.e.} the
environment, are known to the cognitive system. 
For all practical purposes it is useful to postulate 
an environment existing independently of the cognitive system.

\index{cognitive system!abstract identity} It is, however, important
to realize that the cognitive system per se is an abstract identity,
{i.e.} the dynamical activity patterns. The physical
support, i.e. computer chips and brain tissue, are
not part of the cybernetic  or of the human
cognitive system, respectively. We, as cognitive systems, are
abstract identities and the physical brain tissue 
therefore also belongs to our environment!

One may differentiate this statement to a certain
extent, as direct manipulations of our neurons
may change the brain dynamics directly. This
may possibly occur without our external and
internal sensory organs noticing the manipulatory process. 
In this respect the brain tissue is distinct from the rest 
of the environment, since changes in the rest of the 
environment influence the brain dynamics exclusively via 
sensory inputs, which may be either internal, 
such as a pain signal, or external, like an auditory signal.

For practical purposes, when designing an artificial environment for
a cognitive system, the distinction between a directly observable
part of the outside world and the non-observable part becomes
important. Only the observable part generates, per definition,
sensorial stimuli, but one needs to keep in mind that the actions of
the cognitive system may {also influence}
the non-observable environment.

\runinhead{Classification of Learning Procedures} It is customary to
broadly classify possible learning procedures. We discuss {briefly}
the most important cases of learning algorithms{;} for details we
refer to the literature.\vspace*{2pt}
\begin{itemize}
\item[--] Unsupervised Learning: \index{learning!unsupervised}The system learns completely by itself, without any external teacher.
\item[--] Supervised Learning: \index{learning!supervised}Synaptic changes are made \qut{by hand}, by the external teacher
  and not determined autonomously. Systems with supervised learning {in most cases have} distinguished periods for training and
  performance (recall).
\item[--] Reinforcement Learning: \index{learning!reinforcement}Any cognitive system faces the fundamental dilemma of action
  selection, namely that the final success or failure of a series
  of actions may {often be evaluated only at} the end. When playing a
  board game one knows only {at} the end whether one {has won or lost}.

\hspace*{6pt}  Reinforcement learning denotes strategies {that allow one} to employ
  the positive or negative reward signal obtained at the end of a
  series of actions to either rate the actions taken or to reinforce the
  problem solution strategy.

\item[--] Learning from Mistakes: \index{learning!from mistakes}Random action selection will normally result in mistakes and not
  in success. In normal life learning from mistakes
  is therefore by far more important than learning
  from positive feedback.
\item[--] Hebbian Learning: \index{Hebbian learning}\index{learning!Hebbian}Hebbian learning denotes a specific instance of a linear
  synaptic modification procedure in neural networks.
\begin{itemize}
\item[--] Spiking Neurons: For spiking neurons Hebbian learning results
      in a long-term potentiation (LTP) of the synaptic strength when the
      presynaptic neuron spikes shortly before the postsynaptic neuron
      (causality principle). The reversed spiking timing results in
      long-term depression (LTD).
\item[--] Neurons with Continuous Activity: The synaptic strength is increased when both {postsynaptic and presynaptic}
      neurons are active. Normally one assumes the synaptic
      plasticity to be directly proportional to the product
      of {postsynaptic and presynaptic} activity levels.
\end{itemize}
\end{itemize}\vspace*{3pt}

\runinhead{Learning Within an Autonomous Cognitive System} Learning
within an autonomous cognitive system with self-induced dynamics is,
strictly speaking, {} unsupervised. Direct synaptic
modifications by an external teacher are clearly not\break
admissible. But also reinforcement learning is, at its basis,
unsupervised, as the system has to select autonomously what it
accepts as a reward signal.

The different forms of learning are, however, significant when
taking the internal subdivision of the cognitive system into various
modules into account. In this case a diffusive control unit can
provide the reward signal for a cognitive information processing
module. Also internally supervised learning is
conceivable.\vspace*{3pt}

\runinhead{Runaway Synaptic Growth} \index{learning!runaway effect}
Learning rules in a continuously active dynamical system need
careful considerations. A learning rule might foresee fixed
boundaries, viz limitations, for the variables involved in learning
processes and for the parameters modified during metalearning. In
this case when the parameter involved reaches the limit, learning
might potentially lead to saturation, which is \nobreak suboptimal
for \nobreak information storage and processing. With no limits
encoded the continuous learning process might lead to unlimited
synaptic weight growth.
\begin{quotation}{\it Runaway Learning.\enspace} When a
specific learning rule acts over time continuously with the same
sign it might lead to an unlimited growth of the affected variables.
\end{quotation}
Any instance of runaway growth needs to be avoided, as
it will inevitably lead the system out of suitable
parameter ranges. This is an example of the general problem of
working point optimization, see Sect.~\ref{cogSys_principles}.

\runinhead{Optimization vs.\ Maximization}
Biological processes generally aim for optimization and not
for maximization. The naive formulation of Hebbian learning
is an instance of a maximization rule. It can be transformed
into an optimization process by demanding for
the sum of active incoming synaptic strengths
to adapt towards a given value. This procedure
leads to both LTP and LTD; an explicit rule for
LTD is then not necessary.

\runinhead{Biological Memories} Higher mammalian brains are capable
of storing information in several distinct ways. Both experimental
psychology and neuroscience are investigating the different storage
capabilities {and} suitable nomenclatures have been
developed. Four types of biophysical different storing mechanisms
have been identified so far:

\vspace*{6pt}
\begin{enumerate}\leftskip7pt
\item[(i)\phantom{ii}] Long-Term Memory: \index{memory!long-term}The brain is made up by a network of neurons
{that} are
  interconnected via synapses. All long-term information is
  therefore encoded, directly or indirectly, in the
  respective synaptic strengths.

\item[(ii)\phantom{i}] Short-Term Memory: \index{memory!short-term}The short-term memory corresponds to
transient modifications
  of the synaptic strength. These modifications decay after
  a characteristic time, which may be of the order of minutes.

\item[(iii)] Working Memory: \index{memory!working}
The working memory corresponds to firing states of individual
  neurons or neuron ensembles that are kept active for a certain
  period, up to several minutes,
  even after the initial stimulus has subsided.

\item[(iv)] Episodic Memory: \index{memory!episodic}\index{hippocampus}
The episodic memory is mediated by the hippocampus, a subcortical
  neural structure. The core of the hippocampus, called CA3,
  contains only about $3\cdot10^5$ neurons (for humans).
  All daily episodic experiences, from the visit to the
  movie theater to the daily quarrel with the spouse, are kept
  active by the hippocampus. A popular theory of sleep assumes
  that fixation of the episodic memory in the cortex
  occurs during dream phases when sleeping.
\end{enumerate}
\vspace*{6pt}
In Sect.~\ref{cogSys_dhan} we will treat a generalized 
neural network layout illustrating the homeostatic 
self-regulation of long-term synaptic plasticities and
the encoding of memories in terms of local active clusters.

\runinhead{Learning and Memory Representations} The representation
of the environment, via suitable filtering of prominent patterns
from the sensory input data stream, is a basic need for any
cognitive system. We discuss a few important considerations.

\begin{itemize}
\item[--] Storage Capacity: \index{memory!storage capacity}
     Large quantities of new information needs to be stored without
     erasing essential memories.
\begin{quotation}{\it Sparse/Distributed Coding.\enspace} \index{sparse
coding}\index{distributed coding}\index{neural network!sparse vs.\
distributed coding} A network of local computational units in which
only a few units are active at any given time is said to use
\qut{sparse coding}. If on the average half of the neurons are
active, one speaks of \qut{distributed coding}.
\end{quotation}

   Neural networks with sparse coding have a substantially higher
   storage capacity than neural networks with an average activity
   of about 1/2. The latter have a storage capacity scaling only linearly
   with the number of nodes. A typical value for the storage capacity
   is in this case 14\%, with respect to the system
   size.\footnote{This is a standard result for so-called
   Hopfield neural networks, see e.g.\ Ballard (2000).}

  \hspace*{12pt} In the brain only a few percent of all neurons are
   active at any given time. Whether this occurs in order to
   minimize energy consumption or to maximize the storage
   capacity is not known.

\item[--] Forgetting: \index{memory!forgetting}
No system can acquire and store new
information forever. There are very different approaches
to how to treat old information and memories.

\begin{itemize}
\item[--] Catastrophic Forgetting: \index{catastrophic forgetting}
   One speaks of \qut{catastrophic forgetting} if all
   previously stored memories are erased completely whenever
   the system surpasses its storages capacity.
\item[--] Fading Memory:
\index{fading memory}
   The counterpoint is called \qut{fading memory};
   old and seldomly reactivated memories are overwritten
   gradualy with fresh impressions.
\end{itemize}

   \index{neural network!recurrent}
   Recurrent neural networks\footnote{A neural network is
   denoted \qut{recurrent} when loops dominate the network
   topology.} with distributed coding
   forget catastrophically. Cognitive systems can
   only work with a fading memory, when old information is
   overwritten gradualy.\footnote{For a mathematically precise
   definition, a memory is termed fading
   when forgetting is scale-invariant,
   viz having a power law functional time dependence.
                               }

\item[--] The Embedding Problem: \index{learning!embedding problem}
There is no isolated information. Any new information is
   only helpful if the system can embed it into the web of
   existing memories. This embedding, at its basic level,
   needs to be an automatic process, since any search algorithm
   would blast away any available computing power.

   \hspace*{12pt}In Sect.~\ref{cogSys_dhan} we will present a cognitive module
   for environmental data representation, which allows for
   a crude but automatic embedding.

\item[--] Generalization Capability: \index{learning!generalization capability}The encoding used for
memories must allow the system to work
   with noisy and incomplete sensory data. This is a key requirement
   {that} one can regard as a special case of a broader
   generalization capability necessary for universal
   cognitive systems.

   \hspace*{12pt}An efficient data storage format would
   allow the system to automatically find, without
   extensive computations, common characteristics of
   distinct input patterns. If all patterns corresponding
   to ``car'' contain elements corresponding to
   ``tires''  and ``windows'' the data representation should
   allow for an automatic prototyping of the kind
   ``car\,=\,tires\,+\,windows''.

   \hspace*{12pt}Generalization capabilities and noise tolerance are intrinsically
   related. Many different neural network setups have this
   property, due to distributed and overlapping memory storage.
\end{itemize}

\section{Motivation, Benchmarks and Diffusive Emotional Control}
\label{cogSys_motivation}

Key issues to be considered for the general layout
of a working cognitive system are:
\begin{itemize}
\item[--] Cognitive Information Processing:
\index{cognitive information processing}
Cognitive information processing involves the dynamics
of the primary variables, compare Sect.~\ref{cogSys_principles}.
We will discuss a possible modular layout in Sect.~\ref{cogSys_tasks}.
\item[--] Diffusive Control: \index{diffusive control}
Diffusive control is at the heart of homeostatic
self-regulation for any cognitive system. The layout
of the diffusive control depends to a certain extent
on the specific implementation of the cognitive modules.
We will therefore restrict ourselves here
to general working principles.

\item[--] Decision Processes: \index{decision process}
\index{cognitive system!decision processes}
Decision making in a cognitive system depends strongly on
the specifics of its layout. A few general guidelines may
be formulated for biologically inspired cognitive systems;
we will discuss these in Sect.~\ref{cogSys_benchmarks}
\end{itemize}

\subsection{Cognitive Tasks}
\label{cogSys_tasks}

\runinhead{Basic Cognitive Tasks} A rudimentary cognitive system
needs at least three types of cognitive modules. The individual
modules comprise cognitive units for
\begin{enumerate}\leftskip5pt
\item[(a)] environmental data representation via unsupervised learning
           (compare\break Sect.~\ref{cogSys_memory}),
\item[(b)] modules for model building of the environment via
           internal supervised learning, and
\item[(c)] action selection modules via learning by reinforcement
           or learning by error.
\end{enumerate}
We mention here in passing that the assignment of these
functionalities to specific brain areas is an open issue, one
possibility being a delegation to the cortex, the cerebellum and to
the basal ganglia, respectively.

\runinhead{Data Representation and Model Building} In
Sect.~\ref{cogSys_dhan} we will treat in depth
the problem of environmental data representation and automatic
embedding. Let us note here that the problem of model building is
not an all-in-one-step operation. Environmental data representation
and basic generalization capabilities normally go hand in hand, but
this feature falls far short of higher abstract concept generation.

An example of a basic generalization process is,
to be a little more concrete, the generation of
the notion of a ``tree'' derived by suitable
averaging procedures out of many
instances of individual trees occurring in
the visual input data stream.

\runinhead{Time Series Analysis and Model Building} The analysis of
the time sequence of the incoming sensory data has a high biological
survival value and is, in addition, at the basis of many cognitive
capabilities. It allows for quite sophisticated model building and
for the generation of abstract concepts. In
Sect.\ref{cogSys_prediction} we will treat a neural network setup
allowing for universal abstract concept generation, resulting from
the task to predict the next incoming sensory data{;
a} task {that} is independent of the nature of the
sensory data and in this sense universal. When applied to a
linguistic incoming data stream, the network generates, with zero
prior grammatical knowledge, concepts like \qut{verb}, \qut{noun}
and so~on.

\vspace*{-6pt}

\subsection{Internal Benchmarks}
\label{cogSys_benchmarks}

Action selection occurs in an autonomous cognitive system via
internal reinforcement signals. The reward signal can be either
genetically predetermined or internally generated. To give a
high-level example: We might find it positive to win a chess game if
playing against an opponent but we may also enjoy
{losing} when playing with our son or daughter.
Our internal state is involved when selecting the reward signal.

We will discuss the problem of action selection by a
cognitive system first on a phenomenological level
and then relate these concepts to the general layout
in terms of variables and diffusive control units.

\runinhead{Action Selection} Two prerequisites are fundamental to
any action {taken} by a cognitive system:
\begin{enumerate}
\leftskip6pt
\item[($\alpha$)] Objective: No decision can be taken without an objective of what to do.
  A goal can be very general or quite specific. \qut{I am bored,
  I want to do something interesting} would result in a general
  explorative strategy, whereas \qut{I am thirsty and I have a
  cup of water in my hand} will result in a very concrete
  action, namely drinking.
\item[($\beta$)] Situation Evaluation: In order to decide between many possible actions the
  system needs to evaluate them. We define by \qut{situation}
  the combined attributes characterizing the current internal
  status and the environmental conditions.
\begin{center}
\framebox{\parbox{10cm}{\begin{center}
\begin{tabular}{rcl}
Situation &= & (internal status)\ \ +\ \ (environmental conditions) \\
Situation &$\to$ & value \\
\end{tabular}
\end{center}
         }}
\end{center}
The situation \qut{(thirsty)\,+\,(cup with water in my hands)} will
normally be evaluated positively, the situation
\qut{(sleepy)\,+\,(cup with water in my hand)} on the other
{hand} not.
\end{enumerate}

\runinhead{Evaluation and Diffusive Control} The evaluation of a
situation goes hand in hand with feelings and emotions. Not only for
most human {does} the evaluation
{belong} to the domain of diffusive control. The
reason being that the diffusive control units, see
Sect.~\ref{cogSys_vs}, are responsible {for
keeping} an eye on the overall status of the cognitive
system{;} they need to evaluate the internal status
constantly in relation to what {is happening} in
the outside world, viz in the sensory input.

\runinhead{Primary Benchmarks} \index{decision process!primary
benchmarks} Any evaluation needs a benchmark: What is good and what
is bad for oneself? For a rudimentary cognitive system the
benchmarks {and} motivations are given by the fundamental need to
survive: If certain parameter values, like hunger and pain signals
arriving from the body, or more specific signals about protein
support levels or body temperature, are in the \qut{green zone}, a
situation, or a series of events leading to the present situation,
is deemed good. Appropriate corresponding \qut{survival variables}
need to be defined for an artificial cognitive\break system.
\begin{quotation}{\it Survival Parameters.\enspace} \index{decision process!survival
parameters} \index{cognitive system!survival parameters}
\index{survival parameters} We denote the parameters regulating the
condition of survival for a living dynamical system as survival
parameters.
\end{quotation}
The survival parameters are part of the sensory input, compare
Fig.~\ref{cogSys_fig_CS_illust}, as they convene information about
the status of the body, viz the physical support complex for the
cognitive system. The survival parameters affect the status of
selected diffusive control units{;} {generally they do not interact}
directly with the cognitive information processing.\vspace*{3pt}

\runinhead{Rudimentary Cognitive Systems} A cognitive system will
only survive if its benchmarking favors actions
{that} keep the survival parameters in the green
zone.
\begin{quotation}{\it Fundamental Genetic Preferences.\enspace} \index{decision
process!genetic preferences}\index{genetic preferences} The
necessity for biological or \hbox{artificial} cognitive systems to
keep the survival parameters in a given range corresponds to primary
goals, which are denoted \qut{fundamental genetic preferences}.
\end{quotation}
The fundamental genetic preferences are not \qut{instincts} in the
classical sense, as they do not lead deterministically and directly
to observable behavior. The cognitive system needs to learn which of
its actions satisfy the genetic preferences, as it acquires
{information} about the world it is born into only by direct
personal \nobreak experiences.
\begin{quotation}{\it Rudimentary Cognitive {Systems}.\enspace}
\index{cognitive system!rudimentary} A rudimentary cognitive system
is determined fully by its fundamental genetic preferences.
\end{quotation}
A rudimentary cognitive system is very limited with respect to the
complexity level {that} its actions can achieve, since they are all
directly related to primary survival. The next step in benchmarking
involves the diffusive control units.\vspace*{3pt}

\runinhead{Secondary Benchmarks and Emotional Control}
\index{decision process!emotional control} Diffusive control units
are responsible {for keeping} an eye on the overall status of the
dynamical system. We can divide the diffusive control units into
two\vadjust{\pagebreak} classes:
\begin{itemize}
\item[--]Neutral Units: \index{diffuse control!neutral}
     These diffusive control units have no preferred activity
     level.
\item[--] Emotional Units: \index{diffuse control!emotional}
     These diffusive control units have a (genetically
     determined) preferred activity level.
\end{itemize}
Secondary benchmarks involve the emotional diffusive
control units. The system tries to keep the activity
level of those units in a certain green zone.

\vspace{-3pt} \begin{quotation}{\it Emotions.} \quad By emotions we
denote for a cognitive system the goals resulting from the desire to keep
emotional diffusive control units at a preprogrammed level.
\end{quotation}

We note that the term \qut{emotion} is to a certain
extent controversial here. The relation of real
emotions experienced by biological cognitive systems, e.g.\ us
humans, to the above definition from cognitive system theory is
not fully understood at present. It is however known, that there
are no emotions without the concomitant release of appropriate
neuromodulators, viz without the activation of diffusive
control mechanisms.

\runinhead{Diffusive Emotional Control and Lifetime Fitness}
Emotional control is very powerful. An emotional diffusive control
signal like ``playing is good when you are not hungry or thirsty'',
to give an example, can lead a cognitive system to slowly develop
complex behavioral patterns. Higher-order explorative strategies,
like playing, can be activated when the fundamental genetic
preferences are momentarily satisfied. From the evolutionary
perspective emotional control serves to optimize lifetime fitness,
with the primary genetic preferences being responsible for the
day-to-day survival.

\runinhead{Tertiary Benchmarks and Acquired Tastes} The vast
majority of our daily actions is not directly dictated by our
fundamental genetic preferences. A wish to visit a movie theater
instead of a baseball match cannot be tracked back in any meaningful
way to the need to survive, to eat and to sleep.

Many of our daily actions are also difficult to
directly relate to emotional control. The decision to
eat an egg instead of a toast for breakfast involves
partly what one calls acquired tastes or preferences.

\vspace{-3pt} \begin{quotation}{\it Acquired Preferences.\enspace} A
learned connection, or association, between environmental sensory
input signals and the status of emotional control units is denoted
{as} an acquired taste or preference.
\end{quotation}
The term \qut{acquired taste} is used here in a very general
context, it could contain both positive or negative connotations,
involve the taste of food or the artistic impression of a painting.

Humans are able to go even one step further. We can establish
positive/negative feedback relations between essentially every
internal dynamical state of the cognitive system and emotional
diffuse control, viz we can set ourselves virtually any goal and
task. This capability is called \qut{freedom of will} in everyday
language. This kind of freedom of will is an emergent feature of
certain complex but\vadjust{\pagebreak} deterministic \nobreak
dynamical systems and we sidestep here the philosophically rather
heavy question of whether the thus defined freedom of will
corresponds to the true freedom of will.\footnote{From the point of
view of\index{freedom of action} dynamical systems theory effective
freedom of action is conceivable in connection to a true dynamical
phase transition, like the ones discussed in the
Chap.~\ref{chap_networks2}  possibly occurring in a high-level
cognitive system. Whether dynamical phase transitions are of
relevance for the brain of mammals, e.g.\ in relation to the
phenomenon of consciousness, is a central and yet unresolved issue.}

\begin{figure}[t]
\centerline{\includegraphics{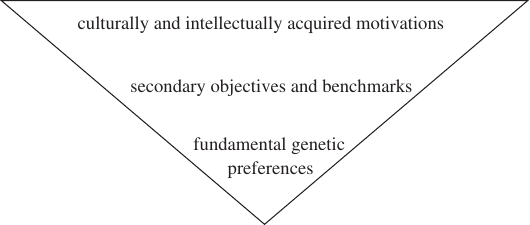}}
\caption{The inverse pyramid for the internal benchmarking of
complex and universal cognitive systems. The secondary benchmarks
correspond to the emotional diffusive control and the culturally
acquired motivations to the tertiary benchmarks, the acquired
preferences. A rudimentary cognitive system contains only the basic
genetic preferences, viz the preferred values for the survival
variables, for action selection} \label{cogSys_fig_invPyramid}
\end{figure}

\runinhead{The Inverse Pyramid} \index{cognitive system!benchmarking
pyramid} An evolved cognitive system will develop complex behavioral
patterns and survival strategies. The delicate balance of internal
benchmarks needed to stabilize complex actions goes beyond the
capabilities of the primary genetic preferences. The necessary fine
tuning of emotional control and acquired preferences is the domain
of the diffusive control system.

Climbing up the ladder of complexity, the cognitive system
effectively acquires a de facto freedom of action. The price for
this freedom is the necessity to benchmark internally any possible
action against hundreds and thousands of secondary and tertiary
desires and objectives, {which is} a delicate balancing
problem.

The layers of internal benchmarking can be viewed as an inverse
benchmarking pyramid, see Fig.~\ref{cogSys_fig_invPyramid} for an
illustration. The multitude of experiences and tertiary preferences
plays an essential role in the development of the inverse
pyramid{;} an evolved cognitive system is more than the
sum of its genetic or computer codes.

\enlargethispage{-6pt}

\section{Competitive Dynamics and Winning Coalitions}
\label{cogSys_dhan} Most of the discussions presented in this
chapter so far were concerned with general principles and concepts.
We will now discuss a functional basic cognitive \nobreak module
implementing illustratively the concepts treated in the preceding
sections. This \nobreak network is useful for environmental data
representation and storage and shows a continuous and self-regulated
transient state dynamics in terms of associative thought processes.
For some of the more technical details we refer to the literature.

\begin{figure}[t]
\centerline{\includegraphics{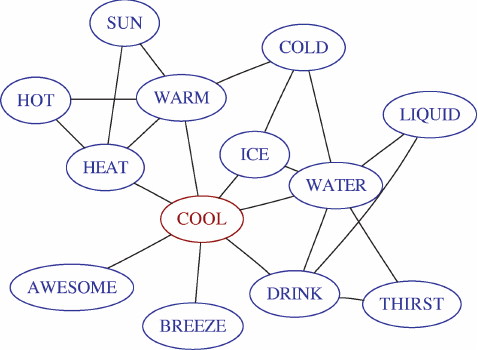}}
\caption{Extract of the human associative database. Test subjects
         were asked to name the first concept coming to their
         mind when presented with a cue randomly drawn
         from a dictionary database. In this graphical representation,
         starting from \qut{cool}, links have been drawn whenever the
         corresponding association was named repeatedly in several trials
         (generated from Nelson et al., 1998)}
\label{cogSys_fig_human_ass_database}
\end{figure}

\subsection{General Considerations}

\runinhead{The Human Associative Database} \index{associative!human
database} The internal representation of the outside world is a
primary task of any cognitive system with universal cognitive
capabilities, i.e. capabilities that are suitable for a certain
range of environments that are not explicitly encoded in genes or in
software. Associations between distinct representations of the
environment play an important role in human thought processes and
may rank evolutionary among the first cognitive capabilities not
directly \nobreak determined by gene expression. Humans dispose of a
huge commonsense knowledge base, organized dominantly via
associations, compare Fig.~\ref{cogSys_fig_human_ass_database}.
These considerations imply that associative information processing
in the form of associative thought processes plays a basic role in
human thinking.\enlargethispage*{6pt}
%

\begin{quotation}{\it Associative Thought Processes.\enspace}
\index{associative!thought process}
An associative thought process is the spontaneous generation
of a time series of transient memory states with a high
associative overlap.
\end{quotation}
Associative thought processes are natural candidates for transient
state \nobreak dynamics (see Sect.~\ref{cogSys_requirements}). The
above considerations indicate that associative thought\break
processes are, at least in part, generated directly in the cognitive
modules responsible for the environmental data representation.
{Below we will define the notion of \qut{associative} overlaps, see
Eqs.~(\ref{cogSys_ass_overlap_zero}) and
(\ref{cogSys_ass_overlap_one})}.\vspace*{3pt}

\runinhead{The Winners-Take-All Network} \index{neural
network!winners-take-all}\index{winners-take-all network}
Networks
in which the attractors are given by finite clusters of active
sites, the \qut{winners}, are suitable candidates for data
storage because (i) they have a very high storage capacity
and (ii) the competitive dynamics is directly controllable when
clique encoding is used.
\vspace{-4pt} \begin{quotation}{\it Cliques.\enspace}
\index{clique!winners-take-all network} A fully connected
subgraph of a network is called a clique, compare
Sect.~\ref{networks1_random_concepts}.
\end{quotation}
\vspace{-4pt}
Cliques are natural candidates for winning coalitions
of mutually supporting local computing units. Examples
for cliques in the human associative database, see
Fig.~\ref{cogSys_fig_human_ass_database}, are
(heat,hot,warm) and (drink,thirst,water).

\runinhead{Data Embedding} \index{learning!embedding problem} Data
is meaningless when not embedded into the context of other, existing
data. When properly embedded, data transmutes to information, see
the discussion in Sect.~\ref{cogSys_memory}.

\begin{figure}[t]
\centerline{\includegraphics{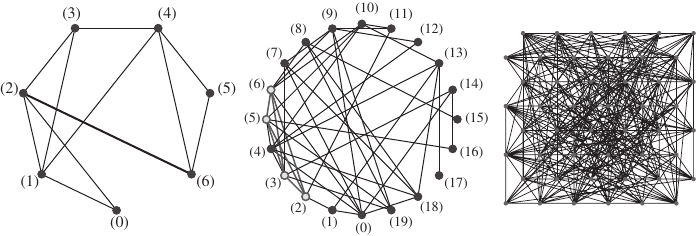}}
\caption{Illustration of winners-take-all networks with clique
encoding. Shown are the excitatory links. Sites not connected by a
line are inhibitorily connected. \textit{Left}: This 7-site network
contains the cliques (0,1,2), (1,2,3), (1,3,4), (4,5,6) and (2,6).
\textit{Middle}: This 20-site network contains 19, 10 and 1 cliques
with 2, 3 and 4 sites. The only 4-site clique (2,3,5,6) is
highlighted. \textit{Right}: This 48-site network contains 2, 166,
66 and 2 cliques (a total of 236 memories) with 2, 3, 4 and 5 sites,
respectively. Note the very high density of links}
\label{cogSys1_fig_memory_states}\vspace*{3pt}
\end{figure}

Sparse networks with clique encoding allow for a crude but automatic
embedding, viz embedding with zero computational effort. Any memory
state added to an existing network in the form of a clique, compare
Fig.~\ref{cogSys1_fig_memory_states}, will normally share nodes with
other existing cliques, viz with other stored memories. It thus
{automatically acquires} an
\qut{associative context}. The notion of associative context or
associative overlap will be defined precisely {}
below, see Eqs.~(\ref{cogSys_ass_overlap_zero}) and
(\ref{cogSys_ass_overlap_one}).

\runinhead{Inhibitory Background} Winners-take-all networks function
on the basis of a strong inhibitory background. In
Fig.~\ref{cogSys1_fig_memory_states} a few examples of networks with
clique encoding are presented. Fully connected clusters, the
cliques, mutually {excite} themselves. The winning
coalition suppresses the activities of all other sites, since there
is at least one inhibitory link between one of the sites belonging
to the winning coalition and any other site. All cliques therefore
form stable attractors.

The storage capacity is very large, due to the sparse coding. The
48-site network illustrated in Fig.~\ref{cogSys1_fig_memory_states}
has 236 stable memory states (cliques). We note for comparison that
maximally $6\approx 1.4*N$ memories could be stored for a $N=48$
network with distributed coding.

\begin{figure}[t]
\centerline{\includegraphics{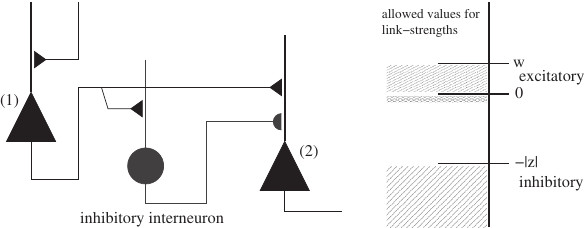}}
\caption{Synaptic strengths might be discontinuous when using
effective neurons. \textit{Left}: A case network of biological
neurons consisting of two neurons with exhibitory couplings (1) and
(2) and an inhibitory interneuron. The effective synaptic strength
(1)$\to$(2) might be weakly positive or strongly negative depending
on the activity status of the interneuron. The \textit{vertical
lines} symbolize the dendritic tree, the \textit{thin lines} the
axons ending with respective synapses. \textit{Right}: The resulting
effective synaptic strength. Weak inhibitory synaptic strengths do
not occur. For the significance of the small negative allowed range
for $w_{ij}$ compare the learning rule
Eq.~(\ref{cogSys_w_L_dot_opt}) (from Gros,\break 2007b) }
\label{cogSys_fig_gaps} 
\end{figure}

\runinhead{Discontinuous Synaptic Strengths} The clique encoding
works when the excitatory links are weak compared to the inhibitory
background. This implies that any given link cannot be weakly
inhibitory; the synaptic strength is discontinuous,
see~Fig.~\ref{cogSys_fig_gaps}. 

Discontinuous synaptic strengths also arise generically when
generating effective neural networks out of biological neural nets.
Biological neurons come in two types, excitatory neurons and
inhibitory interneurons. A biological neuron has either exclusively
excitatory or inhibitory outgoing synapses, never both types. Most
effective neurons used for technical neural networks have, on the
other {hand}, synaptic strengths of both signs. Thus,
when mapping a biological network to a network of effective neurons
one has to eliminate one degree of freedom, e.g.\ the inhibitory
interneurons.

\vspace{-3pt} \begin{quotation}{\it Integrating out Degrees of
Freedom.\enspace}
           A transformation of a model (A) to a model (B) by
           eliminating certain degrees of freedom occurring in
           (A), but not in (B) is called \qut{integrating out a
           given degree of freedom}, a notion of widespread
           use in theoretical physics.
\end{quotation}
This transformation depends strongly on the properties of the
initial model. Consider the small biological network depicted in
Fig.~\ref{cogSys_fig_gaps}, for the case of strong inhibitory
synaptic strength. When the interneuron is active/inactive the
effective (total) influence of neuron (1) on neuron (2) will be
strongly negative/weakly positive.\footnote{ We note that general
$n$-point interactions could be generated additionally when
eliminating the interneurons. \qut{$n$-point interactions}
are terms entering the time evolution of dynamical systems
depending on ($n-1$) variables. Normal synaptic interactions
are 2-point interactions, as they involve two neurons,
the presynaptic and the postsynaptic neuron. When integrating
out a degree of freedom, like the activity of the interneurons,
$n$-point interactions are generated generally.
The postsynaptic neuron is then influenced only when
({$n-1$}) presynaptic neurons are active
simultaneously. $n$-point interactions are normally not considered
in neural networks theory. They complicate the analysis of the
network dynamics considerably.
           }

\runinhead{Transient Attractors} \index{attractor!transient} The
network described so far has many stable attractors,
{i.e.} the cliques. These patterns are memories
representing environmental data found as typical patterns in the
incoming sensory data stream.

It clearly does not make sense for a cognitive system to
remain stuck for eternity in stable attractors. Every attractor
of a cognitive system needs to be a transient
attractor,\footnote{Here we use the term
  \qut{transient attractor} as synonymous with
  \qut{attractor ruin}, an alternative terminology from
   dynamical system theory.}
i.e.\ to be part of the transient state dynamics.

There are many ways in dynamical {systems} theory
by which attractors can become unstable. The purpose of any
cognitive system is cognitive information processing and associative
thought processes constitute the most fundamental form of cognitive
information processing. We therefore discuss here how memories can
take part, in the form of transient attractors, in associative
thought processes.

\runinhead{Associative Overlaps} \index{associative!overlap} Let us
denote by $x_i\in[0,1]$ the activities of the network
($i=1,\ldots,N$) and by
$$
x_i^{(\alpha)},\qquad\quad \alpha=1,\ldots,N^{(m)}
$$
the activation patterns of the $N^{(m)}$ memories, the
stable attractors. In winners-take-all networks {}
$x_i^{(\alpha)}\to 0,1$.

For the seven-site network illustrated in
Fig.~\ref{cogSys1_fig_memory_states} the number of cliques is
$N^{(m)}=5$ and for the clique $\alpha = (0,1,2)$ the activities
approach $x_i^{(0,1,2)}\to1$ (i=0,1,2) for members of the winning
coalition and $x_j^{(0,1,2)}\to0$ ($j=3,4,5,6$) for the
out-of-clique units.
\begin{quotation}{\it Associative Overlap of Order Zero.\enspace}
We define the associative overlap of zero order
\begin{equation}
A_0[\alpha,\beta] \ =\
\sum_{i=0}^N x_i^{(\alpha)}x_i^{(\beta)}
\label{cogSys_ass_overlap_zero}
\end{equation}
for two memory states $\alpha$ and $\beta$ and
for a network using clique encoding.
\end{quotation}
The associative overlap of order zero just counts the
number of common constituting elements.

For the seven-site network shown in
Fig.~\ref{cogSys1_fig_memory_states} we have $A_0[(0,1,2),(2,6)]=1$
and $A_0[(0,1,2),(1,2,3)]=2$.
\begin{quotation}{\it Associative Overlap of Order 1.\enspace}
We define by
\begin{equation}
A_1[\alpha,\beta] \ =\
\sum_{\gamma\ne\alpha,\beta}
 \left(\sum_{i} x_i^{(\alpha)}(1-x_i^{(\beta)})x_i^{(\gamma)}
 \right)
 \left(\sum_{j} x_j^{(\gamma)}(1-x_j^{(\alpha)})x_j^{(\beta)}\right)
\label{cogSys_ass_overlap_one}
\end{equation}
the associative overlap of first order for two memory states
$\alpha$ and $\beta$ and a network using clique encoding.
\end{quotation}
The associative overlap of order 1 is the sum of multiplicative
associative overlap of zero order {that} the
disjunct parts of two memory states $\alpha$ and $\beta$ have with
all third memory states $\gamma$. It counts the number of
associative links connecting two memories.

For the seven-site network shown in
Fig.~\ref{cogSys1_fig_memory_states} we have
$A_1[(0,1,2),(4,5,6)]=2$ and $A_1[(0,1,2),(1,3,4)]=1$.

\runinhead{Associative Thought Processes} \index{associative!thought
process} Associative thought processes convenes maximal cognitive
information processing when they correspond to a time
series of memories characterized by high associative overlaps of
order zero or one.

In Fig.~\ref{cogSys_fig_7sites} the orbits resulting from a
transient state dynamics{, which we will introduce in
Sect.~\ref{cogSys_sec_ass_thinking}} are illustrated.
Therein two consecutive winning coalitions have either an
associative overlap of order zero, such as the transition
$(0,1)\to(1,2,4,5)$ or of order 1, as the transition
$(1,2,4,5)\to(3,6)$.

\subsection{Associative Thought Processes}
\label{cogSys_sec_ass_thinking}
\index{associative!thought process|textbf}

We now present a functioning implementation, in terms of a set of
appropriate coupled differential equations, of the notion of
associative thought processes as a time series of transient
attractors representing memories in the environmental data
representation module.\enlargethispage*{12pt}

\runinhead{Reservoir Variables} \index{reservoir!variable} A
standard procedure, in dynamical system theory, to control the
long-term dynamics of a given variable of interest is to couple it
to a second variable with much longer time scales. This
is the principle of time scale separation.
\index{time scale separation} To be concrete we
denote, as hitherto, by $x_i\in[0,1]$ the
activities of the local computational
units constituting the network and by
$$
\varphi_i\ \in\ [0,1]
$$
a second variable, which we denote {\em reservoir}. The differential
equations\vspace{8pt}
\begin{eqnarray} \label{cogSys_xdot}
\dot x_i &=& (1-x_i)\,\Theta(r_i)\,r_i \,+\,
x_i\,\Theta(-r_i)\,r_i~,
\\[8pt] \label{cogSys_ri}
    r_i &=&
    \sum_{j=1}^N \Big[
    f_w(\varphi_i) \Theta(w_{ij}) w_{i,j}
    + z_{i,j}f_z(\varphi_j)
        \Big] x_j~,
\\[8pt] \label{cogSys_phidot}
\dot\varphi_i & =& \Gamma_\varphi^+\, (1-\,\varphi_i)(1-x_i/x_c)
\Theta(x_c-x_i) \,-\, \Gamma_\varphi^-\,\varphi_i\,\Theta(x_i-x_c)~,
\\[8pt] \label{cogSys_z_t}
z_{ij} & =& -|z|\,\Theta(-w_{ij})\vspace{8pt}
\end{eqnarray}
generate associative thought processes. We now discuss some
properties of Eqs.~(\ref{cogSys_xdot}), (\ref{cogSys_ri}), (\ref{cogSys_phidot}) and (\ref{cogSys_z_t}).
The general form of these differential equations is termed
the \qut{Lotka--Volterra} type.
\index{Lotka--Volterra equations}
\index{differential equation!Lotka-Volterra}
\begin{itemize}

\item[--]Normalization: {Equations
(\ref{cogSys_xdot}), (\ref{cogSys_ri}) and (\ref{cogSys_phidot})} respect the
normalization $x_i,\varphi_i\in[0,1]$, due to the prefactors
$x_i$,$(1-x_i)$, $\varphi_i$ and $(1-\varphi_i)$ in
Eqs.~(\ref{cogSys_xdot}) and (\ref{cogSys_phidot}), for the
respective growth and depletion processes, {and}
$\Theta(r)$ is the Heaviside step function.

\item[--]Synaptic Strength: \index{synaptic!strength}The synaptic strength is split into
excitatory and inhibitory contributions, $\propto w_{i,j}$ and
$\propto z_{i,j}$, respectively, with $w_{i,j}$ being the primary
variable: The inhibition $z_{i,j}$ is present only when the link is
not excitatory{, Eq.~}(\ref{cogSys_z_t}). With
$z\equiv-1$ one sets the inverse unit of time.

\item[--]{The} Winners-Take-All Network: \index{winners-take-all network}Equations (\ref{cogSys_xdot}) and
(\ref{cogSys_ri}) describe, in the absence of a coupling to the
reservoir via $f_{z/w}(\varphi)$, a competitive winners-take-all
neural network with clique encoding. The system relaxes towards the
next attractor made up of a clique of $Z$ sites $(p_1,\dots,p_Z)$
connected excitatory via $w_{p_i,p_j}>0$ ($i,j=1,\ldots ,Z$).

\begin{figure}[t]
\sidecaption[b] \vspace{6pt}
{\includegraphics{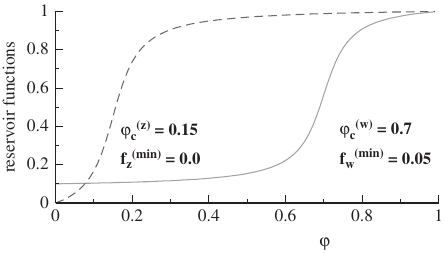}}
\caption{{The} reservoir functions
$f_{w}(\varphi)$ (\textit{solid line}) and $f_{z}(\varphi)$
(\textit{dashed line}), see Eq.~(\ref{cogSys_ri}), of sigmoidal form
with respective turning points $\varphi_c^{(f/z)}$ and width
$\Gamma_\varphi=0.05$}
\label{cogSys_fig_reservoirFunctions}
\end{figure}

\item[--]Reservoir Functions: \index{reservoir!function}The reservoir functions
$f_{z/w}(\varphi)\in[0,1]$ govern the interaction between the
activity levels $x_i$ and the reservoir levels $\varphi_i$. They may
be chosen as washed out step functions of sigmoidal form\footnote{A
possible mathematical implementation for the reservoir functions,
with $\alpha=w,z$, is $ f_\alpha(\varphi)\ =\ f_\alpha^{(\min)}
\,+\, \left(1-f_\alpha^{(\min)}\right) { {\rm
atan}[(\varphi-\varphi_c^{(\alpha)})/\Gamma_\varphi] - {\rm
atan}[(0-\varphi_c^{(\alpha)})/\Gamma_\varphi] \over {\rm
atan}[(1-\varphi_c^{(\alpha)})/\Gamma_\varphi] - {\rm
atan}[(0-\varphi_c^{(\alpha)})/\Gamma_\varphi] } $. Suitable values
are $\varphi_c^{(z)}=0.15$, $\varphi_c^{(w)}=0.7$
$\Gamma_\varphi=0.05$, $f_w^{(\min)}=0.1$ and $f_z^{(\min)}=0$.
 }
with a suitable width $\Gamma_{\varphi}$ and inflection points
$\varphi_c^{(w/z)}$, see Fig.~\ref{cogSys_fig_reservoirFunctions}.

\item[--]Reservoir Dynamics: \index{reservoir!dynamics}The reservoir levels of the winning
clique deplete slowly, see Eq.~(\ref{cogSys_phidot}), and recovers
only once the activity level $x_i$ of a given site has dropped below
$x_c$. The factor $(1-x_i/x_c)$ occurring in the reservoir growth
\hbox{process}, see the {right-hand side} of
{Eq.~}(\ref{cogSys_phidot}), serves {as} a
stabilization of the transition between subsequent memory states.

\item[--]Separation of Time Scales: \index{time scale separation}\looseness1 A separation of time scales is
obtained when {} $\Gamma_\varphi^\pm$ are much smaller
than the average strength of an excitatory link, $\bar w$, leading
to transient state dynamics. Once the reservoir of a winning clique
is depleted, it {loses}, via $f_z(\varphi)$, its
ability to suppress other sites. The mutual intraclique excitation
is suppressed via $f_w(\varphi)$.

\end{itemize}

\begin{figure}[t]
\centerline{\includegraphics{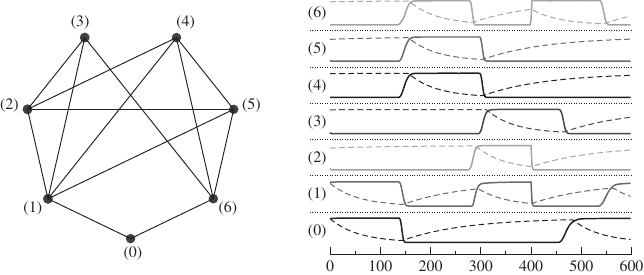}}
\caption{\textit{Left}: A seven-site network{;} shown
are links with $w_{i,j}>0$, containing six cliques, (0,1), (0,6),
(3,6), (1,2,3), (4,5,6) and (1,2,4,5).  \textit{Right}: The
activities $x_i(t)$ (\textit{solid lines}) and the respective
reservoirs $\varphi_i(t)$ (\textit{dashed lines}) for the transient
state dynamics $(0,1)\rightarrow(1,2,4,5) \rightarrow (3,6)
\rightarrow(1,2,4,5)$ \label{cogSys_fig_7sites}
        }\vspace{6pt}
\end{figure}

\vspace{6pt} \runinhead{Fast and Slow Thought Processes}
\index{variable!fast and slow} 
Figure \ref{cogSys_fig_7sites} illustrates the transient state dynamics
resulting from Eqs.\ (\ref{cogSys_xdot}), (\ref{cogSys_ri}), 
(\ref{cogSys_phidot}) and (\ref{cogSys_z_t}), in the
absence of any sensory signal. When the growth/depletion rates
$\Gamma_\varphi^\pm\to0$ are very small, the individual cliques turn
into stable attractors.

The possibility to regulate the \qut{speed} of the associative
thought process arbitrarily by setting {} $\Gamma_\varphi^\pm$ is
important for applications. For a working cognitive system it is
enough if the transient states are just stable for a certain minimal
period, anything longer just would be a \qut{waste of
time}.\vspace*{3pt}

\runinhead{Cycles} \index{cycle!thought process} The system in
Fig.~\ref{cogSys_fig_7sites} is very small and the associative
thought process soon settles into a cycle, since there are no
incoming sensory signals in the simulation of
Fig.~\ref{cogSys_fig_7sites}.

\begin{figure}[t]
\centerline{\includegraphics{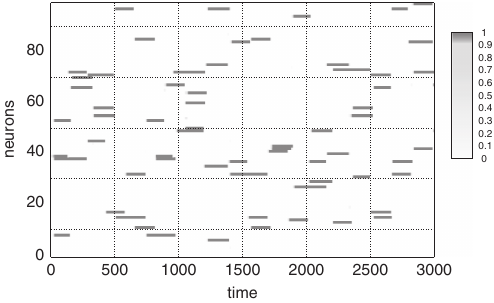}}
\caption{Example of an associative thought process in a network
containing 100 artificial neurons and 713 stored memories. The times
runs horizontally, the site index
vertically ($i=1,\ldots,100$). The neural
activities $x_i(t)$ are color coded
        }
\label{cogSys_fig_100_acti}
\end{figure}

For networks containing a somewhat larger number of sites, see
Fig.~\ref{cogSys_fig_100_acti}, the number of attractors can be very
large. The network will then generate associative thought processes
{that} will go on for very long time spans before
entering a cycle. Cyclic \qut{thinking} will normally not occur for
real-world cognitive systems interacting continuously with the
environment. Incoming sensory signals will routinely interfere with
the ongoing associative dynamics, preempting cyclic activation of
memories.

\runinhead{Dual Functionalities for Memories} \index{memory!dual
functionality} \index{model!dHAN} \index{dHAN model} The network
discussed here is a dense and homogeneous associative network
(dHAN). It is homogeneous since memories have dual functionalities:
\begin{itemize}
\item[--] Memories are the transient states of the associative thought
      process.
\item[--] Memories define the associative overlaps,
      see Eq.~(\ref{cogSys_ass_overlap_one}), between two
      subsequent transient states.
\end{itemize}
The alternative would be to use networks with two kinds
of constituent elements, as in semantic networks.
\index{network!semantic}
\index{semantic network}
The semantic relation\vspace{6pt}
\begin{center}
\framebox{\parbox{3cm}{\begin{center}\large car\end{center}}} \ \ \
\rule{1cm}{1pt}  \ \ \ {\large is} \ \ \ \rule{1cm}{1pt} \  \ \
\framebox{\parbox{3cm}{\begin{center}\large blue\end{center}}}
\end{center}\vspace{6pt}
can be thought to be part of a (semantic) network containing the
nodes \qut{car} and \qut{blue} linked by the relation \qut{is}. Such
a network would contain two kinds of different constituting
elements, the nodes and the links. The memories of the dHAN, on the
other hand, are made up of cliques of nodes and it
is therefore\break homogeneous.

A rudimentary cognitive system knows of no predefined concepts and
cannot, when starting from scratch, {initially classify} data into \qut{links} and
\qut{nodes}. A homogeneous network is consequently the network of
choice for rudimentary cognitive systems.

\vspace{6pt}

\runinhead{Dissipative Dynamics} Interestingly, the phase space
contracts at all times in the absence of external inputs. With
respect to the reservoir variables, we have
$$
\sum_i {\partial \dot\varphi_i\over\partial\varphi_i}
\ =\ -\sum_i\left[\Gamma_\varphi^+(1-x_i/x_c)\Theta(x_c-x_i)
                 +\Gamma_\varphi^-\Theta(x_i-x_c)
            \right] \ \le\ 0~,
$$
$\forall x_i\in[0,1]$, where we have used Eq.~(\ref{cogSys_phidot}).
We note that the diagonal contributions to the link matrices vanish,
$z_{ii}=0=w_{ii}$, and therefore $\partial r_i/\partial x_i =0$. The
phase space {consequently
contracts} also with respect to the activities,
$$
\sum_i {\partial \dot x_i\over\partial x_i}
\ =\ \sum_i \,\Big[\,\Theta(-r_i) -\Theta(r_i)\, \Big]\, r_i
 \ \le\ 0~,
$$
where we have used {Eq.~}(\ref{cogSys_xdot}). The system
is therefore strictly dissipative, compare
Chap.~\ref{chap_chaos1} in the absence of
external stimuli.

\vspace{6pt}

\runinhead{Recognition} \index{recognition} Any sensory stimulus
arriving {in} the dHAN needs to compete with the
on\-going intrinsic dynamics to make an impact. If the sensory
signal is not strong enough, it cannot deviate the autonomous
thought process. This feature results in an intrinsic recognition
property of the dHAN: A background of noise will not influence the
transient state dynamics.

\subsection{Autonomous Online Learning}
\label{cogSys_subsec_online_learning}

It is characteristic to the theory of cognitive
systems, as pointed out in the introduction
(Sect.~\ref{cogSys_introduction}), that the
exact equations used to model a phenomena of
interest are not of importance. There is a
multitude of possible formulations and a range of
suitable modeling approaches may lead to
similar overall behavior -- the principles are
more important than the details of their actual
implementation. In the preceding
section we have discussed a formulation of
transient state dynamics based on competitive
clique encoding. Within this framework we will
now illustrate the basic functioning of
homeostatic regulation.

\runinhead{Local Working Point Optimization}
\index{working point optimization}
Dynamical systems normally retain their functionalities
only when they keep their dynamical properties within certain\break
regimes. They need to regulate their own working point,
as discussed in Sect.~\ref{cogSys_principles}, via
homeostatic regulation. The working point optimization
might be achieved either through diffusive control signals
or via local optimization rules. Here we discuss an example of
a local rule.

\runinhead{Synaptic Plasticities}
\index{synaptic!plasticity}
The inhibitory and the excitatory synaptic strength
have different functional roles in the
dHan formulation of transient state dynamics.
The average strengths $|z|$ and $\bar w$
of the inhibitory and excitatory links
differ substantially,
$$
  |z|\ \gg\  \bar w~,
$$
compare Fig.~\ref{cogSys_fig_gaps}. Homeostatic
regulation is a slow process involving incremental
changes. It is then clear that these gradual changes
in the synaptic strengths will affect dominantly
the excitatory links, as they are much smaller,
since small changes of large parameters (like the
inhibitory links) do not influence substantially,
quite in general, the properties of a dynamical system.
We may therefore consider the inhibitory background
as given and fixed and restrict the effect of
homeostatic regulation to the excitatory $w_{ij}$.

\runinhead{Effective Incoming Synaptic Strength} The
average magnitude of the growth rates $r_i$, see
Eq.~(\ref{cogSys_ri}), determines the time scales
of the autonomous dynamics and thus the working
point. The $r_i(t)$ are, however, quite strongly
time dependent. The effective incoming
synaptic signal\vspace{3pt}
\[
\tilde r_i \,=\, \sum_{j}\Big[ w_{i,j}x_j
  \,+\, z_{i,j}x_jf_z(\varphi_j)\Big]~,\vspace{3pt}
\]
which is independent of the postsynaptic reservoir,
$\varphi_i$, is a more convenient control parameter
for a local homeostatic regulation, since $\tilde r_i$
tends to the sum of active incoming links,\vspace{3pt}
$$
\tilde r_i\,\to\, \sum_{j\in\alpha} w_{i,j}~,\vspace{3pt}
$$
for a transiently stable clique
$\alpha=(p_1,\dots,p_Z)$.

\runinhead{Optimal Growth Rates}
The working point of the dHan is optimal when the effective
incoming signal is, on the average, of comparable
magnitude $r^{(opt)}$ for all sites,
\begin{equation}
\tilde r_i\ \to\ r^{(\mathrm{opt})}~.
\label{cogSys1_r_opt}
\end{equation}
$r^{(\mathrm{opt})}$ is an unmutable parameter, compare
Fig.~\ref{cogSys_fig_parameters}. There is no need to
fulfill this rule exactly for every site $i$. The dHan
network will retain functionality whenever
Eq.~(\ref{cogSys1_r_opt}) is approached slowly and on
the average by suitable synaptic plasticities.

\runinhead{Long-Term Homeostatic Plasticities}
The working point optimization Eq.~(\ref{cogSys1_r_opt})
can be achieved through a suitable local rule:
\begin{eqnarray}
\label{cogSys_w_L_dot_opt} \dot w_{ij}(t) & =&
\Gamma_{L}^{(\mathrm{opt})} \Delta \tilde r_i \Big[\,
\left(w_{ij}-W_L^{(\min )}\right) \Theta(-\Delta \tilde r_i) +
\Theta(\Delta \tilde r_i)
                \,\Big] \\ &&\cdot
\,\Theta(x_i-x_c)\,\Theta(x_j-x_c),
\nonumber \\
& - & \Gamma_L^- \, d(w_{ij})\, \Theta(x_i-x_c)\,\Theta(x_c-x_j)~,
\label{cogSys_w_L_dot_decay}
\end{eqnarray}
with
$$
\Delta \tilde r_i\ =\ r^{(\mathrm{opt})}-\tilde r_i~.
$$

Some comments:
\begin{itemize}
\item[--] Hebbian learning: \index{Hebbian learning}
     The learning rule Eq.~(\ref{cogSys_w_L_dot_opt}) is local
     and of Hebbian type. Learning occurs only when the presynaptic
     and the postsynaptic neurons are active. Weak forgetting, i.e.\
     the decay  of rarely used links, Eq.~(\ref{cogSys_w_L_dot_decay})
     is local too.
\item[--]Synaptic Competition: \index{synaptic!competition}
    When the incoming signal is weak/strong, relative to the
    optimal value $r^{(\mathrm{opt})}$, the active links are
    reinforced/weakened, with $W_L^{(\min )}$ being the minimal
    value for the $w_{ij}$. The baseline $W_L^{(\min )}$ may be
    chosen to be slightly negative, compare Fig.~\ref{cogSys_fig_gaps}.

      The Hebbian-type learning then takes place in the form of
      a competition between incoming synapses -- frequently
      active incoming links will gain strength, on the average,
      on the expense of rarely used links.
\item[--]Asymmetric Decay of Inactive Links:
      The decay term $\ \propto \Gamma_L^->0\ $
      in Eq.~(\ref{cogSys_w_L_dot_decay})
      is taken to be asymmetric,
      viz when the presynaptic neuron is inactive
      with the postsynaptic neuron being active. The
      strength of the decay is a suitable non-linear function
      $d(w_{ij})$ of the synaptic strength $w_{ij}$.
      Note that the opposite asymmetric decay, for which
      $w_{ij}$ is weakened whenever the {presynaptic}/postsynaptic neurons
      are active/inactive, may potentially lead to the
      dynamical isolation of the currently active clique by
      suppressing excitatory out-of-clique synapses.
\item[--] Suppression of Runaway Synaptic Growth:
          \index{learning!runaway effect}
      The link dynamics{, Eq.~}(\ref{cogSys_w_L_dot_opt}) suppresses
      synaptic runaway growth, a general problem common to
      adaptive and continuously active neural networks. It has
      been shown that similar rules for discrete neural networks
      optimize the overall storage capacity.
\item[--] Long-Term Dynamical Stability:
      In Fig.~\ref{cogSys_fig_100_acti} an example for an
      associative thought process is shown for a
      100-site network containing 713 memories.
      When running the simulation
      for very long times one finds that the values of
      excitatory links $w_{ij}$ tend to a steady-state
      distribution, as the result of the continuous
      online learning. The system is self-adapting.
\end{itemize}

\runinhead{Conclusions} In this section we presented and discussed
the concrete implementation of a module for competitive transient
state dynamics within the dHan (dense and homogeneous associative
network) approach. Here we have discussed only the isolated module,
one can couple this module to a sensory input stream and the
competitive neural dynamics will then lead to semantic learning. The
winning coalitions of the dHan module, the cliques, will then
acquire a semantic context, corresponding via their respective
receptive fields to prominent and independent patterns and objects
present in the sensory stimuli.\enlargethispage*{6pt}

The key point is however that this implementation fulfills
all requirements necessary for an autonomous cognitive system,
such as locality of information processing, unsupervised online
learning, huge storage capacity, intrinsic generalization
capacity and self-sustained transient state dynamics in
terms of self-generated associative thought processes.

\section{Environmental Model Building}
\label{cogSys_prediction}
\index{environment!model building|textbf}

The representation of environmental data, as discussed in
Sect.~\ref{cogSys_dhan}, allows for simple associational reasoning.
For anything more sophisticated, the cognitive system needs to learn
about the structure of the environment itself, {i.e.} it
has to build models of the environment.

The key question is then: Are there universal principles
{that} allow for environmental model building
without any a priori information about the environment? Principles
{that} work independently of whether the cognitive
system lives near a {lakeside in a}
tropical rain forest or in an artificial cybernetical
{world.}

Here we will discuss how {\em universal prediction tasks} allow for
such universal environmental model building and for the spontaneous
generation of abstract \nobreak concepts.

\subsection{The Elman Simple Recurrent Network}
\label{cogSys_subsect_SRN}
\index{Elman network|textbf}

\runinhead{Innate Grammar} Is the human brain completely empty at
birth and {can babies} learn with the same ease
any language, natural or artificial, with arbitrary grammatical
organization? Or do we have certain gene determined predispositions
toward certain innate grammatical structures? This issue has been
discussed by linguists for decades.

\begin{figure}[t]
\centerline{\includegraphics{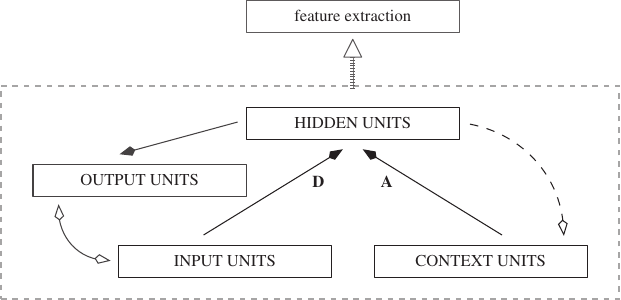}}
\caption{The Elman simple recurrent network (inside the
\textit{dashed box}). The connections ({\bf D}: input$\to$hidden),
({\bf A}: context$\to$hidden) and (hidden$\to$output) are trained
via the backpropagation algorithm.  At every time step the content
of the hidden units is copied into the context units on a one-to-one
basis. The difference between {the} output signal and the new input
signal constitutes the error for the training. The hidden units
generate abstract concepts {that} can be used for further processing
by the cognitive system via standard feature extraction
        }
\label{cogSys_fig_simpleRecurrent}
\end{figure}

In this context {in 1990 Elman
performed} a seminal case study, examining the representation of
time-dependent tasks by a simple recurrent network. This network is
universal in the sense that no information about the content or
structure of the input data stream is used in its layout.

Elman discovered that lexical classes are spontaneously generated
when the network {is} given the task to predict the
next word in an incoming data stream made up of natural sentences
constructed from a reduced vocabulary.

\runinhead{The Simple Recurrent Network} When the task of a neural
network extends into the time domain it needs a memory, otherwise
comparison of current and past states is impossible. For the simple
recurrent network, see Fig.~\ref{cogSys_fig_simpleRecurrent}, this
memory is constituted by a separate layer of neurons denoted {\sl
context units}.

The simple recurrent network used by Elman employs
discrete time updating. At every time step the following
computations are performed:
\begin{enumerate}
\item The activities of the hidden units
  are determined by the activities of the input units
  and by the activities of the context units
  and the respective link matrices.
\item The activities of the output units
  are determined by the activities of the hidden units
  and the respective link matrix.
\item The activities of the hidden units are copied
      one-by-one to the context unit.
\item The next input signal is copied to the input units.
\item The activities of the output units are compared
      to the current input and the difference yields the
      error signal. The weight of the link matrices
      (input$\to$hidden),  (context$\to$hidden) and
      (hidden$\to$output) are adapted such to reduce the
      error signal. This procedure is called the back-propagation
      algorithm.
\end{enumerate}
The Elman net does not conform in this form to the requirements
needed for modules of a full-fledged cognitive system, see
Sect.~\ref{cogSys_requirements}. It employs discrete time
synchronous updating and non-local learning rules based on a global
optimization condition, the so-called back-propagation algorithm.
This drawback is, however, not essential at this point, since we are
interested here in the overall and generic \hbox{properties} of the
simple recurrent network.

\runinhead{The Lexical Prediction Task} \index{Elman network!lexical
prediction task} \index{prediction task!lexical} The simple
recurrent network works on a time series $\veci x(t)$ of inputs
$$
   \veci x(1),\ \veci x(2),\ \veci x(3),\ \ldots
$$
which are presented to the network one after the other.

The network has the task to predict the next input. For the case
studied by Elman the inputs $\veci x(t)$ represented randomly
encoded words out of a reduced vocabulary of 29 lexical items. The
series of inputs corresponded to natural language sentences obeying
English grammar rules. The network {then had} the
task to predict the next word in a sentence.

\runinhead{The Impossible Lexical Prediction Task} \index{prediction
task!impossible} The task to predict the next word of a natural
language sentence is impossible to fulfill. Language is
non-deterministic, communication would otherwise convene no
information.

The grammatical structure of human languages places constraints on
the possible sequence of words, a verb is more likely to follow a
noun than another verb, to give an example. The expected frequency
of possible successors, {implicit} in the set
of training sentences, is, however, deterministic and is reproduced
well by the simple recurrent network.

\begin{figure}[t]
\centerline{\includegraphics{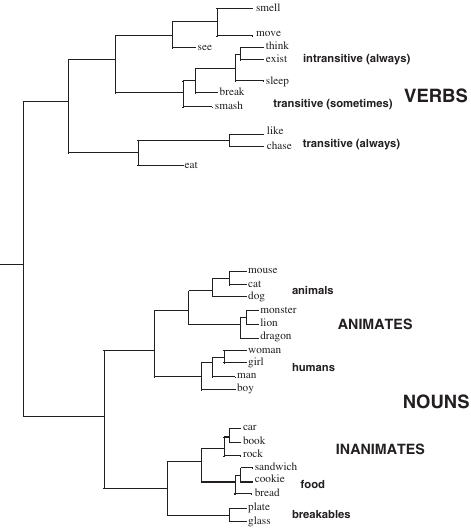}}
\caption{Hierarchical cluster diagram of the hidden units activation
pattern. Shown are the relations and similarities of the hidden unit
activity patterns according to a hierarchical cluster analysis (from
Elman, 2004)
           }
\label{cogSys_fig_elmanHieracy} 
\end{figure}

\runinhead{Spontaneous Generation of Lexical Types} Let us
recapitulate the situation:
\begin{enumerate}\leftskip6pt
\item[i.\phantom{ii}] The lexical prediction task given to the network
           is impossible to fulfill.
\item[ii.\phantom{i}] The data input stream has a hidden grammatical structure.
\item[iii.] The frequency of successors is not random.
\end{enumerate}
{As a
consequence, the network generates} in its hidden layer
representations of the 29 used lexical items, see
Fig.~\ref{cogSys_fig_elmanHieracy}. These representations, and this
is the central result of Elman's 1990 study, have a characteristic
hierarchical structure. Representations of different nouns, e.g.\
\qut{mouse} and \qut{cat}, are more alike than the representations
of a noun and a verb, e.g.\ \qut{mouse} and \qut{sleep}. The network
has generated spontaneously abstract lexical types like verb, nouns
of animated objects and nouns of inanimate objects.

\runinhead{{Tokens} and Types} The network
{actually generated} representations of
the lexical items dependent on the context, the tokens. There is not
a unique representation of the item {\sl boy}, but several, viz $\sl
boy_1$, $\sl boy_2, \ldots$, which are very similar to each other,
but with fine variations in their respective activation patterns.
These {depend} on the context, as in the
following training sentences:\vspace{3pt}
\medskip

\centerline{man\_smell\_BOY,\hspace{4ex}
            man\_chase\_BOY,\hspace{4ex}... }
\medskip\vspace{3pt}
The simple recurrent network is thus able to generate both abstract
lexical types and concrete lexical tokens.

\vspace{6pt}

\begin{figure}[t]
\centerline{\includegraphics{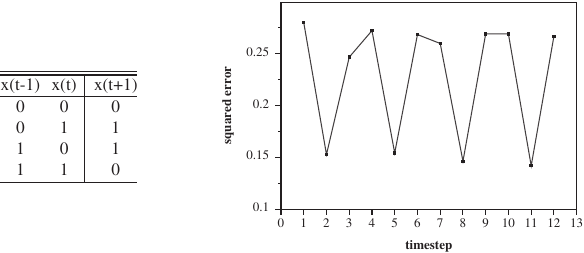}}
\caption{The temporal XOR. \textit{Left}: The prediction task.
\textit{Right}: The performance $(y(t+1)-x(t+1))^2$ ($y(t)\in[0,1]$
       is the activity of the single output neuron{} of a simple recurrent
       network, see Fig.~\ref{cogSys_fig_simpleRecurrent},
       with two neurons in the hidden layer after 600 sweeps
       through a 3000-bit training \nobreak sequence
        }
\label{cogSys_fig_temporal_XOR}
\end{figure}

\runinhead{Temporal XOR} \index{XOR}\index{temporal XOR} The XOR
problem, see Fig.~\ref{cogSys_fig_temporal_XOR}, is a standard
prediction task in neural network theory. In its temporal version
the two binary inputs are presented one after the other to the same
input neuron as $x(t-1)$ and $x(t)$, with the task to predict the
correct $x(t+1)$.

The XOR problem is not linearly decomposable, i.e.\ there
are no constants $a,b,c$ such that
$$
x(t+1)\ =\ a\,x(t)\,+\,b\,x(t-1)\,+\,c~,
$$
and this is why the XOR problem serves as a benchmark for
neural prediction tasks. Input sequences like
$$
\dots\,\underbrace{0\,0\,0}_{}\,
       \underbrace{1\,0\,1}_{}\,
       \underbrace{1\,1\,0}_{}\,\dots
$$
are presented to the network with the caveat that the network does
not know when an XOR-triple starts. A typical result is shown in
Fig.~\ref{cogSys_fig_temporal_XOR}. Two out of three prediction
results are random, as expected but every third prediction is quite
good.\enlargethispage*{10pt}

\vspace*{-3pt}

\runinhead{The Time Horizon} \index{time!horizon} Temporal
prediction tasks may vary in complexity depending on the time scale
$\tau$ characterizing the duration of the temporal dependencies in
the input data $\veci x(t)$. A well known example is the Markov
process.

\begin{quotation}{\it The Markov Assumption.\enspace} \index{time series analysis!Markov
assumption}\index{Markov assumption} The distribution of possible
$\veci x(t)$ depends only on the value of the input at the previous
time step, $\veci x(t-1)$.
\end{quotation}
For Markovian-type inputs the time correlation length of the input
data is {1;} $\tau=1$. For the temporal XOR problem
$\tau=2$. {In
principle, the simple recurrent network is able to handle time
correlations of arbitrary length}. It has been tested with respect
to the temporal XOR and to a letter-in-a-word prediction task. The
performance of the network in terms of the accuracy of the
prediction results{, however, is} expected to
deteriorate with increasing $\tau$.

\vspace*{-6pt}

\subsection{Universal Prediction Tasks}
\index{prediction task!universal|textbf}
\index{universality!temporal prediction task|textbf}

\runinhead{Time Series Analysis} \index{time series analysis!neural
network}\index{neural network!time series analysis} The Elman simple
recurrent network is an example of a neural network layout {that is}
suitable for time series analysis. Given a series of~vectors
$$
\veci x(t),\qquad\quad t=0,\ 1,\ 2,\ \ldots
$$
one might be interested in forecasting $\veci x(t+1)$ when $\veci
x(t),\ \veci x(t-1),\ldots$ are known. Time series analysis is very
important for a wide range of applications and a plethora of
specialized algorithms have been developed.\enlargethispage{12pt}
\runinhead{State Space Models} \index{model!state space}\index{state
space model}\index{time series analysis!state space model} Time
series generated from physical processes can be described by
\qut{state space models}. The daily temperature in Frankfurt is a
complex function of the weather dynamics, which contains a huge
state space of (mostly) unobservable variables. The task to predict
the local temperature {from only} the knowledge
of the history of previous temperature readings constitutes a time
series analysis task.

Quite generally, there are certain deterministic
or stochastic processes generating a series
$$
\veci s(t),\qquad\quad t=0,\ 1,\ 2,\ \ldots
$$
of vectors in a state space, which is mostly unobservable. The
readings $\veci x(t)$ are then some linear or non-linear functions
\begin{equation}
\veci x(t)\ =\ \veci F[\veci s(t)]\,+\, \boldsymbol{\eta}(t)
\label{cogSys_mapping_s_x}
\end{equation}
of the underlying state space, possibly in addition to some noise
$\boldsymbol{\eta}(t)$. Equation (\ref{cogSys_mapping_s_x}) is
denoted a state space model.

\runinhead{The Hidden Markov Process} \index{hidden Markov process}
There are many possible assumptions for the state space dynamics
underlying a given history of observables $\veci x(t)$. For a hidden
Markov process, to give an example, one assumes that
\begin{enumerate}\leftskip4pt
\item[(a)] $\veci s(t+1)$ depends only on $\veci s(t)$
(and not on any previous state space vector, the
{\sl Markov assumption}) and that
\index{Markov assumption}
\item[(b)] the mapping $\veci s(t)\to \veci s(t+1)$ is
stochastic.
\end{enumerate}
The process is dubbed \qut{hidden}, because the state
space dynamics is not directly observable.

\runinhead{The Elman State Space Model}
The dynamics of the Elman simple recurrent
network is given by
\begin{equation}
\veci s(t)\ =\ {\boldsymbol{\sigma}}\Big[{\bf A}\veci s(t-1)+{\bf
D}\veci x(t)\Big], \qquad\quad \sigma[y]={1\over 1+e^{-y}}~,
\label{cogSys_state_space_elman}\vspace{5pt}
\end{equation}
were $\veci x(t)$ and $\veci s(t)$ correspond to the activation
patterns of input and hidden units, respectively. The {\bf A} and
{\bf D} are the link matrices (context$\to$hidden) and
(input$\to$hidden), compare Fig.~\ref{cogSys_fig_simpleRecurrent},
and $\sigma(y)$ is called the {\em sigmoid function}.\index{sigmoid
function} The link matrix (hidden $\to$output) corresponds to the
prediction task $\veci s(t)\to\veci x(t+1)$ given to the Elman
network.

The Elman simple recurrent network extends the classical
state space model. For a normal state space model the readings
$\veci x(t)$ depend only on the current state $\veci s(t)$ of the
underlying dynamical system, compare Eq.~(\ref{cogSys_mapping_s_x}).
Extracting $\veci x(t)$ from Eq.~(\ref{cogSys_state_space_elman}),
one obtains\vspace{5pt}
\begin{equation}
\veci x(t)\ =\ \veci F[\veci s(t),\veci s(t-1)]~,
\label{cogSys1_SRB_state_space}\vspace{5pt}
\end{equation}
which is a straightforward generalization of
Eq.~(\ref{cogSys_mapping_s_x}). The simple recurrent net has a
memory since $\veci x(t)$ {in
Eq.~(\ref{cogSys1_SRB_state_space}) depends} both on $\veci s(t)$
and on $\veci s(t-1)$.

\runinhead{Neural Networks for Time Series Analysis}
The simple recurrent network can be generalized in several ways,
e.g.\ additional hidden layers result in a non-linear state space
dynamics. More complex layouts lead to more powerful prediction
capabilities, but there is a trade-off. Complex neural networks with
lots of hidden layers and recurrent connections need very big
training data. There is also the danger of overfitting the data,
when the model has more free parameters than the\break input.

\vspace{6pt}
\runinhead{Time Series Analysis for Cognitive Systems}
For most technical applications one is interested
exclusively in the time prediction capability of the
algorithm employed and an eventual spontaneous generation
of abstract concepts is not of interest. Pure time
series prediction is, however, of limited use for a
cognitive system. An algorithm allowing the prediction of
future events that at the same time generates
models of the environment is, however, extremely useful
for a cognitive system.

This is the case for state space models, as they generate explicit
proposals for the underlying environmental states describing the
input data. For the simple recurrent network these proposals are
generated in the hidden units. The activation state of the hidden
units can be used by the network for further cognitive information
processing via a simple feature extraction procedure, see
Fig.~\ref{cogSys_fig_simpleRecurrent}, e.g.\ by a Kohonen\break
layer.\footnote{A Kohonen network is an example of a neural
classifier via {one-winner-takes-all} architecture, see e.g.\ Ballard (2000).}
\index{Kohonen network}

\vspace{6pt}

\runinhead{Possible and Impossible Prediction Tasks} A cognitive
system is generally confronted with two distinct types of prediction
tasks.
\begin{itemize}
\item[--]Possible Prediction Tasks: Examples are the prediction of the limb dynamics
    as a function of muscle activation or the
    prediction of physical processes like the motion
    of a ball in a soccer game.
\item[--]Impossible Prediction Tasks: \index{prediction task!impossible}
    When a series of events is unpredictable it is,
    however, important to be able to predict the
    class of the next events. When we drive with a
    car behind another vehicle we automatically
    generate in our mind a set of likely {maneuvers that we} we expect
    the vehicle in front of us to perform next.
    When we listen to a person speaking we
    generate expectancies of what the person
    is likely to utter next.
\end{itemize}

\vspace{6pt}

\runinhead{Universal Prediction Tasks and Abstract Concepts}
\index{prediction task!universal}\index{abstract concept} Impossible
prediction tasks, like the lexical prediction task discussed in
Sect.~\ref{cogSys_subsect_SRN}, lead to the generation of abstract
concepts in the hidden layer, like the notion of \qut{noun} and
\qut{verb}. This is not a coincidence, but a necessary consequence
of the task given to the network. Only classes of future events can
be predicted in an impossible prediction task and not concrete
instances. We may then formulate the key result of this section in
the form of a lemma.

\begin{quotation}
{\it Universal Prediction Task Lemma.\enspace}
The task to predict future events leads to
universal environmental model building for neural networks with
state space layouts. When the prediction task is impossible to
carry out, the network will automatically generate abstract
concepts {that} can be used for further processing
by the cognitive system.
\end{quotation}

\runinhead{Conclusions}
Only a small number of genes, typically a few thousands, are
responsible for the growth and the functioning of mammalian
brains. This number is by far smaller than the information
content which would be required for an explicity encoding
of the myriad of cognitive capabilities of mammalian
brains. All these cognitive skills, apart from a few biologically
central tasks, must result from a limited number of universal
principles, the impossible time prediction task being one of
them.

\section*{Exercises}
\addcontentsline{toc}{section}{Exercises}
\begin{list}{}
\item \hspace*{-15pt}{\sc Transient State Dynamics} \\
Consider a system containing two variables, $x,\varphi\in[0,1]$.
Invent a system of coupled differential equations for which $x(t)$
has two transient states, $x\approx1$ and $x\approx0$. One
possibility is to consider $\varphi$ as a reservoir
and to let $x(t)$ autoexcite/autodeplete itself when the reservoir
is high/low.

The transient state dynamics should be rigorous. Write a code
implementing the differential equations.
\item \hspace*{-15pt}{\sc The Diffusive Control Unit} \\
Given are two signals $y_1(t)\in[0,\infty]$ and
$y_2(t)\in[0,\infty]$. Invent a system of differential 
equations for variables $x_1(t)\in[0,1]$ and 
$x_2(t)\in[0,1]$ driven by the $y_{1,2}(t)$ such 
that $x_1\to1$ and $x_2\to0$ when $y_1>y_2$ and
vice versa. Note that the $y_{1,2}$ are not 
necessarily normalized.
\item \hspace*{-15pt}{\sc Leaky Integrator Neurons} \\
\index{model!leaky integrator}
Consider a two-site network of neurons, having membrane
potentials $x_i$ and activities $y_i\in[-1,1]$, the so-called
\qut{leaky integrator} model for neurons,
$$
\dot x_1 \ =\ -\Gamma x_1 - w y_2,
\qquad\quad
\dot x_2 \ =\ -\Gamma x_2 + w y_1,
\qquad\quad
y_i \ =\  {2\over e^{-x_i} +1}-1~,
$$
with $\Gamma>0$ being the decay rate.
The coupling $w>0$ links neuron one (two)
excitatorily (inhibitorily) to neuron two (one).
Which are the fixpoints and for which parameters
can one observe weakly damped oscillations?
\item \hspace*{-15pt}{\sc Associative Overlaps and Thought Processes} \\
Consider the seven-site network of
Fig.~\ref{cogSys1_fig_memory_states}. Evaluate 
all pairwise associative overlaps of order zero 
and of order one between the five cliques, using 
Eqs.~(\ref{cogSys_ass_overlap_zero}) and (\ref{cogSys_ass_overlap_one}).
Generate an associative thought process of cliques $\alpha_1,\
\alpha_2,\ldots$, where a new clique $\alpha_{t+1}$ is selected
using the following simplified dynamics:
\begin{description}
\item[(1)] $\alpha_{t+1}$ has an associative overlap of order
      zero with $\alpha_{t}$ and is distinct from
      $\alpha_{t-1}$.
\item[(2)] If more than one clique
      satisfies criterium (1), then the clique with
      the highest associative overlap of order
      zero with $\alpha_{t}$ is selected.
\item[(3)] If more than one clique
      satisfies criteria (1)--(2), then one of
      them is drawn randomly.
\end{description}
Discuss the relation to the dHAN model treated in
Sect.\ref{cogSys_sec_ass_thinking}.
\end{list}


\def\refer#1#2#3#4#5#6{\item{\frenchspacing\sc#1}\hspace{4pt}
                       #2\hspace{8pt}#3 {\it \frenchspacing#4} {\bf#5}, #6.}
\def\bookref#1#2#3#4{\item{\frenchspacing\sc#1}\hspace{4pt}
                     #2\hspace{8pt}{\it#3}  #4.}

\addcontentsline{toc}{section}{Further Reading} 
\section*{Further Reading}
\markboth{\thechapter\enspace Elements of Cognitive {Systems}
Theory}{Further Reading}

For a general introduction to the field of artificial 
intelligence (AI), see Russell and Norvig (1995). For a
handbook on experimental and theoretical neuroscience, 
see Arbib (2002). For exemplary textbooks on neuroscience,
see Dayan and Abbott (2001) and for an introduction to 
neural networks, see Ballard (2000).

Somewhat more specialized books for further reading 
regarding the modeling of cognitive processes by
small neural networks is that by McLeod et~al. (1998) 
and on computational neuroscience
that by O'Reilly and Munakata (2000).

For some relevant review articles on dynamical
modeling in neuroscience the following are recommended:
Rabinovich et~al. (2006); on reinforcement 
learning Kaelbling et~al. (1996), and on
learning and memory storage in neural nets Carpenter (2001).

We also recommend to the interested reader to go 
back to some selected original literature dealing 
with \qut{simple recurrent networks} in the context of 
grammar acquisition (Elman, 1990; 2004), 
with neural networks for time series prediction
tasks (Dorffner, 1996), with \qut{learning by error} 
(Chialvo and Bak, 1999), with the assignment of the 
cognitive tasks discussed in Sect.~\ref{cogSys_tasks} 
to specific mammal brain areas (Doya, 1999),
with the effect on memory storage capacity 
of various Hebbian-type learning rules (Chechik et~al. 2001),
with the concept of \qut{associative thought processes} 
(Gros, 2007; 2009a) and with \qut{diffusive emotional
control} (Gros, 2009b).

It is very illuminating to take a look at the 
freely available databases storing human associative 
knowledge (Nelson et~al. 1998; Liu and Singh, 2004).

{\baselineskip=15pt
\begin{list}{}{\leftmargin=2em \itemindent=-\leftmargin%
\itemsep=3pt \parsep=0pt \small}
\refer{Abeles M. et al.}{1995} {Cortical activity flips among
quasi-stationary states.} {Proceedings of the National Academy of
Science, USA}{92}{8616--8620}
\bookref{Arbib, M.A.}{2002} {The Handbook of Brain Theory and Neural
Networks.}{MIT Press, Cambridge, MA}
\refer{Baars, B.J., Franklin, S.}{2003} {How conscious experience
and working memory interact.} {Trends in Cognitive
Science}{7}{166--172}
\bookref{Ballard, D.H.}{2000}{An Introduction to Natural Computation.}
        {MIT Press, Cambridge, MA}
\refer{Carpenter, G.A.}{2001} {Neural-network models of learning and
memory: Leading questions
 and an emerging framework.}
{Trends in Cognitive Science}{5}{114--118}
\refer{Chechik, G., Meilijson, I., Ruppin, E.}{2001} {Effective
neuronal learning with ineffective Hebbian learning rules.} {Neural
Computation}{13}{817}
\refer{Chialvo, D.R., Bak, P.}{1999} {Learning from mistakes.}
{Neuroscience}{90}{1137--1148}
\refer{Crick, F.C., Koch, C.}{2003} {A framework for
consciousness.}{Nature Neuroscience}{6}{119--126}
\bookref{Dayan, P., Abbott, L.F.}{2001}{
         Theoretical Neuroscience: Computational and Mathematical
         Modeling of Neural Systems.}{MIT Press, Cambridge, MA}
\refer{Dehaene, S., Naccache, L.}{2003} {Towards a cognitive
neuroscience of consciousness:
 Basic evidence and a workspace framework.}{Cognition}{79}{1--37}
\refer{Dorffner, G.}{1996} {Neural networks for time series
processing.} {Neural Network World}{6}{447--468}
\refer{Doya, K.}{1999}
{What are the computations of the cerebellum,
 the basal ganglia and the cerebral cortex?}{Neural Networks}
 {12}{961--974}
\bookref{Edelman, G.M., Tononi, G.A.}{2000} {A Universe of
Consciousness.}{Basic Books, New York}
\refer{Elman, J.L.}{1990} {Finding structure in time.}{Cognitive
Science}{14}{179--211}
\refer{Elman, J.L.}{2004}
{An alternative view of the mental lexicon.}
{Trends in Cognitive Sciences}{8}{301--306}
\refer{Gros, C.}{2007}{Neural networks with transient state dynamics.}
{New Journal of Physics}{9}{109}
\refer{Gros, C.}{2009a}{Cognitive computation with autonomously active
neural networks: an emerging field.}
{Cognitive Computation}{1}{77}
\bookref{Gros, C.}{2009b} {{\rm Emotions, diffusive emotional control 
and the motivational problem for autonomouscognitive systems}, in 
Handbook of Research on Synthetic Emotionsand Sociable Robotics: 
New Applications in Affective Computing and Artificial Intelligence,
{\rm J. Vallverdu, D. Casacuberta (Eds)}.}{IGI-Global}
\refer{Kaelbling, L.P., Littman, M.L., Moore, A.}{1996}
{Reinforcement learning: A survey.}{Journal of Artificial
Intelligence Research}{4}{237--285}
\refer{Kenet, T., Bibitchkov, D., Tsodyks, M., Grinvald, A.,
Arieli, A.}{2003}{Spontaneously emerging cortical representations of
visual attributes.}{Nature}{425}{954--956}
\refer{Liu, H., Singh, P.}{2004}{ConcepNet {}
 a practical commonsense reasoning tool-kit.}
 {BT Technology Journal}{22}{211--226}
\bookref{McLeod, P., Plunkett, K., Rolls, E.T.}{1998}
         {Introduction to {Connectionist} Modelling.}
         {Oxford University Press New York}
\bookref{Nelson, D.L., McEvoy, C.L., Schreiber, T.A.}{1998} {The
University of South Florida Word Association, Rhyme,
  and Word Fragment Norms.}{Homepage:
  \url{http://www.usf.edu/FreeAssociation}}
\bookref{O'Reilly, R.C., Munakata, Y.}{2000}
  {Computational Explorations in Cognitive Neuroscience:
   Understanding the Mind by Simulating the Brain.} {MIT Press Cambridge}
\refer{Rabinovich, M.I., Varona, P., Selverston, A.I.
       and Abarbanel, H.D.I.}{2006}
{Dynamical principles in neuroscience.}
{Review of Modern Physics}{78}{1213--1256}
\bookref{Russell, S.J., P Norvig, P.}{1995} {Artificial
Intelligence: A Modern Approach.}{Prentice-Hall, Englewood Cliffs,
NJ}
\end{list}
\par}



\backmatter



\printindex

\end{document}